\documentclass[a4paper,10pt]{report}
\usepackage[inner=3.cm,outer=3.cm,bottom=2cm]{geometry}
\usepackage[utf8]{inputenc}
\usepackage{graphicx}
\usepackage{tikz}
\usepackage{amsmath}
\usepackage{amssymb}
\usepackage{xspace}

\usepackage{bm}
\usepackage{bbm}
\usepackage{amsfonts}

\usepackage[colorlinks=true, linkcolor=blue, bookmarks=true]{hyperref}

\newcommand{\R}{\ensuremath{\mathbbm{R}}}
\newcommand{\C}{\ensuremath{\mathbbm{C}}}

\newcommand{\Rs}{\ensuremath{\mathcal{R}}}
\renewcommand{\L}{\ensuremath{\mathcal{L}}}
\newcommand{\W}{\ensuremath{\mathcal{W}}}

\newcommand{\bu}{\bm{u}}
\newcommand{\bh}{\bm{h}}
\newcommand{\bx}{\bm{x}}
\newcommand{\GN}{G}

\renewcommand{\bu}{u}
\renewcommand{\bh}{h}
\renewcommand{\bx}{x}

\newcommand{\ads}{AdS$_3$\xspace}
\newcommand{\pc}[1]{\tilde{#1}}
\newcommand{\LS}{$\mathcal{L}$\xspace}
\newcommand{\cZ}{\ensuremath{Z}}
\newcommand{\lb}{\linebreak}
\newcommand{\tr}{\mathrm{tr}}

\newcommand{\hf}{\ensuremath{\hat{f}}}
\newcommand{\hh}{\ensuremath{\hat{h}}}
\newcommand{\hB}{\ensuremath{\hat{B}}}
\newcommand{\hb}{\ensuremath{\hat{b}}}
\newcommand{\hz}{\ensuremath{\hat{z}}}

\newcommand{\slr}{\ensuremath{SL(2,\R)}\xspace}
\newcommand{\sln}[1]{\ensuremath{SL(#1,\R)}}

\newcommand{\rd}{\ensuremath{\mathrm{d}}}

\usetikzlibrary{shapes,snakes}
\usetikzlibrary{decorations.pathmorphing}
\usetikzlibrary{decorations.markings}
\graphicspath{ {Figures/} }

\begin{document}
\begin{titlepage}

\begin{tikzpicture}[remember picture,overlay]
   \node[yshift=2cm,xshift=-4cm] at (current page.south east)
              {\includegraphics[scale=0.4]{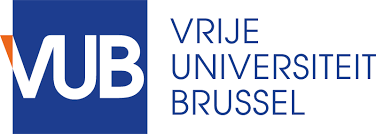}};
  \node[yshift=2cm,xshift=3cm] at (current page.south west)
              {\includegraphics[scale=0.4]{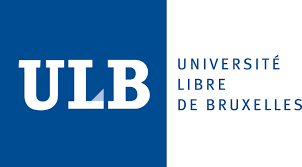}};
\end{tikzpicture}
\begin{center}
{\Huge Black hole formation, holographic thermalization and the AdS/CFT correspondence}
\vspace{4pt}
{\Huge \par}
\begin{center}{\LARGE Erik Jonathan Lindgren}\end{center}{\Huge \par}
\begin{center}
\includegraphics[scale=0.8]{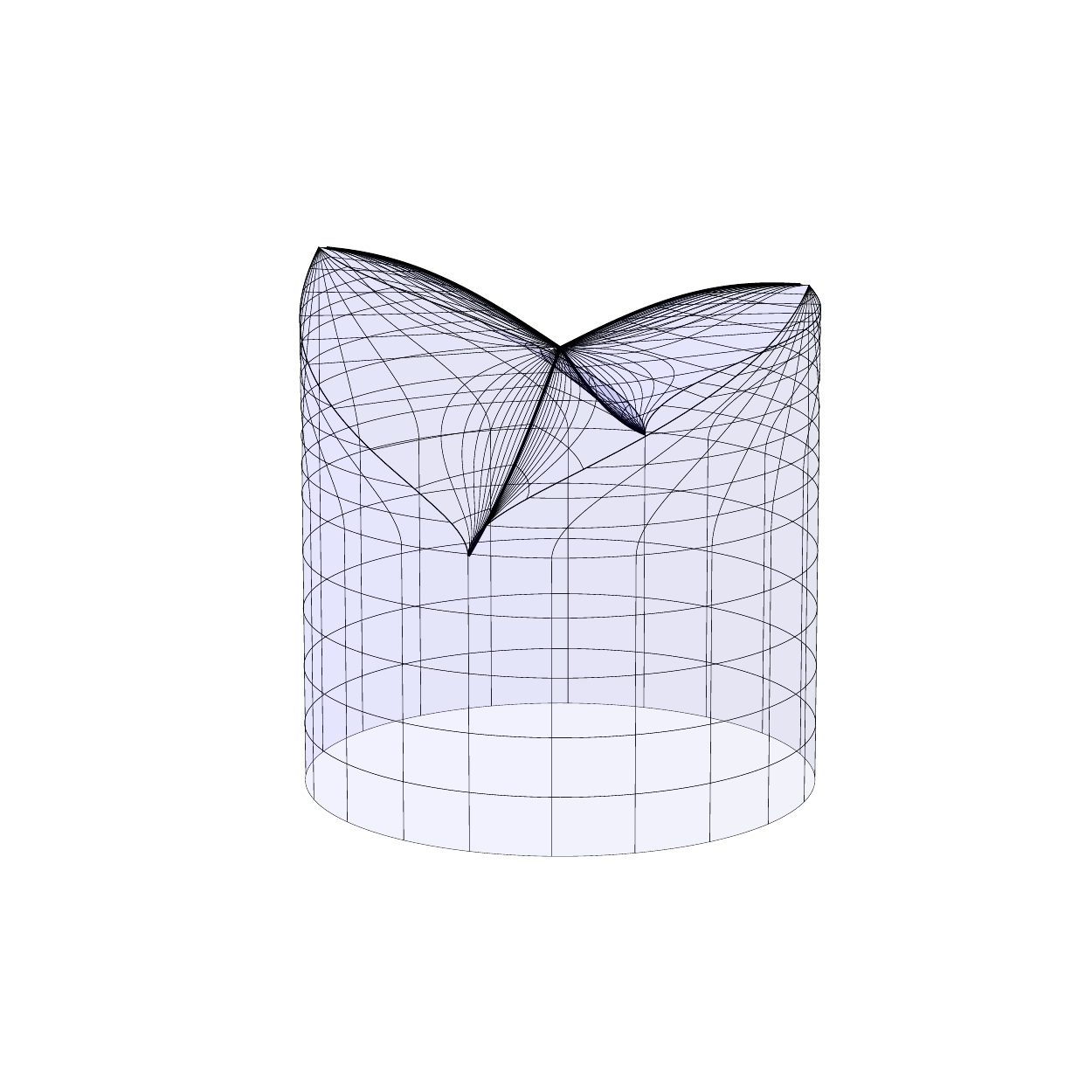} 
\end{center}

\begin{center}{\Large Promotors at Vrije Universiteit Brussel: Prof. Ben Craps and Prof. Alexander Sevrin}\end{center}
\vspace{5pt}
\begin{center}{\Large Promotor at Universit\'e Libre de Bruxelles: Prof. Marc Henneaux}\end{center}
\vspace{10pt}
\begin{center}{ Thesis submitted in fulfillment of the requirements for the degree of}\end{center}
\begin{center}{\it \large Doctor in de Wetenschappen}\end{center}
\begin{center}{at Vrije Universiteit Brussel and}\end{center}
\begin{center}{\it \large Docteur en Sciences}\end{center}
\begin{center}{at Universit\'e Libre de Bruxelles}\end{center}
\end{center}
\end{titlepage}

\newpage
\thispagestyle{empty} 
\null
\newpage
\noindent This thesis was defended behind closed doors at Universit\'e Libre de Bruxelles on June the 9th, 2017, in front of the following jury:
\begin{itemize}
 \item Ben Craps, Vrije Universiteit Brussel
 \item Alexander Sevrin, Vrije Universiteit Brussel
 \item Marc Henneaux, Universit\'e Libre de Bruxelles
 \item St\'ephane Detournay, Universit\'e Libre de Bruxelles
 \item Sophie de Buyl, Vrije Universiteit Brussel
 \item Thomas Van Riet, Katholieke Universiteit Leuven
 \item Jan de Boer, University of Amsterdam
 
\end{itemize}

It was defended publicly at Vrije Universiteit Brussel on July the 5th, 2017.
\newpage

\tableofcontents
\pagebreak
\pagenumbering{roman}
\chapter*{Abstract}

\addcontentsline{toc}{chapter}{Abstract}

The AdS/CFT correspondence is one of the most important discoveries in theoretical physics in recent years. It states that certain quantum mechanical theories can actually be described by classical gravity in one higher dimension, in a spacetime called anti-de Sitter (AdS) space. This means that to compute any measurable quantity in the quantum theory, we can instead do a computation in classical general relativity, and vice versa. What makes this duality so useful is that it relates theories with weak coupling to theories with strong coupling and thus provides a new tool for tackling strongly coupled quantum field theories, which are notoriously difficult to handle using conventional methods. Originally discovered in the context of string theory, this duality has now found a wide range of applications, from condensed matter physics to high temperature plasmas in quantum chromodynamics (QCD).\\
\indent During the course of my PhD I have mostly studied time dependent processes, in particular thermalization processes, in quantum field theories using the AdS/CFT correspondence. On the gravity side, this is dual to dynamical formation of black holes from the collapse of matter fields. By studying the gravitational collapse process in detail, we can then draw conclusions about the dynamical formation of a thermal state in the dual quantum field theory. Certain quantum field theories (such as QCD) enjoy a property called confinement, which in the case of QCD states that quarks can not be isolated. Using mostly numerical methods, I have studied how confinement affects thermalization in quantum field theories. We found that sometimes the system never thermalizes and field theory observables undergo interesting quasiperiodic behaviour. In another line of research, I have studied formation of black holes in three dimensions which due to the simplified nature of three-dimensional gravity can be done using analytical methods. This has led to the discovery of new solutions of three-dimensional gravity corresponding to the formation of black holes without spherical symmetry, which can provide a deeper understanding of thermalization in two-dimensional quantum field theories. In a third line of research, I have studied higher spin gravity in three dimensions, an exotic extension of three-dimensional gravity which includes fields with spin higher than two, and we outline a new method to construct black hole solutions carrying higher spin charge.\\
\pagebreak
\chapter*{Samenvatting}

\addcontentsline{toc}{chapter}{Samenvatting}

De AdS/CFT dualiteit is een van de belangrijkste ontdekkingen in de theoretische natuurkunde van de laatste jaren. Ze zegt dat sommige kwantummechanische theorie\"en kunnen worden beschreven  door klassieke zwaarte\-kracht in een hogere dimensie, in een ruimtetijd genaamd anti-de Sitter (AdS) ruimte. Dit betekent dat om een meetbare grootheid in de kwantumtheorie te berekenen, we ook een berekening in de klassieke algemene relativiteitstheorie kunnen doen. Wat deze dualiteit zo praktisch maakt is dat ze sterk gekoppelde theorie\"en en zwak gekoppelde theorie\"en relateert en daarom een nieuw hulpmiddel biedt voor het behandelen van sterk gekoppelde kwantumveldentheorie\"en, die erg moeilijk met traditionele methoden te behandelen zijn. Hoewel ze oorspronkelijk ontdekt werd in de context van snaartheorie, heeft deze dualiteit nu vele toepassingen gevonden, van de fysica van gecondenseerde materie naar hoge-temperatuursplasma in kwantumchromodynamica (QCD).\\
\indent Tijdens mijn doctoraatsonderzoek heb ik tijdsafhankelijke processen, voornamelijk thermalisatieprocessen, in kwantumveldentheorie\"en bestudeerd door de AdS/CFT dualiteit te gebruiken. In de zwaartekrachtbeschrijving komt dit overeen met de vorming van zwarte gaten door het instorten van materievelden. Door de gravitationele ineenstorting in detail te bestuderen, kunnen we conclusies trekken over de vorming van een thermische toestand in de kwantumveldentheorie. Sommige kwantumveldentheorie\"en (zoals QCD) hebben een eigenschap genaamd confinement, die voor QCD betekent dat quarks niet ge\"isoleerd kunnen worden. Door vooral numerieke methoden, heb ik bestudeerd hoe confinement een invloed heeft op thermalisatieprocessen in kwantumveldentheorie\"en. We hebben ontdekt dat in sommige gevallen het systeem nooit thermaliseert en de observabelen in de kwantumvelden\-theorie quasiperiodiek gedrag ervaren. In een andere onderzoekslijn heb ik de vorming van zwarte gaten in drie dimensies bestudeerd, wat door de vereenvoudigde aard van driedimensionale zwaartekracht mogelijk is met analyti\-sche methoden. Dit heeft geleid tot de ontdekking van nieuwe oplossingen in driedimensionale zwaartekracht, corresponderend met de vorming van zwarte gaten zonder sferische symmetrie, wat tot een dieper begrip kan leiden van thermalisatieprocessen in tweedimensionale kwantumveldentheorie\"en. In een derde onderzoekslijn, heb ik hogere-spin-zwaartekracht bestudeerd, een exo\-tische uitbreiding van driedimensionale zwaartekracht die velden met spin hoger dan twee bevatten, en heb ik nieuwe methoden gevonden om zwarte gaten met hogere-spin-lading te construeren.
\pagebreak
\chapter*{R\'esum\'e}

\addcontentsline{toc}{chapter}{Résumé}

\thispagestyle{empty} 
La correspondence AdS/CFT est parmi les d\'ecouvertes les plus importantes dans la physique th\' eorique des ann\' ees r\'ecentes. Elle declare que quelques th\' eories m\'ecanique quantique peuvent en fait \^etre d\'ecrites par la gravit\'e classique dans une plus haute dimension, dans un espace-temps qui s'appele l'espace anti-de Sitter (AdS). Cela signifie que, pour calculer une quantit\'e mesurable dans la theori\'e quantique, on peut plut\^ot utiliser la th\'eorie de la relativit\'e g\'en\'erale, et vice versa. Ce qui rend cette dualit\'e si utile est qu'elle relie des th\'eories ayant une interaction forte avec les th\'eories ayant une interaction faible. Par cons\'equent, elle fournit un nouveau moyen pour traiter les th\'eories quantiques des champs fortement coupl\'ees, qui sont difficiles \`a resoudre avec les m\'ethodes traditionelles. Bien que cette correspondence ait \'et\'e d\'ecouverte dans le cadre de la th\'eorie des cordes, elle a trouv\'e plusieurs applications dans differents domaines en physique, par example en physique de la mati\`ere condens\'ee et dans des plasmas \`a haute temp\'erature de la chromodynamique quantique (QCD).\\
\indent Au cours de mon doctorat, j'ai \'etudi\'e des processus d\'ependants du temps, en particulier des processus thermalisation, dans des th\'eories quantiques des champs en utilisant la correspondence AdS/CFT. Du c\^ot\'e gravitationnel, ils sont d\'ecrits par la formation d'un trou noir issu du collapsus gravitationnel des champs de mati\`ere. En \'etudiant cette solution en d\'etail on peut en tirer des conclusions sur le processus de thermalisation dans la th\'eorie quantique. En particulierement, j'ai utilis\'e cette m\'ethode pour \'etudier des th\'eories qui poss\`edent une propri\'et\'e appel\'ee confinement. Cette propri\'et\'e dit essentiellement que les quarks (ou d'autre particules elementaires) ne peuvent pas \^etre isol\'es, et QCD est l'example le plus connu. Nous avons trouv\'e que parfois le syst\`eme ne thermalise pas et les observables peuvent subir des oscillations dans une \'echelle de temps longue. Dans une autre ligne de recherche, j'ai \'etudi\'e la formation des trous noirs en trois dimensions par collision de particules ponctuelles. En raison de la nature simplifi\'ee de la gravit\'e tridimensionnelle, on peut l'\'etudier avec des m\'ethodes analytiques. Cette recherche a donn\'e lieu \`a la d\'ecouverte de nouvelles solutions de la gravit\'e tridimensionelle qui correspondent \`a la formation de trous noirs sans sym\'etrie sph\'erique. Ces solutions peuvent donner plus de compr\'ehension \`a thermalisation dans des th\'eories bidimensionnelles des champs conformes. Dans une troisieme ligne de recherche, j'ai \'etudi\'e la gravit\'e de spin \'elev\'e en trois dimensions, une extension exotique de la gravit\'e qui inclut des champs avec un spin sup\'erieur \`a deux, et nous avons trouv\'e une nouvelle m\'ethode pour construire les solutions des trous noirs qui poss\`edent une charge de spin \'elev\'e.

\chapter*{Acknowledgements}
\addcontentsline{toc}{chapter}{Acknowledgements}

First of all I want to thank Ben Craps, for his excellent guidance and support during the course of my PhD, and also for giving me a lot of freedom when carrying out my research in theoretical physics. I am also grateful for his detailed feedback on earlier drafts of this thesis. Without his guidance this thesis would certainly not have been possible. I also want to thank my other two promotors, Marc Henneaux and Alexander Sevrin, for their support and encouragement and for allowing me to carry out my PhD in Brussels. I also want to express my gratitude to all the other members of my PhD jury, namely Jan de Boer, Thomas Van Riet, St\'ephane Detournay and Sophie de Buyl for taking their time to read my thesis and asking interesting and challenging questions during my PhD defense. I also want to thank Ben, Hongbao, Anastasios, Joris, Ioannis, Tim, Oleg, Andrea and Xavier for our interesting collaborations, as well as my other colleagues and friends in physics both in Belgium and abroad. In particular I want to thank Amaury Leonard for his unwavering friendship during this period in Brussels. I also thank my family for always being there for me and supporting my choice to pursue a PhD in Belgium. Lastly, I want to thank Mengshu Tang for her unconditional love and support which have been crucial for my well-being during the more difficult times during this PhD.\\
\newline
I also thank the Fonds Wetenschappelijk Onderzoek (FWO) and the European Research Council (ERC) for funding, as well as the HPC Computing Center where some numerical calculations in this thesis were carried out.

\newpage

\chapter*{List of publications}
\addcontentsline{toc}{chapter}{List of publications}
During the course of my PhD, I was involved in the following publications:
\begin{enumerate}
\item Collisions of massive particles, timelike thin shells and formation of black holes in three dimensions\\
J. Lindgren\\
{\it JHEP 1612 (2016) 048}
\item Black hole formation from pointlike particles in three-dimensional anti-de Sitter space\\
E. J. Lindgren\\
{\it Class.Quant.Grav. 33 (2016) no.14, 145009}
\item Holographic thermalization in a top-down confining model\\
B. Craps, E. J. Lindgren, A. Taliotis\\
{\it JHEP 1512 (2015) 116}
\item Holographic hall conductivities from dyonic backgrounds\\
J. Lindgren, I. Papadimitriou, A. Taliotis, J. Vanhoof\\
{\it JHEP 1507 (2015) 094}
\item Holographic gravitational infall in the hard wall model \\
B. Craps, E. J. Lindgren, A. Taliotis, J. Vanhoof and H. Zhang\\
{\it Phys. Rev. D 90, 086004 (2014)}
\end{enumerate}
This thesis is based on publications 1, 2, 3, 5 and ongoing unpublished work.
\newpage

\chapter*{Introduction}\label{intro}
\addcontentsline{toc}{chapter}{Introduction}
\pagestyle{plain}

One of the most important unsolved problems in theoretical physics is that of understanding how the theory of gravity fits in with the formalism of quantum mechanics, a field known as quantum gravity. They are both highly successful theories in their regimes of validity. Quantum mechanics, and quantum field theory, has been remarkably successful in describing phenomena in condensed matter physics, atomic physics, nuclear physics and high energy collisions like those in the Large Hadron Collider (LHC). In particular, quantum field theory can explain three of the four fundamental forces: the strong, the weak and the electromagnetic force, and describes all measured elementary particles in a framework known as the standard model. The common denominator is that these phenomena all take place at very small length scales. General relativity deals with the fourth fundamental force, gravity. It is also a remarkably successful theory, explaining deviations from its predecessor Newtonian gravity to planetary motion and gravitational lensing, providing accurate descriptions of large scale features of the universe and predicting the existence of gravitational waves. The phenomena typically described by general relativity happen on very large length scales.\\
\lb
However, at extremely high energies, or equivalently at extremely small length scales (approximately $\sim 10^{-35}$ meters, a length scale known as the Planck length), the quantum nature of gravity is expected to become important. One of the most successful candidates for describing quantum gravity is {\it string theory}, a theory which postulates that all elementary particles are composed of small string-like objects. String theory has been very successful in putting all fundamental forces on an equal footing and both general relativity (the theory of gravity) and gauge theories (which are the basic building blocks of the standard model) follow naturally from the spectrum of these quantum strings. However, there are still many outstanding problems, such as the lack of a fundamental definition of the theory (essentially only the perturbation theory is known), the prediction of the existence of supersymmetry (a symmetry which has not been verified experimentally) and the fact that we are not able to carry out experiments at the Planck scale and thus it is very hard to verify or falsify the theory. The sheer number of different theories that can be constructed from string theory (estimates range between $10^{10}$ and $10^{500}$ different four-dimensional theories obtained from string theory) is also troubling since it reduces the predictive power of the theory and it seems impossible to derive the parameters in the Standard Model without making additional assumptions.\\
\lb
However, in the late 1990s, a new surprising application of string theory was discovered \cite{Maldacena:1997re}. By considering different limits of string theory close to certain extended objects known as branes, a remarkable duality between a string theory and a certain four dimensional supersymmetric Yang-Mills theory could be derived. This duality, known as the AdS/CFT correspondence, has been the focus of an intense research program over the last two decades. The name comes from the fact that the string theory lives in a space known as anti-de Sitter (AdS) space, which is a solution to Einstein's equations with a negative cosmological constant, and the dual Yang-Mills theory is a conformal field theory (CFT), meaning it enjoys (super) conformal invariance. One of the most important features of this duality is the fact that it relates strong coupling on one side to weak coupling on the other side which makes it useful since it may relate hard problems on one side to easier problems on the other. The duality has been mostly used in the limit where the string theory has weak coupling and high string tension, for which it reduces to classical (super) gravity. This is then dual to a Yang-Mills theory in the strong coupling regime, which is famous for being a very difficult regime to deal with using traditional methods in quantum field theory. After the initial example by Maldacena, many other similar dualities have been conjectured. \\
\lb
One of the most difficult questions to study in quantum field theory is that of dynamics and time-dependent processes. In particular, at finite temperature it is not even clear how to formulate time-dependent setups. However, these problems become conceptually very easy in view of the AdS/CFT correspondence, where dynamics just amounts to solving the time-dependent Einstein's equations and dynamics at finite temperature corresponds to dynamical solutions including black holes. In practice however, this is easier said than done since Einstein's equations (being non-linear partial differential equations) can be quite cumbersome to solve and one must often resort to sophisticated numerical tools to handle any problems that do not enjoy a large degree of symmetry. The main focus of this thesis will be to study dynamical processes (in particular the formation of black holes) in AdS, with applications to dynamics in quantum field theories. In particular, we will study dynamics in confining field theories by using the AdS/CFT correspondence. Although we will not make comparisons with experiments, this can have potential applications to quantum chromodynamics (QCD), the theory for the strong nuclear force, since it is a confining theory at low temperatures. We will also study black holes and black hole formation in three-dimensional gravity theories with potential applications to dynamics in two-dimensional conformal field theories. The discussions will mostly focus on constructing solutions corresponding to black hole formation in AdS (and other asymptotically AdS spacetimes), but we will also relate our gravity computations to field theory quantities by using the AdS/CFT correspondence.\\
\lb
The outline of the thesis is as follows. In Chapter \ref{adscft} we review the AdS/CFT correspondence and previous well known results, focusing on topics relevant for this thesis such as confining models and time dependence. This chapter barely contains any new results. In Chapter \ref{confinement} we study dynamics in confining field theories using holographic methods. This chapter is based on \cite{Craps:2014eba} and \cite{Craps:2015upq}. In Chapter \ref{threed} we study black hole formation in three-dimensional gravity, focusing on the interesting process where a black hole forms from collisions of pointlike particles. This chapter is based on \cite{Lindgren:2015fum} and \cite{Lindgren:2016wtw}. In Chapter \ref{higher} we study black holes (and other solutions) in an extension of gravity known as higher spin gravity, which in our case will be a theory of gravity coupled to a spin-3 field. This chapter is based on work in progress together with Andrea Campoleoni and Xavier Bekaert. We end with some general conclusions in Chapter \ref{conclusions}. The different chapters can be read more or less independently and references between different chapters will be minor and not crucial for understanding the content.

\pagenumbering{arabic}
\chapter{Aspects of the AdS/CFT correspondence}\label{adscft}

The AdS/CFT correspondence is one of the most successful applications of string theory, and arguably one of the most important discoveries in theoretical physics in recent years. It states the equivalence of two very different physical theories, a gravitational theory and a quantum field theory, and provides a framework for using one of them to make predictions about the other. Moreover, the duality between the two theories is strong-weak, meaning that it relates the strong coupling sector of one theory to the weak coupling sector of the other, and vice versa. This is extremely useful, since this means in general that difficult calculations in one theory are mapped to easy calculations in the other theory. In particular, it allows us to make statements about strongly coupled quantum field theories by doing calculations in classical gravity.\\
\linebreak
To be more precise, the AdS/CFT correspondence relates (quantum) gravity in anti-de Sitter (AdS) space in $d+1$ dimensions, to a certain (conformal) field theories in $d$ dimensions on the boundary of AdS. The first  example of such a duality was discovered by Juan Maldacena \cite{Maldacena:1997re}, and related supergravity in $AdS_5\times S^5$ to a supersymmetric Yang-Mills theory in 3+1 dimensions. Although the most well established examples of the correspondence include high degrees of super and conformal symmetry, dualities between theories that break these symmetries have also been constructed. Moreover, the duality is also often used to construct phenomenological models to describe condensed matter systems or hydrodynamical processes, even though the theoretical foundations of the applicability of the duality in these cases is less reliable. \\
\linebreak
We will in this section review the basics of the AdS/CFT correspondence, focusing on aspects relevant for the other chapters in this thesis. We will first quickly review anti-de Sitter space and conformal field theory. Then we will sketch the derivation of the first example of the duality. After that we will explain how to relate the different quantities on the two sides of the duality. We will then go to more specialized topics of relevance to this thesis, such as confining holographic models as well as time dependence and black hole formation.
\section{Anti-de Sitter space and black hole solutions}\label{sec_adsbh}
Anti-de Sitter (AdS) space is the maximally symmetric solution to Einstein's equations with a cosmological constant. It can be defined, in $d+1$ dimensions, by an embedding in a $d+2$ dimenssional space with two time coordinates, via the equation
\begin{equation}
-X_1^2-X_2^2+\sum_{i=3}^{d+2}X_i^2=-L^2,
\end{equation}
where $L$ is called the AdS radius. The resulting spacetime is a solution to Einstein's equations with cosmological constant given by
\begin{equation}
\Lambda=-\frac{d(d-1)}{2L^2}.
\end{equation}
There are two important coordinate systems used to parametrize AdS, called respectively {\it global coordinates} and {\it Poincar\'e coordinates}. The global coordinate system, as the name suggests, covers all of AdS, while Poincar\'e coordinates only cover a submanifold called the {\it Poincar\'e patch}. We will review both coordinate systems below.\\
\linebreak
Another important family of solutions are spacetimes that are {\it asymptotically} AdS. This essentially means that they have a boundary that looks like the boundary of AdS, although they are not necessarily equivalent to AdS everywhere. Asymptotically AdS spacetimes are really the type of spacetimes one considers in the AdS/CFT correspondence, not just pure AdS geometries. With further abuse of notation, we will frequently refer to the boundary of an asymptotically AdS boundary just as the {\it AdS boundary}, even though the spacetime is not equivalent to AdS globally. A solution can be asymptotic to either global AdS or to Poincar\'e AdS, and it will be obvious from context which one is meant.\\
\linebreak
A notable example of asymptotically AdS spacetimes are AdS black branes and AdS black holes, which we will also describe in this section. These are asymptotically AdS solutions which contain a region which is causally disconnected from the boundary, and the boundary of this region is referred to as the event horizon. We may have black hole solutions that are either asymptotically Poincar\'e AdS or asymptotically global AdS. The former are often referred to as black branes since they are translationally invariant solutions (but note that sometimes they are also referred to as just black holes), while the latter have event horizons enjoying spherical symmetry. For simplicity we will here only look at static black holes without angular momentum. Note also that there exists many other black hole solutions, for example black holes charged under some other gauge field, which would then solve Einstein's equations coupled to matter and not the vacuum Einstein equations.\\
\linebreak
Black holes can be characterized in terms of the mass, or equivalently in terms of the {\it temperature}. The concept of a temperature comes originally from the discovery by Stephen Hawking \cite{Hawking:1974sw} that black holes radiate with a spectrum equivalent to that of a black body (thus the temperature is often referred to as the {\it Hawking} temperature). This computation is relatively involved and requires the quantization of a field in curved spacetime. There is however a much easier way to obtain the temperature of black holes. One simply considers the black hole in Euclidean signature (replacing the time coordinate $t$ via the relation $\tau=it$), and enforces regularity at the horizon. This will force the time coordinate $\tau$ to be periodic as $\tau\sim\tau+\beta$ (or otherwise there will be a conical singularity at the horizon), and one simply identifies the temperature as $T=1/\beta$. This identification is even more natural from the interpretation of the AdS/CFT correspondence, where it corresponds to looking at the boundary field theory in Euclidean signature, for which it is natural that periodicity in Euclidean time should correspond to a thermal state.\\
\linebreak
{\bf Poincar\'e coordinates}\\
The Poincar\'e coordinates are obtained by parametrizing AdS by
\begin{align}
X_1&=\frac{L^2}{2r}(1+\frac{r^2}{L^4}(L^2+\vec{x}^2-t^2)),\nonumber\\
X_2&=\frac{r}{L}t,\nonumber\\
X_i&=\frac{r}{L}x_i,\quad i\in \{3,\ldots,d+1\},\nonumber\\
X_{d+2}&=\frac{L^2}{2r}(1-\frac{r^2}{L^4}(L^2-\vec{x}^2+t^2)).
\end{align}
The metric takes the form
\begin{equation}
\rd s^2=-\frac{r^2}{L^2}\rd t^2+\frac{L^2}{r^2}\rd r^2+\frac{r^2}{L^2}\rd \vec{x}^2,
\end{equation}
with the AdS boundary located at $r\rightarrow\infty$. The boundary takes the form $\R ^d$. Another convenient choice of coordinates can be obtained by defining $z=L^2/r$, for which the metric becomes
\begin{equation}
\rd s^2=\frac{L^2}{z^2}\left(-\rd t^2+\rd z^2+\rd\vec{x}^2\right),\label{poincaremetric}
\end{equation}
for which the boundary is at $z\rightarrow0$. The AdS radius $L$ in these coordinates plays no significant role and can often be set to one. What we mean by this statement is that if we solve Einstein's equations (coupled to matter) with Poincar\'e AdS as boundary condition for the metric, the AdS radius can be scaled out by rescaling the metric (and rescaling eventual coupling constants for the matter fields). Thus it is enough to solve the equations for $L=1$ and then all solutions with different $L$ can be obtained from this result. This statement is not true for solutions with global AdS as boundary condition.\\
\linebreak
The black brane solutions are black holes that extend in $d-1$ dimensions and are asymptotically AdS. They also solve the vacuum Einstein equations with negative cosmological constant and can described by the metric
\begin{equation}
\rd s^2=-\frac{r^2}{L^2}\rd t^2(1-\frac{M}{r^d})+\frac{L^2}{r^2}\rd r^2(1-\frac{M}{r^d})^{-1}+\frac{r^2}{L^2}\rd\vec{x}^2,\label{poincareblackbrane}
\end{equation}
for a constant $M$. These solutions have an event horizon at $r=M^{1/d}$, which is the reason why they are called {\it black} branes. Note that we can now rescale our coordinates such that the black hole horizon is equal to one, and thus in Poincar\'e coordinates there does not exist any classification of black branes as {\it small} or {\it large} (note that there is even no meaning in comparing the event horizon to the AdS radius). An implication of this fact is that arbitrarily small spherically symmetric time dependent perturbation of AdS will always result in the formation of a black brane. The temperature can be computed by considering the metric around the horizon in Euclidean signature. By defining $\rho^2=(r-r_h)dr_h/L^2$ the metric around the horizon takes the form
\begin{equation}
\rd s^2\approx \frac{4L^4}{d^2r_h^2}\rd\rho^2+\rho^2\rd\tau^2+\ldots.
\end{equation}
This is essentially the same as flat space in polar coordinates, has a conical singularity unless $\tau$ is periodic with a specific periodicity. The periodicity $\beta$ then gives the temperature as
\begin{equation}
T=\frac{dr_h}{4\pi L^2}.\label{temppoincare}
\end{equation}
\linebreak
Many interesting applications of the AdS/CFT correspondence are to translationally invariant systems and thus asymptotically Poincar\'e AdS spacetimes are most often used for applications (such as to condensed matter systems or the quark gluon plasma). Will use these coordinates when studying dynamics in confining holographic models in Chapter \ref{confinement}. On the other hand, if one is interested in applications to field theories defined on a sphere, one should consider spacetimes that asymptotes to global AdS as described next.\\
\linebreak
{\bf Global coordinates}\\
The global coordinates are obtained by for instance the parametrization
\begin{align}
X_1=L\cosh\chi\cos t,\nonumber\\
X_2=L\cosh\chi\sin t,\nonumber\\
X_i=Lx_i\sinh\chi,\quad\sum_{i=3}^{d+2}x_i^2=1.
\end{align}
The $x_i$ parametrize a sphere of dimension $d-1$ with line element $\rd\Omega_{d-1}^2$. The metric then takes the form
\begin{equation}
\rd s^2=L^2(-\cosh^2\chi \rd t^2+\rd\chi^2+\sinh^2\chi \rd\Omega_{d-1}^2).
\end{equation}
The (conformal) boundary, located at $\chi\rightarrow\infty$, takes the form $\R\times S^{d-2}$. The time coordinate $t$ is now periodic with period $2\pi$, but to avoid closed timelike curves we can ``unwrap'' this coordinate and AdS is usually defined as such. By defining the coordinate $r=L\sinh\chi$, the metric can be written in the form
\begin{equation}
\rd s^2=-(1+\frac{r^2}{L^2})\rd t^2+\frac{1}{1+\frac{r^2}{L^2}}\rd r^2+r^2\rd\Omega_{d-1}^2.
\end{equation}
Note also that for large $r$ the metric approaches that of Poincar\'e patch AdS. Contrary to the Poincar\'e coordinates, in global coordinates the AdS radius plays an important role and sets a relevant length scale. Thus black hole solutions, which are now asymptotically {\it global} AdS, can be classified as {\it small} or {\it large}, depending on how the size (event horizon) compares to the AdS radius (recall that this is not possible for the AdS black branes). The black hole solutions have the metric
\begin{equation}
\rd s^2=-\left(1+\frac{r^2}{L^2}-\frac{M}{r^{d-2}}\right)\rd t^2+\frac{1}{1+\frac{r^2}{L^2}-\frac{M}{r^{d-2}}}\rd r^2+r^2\rd\Omega_{d-1}^2,
\end{equation}
with an event horizon located at the solution of $1+\frac{r^2}{L^2}-\frac{M}{r^{d-2}}=0$. It is possible to obtain the Poincar\'e patch AdS black branes by sending $r\rightarrow\infty$ while rescaling $M$ appropriately. Note also that an AdS black hole looks like a standard Schwarzschild black hole in flat space. The temperature can be computed in exactly the same way as for the black brane. By considering the metric close to the horizon $r_h$, we obtain
\begin{equation}
\rd s^2\approx(r-r_h)\left(d\frac{r_h}{L^2}+(d-2)\frac{1}{r_h}\right)\rd\tau^2+\frac{1}{(r-r_h)\left(d\frac{r_h}{L^2}+(d-2)\frac{1}{r_h}\right)}\rd r^2+\ldots,
\end{equation}
and by identifying the periodicity of $\tau$ that results in a smooth geometry without a conical singularity, we obtain the temperature as
\begin{equation}
T=\frac{d\frac{r_h}{L^2}+(d-2)\frac{1}{r_h}}{4\pi}.
\end{equation}
Note that this results in \eqref{temppoincare} when $r_h\rightarrow\infty$. It is interesting to note that there are qualitative differences between small and large black holes in global AdS, for instance the temperature has a minimum at $r_h=L\sqrt{(d-2)/d}$ and thus small black holes will have a negative specific heat (defined as the rate of change of the temperature as a function of the mass $M$). Interestingly, it has also been shown that small black holes suffer from a so-called Gregory-Laflamme instability \cite{Gregory:1993vy}, if we consider them as solutions of 10-dimensional supergravity \cite{Banks:1998dd,Horowitz:1999uv}. This is related to the so-called correlated stability conjecture \cite{Gubser:2000ec,Gubser:2000mm}, which states precisely that thermodynamic instabilities and dynamical instabilities should be correlated (but there are counterexamples to this conjecture, see for instance \cite{Friess:2005zp,Buchel:2010wk,Buchel:2011ra}). Dynamical instabilities of higher dimensional black holes is a whole research in itself, and we will not comment further on this topic.

\section{Conformal field theory}\label{sec_cft}
\begin{figure}
 \includegraphics[scale=0.7]{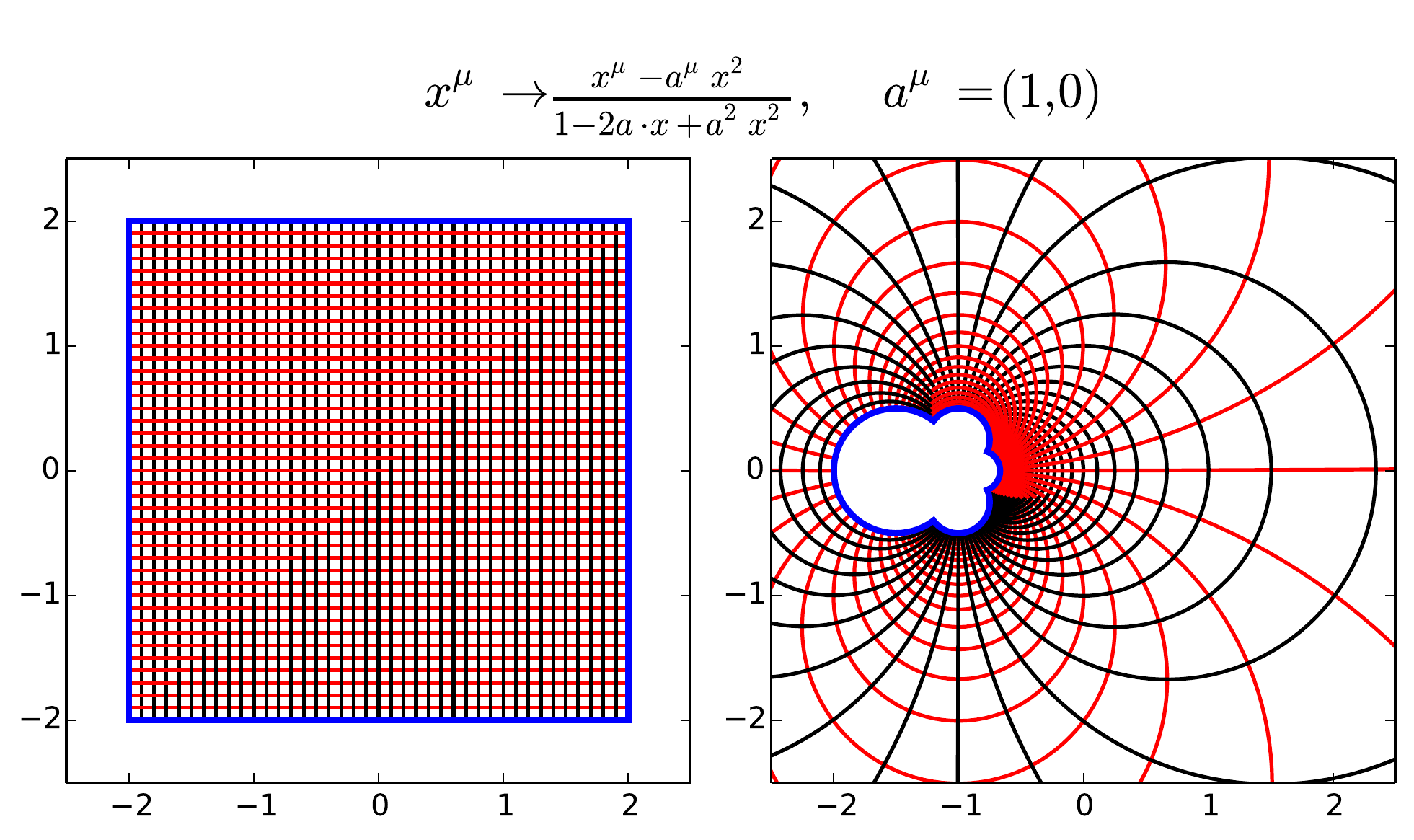}
 \caption{\label{specconf} A special conformal transformation in two dimensions.}
\end{figure}

In this section we will briefly review some of the aspects of conformal field theory that will be relevant for this thesis. Conformal symmetry is a symmetry that generalizes scale invariance. More precisely, a theory possessing scale invariance is invariant if we rescale all our coordinates as $\vec{x}\rightarrow\lambda^2\vec{x}$. This would obviously rescale the metric of the theory as $g_{\mu\nu}\rightarrow\lambda^2g_{\mu\nu}$. A conformal transformation on the other hand is a coordinate transformation that will rescale the metric by a factor that depends on location, namely by
\begin{equation}
g_{\mu\nu}(x)\rightarrow \lambda^2(x)g_{\mu\nu}(x).
\end{equation}
In particular, conformal transformations will preserve angles, and in two dimensions (in Euclidean signature) they coincide with the holomorphic (analytic) maps in complex analysis. Conformal invariance obviously implies scale invariance, and while the converse is not true in general, in many cases scale invariance plus some additional information (for example unitarity and Lorentz invariance) implies conformal invariance. The conformal group (for $d>2$) is generated by the generators
\begin{align}
M_{\mu\nu}&\equiv i(x_\mu\partial_\nu-x_\nu\partial_\mu),\nonumber\\
P_{\mu}&\equiv -i\partial_\mu,\nonumber\\
D&\equiv -ix_\mu\partial^\mu,\nonumber\\
K_\mu&=i(x^2\partial_\mu-2x_\mu x_\nu\partial^\nu).
\end{align}
where $M_{\mu\nu}$ generate Lorentz transformations, $P_\mu$ generate translations, $D$ generates dilatations (scalings) and $K_\mu$ generates so-called special conformal transformations (which consists of an inversion, followed by a translation and then an inversion again). An illustration of a special conformal transformation is shown in Figure \ref{specconf}
This group is isomorphic to $SO(2,d)$ (in Euclidean signature it would instead be isomorphic to $SO(1,d+1)$). In two dimensions, these generators only generate a part of the total symmetry group. This subgroup is called the {\it global} subgroup since it is formed by generators that are well-defined in the whole plane (including the point at infinity). In two dimensions, the symmetry group is larger and consists of (two copies of) the Virasoro group which is infinite dimensional. In Euclidean signature, we can formulate the theory in terms of two complex coordinates $z$ and $\bar z$ and the conformal group in two dimensions coincides with the set of holomorphic and antiholomorphic transformations $z\rightarrow f(z),$ $\bar z\rightarrow \bar{f}(\bar z)$. The infinite number of generators of the Virasoro group are related to the expansion coefficients of the Laurent expansion of the functions $f$ and $\bar f$. The first hint for the AdS/CFT correspondence comes from matching the symmetries on both sides, and for $d>2$ the conformal symmetry group is exactly the same as the isometry group of AdS$_{d+1}$. For three-dimensional asymptotically AdS spaces, the isometry group can be extended to an asymptotic symmetry group (the group of transformations that preserve the asymptotic \ads form of the metric) which turns out to be exactly the same as the full conformal symmetry group in two dimensions \cite{Brown:1986nw}. This will be explained in further detail in Section \ref{ads3asa}.\\
\lb
A conformal field theory (CFT) is a (quantum) field theory (QFT) that is invariant under conformal transformations. Conformal field theory has numerous applications, not only in high energy physics and string theory where for instance the two-dimensional worldsheet in string theory is described by a CFT, but also in condensed matter systems close to phase transitions. CFTs are often specified and studied by the set of operators present in the theory without necessarily refering to an action principle. The operators are then related to each other via the {\it operator product expansion} (OPE), which dictates how the product of two fields behaves when they are brought close to each other. Due to the high degree of symmetry, many questions have universal answers. In particular, two- and three-point functions are almost completely fixed by conformal invariance. For example, the two-point function of two operators $\mathcal{O}_1$ and $\mathcal{O}_2$ are given by
\begin{equation}
\langle \mathcal{O}_1(x_1)\mathcal{O}_2(x_2)\rangle= C\frac{\delta_{\Delta_1,\Delta_2}}{|x_1-x_2|^{2\Delta_1}},
\end{equation}
where $C$ is an overall constant and $\Delta_1$ and $\Delta_2$ are the {\it scaling dimensions} of the operators. The scaling dimension $\Delta$ of an operator $\mathcal{O}$ is determined by its transformation under a dilatation (each operator will transform according to some representation of the conformal group which can then be labeled by its eigenvalue under the dilatation operator). Under a scaling $x'=\lambda x$, the operator transforms as
\begin{equation}
\mathcal{O}'(x')=\lambda^{-\Delta}\mathcal{O}(x).
\end{equation}
The scaling dimension of an operator will be related to the boundary behaviour of bulk fields in the AdS/CFT correspondence.\\
\lb
In a CFT there is also an important class of operators called {\it primary} operators. They are defined by the fact that they are annihilated by all lowering operators of the conformal algebra. Operators obtained by acting with raising operators on a primary operator are known as {\it descendants}, and any operator in a (unitary) CFT can then be formed as linear combinations of descendants and primaries. In two dimensions, under a conformal transformation characterized by two functions $f$ and $\bar f$, a primary operator transforms as
\begin{equation}
 \mathcal{O}'(f(z),\bar{f}(\bar z))=(f'(z))^{-h}(\bar{f}'(\bar z))^{-\bar h}\mathcal{O}(z,\bar z),
\end{equation}
where $h$ and $\bar h$ are called the conformal weights of the operator. Thus we see that the scaling dimension is given by $\Delta=h+\bar h$, while $s=h-\bar h$ is called the {\it spin} of the operator (since it is the eigenvalue under a rotation).\\
\lb
In many examples of the AdS/CFT correspondence, the dual field theory has {\it super}conformal symmetry, which means that on top of conformal symmetry there is also supersymmetry, a symmetry that mixes bosonic and fermionic degrees of freedom. This symmetry is a natural consequence of the fact that all consistent string theories are supersymmetric and supersymmetry has been proposed as a candidate for solving several problems in the standard model of particle physics (although it has yet to be discovered in nature). We will not discuss supersymmetry in any detail in this thesis.

\section{Obtaining the duality from string theory}\label{stringtheory}
In this section we will review the original example of the AdS/CFT correspondence, which was first discovered by Juan Maldacena \cite{Maldacena:1997re,Aharony:1999ti}. It states that {\it type IIB superstring theory in AdS$_5\times S^5$ is dual to $\mathcal{N}=4$ $SU(N)$ SYM theory in 3+1 dimensions}, in certain limits that we will clarify below. $\mathcal{N}=4$ $SU(N)$ SYM stands for a supersymmetric Yang-Mills theory with gauge group $SU(N)$, in four dimensions and with maximal supersymmetry. As the name suggests, it has four supercharges\footnote{There are four supercharges if we count the number of spinors, but the number of {\it real} supercharges (spinor components) is 16.} and supersymmetry completely determines the action of the theory. In the strongest version of the correspondence, the duality holds for all parameters of the two theories. In a weaker version, the duality only holds in the limit of large $N$ (which corresponds to small string coupling $g_s$). The weakest version of the correspondence states that the two theories are equal for large $N$ and large 't Hooft coupling $\lambda=g_{YM}^2 N$ where $g_{YM}$ is the coupling of the gauge theory (this corresponds to small string coupling $g_s$ and the high tension limit $T\sim1/\sqrt{\ell_s}\rightarrow \infty$). This is the version that is most commonly used (and will be in this thesis) since the string theory side reduces to classical gravity. The derivation of this duality relies on taking a certain low energy limit of string theory in the presence of so-called D$p$-branes, and we will review all relevant concepts below. There are now many more examples of similar dualities derived by taking limits of different setups in string theory. For more information about string theory in general, see for instance \cite{Green:1987sp,Green:1987mn,Polchinski:1998rq,Polchinski:1998rr,Becker:2007zj}.
\subsection{String theory}
String theory is based on the idea that elementary particles are actually not pointlike particles, but rather tiny vibrating strings. Quantizing the vibrations of the string then gives rise to a discrete spectrum of vibrations, and these different modes will then correspond to different elementary particles (or different quantum fields) in the low energy effective description where the string theory reduces to a field theory and the strings look pointlike. Replacing fundamental particles by strings is motivated partly by the goal to reconcile quantum physics and the standard model with general relativity, and one of the most successful aspects of string theory is that the graviton naturally appears in the spectrum of string vibrations, along with other basic building blocks in the standard model such as spin-1/2 fermions and spin-1 vector bosons. The perturbative expansions in Feynman diagrams in (Euclidean) quantum field theory are replaced in string theory as a sum over different topologies of a string's worldsheet, which is one of many examples where a concept in quantum field theory or gravity attains a new interesting geometrical description in string theory. The different worldsheet topologies are weighted by $g_s^{-\chi}$ where $g_s$ is the string coupling constant (determining the strength of interactions between strings) and $\chi$ is the Euler characteristic of the worldsheet (related to the genus by $\chi=2-2g$). Thus the lowest order corresponds to a string worldsheet with the topology of a sphere and the next order consists of worldsheets with the topology of a torus. In the quantum field theory limit, the number of handles $g$ corresponds to the number of loops in the Feynman diagrams.\\
\linebreak
The dynamics of a (bosonic) string is governed by the Polyakov action
\begin{equation}
S=\frac{\mathcal{T}}{2}\int \rd^2x \sqrt{-h}h^{ab}g_{\alpha\beta}\partial_aX^\alpha \partial_bX^\beta,
\end{equation}
where $a,b$ are indices for the coordinates parametrizing the string's two-dimensional worldsheet, and $X^\alpha$ are the coordinates of the spacetime where the string propagates. The dynamical fields living on the string worldsheet are the metric $h_{ab}$ and the scalars $X^\alpha$, and $\mathcal{T}=1/2\pi\alpha'$ is a parameter known as the string tension. The parameter $\alpha'$ is the only dimensionful parameter in string theory, and is often instead written in terms of the string length $\ell_s=\sqrt{\alpha'}$. By computing the equations of motion and solving for $h_{ab}$, we obtain the Nambu-Goto action
\begin{equation}
S=\frac{\mathcal{T}}{2}\int \rd^2x \sqrt{-\det h_{ab}},
\end{equation}
where $h_{ab}=\partial_aX^\alpha\partial_bX^\beta g_{\alpha\beta}$ is now the induced metric on the worldsheet. This action might look more familiar since it just computes the area of the string's worldsheet. Classically, these two actions are the same, but at a quantum level this is no longer guaranteed and the Polyakov action is preferred since it is much easier to quantize. Quantizing the Polyakov action then gives rise to different quantum states of the string, which are interpreted as different elementary particles. Note that these actions describe {\it bosonic} string theory, but for consistency one must add fermionic coordinates to the worldsheet and such theories are referred to as {\it super} string theories. There are currently five known consistent string theories, which all include some degrees of supersymmetry and only exist in 10 dimensions (a bosonic string theory exists in 26 dimension but suffers from an instability triggered by negative mass modes in the string spectrum). The particular string theory that we will focus on, which is often used in model building in the context of AdS/CFT and cosmology, is type IIB string theory. Another important property distinguishing different strings is if they are open or closed. Open strings will have either Neumann or Dirichlet boundary conditions at the endpoints of the string, and closed strings will have periodic boundary conditions, and they result in different quantum fields (different particles) in the low energy effective description.
\subsection{D-branes}
Another important object in string theory are so-called D$p$-branes, which are objects extending in $p$ (spatial) dimensions. The defining property of D-branes are that these are surfaces on which the open strings can end. The open string will then have Dirichlet boundary conditions in the directions perpendicular to the brane, and Neumann boundary conditions in the directions parallel to the brane (in which the endpoints move freely). The open strings attached to the brane may interact with closed strings which are not confined to the brane, for instance two open strings may collide and merge to a single closed string which then is free to leave the world volume of the brane. Equivalently, closed strings may be absorbed on the brane and decay into open strings. We may also have open strings that stretch between different D-branes. The strong coupling low energy description of D$p$-branes is believed to be extended solitonic solutions in supergravity, called extremal $p$-branes, which we will describe next.


\subsection{Relation to supergravity}
The closed string spectrum consists of a set of massless and massive modes. All the massive modes have a mass proportional to $1/\sqrt{\alpha'}$, and thus when taking the low energy limit, equivalent to taking $\alpha'\rightarrow0$, all massive modes will decouple and the effective theory thus corresponds only to the massless part of the spectrum. When taking this limit, the strings (which are extended objects) will reduce to pointlike objects and the effective theory one obtains from the five different consistent string theories coincide exactly with the five known anomaly free supergravity theories in 10 dimensions. The extra fermionic fields will not be important in this thesis, but supersymmetry also constrains the bosonic part of the supergravity action. Likewise, when taking the low energy limit of open strings attached to a D-brane, we end up with a supersymmetric field theory living on the brane. Moreover, if we consider a string propagating in a curved background, it can be shown that the beta function vanishes if and only if the background satisfies the supergravity equations of motion. The beta function must vanish for a theory with conformal symmetry, and since the conformal symmetry on the worldsheet is interpreted as a gauge symmetry and thus has to hold at the quantum level we conclude that the background must satisfy the supergravity equations of motion for consistency.\\
\linebreak
For type IIB supergravity (in ten dimensions), which is the $\alpha'\rightarrow0$ limit of type IIB string theory, the bosonic field content is the metric $g_{\mu\nu}$, a scalar $\phi$, an antisymmetric tensor $B_{\mu\nu}$ and three gauge potentials $C^{(0)}$, $C^{(2)}_{\mu\nu}$ and $C^{(4)}_{\mu\nu\rho\sigma}$ which are tensors of rank 0, 2 and 4 respectively. Supersymmetry requires that $C^{(4)}$ is self dual, namely that $F=*F$ where $F_{\mu\nu\rho\sigma\delta}=\partial_{[\mu}C^{(4)}_{\nu\rho\sigma\delta]}$. There are natural objects in supergravity that couple to these gauge potentials, known as black $p$-branes. These are solutions of the supergravity equations of motion and are natural extensions of black holes that extend in $p$ spatial dimensions. A black $p$-brane can carry charge under a ($p+1$)-form, and thus in type IIB supergravity there are black 1-branes and 3-branes. This is analogous to the standard electric charge for black holes in four dimensions. There is generically a bound on the charge in terms of the mass of the black brane derived from enforcing absence of a naked singularity, and black branes saturating this inequality are called {\it extremal}. One of the fundamental hypotheses in string theory (and crucial for the derivation of the AdS/CFT correspondence) is that extremal black $p$-branes in supergravity are the same as D$p$-branes in string theory. 

\subsection{Obtaining the correspondence}
We are now ready to sketch the derivation of the correspondence between $\mathcal{N}=4$ SYM and supergravity in $AdS_5\times S^5$. The starting point is a stack of $N$ D3-branes in ten-dimensional type IIB super string theory. By describing the low energy limit of this setup both from the point of view of D3-branes in string theory, and from the point of view of extremal 3-branes in supergravity, one may motivate the desired duality.\\
\linebreak
We thus start with the viewpoint of $N$ D3-branes extending in the directions $x_1,x_2,x_3$ as well as time $t$, while the other transverse coordinates are $y_1,\ldots,y_6$. The action of the full system can be described as
\begin{equation}
S=S_{bulk}+S_{brane}+S_{int},
\end{equation}
where $S_{bulk}$ is the action for the closed strings outside the brane, $S_{brane}$ is the action describing the dynamics of the open strings on the branes (where an open string generically stretch between different D-branes) and $S_{int}$ is the interaction between the two. When taking the low energy limit, $\alpha'\rightarrow0$, the bulk theory reduces to type IIB supergravity and the brane theory reduces to $\mathcal{N}=4$ SYM with gauge group $SU(N)$. Moreover, for the interaction term we have $S_{int}\sim g_s\alpha'^2$ where $g_s$ is the string coupling constant. Thus in the low energy limit the two theories decouple and we obtain both ten-dimensional type IIB supergravity as well as the $\mathcal{N}=4$ SYM theory.\\
\linebreak
From the other point of view, the $N$ D3-branes can be interpreted as an extremal 3-brane charged under the potential $C^{(4)}$. The solution takes the form \cite{Horowitz:1991cd}
\begin{equation}
\rd s^2=H^{-1/2}\rd x_{\|}^2+H^{1/2}(\rd r^2+r^2\rd\Omega_5^2),
\end{equation}
\begin{equation}
C^{(4)}_{0123}=H,
\end{equation}
\begin{equation}
H=1+\frac{L^4}{r^4},\quad L^4=4\pi g_s N(\alpha')^2.
\end{equation}
The low energy excitations of this background consist of two sectors. First of all, we have excitations deep inside the throat, namely at small $r$. For an observer at $r\rightarrow\infty$, the energy $E(r)$ of an object at position $r$ is seen as having energy $E=H(r)^{-1/4}E(r)\sim r E(r)/L$ and thus goes to zero when $r\rightarrow0$. Moreover, small excitations deep inside the throat will not be able to climb up from the throat and are thus confined at small $r$ and decoupled from the region at large $r$. The other sector consists of large wavelength excitations at large length scales. They can be viewed as living in ten-dimensional flat space since the wavelengths are too long to resolve the 3-brane, and thus do not affect (and are not affected by) the dynamics of and inside the brane. Thus these two sectors are completely decoupled and do not interact with each other, and we conclude that the low energy dynamics from this point of view consists of supergravity deep inside the throat of the 3-brane as well as ten-dimensional supergravity in flat space. Note also that $r\rightarrow0$ implies that $H\sim L^4/r^4$, thus the metric becomes
\begin{equation}
\rd s^2=\frac{r^2}{L^2}\rd x_{\|}^2+\frac{L^2}{r^2}\rd r^2+L^2\rd\Omega^2_5,
\end{equation}
which is that of $AdS_5\times S^5$ in Poincar\'e coordinates where the radius of the five sphere as well as the AdS radius are both equal to $L$.\\
\linebreak
Since both viewpoints include a sector with ten-dimensional supergravity, it is tempting to conclude that the other two theories are equal, namely that {\it supergravity in $AdS_5\times S^5$ is equivalent to $\mathcal{N}=4$ SYM in 3+1 dimensions with gauge group $SU(N)$, when $\alpha'\rightarrow0$}. Note that if we want the gravity sector to be governed by classical supergravity, we need both $g_s\rightarrow0$ (no quantum corrections) and $\ell_s/L\rightarrow0$ (such that the strings are approximately pointlike), which on the field theory side translates into $N\rightarrow\infty$ (a large gauge group) and $\lambda=g_{YM}^2N=g_s N\rightarrow\infty$ (strong coupling). For a more in-depth analysis of the different limits we refer the reader to for instance \cite{Aharony:1999ti,Nastase:2007kj}.\\
\linebreak
We have in this section very roughly sketched the original motivation for the correspondence, but right now it is just a very vague statement and exactly how the quantities are related on the two sides of the duality is still unclear. In the next section we will explain how to relate the gravity quantities to the field theory observables, a prescription usually called the {\it holographic dictionary}, which was developed after Maldacena's original paper.


\section{The dictionary and holographic renormalization}\label{sec_holrenorm}
In this section we will explain the map between observables on the CFT side and quantities on the gravity side. The program of identifying sources and expectation values of field theory observables in terms of bulk quantities, known as the {\it holographic dictionary}, involves a procedure called {\it holographic renormalization} \cite{Skenderis:2002wp} which is necessary to render the observables finite. We will first describe the general formalism and then illustrate the technique with a massless scalar field coupled to gravity, which will be the main model we will use in the rest of the thesis.\\
\linebreak
The map between bulk and boundary quantities follows from the so called GKPW \cite{Witten:1998qj,Gubser:1998bc} formula, named after Gubser, Klebanov, Polyakov and Witten, which states that
\begin{equation}
Z_{sugra}[J]\equiv\int_{\Phi\sim J} D\Phi \exp(-S[\Phi])=\langle \exp\left(-\int_{\partial AAdS}J\mathcal{O}\right)\rangle\equiv Z_{QFT}[J].
\end{equation}
Here the left hand side is the supergravity partition function in anti-de Sitter space, and the right hand side is the partition function of the dual QFT on the boundary of the asymptotically AdS spacetime we are considering. $\Phi$ can be interpreted as denoting all relevant fields in the supergravity theory (although we will simply treat it as a single field for simplicity), and $\Phi\sim J$ means that the boundary condition at the AdS boundary is given by the source $J$ in a way that will be made precise later. On the field theory side, $J$ is a source coupled to an operator $\mathcal{O}$ which we say is {\it dual} to $\Phi$. This equality can be taken as the fundamental hypothesis underlying the AdS/CFT correspondence, and in the strongest version of the duality this holds for arbitrary parameters on both sides (in which the left hand side should really be replaced by the partition function of string theory). In the limit of small string coupling and large string tension, $g_s\rightarrow0$ and $\ell_s<<L$ (corresponding to the large $N$ and large $\lambda$ limit in the dual field theory), the correspondence reduces to an equivalence between the partition function of the QFT and the partition function of supergravity in the saddle point approximation as
\begin{equation}
\exp(-S_{on-shell}[J])=\langle \exp\left(-\int_{\partial AAdS}J\mathcal{O}\right)\rangle_{N\rightarrow\infty, \lambda\rightarrow \infty},
\end{equation}
where the right hand side is the QFT partition function in the large $N$ and strong ('t Hooft) coupling limits and the left hand side involves the on-shell supergravity action. This is the form of the GKPW formula that we will use throughout this thesis, and is the form that is used in the vast majority of applications of the AdS/CFT correspondence. By taking functional derivatives of this relation we can compute expectation values of CFT observables by using techniques from classical (super) gravity. We have in particular that correlators when the source is turned off can be computed as
\begin{equation}
\langle \mathcal{O}(x)\rangle \sim\frac{\delta S_{on-shell}}{\delta J(x)}|_{J=0}.\label{1pfctsketch}
\end{equation}
\begin{equation}
\langle \mathcal{O}(x_1)\mathcal{O}(x_2)\rangle \sim\frac{\delta^2 S_{on-shell}}{\delta J(x_1)\delta J(x_2)}|_{J=0}.\nonumber
\end{equation}
\begin{equation}
\ldots\nonumber
\end{equation}
We will make the $\sim$ more precise later.
We will now focus on the one-point functions and not concern ourselves with higher order correlation functions. Higher order correlation functions are more sensitive to the signature of spacetime and there are subleties in Lorentzian signature which we do not want to get into (see for instance \cite{Skenderis:2008dh,Skenderis:2008dg,Son:2002sd}). For one-point functions, the result is less sensitive to these issues and we will use Lorentzian signature throughout.\\
\linebreak
Let us now make \eqref{1pfctsketch} more precise. First of all, the boundary behaviour of the fields $\Phi$ will depend on the scaling dimension (see Section \ref{sec_cft}) of the dual operator in the form
\begin{equation}
\Phi=Jz^{d-\Delta}+\ldots+\Phi_\Delta z^{\Delta}+\ldots,\label{Phiasympt}
\end{equation}
where $Jz^{d-\Delta}$ and $\Phi_\Delta z^{\Delta}$ are the two independent modes and the metric \eqref{poincaremetric} has been used for the asymptotically AdS spacetime. This is what we meant by the boundary condition $\Phi\sim J$ which thus depends on the parameter $\Delta$ which coincides with the scaling dimension of the operator $\mathcal{O}$. Moreover, the terms between $J$ and $\Phi_\Delta$ will be local functions of $J$, while the terms after $\Phi_\Delta$ will be functions of both $J$ and $\Phi_\Delta$. As we will see later, $\Phi_\Delta$ will be related to the expectation value of the operator $\mathcal{O}$. Given the boundary condition $J$, $\Phi_\Delta$ can in principle be determined by solving the full equations of motion in the bulk (i.e.\ not just the asymptotic expansion). Furthermore, when evaluating the on-shell action, it will typically diverge and it is necessary to regularize it. We will thus not integrate the radial coordinate all the way to the boundary, but stop at some finite position $z=\epsilon$. To cancel to divergence at $\epsilon\rightarrow0$ we need to add counterterms that make the action finite. We thus define a renormalized action by
\begin{equation}
S_{ren}=S+S_{ct},
\end{equation}
where $S_{ct}$ includes the necessary counterterms and $S_{ren}$ on-shell is finite when $\epsilon\rightarrow0$. Moreover, even though the metric does not define an induced metric on the boundary, it can define a boundary metric up to a conformal transformation (which is known as a {\it conformal structure}). Thus we can construct a metric for the boundary theory by $\hat{g}_{ij}^{(0)}$, which is the leading term in $\hat{g}_{ij}$ defined by writing the metric in the form
\begin{equation}
\rd s^2=\frac{L^2}{z^2}\left(\rd z^2+\hat{g}_{ij}\rd x^i\rd x^j\right).\label{defhatg}
\end{equation}
$\hat{g}_{ij}^{(0)}$ is usually just flat Minkowski space. Thus the partition function should really read
\begin{equation}
Z_{QFT}[J]=\langle \exp\left(-\int_{\partial AAdS}\sqrt{-\hat{g}^{(0)}}J\mathcal{O}d^dx\right)\rangle.
\end{equation}
We now define our renormalized one-point functions as
\begin{equation}
\langle \mathcal{O}(x)\rangle\equiv\lim_{\epsilon\rightarrow0}\frac{1}{\sqrt{|\hat{g}^{(0)}|}}\frac{\delta S_{ren}}{\delta J(x)}|_{J=J_0},
\end{equation}
where we have also included the possibility of having a non-zero source $J_0$ turned on. Moreover, we can rewrite this in a covariant way in terms of the field $\Phi$ as
\begin{equation}
\langle \mathcal{O}(x)\rangle\equiv\lim_{\epsilon\rightarrow0}\epsilon^{d-\Delta}\frac{1}{\sqrt{|\hat{g}^{(0)}|}}\frac{\delta S_{ren}}{\delta \Phi}|_{J=J_0},
\end{equation}
since $\Phi\rightarrow \epsilon^{d-\Delta}J$ when $\epsilon\rightarrow0$. This renormalization procedure is reminiscent of the renormalization one does when computing QFT observables in perturbation theory. Sometimes these counterterms are not uniquely determined from the requirement that the on-shell action is finite, but instead there is room for adding finite counterterms which are not fixed by this requirement but will still affect the boundary theory observables. In these cases there is a scheme dependence in the boundary quantities and they are thus not uniquely defined. Typically this only happens while the source is non-zero and when $d$ is even. Often we will only study the boundary observables after the source has been turned off and thus this sublety will not play a major role, but we will demonstrate this phenomenon explicitly for a massless scalar with $d=4$ below.\\
\linebreak
Another important concept are so-called holographic Ward identities. These are relations between the sources and expectation values of operators in the boundary theory that can be obtained purely from the asymptotic boundary expansion, and follow from the diffeomorphism invariance of the bulk theory. These should be interpreted in the boundary theory as the traditional Ward identities in QFT which are derived from global symmetries on the boundary. We will encounter such relations in Chapter \ref{confinement}, although we will obtain them directly from the boundary expansion of equations of motion instead of considering diffeomorphism invariance.\\
\linebreak
In the next two sections we will discuss how to obtain the boundary stress-energy tensor as well as go through a very simple example to illustrate the holographic renormalization procedure, namely that of a massless scalar field in the probe limit in four-dimensional AdS. In Chapter \ref{confinement} we will consider more complicated setups with less symmetry and beyond the probe limit.

\subsection{The stress-energy tensor}
The formula for the stress-energy tensor in the dual CFT is given by
\begin{equation}
\langle T_{ij}\rangle\equiv\lim_{\epsilon\rightarrow0}\frac{2}{\sqrt{|\hat{g}^{(0)}|}}\frac{\delta S_{ren}}{\delta \hat{g}^{ij,(0)}}=\lim_{\epsilon\rightarrow0}\frac{2}{\sqrt{|\gamma|}}\left(\frac{L}{\epsilon}\right)^{d-2}\frac{\delta S_{ren}}{\delta \gamma^{ij}},
\end{equation}
where $\gamma_{ij}$ is the induced metric at $z=\epsilon$ and $\hat{g}_{ij}^{(0)}$ is the boundary metric defines as the boundary value of $\hat{g}_{ij}$ in equation \eqref{defhatg}. $S_{ren}$ is given by
\begin{equation}
S_{ren}=S_{EH}+S_{GHY}+S_{ct},
\end{equation}
where $S_{EH}$ is the Einstein-Hilbert action given by
\begin{equation}
S_{EH}=\frac{1}{2\kappa^2}\int_{\mathcal{M}} \rd^{d+1}x \sqrt{-g}\left(R-2\Lambda\right),
\end{equation} 
and $S_{GHY}$ is the Gibbons-Hawking-York term that must be present to render the variational principle well defined\cite{Gibbons:1976ue}. It does not affect the equations of motion, but it will affect the holographic renormalization at the boundary. It is given by
\begin{equation}
S_{GHY}=\frac{1}{\kappa^2}\int_{\partial \mathcal{M}} \rd^dx\sqrt{-h}K,
\end{equation}
where $K$ is the trace of the extrinsic curvature. $S_{ct}$ is the necessary counterterm action to render the total renormalized action (and the field theory observables) finite.
The counterterm action will depend on the dimension of the spacetime, and there is no closed form expression for general dimension. The first few terms in the counterterm action are \cite{Emparan:1999pm}
\begin{equation}
S_{ct}=-\frac{1}{\kappa^2}\int_{z=\epsilon}\rd^dx\sqrt{-\gamma}\left(\frac{d-1}{L}+\frac{L}{2(d-2)}\hat{R}+\frac{L^3}{2(d-4)(d-2)}\left(\hat{R}_{ab}\hat{R}^{ab}-\frac{d}{4(d-1)}\hat{R}^2\right)+\ldots\right),
\end{equation}
where $\ldots$ are higher order terms in the Ricci tensor $\hat{R}_{ab}$ computed from the induced metric $\gamma_{ab}$. The number of terms in this expansion that should be included depends on the dimension of the spacetime, and only terms up to the $\lfloor (d+1)/2\rfloor$'th term should be included (in particular terms with an overall coefficient $1/(d-k)$ are obviously not included for $d=k$). The stress-energy tensor is then given by the formula
\begin{equation}
\langle T_{ij} \rangle\equiv\lim_{\epsilon\rightarrow0}\frac{2}{\sqrt{|\hat{g}^{(0)}|}}\frac{\delta S_{ren}}{\delta \hat{g}^{ij}_{(0)}}.
\end{equation}
The final expression for the stress-energy tensor in terms of the boundary expansion of the metric is
\begin{equation}
\langle T_{ij} \rangle = \frac{L^{d-1}d}{16\pi G}\hat{g}^{(d)}_{ij}+X_{ij}[\hat{g}^{(0)}_{ij}],\label{Tijformulev}
\end{equation}
where $X_{ij}=0$ for odd $d$ but otherwise is a complicated function of the boundary metric and its derivatives. The explicit formula for $X_{ij}$ is scheme dependent and depends strongly on dimension (see \cite{Skenderis:2000in}). Here $\hat{g}_{ij}^{(0)}$ and $\hat{g}_{ij}^{(d)}$ are the 0th and $d$th order coefficients in the expansion of $\hat{g}_{ij}$ defined by equation \eqref{defhatg}. However, if the boundary metric $\hat{g}_{ij}$ can be written as $\hat{g}_{ij}=\eta_{ij}+z^d\hat{g}_{ij}^{(d)}+O(z^{d+1})$, namely if the boundary metric is the flat metric $\eta_{ij}$ and the intermediate terms in the expansion vanish, the situation simplifies and $X_{ij}=0$ so that the result is
\begin{equation}
\langle T_{ij} \rangle = \frac{L^{d-1}d}{16\pi G}\hat{g}^{(d)}_{ij}.\label{Tijformul}
\end{equation}
We want to emphasize that \eqref{Tijformul} is valid for odd $d$ even when the boundary metric is not flat. When there is no matter present, the criteria of having flat boundary metric is enough to guarantee that the intermediate coefficients $g^{(n)}_{ij}$ with $0<n<d$ vanish. However, if there is additional matter present, the intermediate terms can be non-zero and depend on the matter fields. At least for the matter content we discuss in this thesis (a massless scalar field), it can be easily seen that if the source is turned off, the matter part of the action will go to zero at the boundary and not contribute to the stress-energy tensor. Thus the formula \eqref{Tijformul} holds also for the case when a massless scalar with vanishing boundary source is present (this argument should hold for a large class of matter actions but we will not discuss that further).\\
\linebreak
As an example, for the black brane metric \eqref{poincareblackbrane}, we obtain
\begin{equation}
\langle T_{tt} \rangle =(d-1)\langle T_{x_ix_i} \rangle = \frac{(d-1)ML^{d-1}}{16\pi G},
\end{equation}
so we see that $M$, which is usually called the mass of the black brane, is indeed proportional to the energy density of the boundary theory. In the above equation we used $x_i$ to denote the spatial coordinates in the boundary theory. The off-diagonal components vanish.

 
\subsection{Example: Massless scalar field}
As an illustrative example, we now consider a free massless scalar field in the probe limit of AdS. We will consider the case of $d=3$ in detail before quoting the general result in the absence of sources, and we will restrict to the case where the scalar only depends on time and the radial coordinate. In Section \ref{hardwallmodel} we will study the full system (including backreaction) for both $d=3$ and $d=4$ and we refer to that section for a more in-depth analysis. The action for the scalar is
\begin{equation}
S=\frac{1}{2\kappa^2}\int \rd^{d+1}x \sqrt{-g}\left(-\frac{1}{2}\partial_\mu\phi g^{\mu\nu}\partial_\nu\phi\right),
\end{equation}
where $2\kappa^2=16\pi G$ where $G$ is Newton's constant. We will in this section only be concerned with the matter part since we assume that the scalar field is in the probe limit (no backreaction to the metric). The equation of motion for the scalar field in the background \eqref{poincaremetric} is thus
\begin{equation}
\square \phi=\frac{1}{\sqrt{-g}}\partial_\mu(\sqrt{-g}g^{\mu\nu}\partial_\nu\phi)=\frac{z^2}{L^2}\left(-\partial_t^2\phi+\partial^2_z\phi-(d-1)z^{-1}\partial_z\phi\right)=0.
\end{equation}
The expansion close to the boundary is
\begin{equation}
\phi(z,t)=J(t)+\phi_1(t)z+\ldots+\phi_lz^d\log z+\phi_d(t)z^{d}+\ldots.
\end{equation}
The first coefficient $\phi_0\equiv J$ as well as the coefficient $\phi_d$ are undetermined by the boundary expansion. The term proportional to $z^d\log z$ only exists for even $d$. Note that the scaling dimension $\Delta$ is equal to $d$. Usually one specifies the source $J$ and then computes the coefficient $\phi_d$ which will essentially be the expectation value of the dual operator. Note that although $\phi_d$ is undetermined by the boundary analysis, solving the full equations will generically determine $\phi_d$ in terms of $J$. The other coefficients $\phi_i$, $i\neq0,d$ are determined in terms of $J$ and $\phi_d$. For a massive scalar field with mass $m$, we would have $\Delta$ determined by the equation $\Delta(d-\Delta)=m^2$, and the field would not go to a constant at the boundary (it would have the general behaviour given by \eqref{Phiasympt}). We will not discuss massive scalars in this thesis but refer to\cite{Skenderis:2002wp} for more information.\\
\linebreak
For $d=3$ the boundary expansion reads
\begin{equation}
\phi_1=0,\quad\quad\phi_2=-\frac{1}{2}\ddot{J},
\end{equation}
with $\phi_3$ unconstrained. 
We can now evaluate the on-shell action (for the scalar field) with the cutoff $\epsilon$ in the $z$ coordinate. For $d=3$, after carrying out the $z$ integral, we obtain that
\begin{equation}
S=\frac{1}{2\kappa^2}\int \rd^3x\left( \frac{1}{2}L^2\dot{J}^2\epsilon^{-1}+\ldots\right),
\end{equation}
where $\ldots$ are finite terms. As we can see, this action diverges when $\epsilon\rightarrow0$ so we must add counterterms to the action. It turns out that it is enough to add a term of the form
\begin{equation}
S_{ct}=\frac{L}{2\kappa^2}\int_{z=\epsilon} \rd^3x\sqrt{-\gamma}\frac{1}{2}\partial_i\phi\gamma^{ij}\partial_j\phi,
\end{equation}
where $\gamma_{ij}$ is the induced metric at $z=\epsilon$. The resulting renormalized on-shell action, in the limit $\epsilon\rightarrow0$, is then
\begin{equation}
S_{ren}=S+S_{ct}=\frac{1}{2\kappa^2}\left[-\int \rd^4x \sqrt{-g}\frac{1}{2}\partial_\mu\phi g^{\mu\nu}\partial_\nu\phi+L\int_{z=\epsilon} \rd^3x\sqrt{-\gamma}\frac{1}{2}\partial_i\phi\gamma^{ij}\partial_j\phi\right].
\end{equation}
The variation of the renormalized on-shell action with respect to $\phi$ can now be evaluated, by using the equations of motion and integration by parts, as
\begin{align}
\delta S_{ren, on-shell}=&\frac{1}{2\kappa^2}\int_{z=\epsilon} \rd^3x\left(\delta\phi\sqrt{-g}g^{zz}\partial_z \phi-L\delta\phi \partial_i(\sqrt{-\gamma}\gamma^{ij}\partial_j \phi)\right)\nonumber\\
=&\frac{1}{2\kappa^2}\int_{z=\epsilon} \rd^3x\delta\phi\left(3L^2\phi_3+\ldots\right)
\end{align}
which results in the 1-point function
\begin{equation}
\langle\mathcal{O}\rangle=\frac{3L^2}{16\pi G}\phi_3.\label{vevd3}
\end{equation}
In practice, this complicated method of adding counterterms is only relevant when the source of the field is non-zero, because all the extra contributions (including the contributions from the finite counterterms) vanish when the source is turned off. Remember also that for even $d$ the contributions from finite counterterms are scheme dependent, so in these cases the result when the source is turned on is ambiguous anyway. This is why we will not always carry out this procedure and instead just consider boundary observables after the relevant time-dependent sources have been turned off, which will be sufficient for answering many of the physical questions we are interested in. The general result for a massless scalar field (also when including backreaction) for odd $d$, as well as for even $d$ in the absence of sources, is given by
\begin{equation}
\langle\mathcal{O}\rangle=\frac{L^{d-1}d}{16\pi G}\phi_d.
\end{equation}
We will look at both $d=3$ and $d=4$ in more detail, including backreaction, in Section \ref{hardwallmodel}.

\section{Confining holographic models}\label{sec_conf}
Since confining holographic models will play a prominent role in Chapter \ref{confinement}, we will here review the basic properties of confining theories from a holographic viewpoint. The most well known example of a confining quantum field theory is quantum chromodynamics (QCD), the theory of the strong interaction in nuclear physics. The intuitive picture of confinement is that some elementary particles of the theory can not be separated by an arbitrary large distance (they are {\it confined} to stay close to each other). In the case of QCD, these elementary particles are the quarks (and gluons), and we will call them quarks from now on. It is instead more energetically favorable for the theory to spontaneously create new quarks and form bound states. In other words, the potential energy between a quark and an antiquark will grow indefinitely with distance and thus it would require an infinite amount of energy to separate the two quarks. We will restrict the definition of confinement to a {\it linearly} growing potential, although in principle the growth of the potential could have any monotonic behaviour.\\
\subsection{Confinement in field theory}
Confinement can be defined in terms of the Wilson loop operator. Given a non-abelian gauge theory, the Wilson loop is a gauge invariant observable. Moreover, the space of Wilson loops spans the whole space of gauge invariant observables. The Wilson loop is defined as
\begin{equation}
W(C)=\textrm{Tr} \left(\mathcal{P}\textrm{exp}\ i\oint_C A_\mu \rd x^\mu\right),
\end{equation}
where Tr is the trace, $\mathcal{P}$ means path-ordering, $C$ is a closed loop and $A_\mu$ is the gauge field. If we consider the field theory in Euclidean signature, and choose the curve $C$ to be a rectangular loop with length $R$ in a spatial direction and length $T$ in the time direction, the expectation value of the Wilson loop can be related to the quark-antiquark potential $V(R)$ of two stationary quarks separated by a distance $R$ as
\begin{equation}
W(C)=\textrm{Tr} \left(\mathcal{P}\textrm{exp}\ i\oint_\square A_\mu \rd x^\mu\right)\sim e^{-T V(R)}.
\end{equation}
Thus wanting the quark-antiquark potential to grow linearly for large distance $R$ is equivalent to saying that the logarithm of the rectangular Wilson loop grows linearly with area.

\subsection{Confinement in holography}
The Wilson loop operator can be computed using the AdS/CFT correspondence by using an extremal surface in the bulk (see \cite{Maldacena:1998im} for a derivation and a more rigorous treatment). We first define a two-dimensional surface $\mathcal{A}$ in the bulk spacetime, which is attached to the boundary such that $\partial \mathcal{A}=C$, where $C$ is the Wilson loop on the boundary. The surface should also be a minimal surface, meaning that it extremizes the Nambu-Goto action
\begin{equation}
S=\frac{1}{2\pi\alpha'}\int_{\mathcal{A}} \sqrt{|\textrm{det}(h_{ab})|},\label{wilsonloopaction}
\end{equation}
where $h_{ab}$ is the induced metric on the worldsheet surface. The Wilson loop is then given by the exponential of (minus) the Nambu-Goto action. Note that this action diverges and must be regularized. The divergence can be interpreted as coming from the rest mass of the two quarks, which in this approximation is infinite (see \cite{Berenstein:1998ij} for more information on this point). We will thus use the regularization scheme where we subtract the masses of two {\it isolated} quarks as computed in AdS by two strings hanging straight into the bulk.\\
\begin{figure}[t]
 \includegraphics[scale=0.6]{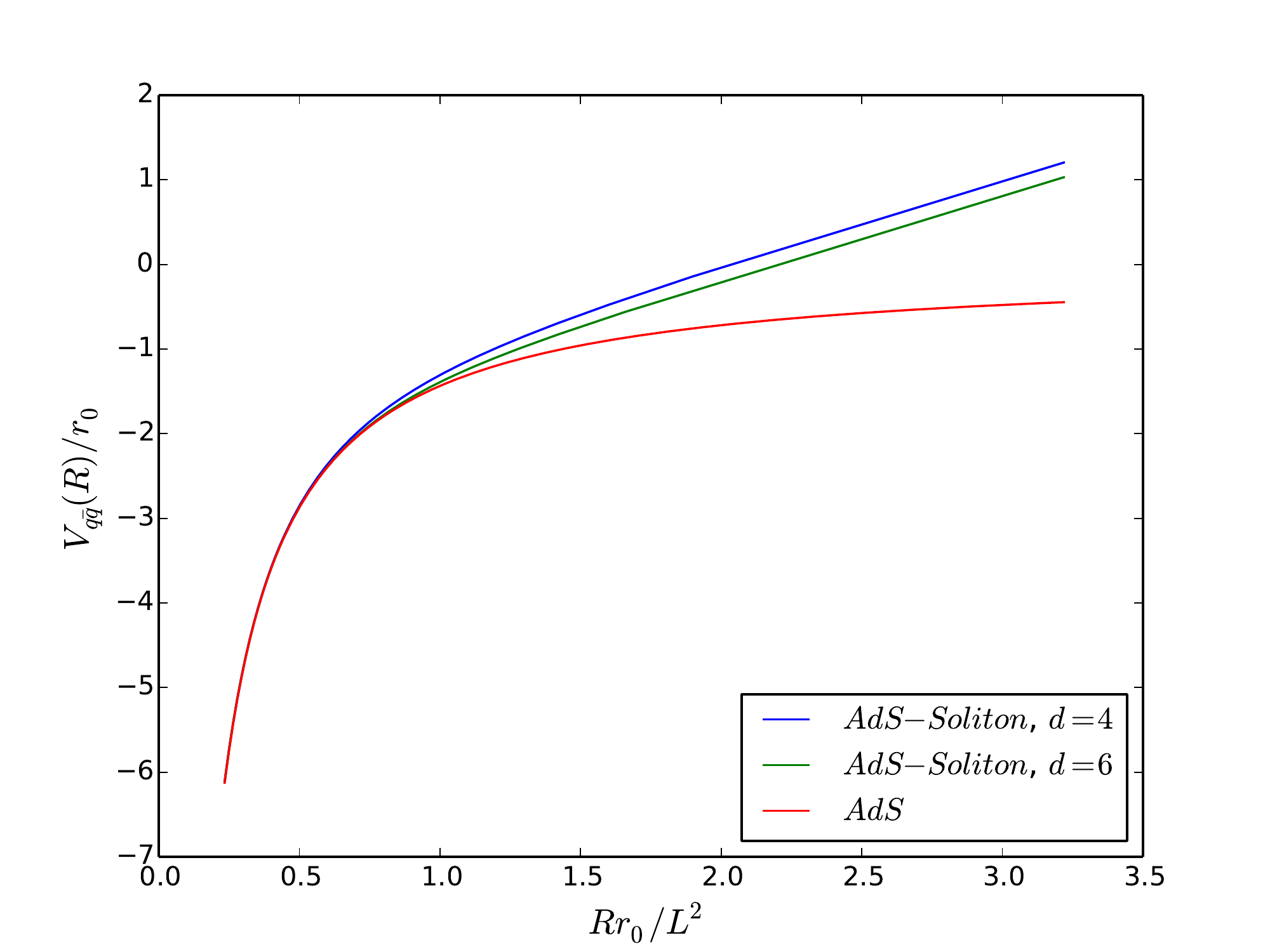}
 \caption{\label{qqpot} The quark-antiquark potential for a confining model (AdS Soliton) compared to that of AdS. For the confining model the potential grows linearly for large separation, while for small separation it coincides with that of AdS.}
\end{figure}
\linebreak
We will now present an argument for when a holographic model is confining. Consider an asymptotically AdS geometry with some radial coordinate $r$ that goes to $\infty$ at the boundary. Now assume that the geometry ends at some coordinate $r_0$, or in other words there is some impenetrable obstruction beyond which nothing can propagate. If we now consider a static string hanging down in this geometry, for large boundary separation $R$, the string will fall along the obstruction at $r=r_0$ and this will give a linear contribution to the potential. Thus for large separate $R$, for a diagonal metric, we expect that the potential will grow as
\begin{equation}
V(R)\sim R\frac{1}{2\pi\alpha'}\sqrt{|g_{tt}g_{xx}|},
\end{equation}
where $x$ is the spatial direction of the rectangular Wilson loop and $t$ is the time direction. This can be shown by parametrizing the surface by the coordinates $x$ and $t$ and using \eqref{wilsonloopaction}. Note that in a black brane background, the string would hang along the horizon, but $g_{tt}\rightarrow0$ at the horizon and thus the geometry is not confining, as expected. Before carrying out the actual computations, let us first introduce two confining holographic models. The simplest confining model one can think of is the so-called {\it hard wall model}. This spacetime is defined by just taking AdS (in Poincar\'e coordinates) and inserting a hard cutoff at some position $z_c$. This is a quite ad hoc way of obtaining the confining property and this model can not be embedded in string theory, thus one can question how realistic the results can be. Moreover, the setup is not well defined unless some boundary conditions are imposed on the hard wall. These boundary conditions would also affect how one computes Wilson loops, and we will thus not consider this geometry in this section. In Chapter \ref{confinement} we will study dynamics in the hard wall model.\\
\linebreak
Another confining model is the so-called AdS Soliton. It is given by the metric
\begin{equation}
ds^2=\frac{r^2}{L^2}(-f(r)\rd t^2+\rd\vec{x}^2_{d-2})+f(r)^{-1}\rd\theta^2+\frac{L^2}{r^2}\rd r^2,\label{solitonmetric0}
\end{equation}
where $f(r)=1-(r/r_0)^{d}$. This metric can be obtained via a double Wick rotation from an AdS black brane. It has a smooth cutoff at $r=r_0$ and the geometry can be thought of as a cigar. To avoid a conical singularity the coordinate $\theta$ must be periodic. This geometry can be embedded in string theory \cite{Witten:1998zw}, and no additional ad hoc boundary conditions have to be imposed at $r=r_0$. The field theory dual of this model is believed to be a non-supersymmetric $SU(N)$ gauge theory at strong coupling and large $N$. This duality can also be derived from string theory \cite{Witten:1998zw}, and thus this model has a much more solid theoretical foundation than for instance the hard wall model. We will now briefly sketch how this duality can be derived for $d=6$. One starts with M-theory, which is a hypothetical (quantum) theory of which 11-dimensional supergravity is the classical low energy limit. In M-theory there are objects known as M5-branes, whose existence are supported by for instance 5-brane solutions of 11-dimensional supergravity. By considering the low energy limit of a stack of M5-branes and using similar techniques as in Section \ref{stringtheory}, one can derive a duality between a certain exotic six-dimensional supersymmetric theory with $(2,0)$ supersymmetry and M-theory in AdS$_7\times S^4$ \cite{Maldacena:1997re}. After compactifying first on a circle with supersymmetry preserving boundary conditions (antiperiodic fermions), we obtain a five-dimensional supersymmetric Yang-Mills theory which is dual to type IIA string theory on a background related to AdS$_6\times S^4$ by a Weyl transformation \cite{Kanitscheider:2008kd} (the M5-branes in M-theory become D4-branes in type IIA string theory). After this we compactify on another circle with radius $R_2$ and with supersymmetry breaking boundary conditions (periodic fermions). The breaking of supersymmetry results in the fermions attaining masses of order $\sim1/R_2$, and it is believed that also the scalars receive masses from quantum corrections. Thus we are left with a pure non-supersymmetric $SU(N)$ gauge theory, and the dual spacetime can be written in the form \eqref{solitonmetric0} with $d=6$.\\
\linebreak
We will now compute the quark-antiquark potential in this geometry and compare it to the result of AdS. We will start directly with the AdS Soliton, and the result for AdS can then be obtained by setting $r_0=\infty$.\\
\linebreak
We will parametrize the surface $\mathcal{A}$ by $t$ and the coordinate $x$, thus the action \eqref{wilsonloopaction} becomes
\begin{equation}
S=2T\int_{r_m}^{r_c}\sqrt{f(r)^{-1}+\frac{r^4}{L^4}(x')^2}\rd r.\label{wilsonactionsol}
\end{equation}
where $r_m$ is the minimal value of $r$ on the surface $\mathcal{A}$ and $'$ denotes derivative with respect to $r$, and we have for simplicity set $2\pi\alpha'=1$. We have also regularized the action by integrating up to a maximal value $r_c$ instead of infinity. Since the action is independent of $x$, Noether's theorem gives that the canonical momentum $p_x=\partial S/\partial x'$ is constant, and from this we can solve for $x'$ as
\begin{equation}
(x')^2=\left(\frac{L}{r}\right)^4f(r)^{-1}\frac{1}{\left(\frac{r}{r_m}\right)^4-1}.
\end{equation}
The value of the action is then
\begin{equation}
S=2T\int_{r_m}^{r_c}\frac{r^2}{r_m^2\sqrt{f(r)\left(\left(\frac{r}{r_m}\right)^4-1\right)}}\rd r.
\end{equation}
\begin{figure}[t]
 \includegraphics[scale=0.7]{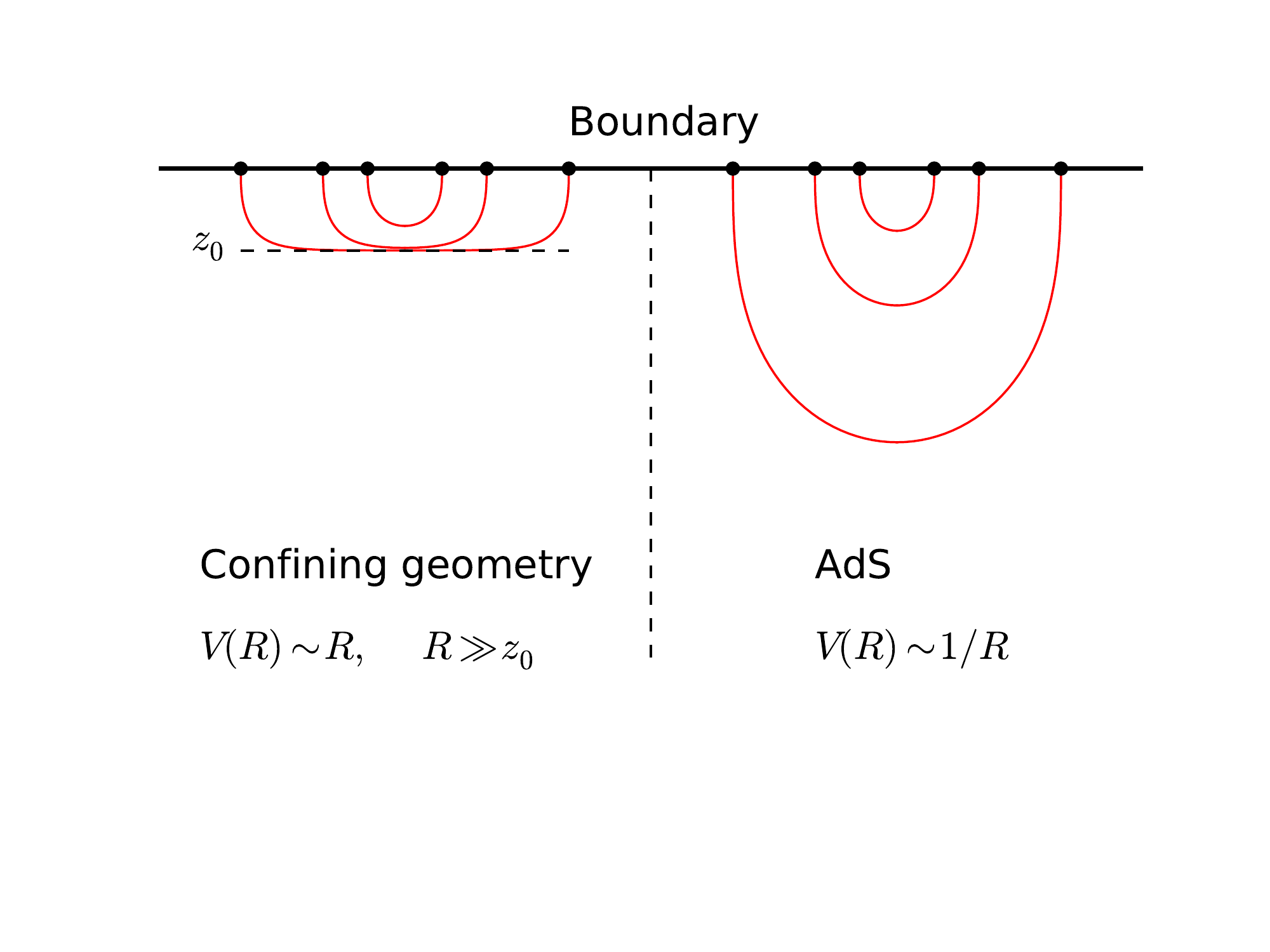}
 \caption{\label{wlillustr} Illustration of the quark-antiquark potential computation in AdS/CFT for two different boundary intervals, both in AdS and in the confining AdS Soliton model \eqref{solitonmetric0} with parameter $z_0=L^2/r_0$. }
\end{figure}
To obtain a finite answer when $r_c\rightarrow\infty$, we will subtract the action of two isolated quarks, computed as two worldsheets that hang straight into the bulk in AdS. This might seem like an arbitrary regularization scheme, but for our purposes of just illustrating the confining property it is enough. The final regularized action is thus
\begin{equation}
S_{reg}=\lim_{r_c\rightarrow \infty} (S-2T\int_0^{r_c}\rd r).
\end{equation}
This will give us the Wilson loop as a function of $r_m$. To obtain it as a function of $R$, we have to invert the relation
\begin{equation}
R=2\int_{r_m}^\infty x' \rd r=2\int_{r_m}^\infty x' \rd r=2\int_{r_m}^\infty\frac{L^2}{r^2\sqrt{f(r)\left(\left(\frac{r}{r_m}\right)^4-1\right)}}\rd r,
\end{equation}
which has to be done numerically. Let us now consider empty AdS, obtained by setting $f(r)\rightarrow1$. In this case we can compute $R$ analytically as
\begin{equation}
R=\frac{2L^2}{r_m}\int_1^{\infty}\frac{1}{y^2\sqrt{y^4-1}}\rd y=\frac{2\sqrt{\pi}L^2\Gamma(\frac{3}{4})}{r_m\Gamma(\frac{1}{4})}.
\end{equation}
It is also possible to compute $S_{reg}$, and we obtain the result
\begin{align}
S_{reg}=&2Tr_m\left[\int_{1}^{\infty}\left(\frac{y^2}{\sqrt{y^4-1}}-1\right)\rd y-1\right]=Tr_m\frac{\sqrt{\pi}\Gamma(-\frac 1 4)}{2\Gamma(\frac 1 4)}\nonumber\\
=&-T\frac{8\pi^3L^2}{\Gamma(\frac 1 4)^4R},
\end{align}
from which we see that the potential is behaving like a Coulomb potential. In Figure \ref{qqpot} we show the behaviour of the quark-antiquark potential in AdS and in the AdS Soliton. We see indeed that for small separations they concide, while for larger separations the potential grows linearly for the AdS Soliton. The linear growth for large separation is independent of dimension and equal to $r_0^2 R/L^2$. This can also be inferred analytically from \eqref{wilsonactionsol} by considering a part of the worldsheet that stretches along $r=r_0$ with $\rd r/\rd x=0$ which is expected to be the dominant contribution for large $R$. An illustration of the dual picture of the quark-antiquark potential is shown in Figure \ref{wlillustr}.


\section{Dynamics and holographic thermalization}\label{timedep}
\begin{figure}[t]
\resizebox{15cm}{!}{
\begin{tikzpicture}
\def\s{0}
    \def\y{-2}

      \def\a{4}
      \draw[color=black] ({\s},{\y})--({10},{\y});
    \draw[color=black] ({\s},{\y})--({\s},{\y+4});
    \draw[color=black,dashed] ({\s+1},{\y+1})--({\s+2},{\y+1}) node[right] {$J(t)$};
    \draw[color=black] ({\s+1},{\y+.5})--({\s+2},{\y+.5}) node[right] {$\langle \mathcal{O}' \rangle$};
    
    \draw[color=black] ({0},{\y+2})--({10},{\y+2}) node[right] {};
    \node[rotate=90] at (-.5,\y+3) {Gravity};
    \node[rotate=90] at (-.5,\y+1) {Field Theory};
    \draw[color=black,dashed] ({\s+3},{\y+2})--({\s+3},{\y+4});
    \draw[color=black,dashed] ({\s+6},{\y+2})--({\s+6},{\y+4});
    
    \node at (1.5,\y+3) {$g_{\mu\nu}=g_{\mu\nu}^{(AdS)}$};
    \node at (1,\y+2.5) {$\phi=0$};

    \node at (4,\y+3) {$g_{\mu\nu}(t)$};
    \node at (4,\y+2.5) {$\phi(t)\neq0$};
    
    \node at (7.5,\y+3) {$g_{\mu\nu}\approx g_{\mu\nu}^{(BH)}$};
    \node at (7,\y+2.5) {$\phi\approx0$};

     \draw[domain=0:8,samples=200,dashed,color=black] plot (\x,{\y+exp(-(\x-\a)*(\x-\a)*10)}) node[left] {};
     
     \draw[domain=2:3,samples=200,color=black] plot (\x,{\y+(10*(\x-\a)^2*exp(-(\x-\a)*(\x-\a)*10)+10*(\x-\a-1.2)^3*exp(-(\x-\a-.9)*(\x-\a-1.4)*10)+1.5*(1/2+tanh((\x-\a))/2)*0)}) node[left] {};
     \draw[domain=3:7,samples=200,color=black] plot (\x,{\y+(10*(\x-\a)^2*exp(-(\x-\a)*(\x-\a)*10)+10*(\x-\a-1.2)^3*exp(-(\x-\a-.9)*(\x-\a-1.4)*10)+1.5*(1/2+tanh((\x-\a))/2)*(1/2+tanh((\x-\a)*2.4)/2))}) node[left] {};
     \draw[domain=7:10,samples=200,color=black] plot (\x,{\y+(10*(\x-\a)^2*exp(-(\x-\a)*(\x-\a)*10)+10*(\x-\a-1.2)^3*exp(-(\x-\a-.9)*(\x-\a-1.4)*10)+1.5*(1/2+tanh((\x-\a))/2))}) node[left] {};
     
   
     \draw [decorate,decoration={brace,amplitude=5pt},xshift=0pt,yshift=-4pt]
 ({\a+0.4},{\y}) -- ({\a-0.4},{\y}) node [black,midway,yshift=-15pt] {\footnotesize
 $\delta t$};
 \draw [decorate,decoration={brace,amplitude=5pt},xshift=0pt,yshift=-4pt]
 ({\a+6},{\y}) -- ({\a+2.45},{\y}) node [black,midway,yshift=-15pt] {\tiny
Linear regime/approach to equilibrium};
 
 \baselineskip=6pt
 \draw [decorate,decoration={brace,amplitude=5pt},xshift=0pt,yshift=-4pt]
 ({\a+2.4},{\y}) -- ({\a+0.45},{\y}) node [black,midway,yshift=-15pt,align=left] {\tiny
 Complicated\\\tiny dynamics};
 
 \draw[->] (1,{\y-.3})--(2,{\y-.3}) node[black,midway,yshift=-8pt] {$t$};

\end{tikzpicture}
}
\caption{\label{thermsketch} Schematic illustration of how a thermalization process in a quantum field theory with a gravity dual, after quenching some operator $J$, could look like. The upper panel shows the dual gravity picture, where some matter field $\phi$ is sourced at the boundary and then collapses to a black hole.}
\end{figure}
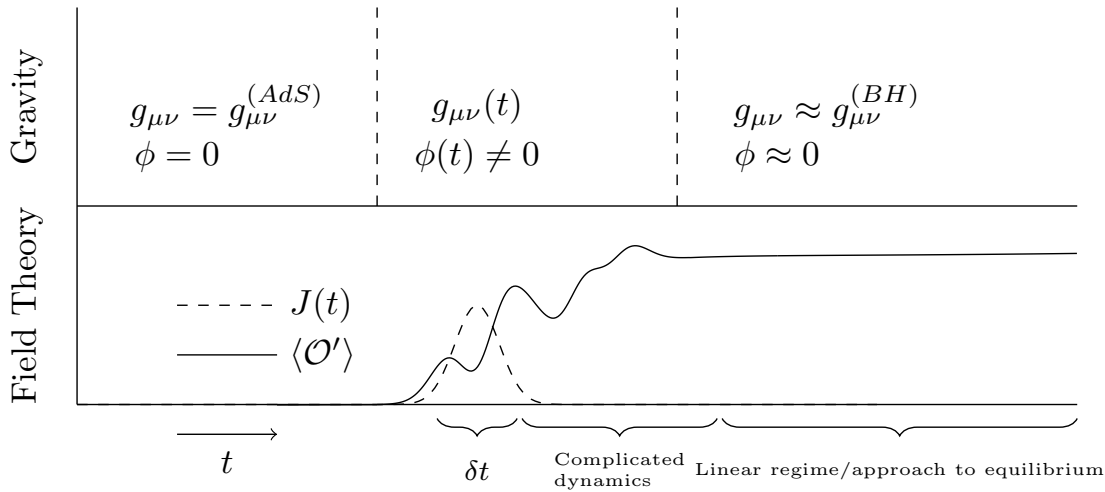

In this section we will discuss dynamical processes in AdS/CFT, in particular the process of matter collapsing to form a black hole after injecting energy at the boundary of AdS. Since a black hole is dual to a thermal state in the field theory, the formation of a black hole solution in the bulk is dual to a thermalization process on the boundary. Such a process is thus called a {\it holographic thermalization} process. We will now describe the particular mechanism for triggering a thermalization process that we will focus on.\\
\linebreak
Consider a quantum field theory governed by some Lagrangian $\mathcal{L}$, prepared in the vacuum state $|\Psi\rangle$. We now couple the Lagrangian to a source $J$ via some operator $\mathcal{O}$, as $\mathcal{L}\rightarrow \mathcal{L}+J\mathcal{O}$. The source is then turned on for a brief period of time $J\rightarrow J(t)$. The shape of $J(t)$ will be that of a short pulse (typically in the form of a gaussian function), and we will refer to this procedure as a {\it quench}. Note that the source $J$ is viewed as a classical parameter, or in other words it is a function taking values in $\R$, while $\mathcal{O}$ is a quantum mechanical operator. This procedure will put the system out of equilibrium, and note also that the time dependent source breaks the time translation invariance of the Lagrangian. Thus energy is not conserved, and a reasonable physical assumption is that energy is injected into the system. The system will then undergo some non-trivial dynamical evolution, but at some later time it will typically equilibrate to a static state. This static state, although being a pure state, will generically behave like a thermal state with some temperature $T$. This fits well with the intuition that if we inject energy into a physical system, the temperature will rise. This equilibration process to a thermal state is what we call {\it thermalization}, and it can be probed by looking at expectation values of different operators. This process is illustrated in Figure \ref{thermsketch}. \\
\linebreak
Now let us consider what this is dual to on the gravity side. The initial state will be some static solution of Einstein's equations (in the simplest example, it will just be AdS), which is dual to the boundary vacuum state $|\Psi\rangle$. The operator $\mathcal{O}$ will be dual to some field $\phi$ in the bulk. As an illustrative example, we can take a massless scalar field $\phi$ which is thus dual to a dimension $d$ operator on the boundary. According to the AdS/CFT dictionary, the asymptotic behaviour at the boundary is given by
\begin{equation}
\phi(t,z)=J(t)+\ldots+\phi_d z^d+\ldots,
\end{equation}
where $\phi_d$ is related to the vacuum expectation value of the dual operator $\mathcal{O}$. Thus quenching the source $J$ coupled to the operator $\mathcal{O}$ is dual to perturbing the boundary condition of $\phi$. This perturbation will then propagate into the bulk with dynamics governed by Einstein's equations (plus the equations of motion for the matter fields). Thermalization of the dual field theory is then dual to the statement that the field $\phi$ collapses to form a black hole, which happens in many situations. The temperature of the final state is then the Hawking temperature of the final black hole.\\
\linebreak
To model this process in full generality, one will usually have to solve Einstein's equations numerically. For setups with a large number of symmetries (namely where the fields only depend on time and the radial bulk coordinate $z$), this is a relatively manageable problem since the full dynamics is completely specified by the {\it constraints} of Einstein's equations and the partial differential equations are only two-dimensional. However, for setups without translational symmetry, this problem becomes much more difficult and requires much more sophisticated methods than what we discuss in this thesis (see for instance \cite{Lehner:2001wq,Chesler:2013lia} for discussions on more advanced techniques in numerical relativity). We will discuss numerical methods in Section \ref{ssec_num}. We will also discuss two other special cases that allow us to make progress analytically. One is to consider a certain limit where the source that is turned on is weak (or equivalently when the width of the source goes to zero), for which one can make progress using perturbative methods. This also allows us to make some progress also in the case without translational symmetry. The other is to consider the width of the source being strictly equal to zero (namely a delta function in time), for which the matter content in the bulk takes the form of an infinitesimally thin shell of matter. 

\subsection{Quasinormal modes of black holes}\label{sec_qnm}
One concept that is very relevant for that of dynamics and black hole formation are the so-called {\it quasinormal modes} of a black hole or black brane. These are the discrete frequencies one can obtain by solving for an oscillating perturbation in a black brane background. Typically we would thus assume that the metric (or in principle some other field) is perturbed as $g_{\mu\nu}=g_{\mu\nu}^{BH}+\epsilon e^{i\omega t} h_{\mu\nu}$ where $h_{\mu\nu}$ is a metric independent of time and $\epsilon$ is a small parameter. One then solves the Einstein equations for this perturbation by enforcing infalling boundary conditions at the horizon, and this will result in a discrete set of complex frequencies $\omega_i$. Thus the perturbations will be oscillations modulated with a decaying exponential. The physical mechanism for this decay is dissipation resulting from energy falling behind the event horizon. Any small perturbation around a black hole solution can then be expanded in a basis of the quasi normal modes, and for late times the slowest decaying (lowest lying) quasi normal mode will be the most long lived (assuming a generic perturbation which has a non-zero overlap with the lowest mode). Thus it is expected that the decay of any perturbation on a black brane background will be governed by the lowest lying quasi normal mode for late times, which can indeed be seen for the results of this section, as is illustrated in for instance Figure \ref{BMvsqnm}. We will not compute the frequencies here but instead just quote the result which was obtained in for instance \cite{Horowitz:1999jd}. For AdS black branes, the quasinormal modes are always proportional to the temperature and the lowest values (of the imaginary part, which sets the decay constant) are $\omega_0=11.16T$, $\omega_0=8.63T$, $\omega_0=5.47T$ for $d=3$, $d=4$ and $d=6$ respectively.

\subsection{Numerical treatment}\label{ssec_num}
\begin{figure}
 \includegraphics[scale=.7]{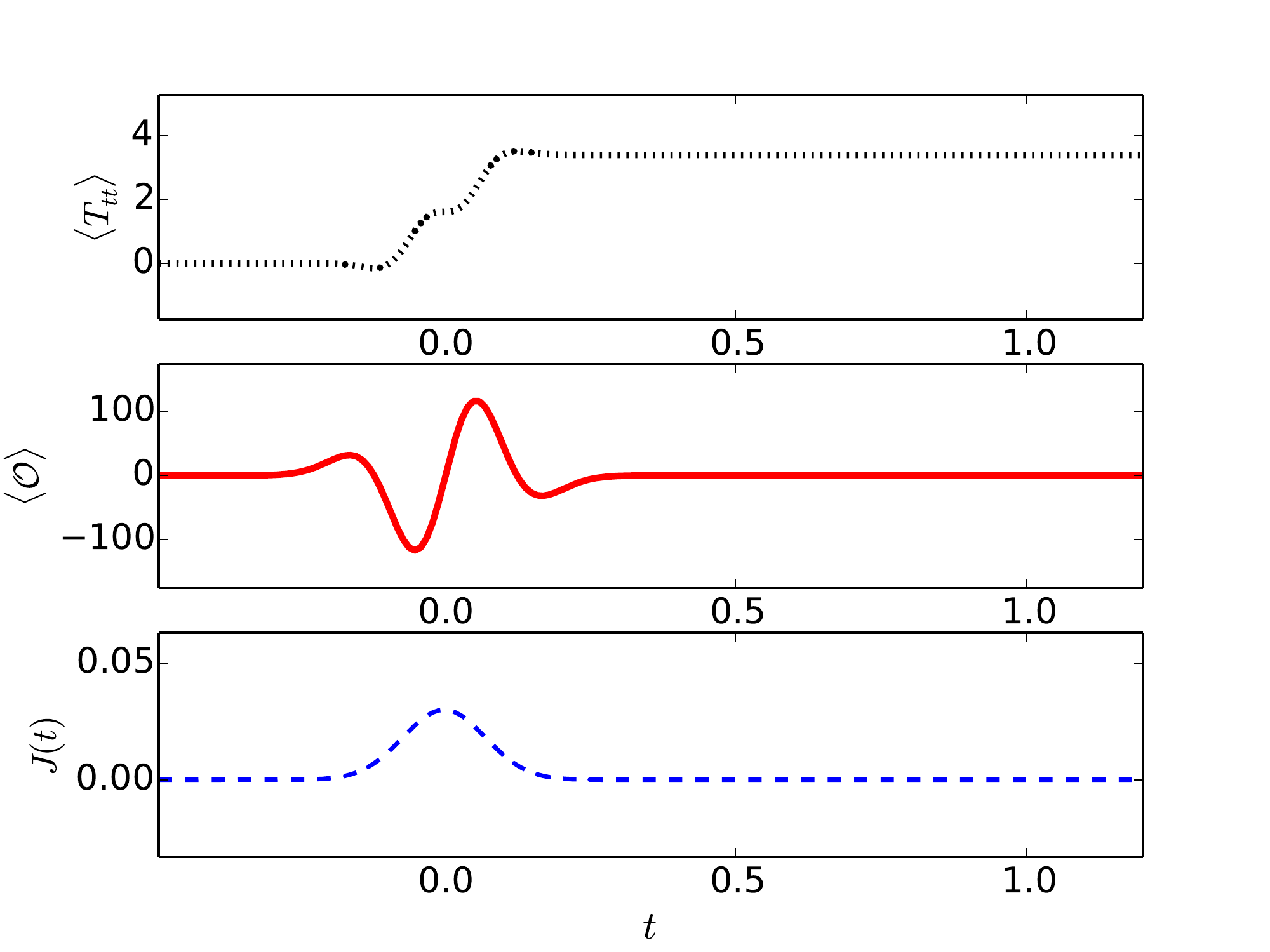}
 \caption{\label{adsnumbh}Evolution of field theory observables when a black hole forms after quenching AdS$_4$. The source is given by \eqref{numsource} with parameters $\epsilon=0.03$ and $\delta t=0.1$. The expectation values are in units of $L^2/16\pi G$.}
\end{figure}

In this section we will use numerical methods to model black brane formation resulting from quenching the boundary condition of a bulk field. We will only consider the case where the setup depends only on time and the radial bulk coordinate. This is dual to a field theory setup which is translationally invariant and only depends on time and the energy will be injected via a massless scalar field minimally coupled to gravity. The initial state will be taken as empty AdS (but see Chapter \ref{confinement} where we will investigate what happens when the initial state is a confining solution). The action we will consider is thus
\begin{equation}
S=\frac{1}{2\kappa^2}\int \rd^{d+1}x\sqrt{-g}\left(R-2\Lambda-\frac{1}{2}(\partial\phi)^2\right).
\end{equation}
Here $(\partial\phi)^2$ is shorthand notation for $\partial_\mu\phi g^{\mu\nu}\partial_\nu\phi$. The equations of motion are the Einstein's equations
\begin{equation}
R_{\mu\nu}-\frac{1}{2}R g_{\mu\nu}+\Lambda g_{\mu\nu}=\frac{1}{2}\partial_\mu\phi\partial_\nu\phi-\frac{1}{4}(\partial\phi)^2g_{\mu\nu},
\end{equation}
and the wave equation for the scalar field
\begin{equation}
\partial_\mu(\sqrt{-g}g^{\mu\nu}\partial_\nu\phi)=0.
\end{equation}
The cosmological constant $\Lambda$ is related to the AdS radius by $\Lambda=-\frac{d(d-1)}{L^2}$. It can be shown that by using our symmetry restrictions, the metric can be brought to the form
\begin{equation}
\rd s^2=\frac{L^2}{z^2}\left(-f(z,t)e^{-2\delta}\rd t^2+\frac{\rd z^2}{f(z,t)}+\rd\vec{x}^2\right),\label{adsnummetric}
\end{equation}
where $f$ and $\delta$ are functions to be determined by the equations of motion. At the initial conditions, where the metric equals that of AdS, we thus have $f=1$ and $\delta=0$. Moreover, the metric \eqref{adsnummetric} has a residual gauge invariance corresponding a shift in $\delta$ by $\delta(z,t)\rightarrow\delta(z,t)+p(t)$. We will fix this gauge freedom by requiring that $\delta(0,t)=0$, so that $t$ will coincide with the boundary time. We will denote derivative with respect to $z$ by $'$ and use $\dot{}$ to denote derivative with respect to $t$. In some applications of numerical holography it is more useful to use so-called {\it infalling} coordinates (see for instance \cite{Chesler:2013lia}), where one uses a infalling time coordinate $v$ that parametrizes a null-direction such that infalling lightrays move along constant $v$. However, for our setup where a black brane is forming from a time dependent process in AdS, such coordinates are not as convenient as the type of time coordinates we use here. This choice of coordinates are even more crucial in Chapter \ref{confinement} where there will be solutions including both ingoing and outgoing waves of matter and it would not make any sense to single out only infalling waves by working in infalling coordinates. However, we will use infalling coordinates in \ref{ssec_weak} where we study black brane formation using perturbative methods.\\
\linebreak
Defining $\Phi=\phi'$ and $\Pi=e^{\delta}\dot{\phi}/f$, the equations of motion take the form
\begin{subequations}
\begin{align}
\dot{\Phi}&=(fe^{-\delta}\Pi)',\quad\quad \quad \dot{\Pi}=z^{d-1}\left(\frac{fe^{-\delta}\Phi}{z^{d-1}}\right)',\label{adsnumscal}\\
\dot{f}&=\frac{z}{d-1}f^2e^{-\delta}\Phi\Pi,\quad\quad \delta'=\frac{z}{2(d-1)}(\Phi^2+\Pi^2),\label{adsnumdyn}\\
&\quad f'=f\delta'+\frac{d}{z}(f-1).\label{adsnumconstr}
\end{align}
\end{subequations}
We will use \eqref{adsnumscal} and \eqref{adsnumdyn} to evolve our system, while \eqref{adsnumconstr} is a constraint. \eqref{adsnumconstr} follows from the other equations, and can thus be followed during the evolution to check the quality of the numerics. Note the confusing double usage of the word ``constraint equations''. From our point of view, equations \eqref{adsnumscal} and \eqref{adsnumdyn} are dynamical equations used to evolve our system in time, while \eqref{adsnumconstr} is a constraint equations which only constrains the initial data. However, from the point of view of the Hamiltonian formulation (or commonly known as the ADM formalism\cite{Arnowitt:1959ah,Misner:1974qy}) of general relativity, they are all constraint equations arising from diffeomorphism invariance of the theory (recall that in Hamiltonian formulations of dynamical systems, each gauge symmetry corresponds to a constraint on phase space). 
Note that the equations for the metric \eqref{adsnumdyn} and \eqref{adsnumconstr} are all first order in time, so they can indeed be interpreted as constraints on the metric components and the scalar field and their respective canonical momenta. The dynamical equations in general relativity are generically second order in both space and time and will for instance be relevant in Section \ref{adssoliton} where we have less symmetry.\\
\linebreak
The dynamical evolution can now be triggered by quenching the source for the scalar, as explained in the beginning of this section. We will now for simplicity focus on the special case of $d=3$. The boundary expansion of the scalar field can be computed from the equations of motion as
\begin{equation}
\phi(z,t)=J-\frac{1}{2}\ddot{J}z^2+\phi_3 z^3+O(z^4),
\end{equation}
where $J(t)$ is the dynamical source. The boundary expansion of the metric coefficients are
\begin{equation}
f=1-z^2\frac{\dot{J}^2}{4}+z^3f_3+O(z^4),
\end{equation}
\begin{equation}
\delta=\frac{\dot{J}^2}{8}z^2+O(z^4).
\end{equation}
$\phi_3$ and $f_3$ are determined by solving the full non-linear equations, and are related to the vacuum expectation value of the dual operator and the boundary stress energy tensor as
\begin{equation}
\langle\mathcal{O}\rangle=\frac{3L^2}{16\pi G}\phi_3,\quad\quad\langle T_{tt}\rangle=2\langle T_{xx} \rangle=2\langle T_{yy} \rangle = \frac{L^{2}}{8\pi G}f_3.
\end{equation}
This result follows from the holographic renormalization procedure explained in Section \ref{sec_holrenorm}. The formula for the stress-energy tensor \eqref{Tijformul} remains valid even when there is a massless scalar field present, and the formula for the expectation value \eqref{vevd3} computed in the probe limit is also still valid. Even if it is easy to convince oneself that this is true, we will prove it explicitly in Section \ref{hardwallmodel} when we consider both $d=3$ and $d=4$ in detail. For simplicity we will consider a source on the form
\begin{equation}
J(t)=\epsilon e^{-\frac{t^2}{\delta t^2}},\label{numsource}
\end{equation}
which corresponds to a localized pulse with approximate width $\delta t$ in time. Note that since we are working with $d=3$, there are no extra contributions from the source while it is turned on and we can ignore the counterterms. We will do the full holographic renormalization for $d=3$ and $d=4$ in Section \ref{hardwallmodel}. \\
\linebreak
The equations of motion can be solved by discretizing the radial coordinate $z$, by replacing it with a discrete set of points $z_i$. The partial differential equations will then turn into a system of ordinary differential equations in time for the functions $f$, $\Phi$ and $\Pi$ evaluated at the points $z_i$. Derivatives with respect to $z$ will be replaced by matrix multiplications, and we can use some standard ODE library for evolving the functions $f$, $\Phi$ and $\Pi$. The function $\delta$ can be obtained at every time step from \eqref{adsnumconstr}. Note that we could also have used equation \eqref{adsnumconstr} to solve for $f$ at every time step, but it turns out that evolving $f$ in time is more stable numerically.\\
\linebreak
There are two main methods for discretizing the radial direction. The simplest one is the {\it finite difference method}. In this method the grid points $z_i$ are located uniformly, meaning that $z_{i+1}-z_i=h$ for some small number $h$. For a generic function $F$, a derivative can then be approximated as
\begin{equation}
F'(z_i)\approx\frac{F(z_{i-1})-F(z_i)}{2h}.
\end{equation}
This is called a {\it second order symmetric} finite difference scheme, as the order of the error is $O(h^2)$. For more accurate results, one can use higher order schemes which use more points. For instance, the fourth order symmetric scheme looks like
\begin{equation}
F'(z_i)\approx\frac{\frac{1}{12}F(z_{i-2})-\frac{2}{3}F(z_{i-1})+\frac{2}{3}F(z_{i+1})-\frac{1}{12}F(z_{i+2})}{h}. 
\end{equation}
The derivatives can then be formulated as matrix multiplications acting on the vector $F_i\equiv F(z_i)$. Note that we have to restrict ourselves to a finite domain $(0,z_c)$ of $z$, and $z_c$ must be chosen adequately large. The numerical scheme will then only work until any degrees of freedom has propagated up to $z_c$. In particular, if we want to study formation of black branes, we must make sure that the event horizon of the final black brane is located inside the interval $(0,z_c)$.\\
\linebreak
The other numerical scheme that is widely used, and which we have used in this section, is called the {\it Chebyshev pseudospectral method}, which is a method based on a set of orthogonal polynomials called Chebyshev polynomials. To derive this scheme, consider a set of orthogonal polynomials $P_i(z)$ defined on the interval $(0,z_c)$. We can then expand a function $f(z)$ as
\begin{equation}
f(x)\approx\sum_{i=0}^N C_iP_i(z),
\end{equation}
where the expansion has been truncated at order $N$. Derivatives are then easy to compute as
\begin{equation}
f'(x)\approx\sum_{i=0}^N C_iP_i'(z).
\end{equation}
A spectral method would rely on reformulating the system as an algebraic equation for the coefficients $C_i$. However, for non-linear problems, this is not very convenient and the truncation in the series may pose problems. What one does instead is define a grid of $N+1$ points $z_i$, and then demand that the differential equation only holds on these $z_i$ points. It turns out that there exists an optimal set of $z_i$ that minimizes the error, for our problems given by
\begin{equation}
z_i=\frac{1+\cos\frac{i\pi}{N}}{2}.\quad i=0,\ldots,N.
\end{equation}
The derivatives, just like in the finite difference method, can then be formulated as a matrix acting on the vector $f_i=f(z_i)$ as $f'_i=D_i^{\;j}f_j$. For the Chebyshev pseudospectral method in our case it takes the form
\begin{equation}
D_i^{\;j}=\left\{\begin{array}{cc}
       \frac{2N^2+1}{3z_c}&i=j=0,\\
       \frac{-z_j}{z_c(1-z_j^2)}&0<i=j<N,\\
       \frac{c_i}{c_j}\frac{2(-1)^{i+j}}{z_c(z_i-z_j)}&i\neq j,\\
       -\frac{2N^2+1}{3z_c}&i=j=N,\\
      \end{array}\right.
\end{equation}
where $c_0=c_N=2$ and $c_i=1$ otherwise. The Chebyshev pseudospectral method can be viewed as a maximal order finite difference scheme and has an error that converges exponentially. However, the fact that the differentiation matrix uses all points on the grid to compute the derivative at one point can cause errors from one part to quickly propagate over the whole grid, and sometimes a finite difference method can be more stable. Moreover, since the differentiation matrices are not sparse, it may take longer time for a computer to evaluate the matrix multiplications (this is however usually overcome by the fact that a pseudospectral grid uses much fewer grid points than a finite difference scheme). Sometimes it may be difficult to impose the boundary conditions directly at $z=0$ and the diverging terms in the equations of motion may lead to numerical instabilities. One can then for example introduce a small cutoff $\epsilon$ and impose the boundary conditions at $z=\epsilon$ as was done in \cite{Wu:2012rib}. We have found that a different route is better, namely introducing a new variable by $\Phi(z,t)=z\tilde \Phi(z,t)$ and rewriting the equations of motion in terms of this varible. This turns out to be more stable numerically and the boundary conditions can now be imposed exactly at $z=0$ by setting $\tilde{\Phi}(0,t)=-\ddot{J}(t)$.\\
\linebreak
In Figure \ref{adsnumbh} we show the typical evolution of the field theory observables for $d=3$. The energy density goes to a constant and is then preserved in time (which is guaranteed by energy conservation). The expectation value of the scalar operator has some non-trivial behaviour when the source is turned on and then goes to zero. It typically goes to zero on a timescale set by the lowest quasinormal mode of the final black brane but it is difficult to resolve this decay due to numerical instabilities when the black brane forms.\\
\linebreak
In Chapter \ref{confinement}, we will use numerical techniques to study dynamics and holographic thermalization in confining holographic models.

\subsection{Weak field black hole formation}\label{ssec_weak}
In this section we will describe black hole formation in the weak field regime (meaning the amplitude of the source is small), where we can use perturbative methods to predict the time evolution. We will again assume that the matter content we use to inject energy into the system is that of a massless scalar field, with a source that is turned on during some small time $\delta t$. We will restrict attention to the case of black hole formation in AdS$_4$, although in principle the methods can be extended at least to any even dimension. We will first describe black hole formation from a homogeneous quench on the boundary, based mainly on \cite{Bhattacharyya:2009uu}, and then move on to study the setup where the boundary source also depends on a spatial coordinate, based on \cite{Balasubramanian:2013oga} and some unpublished results. We will for simplicity set $L=1$.
\subsubsection{Homogeneous black hole formation}
We will work in infalling coordinates, which is a different gauge than what was used in Section \ref{ssec_num}. In these coordinates, the metric ansatz can be taken as
\begin{equation}
\rd s^2=-g(v,r)\rd v^2+2\rd v\rd r+f(v,r)^2(\rd x^2+\rd y^2),
\end{equation}
and thus only depends on two free functions $f(r,v)$ and $g(r,v)$. In these coordinates, infalling lightrays (meaning lightrays that start at the boundary of AdS) travel along trajectories of constant $v$, which is why these coordinates are suitable for studying a {\it massless} scalar field coupled to gravity. The equations of motion can be shown to take the form
\begin{equation}
\partial_r(f^2g\partial_r\phi)+\partial_v(f^2\partial_r\phi)+\partial_r(f^2\partial_v\phi)=0,
\end{equation}
\begin{align}
(\partial_r\phi)^2&=-\frac{4\partial_r^2f}{f},\nonumber\\
\partial_r(fg\partial_rf&+2f\partial_vf)=3f^2.\label{eomweak}
\end{align}
The boundary conditions at $r\rightarrow\infty$ are
\begin{equation}
\phi=\epsilon J(v)+O(r^{-1}),\quad\quad g=r^2+O(1),\quad\quad f=r+O(r^{-1}),
\end{equation}
where $\epsilon$ is a small parameter enforcing that the amplitude of the source is small. The source will be turned on for a short period of time $\delta t$, and we will assume for simplicity that this window is in $0<t<\delta t$. We will now solve the equations of motion perturbatively in $\epsilon$, namely we expand the fields as
\begin{equation}
f=\sum_{i\geq0} f_{2i}\epsilon^{2i},\quad g=\sum_{i\geq0} g_{2i}\epsilon^{2i},\quad \phi=\sum_{i\geq0} \phi_{2i+1}\epsilon^{2i+1}.
\end{equation}
At first order in $\epsilon$, the equations of motion for the scalar field takes the form
\begin{equation}
\partial_r(r^4\partial_r\phi_1)+2r\partial_r(r\partial_v\phi_1)=0,
\end{equation}
which, with our boundary conditions, has the solution
\begin{equation}
\phi_1(v,r)=J(v)+\frac{\dot{J}(v)}{r}.
\end{equation}
At second order in $\epsilon$ there is a contribution to the metric. By solving \eqref{eomweak} we obtain 
\begin{equation}
f_2=-\frac{\dot{J}^2}{8r},\label{weakeqf2}
\end{equation}
\begin{equation}
g_2=-\frac{3}{4}\dot{J}^2+\frac{1}{2r}\dot{J}\ddot{J}-\frac{1}{2r}\int_{-\infty}^v\ddot{J}^2\rd t.\label{weakeqg2}
\end{equation}
Since we assume that the source $J$ vanishes for $t>\delta t$, it can be seen that the metric for times $t>\delta t$ at this order takes the form of a black brane with mass given by
\begin{equation}
M=\epsilon^2\frac{1}{2}\int_{-\infty}^\infty \ddot{J}^2\rd t.\label{weakeqM}
\end{equation}
At third order there is a correction to the scalar field, and the result is
\begin{equation}
\phi_3=\frac{1}{8r^3}\int_{-\infty}^v \dot{J}\left[-\int_{-\infty}^t\ddot{J}^2dt'+3\dot{J}\ddot{J}\right]\rd t. \label{phi3}
\end{equation}
Note that when the source is turned on, $\phi_3$ is not exactly equal to the vacuum expectation value $\langle \mathcal{O}\rangle$ of the dual operator. There will be additional terms directly proportional to the source and its derivatives, and $\langle \mathcal{O}\rangle$ will in general be much larger than $O(\epsilon^3)$, which could have been incorrectly concluded from \eqref{phi3}. However, for the dynamics after the source has been turned off, which we will focus on in this section, $\langle \mathcal{O}\rangle$ will be proportional to $\phi_3$.\\
\linebreak
We will not go to higher orders in $\epsilon$. In \cite{Bhattacharyya:2009uu}, higher order contributions were computed and it was discovered that the perturbation theory breaks down for late times, namely that higher order terms dominate lower order terms. Such terms that invalidate perturbation theory are called {\it secular} terms, and we will come back to this concept later. There are many ways to understand why this happens. The intuitive picture is that after the quench a black brane has formed, which is intrinsically a non-perturbative object. As soon as the black brane has formed, we could rescale our coordinates such that the mass becomes order 1, thus it is invalid to treat the term $g_2$ as order $\epsilon^2$ which we have done and it must be treated as a correction to $g_0$. In other words, there is no such thing as a ``black brane with small mass'', thus any dynamics solved in the black brane background must be manifestly non-perturbative. Another way to see it is to look close to the horizon at $r^3\sim M$, and we see that at this radial position $g_2$ becomes of the same magnitude as $g_0$ and the perturbation theory is not well-defined for small values of $r$. \\
\begin{figure}[t]
 \includegraphics[scale=0.7]{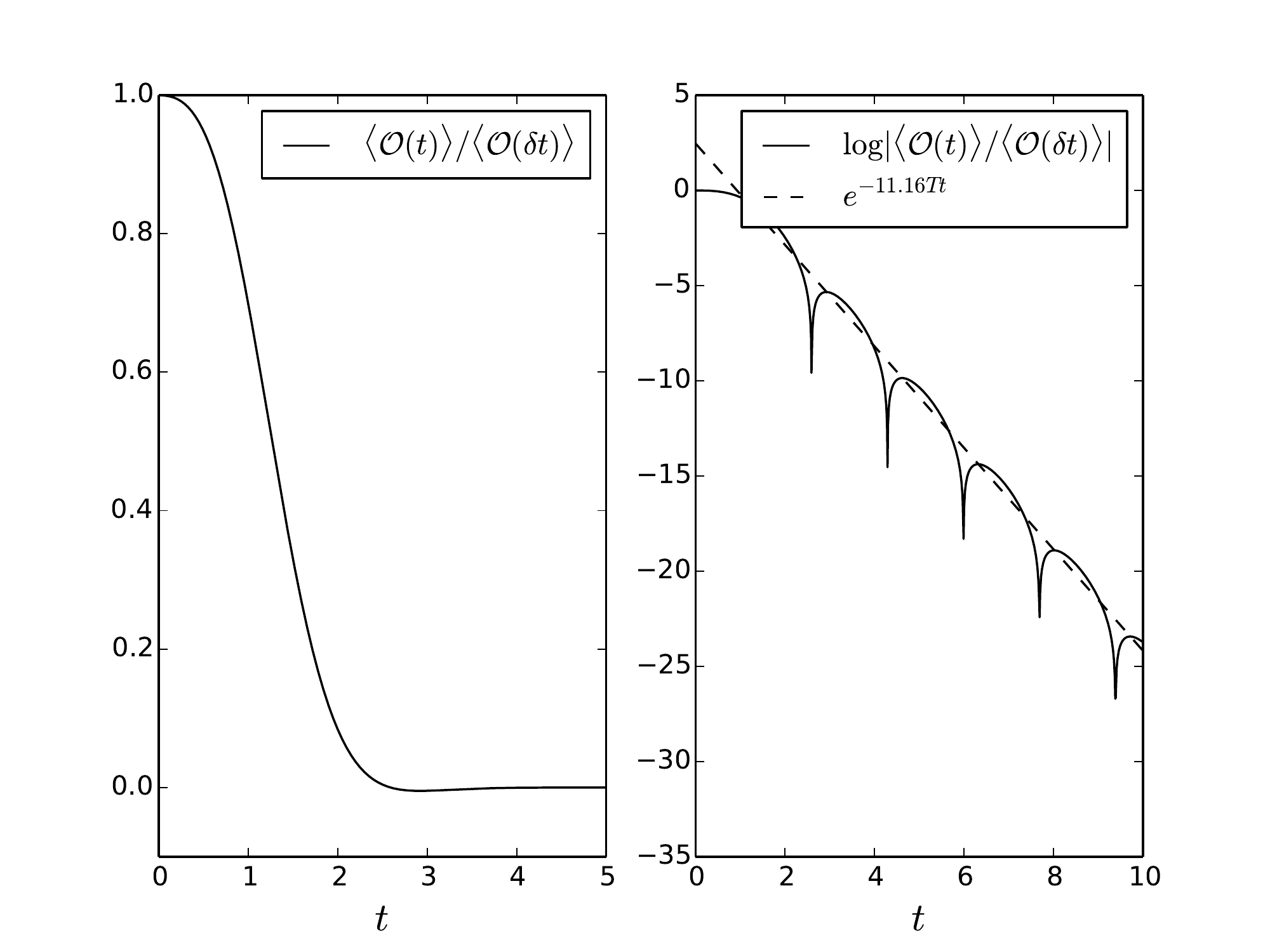}
 \caption{\label{BMvsqnm} Dynamics of a massless scalar field perturbation after a black brane has formed from quenching the scalar, computed using the perturbative methods in this section. By fitting a decaying exponential function to the maxima it is seen that the decay constant is $11.16T$, which is exactly the value of the lowest quasi normal mode of the black brane as quoted in \cite{Horowitz:1999jd}.}
\end{figure}
\linebreak
The correct way to deal with this issue is to treat the mass of the black brane as an $O(1)$ quantity, and solve the long-time dynamics in a black brane background instead. Note that in the naive perturbation theory, the term $\phi_3$ is constant for all times. This is another way to see that the perturbation theory should not be valid for late times since typically we would expect perturbations to dissipate behind the horizon of the black brane. Since the naive perturbation theory is valid for short times, we can use it as an initial condition. We thus use the profile \eqref{phi3} for $\phi_3$ as an initial condition for the wave equation in a black brane background, which takes the form 
\begin{equation}
\partial_r(r^4\partial_r\phi)+2r\partial_r(r\partial_v\phi)=M(r\partial^2_r\phi+\partial_r\phi), \label{BMwaveeqnum}
\end{equation}
Solving this equation with the initial condition given by $\phi(v=\delta t,r)=\epsilon^3\phi_3(v=\delta t,r)$ will turn the constant $\phi_3$ term into an oscillating decaying function. This is shown in Figure \ref{BMvsqnm}. To solve equation \eqref{BMwaveeqnum} numerically we first introduce the new variables $z=M^{1/3}/r$ and $t=M^{1/3}v$ and rewrite \eqref{BMwaveeqnum} as
\begin{equation}
I\dot{\phi}=\frac{1}{z}\phi''(1-z^3)-z\phi'-\frac{2}{z}\phi',
\end{equation}
where the operator $I$ is
\begin{equation}
I=2\partial_z \frac{1}{z}.
\end{equation}
Now $'$ means derivative with respect to $z$ and $\dot{}$ is derivative with respect to $t$, and the initial condition will be $\phi\sim z^3$. We can now discretize the radial direction with a standard finite difference scheme, such that the derivatives become matrices $D$ acting on the vectors $\phi_i(t)\equiv\phi(t,z_i)$. We can then invert the matrix $I_{ij}$ corresponding to the discretized version of the operator $I$ and write the system as a set of coupled ordinary differential equations for the functions $\phi_i(t)$ which can then easily be solved by standard computing libraries. We can then read off the expectation value of the operator corresponding to the field $\phi$ as the coefficient of $z^3$ in a taylor expansion of $\phi$ at the boundary, which is what we plot in Figure \ref{BMvsqnm}. Note that in the original paper \cite{Bhattacharyya:2009uu} a different numerical scheme was used to solve this equation.\\
\linebreak
This result can be cross checked with the quasi normal modes of the black brane. As explained in Section \ref{sec_qnm}, any small perturbation around a black brane can be decomposed in the quasi normal modes of the black branes, and for late times the slowest decaying mode will be dominant. The imaginary part of the lowest lying quasi normal mode for a black brane in four dimensions is given in terms of the temperature as $\omega_0\approx11.16T$, and by fitting a decaying exponential to the numerical results obtained above we obtain that value (see Figure \ref{BMvsqnm}).

\subsubsection{Inhomogeneous black hole formation}

\begin{figure}[t]
\includegraphics[scale=0.7]{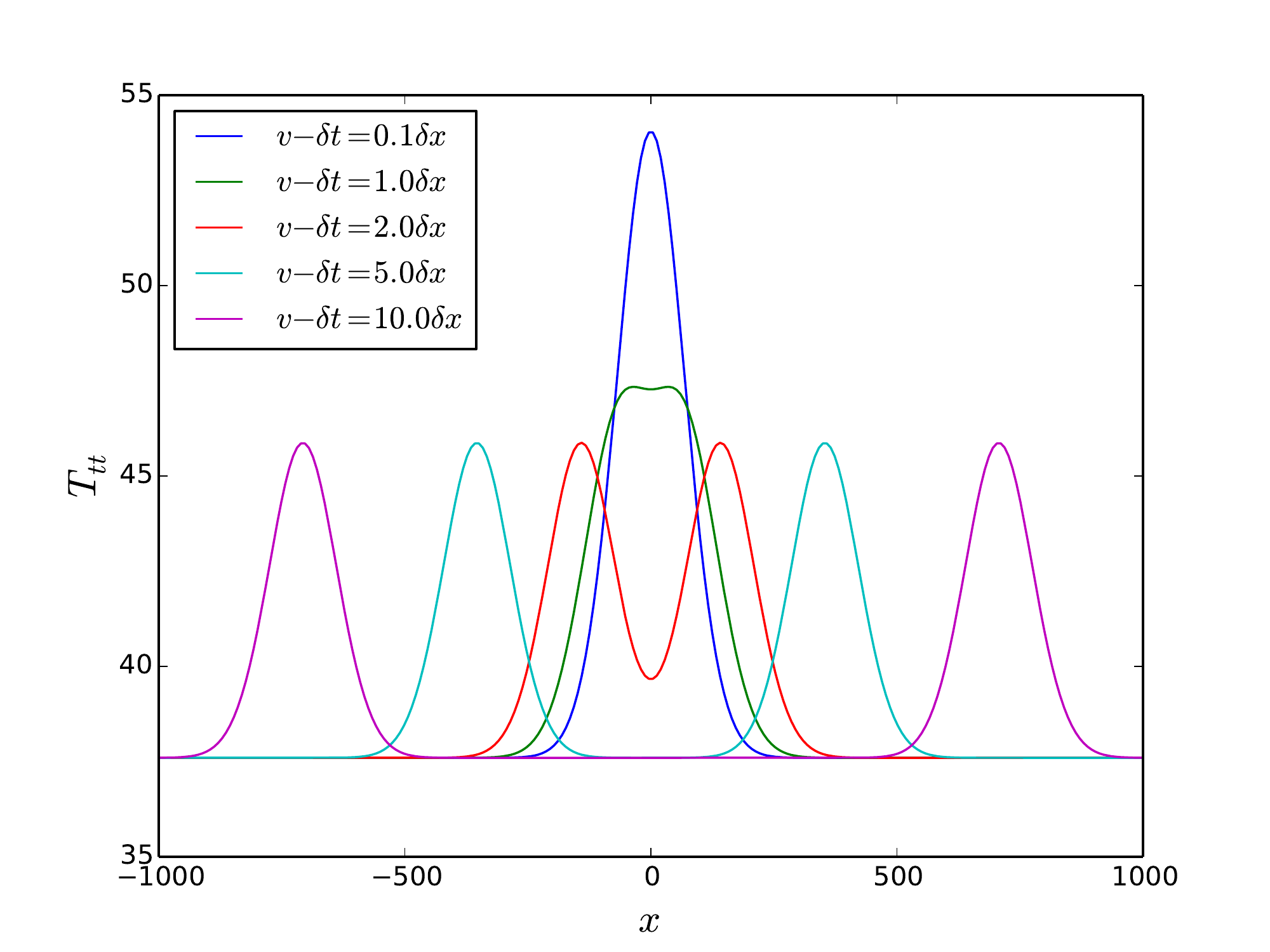}
\caption{\label{Ttt_weak_inhom} Stress-energy tensor after an inhomogeneous quench, in the limit where the amplitude of the quench is small and the amplitude and gradients of the inhomogeneities are both small compared to the background temperature. The energy density splits up into left- and right-moving modes, as expected from intuition from linear hydrodynamics. In this approximation, dissipation is not included and is expected to show up at higher orders. The boundary source is given by \eqref{inhomsource} with parameters $\delta t=0.1$, $a=0.2$, $\epsilon=0.1$ and $\delta x=100$, and the energy density is given in units of $L^2/16 \pi G$.}
\end{figure}

We will now consider a setup where the source we use to quench the system is not translationally invariant. This section is based on \cite{Balasubramanian:2013oga} and some unpublished results (together with Ben Craps, Tim De Jonckheere and Oleg Evnin) and can be skipped at a first reading unless the reader is specifically interested in this topic.\\
\lb
We thus consider AdS$_4$ quenched by a source that is not translationally invariant, but instead has a dependence on an $x$-coordinate as $J=J(t,x)$. The source will still be localized in time (meaning that the pulse length $\delta t$ is small compared to the inverse temperature of the formed black brane), but now the amplitude at any given time will depend on $x$. In general this is a very hard problem to solve. Even if we assume that the amplitude $\epsilon$ is small, computing the coefficients in the amplitude expansion at each order in $\epsilon$ can not be done analytically. We will thus make another assumption, namely that derivatives with respect to $x$ are small, or in other words that the dependence in the $x$ direction is weak. We can then also do an expansion in gradients, which can be counted by introducing a formal derivative counting parameter $x\rightarrow \mu x$. We will denote the characteristic length scale for the $x$-dependence as $\delta x$. $\mu$ can then either be viewed as a formal parameter that is set to one in the end, or as a small parameter characterizing the length scale for the $x$-dependence as $\delta x=1/\mu$.  
\linebreak
The metric ansatz must also be generalized such as to incorporate the broken translational invariance in the $x$ direction. One convenient choice, which was used in \cite{Balasubramanian:2013oga}, is
\begin{equation}
\rd s^2=-g\rd v^2+2\rd v(\rd r+k\rd x)+f^2(e^{B}\rd x^2+e^{-B}\rd y^2),
\end{equation}
where $f$, $g$, $k$ and $B$ are all functions of $r$, $v$ and $x$. All quantities are then expanded in a double expansion in $\mu$ and $\epsilon$ as
\begin{equation}
f=\sum_{i,n\geq0} f_{2i,n}\epsilon^{2i}\mu^{n},\quad\quad g=\sum_{i\geq0} g_{2i,n}\epsilon^{2i}\mu^n,\quad\quad k=\sum_{i\geq0} k_{2i,n}\epsilon^{2i}\mu^n,
\end{equation}
\begin{equation}
B=\sum_{i\geq0} B_{2i,n}\epsilon^{2i}\mu^n,\quad\quad\phi=\sum_{i\geq0} \phi_{2i+1,n}\epsilon^{2i+1}\mu^n.
\end{equation}
At zeroth order in the gradient expansion, the solution is just the same as in the homogeneous case, but where all quantities are modulated with a dependence on $x$, namely
\begin{equation}
\phi_{1,0}(v,r,x)=J(v,x)+\frac{\dot{J}(v,x)}{r}.
\end{equation}
\begin{equation}
f_{2,0}(v,r,x)=-\frac{\dot{J}^2(v,x)}{8r},\label{weakeqinhomf2}
\end{equation}
\begin{equation}
g_{2,0}(v,r,x)=-\frac{3}{4}\dot{J}^2(v,x)+\frac{1}{2r}\dot{J}(v,x)\ddot{J}(v,x)-\frac{1}{2r}\int_{-\infty}^v\ddot{J}^2(t,x)\rd t.\label{weakeqinhomg2}
\end{equation}
Thus after the source has been turned off, we can define an $x$-dependent local mass density $M(x)$ by
\begin{equation}
M(x)=\frac{1}{2}\int_{-\infty}^\infty\ddot{J}^2(t,x)\rd t,\label{inhomweakmass}
\end{equation}
from which we can define a local temperature as $T(x)=\frac{3}{4\pi}M^{1/3}(x)$. At first order in the gradient expansion we obtain a correction to the function $k$ as
\begin{equation}
k_{2,1}(v,r,x)=\frac{1}{2}\dot{J}(v,x)J'(v,x)+\frac{1}{3r}\int_{-\infty}^v\rd\tau\left[-\dot{J}(\tau,x)\ddot{J}'(\tau,x)+\int_{-\infty}^\tau \rd t\ddot{J}'(t,x)\ddot{J}(t,x)\right].\label{k21eq}
\end{equation}
We will not consider higher orders in neither the gradient expansion nor the amplitude expansion, and refer to \cite{Balasubramanian:2013oga} for more details. Note that the first order correction in the gradient expansion results in terms that grow with time and invalidate the perturbative expansion for times of order $v\sim \delta x$. Thus for late times these solutions are not valid. However, we have already seen in the homogeneous case that the amplitude expansion already invalidates the results at a time scale of order $1/T$, which is assumed to be shorter than the time scale set by $\delta x$. However, these secular terms are not expected to affect the long-time dynamics of the metric which we are interested in here. As we saw in the homogeneous case, by renormalizing the mass of the black brane (namely treating the mass term as an $O(1)$ quantity) we can resum the secular terms showing up in the scalar field. The result was that the scalar field decays very fast (fast compared to the time scale $\delta x$ which we are interested in), and for long time the dynamics of the scalar field is not expected to affect the dynamics of the metric in any significant way.\\
\linebreak
Thus after the source has been turned off ($v\gtrsim \delta t$), the metric functions can be written as
\begin{equation}
f_{2,0}=0,\quad g_{2,0}=-\frac{M(x)}{r},\quad k_{2,1}=\frac{1}{r}\left(K+(v-\delta t)\frac{1}{3}M'(x)\right).\label{metricfunctionsinhom}
\end{equation}
From this point of view $K$ and $M$ are just integration constants that are fixed from the initial conditions (which in turn are determined in terms of $J$ by solving the full system including the scalar field while the source is turned on). The formulas for $M$ and $K$ are given in terms of $J$ by equations \eqref{inhomweakmass} and \eqref{k21eq}, namely by
\begin{equation}
K=\frac{1}{3}\int_{-\infty}^\infty \rd\tau\left[-\dot{J}(\tau,x)\ddot{J}'(v,x)-(\tau-\delta t)\ddot{J}'(t,x)\ddot{J}(t,x)\right],\quad M(x)=\frac{1}{2}\int_{-\infty}^\infty\ddot{J}^2(t,x)\rd t
\end{equation}
It can be shown from the linearized Einstein equations that $k_{2,1}$ satisfies the equations
\begin{equation}
r^2\partial_r^2k_{2,1}-2k_{2,1}=0,\quad\quad 2\partial_vk_{2,1}-r\partial_r\partial_vk_{2,1}=\frac{M'(x)}{r},\label{k21eom}
\end{equation}
from which the solution \eqref{metricfunctionsinhom} for $k_{2,1}$ follows (note that we also impose the boundary condition that $k_{2,1}\sim O(1/r)$). The equations of motion for $g_{2,1}$ resulting from the Einstein equations are
\begin{equation}
\partial_vg_{2,1}=0,\quad\quad r\partial_rg_{2,1}+g_{2,1}=0,\label{g21eom}
\end{equation}
from which we obtain $g_{2,1}=0$, and it can be shown that $f_{2,0}$ and $f_{2,1}$ vanish. We will take this metric, which is the form of the metric right after the source has been turned off, as initial conditions for computing the long-time dynamics. The initial conditions can thus, at this order, be specified by just the two functions $M(x)$ and $K(x)$ and we do not need to explicitly refer to $J(x,t)$. For higher order corrections there will of course be corrections that depend on $J(x,t)$ (constructed by taking integrals of expressions involving $J(x,t)$ and its derivatives that are algebraically independent of $M(x)$ and $K(x)$). The second term in $k_{2,1}$, which grows with time, is called a secular term and is the reason why the perturbative expansion is not valid for late times. There exists a standard technique for dealing with secular terms in perturbation theory, a resummation method similar to the renormalization procedure one employs to deal with divergences in quantum field theory \cite{Chen:1995ena}. One essentially absorbs the secular term into a lower order integration constant, by making this lower order integration constant depend on time slowly. In other words, the perturbative expansion is carried out around a slowly moving zeroth order solution instead around a constant solution. To determine the exact time dependence of the integration constants, we solve a set of differential equations arising from a self consistency requirement when absorbing the secular term into the integration constants (these differential equations are similar to the equation for the running coupling constants in QFT, and we will refer to them as renormalization group (RG) equations). For this method to be convenient, the secular term should have a form that is suitable for being absorbed into the integration constants, otherwise the resulting RG equations might be too complicated to be of any use. This method has been recently used to study instability in global AdS\cite{Craps:2014vaa,Craps:2014jwa,Craps:2015jma} and see also \cite{Chen:1995ena} for other applications.\\
\linebreak
We thus generalize our setup to also include an integration constant in $k_{2,0}$, the zeroth order coefficient in the gradient expansion of $k$. Recall that the coefficient $k_{2,1}$ was obtained by solving equation \eqref{k21eom}. The integration constant in $k_{2,0}$ we are looking for arises when we consider a general solution to the {\it homogeneous} linearized equations, namely \eqref{k21eom} with the source (the right hand side) omitted. As can be shown from equations \eqref{k21eom}, the most general solution of $k_{2,0}$ satisfying the boundary conditions takes the form
\begin{equation}
k_{2,0}=\frac{C(x)}{r},
\end{equation}
for some $x$-dependent integration constant $C(x)$. This homogeneous solution has exactly the right $r$-dependence to absorb the secular term in $k_{2,1}$. However, introducing this integration constant in $k_{2,0}$ results in a correction at the leading gradient expansion to the equations \eqref{g21eom} for $g_{2,1}$. Now $g_{2,1}$ must satisfy
\begin{equation}
\partial_vg_{2,1}=-\frac{3C'(x)}{2r},\quad\quad r\partial_rg_{2,1}+g_{2,1}=\frac{C'(x)}{2r^2},
\end{equation}
which has the solution
\begin{equation}
g_{2,1}=-\frac{C'(x)}{2r^2}-(v-\delta t)\frac{3C'(x)}{2r}.
\end{equation}
which also includes a secular term. We thus also need to absorb this secular term into the integration constant in the zeroth order coefficient $g_{2,0}$. We thus replace the mass $M(x)$ by an integration constant $G(x)$. We will then renormalize $G(x)$ and $C(x)$ and replace them by {\it renormalized} integration constants $G_R(x,v)$ and $C_R(x,v)$. This renormalization is done by solving RG equations and has the effect that the secular terms are absorbed into the renormalized integration constants (there will however be a non-secular correction in $B$ that we will not treat here). The initial conditions are then imposed as $G_R(x,\delta t)=M(x)$ and $C_R(x,\delta t)=K(x)$.\\
\linebreak
Let us now carry out this procedure. The part of $g$ and $k$ including the secular terms and the relevant homogeneous solutions are
\begin{equation}
g_{sec}=-\frac{G(x)}{r}-\mu\frac{3vC'(x)}{2r},
\end{equation}
\begin{equation}
k_{sec}=\frac{C(x)}{r}+\mu\frac{vG'(x)}{3r}.
\end{equation}
The usual method \cite{Craps:2014vaa,Chen:1995ena} for carrying out this renormalization method is to pick an arbitrary parameter $\tau$, and rewrite the time dependence in the secular terms as $v=(v-\tau)+\tau$. We then define renormalized integration constants by
\begin{equation}
G_R(x,\tau)\equiv G(x)+\tau \frac{3}{2}\mu C'(x),\quad\quad C_R(x,\tau)\equiv C(x)+\tau \frac{1}{3}\mu G'(x),
\end{equation}
which are such that the secular terms are cancelled exactly at time $v=\tau$. Now we replace the occurence of $G$ and $C$ at order $\mu$ by $G_R$ and $C_R$ (since they only differ by higher order terms), and demand that the resulting expansion is independent of $\tau$ (by demanding that a derivative with respect to $\tau$ vanishes). This results in the differential equations
\begin{equation}
\partial_\tau G_R(x,\tau)=\frac{3C_R'(x,\tau)}{2},
\end{equation}
\begin{equation}
\partial_\tau C_R(x,\tau)=\frac{G_R'(x,\tau)}{3},
\end{equation}
which are the RG equations determining the behaviour of $G_R$ and $C_R$. To obtain an optimal solution, which in some sense cancels the secular terms at every time $v$, we simply set $\tau=v$. This last step is analogous to the step in QFT where one sets the renormalization scale equal to the energy scale one is interested in. The intuitive explanation of this procedure is that we slowly update our integration constants to eliminate the secular terms at all times. The differential equations we obtained are equivalent to the wave equation (which can be seen by differentiating with respect to $\tau$ and $x$). The solutions, with the initial conditions $C_R(x,\delta t)=K(x)$ and $G_R(x,\delta t)=M(x)$, are
\begin{equation}
 G_R(x,v)=\frac{1}{2}(M-\frac{3}{\sqrt{2}}K)(x-\frac{1}{\sqrt{2}}(v-\delta t))+\frac{1}{2}(M+\frac{3}{\sqrt{2}}K)(x+\frac{1}{\sqrt{2}}(v-\delta t)).
\end{equation}
\begin{equation}
 C_R(x,v)=-\frac{1}{2}(\frac{\sqrt{2}}{3}M-K)(x-\frac{1}{\sqrt{2}}(v-\delta t))+\frac{1}{2}(\frac{\sqrt{2}}{3}M+K)(x+\frac{1}{\sqrt{2}}(v-\delta t)).
\end{equation}
We thus see that the inhomogeneities split up into right- and left-moving modes with the speed of $1/\sqrt{2}$, which is the speed of sound in a conformal field theory in 2+1 dimensions (which is given in general dimension $n+1$ by the formula $1/\sqrt{n}$ \cite{Bhaseen:2013ypa}). From this we can now obtain the stress-energy tensor by the prescription outline in Section \ref{sec_holrenorm}, and the energy density is given by
\begin{equation}
T_{tt}(x,t)=\frac{1}{8\pi G} C_R(x,t).
\end{equation}
The pressures can also be computed, for which a non-secular correction to $B$ must first be obtained, but we will not do that here. The explicit formulas for the full stress-energy tensor in terms of the boundary expansion can be found in \cite{Balasubramanian:2013oga}. The evolution of the energy density is illustrated in Figure \ref{Ttt_weak_inhom}, for a source of the form
\begin{equation}
J(t,x)=\epsilon e^{-t^2/\delta t^2}(1+ae^{-x^2/\delta x^2}).\label{inhomsource}
\end{equation}
As we see, right after the source has been turned off, the energy density has a profile corresponding precisely to the energy profile of the source. This profile then splits up into right- and left-moving modes (see \cite{Bhaseen:2013ypa,Lucas:2015hnv} for similar results).\\
\lb
One should be careful about the regime of validity of these solutions. When we absorbed the higher order corrections into the lower order integration constants, the homogeneous solutions at zeroth order solve the {\it linearized} homogeneous equations. These solutions are formally second order in $\epsilon$, but we know that these should really be treated as zeroth order quantities for the time scales we are considering here (which are of the order $\delta x\gg 1/T$). Although the homogeneous solution in $g$ is still a solution at the non-linear level, the homogeneous solution for $k$ is not. Thus it is imperative that the deviations from a homogeneous background really are small. This means that non only do we need small gradients, we also need to be able to define a background homogeneous solution with non-zero local temperature such that the deviations from this solution are everywhere small. For the particular source \eqref{inhomsource}, we must have $a\ll 1$. The conclusion is that the result in this section is really just a way to obtain and solve the equations of linearized hydrodynamics for small perturbations around a formed black brane with initial conditions given by equations \eqref{weakeqinhomf2}-\eqref{k21eq}. This is the reason why there is no dissipation in Figure \ref{Ttt_weak_inhom}, i.e. the two modes will move undisturbed for all times. When including higher order hydrodynamical corrections, dissipation is expected to arise.\\
\linebreak
In this section we have obtained a solution for the formation of an inhomogeneous black brane from quenching a boundary source, which is valid in certain limits. In Chapter \ref{threed}, we will look at a similar setup in three dimensions and show how to construct an exact solution of an inhomogeneous black hole being formed from a boundary source. In that example, the stress-energy tensor will also split up into left- and right-moving modes, but contrary to the results in this section, that is an exact result. The reason why there is no dissipation in that case is due to the enhanced symmetry group in the two-dimensional conformal field theory on the boundary which results in an infinite number of conserved charges.

\subsection{AdS Vaidya spacetimes}\label{ssec_vaidya}

Another analytical model, the Vaidya spacetime, describing black hole formation can be obtained by assuming that the matter content is composed of pressureless spherically (or translationally) symmetric null dust. We will only consider the Vaidya spacetimes for formation of AdS black holes, although flat space versions also exist and have been used to study black hole formation in spacetimes more similar to our own. In global AdS, with AdS radius $L=1$, the metric takes the form
\begin{equation}
\rd s^2=-\left(r^2+1-\frac{M(v)}{r^{d-2}}\right)\rd v^2+2\rd v\rd r+r^2\rd\Omega^2_{d-1}.\label{globalvaidya}
\end{equation}
Note that for constant $M$, this is just the black hole metric in global coordinates. For a time-dependent mass $M$, this spacetime requires a non-zero stress-energy tensor of the form
\begin{equation}
8\pi G T_{vv}=\frac{d-1}{2r^{d-1}}M'(v),\label{stressenergyvaidya}
\end{equation}
which corresponds to a continuous creation of spherically symmetric infalling lightlike matter from the boundary of the spacetime at $r\rightarrow\infty$. We can define a similar toy model in Poincar\'e coordinates for the formation of a translationally invariant black brane, and the metric can be written in the form
\begin{equation}
\rd s^2=-r^2\left(1-\frac{M(v)}{r^{d}}\right)\rd v^2+2\rd v\rd r+r^2\rd\vec{x}^2_{d-1},\label{translvaidya}
\end{equation}
with the stress-energy tensor again given by \eqref{stressenergyvaidya}. This spacetime corresponds to a continuous creation of translationally invariant infalling lightlike matter from the boundary. Via the AdS/CFT correspondence, the infalling matter in both versions of the AdS Vaidya spacetime can be interpreted as being created by a time dependent source in the dual CFT.\\
\linebreak
A very useful special case of these spacetimes can be obtained by assuming that the matter content is an infinitesimally thin shell. The mass profile then takes the form of a step function and such spacetimes thus only depend on the final mass of the black hole. Studying thin shell spacetimes can be useful since there is only one free parameter (the mass $M$), compared to if we assumed some arbitrary profile for the infalling matter content (which would give us a continuous function as a free parameter). In the boundary CFT these spacetimes are dual to thermalization processes triggered by an {\it instantaneous} quench, and non-local observables in a thermalizing CFT have been studied using these spacetimes \cite{Balasubramanian:2011ur,Ziogas:2015aja}. Note also that the translationally invariant Vaidya spacetime \eqref{translvaidya}, for $d=3$, can be obtained by taking the thin shell limit of the thermalization process from a massless scalar field considered in Section \ref{ssec_weak}, as can be seen from equations \eqref{weakeqf2}-\eqref{weakeqM}. In Chapter \ref{threed}, we will in particular show that the spherically symmetric Vaidya spacetime \eqref{globalvaidya} (and generalizations thereof), for $d=2$, can be obtained from a limit of an infinite number of pointlike particles falling in from the boundary.

\subsection{Dynamics in global AdS}
Dynamics in global AdS has been shown to exhibit much more complicated and interesting structure than Poincar\'e patch AdS, and for instance it is very difficult to know if small perturbations will result in black hole formation or not \cite{Bizon:2011gg} (as opposed to in Poincar\'e coordinates where a black brane always forms for translationally invariant perturbations). This is related to the fact that the AdS radius in global coordinates plays an important role and sets a scale to which one can compare the injected energy, so that there can be qualitatively different solutions depending on the energy of the perturbation. The typical behaviour is that the matter wave packet will fall in to the center of AdS and bounce back to the boundary before a black hole has formed. The wave packet then becomes more and more localized the more times it scatters at the origin and after a number of bounces it is localized enough to collapse to a black hole. We will find similar behaviour in Chapter \ref{confinement} when we study dynamics in confining holographic models. Dynamics in global AdS has also been studied perturbatively \cite{Craps:2014jwa,Craps:2015iia,Craps:2015xya,Craps:2014vaa,Evnin:2015wyi}. When solving the equations of a scalar field coupled to gravity perturbatively there will be secular terms showing up in the perturbative expansion, which can be resummed using the techniques we used in \ref{ssec_weak}. This resummed perturbation theory reveals a lot of interesting features, such as hidden conservation laws for the transfer of energy between the normal modes that can be related to the isometries of AdS. Although these are very interesting research questions, we will not discuss them in this thesis. In Chapter \ref{threed}, we will however look at black hole formation in {\it three-dimensional} global AdS, but from a different mechanism than what has been the focus of the above mentioned lines of research. In three dimensions there is another interesting dynamical process that can result in formation of a black hole \cite{Matschull:1998rv}, namely that of colliding pointlike particles (in three dimensions, pointlike particles coincide with conical singularities, contrary to higher dimensions where a pointlike massive solution of Einstein's equations can not exist since it must necessarily be covered by an event horizon). It turns out that one can consider collisions of pointlike particle where the initial particles merges into one single resulting object. If the energy is low, the resulting object will just be another pointlike particle, but if the energy is large enough the resulting object is instead a black hole.


\chapter{Holographic thermalization in confining theories}\label{confinement}

In this chapter we will study dynamical processes in confining theories using the AdS/CFT correspondence. After some brief motivation for studying these processes we will then study in detail dynamics in the hard wall model and in the AdS Soliton model which were both introduced in Section \ref{sec_conf}. The setup will be similar to what we did in Section \ref{timedep}, except that we will use a confining geometry as initial condition instead of AdS. We end with some general conclusions and directions for future research. This chapter is based on the papers \cite{Craps:2014eba} and \cite{Craps:2015upq}.

\section{Introduction and motivation}
One of the most successful applications of the AdS/CFT correspondence to real world systems, has been the application to high temperature strongly coupled plasmas of $SU(N)$ gauge theories, in particular to quantum chromodynamics (QCD). This is particularly relevant due to the experimental discovery of the quark-gluon plasma (QGP), which is a high temperature state of QCD where the quarks and gluons are deconfined. The essential idea for this application is to first consider the supersymmetric $\mathcal{N}=4$ $SU(N)$ SYM theory in four dimensions, which is one of the sides in the original AdS/CFT duality. This theory is obviously very different from a (non-supersymmetric) $SU(N)$ gauge theory. However, if we put the theories at finite temperature, the supersymmetry will be broken and in this state it is believed that the two theories are more similar. Thus by studying the supersymmetric $SU(N)$ gauge theory at finite temperature by using the AdS/CFT correspondence we may be able to deduce qualitative behaviour for high temperature non-supersymmetric gauge theories. For the particular application to real QCD, there is another subtlety which is that the gauge theory is $SU(3)$ while the only tractable version of the AdS/CFT correspondence requires that the number of colors, $N$, is large. Thus for applications to QCD we also rely on the approximation that $3\gg1$. The validity of this approximation can obviously be seriously questioned, but comparisons to experiments indicate that we are not too far off. Moreover, one can argue that AdS/CFT applications to QCD are better motivated than in for example several condensed matter applications where it is even difficult to identify any parameter that corresponds to the parameter $N$ in the AdS/CFT duality. \\
\linebreak
The AdS/CFT correspondence has been used to study many different properties of strongly coupled plasmas at high temperature. One of the most famous result emanating from applying AdS/CFT to strongly coupled plasmas concerns the value of the shear viscosity $\eta$, and it has been shown that \cite{Kovtun:2004de,Cremonini:2011iq} that it takes the
\begin{equation}
\frac{\eta}{s}=\frac{1}{4\pi}\frac{\hbar}{k_B},
\end{equation}
for a large class of field theories (namely those that have a gravity dual described by a black brane). Here $s$ is the entropy density. Comparing to experiments, this is of the same magnitude as what has been measured in the quark-gluon plasma formed in particle colliders \cite{PhysRevC.78.034915}, and the AdS/CFT description works better than perturbative QCD calculations providing more evidence that the quark-gluon plasma is strongly coupled. A more general result building on this, the famous so-called fluid-gravity correspondence, showed that the long wavelength dynamics of a black brane in AdS can be described by a hydrodynamics theory involving boundary variables \cite{Bhattacharyya:2008jc}. Another strong advantage of using AdS/CFT is that formulating non-perturbative dynamics at finite temperature becomes tractable. Studying dynamics in a quantum field theory, especially at strong coupling, is a very difficult task, and even formulating what one means with dynamics (away from linear fluctuations) in a thermodynamical context at finite temperature is not straightforward. In AdS/CFT, it all boils down to solving Einstein's equations in the bulk and it thus becomes conceptually trivial to formulate solution strategies for problems that would on the field theory side require techniques from non-equilibrum statistical physics and time-dependent lattice gauge theory. Many out-of-equilibrium problems of strongly coupled plasmas have been studied using holography, see for instance \cite{Chesler:2013lia,Chesler:2015wra,Chesler:2008hg,vanderSchee:2013pia,Balasubramanian:2013oga}. The overall qualitative features, such as a fast thermalization time and good agreement with hydrodynamics, agree with measurements of the QGP.\\
\linebreak
The applications of AdS/CFT to QCD described above have been in the high temperature deconfined phase, but it is also interesting to ask what happens in the confined phase. Corrections from the confinement property of QCD can also be relevant in the deconfined phase near the phase transition. With applications to dynamics of confining QCD in mind, we will in the remainder of this chapter study dynamics in two confining theories, but we will keep the discussion at a formal level and not compare with experiments or with real applications of QCD. The first model we will study is the hard wall model. As mentioned in Section \ref{sec_conf}, this model is essentially AdS but with a hard cutoff that has been put in by hand. This model can not be obtained from a string theory embedding and thus it is not clear how realistic results obtained using such a model will be. It will however will serve as a useful toy model to build some intuition of what behaviour we can expect in more well founded confining models. The second example we will study is the AdS Soliton. This model has a much more well defined theoretical motivation, and as we explained in Section \ref{sec_conf} can be obtained by considering compactifications in string theory. This model also underlies the Sakai-Sugimoto model \cite{Sakai:2004cn,Sakai:2005yt}, which is an extension that adds quarks and chiral symmetry breaking to try to capture more properties of QCD. In both models, the action will be that of a massless scalar field coupled to gravity, given by
\begin{equation}
S=\frac{1}{16\pi G}\int \rd^{d+1}x\sqrt{-g}\left(R-2\Lambda-\frac{1}{2}(\partial\phi)^2\right)+S_{GYH},\label{actionconf}
\end{equation}
where $S_{GYH}$ is the necessary boundary term to render the variational principle well defined\cite{Gibbons:1976ue}, but which does not affect the equations of motion. This is the same action we used in Chapter \ref{adscft}. The equations of motion following from this action are
\begin{align}
R_{\mu\nu}&-\frac{1}{2}R g_{\mu\nu}+\Lambda g_{\mu\nu}=\frac{1}{2}\partial_\mu\phi\partial_\nu\phi-\frac{1}{4}(\partial\phi)^2g_{\mu\nu},\nonumber\\
&\partial_\mu(\sqrt{-g}g^{\mu\nu}\partial_\nu\phi)=0.\label{eomconf}
\end{align}
The setup will be similar to the setup in Section \ref{timedep}, namely we will quench the boundary condition $J$ of the scalar field $\phi$ by giving a time dependence in the form of a short (Gaussian) pulse. This corresponds to turning on a source in the boundary field theory coupled to the operator $\mathcal{O}$ that is dual to the scalar field $\phi$. For the AdS Soliton, we will also consider turning on a source coupled to the boundary stress-energy tensor, which is equivalent to giving a time dependence to components of the boundary metric. The most important question we will ask is whether or not turning on a time dependent source on the boundary results in the formation of a black hole (dual to a thermalization process on the boundary), as well as what the qualitative behaviour of the resulting dynamics is both in the case of black hole formation but also in the case when no black hole forms.  
\pagebreak

\section{The hard wall model}\label{hardwallmodel}
In this section we will describe the process of holographic thermalization in the hard wall model. As we have emphasized previously, this is a very crude model of a gravitational background that enjoys the confining property. However, it will be a useful toy model to get some intuition for the behaviour we might expect from more well established models and as we will see many features will indeed also be present for the AdS Soliton.
\subsection{Holographic setup}
Our bulk setup is based on Einstein gravity in $d+1$ dimensions with a negative cosmological constant, minimally coupled to a massless scalar field $\phi$. The action will be that of a massless scalar field coupled to gravity given by \eqref{actionconf}. The metric ansatz will be the same as in \ref{ssec_weak} and given by
\begin{equation} \label{ds}
\rd s^{2}=\frac{L^2}{z^{2}}\left(-f(z,t)e^{-2\delta(z,t)}\rd t^{2}+\frac{\rd z^{2}}{f(z,t)}+\rd\vec{x}^{2}\right),
\end{equation}
where we fix the residual gauge freedom $\delta(z,t)\mapsto\delta(z,t)+p(t)$ by requiring the boundary condition $\delta(0,t)=0$. At early times, we start from the AdS metric, $f=1$ and $\phi=\delta=0$, and then we inject energy into the system by turning on the source $J(t)\equiv\phi(z=0,t)$ for the scalar field $\phi$. The source will be a Gaussian of the form
\begin{equation}
J(t)=\epsilon e^{-\frac{t^2}{\delta t^2}}. \label{hwsource}
\end{equation} 
Writing prime and dot for differentiation with respect to $z$ and $t$, respectively, and introducing $\Phi\equiv \phi'$ and $\Pi \equiv e^{\delta}\dot{\phi}/f$, the equations of motion reduce to
\begin{subequations}\label{eom}
\begin{align}
\dot{\Phi}=&\left(fe^{-\delta}\Pi\right)',\label{Ad} \\
\dot{\Pi}=&z^{d-1}\left(\frac{fe^{-\delta}\Phi}{z^{d-1}}\right)',\label{Pid}\\
\dot{f}=&\frac{z}{d-1}f^{2}e^{-\delta}\Phi\Pi, \label{fd}\\
f'=&\frac{z}{2(d-1)}f\left(\Phi^{2}+\Pi^{2}\right)+\frac{d}{z}(f-1),\label{fp}\\
\delta'=&\frac{z}{2(d-1)}\left(\Phi^{2}+\Pi^{2}\right)\label{dp}.
\end{align}
\end{subequations}
So far everything is the same as in Section \ref{ssec_num} where we studied dynamics and black hole formation in AdS. Now we introduce the hard wall  by restricting the range of the $z$ coordinate to $0<z<z_0$, where the location of the hard wall is inversely proportional to the confinement scale $\Lambda$ of the boundary theory, $\Lambda \sim 1/z_0$. The physical interpretation of the confinement scale is that the larger it is, the sooner the linear behaviour of the quark-antiquark potential in Figure \ref{qqpot} kicks in (and the linear potential grows faster). At the hard wall, we will mainly consider two possible boundary conditions on the scalar field: Dirichlet boundary conditions $\left.\phi\right|_{z=z_{0}}=0$, corresponding to $\Pi=0$, or Neumann boundary conditions $\left.\phi'\right|_{z=z_{0}}=0$, corresponding to $\Phi=0$. 
\subsection{Boundary expansion and holographic renormalization}\label{hwholrenorm}
The boundary quantities can be extracted by using the holographic renormalization procedure explained in Section \ref{sec_holrenorm}. We will go through the procedure in detail here for the cases of $d=3$ and $d=4$. One has to evaluate the bulk action
\begin{align}
S=&\frac{1}{16\pi G}\int_{\mathcal{M}}\rd^{d+1}x\sqrt{-g}\left(R+\frac{d(d-1)}{L}-\frac{1}{2}(\partial \phi)^2\right)+\frac{1}{8\pi G}\int_{\partial\mathcal{M}}\rd^{d}x\sqrt{-\gamma}\,K+S_{ct}
\end{align}
for the on-shell field solutions where the counterterm action $S_{ct}$ has been determined such that the total action is finite. Here $\gamma_{ij}$ is the induced metric at the boundary $\partial\mathcal{M}$ and is used for raising and lowering indices on $\partial\mathcal{M}$ (note that this is different from the boundary metric $g^{(0)}_{ij}$ with respect to which the functional derivaties should be taken when computing the stress-energy tensor, and the two are related by $\gamma_{ij}=L^2g^{(0)}_{ij}/z^2$). The boundary theory expectation values are then given by functional derivatives with respect to the sources.\\
\linebreak
The results for $d=3$ can in principle be obtained from Section \ref{sec_holrenorm}, but we will carry out the procedure again explicitly for the full theory (both the scalar field and gravity). The boundary expansion, which is the same as the one in Section \ref{ssec_num}, is
\begin{equation}
\phi=J-\frac{1}{2}\ddot{J}z^2+\phi_3 z^3+O(z^4),
\end{equation}
\begin{equation}
f=1-z^2\frac{\dot{J}^2}{4}+z^3f_3+O(z^4),
\end{equation}
\begin{equation}
\delta=\frac{\dot{J}^2}{8}z^2+O(z^4).
\end{equation}
We will also here state the expansions of the extrinsic curvature $K_{ij}$ and its trace $K$, which are
\begin{equation}
z^2K_{tt}=-L+\frac{L}{8}\dot{J}^2z^2+O(z^4),\quad z^2K_{x_ix_i}=L-\frac{L}{8}\dot{J}^2z^2+\frac{L}{2}f_3z^3+O(z^4),\quad K=\frac{3}{L}+\frac{1}{8L}\dot{J}^2z^2+O(z^4),
\end{equation}
where $x_i$ denotes one of the two spatial coordinates. By evaluating the on-shell action, it can be seen that the counterterms are given by
\begin{equation}
S_{ct}=\frac{1}{16\pi G}\int_{\partial\mathcal{M}}\rd^{3}x\sqrt{-\gamma}\left(-\frac{4}{L}-L\hat{R}+\frac{L}{2}(\hat{\partial}\phi)^{2}\right),
\end{equation}
where the hat signifies quantities evaluated using the induced metric $\gamma_{ij}$ at $z=\epsilon$. Note that this is just the same counterterm discussed in Section \ref{sec_holrenorm}. The variation of the renormalized on-shell action can now be written as
\begin{align}
\delta S_{ren,on-shell}=\frac{1}{16\pi G}\int_{z=\epsilon} \rd^{3}x&\Big(\sqrt{-\gamma}(K_{ij}-K\gamma_{ij}+\frac{2}{L}\gamma_{ij}+\frac{L}{2}\hat{R}\gamma_{ij}-L\hat{R}_{ij}-\frac{L}{4}(\hat{\partial}\phi)^2\gamma_{ij}\nonumber\\
&+\frac{L}{2}\partial_i\phi\partial_j\phi)\delta \gamma^{ij}+(\sqrt{-g}g^{zz}\partial_z \phi-L\partial_i(\sqrt{-\gamma}\gamma^{ij}\partial_j \phi))\delta\phi\Big),
\end{align}
where some boundary terms have been ignored. The Ricci tensor and scalar will both vanish identically (since the boundary metric is flat). By using the expansions of the extrinsic curvature, the scalar fields and the metric components one can see that the divergent part cancels out (as expected), and the result can be written as
\begin{equation}
\delta S=\frac{L^2}{16\pi G}\int_{z=\epsilon} \rd^{3}x\left(\left(\frac{L^2}{\epsilon^2}\right)\left(f_3\delta \gamma^{tt}+\frac{1}{2}f_3\delta \gamma^{xx}+\frac{1}{2}f_3\delta \gamma^{yy}\right)+3\phi_3\delta \phi+O(\epsilon)\right).
\end{equation}
From this we obtain the expectation values as 
\begin{equation}
\langle{\cal O}\rangle=\frac{3L^2}{16\pi G}\phi_3,\quad \langle T_{tt}\rangle=2\langle T_{x_{i}x_{i}}\rangle=-\frac{L^2}{8\pi G}f_3.  
\end{equation}
The trace of the stress-energy tensor is identically zero, $\langle\text{Tr}\left(T_{\mu\nu}\right)\rangle=-\langle T_{tt}\rangle+2\langle T_{x_{i}x_{i}}\rangle=0$. By solving the equations of motion asymptotically near the boundary, one can deduce that $\dot{f}_3(t)=(3/2)\dot{J}(t)\phi_3(t)$ which can also be obtained as a Ward identity as discussed in Section \ref{sec_holrenorm}, and can be reformulated in terms of the boundary expectation values and sources.\\
\linebreak
For $d=4$ the situation is a bit more involved, since there will be finite contributions from the source to the expectation values. We also have the sublety of having to deal with finite counterterms. The expansion of the metric components and the scalar field reads
\begin{equation}
\phi=J-\frac{\ddot{J}}{4}z^2+\left(\frac{\dot{J}^2\ddot{J}}{8}-\frac{\ddddot{J}}{16}\right)z^4\log z+\phi_4z^4+\ldots,\label{d4phiexp}
\end{equation}
\begin{equation}
f=1-\frac{\dot{J}^2}{12}z^2+\left(\frac{\dot{J}^4}{24}-\frac{\dot{J}\dddot{J}}{12}+\frac{\ddot{J}^2}{24}\right)z^4\log z+f_4z^4+\ldots,\label{d4fexp}
\end{equation}
\begin{equation}
\delta=\frac{\dot{J}^2}{12}z^2+\left(\frac{\dot{J}^4}{72}-\frac{\dot{J}\dddot{J}}{48}+\frac{\ddot{J}^2}{96}\right)z^4+\ldots.\label{d4deltaexp}
\end{equation}
For the extrinsic curvatures we have
\begin{align}
z^2K_{tt}&=-L+\frac{L}{24}\dot{J}^2z^2+L\left(\frac{\dot{J}^4}{48}-\frac{\dot{J}\dddot{J}}{24}+\frac{\ddot{J}^2}{48}\right)z^4\log z+L\left(\frac{f_4}{2}+\frac{25}{1152}\dot{J}^4\right)z^4+ O(z^5),\nonumber\\
z^2K_{xx}&=L-\frac{L}{24}\dot{J}^2z^2+L\left(\frac{\dot{J}^4}{48}-\frac{\dot{J}\dddot{J}}{24}+\frac{\ddot{J}^2}{48}\right)z^4\log z+L\left(\frac{f_4}{2}-\frac{1}{1152}\dot{J}^4\right)z^4+ O(z^5),\nonumber\\
K&=\frac{4}{L}+\frac{1}{12L}\dot{J}^2z^2+\frac{1}{L}\left(\frac{\dot{J}^4}{36}-\frac{\dot{J}\dddot{J}}{24}+\frac{\ddot{J}^2}{48}\right)z^4+O(z^5).
%
\end{align}
It turns out that the counterterms are
\begin{equation}
S_{ct}=\frac{1}{16\pi G}\int_{\partial\mathcal{M}}\rd^{4}x\sqrt{-\gamma}\left(-\frac{6}{L}+\frac{L}{4}(\hat{\partial}\phi)^{2}+L^3\log(\epsilon)\left(\frac{1}{8}(\hat{\Box}\phi)^{2}+\frac{1}{24}(\hat{\partial}\phi)^{4}-\frac{1}{24}\hat{R}^{ij}\partial_i\phi\partial_j\phi\right)\right).
\end{equation}
Note that even though $\hat{R}_{ij}$ vanishes, its variation does not. These counterterms are not unique and there is an ambiguity up to finite counterterms which we will parametrize as
\begin{equation}
\tilde{S}_{ct}=\frac{L^3}{16\pi G}\int_{\partial\mathcal{M}}\rd^{4}x\sqrt{-\gamma}\left(\frac{\alpha}{8}(\hat{\Box}\phi)^{2}+\frac{\beta}{24}(\hat{\partial}\phi)^{4}-\frac{\sigma}{24}\hat{R}^{ij}\partial_i\phi\partial_j\phi\right),
\end{equation}
with arbitrary constants $\alpha$, $\beta$ and $\sigma$. The variation of the renormalized on-shell action yields a quite tedious result which can be written as
\begin{align}
\delta S_{ren,on-shell}=&\frac{1}{16\pi G}\int_{z=\epsilon} \rd^{4}x\Big(\sqrt{-\gamma}\left[K_{ij}-K\gamma_{ij}+\frac{3}{L}\gamma_{ij}-\frac{L}{8}(\hat{\partial}\phi)^2\gamma_{ij}+\frac{L}{4}\partial_i\phi\partial_j\phi\right]\delta \gamma^{ij}\nonumber\\
&+\sqrt{-\gamma}L^3\log(\epsilon)\left[-\frac{1}{16}\gamma_{ij}(\hat{\Box}\phi)^{2}+\frac{1}{4}\partial_i\partial_j\phi\hat{\Box}\phi-\gamma_{ij}\frac{1}{48}(\hat{\partial}\phi)^4+\frac{1}{12}\partial_i\phi\partial_j\phi(\hat{\partial}\phi)^2\right]\delta \gamma^{ij}\nonumber\\
&+\sqrt{-\gamma}L^3\log(\epsilon)\left[\frac{1}{8}\gamma_{ij}\gamma^{kl}\partial_k(\partial_l\phi\hat{\Box}\phi)-\frac{1}{4}\partial_i(\partial_j\phi\hat{\Box}\phi)\right]\delta \gamma^{ij}\nonumber\\
&+\sqrt{-\gamma}L^3\log(\epsilon)\left[\frac{1}{24}\left(2\partial_k\partial_i(\partial^k\phi\partial_j\phi)-\partial^2(\partial_i\phi\partial_j\phi)-\gamma_{ij}\partial_k\partial_l(\partial^k\phi\partial^l\phi)\right)\right]\delta \gamma^{ij}\nonumber\\
&+\sqrt{-\gamma}L^3\left[-\frac{\alpha}{16}\gamma_{ij}(\hat{\Box}\phi)^{2}+\frac{\alpha}{4}\partial_i\partial_j\phi\hat{\Box}\phi-\gamma_{ij}\frac{\beta}{48}(\hat{\partial}\phi)^4+\frac{\beta}{12}\partial_i\phi\partial_j\phi(\hat{\partial}\phi)^2\right]\delta \gamma^{ij}\nonumber\\
&+\sqrt{-\gamma}L^3\left[\frac{\alpha}{8}\gamma_{ij}\gamma^{kl}\partial_k(\partial_l\phi\hat{\Box}\phi)-\frac{\alpha}{4}\partial_i(\partial_j\phi\hat{\Box}\phi)\right]\delta \gamma^{ij}\nonumber\\
&+\sqrt{-\gamma}L^3\left[\frac{\sigma}{24}\left(2\partial_k\partial_i(\partial^k\phi\partial_j\phi)-\partial^2(\partial_i\phi\partial_j\phi)-\gamma_{ij}\partial_k\partial_l(\partial^k\phi\partial^l\phi)\right)\right]\delta \gamma^{ij}\nonumber\\
&+L^3\log(\epsilon)\left[\frac{1}{4}\partial_i\partial_j(\gamma^{ij}\sqrt{-\gamma}\hat{\Box}\phi)-\frac{1}{6}\partial_i(\sqrt{-\gamma}\gamma^{ij}\partial_j\phi(\hat{\partial}\phi)^2)\right]\delta \phi\nonumber\\
&+L^3\left[\frac{\alpha}{4}\partial_i\partial_j(\gamma^{ij}\sqrt{-\gamma}\hat{\Box}\phi)-\frac{\beta}{6}\partial_i(\sqrt{-\gamma}\gamma^{ij}\partial_j\phi(\hat{\partial}\phi)^2)\right]\delta \phi\nonumber\\
&+\left[\sqrt{-g}g^{zz}\partial_z \phi-\frac{L}{2}\partial_i(\sqrt{-\gamma}\gamma^{ij}\partial_j \phi)\right]\delta\phi\Big).
\end{align}
It can be checked explicitly by using the expansions \eqref{d4phiexp}-\eqref{d4deltaexp} that all divergent terms cancel (just as in the $d=3$ case, the Ricci scalar and tensor vanish). The result can then be written as
\begin{align}
\delta S_{ren,on-shell}=\frac{L^3}{16\pi G}&\int_{z=\epsilon} \rd^{4}x\Bigg(\left[\frac{3}{2}f_4+\frac{(11-24\beta)}{384}\dot{J}^{4}+\frac{(2\alpha-1)}{16}\dot{J}\dddot{J}-\frac{\alpha}{16}\ddot{J}^{2}\right] \left(\frac{z^2}{L^2}\right)\delta\gamma^{tt}\nonumber\\
&+\left[\frac{1}{2}f_4+\frac{(1-8\beta)}{384}\dot{J}^{4}+\frac{9\alpha-1}{48}\ddot{J}^{2}+\frac{6\alpha-1}{48}\dot{J}\dddot{J}-\frac{\sigma}{24}\partial_t^2(\dot{J}^2)\right] \times\nonumber\\
&\left(\frac{z^2}{L^2}\right)(\delta\gamma^{x_1x_1}+\delta\gamma^{x_2x_2}+\delta\gamma^{x_3x_3})\nonumber\\
&+\left[4\phi_4+\frac{(19-24\beta)}{48}\dot{J}^{2}\ddot{J}-\frac{(3-4\alpha)}{16}\ddddot{J}\right]\delta\phi+O(\epsilon)\Bigg).
\end{align}
Thus we can identify the boundary field theory quantities as
\begin{equation}
\langle{\mathcal O}\rangle=\frac{L^3}{16\pi G}\left(4\phi_4+\frac{(19-24\beta)}{48}\dot{J}^{2}\ddot{J}-\frac{(3-4\alpha)}{16}\ddddot{J}\right),
\end{equation}
\begin{equation}
\langle T_{tt}\rangle=\frac{L^3}{16\pi G}\left(-3f_4-\frac{(11-24\beta)}{192}\dot{J}^{4}-\frac{(2\alpha-1)}{8}\dot{J}\dddot{J}+\frac{\alpha}{8}\ddot{J}^{2}\right),
\end{equation}
\begin{equation}
\langle T_{x_{i}x_{i}}\rangle=\frac{L^3}{16\pi G}\left(-f_4-\frac{(1-8\beta)}{192}\dot{J}^{4}-\frac{9\alpha-1}{24}\ddot{J}^{2}-\frac{6\alpha-1}{24}\dot{J}\dddot{J}+\frac{\sigma}{12}\partial_t^2(\dot{J}^2)\right),
\end{equation}
For simplicity we pick the scheme where $4\alpha=3$ and $24\beta=19$ so that $\langle \mathcal{O}\rangle$ simplifies. 
 By solving the equations of motion asymptotically near the boundary, one can deduce that
\begin{align}
\dot{f}_4=&\frac{4}{3}\dot{J}\phi_4+\frac{1}{18}\dot{J}^{3}\ddot{J}+\frac{1}{24}\ddot{J}\dddot{J}-\frac{1}{48}\dot{J}\ddddot{J},
\end{align}
which is the Ward identity and can be reformulated in terms of the boundary expectation values and sources. Note that the trace of the stress-energy tensor is non-zero, signifying the breaking of the conformal symmetry by the source $J$.


\subsection{Numerical methods}
To solve the system numerically, we discretized the equations in the bulk coordinate $z$ using a pseudospectral method based on Chebychev polynomials, see Section \ref{ssec_num} for more details. It will turn out that the trick to define a new variable by $\Phi(z,t)=z\tilde{\Phi}(z,t)$ is crucial for long stable evoution, in particular in the $d=4$ case. Note that this does not affect the boundary conditions at the hard wall, and Dirichlet boundary conditions are given by $\tilde\Phi|_{z=z_0}=0$. We will also use our rescaling freedom of the coordinates to set $z_0=1$ and then restore it by dimensional arguments afterwards.
\subsection{The black hole phase and the scattering phase}
\begin{figure}
 \includegraphics[scale=.7]{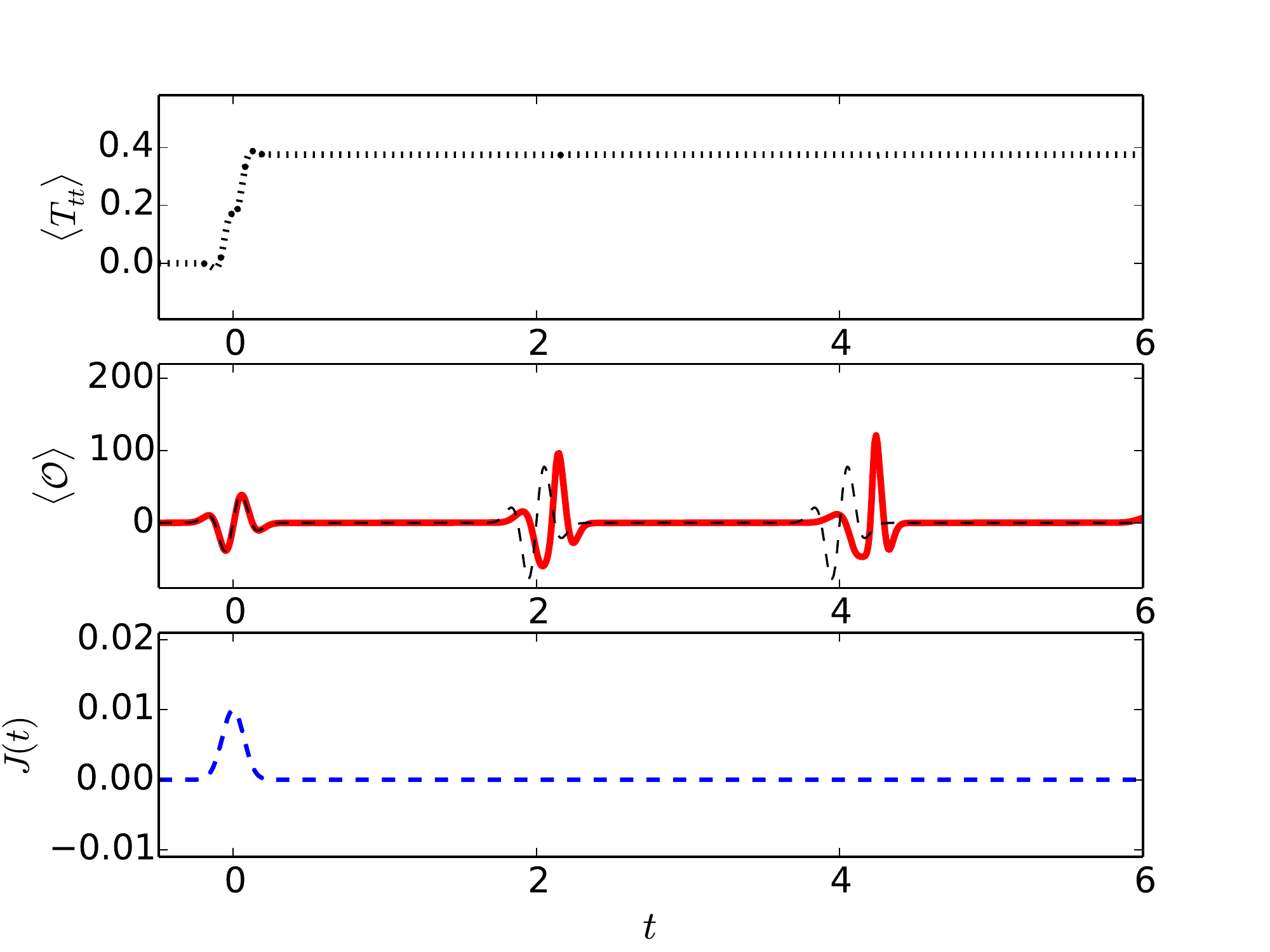}
 \caption{\label{evolhw2}Evolution of field theory observables for a scattering solution in $d=3$ and Neumann boundary conditions. The source is given by \eqref{hwsource} with parameters $\epsilon=0.01$ and $\delta t=0.1z_0$. The expectation values are in units of $L^2/16\pi G$. The dashed black line in the center panel is the expectation from a weak field probe limit calculation as explained in Section \ref{hwweak}.}
\end{figure}

\begin{figure}[t]
\centering 
\includegraphics[scale=0.60]{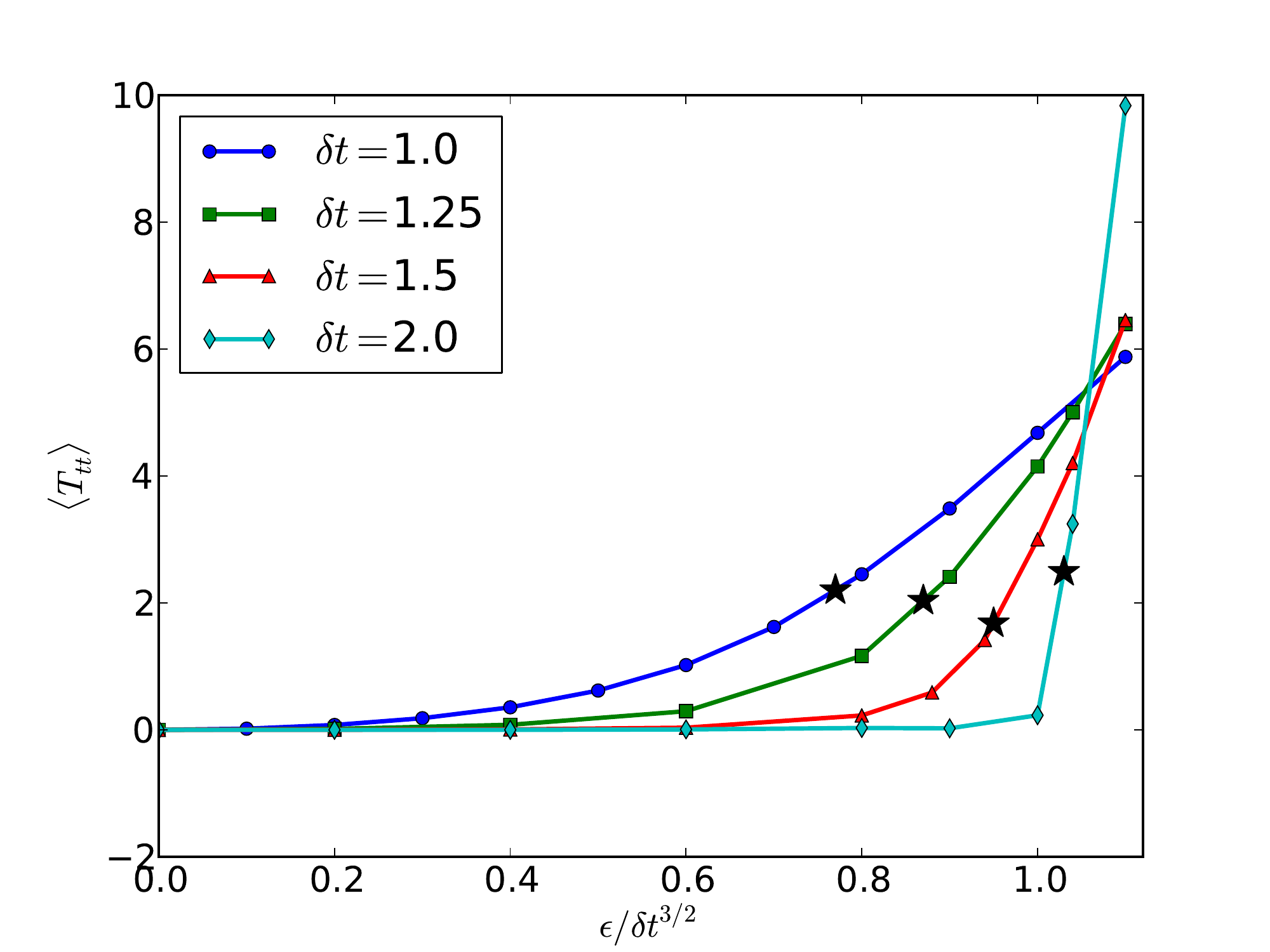}
\caption{\label{T_eps} Total injected energy as a function of $\epsilon$ for $d=3$ and Neumann boundary conditions. The black stars mark the critical value for black hole formation. The scattering regime is on the left of these markers and the black brane formation regime is on the right.}
\end{figure}

If the energy is large enough, the situation is very similar to that of quenching AdS and a black brane forms. This happens when the black hole horizon that would be formed in AdS is sufficiently large, namely when the would be horizon $z_h$ satisfies $z_h\lesssim z_0$. This is only an intuitive argument that becomes exact in the $\delta t\rightarrow0$ limit, and the details depends on the full dynamics and the particular source profile used. If the source is not localized (if $\delta t$ is not sufficiently small), the total injected energy can not be computed analytically and depends on the full dynamics in the bulk. Black hole formation is signalled by the formation of an apparent horizon, where the blackening factor $f$ vanishes. Since in the coordinate system (\ref{ds}) an apparent horizon is only reached at infinite time, we declare that a black brane has been formed whenever the minimum of $f$ goes below a cutoff we choose to be $0.02$. When a black brane forms, the evolution will typically just look like that of pure AdS (see Figure \ref{adsnumbh}), since the hard wall will be screened by formation of the horizon. However, it is possible that if the injected energy is very close to the threshold ($z_h\sim z_0$), the dynamics might still be affected by the hard wall even though a black brane still forms. We will not look at this in more detail here, but we will discuss such critical behaviour more in the AdS Soliton model.\\
\linebreak
If the energy is not large enough, a black brane will not form because the wave packet will reach the hard wall before it has time to collapse (in other words, the black brane that would have formed in AdS would have a horizon $z_h$ satisfying $z_h\gtrsim z_0$). The wave packet will then bounce on the hard wall and return to the boundary. This results in what we call {\it scattering solutions}, where the matter bounces back and fourth between the hard wall and the AdS boundary, and the system never equilibrates/thermalizes. Every time the matter bounces on the boundary, there is a reaction in the boundary observables and thus the boundary observables will have a quasi periodic behaviour. In Figure \ref{evolhw2} we illustrate this behaviour with a short pulse with $\delta t=0.1 z_0$, $\epsilon=0.01$ and Dirichlet boundary conditions. As we can see, the expectation value of the scalar undergoes a quasi periodic behaviour (even though the energy density goes to a constant).\\
\linebreak
In Figure \ref{evol} we show examples where the pulse width is of the same order of magnitude as the confinement scale (location of the hard wall). We illustrate the evolution of the minimum of the metric function $f$, as well as the evolution of the expectation value of the scalar field's dual operator, for $d=4$ field theory dimensions, Neumann boundary conditions, $\delta t=z_0$ and several values of $\epsilon$ (including both scattering solutions and black hole formation). Panel (g) shows the formation of a black brane signalled by $f$ going to zero. As we see, with increasing $\epsilon$ and thus increased backreaction, the oscillations of the scalar becomes more and more distorted.\\
\linebreak
One may also ask the question what exactly the energy density of the final state is as a function of the source amplitude $\epsilon$ and the injection time $\delta t$. As we already mentioned, for small $\delta t$ the total energy can be computed analytically (see the next section), while for larger $\delta t$ the whole non-linear dynamics must be taken into account. Figure~\ref{T_eps} shows that for fixed $\delta t$, the injected energy increases as a function of $\epsilon$. While for small $\delta t$ the increase is gradual, for large $\delta t$ the injected energy density is very small in the scattering phase (as can be expected for a source that is turned on and off almost adiabatically), but increases very sharply when the threshold for black brane formation is crossed. 

\subsection{Weak field solution in the scattering phase}\label{hwweak}
For very weak quenches in $d=3$, it is possible to find the scattering solutions analytically. The general solution of the Klein-Gordon equation for the scalar field is
\begin{equation}
\phi(z,t)=A(t-z)+zA'(t-z)+B(t+z)-zB'(t+z),
\end{equation}
for two arbitrary functions of time $A$ and $B$. $A$ corresponds to infalling waves and $B$ corresponds to outgoing waves. In pure AdS our setup would require $B=0$ and $A$ would then be equal to the boundary source $J$ (as was done in Section \ref{ssec_weak}). However, the presence of the hard wall will force us to also consider outgoing waves resulting from the infalling wave bouncing on the wall. By imposing the boundary condition $\phi(0,t)=J(t)$, we obtain that $A(t)+B(t)=J(t)$ and thus the solution reads
\begin{equation}
\phi(z,t)=J(t-z)+zJ'(t-z)+B(t+z)-B(t-z)-z(B'(t+z)+B'(t-z)),
\end{equation}
By imposing Neumann boundary conditions, we obtain the relation $J'(t-z_0)=B'(t-z_0)+B'(t-z_0)$ which has the solution
\begin{equation}
B(t)=\sum_{m=1}^{\infty}J(t-2z_0m)(-1)^{m+1}.
\end{equation}
The solution with Dirichlet boundary conditions is more complicated and we refer to \cite{Craps:2013iaa} for more details. From this it is then possible to extract the boundary field theory quantities. The full solution thus takes the form
\begin{align}
\phi(z,t)=&J(t-z)+zJ'(t-z)+\sum_{m=1}^\infty\Big[J(t+z-2mz_0)-J(t-z-2mz_0)\nonumber\\
&-z(J'(t+z-2mz_0)+J'(t-z-2mz_0))\Big](-1)^{m+1}.
\end{align}
From this the vacuum expectation value of the dual operator can be obtained as
\begin{equation}
\langle \mathcal{O}\rangle=\frac{L^2}{16\pi G}\left(\dddot{J}(t)-\sum_{m=1}^\infty(-1)^{m+1}\dddot{J}(t-2mz_0)\right).
\end{equation}
This result is shown in Figure \ref{evolhw2} compared to the numerical simulation.

\subsection{The phase diagram}
\begin{figure}[t]
\centering 
\includegraphics[scale=0.70]{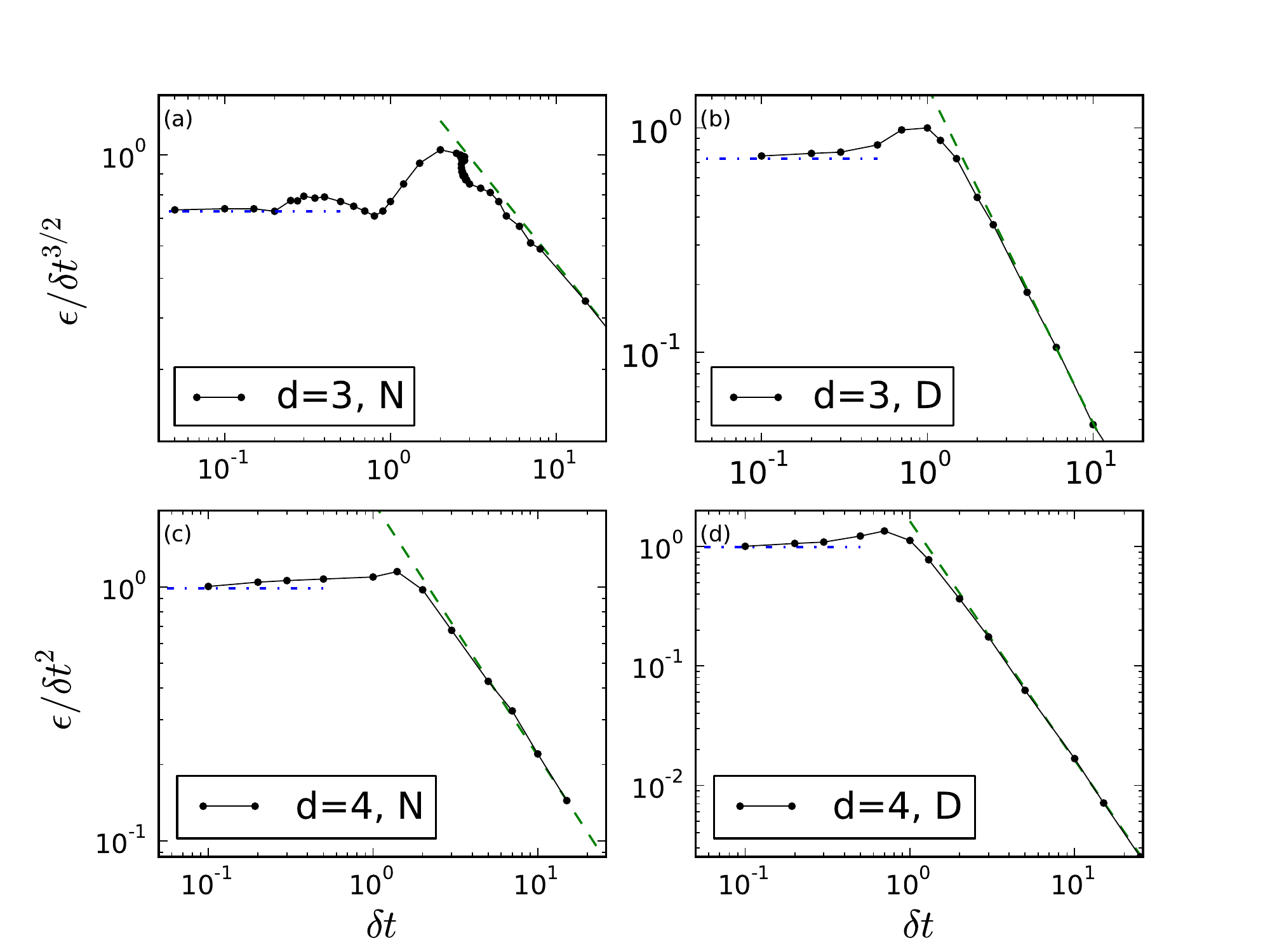}
\caption{\label{eps_delta_hw} Dynamical phase diagrams showing the critical parameters that separate the scattering regime from the black brane regime. The different panels show different dimensions and different boundary conditions, with panels (a,b,c,d) representing d=3 Neumann, d=3 Dirichlet, d=4 Neumann and d=4 Dirichlet, respectively. The dots are the numerically computed critical parameters for gravitational collapse. Above (below) the dots we have the black brane (scattering) phase. The straight lines correspond to analytically computed asymptotic behaviours, as explained in the text. Note the logarithmic scaling of the axes.}
\end{figure}

It is interesting to ask for what values of the parameters $\delta t$ and $\epsilon$ the system thermalizes. This is of course a question that depends on the particular source involved and may not give universal answers, but it is nevertheless an interesting question and as we will see there are several analytical checks. With {\it phase diagram} we specifically mean the separation between the parameter space that results in the formation of a black brane and the parameter space that results in a scattering solution. This depends on the boundary conditions (Neumann or Dirichlet) as well as the dimension and is shown in Figure \ref{eps_delta_hw} for $d=3$ and $d=4$. In the limit of small and large $\delta t$, analytic understanding can be obtained of the phase diagram which we will describe below.\\
\linebreak
{\bf Small $\delta t$ behaviour}\\
When $\delta t\ll z_0$, the infalling matter approximately takes the form of a thin shell. As explained earlier, a black brane will form in the hard wall model if the brane that would have formed in the absence of the hard wall. Moreover, in the limit of small $\delta t$, it is possible to analytically compute the total energy injected from the source. For $d=3$, it is given by equation \eqref{weakeqM}, and for the profile \eqref{hwsource} the mass is $M=\sqrt{\frac{9\pi}{8}}\frac{\epsilon^2}{\delta t^3}$. Since the event horizon is given by $z_h=M^{-1/3}$, comparing the event horizon $z_h$ with the location of the hard wall $z_0$ we obtain that the critical value for $\epsilon$ is given by
\begin{equation}
\epsilon_{c}=\left(8/(9\pi)\right)^{1/4}\delta t^{3/2}z_0^{-3/2}.
\end{equation}
For $d=4$, the formula for the total injected energy is given by formula (B.14) in \cite{Bhattacharyya:2009uu} and for our source \eqref{hwsource} the mass is then $M=\frac{\pi\epsilon^2}{3\delta t^4}$. The event horizon is then given by $z_h=M^{-1/4}$ which results in the critical $\epsilon$ value
\begin{equation}
\epsilon_c=\left(3/\pi\right)^{1/2}\delta t^{2}z_0^{-2}.
\end{equation}
These predictions are given by the blue dashed lines in Figure \ref{eps_delta_hw} and agree with the numerical results.\\
\linebreak
{\bf Large $\delta t$ behaviour}\\
The regime where $\delta t\gg z_0$ can also be treated analytically. This is in essence an {\it adiabatic} approximation, and we will thus neglect time derivatives in the equations of motion and consider quasi-static solutions with $\dot{f}=\dot{\Pi}=\dot{\Phi}$. For Neumann boundary conditions we have $\Phi=0$ at the hard wall. From equation \eqref{Pid} we then find that $\Phi=Cz^{d-1}f^{-1}e^{\delta}$, where $C$ is an integration constant, and using the Neumann boundary condition we must have $C=0$ and so $\Phi=0$ throughout the bulk (which then also satisfies equation \eqref{fd} since $\dot f=0$). Assuming the boundary conditions $\Pi(z=0)=\lambda$, \eqref{Ad} now gives $\Pi=f^{-1}e^{\delta}\lambda$ since $f(z=0)=1$ and $\delta(z=0)=0$. Using these solutions in \eqref{fp} and \eqref{dp}, we obtain an ordinary differential equation (ODE), conveniently written in terms of $S\equiv fe^{-\delta}$, as
\begin{equation}
 S'=\frac{d}{z}\left(S-e^{-\frac{\lambda^2}{2(d-1)}\int_0^z z'S(z')^{-2} \rd z' } \right),\label{oden}
\end{equation} 
where we have again used the boundary condition $\delta(z=0)=0$. This ODE can be solved numerically by scanning over possible boundary values of $S$ at the hard wall. However, it turns out there is a critical $\lambda_c$ such that for $|\lambda|\geq\lambda_c$ this ODE is not solvable anymore and this indicates that the adiabatic approximation breaks down and a black brane must form instead. For $d=3$ we have $\lambda_c\approx 1.47$ and for $d=4$ we have $\lambda_c\approx 1.85$. Going back to our original setup, $\lambda$ will be (slowly) time-dependent and equal to the time derivative of the boundary condition of the scalar field. Thus we draw the conclusion that for large injection times, a black hole is formed if we have $\textrm{max}\{|\dot{J}(t)|\}\geq \lambda_c$. For the profile given in equation \eqref{hwsource}, we obtain the relation $\epsilon_c=\lambda_c \sqrt{e}\delta t/\sqrt{2}$.\\
\linebreak
The story is very similar for Dirichlet boundary conditions. In this case, we have $\Pi=0$ at the hard wall which then implies from \eqref{Ad} that $\Pi=0$ everywhere in the bulk (and also in this case, this implies that \eqref{fd} is automatically satisfied). We can then solve \eqref{Pid} to obtain $\Phi=\alpha z^{d-1}S^{-1}$, where $S=fe^{-\delta}$ and $\alpha$ is a constant, related to $J$ by the requirement that $\phi=J+\int_0^z\Phi(z') \rd z'$ should vanish on the hard wall. Following the same argument as for the Neumann boundary condition, we obtain the ODE
\begin{equation}
 S'=\frac{d}{z}\left(S-e^{-\frac{\alpha^2}{2(d-1)}\int_0^z S(z')^{-2} z'^{2d-1}\rd z' } \right)\label{oded}.
\end{equation}
We can now numerically find critical parameters, $J_{c}\approx 1.53$ for $d=3$ and $J_{c}\approx 1.63$ for $d=4$, such that for $|J|>J_c$ no solution exists. For the profile \eqref{hwsource}, this leads to the critical amplitude $\epsilon_c=J_{c}$. A technical comment here is that when solving the equation \eqref{oded}, boundary values $S_{hw}$ of $S$ on the hard wall and values of $J$ are not in one-to-one correspondence, thus one must be careful when scanning through the parameter space to find which values of $J$ a solution exists. Since the map from $S_{hw}$ to $J$ is multivalued, if we naivelly just scan monotonically over values of $S_{hw}$ until a solution does not exist anymore and thus finds a critical $S_{hw}^c$, the corresponding value $J(S_{hw}^c)$ will {\it not} be the critical value for $J$. It is thus important to scan over the parameter $J$ and consider all solutions that satisfy this boundary condition, contrary to the case of Neumann boundary conditions where it is perfectly fine (and slightly easier) to just scan over values of $S_{hw}$.\\
\linebreak
The predictions in the adiabatic regime are shown in Figure \ref{eps_delta_hw} by the green dashed lines, and are in excellent agreement with the numerical results.
\begin{figure}[t]
\centering 
\includegraphics[scale=0.60]{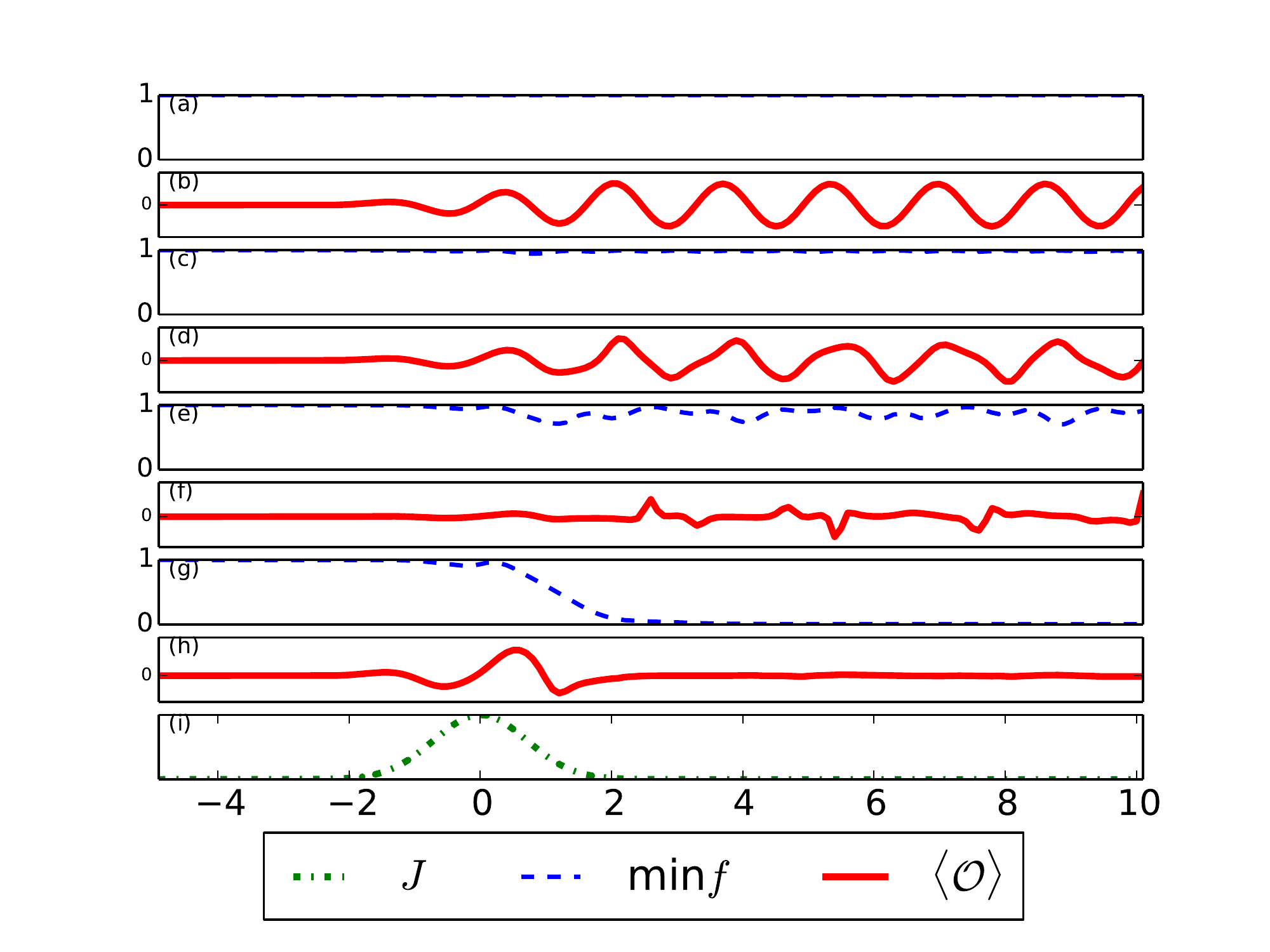}
\caption{\label{evol} Some examples of time evolution for $d=4$ and different $\epsilon$, fixed $\delta t=1$ and Neumann boundary conditions. Panels (a,c,e,g) show the evolution of the minimum of $f$, for three scattering solutions ($\epsilon=0.1$, $\epsilon=0.6$ and $\epsilon=1$) and black brane formation ($\epsilon=1.15$), respectively. Panels (b,d,f,h) contain the corresponding time evolution of the expectation values of the scalar operator in the dual field theory. The last panel shows the profile of the scalar source.}
\end{figure}

\subsection{Late time behaviour}
\begin{figure}[t]
\centering 
\includegraphics[scale=0.60]{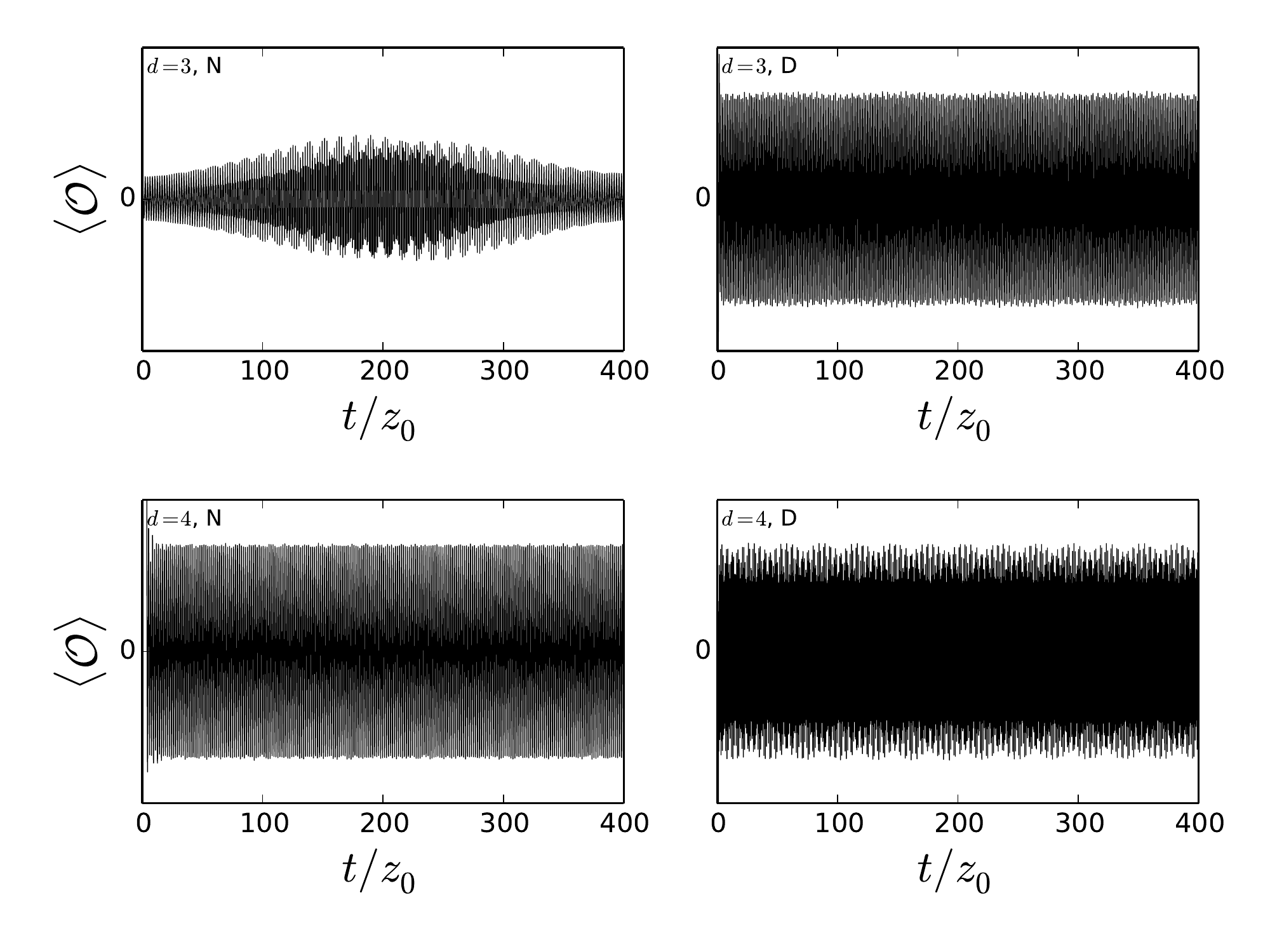}
\caption{\label{longtime}  Long-time evolution and the envelop of $\langle\mathcal{O}\rangle$ in the scattering phase for Neumann (N) and Dirichlet (D) boundary conditions and $d=3$, $4$. Note that only for $d=3$ and Neumann boundary conditions is there an amplitude modulation showing up at a large timescale, which can be explained by a resonant spectrum.}
\end{figure}

One of the most interesting features of our results is the oscillating behaviour of $\langle\mathcal{O}\rangle$ in the scattering phase. If this behaviour persists for all times, it indicates that the out-of-equilibrium state created by the energy injection never thermalizes. While analogous solutions have been found before for field theories in finite volume (dual to asymptotically global AdS spacetimes), this would be the first such example in infinite volume. We have therefore investigated the behaviour of $\langle\mathcal{O}\rangle$ in our scattering solutions for much later times than those displayed in Figure~\ref{evol}. In Figure~\ref{longtime} we show that the oscillations seem to continue regularly for as long as we have followed the evolution. For $d=3$ and Neumann boundary conditions, the amplitude seems to undergo modulation on a larger timescale. The timescale scales with $\epsilon$ roughly like $1/\epsilon^2$. This can be explained with the fact that the spectrum of normal modes is resonant {\it only} for $d=3$ and Neumann boundary conditions as we will explain in \ref{weaklynl}.\\
\linebreak
There is actually a very simple argument why the scattering solutions can not equilibrate to a static solution. First of all, the scattering solutions do not have large enough energy to form a black brane with event horizon $z_h<z_0$ (which we will refer to as a {\it large} black brane). Thus the only possible static solution one can imagine are what we will call small black branes, which are solutions with a blackening factor $f(z)$ corresponding to a black brane that has an event horizon that is outside the physical part of the spacetime (namely with $z_h>z_0$). Such solution do not have an event horizon in the physical part of the spacetime (and are thus not really black holes), but are perfectly allowed by the static vacuum Einstein equations (\eqref{fd}-\eqref{dp} with all time derivatives and matter fields set to zero). However, when we also consider the dynamical equations, a simple argument shows that such solutions can not be formed dynamically. When the scalar field satisfies either Neumann or Dirichlet boundary conditions, equation \eqref{fd} implies that $f$ is constant at the hard wall. Thus since the initial conditions have $f=1$, we conclude that we must have $f=1$ at the hard wall for all times. Since the small black branes have $f<1$, we conclude that such solutions can not be formed dynamically. In fact, this conclusion also holds for much more general boundary conditions of the form $n^\mu\partial_\mu\phi=z_0\sqrt{f}\phi'=F(a)$ where $F(a)$ is an arbitrary function. In that case \eqref{fd} implies that at the location of the hard wall, we have
\begin{equation}
\sqrt{f}=\frac{1}{2d-2}\int_{0}^{\phi}F(\phi')\text{d}\phi'+C,
\end{equation}
where $C$ is an arbitrary constant. Again, if initially we have $f=1$ and $\phi=0$, then at late times we cannot have $f<1$ and $\phi=0$, as would have been the case for a small black brane. While for these more general boundary conditions we cannot exclude that the system might approach another static solution than a black brane (with some non-trivial profile for the scalar), we have seen no hints of this in numerical simulations. 

\subsection{Weakly non-linear perturbation theory}\label{weaklynl}
In global AdS it is known that small perturbations may lead to instabilities\cite{Craps:2014jwa,Craps:2014vaa} on timescales of order $\sim1/\epsilon^2$, which can be explained by resonances in the normal mode spectrum. Thus it is interesting to see what can be said about the normal mode spectrum in the hard wall model, and what cocnlusions can be drawn from it.\\
\linebreak
Thus let us consider a massless scalar field $\phi$ in the probe limit in the hard wall model. The Klein-Gordon equation can be written as
\begin{equation}
\ddot{\phi}-\phi''+(d-1)\phi'/z=0.
\end{equation}
To look for the normal modes, we assume that the scalar field takes the form $\sin(\omega_n t)f_n(z)$, where $\omega_n$ are the different normal mode frequencies, which leads to the radial ODE
\begin{equation}
-\omega_n^2f_n-f_n''+(d-1)f_n'/z=0,
\end{equation}
with the solutions
\begin{equation}
f_n(z)=Az^{d/2}J_{d/2}(\omega_n z)+Bz^{d/2}Y_{d/2}(\omega_nz),
\end{equation}
where $A$ and $B$ are integration constants and $J_{d/2}$ and $Y_{d/2}$ are the Bessel functions of first and second kind, respectively. At the boundary we impose normalizability, $f(0)=0$, and thus $B=0$. Now for Dirichlet boundary conditions we want $f(z_0)=0$, and thus $\omega_n=\gamma_n^{d/2}/z_0$ where $\gamma_n^{d/2}$ are the different zeroes of the Bessel function $J_{d/2}$. A frequency spectrum would typically be resonant if some frequencies are integer multiples of other frequencies, and this does not seem to be the case for the zeroes of the Bessel function. Thus the spectrum is not obviously resonant.\\
\linebreak
For Neumann boundary conditions, we instead want $f_n'(\omega_n z)=0$. For $d=4$ these frequencies also do not have any simple analytical form and the spectrum is thus not expected to be resonant. However, for $d=3$, the situation is different. Using the recursive relation
\begin{equation}
\frac{2\alpha}{z}J_\alpha(z)=J_{\alpha-1}(z)+J_{\alpha+1}(z),
\end{equation}
as well as $J_{1/2}(z)=\sqrt{2/\pi}z^{-1/2}\sin z$ and $J_{-1/2}(z)=\sqrt{2/\pi}z^{-1/2}\cos z$ we can compute $J_{3/2}$ as
\begin{equation}
J_{3/2}(z)=\sqrt{\frac{2}{\pi}}(z^{-3/2}\sin z-z^{-1/2}\cos z).
\end{equation}
By evaluating the derivative of $f_n$, it can be shown that $\omega_n=n \pi /z_0$. Thus the spectrum is resonant, which explains the long time amplitude modulation shown in Figure \ref{longtime} that showed up only for $d=3$ and Neumann boundary conditions. Note also that the analytic result for $J_{3/2}$ still does not give a simple expression for the frequencies in the case of Dirichlet boundary conditions.

\subsection{Summary}
In this section we have  studied time dependent processes in the hard wall model, a very simple model of a confining holographic gauge theory. The main purpose of studying this model was to build some intution of what features confining models might possess, what numerical tools we can use and what analytical tests we can do to check the numerical results. This intuition will be very useful when we go on to study dynamics in the AdS Soliton model and we will see that some features, like the existence of scattering solutions and the qualitative form of the phase diagram, will also be present in that case.

\pagebreak
\section{The AdS Soliton model}\label{adssoliton}
In this section we will study dynamics in the AdS Soliton model. The AdS Soliton was introduced in Section \ref{sec_conf}, and is also a solution to Einstein's equation with negative cosmological constant that is confining according to the critera in Section \ref{sec_conf}. As we explained there, this model can be obtained from string theory and has a much more well understood field theory dual than the hard wall model. We will study the same setup where we inject energy into the system via a massless scalar field, and then we will study the response.
\subsection{Holographic setup}
The model we will consider is again a minimally coupled massless scalar field coupled to gravity. However, instead of using the hard wall geometry as initial condition, we will start with the AdS Soliton which we will now briefly review. The metric can be obtained by taking a Euclidean black brane and Wick rotating one of the spatial coordinates, such the the resulting spacetime still has Lorentzian signature. It can be described by the metric
\begin{equation}
\rd s^2=\frac{L^2}{z^2}\left(-\rd t^2+\frac{\rd z^2}{1-\frac{z^d}{z_0^d}}+(1-\frac{z^d}{z_0^d})\rd \theta^2+\rd \vec{x}_{d-2}^2\right).\label{solitonmetric}
\end{equation}
We will henceforth work in units with $L=1$. The AdS boundary is located at $z=0$ and the confinement scale is set by $z_0$ (which would correspond to the horizon if we Wick rotated back the above to a black brane). Note that the $\theta$ coordinate is compact in order to avoid a conical singularity, and this metric breaks rotational invariance between the $\vec{x}$ coordinates and the $\theta$ coordinate. This will have the implication that in order to solve for the time-dependence of the metric, we will need the second order dynamical Einstein equations. This should be contrasted with rotationally invariant setups, for which the metric can be determined using first order constraint equations alone (as was the case for pure AdS or the hard wall model).\\
\linebreak
We will for most of the time use the same procedure as before to quench the system, namely by using the source $J$, which is the boundary value of $\phi$ (but we will also sometimes consider quenching the metric components as explained below). Just like in our previous thermalization setups, we will use a source on the form
\begin{equation}
J(t)=\epsilon e^{-\frac{t^2}{\delta t^2}}.\label{source}
\end{equation}
After the source has been turned off, the system's energy will have increased and the gravitational bulk solution will have a nontrivial time dependence, governed by the action \eqref{actionconf} and the equations of motion \eqref{eomconf}. The main question is how this time-dependent solution behaves, in particular if it collapses into a black brane solution or not. To avoid having to invoke the extra scalar field, we will also consider quenching the metric, and compare this to the case of perturbing the scalar. We can inject energy by turning on a short time dependence for $\eta_{\theta\theta}/\eta_{x_jx_j}$, where $\eta$ is the boundary metric, which breaks the isotropy of the boundary metric between the $\theta$ and $\vec{x}$ coordinates in the boundary theory (see \eqref{zansatz} and \eqref{bsource}). This can also be interpreted as quenching the size of the compactified dimension. In this case, only the gravitational mode will be turned on, and the dynamics will be qualitatively different from the case where both the scalar and the metric modes are turned on. In pure AdS (or the hard wall model), it is also possible to quench the metric components but that would result in exactly the same dynamical system as quenching the scalar (the equations of motion will be the same). In the AdS Soliton case the two setups are qualitatively different, and quenching the metric can be done without breaking any more symmetries, which would have been the case in pure AdS.\\
\linebreak
Now to solve the Einstein equations we need to choose specific coordinates, as we did for AdS and the hard wall model. However, now we have less rotational symmetries and we must be more careful when specifying our metric ansatz, which now must involve more free functions than in the case of quenching AdS (even though our metric components will still only depend on time and the radial bulk coordinate). We can constrain the form of the metric by using the symmetries of the problem and by suitable gauge transformations (diffeomorphisms), but otherwise the metric will be completely general. Also, note in particular that due to parity invariance in the $\theta$ and $\vec{x}$ coordinates, we can set all off-diagonal terms involving these coordinates to zero. We may then use our gauge transformation to bring the metric to a diagonal form with three free functions. Note that in the AdS and hard wall cases, and in many other setups in the literature which enjoy maximal rotational or translational symmetries \cite{ Bhattacharyya:2009uu, Wu:2012rib,Ishii:2015gia}, there will only be two free functions in the metric, which are then completely determined by the constraint equations of Einstein's equations. We have found that the following ansatz is useful:
\begin{equation}
\rd s^2=\frac{L^2}{z^2}\left(-h(z,t)^2\rd t^2+ \frac{f(z,t)^2}{1-\frac{z^d}{z_0^d}}\rd z^2+ (1-\frac{z^d}{z_0^d})e^{(d-2)b(z,t)}\rd \theta^2+e^{-b(z,t)}\rd \vec{x}_{d-2}^2\right),\label{zansatz}
\end{equation}
with the initial conditions that $f=1$, $h=1$ and $b=0$ before the injection. We will refer to the boundary at $z=0$ as the UV or AdS boundary and the point $z=z_0$ as the IR. The coordinate $\theta$ is periodic with period $4\pi z_0/d$ to avoid a conical singularity at $z_0$. This form of the metric has a remaining gauge symmetry corresponding to rescaling of all coordinates. In the numerics, we will use this to set $z_0=1$, but for now we will keep $z_0$ explicit. The coordinates in (\ref{zansatz}) are very badly behaved at $z=z_0$, however, so for numerics we will use a different ansatz, see Section~\ref{numerics} for more details. \\
\linebreak
We will usually inject energy into the system by quenching the scalar with the source \eqref{source}, but as we previously mentioned we will also do this by quenching the metric. We can do this via the function $b(z,t)$, namely we can assume a boundary profile on the form
\begin{equation}
b(0,t)=\epsilon e^{-\frac{t^2}{\delta t^2}},\label{bsource}
\end{equation}
without turning on the source for the scalar. We will thus either use this source for $b$ and keep $\phi=0$, or source the scalar as $\phi(0,t)=J(t)$ and keep $b(0,t)=0$. \\
\linebreak
To decouple the equations of motion, it turns out to be convenient to define the following variables
\begin{equation}
\begin{array}{ccc}\label{PPi}
 \Pi=\dot{\phi}\frac{f}{h},  &\hspace{20pt} P=\dot{b}\frac{f}{h},\\
\Phi=\phi', &\hspace{20pt} B=b',\\
\end{array}
\end{equation}
where $'$ means derivative with respect to the $z$ coordinate. This decoupling is similar to what we did for pure AdS and essentially amounts to identifying the canonical momenta $\Pi$ and $P$. 
Introducing
\begin{equation}
G(z)=1-\frac{z^d}{z_0^d},
\end{equation}
the equations of motion following from the ansatz \eqref{zansatz} can now be written as
\begin{align}
\dot{f}=&z^{1-2d}\frac{(d-2)(d-1)Gh}{2(G z^{-2(d-1)})'}(PBz+2P)+\frac{z^{2-2d}\Phi\Pi Gh}{(G z^{-2(d-1)})'}+\frac{d-2}{2}Ph\label{zdotf},\\\nonumber
\frac{h'}{h}=&\frac{1}{z^{2(d-1)}(Gz^{-2(d-1)})'}\left(\frac{(d-1)(d-2)}{2}(P^2+GB^2+\frac{4GB}{z})+G\Phi^2+\Pi^2\right)+\\
&+(d-2)B-\frac{f'}{f},\\
\frac{h'}{h}=&\frac{2d(d-1)(f^2-1)}{z^{2d}(Gz^{-2(d-1)})'}+\frac{f'}{f},\\
\dot{P}=&\frac{1}{d-1}\left(\frac{(Ge^{(d-1)b})'he^{-(d-1)b}}{fz^{d-1}}\right)'z^{d-1},\\
\dot{\Pi}=&\left(\frac{hG\Phi}{fz^{d-1}}\right)'z^{d-1},\\
\dot{B}=&\left(\frac{Ph}{f}\right)',\\
\dot{\Phi}=&\left(\frac{\Pi h}{f}\right)'.
\end{align}
Note that in this gauge, only derivatives of $b$ appear in the equations of motion, so we do not need to integrate at every time step to obtain $b$. This is the reason for the particular parametrization in \eqref{zansatz}, which is a similar ansatz to what was used in for instance \cite{Chesler:2008hg}.\\
\linebreak
Evaluating equation \eqref{zdotf} at the point $z=z_0$, and using (\ref{PPi}), we obtain
\begin{equation}
\dot{f}_{z=z_0}=(\frac{d-2}{2}P h)_{z=z_0}=(\frac{d-2}{2}\dot{b}f)_{z=z_0},
\end{equation}
so that
\begin{equation} \label{fB}
f_{z=z_0}=C {e^{\frac{d-2}{2}b}}{\Large |}_{r=0}
\end{equation}
for some constant $C$. Since initially $f=1$ and $b=0$, we have that $C=1$. Thus we can state this result as
\begin{equation}
\left(fe^{-\frac{d-2}{2}b}\right)_{z=z_0}=1,\label{feB2}
\end{equation}
which will be crucial for the analysis in Section~\ref{static}. This condition actually is the statement that the regularity (absence of a conical singularity) at $z=z_0$ is preserved in time. This can be easily seen in the ansatz \eqref{ransatz}, which is used in the numerical analysis, and we refer the reader to Section~\ref{numerics} for further discussion.

\subsection{Boundary expansion and holographic renormalization}
To compute field theory observables, we will again use the procedure known as holographic renormalization, which was explained in Section \ref{sec_holrenorm}. As we have explained, this method requires one to add counterterms to the action that render it finite before extracting the boundary observables as functional derivatives with respect to boundary sources. These counterterms, which in odd dimensions give finite contributions to the various one-point functions, must be evaluated explicitly for every dimension and quickly become quite involved for increasing dimension. In addition, these contributions make the one-point functions scheme-dependent. However, when the source is turned off, the first non-trivial term in the boundary expansion is of order $z^d$ and no counterterms are needed. Since we will be interested in the evolution of the one-point functions after the source has been turned off, we will therefore mostly ignore the counterterms. In even dimensions the counterterms typically do not give finite contributions to the one-point functions even when the source is nonzero and thus in this case the counterterms can be ignored, as we saw in particular for $d=3$ in Section \ref{hwholrenorm}. In the end of this section we will provide the full asymptotic boundary expansion for $d=3$.\\
\linebreak
The asymptotic behaviour of the various fields after the source has been turned off is given by
\begin{align}
f=&1+\frac{E}{2(d-1)}z^d+\ldots,\label{fUV}\\
h=&1-\frac{E}{2(d-1)}z^d+\ldots,\\
b=&b_dz^d+\ldots,\\
\phi=&\phi_dz^d+\ldots,
\end{align}
where the $z^d$ coefficients of $f$ and of $h$ have been related by the equations of motion and $E$, $b_d$ and $\phi_d$ are so far undetermined functions that will depend on the full non-linear dynamics in the bulk. We will see later that $E$ will be the total injected energy, while the coefficient $\phi_d$ is related to the vacuum expectation value of the dual operator. We also have that $\dot{E}=0$, which follows from the equations of motion or from holographic Ward identities.\\
\linebreak
To identify the stress energy components at the boundary, we want to write the metric in the Fefferman-Graham gauge
\begin{equation}
\rd s^2=\frac{\rd \zeta^2}{\zeta^2}+\frac{1}{\zeta^2}g_{\alpha\beta}\rd x^\alpha \rd x^\beta.
\end{equation}
Doing this asymptotically, we can identify $z=\zeta-\zeta^{d+1}\frac{1}{2d}(\frac{E}{d-1}+\frac{1}{z_0^d})+O(\zeta^{d+2})$, which gives us the metric
\begin{align}
\rd s^2=\frac{\rd \zeta^2}{\zeta^2}-\frac{1}{\zeta^2}{\Bigg[}& \left(1-(E-\frac{1}{z_0^d})\frac{\zeta^d}{d}+O(\zeta^{d+1})\right)\rd t^2\nonumber\\&+\bigg(1+(\frac{E}{d-1}+\frac{1-d}{z_0^d})\frac{\zeta^d}{d}+(d-2)b_d\zeta^d+O(\zeta^{d+1})\bigg)\rd \theta^2\nonumber\\&+\bigg(1+(\frac{E}{d-1}+\frac{1}{z_0^d})\frac{\zeta^d}{d}-b_d\zeta^d+O(\zeta^{d+1})\bigg)\rd \vec{x}^2 {\Bigg]}. \label{asymptmetric}
\end{align}
Now it is easy to read off the non-zero stress energy components of the boundary field theory following the formulas in Section \ref{sec_holrenorm} as
\begin{subequations}\label{Ttt}
\begin{align}
\langle T_{tt}\rangle=& \frac{1}{16\pi \GN }(E-\frac{1}{z_0^d}),\label{eqTttsoliton}\\
\langle T_{\theta\theta}\rangle=&\frac{1}{16\pi \GN }(\frac{E}{d-1}+\frac{1-d}{z_0^d}+d(d-2)b_d),\\
\langle T_{xx}\rangle=& \frac{1}{16\pi \GN }(\frac{E}{d-1}+\frac{1}{z_0^d}-db_d),
\end{align}
\end{subequations}
from which we see that $E$ is indeed the total injected energy and $-\frac{1}{z_0^d}$ is the  initial AdS soliton energy density (up to a factor that can be set to one by a convenient choice of units). Note also that $\langle T_\mu^\mu \rangle=0$, which is a consequence of conformal invariance and is in general not true while sources are turned on. The vacuum expectation value of the scalar is
\begin{equation}
\langle \mathcal{O} \rangle=\frac{d}{16\pi \GN }\phi_d.\label{eqvevsoliton}
\end{equation}
Note that taking the difference $\langle T_{\theta\theta}\rangle-\langle T_{xx}\rangle$ cancels the total injected energy $E$ and isolates the dynamical mode $b$, which is why we will prefer to plot this quantity instead of the individual pressure components. \\
\linebreak
{\bf Full boundary asymptotics for $d=3$}\\
We will now provide the complete asymptotic expansion for $d=3$, including the time window when the source is turned on. All figures that focus on this time window will be for $d=3$ since in that case the boundary expansion and holographic renormalization is simpler and the boundary field theory quantities are not scheme dependent. To be completely general, we will also assume both a source $J_b$ for the function $b$ and a source $J_\phi$ for the scalar field $\phi$. The asymptotic expansions for the various functions, following from the equations of motion, are then
\begin{equation}
f(z,t)=1+\frac{1}{8}(\dot{J}_b^2+\dot{J}_\phi^2)z^2+\frac{E}{4}z^3+O(z^4),
\end{equation}
\begin{equation}
h(z,t)=1-\frac{1}{4}(\dot{J}_b^2+\dot{J}_\phi^2)z^2-\frac{E}{4}z^3+O(z^4),
\end{equation}
\begin{equation}
b=J_b-\frac{1}{2}\ddot{J}_bz^2+b_3z^3+O(z^4),
\end{equation}
\begin{equation}
\phi=J_\phi-\frac{1}{2}\ddot{J}_\phi z^2+\phi_3z^3+O(z^4).
\end{equation}
We also note the relation
\begin{equation}
3(2b_3-\frac{1}{z_0^3})\dot{J}_b+6\dot{J}_\phi \phi_3+2\dot{E}=0.\label{ward1}
\end{equation}
When going to the Fefferman-Graham gauge, the intermediate $z^2$ terms will not affect the $z^3$ terms. Moreover, as we mentioned in \ref{sec_holrenorm}, in even dimensional AdS spaces the stress-energy tensor does not get any extra contributions even when the boundary metric is not flat. It is also true that the source for the scalar does not give any contributions to the stress-energy tensor (we saw this in Section \ref{hwholrenorm}, and it can be argued that it also holds in the AdS Soliton case). The formula for the vacuum expectation value of the scalar is also unchanged, so we can use equations \eqref{eqTttsoliton} and \eqref{eqvevsoliton} for $d=3$ even when the sources are on. The relation \eqref{ward1} can be written in the form
\begin{equation}
\left(\langle T_{\theta\theta}\rangle-\langle T_{xx}\rangle\right)\dot{J}_b+2\langle \mathcal{O}\rangle\dot{J}_\phi+2\langle \dot T_{tt} \rangle=0,\label{ward2}
\end{equation}
which is a Ward identity which we also saw in Section \ref{hwholrenorm}, which can also be derived from diffeomorphism invariance in the bulk as we mentioned in Section \ref{sec_holrenorm}.\\
\linebreak
{\bf The temperature of the black brane}\\
As seen in \eqref{Ttt}, the energy density will be positive for energies $E>1/z_0^d$, and we expect that black branes will form if this is the case. A black brane can be written as the metric
\begin{equation}
\rd s^2=\frac{1}{\xi^2}\left(-\rd t^2(1-\frac{\xi^d}{\xi_h^d})+\frac{\rd \xi^2}{1-\frac{\xi^d}{\xi_h^d}}+\rd\theta^2+\rd \vec{x}_{d-2}^2\right).\label{bhmetric}
\end{equation}
Note in particular that in the case of dynamically evolving from the AdS soliton background into the black brane \eqref{bhmetric}, isotropy between the $\vec{x}$ and $\theta$ coordinates must be restored. From \eqref{Ttt} this means that we must have $b_d=\frac{1}{(d-1)z_0^d}$ and this is indeed verified numerically. The temperature of such a black brane, obtained by the standard procedure of requiring the absence of a conical singularity for the Euclidean version of \eqref{bhmetric} (see Section \ref{sec_adsbh}), is given by $T=d/4\pi\xi_h$. Asymptotically, the radial coordinates $\xi$ and $\zeta$ are related by $\xi=\zeta-\zeta^{d+1}/2d\xi_h^d$, from which, by comparing with \eqref{asymptmetric}, we can obtain the temperature of the black brane as
\begin{equation}
T=\frac{d}{4\pi \xi_h}=\frac{d}{4\pi}\left[\frac{E-\frac{1}{z_0^d}}{d-1}\right]^{1/d}.
\end{equation}

\subsection{Numerical methods}\label{numerics}
In this section we will list some important tricks that we had to employ to achieve stable numerical evolution. It turns out that it is difficult to use the Chebyshev pseudospectral method, which was used for the AdS and hard wall model, due to numerical instabilities. Instead we used a fourth order finite difference method to discretize the radial direction, and then evolved the resulting system of ordinary differential equations in time. We have as initial conditions $f=1$, $h=1$, $b=0$ and $\phi=0$, corresponding to the AdS soliton geometry. The boundary conditions we impose in the UV are $f(0,t)=1$ and $h(0,t)=1$ as well as $\phi(0,t)=J(t)$ and $b(0,t)=0$ ($b(0,t)=J(t)$ and $\phi\equiv0$ if we quench the metric instead of the scalar), and the source is always taken as a Gaussian $J(t)=\epsilon e^{-t^2/\delta t^2}$. In the IR, we do not have to impose any boundary conditions, since regularity already follows from the equations of motion. However, there are some potential sources for numerical instability and inaccuracy, the coordinate singularity in the IR and the AdS boundary being two examples.

\subsubsection{The coordinate singularity}
The $z$ coordinate ansatz \eqref{zansatz} is very inconvenient in the IR. The reason is that at this point the geometry looks locally like Minkowski space in cylindrical coordinates, with rotational invariance in the $(z,\theta)$ plane. However, the relation to the radial coordinate in this locally flat space is $z=z_0(1-r^2)$, and thus $\rd z=-2rz_0 \rd r$. This means that a small grid spacing in $z$ will be mapped to a very large grid spacing in $r$ (which is the natural coordinate around the point $z=z_0$), so a linearly spaced discretization in the $z$ coordinate will become incredibly bad at this point. We thus found it convenient to instead work with the coordinate $r=\sqrt{1-z/z_0}$, and use the metric ansatz
\begin{equation}\label{ransatz}
\rd s^2=\frac{1}{s(r)^2}(-h(r,t)^2\rd t^2+\frac{4f(r,t)^2}{dg(r)} \rd r^2+ r^2g(r)e^{(d-2)b(r,t)}\rd \tilde{\theta}^2+e^{-b(r,t)}\rd \vec{x}_{d-2}^2),
\end{equation}
where $s(r)=z_0(1-r^2)$ and $g(r)=(1-(1-r^2)^d)/(z_0^2r^2d)$. The advantage of this parametrization of the metric is that now $g(r)$ is a finite slowly varying non-zero function and $g(0)=1/z_0^2$. The new periodic coordinate $\tilde{\theta}$ has period $4\pi z_0^2/d^{3/2}$. While the coordinate system \eqref{zansatz} is convenient to derive analytic results and to extract the boundary field theory observables, the coordinate system \eqref{ransatz} will be used for the numerical evolution. It is also clear in these coordinates, that the regularity condition (absence of conical singularity) means that $fe^{(d-2)b/2}$ remains constant in time, which is exactly the statement \eqref{feB2}. For completeness, we here list the equations of motion resulting from the ansatz \eqref{ransatz}, which are 
\begin{align}
\frac{h'}{h}=&\frac{r^2}{s^{2d-2}(\frac{gr^2}{s^{2d-2}})'}\Big(\frac{2(d-1)(d-2)}{d}P^2+\frac{4}{d}\Pi^2+g\Phi^2+\\\nonumber
&+\frac{(d-1)(d-2)}{2}gB^2+\frac{2(d-2)(d-1)gB s'}{s}\Big)+(d-2)B-\frac{f'}{f},\label{heq}\\
\dot{P}=&\frac{d}{4(d-1)r}\left(\frac{\left(ge^{(d-1)b}r^2\right)'he^{-(d-1)b}}{fs^{d-1}r}\right)'s^{d-1},\\
\dot{f}=&\frac{(d-2)(d-1)gr^2P h B s + 2(d-2)(d-1) gr^2 P h s'+2gr^2\Pi\Phi hs}{2(\frac{gr^2}{s^{2d-2}})'s^{2d-1}}+\frac{d-2}{2}P h,\\
\frac{h'}{h}=&\frac{8(d-1)(f^2-1)r^2}{(r^2gs^{-2(d-1)})'s^{2d}}+\frac{f'}{f},\label{constraint}\\
\dot{B}=&\left(\frac{P h}{f}\right)',\\
\dot{\Pi}=&\frac{d}{4r}\left(\frac{hg\Phi r}{fs^{d-1}}\right)'s^{d-1},\\
\dot{\Phi}=&(\frac{\Pi h}{f} )',
\end{align}
where $s(r)=z_0(1-r^2)$ and $g(r)=(1-(1-r^2)^d)/(z_0^2r^2d)$ and we have defined
\begin{equation}
\begin{array}{ccc}
P=\dot{b}\frac{f}{h}, &\hspace{20pt} \Pi=\dot{\phi}\frac{f}{h},\\
\Phi=\phi', &\hspace{20pt} B=b'.\\
\end{array}
\end{equation}
Note that $'$ now indicates derivative with respect to $r$ instead of $z$, and thus $\Phi$ and $B$ are not the same as in the rest of the text. The functions $f$, $P$, $\Pi$, $\Phi$ and $B$ are then evolved in time, while the function $h$ is solved for at each time step using equation \eqref{heq}, and equation \eqref{constraint} is checked for consistency during the time 
evolution.\\
\linebreak
There is also a convenient trick that can be employed to compute derivatives close to an origin of a polar coordinate grid. Usually if one were to employ finite differences close to a boundary, one would have to resort to non-symmetric stencils which can induce instabilities or numerical inaccuracies. However, at $r=0$ we do not have a boundary, and we can imagine continuing $r$ past $r=0$ to negative values and thus it is possible to still use central difference schemes when computing derivatives close to $r=0$. An equivalent way of reaching the same result is to use the fact that all (scalar) functions must be even functions of $r$ when computing the derivatives.

\subsubsection{Radial discretization}
We have found that high order finite difference discretization has worked well. However, to avoid high frequency spurious oscillations, we have found that it is convenient to put different functions on two different grids. To motivate this, consider first a function $v(t,z)$ satisfying the free wave equation $\ddot{v}=v''$. Defining $V=v'$ and $P=\dot{v}$ we obtain
\begin{align}
\dot{V}=P',\hspace{20pt}\dot{P}=V',
\end{align}
which should be compared to the equations of motion for the scalar field and the metric component $b$. Now, if we discretize the $z$ coordinate by $\{z_j\}_{j=0}^n$, and consider the derivative approximation $(z_{j+1}-z_{j})/\Delta z$, this will compute an approximation to the derivative at the point $(z_j+z_{j+1})/2$. We thus see that it might be convenient to put $V$ and $P$ on two different grids, one on $\{z_j\}_{j=0}^n$, and one one $\{x_j\}_{j=0}^n$ where $x_j=(z_j+z_{j+1})/2$, to improve the accuracy of the derivative approximations. If one were to use a central difference scheme, we find that it typically induces high frequency noise. This high frequency noise is still present when using higher order central difference schemes, but disappears when putting $V$ and $P$ on different grids (also when we use a higher order finite difference scheme).\\
\linebreak
In our more complicated setup, the same reasoning holds for the free wave equation in AdS, and we have found it very useful to employ the same trick even when including backreaction. Thus we have put $\Phi$ and $B$ on one grid, and $\Pi$, $P$ and $f$ on the other. Function values are then interpolated to the other grid when necessary. This proves to result in very stable evolution and the high frequency noise that is present when using central difference schemes with all functions on the same grid disappears, although we do not have a theoretical explanation of this phenomenon.

 \subsubsection{Extracting boundary data}
 To extract the boundary data, we will have to compute quantities like $(f(z)-f(0))/z^d$ when $z\rightarrow0$. This becomes increasingly difficult when the dimension increases, since we are taking the ratio of two very small numbers. In particular for $d=6$, there is high frequency noise which makes it difficult to extract the observables. For the simulations of black hole formation (Figure~\ref{qnm}), we therefore found it appropriate to use a Savitzky-Golay \cite{citeulike:4226570} filter to get rid of this noise and to make the boundary observables more smooth in time. The fact that the final result is insensitive to grid spacing and that it agrees with the quasinormal modes of the formed black brane indicates that the result after applying this filter can be trusted.


\subsection{Phase diagram}\label{phasediagram}
\begin{figure}[t]
\centering
\includegraphics[scale=0.3]{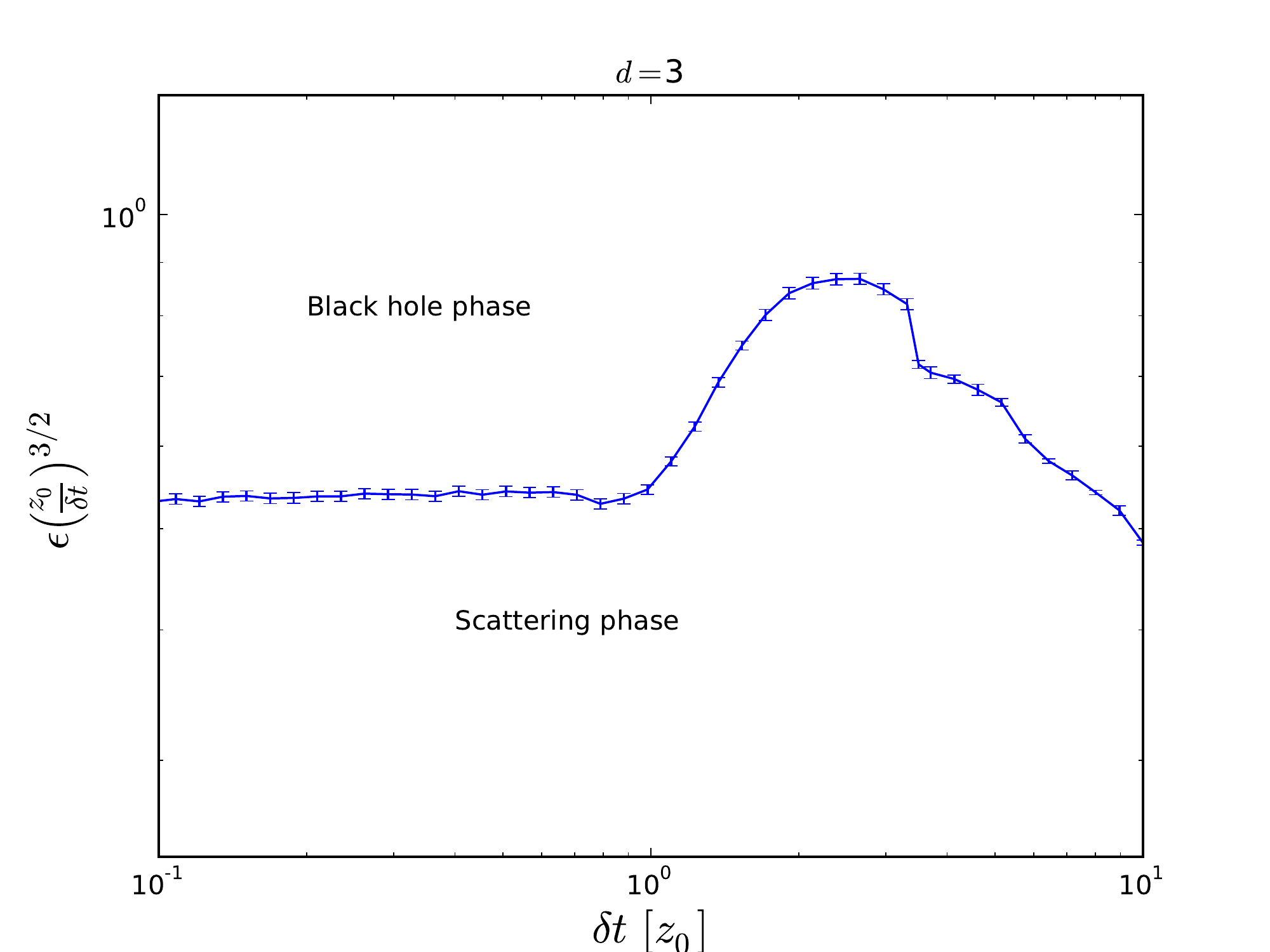}
\includegraphics[scale=0.3]{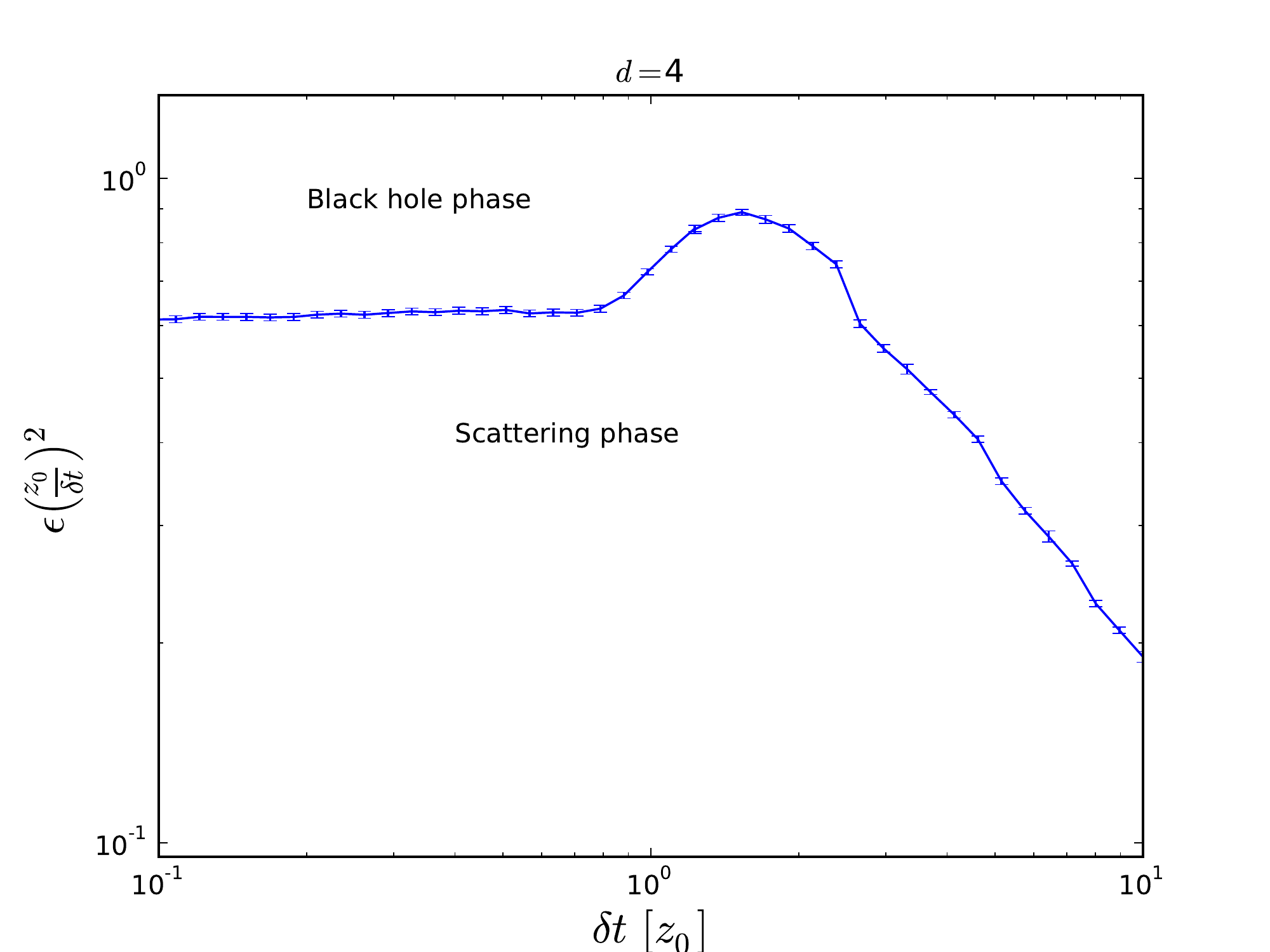}
\includegraphics[scale=0.3]{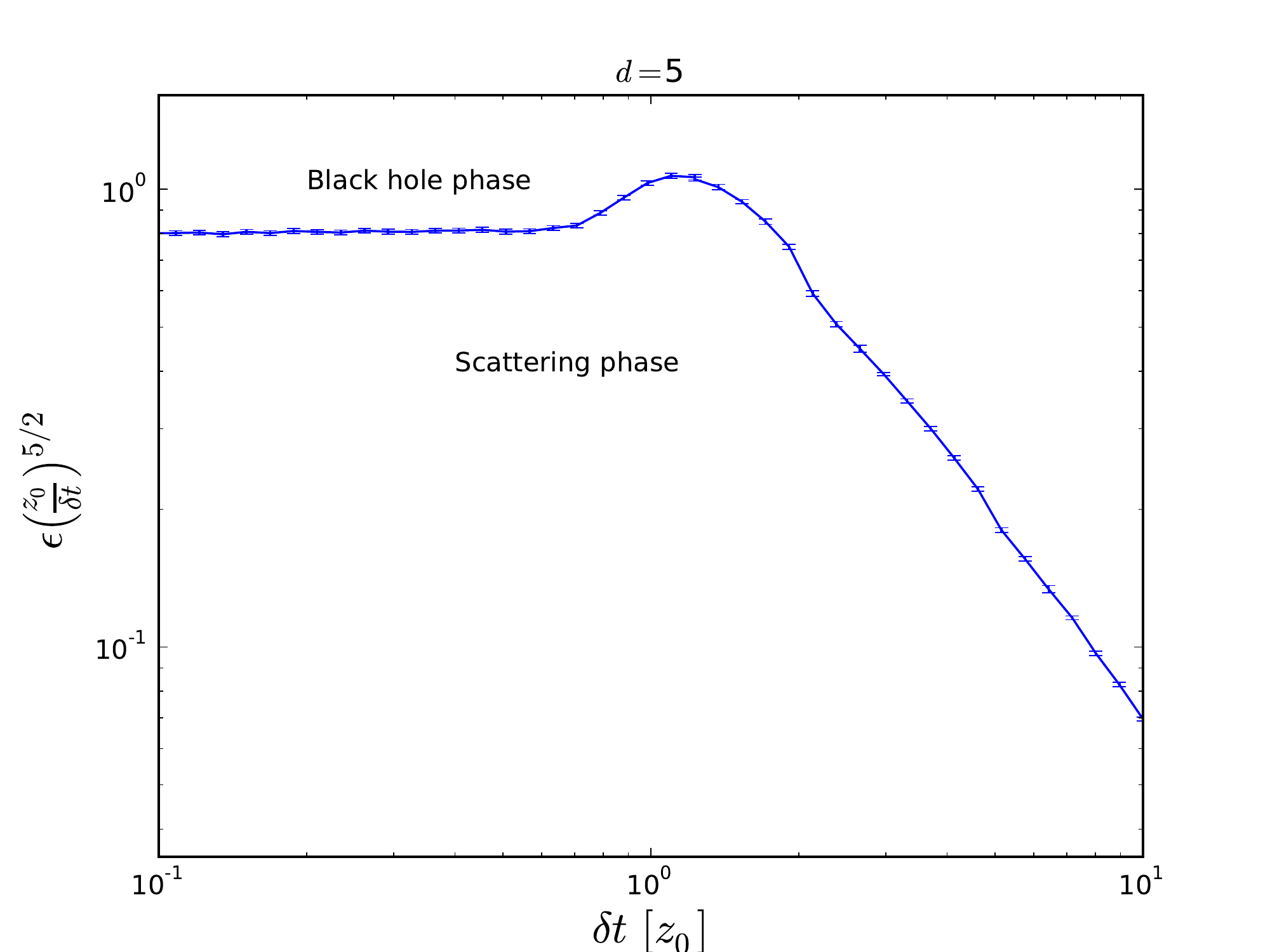}
\includegraphics[scale=0.3]{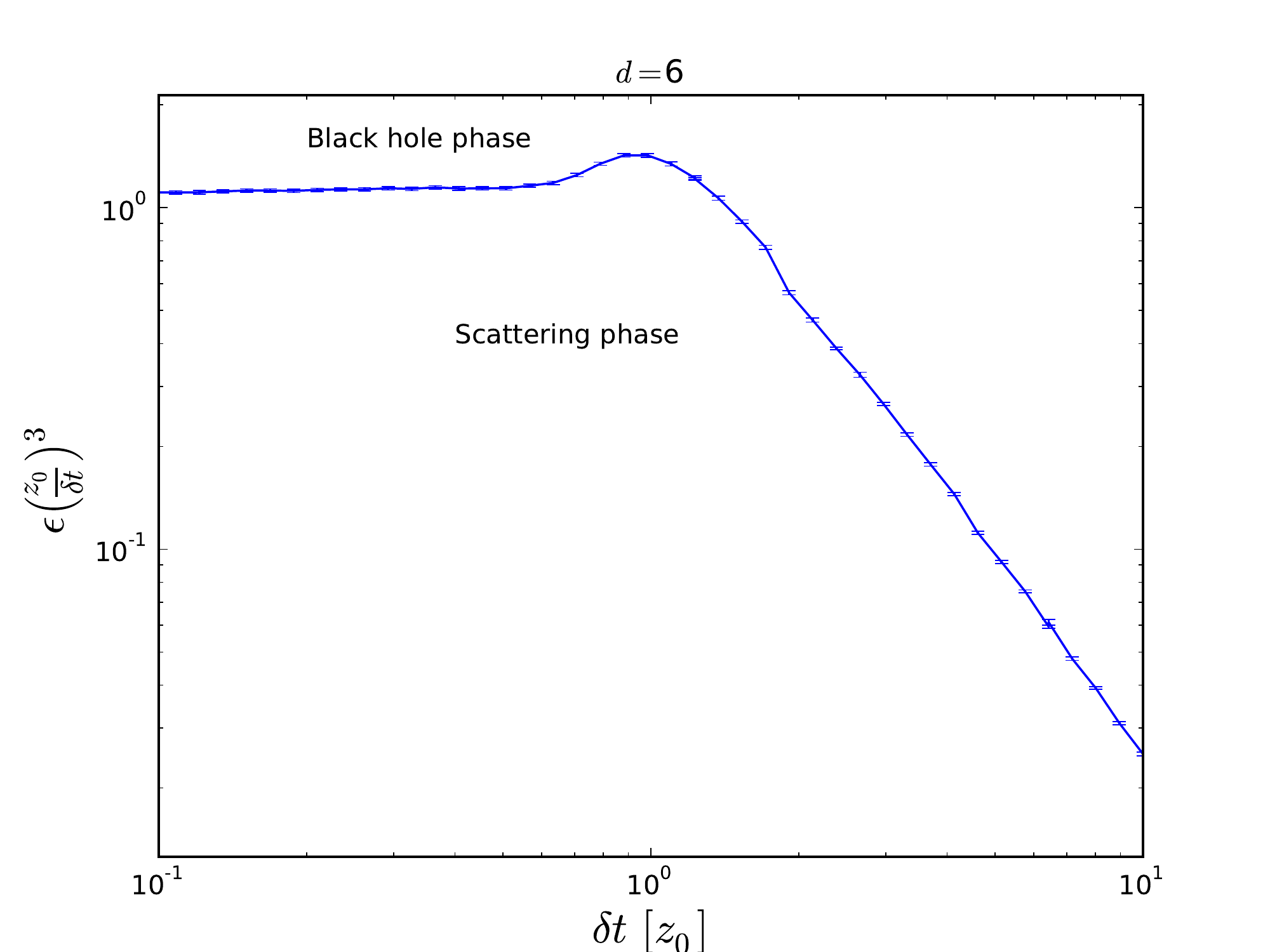}
\caption{The separation between the black hole phase and the scattering phase. For small $\delta t$, we see that $\epsilon\sim\delta t^{d/2}$, which is expected since the total injected energy is expected to go like $E\sim\epsilon^2/\delta t^d$  \cite{Bhattacharyya:2009uu}. For large $\delta t$, we find the relation $\epsilon\sim\delta t$. }
\label{phase_diag} 
\end{figure}

\begin{figure}[t]
\centering
\includegraphics[scale=0.7]{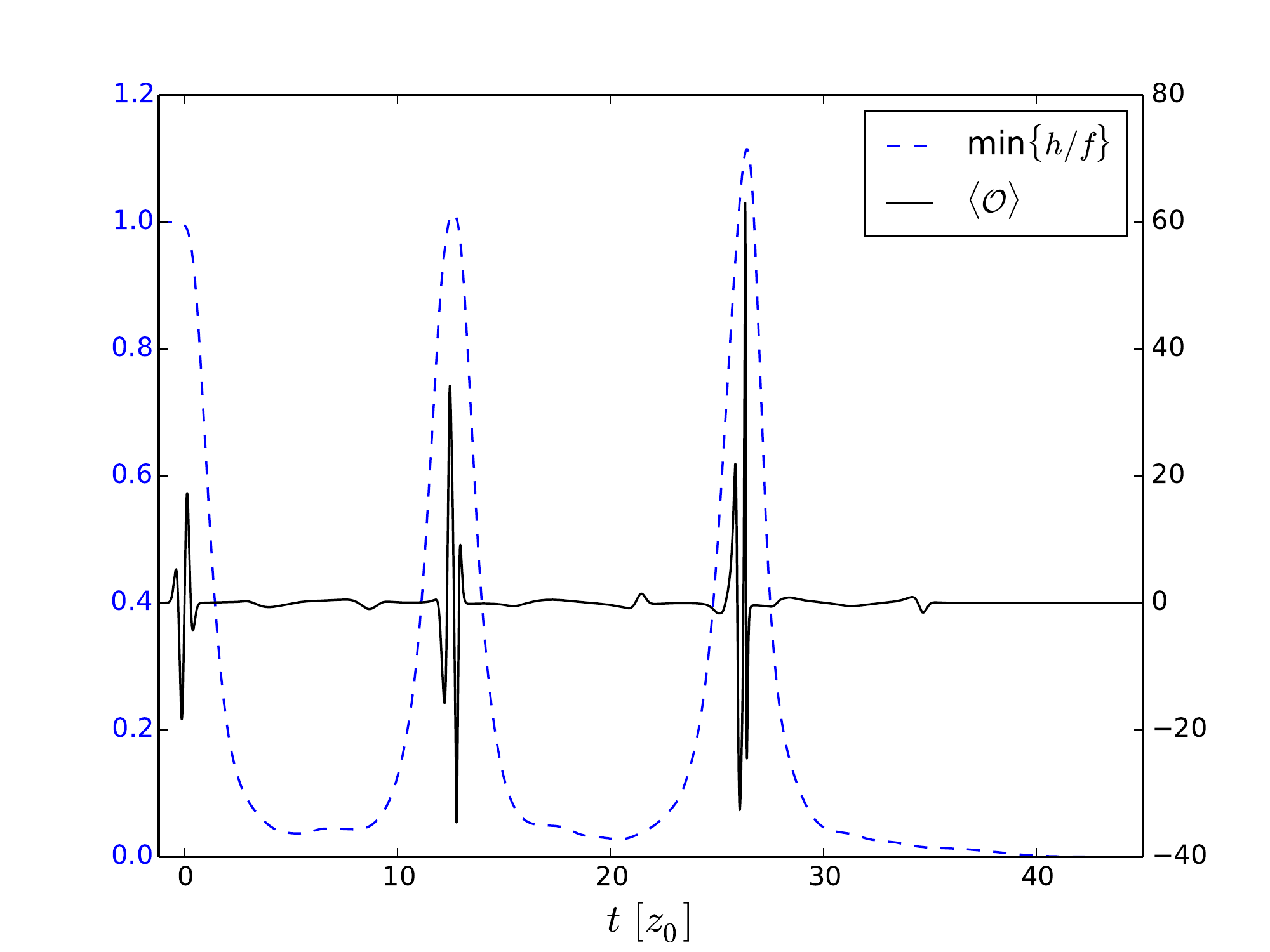}
\caption{Example of a quench where the scalar wave packet reflects twice at the boundary before collapsing into a black brane. Time is here in units of $z_0$. The parameters are $d=3$, $\epsilon=0.06305472$ and $\delta t=0.24 z_0$. The left axis is for $h/f$ and the right axis is for $\langle \mathcal{O} \rangle$ in units of $1/(16\pi \GN  z_0^d)$. Vanishing of $h/f$ signals the formation of an apparent horizon.}
\label{reflections} 
\end{figure}

\begin{figure}[t]
\centering
\includegraphics[scale=0.7]{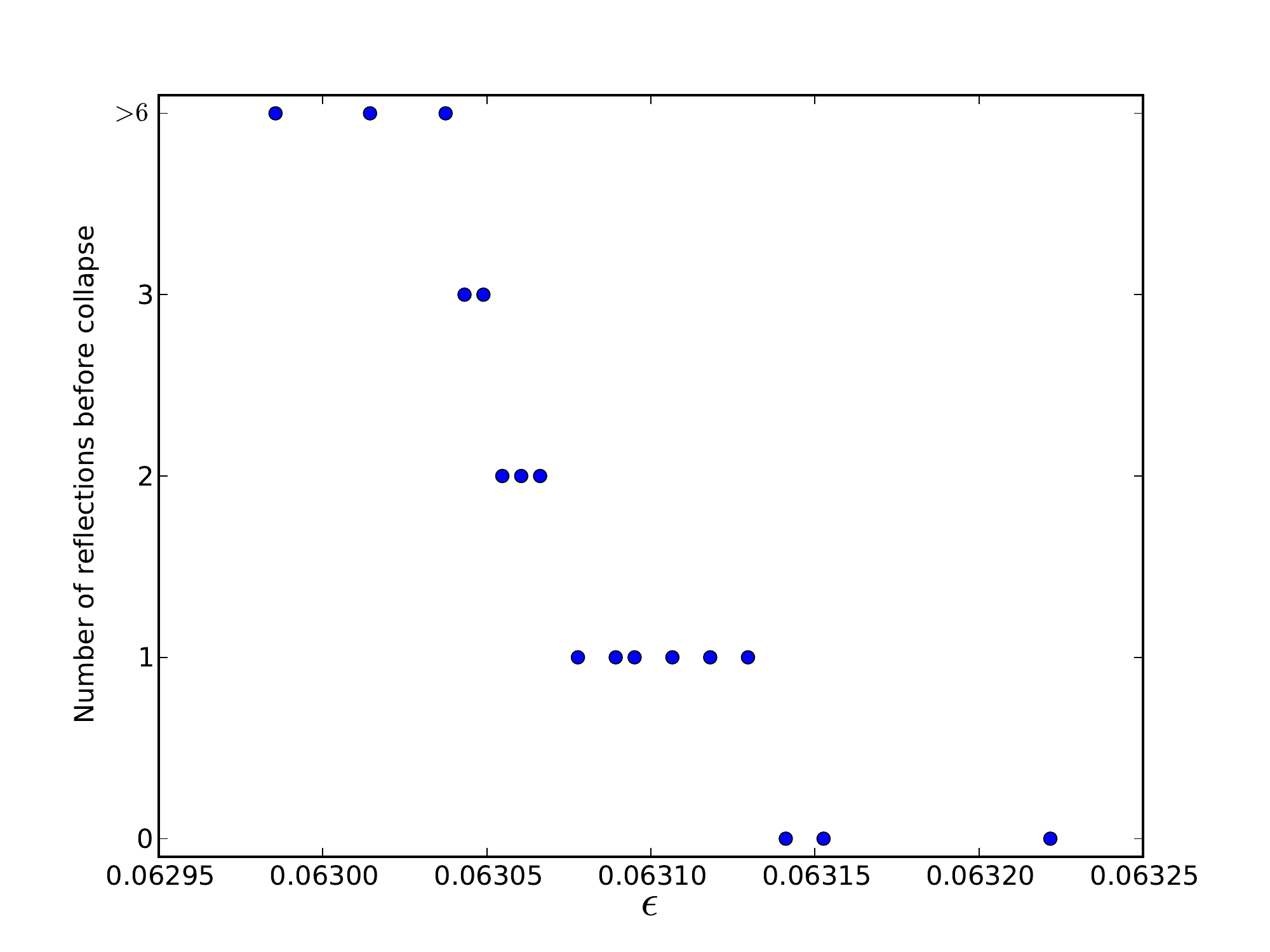}
\caption{Number of reflections at the boundary of the scalar field wave packet before black brane formation, as a function of $\epsilon$ for $d=3$ and $\delta t=0.24 z_0$. Note that for $\epsilon$ smaller than $0.06304$, there is a parameter region where the injected energy density is above the black brane threshold, but nevertheless the solutions seem to scatter for as long as we have been able to follow them. This region is relatively large, since the threshold where the energy density becomes negative is $\epsilon\approx0.0607$.}
\label{nrofbounces} 
\end{figure}

As we have seen, when injecting energy into (the Poincar\'e patch of) vacuum AdS, we always form a black brane. However, since the energy density of the AdS soliton is negative, and any black brane has positive energy density, there should be a threshold for black hole formation when injecting energy into the AdS soliton. This is similar to the hard wall model, and we may ask what solutions we expect if the energy is below the threshold. In the probe limit, the scalar field will just bounce forever between the boundary and the IR, so one could ask if this behaviour will still remain when turning on backreaction, or if the system will equilibrate into some other static solution after a long time. In Section~\ref{static}, we prove that the system cannot equilibrate into any static solution (the proof is similar to that for the hard wall model but more involved). We will thus use the same terminology as in the hard wall model and refer to these solutions as the ``scattering phase'', and the solutions that thermalize into black holes as the ``black hole phase''. In Figure \ref{phase_diag}, we show the separation between the two different phases, in terms of the  parameters $\epsilon$ and $\delta t$. For small $\delta t$ we have the relation $\epsilon\sim\delta t^{d/2}$, which is expected since the injected energy (which is the only parameter associated to the shell in the thin shell limit) is expected to go like $E\sim\epsilon^2/\delta t^d$\cite{Bhattacharyya:2009uu,Ishii:2015gia} (although the overall factor will not be the same as in AdS). The shapes of the phase diagrams resemble those found for the hard wall model (see Figure \ref{eps_delta_hw}). In particular, for large $\delta t$ we find numerically the relation $\epsilon\sim\delta t$, which is the same as in the hard wall model with Neumann boundary conditions.\\
\linebreak
Another interesting question, which we did not explore for the hard wall model, is if we can find scattering solutions above the energy threshold. Intuitively, right above the threshold, a wave packet should bounce before collapsing into a black brane due to the finite width of the wave packet, and this is indeed what we find. Namely, right above the threshold when black brane formation is possible (the energy density is positive), there is a region where solutions reflect many times against the boundary without collapsing (although we are not able to say whether they eventually collapse, due to numerical difficulties in following the solutions for a long time). We have also found solutions that bounce a few times and then collapse into a black hole. In Figure \ref{nrofbounces} we plot the number of scatterings before collapse, as a function of amplitude $\epsilon$ for fixed $\delta t=0.24 z_0$. We see that when decreasing $\epsilon$ the number of reflections against the boundary before collapse varies between 0 and 3, and then for smaller $\epsilon$ there is a large region where the solutions do not seem to collapse. This is similar to what was found in global AdS \cite{Bizon:2011gg}, although in global AdS the behaviour seems to be more regular and easier to understand.\\
\linebreak
In Figure~\ref{reflections}, we show the vacuum expectation value of the scalar operator and min$\{f/h\}$ as a function of time, for a solution that bounces twice before collapsing into a black hole. After two reflections (identified by the sharp peaks in the vacuum expectation value) we see that min$\{f/h\}$ approaches zero, which indicates the formation of an apparent horizon. If the wave packet is very close to collapsing to a black hole while it scatters in the IR, the wave packet usually becomes very squeezed and comes out almost like a shock wave, resulting in the very sharp peaks in the expectation value $\langle \mathcal{O} \rangle$. 

\subsection{Black hole phase}\label{blackholephase}

\begin{figure}[t]
\centering
\includegraphics[scale=0.6]{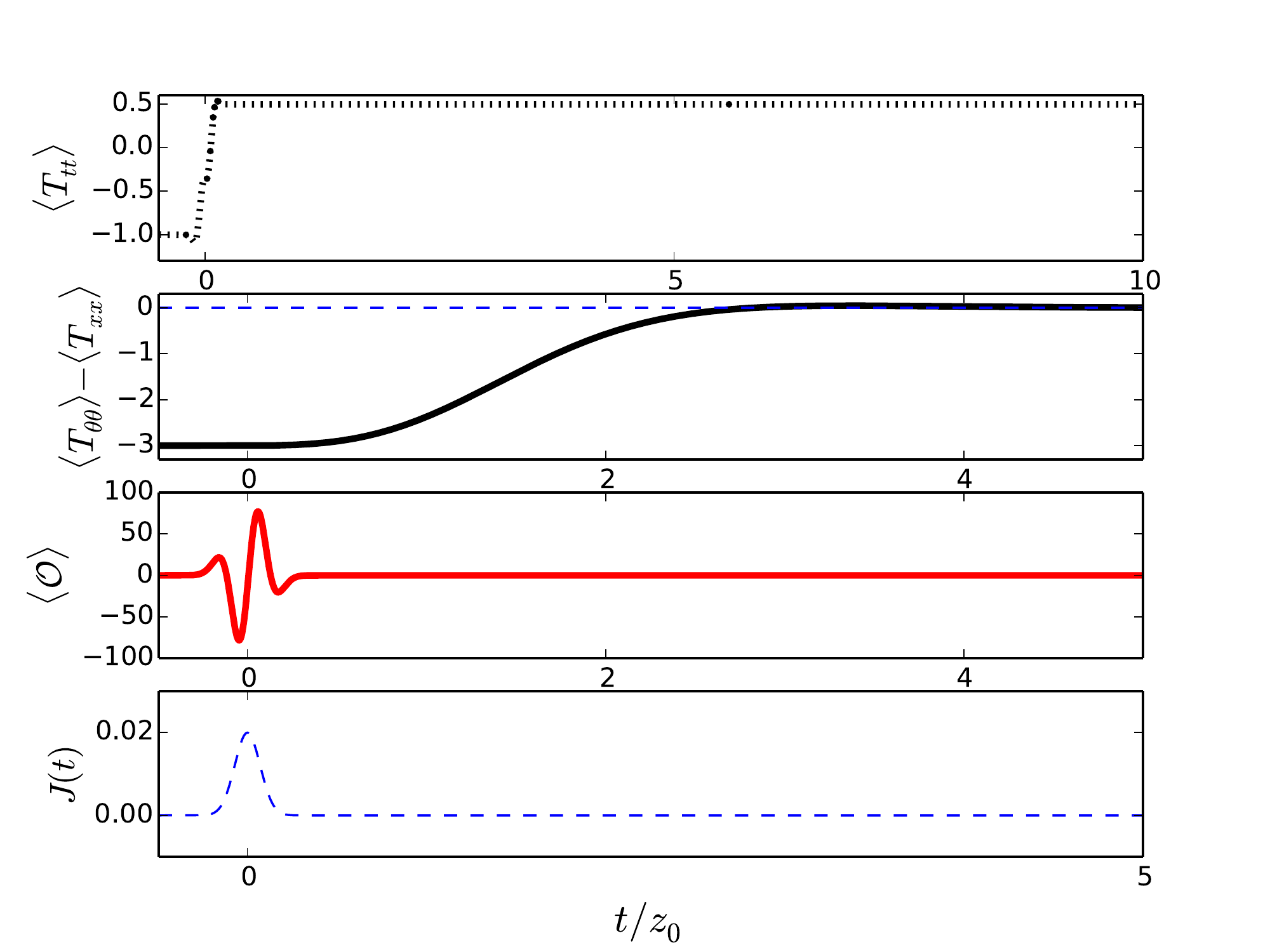}
\caption{\label{blackholesol} Evolution of the difference in the pressure components, the vacuum expectation value of the operator dual to the scalar field and the boundary energy density when a black hole forms (in units of $1/(16\pi \GN z_0^d)$). The parameters are $\epsilon=0.02$ and $\delta t=0.1z_0$. The temperature of the black brane in this example is $T\approx0.15/z_0$.}
\end{figure}

\begin{figure}[t]
\centering
\includegraphics[scale=0.33]{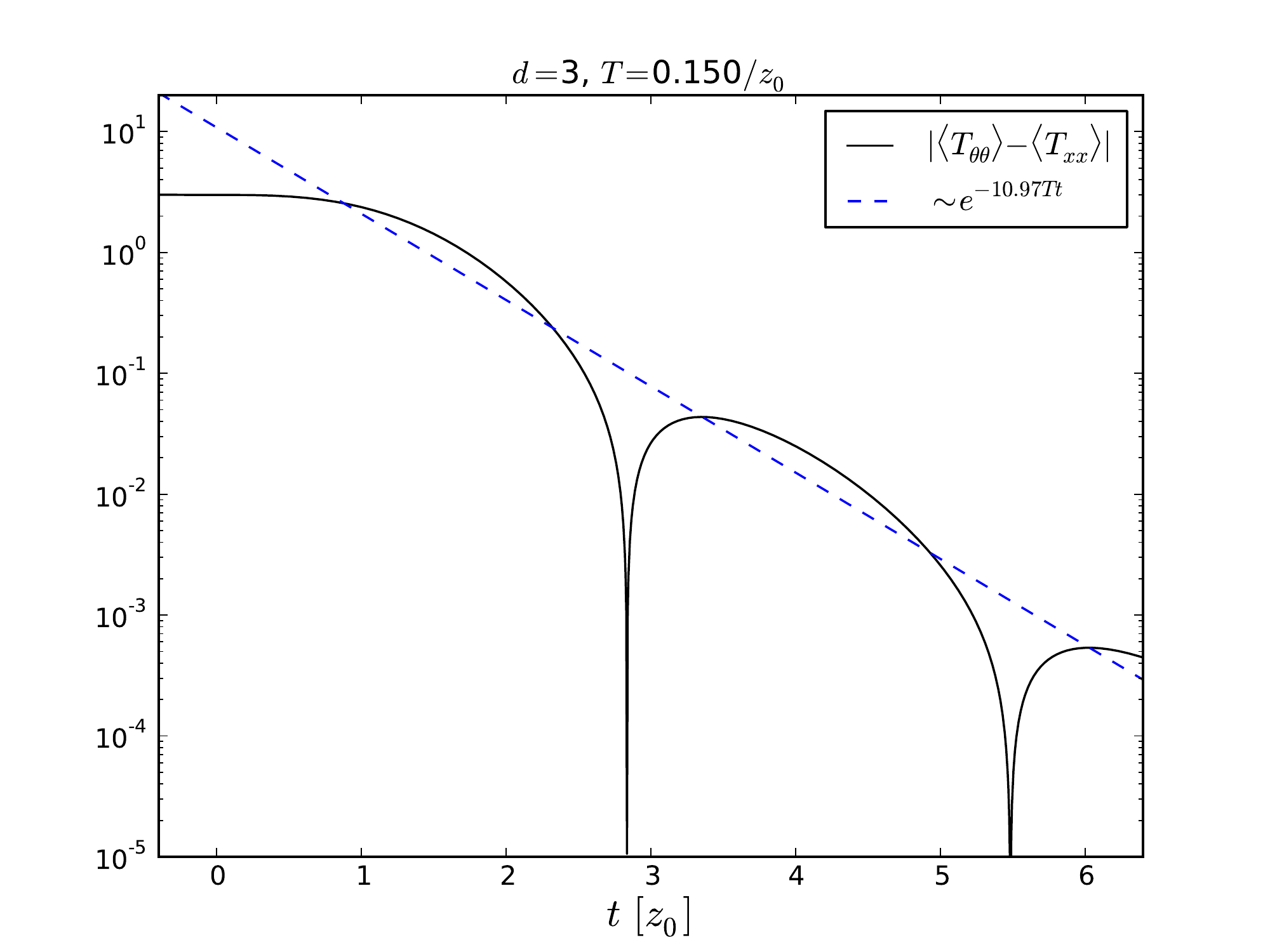}
\includegraphics[scale=0.33]{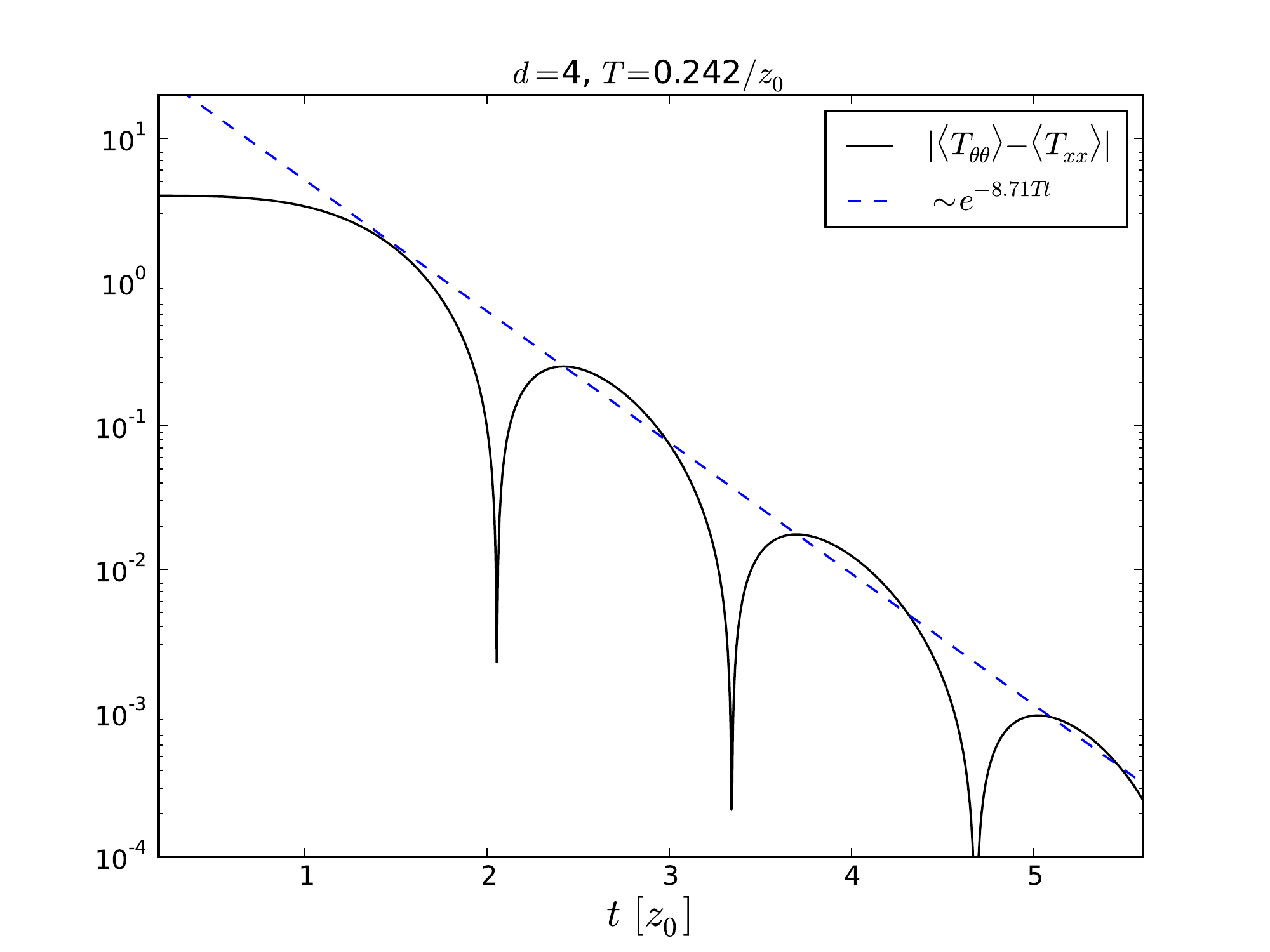}
\includegraphics[scale=0.33]{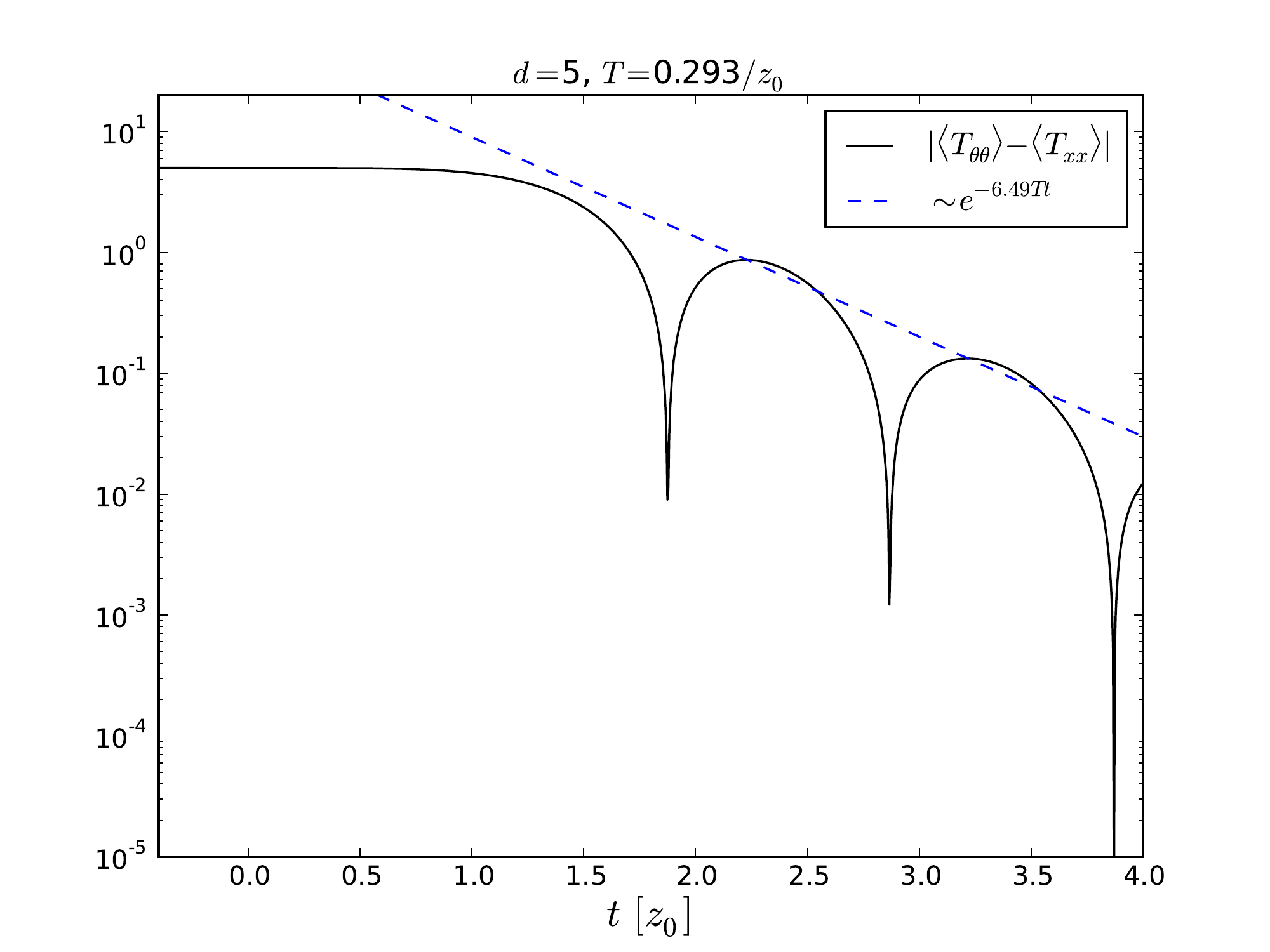}
\includegraphics[scale=0.33]{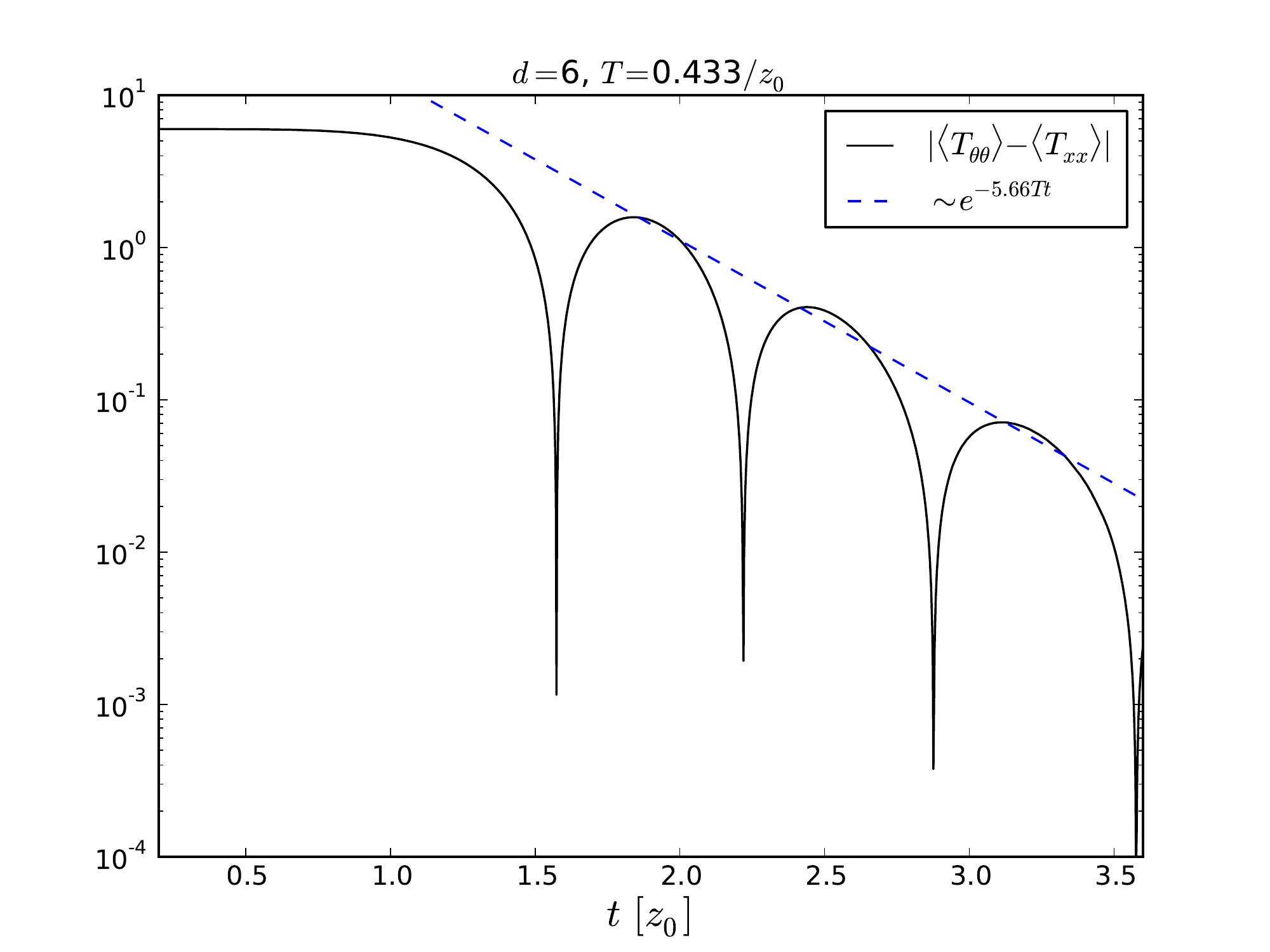}
\caption{Log-lin plots of the absolute deviation from zero of the pressure differences in the late time regime of some black hole collapse processes for various dimensions. Time is expressed in units of $z_0$, $\delta t=0.1z_0$, and the vacuum expectation values are given in units of $1/(16\pi \GN  z_0^d)$. The decay constants $10.97T$, $8.71T$ and $5.66T$, where $T$ is the temperature, are in good agreement with the lowest quasinormal modes ($11.16T$, $8.63T$ and $5.47T$, respectively) quoted in \cite{Horowitz:1999jd}.}
\label{qnm} 
\end{figure}

In the black hole phase, the spacetime will collapse into a black brane, and a horizon will form. The resulting solution will be an AdS$_{d+1}$ black brane. This in particular means that isotropy between the $\theta$ coordinate and the $\vec{x}$ coordinates will be restored, which implies from equation \eqref{Ttt} that $b_d=\frac{1}{(d-1)z_0^d}$, and this is indeed verified numerically. Thus the pressure anisotropy $\langle T_{\theta\theta}-T_{xx}\rangle$ will dynamically evolve from $-d/16z_0^d\pi \GN $ to 0. A relevant question is what this isotropization process looks like. In Figure~\ref{blackholesol}, we show a typical evolution of the energy density, pressure anisotropy and vacuum expectation value for the scalar for $d=3$. The situation looks similar to that in Section \ref{adsnumbh} in the sense that the energy density quickly goes to a constant (as dictated by energy conservation) and the scalar vacuum expectation value is only significant while the source is on. However, we now also have an isotropy in the boundary field theory pressure components which goes to zero on a slower time scale.\\
\lb
As we explained in Section \ref{sec_qnm}, the late-time dynamics should be decomposable into the quasinormal modes of the black brane, and thus field theory observables are expected to be governed by the lowest quasinormal mode. Our numerics will not allow us to follow the evolution for very long times after a black hole has formed, but long enough to see the quasinormal mode behaviour. A standard way to illustrate this behaviour is to plot the logarithm of the absolute value of the deviation of some observable from its final value, as in Figure \ref{BMvsqnm}. In Figure~\ref{qnm}, a few examples of the deviation of the pressure difference $\langle T_{\theta\theta} \rangle-\langle T_{xx} \rangle$ from 0 are shown. We see that, as expected, the decay time is set by the lowest quasinormal mode, since the decay constants $10.97T$ (for $d=3$), $8.71T$ (for $d=4$) and $5.66T$ (for $d=6$) are in good agreement with the values for the lowest quasinormal mode frequencies of AdS Schwarzschild black branes obtained in \cite{Horowitz:1999jd}, namely $11.16T$, $8.63T$ and $5.47T$, respectively. 

\subsection{Scattering phase}\label{scatteringphase}
\begin{figure}[t]
\centering
\includegraphics[scale=0.6]{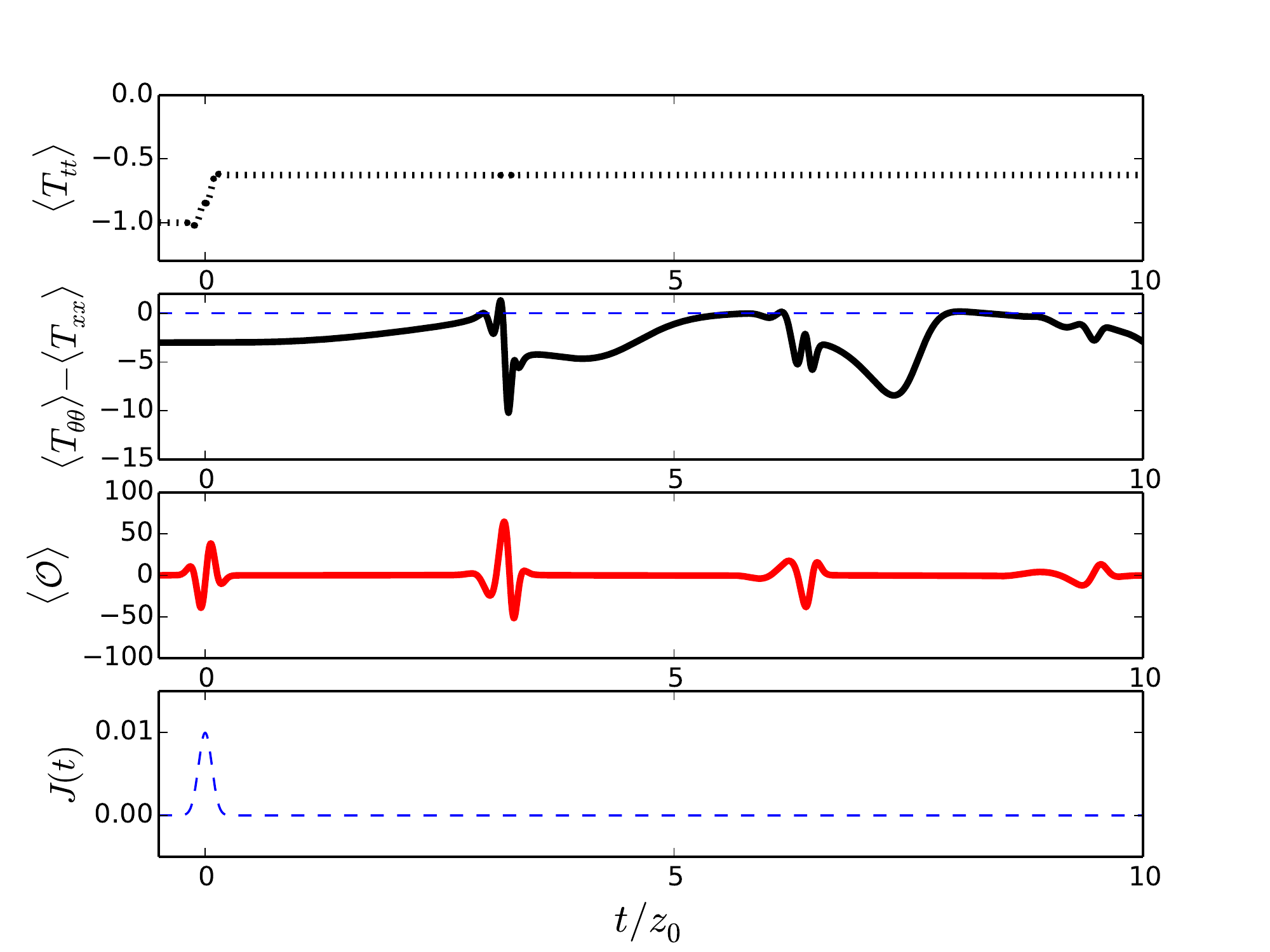}
\caption{Evolution of the difference in the pressure components, the vacuum expectation value of the operator dual to the scalar field and the boundary energy density in a scattering solution (in units of $1/(16\pi \GN z_0^d)$). The parameters are $\epsilon=0.01$ and $\delta t=0.1z_0$.}
\label{scattering}
\end{figure}

\begin{figure}[t]
\centering
\includegraphics[scale=0.6]{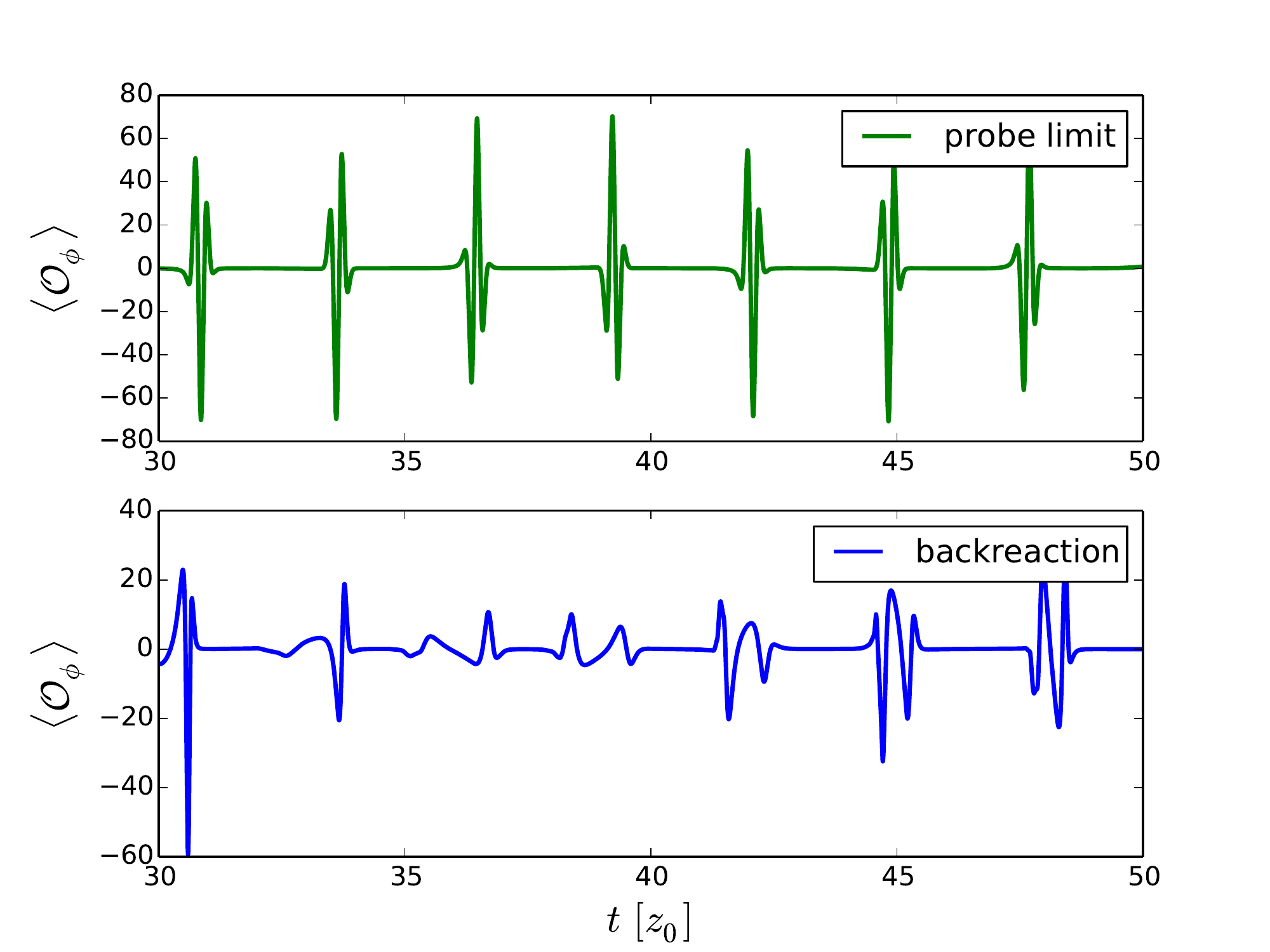}
\caption{The scalar expectation value compared with the probe limit result (in units of $1/(16\pi \GN z_0^d)$). There is a clear distortion of the wave packet which is due to the non-trivial dynamics in the full Einstein equations and can not be seen in the probe limit, although the wavepacket remains fairly localized. This example is for $d=3$, with parameters $\epsilon=0.01$ and $\delta t=0.1z_0$, and time is in units of $z_0$. Note that this is already quite far into the non-linear regime since black hole formation occurs around $\epsilon\approx0.016$.}
\label{waveform} 
\end{figure}

\begin{figure}[t]
\centering
\includegraphics[scale=0.6]{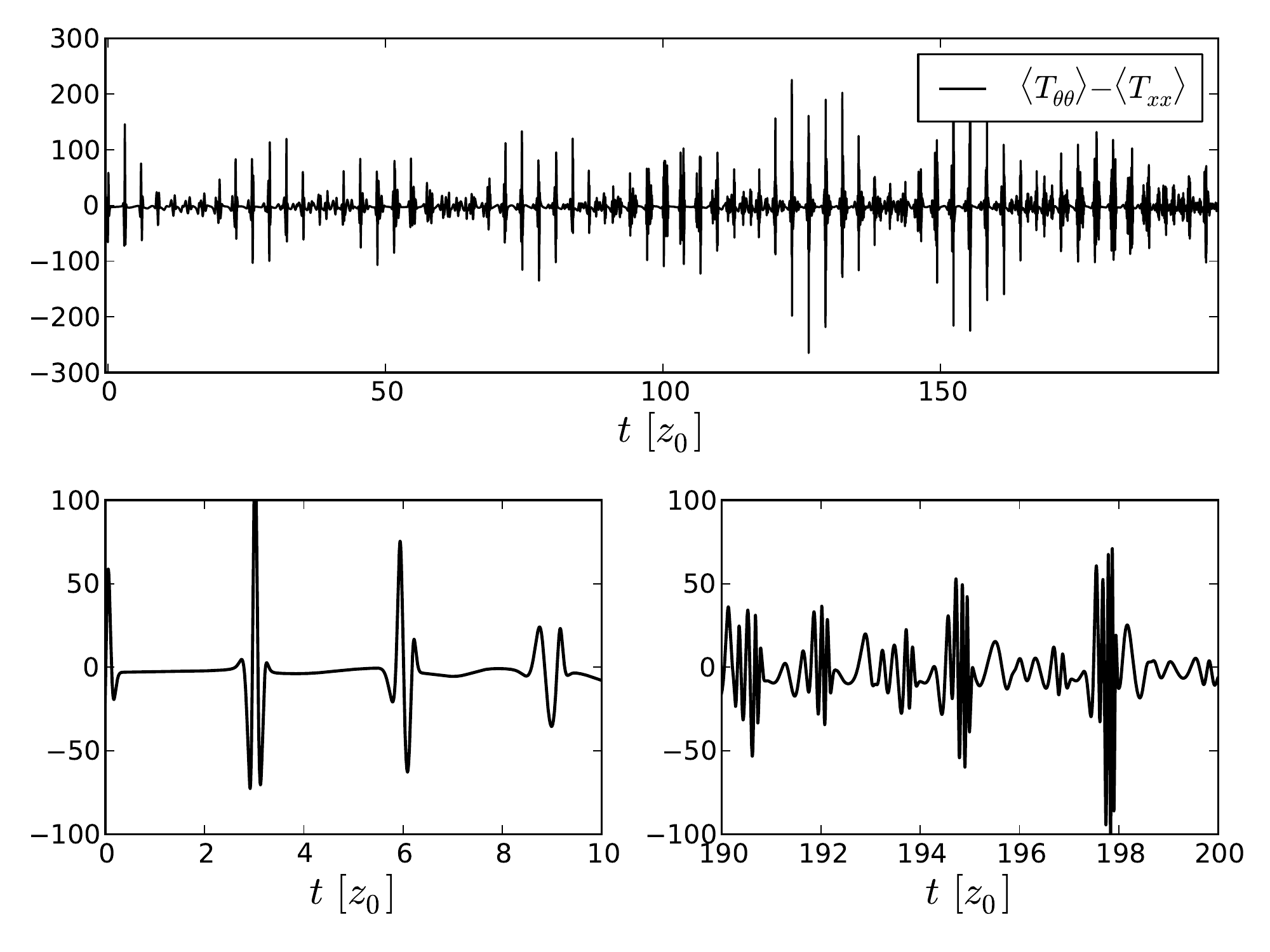}
\caption{The pressure difference after quenching the metric (in units of $1/(16\pi \GN z_0^d)$), with $\epsilon=0.008$ and $\delta t=0.1z_0$ for $d=3$. Time is in units of $z_0$. We see signs of transfer to high frequency modes for late times.}
\label{gwscattering} 
\end{figure}

\begin{figure}[t]
\centering
\includegraphics[scale=0.6]{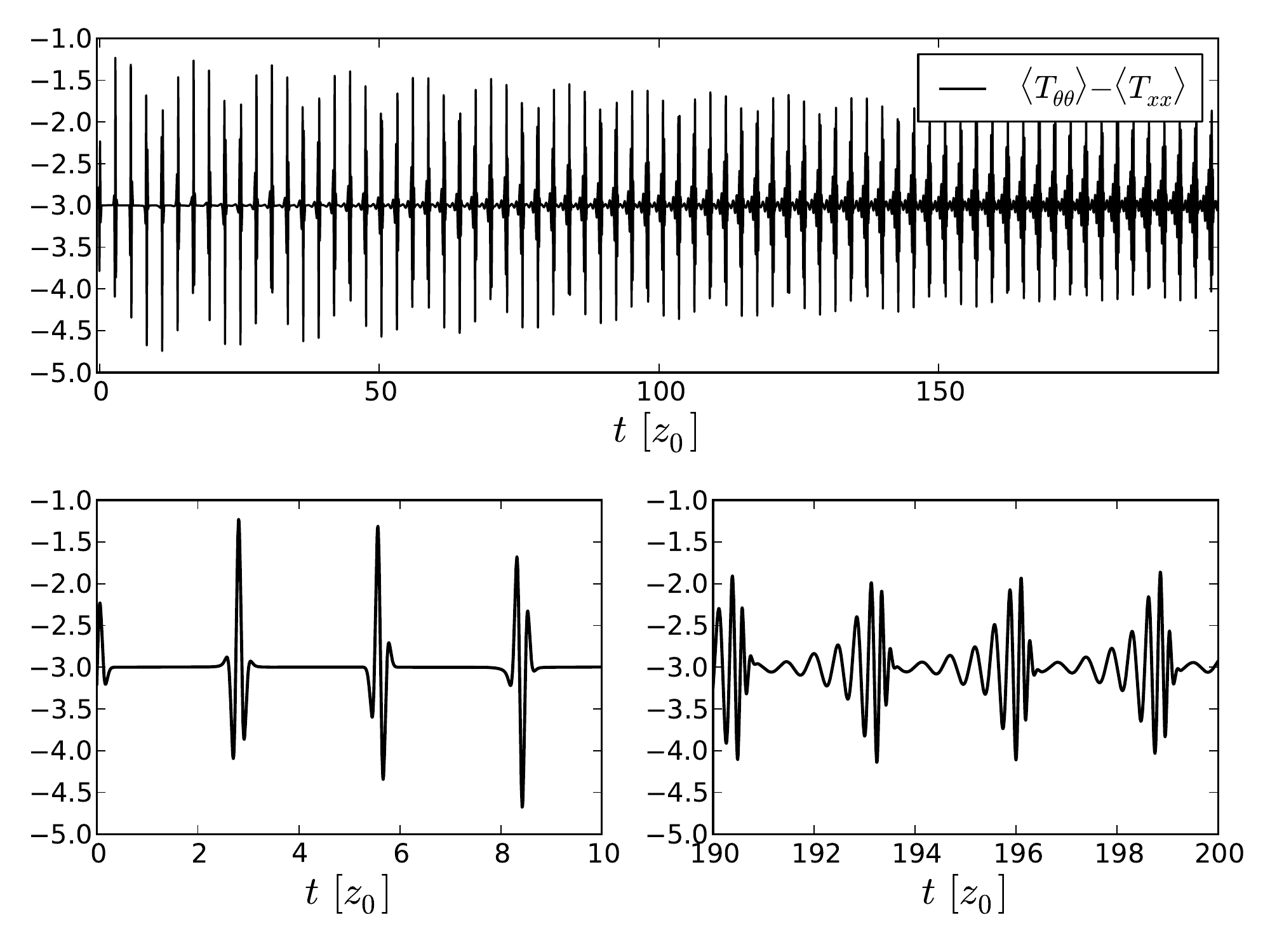}
\caption{The pressure difference after quenching the metric (in units of $1/(16\pi \GN z_0^d)$), with $\epsilon=10^{-4}$ and $\delta t=0.1z_0$ for $d=3$. Time is in units of $z_0$. This figure is unchanged (except for an overall rescaling of the deviations from $-3$), when $\epsilon$ is decreased further. For these small amplitudes, we find no significant transfer to high frequency modes.}
\label{gwscattering001} 
\end{figure}

In the scattering phase, the scalar field wave packet that falls from the boundary will bounce in the deep IR and return to the boundary. When it reaches the boundary there will typically be some excitation of the boundary observables. The wave packet then reflects from the boundary and the scattering repeats. There will be a similar quasiperiodic behaviour in the metric, since due to the broken rotational symmetry between the $\vec{x}$ and $\theta$ coordinates the metric has dynamical degrees of freedom of its own. For all figures we have varied the grid spacing to make sure that the results are not numerical artifacts.\\
\linebreak
In Figure~\ref{scattering}, we show a typical scattering solution. As we can see, every time the scalar field wave packet reaches the boundary, there is a bump in the expectation value, and this oscillation goes on forever as far as we know. We can also see that the dynamical degrees of freedom in the metric are excited, as expected, leading to non-trivial behaviour in the boundary pressure components. This figure should be contrasted to its analog in the hard wall model, Figure \ref{evolhw2}.\\
\linebreak
One interesting feature is that the interpretation of the scattering solution as a localized wave packet persists for very long times, even for solutions where the non-linearities play a significant role. This is definitely not obvious since one could have imagined that the wave packet would broaden and that at late times we would have seemingly random fluctuations, but instead we see that the wave packet remains approximately localized for long times. However, the shape of the wave packet can change with time due to the non-trivial dynamics of the full Einstein equations, as is shown in Figure~\ref{waveform} where we compare a scattering solution close to the black hole threshold with a probe limit computation.\\
\linebreak
In Figure~\ref{gwscattering} we show a typical long time scattering solution when the metric is quenched according to equation \eqref{bsource}. For late times, it looks like the pressure anisotropy develops increasingly sharp features, which could suggest transfer of energy from low to high frequency modes and could be the sign of an instability. If that is the case it would be reminiscent of what happens in global \ads after a rotationally symmetric small-amplitude perturbation where transfer of energy to high frequency modes was indeed interpreted as an instability of \ads \cite{Bizon:2013xha}. To see if this persists for smaller amplitudes, we repeat the analysis for a weaker source in Figure \ref{gwscattering001}, and also in this case it looks like (in the upper panel) that there are sharper and sharper features developing. If the dynamical behaviour is similar to that in \cite{Bizon:2013xha}, namely resulting from secular terms showing up in the perturbative expansion in the source, the transfer of energy to high frequency modes should occur on a time scale that goes to infinity as $1/\epsilon$. However, the behaviour we see in Figure \ref{gwscattering001} is independent of amplitude and lowering the amplitude further just rescales the y axis. Moreover, while a Fourier analysis of \ref{gwscattering} might indicate some transfer of energy to high frequency modes, there is no hint of this in a Fourier analysis of Figure \ref{gwscattering001} and thus the sharper features for later times have nothing to do with transfer of energy to higher frequencies. The Fourier spectrum of \ref{gwscattering001} is actually completely dominated by the normal modes (it has peaks at the normal mode spectrum) with similar amplitudes for all times and thus the metric perturbations are just a completely regular linear superposition of the normal modes. Of course we can not exclude that there are instabilities showing up on longer time scales than what we have studied, but we do not expect that to be the case due to the non-resonant spectrum of the normal modes (note that the spectrum of scalar perturbations in \ads is resonant, which explains the instability in that case).\\
\linebreak
One important question is whether or not these scattering solutions will go on forever, or whether the system will approach some static solution. We will now show that, if the injected energy density is below the black brane threshold, the system can not equilibrate to a static solution.

\subsubsection{Static solutions and non-thermalization}\label{static}
 From equation \eqref{Ttt} we see that we are not able to form a black brane if $E<1/z_0^d$. However, one could imagine that there are other static solutions that the system can end up in. In this section we will show that if $E<1/z_0^d$ there are no static solutions that can be obtained through time evolution. The argument is similar to the argument we gave for the hard wall model but much more involved. To summarize the argument, the key information we get from the dynamical equations is the relation \eqref{feB2}. This condition is essentially the requirement that the spacetime should be regular at $z=z_0$ (such that a conical singularity can not be formed at this point during time evolution). We will then consider static solutions, by looking at the static equations of motion, and show that any possible static solution is incompatible with \eqref{feB2}. Actually, most of the solutions have a completely different asymptotic behaviour at $z=z_0$ and are trivially excluded. The only solutions for which $fe^{-\frac{d-2}{2}b}$ goes to a constant, turn out to be the AdS soliton solutions. However, as we will see, all AdS solitons except our initial condition soliton will have $fe^{-\frac{d-2}{2}b}$ approaching a different constant than 1, violating \eqref{feB2}, so if the injected energy is non-zero, no static solutions can form. The intuitive picture is that we can not form any other AdS Soliton solutions since that would require changing the periodicity of the compactified coordinate.\\
\linebreak
To investigate this we will start by considering a different coordinate system, such that the metric takes the form
\begin{equation}
\rd s^2=\frac{1}{\hz^d}(-\hat{h}^2\rd t^2+\rd \hat{z}^2 \hat{f}^2+\rd \theta^2 e^{(d-2)\hat{b}}+\rd \vec{x}_{d-2}^2e^{-\hat{b}}). \label{newansatz}
\end{equation}
This can be obtained by the coordinate transformations and field redefinitions given by
\begin{equation}
\begin{array}{ccc}
b=\hb-\frac{\log G}{d-1}, &\hspace{10pt} z=\hz G^{\frac{1}{2(d-1)}},\\
\hf^2=\frac{f^2}{G(1-z \frac{G'}{2(d-1)G})^2}, &\hspace{10pt} \hh^2=h^2G^{-\frac{1}{d-1}},\\
\end{array}\label{transformations}
\end{equation}
 where $G(z)=1-z^d/z_0^d$ and $G'(z)=-dz^{d-1}/z_0^d$. The $\hz$ coordinate now ranges from $0$ to $\infty$. The (static) equations of motion for such an ansatz are
\begin{equation}
\frac{\hh'}{\hh}=\frac{\hf'}{\hf}-\frac{d}{\hz}(\hf^2-1),\label{hfeq1}
\end{equation}
\begin{equation}
\frac{\hh'}{\hh}=-\frac{\hf'}{\hf}-\frac{d-2}{4}\hz\hat{B}^2-\frac{1}{2(d-1)}\hz\hat{\Phi}^2,\label{hfeq2}
\end{equation}
\begin{equation}
(\frac{\hat{B}\hat{h}}{\hat{f}\hz^{d-1}})'=0,\label{Bseq}
\end{equation}
\begin{equation}
(\frac{\hat{\Phi}\hat{h}}{\hat{f}\hz^{d-1}})'=0,\label{Phiseq}
\end{equation}
where $\hB=\hat{b}'$, $\hat{\Phi}=\phi'$ and prime now denotes derivative with respect to $\hz$. We can integrate \eqref{Bseq} and \eqref{Phiseq} to obtain
\begin{equation}
\hat{\Phi}=C_\phi \frac{\hat{f}}{\hat{h}} \hz^{d-1},
\end{equation}
\begin{equation}
\hat{B}=C_b \frac{\hat{f}}{\hat{h}} \hz^{d-1}.
\end{equation}
where $C_b$ and $C_\phi$ are integration constants, tuning the UV behaviour. From \eqref{hfeq1} we have $\hat{f}/\hat{h}=e^{\int_0^{\hz} \frac{d}{\hz'}(\hat{f}^2-1)}$, so that we obtain the following formulas for $\hat{B}$ and $\hat{\Phi}$
\begin{equation}
\hat{\Phi}=C_\phi e^{\int_0^{\hz} \frac{d}{\hz'}(\hat{f}^2-1)} \hz^{d-1},\label{Phisol}
\end{equation}
\begin{equation}
\hat{B}=C_b e^{\int_0^{\hz} \frac{d}{\hz'}(\hat{f}^2-1)} \hz^{d-1}.\label{Bsol}
\end{equation}
By eliminating $\hat{h}'/\hat{h}$ from \eqref{hfeq1} and \eqref{hfeq2} and substituting the expressions in \eqref{Bsol} and \eqref{Phisol} for $\hat{B}$ and $\hat{\Phi}$ we obtain that $\hat{f}$ must satisfy
\begin{equation}
2\frac{\hf'}{\hf}-\frac{d}{\hz}(\hf^2-1)=-\frac{d-2}{4}C_b^2\hz^{2d-1}e^{\int_0^{\hz}\frac{2d}{\hz'}(\hf^2-1)\rd \hat{z}'}-\frac{1}{2(d-1)}C_\phi^2\hz^{2d-1}e^{\int_0^{\hz}\frac{2d}{\hz'}(\hf^2-1)\rd \hz'}.\label{staticsolutions}
\end{equation}
With the boundary expansion of $f$ being $f=1+\frac{E}{2(d-1)}z^d+\ldots$ (see \eqref{fUV}), we obtain the boundary expansion of $\hf$ as $\hf=1+\frac{(E-1/z_0^d)}{2(d-1)}\hz^d+\ldots$. We expect that black branes will form when the total energy density is positive, which from \eqref{Ttt} corresponds to $E-1/z_0^d>0$. Here we will now consider the case $E<1/z_0^d$ (negative energy density) and show that any possible static solutions with negative energy density cannot be obtained dynamically. We emphasize that some solutions of \eqref{staticsolutions} might have singular behaviours and should be excluded as relevant solutions by other arguments, but we will not care about such issues here, and just directly show that any static solutions, which must satisfy \eqref{staticsolutions}, cannot be formed dynamically with the AdS soliton as initial condition. Recall the relation \eqref{feB2}, which says that we must have $fe^{-\frac{d-2}{2}b}=1$ in the IR when $\hat{z}\rightarrow\infty$ or equivalently $z\rightarrow z_0$. The idea is now to show that all solutions obtained by solving \eqref{staticsolutions} are inconsistent with this requirement. We will use the notation $\approx$ to mean that two quantities are equal asymptotically, while $\sim$ means that they are equal asymptotically up to an overall constant.\\
\linebreak
To show this we will have to compute the IR asymptotic behaviour of the solutions \eqref{staticsolutions}. We first note that the derivative $\hat{f}'$ in \eqref{staticsolutions} is negative if $0<\hf<1$. Since $\hf=1+(E-1/z_0^d)\hz^d/2(d-1)+\ldots<1$ close to the boundary, we obtain that $\hf<1$ for all $\hz$. \\
\linebreak
It is also easy to see that $\hf$ cannot become negative (because \eqref{staticsolutions} implies that if $\hat{f}=0$ then around that point $\hat{f}\sim \hat{z}^{-\alpha}$ for an $\alpha>0$ so $\hat{f}$ can only go to zero asymptotically). Also, $\hat{f}$ can not asymptote to any other constant than zero. This can be seen by assuming that $\hat{f}\rightarrow c>0$, and then \eqref{staticsolutions} implies that $\hat{f}\sim e^{-\alpha\hz^{\beta}}$ for some $\alpha,\beta>0$ which is inconsistent with $\hat{f}\rightarrow c$ (unless if $C_\phi=C_b=0$, in which case $\hat{f}\sim \hat{z}^{-\alpha}$ for $\alpha>0$, which is also inconsistent, or if $\hat{f}\equiv1$). Since $\hat{f}$ is strictly decreasing, it thus follows that we must have $\hf\rightarrow0$ when $\hz\rightarrow\infty$. When $\hz\rightarrow\infty$ we thus have that
\begin{equation}
e^{\int_0^{\hz}\frac{d}{\hz'}(\hf^2-1)\rd \hat{z}'}\approx C' \hz^{-2d}
\end{equation}
for some constant $C'$. For simplicity of notation we can thus redefine $C_b C'=C_{b, IR}$ and $C_\phi C'=C_{\phi, IR}$. The asymptotic behaviour of $\hat{b}$ is then
\begin{equation}
\hat{b}\approx C_{b,IR}\log \hz. \label{bIR}
\end{equation}
Equation \eqref{staticsolutions} now becomes in the IR
\begin{equation}
 2\frac{\hf'}{\hf}\approx-\left(\frac{d-2}{4}C_{b,IR}^2+\frac{1}{2(d-2)}C_{\phi,IR}^2+d\right)\hz^{-1},
\end{equation}
from which we can obtain the asymptotic behaviour
\begin{equation}
\hat{f}\sim \hat{z}^{-\frac{d-2}{8}C_{b,IR}^2-\frac{1}{4(d-2)}C_{\phi,IR}^2-\frac{d}{2}}.\label{hatfIR}
\end{equation}
We must now compute the asymptotic relations between ($f$, $b$) and ($\hat{f}$, $\hat{b}$), by using the expressions in \eqref{transformations}. We have that $\hz\approx z_0G^{-1/(2(d-1))}$, which directly implies the asymptotic relations
\begin{equation}
b\approx \hat{b}+2\log\frac{\hz}{z_0}\ \ 
\end{equation}
and
\begin{equation}
f\approx \frac{d}{2(d-1)}\hat{f}\left(\frac{\hz}{z_0}\right)^{d-1},\label{IRfehatf}
\end{equation}
which imply
\begin{equation}
fe^{-\frac{d-2}{2}b}\approx \frac{d}{2(d-1)}\frac{\hz}{z_0}\hat{f}e^{-\frac{d-2}{2}\hat{b}} .\label{IRfeb}
\end{equation}
From the above relations and \eqref{bIR} and \eqref{hatfIR} we now obtain the asymptotic behaviour
\begin{equation}
b\approx(C_{b,IR}+2)\log \hz,\label{BIR}
\end{equation}
\begin{equation}
f^2\sim \hz^{\left(2(d-1)-\frac{d-2}{4}C_{b,IR}^2-\frac{1}{2(d-2)}C_{\phi,IR}^2-d\right)},
\end{equation}
so that we finally obtain
\begin{equation}
fe^{-\frac{d-2}{2}b}\sim\hz^{-\frac{d-2}{2}(\frac{C_{b,IR}}{2}+1)^2-\frac{1}{4(d-2)}C_{\phi,IR}^2}.\label{IRfe}
\end{equation}
We thus see that $fe^{-\frac{d-2}{2}b}$ will generically vanish in the IR, trivially violating the condition that $fe^{-\frac{d-2}{2}b}\rightarrow1$. The only way to have $fe^{-\frac{d-2}{2}b}$ approach a constant in the IR, is when the power in \eqref{IRfe} vanishes. This only happens when $C_{\phi,IR}=0$ and $C_{b,IR}=-2$, which in particular implies that the scalar identically vanishes. Only for these particular IR parameters will $fe^{-\frac{d-2}{2}b}$ go to a constant. We will now show, however, that it will go to a constant different from 1.\\
\linebreak
To specify a solution in the bulk, it would be customary to specify the UV behaviour, meaning that we specify $E$ and $C_{b}$ and then integrate to the IR, which should give a unique solution. Specializing to $C_{b,IR}=-2$ should leave us a one parameter family of static solutions. Below we will construct this one parameter family of solutions, which turns out to be all the AdS solitons.\\
\linebreak
An AdS soliton solution with a general confinement scale $z_1$ can be given by the metric \eqref{zansatz} with $b=0$ and $f=h=1$ with $z_0$ replaced by $z_1$. After transforming to the metric \eqref{newansatz}, by using the transformations in \eqref{transformations}, we can obtain the asymptotic behaviour for $\hf$ and $\hat{b}$ as $\hat{b}\approx-2 \log \frac{\hz}{z_1}$ and $\hf\approx\frac{2(d-1)}{d} \left(z_1/\hz\right)^{d-1}$. 
We can now easily obtain from \eqref{IRfeb} that $fe^{-\frac{d-2}{2}b}\rightarrow z_1/z_0$. Thus, the only possible solution that can be obtained is $z_1=z_0$ which corresponds to our initial condition, and which corresponds in the UV to $E=0$. \\
\linebreak
To conclude, when the total energy density lies between that of the AdS soliton we use as initial condition and zero (the threshold for black brane formation), no static solutions exist that can be obtained from dynamical evolution.

\subsubsection{Long time amplitude modulation} \label{sec:long}

\begin{figure}[t]
\centering
\includegraphics[scale=0.33]{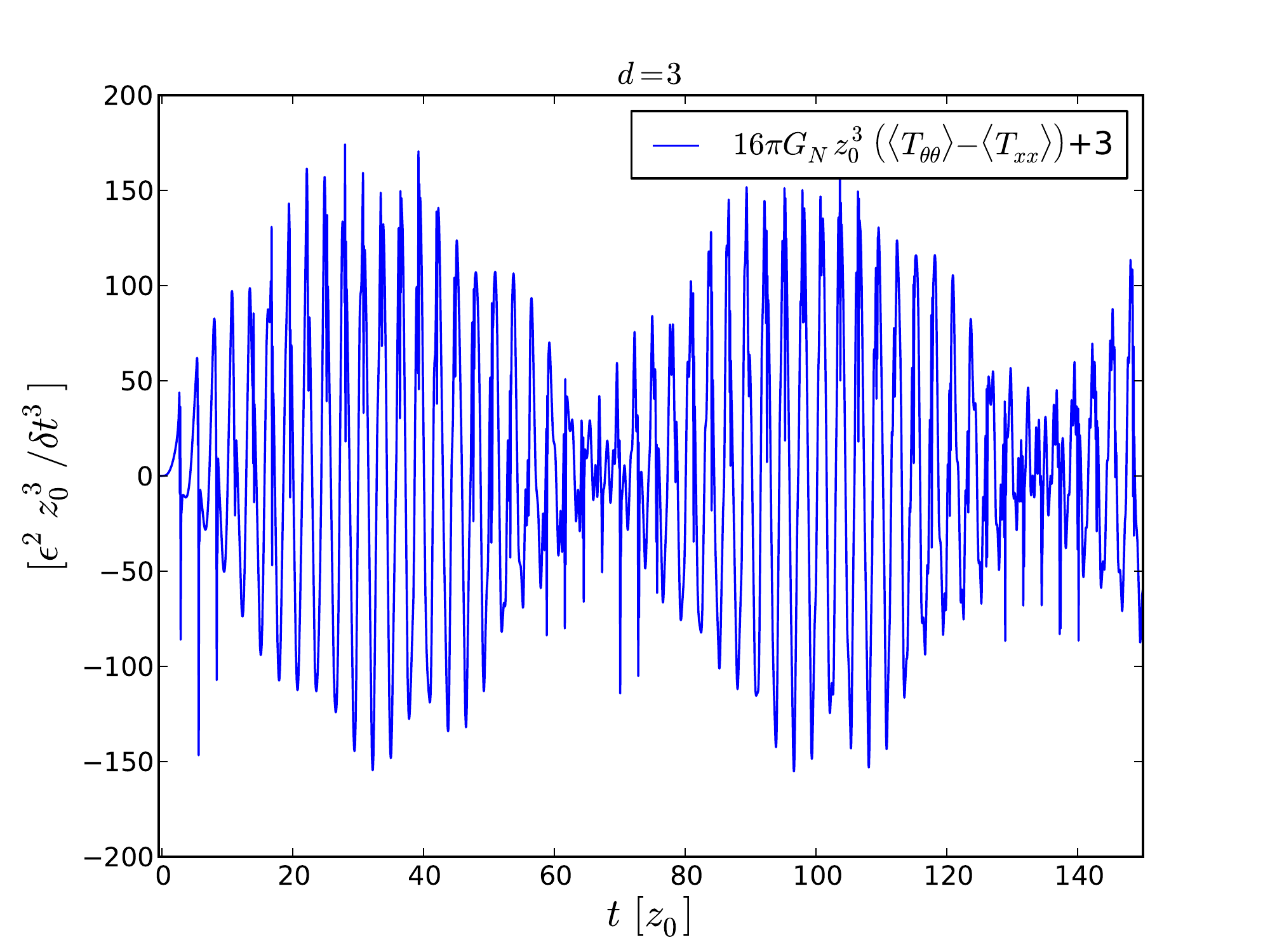}
\includegraphics[scale=0.33]{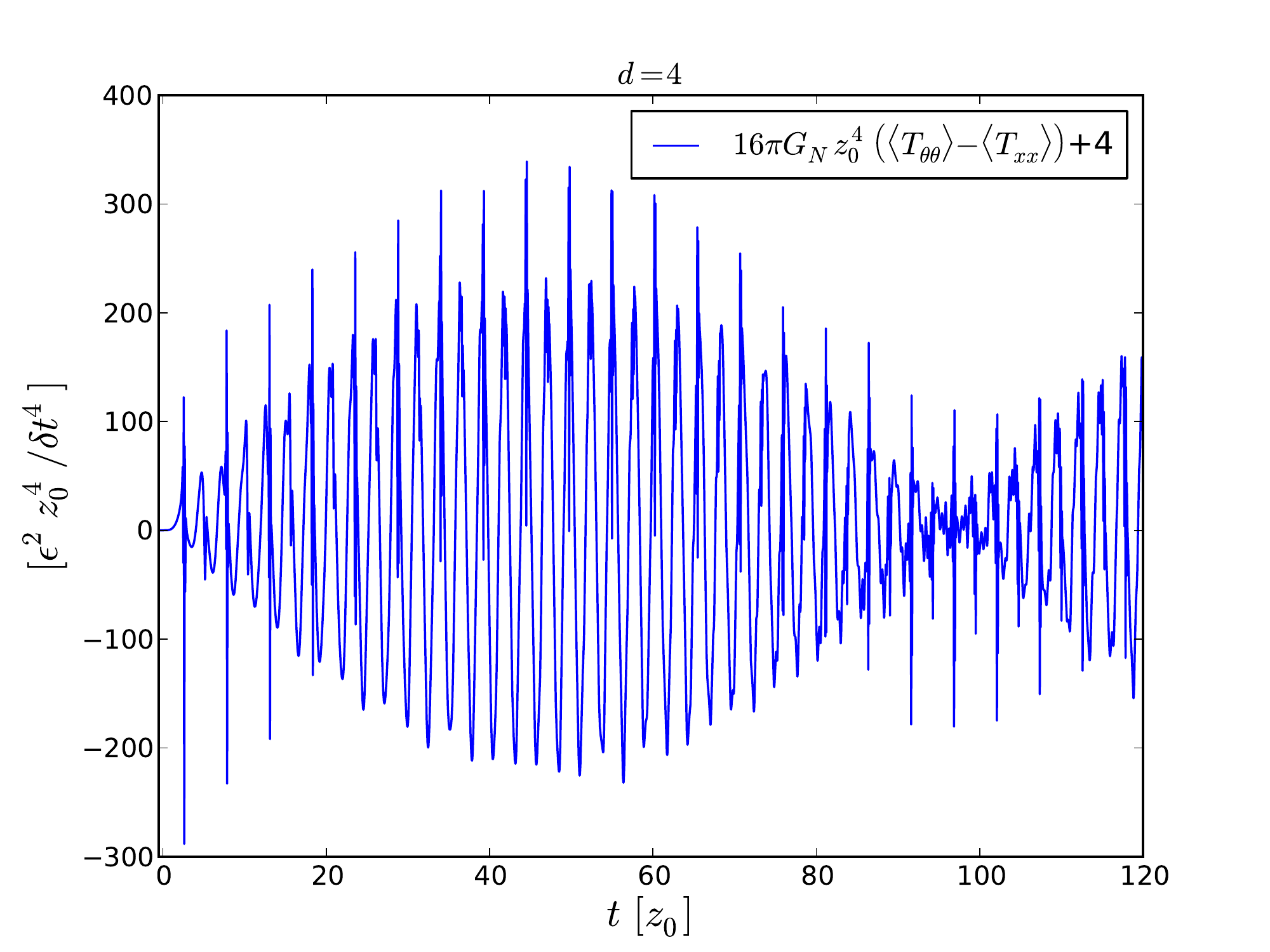}
\includegraphics[scale=0.33]{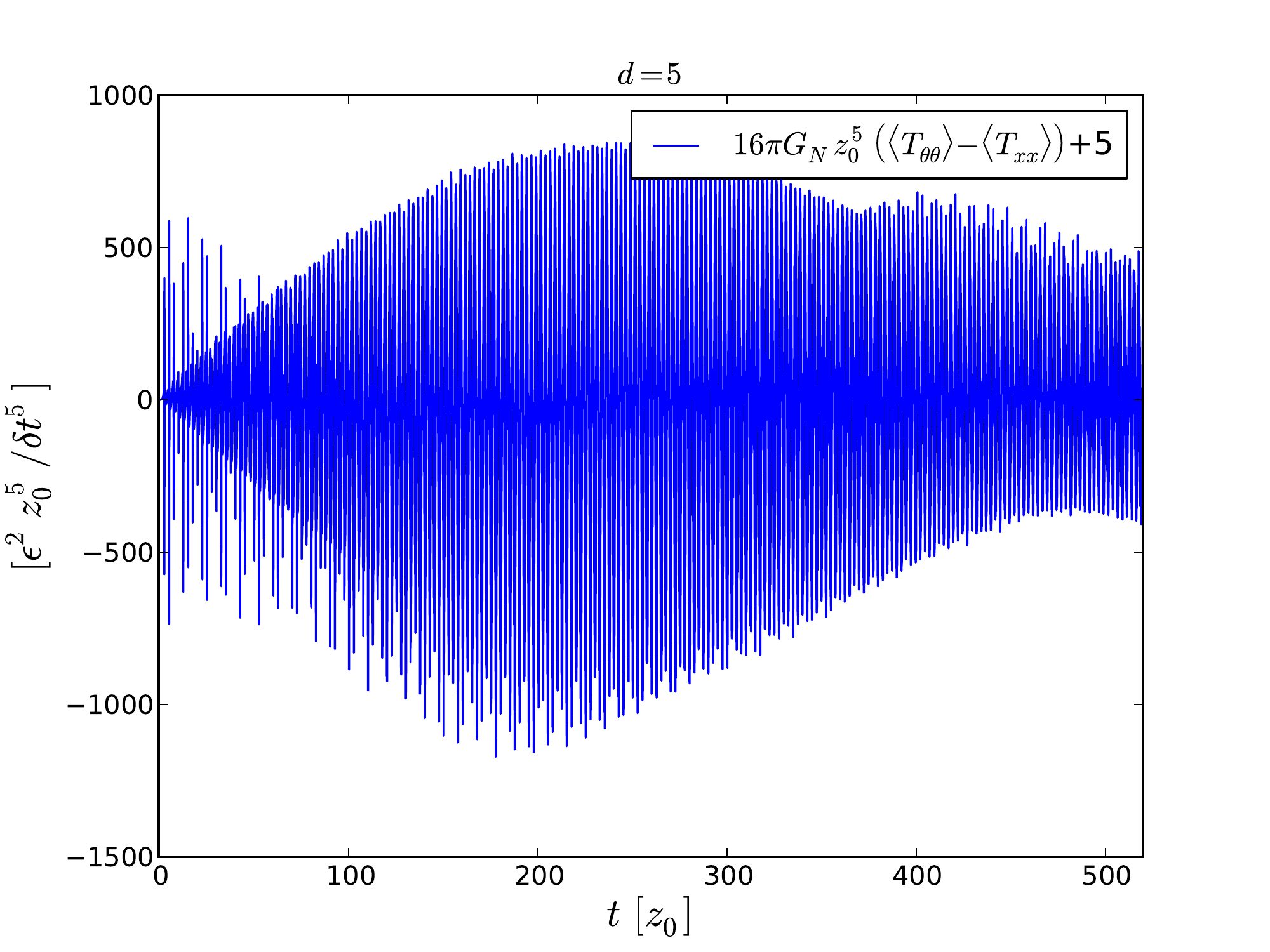}
\includegraphics[scale=0.33]{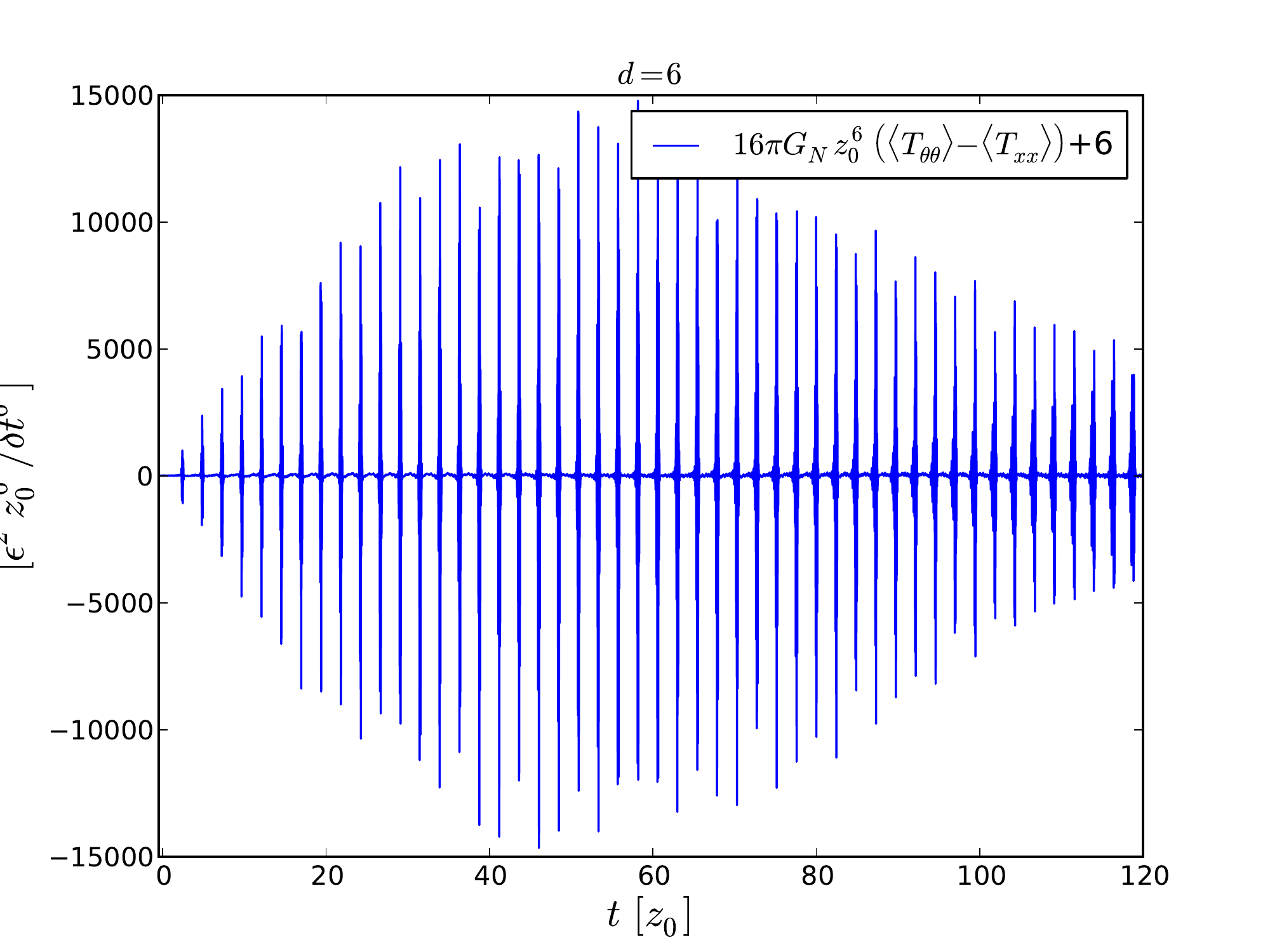}
\caption{The pressure anisotropy after a weak scalar perturbation with small $\epsilon$ and $\delta t=0.1z_0$. Time is measured in units of $z_0$, and the vertical axis has been rescaled by $\epsilon^2/\delta t^d$, which is the expected dependence of the total energy of the system for small $\epsilon,\delta t$. For $d=3,4,5$ we see that the amplitude undergoes an amplitude modulation on a much longer time scale which is in excellent agreement with the result in table \ref{modtable}. For $d=6$, the modulation due to the first normal mode is hidden by the peaks from the bouncing wave packet part, however it is clearly visible in the normal mode decomposition in Figure~\ref{normalmode_decomposition}.}
\label{amplitudemod} 
\end{figure}

\begin{figure}[t]
\centering
\includegraphics[scale=0.33]{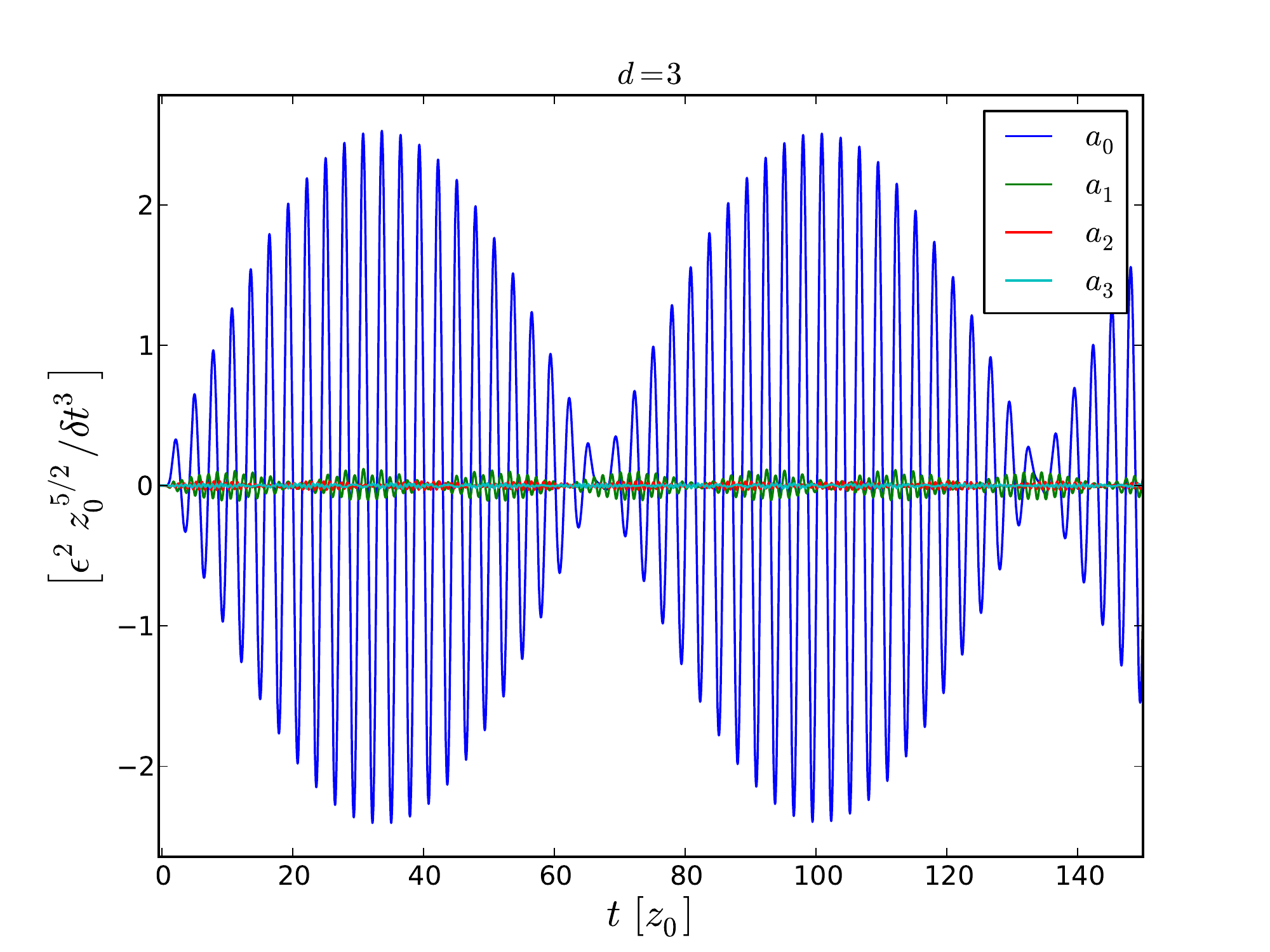}
\includegraphics[scale=0.33]{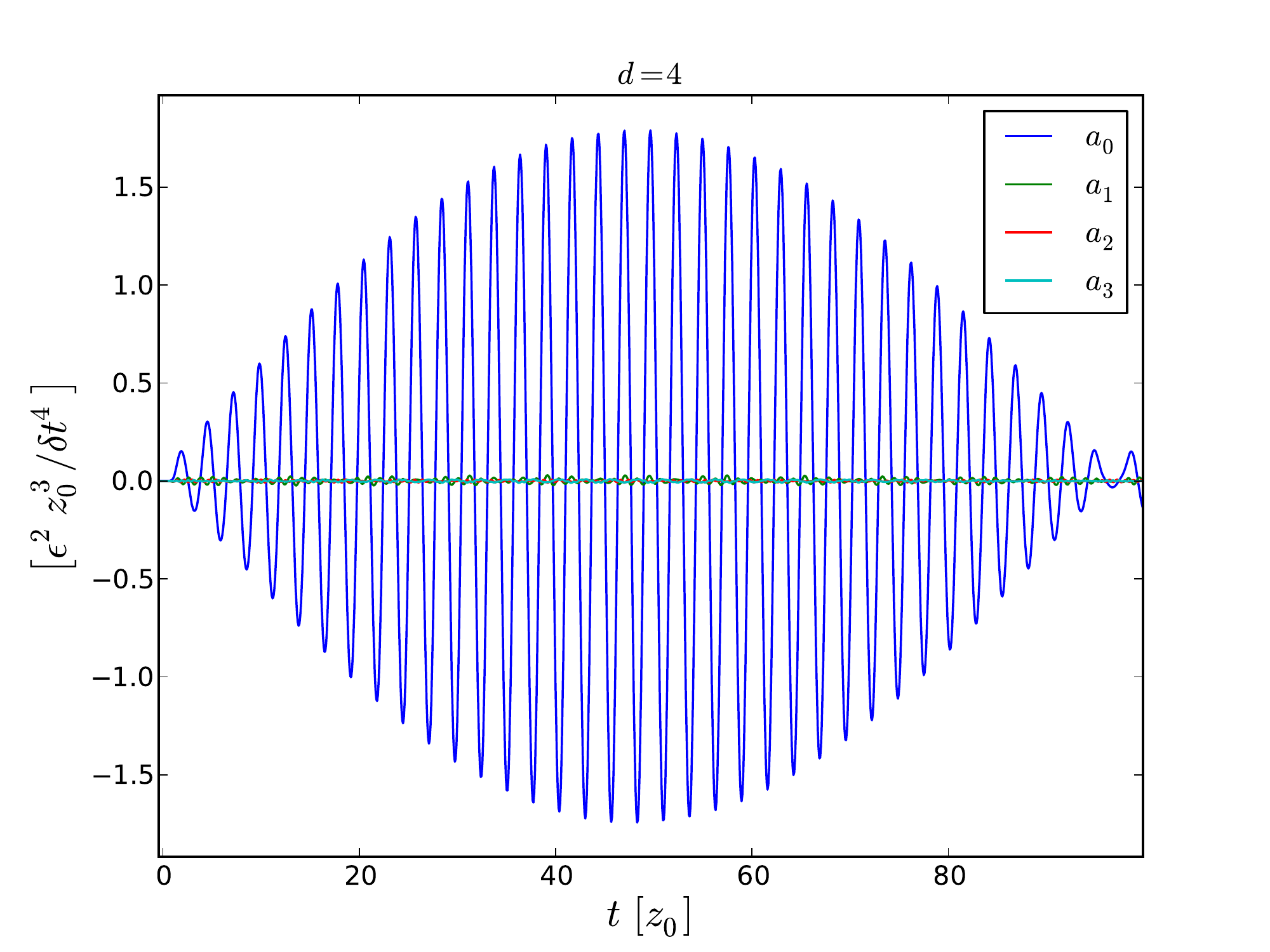}
\includegraphics[scale=0.33]{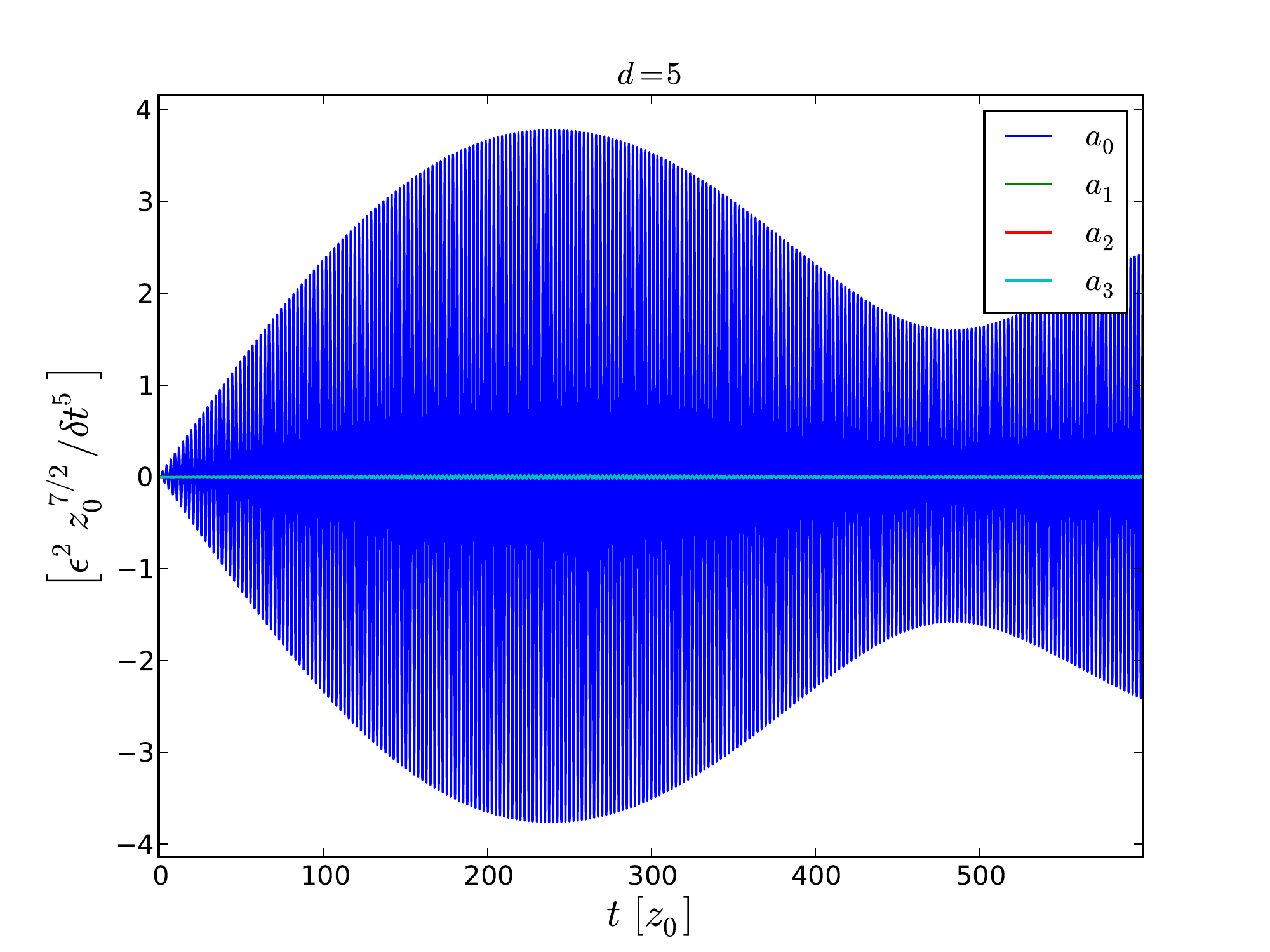}
\includegraphics[scale=0.33]{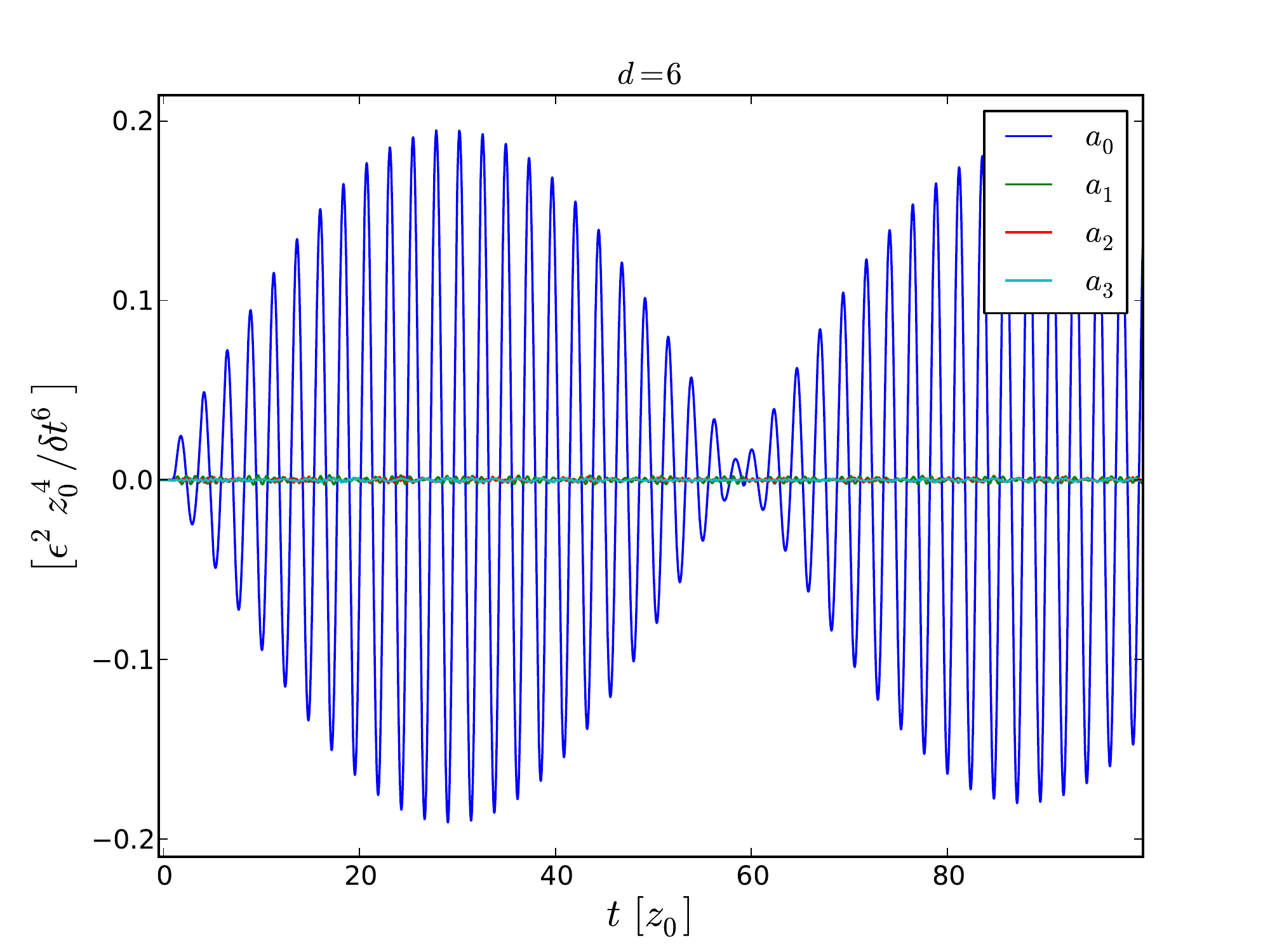}
\caption{The metric function $b$ decomposed in normal modes, after a weak scalar perturbation with small $\epsilon$ and $\delta t=0.1z_0$. Time is measured in units of $z_0$. We see that, as expected, the lowest mode is more excited than higher modes, and undergoes an amplitude modulation which agrees with the result in Table~\ref{modtable}.}
\label{normalmode_decomposition} 
\end{figure}

For small-amplitude scattering solutions (small $\epsilon$), we observe an amplitude modulation in the pressure anisotropy on a long time scale, see Figure~\ref{amplitudemod}. The relevant timescale is actually hard to see for $d=6$, for reasons we will explain below. The time scale can be seen to be independent of $\epsilon$ and $\delta t$ as long as both parameters are sufficiently small. This is different from the $1/\epsilon^2$ modulation time scale observed in the hard wall model for the $d=3$ with Neumann boundary conditions and thus it does not seem to originate from the same mechanism (a resonant normal mode spectrum). As we will now explain, in the present case the modulation is instead due a near-resonance between a metric mode and the frequency of the bouncing scalar shell.\\
\linebreak
In the small $\epsilon$ regime, the scalar field is of order $\mathcal{O}(\epsilon)$. Thus the metric is of order $\mathcal{O}(\epsilon^2)$ and the next order corrections to the scalar are of the order $\mathcal{O}(\epsilon^3)$. Working to order $\mathcal{O}(\epsilon^2)$, we can therefore consider $\phi$ as a probe scalar acting as an external source on the metric. Since the scalar $\phi$ bounces back and forth between the IR and the boundary, the source for the metric backreaction can be characterized by a frequency\footnote{Not to be confused with the metric component $f$ from previous sections.} $f_\phi$. In the limit of small $\delta t$ (the thin shell limit), this frequency will be the same as for a lightlike particle (following lightlike geodesics) that bounces between the boundary and the interior. In particular, for small $\delta t$ the bounce frequency becomes independent of $\delta t$. \\
\linebreak
However, the metric also has some intrinsic frequencies $f_i$ which are the normal mode frequencies of small perturbations of the metric which we compute in the end of this section. Every time the scalar wave packet crosses the spacetime it kicks the metric. It is useful to decompose the metric fluctuation into its normal modes. If $f_\phi=f_j$ for some $j$, we would expect a resonance, such that the amplitude of the $j$'th normal mode will increase linearly with time. But if $f_\phi\approx f_j$, such that we are close to resonance, it would be natural to expect another time scale showing up, namely $T=1/|f_\phi-f_j|$, and this time scale can then be compared to the time scale of the amplitude modulation in the pressure anisotropy. The results are summarized in table \ref{modtable}, and can be compared with the numerical results in Figure~\ref{amplitudemod} and Figure~\ref{normalmode_decomposition}. The latter figure shows the decomposition of $b$ in its normal modes, where the decomposition is of the form $b(z,t)=\sum_{n\geq0} a_n(t)Q_n(z)$. The functions $Q_n$ (corresponding to the $n$th normal mode with frequency $\omega_n$) satisfy the equation \eqref{Qeqbis} with $\omega=\omega_n$, which constitutes a Sturm-Liouville problem (which makes this decomposition possible), and they are normalized with respect to the inner product $\int_0^{z_0} Q_n(z)Q_m(z)z^{-(d-1)}dz=\delta_{mn}$. Note that replacing the frequency of the scalar wave packet by that of a lightlike thin shell still gives decent result, but using the true frequency is required to get accurate results, especially for $d=5$ (where the system is extremely close to resonance). To summarize, the modulation can be traced back to a near resonance between the lowest normal mode frequency of metric perturbations and the frequency of a bouncing scalar shell. As expected from this picture, this type of modulation does not show up when we quench the metric instead of the scalar field, as can be seen in Figure~\ref{gwscattering001}.\\
\linebreak
One can see from the numerics, however, that the metric perturbation consists not only of a few normal modes: it has a slowly moving normal mode part and a rapidly moving wave packet part. The wave packet part is in general smaller than the normal mode part. However, close to the boundary, the wave packet part can still give large contributions to the boundary observables. The intuitive explanation is as follows. Close to the boundary a wave packet looks typically like $~\psi((z-t)/\delta t)$, where $\psi$ is some localized profile and $\delta t$ is the width. Thus when extracting the $z^d$ coefficient when computing the boundary observables, this will be proportional to $\partial_z^d\psi((z-t)/\delta t)\sim 1/\delta t^d$, while the derivatives of the normal modes are of $\mathcal{O}(1)$. We thus see that for larger dimensions, the wave packet part is expected to become more important. These are exactly the sharp peaks one can see in  Figure~\ref{amplitudemod}, and indeed they become larger for larger dimensions. For $d=6$ they completely dominate and this is the reason why we cannot see the modulation due to the first normal mode in the vacuum expectation value in $d=6$. However, it can still be seen in the normal mode decomposition in Figure~\ref{normalmode_decomposition}, since here the contribution from the wave packet part is still small.

\begin{table}[t]
\begin{center}
\begin{tabular}{ |c|c|c|c|c| } 
 \hline
 $d$ & 3 & 4 & 5 & 6\\ 
 \hline
 $z_0f_0$ & 0.34195 &0.37177  &0.40151&0.43004\\ 
 $z_0f$ & 0.35682(19) & 0.38190(36) &0.39944(16)&0.41263(43)\\ 
 $z_0f'$ & 0.3564 & 0.3807 &0.3986&0.4117\\ 
 $T/z_0$ & 67.2(8) & 98.7(35) &484(37)&57.4(14)\\ 
 $T'/z_0$ & 69.2 & 111.9 &344&54.5\\ 
 \hline
\end{tabular}
\end{center}
\caption{\label{modtable} The lowest normal mode frequency $f_0$, the oscillation frequency of the scalar wave packet $f$ for $\delta t=0.1z_0$, the oscillation frequency of a lightlike thin shell $f'$ and the expected modulation times $T=1/|f_0-f|$ and $T'=1/|f_0-f'|$ using $f$ respectively $f'$. Note that the frequency of a thin shell is extremely close to the frequency of the bouncing scalar field. However, note also that in $AdS_6$ ($d=5$) we are extremely close to resonance, and to get an accurate modulation time we must use the true frequency of the scalar wave packet (compare with Figure~\ref{amplitudemod} and Figure~\ref{normalmode_decomposition}). The estimated error comes from reading off the oscillation frequencies of the wave packet from the numerical simulations, while the errors of $f_0$ and $f'$ are negligible.}
\end{table}

\subsubsection{Harmonic oscillator toy model}\label{oscillator_example}
\begin{figure}[t]
\centering
\includegraphics[scale=0.7]{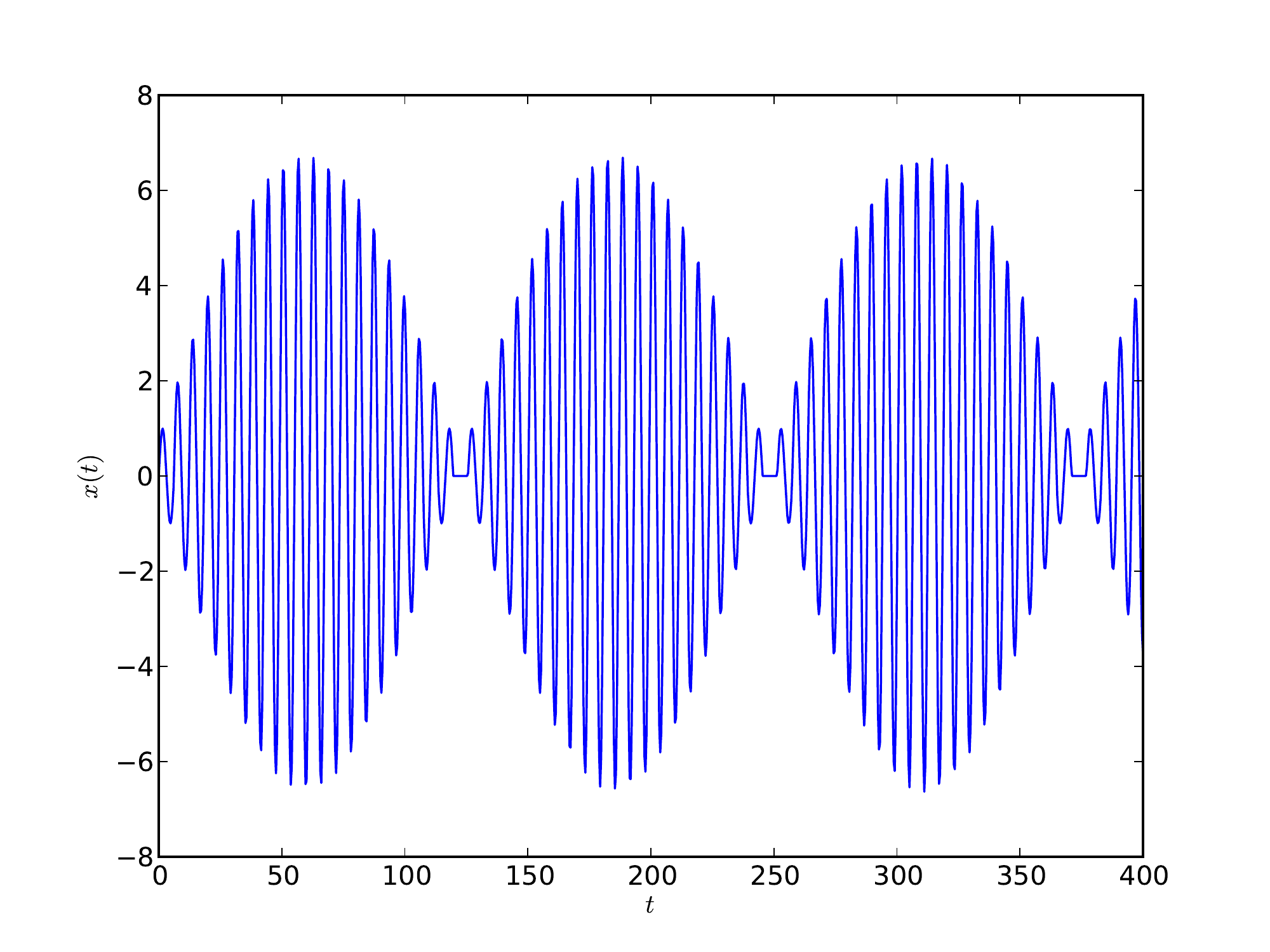}
\caption{Evolution of the sourced harmonic oscillator given by equation \eqref{oscillator}, close to resonance with parameters $\omega=1$ and $T=2\pi/1.05$. We see that the result is a rapidly oscillating solution with a long time modulation with period $2\pi/0.05\approx125.7$.}
\label{oscillator_modulation}
\end{figure}

To develop a better understanding of the modulations we have just described, we now study a sourced harmonic oscillator which is conceptually similar to our gravitational setup (in the small-amplitude scattering phase) and which experiences a similar amplitude modulation phenomenon. This toy model can be solved analytically and provides an analytical understanding of the long time amplitude modulation that shows up when the system is near resonance.\\
\linebreak
Consider a harmonic oscillator with angular frequency $\omega$, sourced by a sequence of local kicks (modelled by delta functions) with period $T$. (We denote the frequency of the kicks by $f=1/T$.) The equation of motion is
\begin{equation}
\ddot{x}+\omega^2x=\sum_{n\geq0} \delta(t-nT),\label{oscillator}
\end{equation}
subject to the initial condition that $x(t)$ should vanish for $t<0$. To compare with our gravitational setup, $x$ is the analogue of $B$ (the metric backreaction), the delta functions are analogous to the stress energy tensor for the scalar $\phi$ which sources the metric, the frequency $f=1/T$ is analogous to the oscillation frequency $f_\phi$ of $\phi$, and $\omega$ is analogous to the lowest normal mode frequency of the metric perturbations. We can solve \eqref{oscillator} by performing a Laplace transform. For the Laplace transformed field $X$ we have
\begin{equation}
s^2X(s)+\omega^2X(s)=\sum_{n\geq0} e^{-nTs}\Rightarrow X(s)=\frac{1}{2i\omega}\sum_{n\geq0}\left(\frac{e^{-nTs}}{s-i\omega}-\frac{e^{-nTs}}{s+i\omega}\right).
\end{equation}
It is now easy to do the inverse Laplace transform, to obtain
\begin{equation}
x(t)=\sum_{n\geq0}\frac{\sin(\omega(t-nT))}{\omega}\theta(t-nT),
\end{equation}
where $\theta(t)$ is the Heaviside step function. By letting $N=\lfloor t/T \rfloor$ (the largest integer less than or equal to $t/T$), we can write this as
\begin{equation}
\omega x(t)=\sum_{n=0}^N\sin(\omega(t-nT))=\mathrm{Im }\enspace e^{i\omega t}\sum_{n=0}^Ne^{-i\omega nT}=\mathrm{Im }\enspace e^{i\omega t}\frac{1-e^{-i\omega (N+1)T}}{1-e^{-i\omega T}},\label{geosum}
\end{equation}
under the assumption that $T\omega\not\equiv 0 \pmod{2\pi}$. Extracting the imaginary part and using some trigonometric identities, we obtain
\begin{equation}
x(t)=\frac{2\sin\left[\frac{\omega T}{2}\right]\sin\left[\omega(t-\frac{NT}{2})\right]\sin\left[\omega\frac{N+1}{2}T\right]}{\omega(1-\cos\left[\omega T\right])}.\label{3sine}
\end{equation}
If the system is almost at resonance, $T\omega\approx2\pi$, the middle factor in \eqref{3sine} will give rise to fast oscillations, while the last factor gives rise to slow amplitude modulations. To see this, we write $f-\omega/2\pi=\epsilon\ll f$ (so $\epsilon$ is the difference between the source frequency and the oscillator frequency). The third factor in \eqref{3sine} now becomes
\begin{align}
\sin\left[\omega\frac{N+1}{2}T\right]&=\sin\left[\pi(1-\epsilon T)(N+1)\right]=\pm \sin \left[\pi \epsilon T (N+1)\right]\nonumber\\
&\approx\pm \sin \left[\pi \epsilon (t+T)\right],
\end{align}
where in the last step we approximated $N=\lfloor t/T \rfloor\approx t/T$, and we see that we indeed obtain an overall amplitude modulation with period $1/\epsilon$. An example with $\omega=1$ and $\epsilon=0.05/2\pi$ is shown in Figure~\ref{oscillator_modulation}. Further, we note that it is the small denominator in (\ref{3sine}) that causes a near-resonant normal mode to dominate the other normal modes.\\
\linebreak
If $T\omega =2\pi k$, for some non-zero integer $k$, the summation of the geometric series in \eqref{geosum} yields instead
\begin{equation}
x(t)=\frac{N+1}{\omega}\sin\omega t\approx \frac{t+T}{2\pi k}\sin\omega t,
\end{equation}
which is a sine function with a (step-wise) increasing amplitude, as expected when we are at resonance. This can also be obtained as a limit $T\omega\rightarrow 2\pi k$ in \eqref{3sine}.\\
\linebreak
This toy model provides a nice analytical example for why a near resonance should result in an amplitude modulation, but it does of course not represent the full dynamics of our gravitational setup. It would of course be more satisfactory to provide a more rigorous analytical calculation for our gravitational problem and we will leave that interesting question as an open problem.

\subsubsection{Normal modes}\label{normalmodes}
Here we will explain how to compute the normal modes which we have used in this section to explain the phenomena in the numerical simulations. Essentially the normal modes are solutions of the linearized equations of motion (small perturbations around the AdS Soliton background) that can be written as a product of a radial function and a harmonically oscillating function of time. The allowed frequencies for such solutions will be a discrete set, which is expected to be the case in confining theories.\\
\linebreak
To find such solutions, we assume that $b=\mathcal{O}(\mu)$, for some small parameter $\mu$, and that the scalar field vanishes. We then solve the equations to linear order in $\mu$. So letting $f=1+\mu f_1+\ldots$, $h=1+\mu h_1+\ldots$, $B=1+\mu B_1+\ldots$ and $P=1+\mu P_1+\ldots$ we obtain
\begin{align}
\dot{f}_1&=z^{1-2d} \frac{(d-2)(d-1)}{(G z^{-2(d-1)})'}P_1G+\frac{d-2}{2}P_1,\\
\dot{P}_1&=\frac{1}{d-1}(G'(h_1-f_1)z^{-(d-1)}+(d-1)B_1Gz^{-(d-1)})'z^{d-1},\\
h_1'&=(d-2)B_1-f_1'+z^{1-2d}\frac{2(d-2)(d-1)GB_1}{(Gz^{-2(d-1)})'},\\
\dot{B}_1&=P_1',\\
h_1'&=\frac{4d(d-1)f_1}{z^{2d}(Gz^{-2(d-1)})'}+f_1',
\end{align}
where $G(z)=1-z^d/z_0^d$. To look for normal modes, we make the ansatz $P_1=Q(z)\cos \omega t$. This implies that $B_1=Q'(z) \sin \omega t /\omega$, $f_1=F(z)\sin\omega t$ and $h_1=H(z)\sin\omega t$, and the functions $Q$, $F$ and $H$ satisfy the ordinary differential equations
\begin{align}
\omega F &= z^{1-2d} \frac{(d-2)(d-1)}{(G z^{-2(d-1)})'}QG+\frac{d-2}{2}Q,\label{Feq}\\
-\omega Q&=\frac{1}{d-1}(G'(H-F)z^{-(d-1)}+\frac{(d-1)Q'G}{\omega }z^{-(d-1)})'z^{d-1},\label{Qeq}\\
H'&=(d-2)\frac{Q'}{\omega}-F'+z^{1-2d}\frac{2(d-2)(d-1)GQ'}{\omega(Gz^{-2(d-1)})'},\\
H'&=\frac{4d(d-1)F}{z^{2d}(Gz^{-2(d-1)})'}+F'.\label{FpHp}
\end{align}
Actually it is possible to extract from these equations a single ordinary differential equation for $Q$. Since $(G'(H-F)z^{-(d-1)})'=G'z^{-(d-1)}(H'-F')$, we can eliminate $H'-F'$ from equation \eqref{Qeq} by using equation \eqref{FpHp}, and then use \eqref{Feq} to eliminate the remaining $F$ such that we end up with
\begin{equation}
-\omega^2 Q=G'\frac{4d}{z^{2d}(Gz^{-2(d-1)})'}\left(z^{1-2d} \frac{(d-2)(d-1)}{(G z^{-2(d-1)})'}QG+\frac{d-2}{2}Q\right)+\left(\frac{Q'G}{z^{d-1}}\right)'z^{d-1},\label{Qeqbis}
\end{equation}
which is a second order ordinary differential equation for $Q$. Demanding regularity in the IR, the only free parameter is $\omega$ (since we can set $Q(z_0)=1$ by an overall rescaling). Then demanding that $Q$ should be normalizable at the boundary (equivalent to $Q(0)=0$), gives us a discrete set of allowed frequencies $\omega$. These are the frequencies of the normal modes, and can be seen in Table \ref{freqtable}.
\begin{table}[t]
\begin{center}
\begin{tabular}{ |c|c|c|c|c| } 
 \hline
 $d$ & 3 & 4 & 5 & 6\\ 
 \hline
 $z_0\omega_0$ & 2.14853 &2.33587  &2.52274&2.70203\\ 
 $z_0\omega_1$ & 4.790 & 5.517 &6.200&6.854\\ 
 $z_0\omega_2$ & 7.116 & 8.069 &8.925&9.719\\ 
 \hline
\end{tabular}
\end{center}
\caption{\label{freqtable} The lowest normal mode frequencies for metric perturbations in various dimensions. The normal mode frequencies are inversely proportional to the confinement scale $z_0$.}
\end{table}
\linebreak
Although not very relevant for this work, it is interesting to note that the normal modes of the scalar field satisfy \eqref{Qeq} but with the first term in the RHS, which is proportional to $Q$, removed, namely
\begin{equation}
-\omega^2 Q=\left(\frac{Q'G}{z^{d-1}}\right)'z^{d-1},\label{Qeqscalar}
\end{equation}
if we assume that $\phi=Q\cos\omega t$. Given that the omitted term is proportional to $Q$, and therefore combines with the LHS, the normal modes of the scalar and the metric fluctuations can be expected to approach each other for large $\omega$. In addition, we have observed numerically that the spectrum becomes linear for large $\omega$. Moreover, global AdS$_3$ is actually identical to the AdS$_3$ Soliton, which leads us to expect that \eqref{Qeqscalar} must give a linear spectrum for $d=2$. Indeed, in this case if we redefine $Q=z^2q$ and $z^2/z_0^2=(x+1)/2$, we obtain the equation
\begin{equation}
(1-x^2)q''+(1-3x)q'+(\frac{\omega^2}{4}-1)q=0,
\end{equation}
which is solved by the Jacobi polynomials $q(x)=P_n^{(\alpha,\beta)}(x)$ with $\alpha=0,\beta=1$ and with $\omega=2(n+1)$ for $n=0,1,2,\ldots$. Note that the normal mode computation of the metric discussed in this section is not well defined in three dimensions.

\subsection{Summary}
In this section we have studied extensively dynamics in the AdS Soliton model. We used mostly numerical methods, but performed several checks on the numerics with analytic calculations. We found similar behaviour as in the hard wall model, namely that a black brane forms when the energy of the injected energy is large enough and otherwise we find perpetual scattering solutions (where the matter bounces back and forth between the boundary and the interior). In the black hole phase we find that the late-time dynamics is governed by the lowest quasinormal modes. In the scattering phase we found a modulation of the amplitude in the boundary pressure anisotropy that can be explained by a near resonance between the lowest normal mode frequency and the frequency of the bouncing scalar field. Note that the origin of this amplitude modulation is completely different to what was found in the hard wall model. We also provided an analytical proof that the scattering solutions can not equilibrate.

\pagebreak
\section{Conclusions and outlook}
In this chapter we have studied dynamical processes in two confining holographic models, the hard wall model and the AdS Soliton, with the aim to try to say something about dynamics in confining quantum field theories. We found that in both models, there is a threshold on the injected energy below which a black hole solution does not form. In this phase we instead obtain a solution that is eternally time dependent and corresponds to a field theory state that never thermalizes. In both solutions we found in some cases long time amplitude modulations in boundary observables, although arising from very different mechanisms in the two models. If these solutions mean anything for real life systems such as the QGP, or if they are artifacts of the AdS/CFT correspondence, is an interesting open question. \\
\linebreak
Our study only provides the very first steps to understanding time dependent processes in confining holographic models and there are many interesting directions for future research. One interesting venue would be to look at more ``realistic'' models, by which we mean dual gravity theories that have been modified with additional matter content to further try to mimic the boundary theory one is interested in. One popular method of constructing holographic duals is to use a scalar field with a some non-trivial scalar potential, and then tune this potential such that the boundary dual satisfies some desired properties. This is the basis of the model called {\it improved holographic QCD}\cite{Gursoy:2007cb,Gursoy:2007er}, which reproduces the qualitative behaviour of many observables of QCD. In \cite{Ishii:2015gia}, the authors studied a similar setup to what we studied in this chapter but where the model was a simplified version of improved holographic QCD, where the scalar potential takes a much simpler form. In this model the field theory observables do not agree as well with QCD observables, but the model is still confining and the simpler boundary behaviour makes the model more tractable for numerics. However, studying time dependence in the full improved holographic QCD model still remains an open interesting problem. There are also interesting extensions of the AdS Soliton model. One is the Sakai-Sugimoto model\cite{Sakai:2004cn,Sakai:2005yt}, which implements quarks and chiral symmetry breaking by adding probe branes to make it a more realistic holographic model of QCD. Another interesting extension are holographic superconductors\cite{Hartnoll:2008kx,Horowitz:2010gk,Cai:2015cya}, which have been constructed with the AdS Soliton as a basis with additional matter fields\cite{Nishioka:2009zj}. Studying time dependence in such extensions would also be a interesting direction for future research.\\
\linebreak
There also remain many fundamental questions about the dynamics in the AdS Soliton (as well as in the hard wall model, but there really is no reason to continue studying the hard wall model instead of the AdS Soliton). For instance one can study time evolution of non-local quantities, such as entanglement entropies. These can be computed as the (renormalized) area of extremal surfaces in the bulk\cite{Ryu:2006bv,Hubeny:2007xt}. Information about non-local observables would give much more information about the dynamical evolution of the boundary field theory. Obtaining a better analytical understanding of the results in this chapter would also be interesting. For instance, in the hard wall model we had analytic control over the asymptotic behaviour of the phase diagram while in the AdS Soliton model we were not as successful. Another interesting technique that has been used recently is to consider gravity in a large $D$ (dimension) expansion. This has been used to study dynamics and instabilities of (higher dimensional) black holes\cite{Emparan:2013moa}, and it would be interesting to see if such techniques can be applied to the AdS Soliton spacetime. In particular, the hard wall model may be viewed as the AdS Soliton where the dimension goes to infinity, and understanding this connection in detail (in particular which hard wall boundary condition corresponds to the large $D$ soliton) would be very interesting. The seemingly accidental near resonance between a bouncing lightray and the lowest normal mode in the AdS Soliton (which is the underlying mechanism for the long time amplitude modulation in the field theory observables) is also a mystery where more analytic understanding is required.\\
\linebreak
Another interesting question is if the scattering solutions we encountered are stable versus spatial inhomogeneities. One can imagine adding a spatial dependence to the source $J(t)\rightarrow J(t,x)$ and then investigate if small such inhomogeneities result in unstable evolution. This is an interesting question since that could mean that the very long lived scattering solutions are not physical as the evolution would look very different if we did not enforce the translational symmetry. In particular, the questions about the long time amplitude modulation would become less relevant. This would be a very interesting future research direction but would require using more sophisticated numerical relativity tools to deal with systems that depend on more than two variables.



\chapter{Collisions of pointlike particles in three-dimensional gravity}\label{threed}

In this chapter we will study a very different process for forming black holes in anti-de Sitter space, which only exists in three dimensions, namely that of formation of a black hole from a collision of pointlike particles. In higher dimensional gravity theories, a pointlike object interacting classically with gravity will be hidden by an event horizon and can thus not be considered pointlike (it will instead be a black hole). However, in three dimensions, pointlike particles instead take the form of conical singularities and are truly pointlike since they do not have an event horizon. Moreover, due to the topological nature of gravity in three dimensions (absence of gravitational waves), pointlike particle solutions can (in a sense that will be specified later) be superimposed and it is thus much easier to construct solutions with many particles than it is for instance to construct solutions with many black holes in higher dimensions. In this Chapter we will consider a process where two or more pointlike particles collide at a single spacetime point in three-dimensional anti-de Sitter space. This process is analogous to an inelastic collision of objects in Newtonian mechanics, and just as in Newtonian mechanics additional information must be provided for the dynamics at the collision point, for instance how many resulting objects there are after the collision. We will restrict to the case where all objects merge to a single object, and for this case no more additional information is needed. It turns out that for small energies, the resulting object is just another (massive) pointlike particle while for energies above a certain threshold the resuling object is a black hole. We will also consider the limit of an infinite number of colliding pointlike particles which can be used to construct spacetimes consisting of a thin shell of matter collapsing to a black hole. The motivation for studying the formation of black holes in three-dimensional anti-de Sitter space again comes from the AdS/CFT correspondence and the fact that such processes are dual to the formation of a thermal state in the boundary two-dimensional field theory. Some of the derivations are carried out in detail in Appendix \ref{appendix}, while for the most involved computations we just refer to calculations done using a symbolic manipulation software. The new results in this chapter are based on the papers \cite{Lindgren:2015fum} and \cite{Lindgren:2016wtw}.

\section{Three-dimensional anti-de Sitter space}\label{adssec}
We will now again quickly review Anti-de Sitter space (AdS), but focusing on the special case of three dimensions and features that are only present in that case. We remind the reader that AdS can be constructed as an embedding in a higher dimensional Minkowski space with two time directions, see Section \ref{sec_adsbh}. For three-dimensional AdS (\ads), the ambient space is four-dimensional Minkowski space with signature $(-,-,+,+)$, and the embedding equation takes the form
\begin{equation}
x_3^2+x_0^2-x_1^2-x_2^2=L^2,\label{embedding_eq}
\end{equation}
where $L$ is the AdS radius, related to the cosmological constant by $\Lambda=-1/L^2$. We will from now on set $L=1$ throughout this chapter. The ambient space has the metric
\begin{equation}
\rd s^2=\rd x_1^2+\rd x_2^2-\rd x_3^2-\rd x_0^2,
\end{equation}
which induces a metric on the submanifold. To parametrize this manifold, one can use the coordinates $(t,\chi,\phi)$ defined by
\begin{equation}
\begin{array}{cc}
x^3=\cosh \chi \cos t, & x^0=\cosh \chi \sin t,\\
x^1=\sinh \chi \cos \phi, & x^2=\sinh \chi \sin \phi, \label{adscoord}
\end{array}
\end{equation}
where $t$ is interpreted as a time coordinate, $\chi$ is a radial coordinate and $\phi$ is an angular coordinate. The spacetime so defined has closed timelike curves, and to resolve this issue one ``unwinds'' the time coordinate, effectively dropping the periodicity of $t$. The ranges of the coordinates are then $0\leq\chi\leq\infty$, $0\leq\phi<2\pi$ and $-\infty<t<\infty$. The boundary of AdS is located at $\chi\rightarrow\infty$ and we will refer to the point $\chi=t=0$ as the {\it origin}. The metric is 
\begin{equation}
\rd s^2=-\cosh^2\chi \rd t^2+\rd \chi^2+\sinh^2\chi \rd\phi^2.\label{adsmetric}
\end{equation}
For figures in this chapter, we will use a different parametrization of the radial coordinate, namely $u=\tanh(\chi/2)$ with $0\leq u\leq1$. With this parametrization, the metric at constant time $t$ takes the form of a Poincar\'e disc, where in particular spacelike geodesics take the form of circle sectors. Explicitly we have
\begin{equation}
\rd s^2=\left(\frac{2}{1-u^2}\right)^2(\rd u^2+u^2\rd \phi^2)-\left(\frac{1+u^2}{1-u^2}\right)^2\rd t^2.\label{figurecoord}
\end{equation}
\linebreak
A very useful property of AdS$_3$ is that it is locally\footnote{\slr is isomorphic to AdS$_3$ {\it before} unwinding the time coordinate. This will not be an issue for any of the computations in this chapter.} isomorphic to \slr, the group manifold of real $2\times2$ matrices with unit determinant. For the Lie algebra $\mathfrak{sl}(2,\R)$ we can define the basis
\begin{equation}
\gamma_0=\left(\begin{array}{ccc} 0 & 1 \\ -1 & 0 \\ \end{array}\right),\hspace{23pt}\gamma_1=\left(\begin{array}{ccc} 0 & 1 \\ 1 & 0 \\ \end{array}\right),\hspace{22pt} \gamma_2=\left(\begin{array}{ccc} 1 & 0 \\ 0 & -1 \\ \end{array}\right),
\end{equation}
which together with the identity matrix $\mathbf{1}$ provides a basis for any $2\times2$ matrix. We can thus expand an arbitrary $2\times2$ matrix $\bx$ as $\bx=x_3\bm 1+x^a\gamma_a$, where raising and lowering indices are done with the metric $\eta_{ab}=\textrm{diag}(-1,1,1)$ with the indices $a,b$ taking values $0,1,2$. The condition of unit determinant yields the embedding equation \eqref{embedding_eq} and the Killing metric coincides exactly with the metric \eqref{embedding_eq}. The isometries of AdS$_3$ can now be implemented by left and right multiplications as
\begin{equation}
\bx\rightarrow g^{-1}\bx h, \quad g,h\in \textrm{\slr}.\label{isom}
\end{equation}
We will use this technique repeatedly to generate isometries of \ads. Note that in the coordinates \eqref{adscoord}, the matrix $\bx$ takes the form
\begin{align}
\bx=&\cosh\chi\cos t+\cosh\chi\sin t\gamma_0+\sinh\chi\cos\phi\gamma_1+\sinh\chi\sin\phi\gamma_2\nonumber\\
&=\cosh\chi\omega(t)+\sinh\chi\gamma(\phi),\label{matrixembads}
\end{align}
where we have defined the convenient matrices
\begin{equation}
\omega(\alpha)=\cos \alpha + \sin\alpha\gamma_0, \hspace{20pt}\gamma(\alpha)=\cos \alpha\gamma_1 + \sin\alpha\gamma_2.\label{omegagamma}
\end{equation}
\subsection{Geodesics}\label{geodesicssec}
We will now illustrate the power of the isomorphism of AdS$_3$ and the group manifold \slr by computing timelike and lightlike geodesics. Starting with the static geodesic located at $\chi=0$, we can apply isometries of AdS$_3$ to generate any timelike geodesic. The isometries we will use (which result in radial timelike geodesics passing the origin) are generated using equation \eqref{isom} with $g=\bh=\bu$, where the group element $\bu=e^{-\frac{1}{2}\zeta\gamma(\psi-\pi/2)}=\cosh\frac{1}{2}\zeta-\gamma(\psi-\pi/2)\sinh\frac{1}{2}\zeta$, and $\gamma$ is given by \eqref{omegagamma}. Computing $\pc\bx=\bu^{-1}\bx \bu$ results in the equations
\begin{align}
\begin{split}
\cos \pc{t}\cosh \pc\chi&=\cos t\cosh \chi,\\
\sin \pc t\cosh \pc\chi&=\cosh\chi\sin t \cosh \zeta - \sinh\chi\sinh\zeta\cos(\phi-\psi),\\
\sinh\pc\chi\cos\pc\phi&=-\cosh\chi\sinh\zeta\sin t\cos\psi+\sinh\chi(-\sin\psi\sin(\phi-\psi)+\cosh\zeta\cos\psi\cos(\phi-\psi)),\\
\sinh\pc\chi\sin\pc\phi&=-\cosh\chi\sinh\zeta\sin t\sin\psi+\sinh\chi(\cosh\zeta\cos(\phi-\psi)\sin\psi+\sin(\phi-\psi)\cos\psi).\label{boosteqs}
\end{split}
\end{align}
It can now be shown that the static geodesic at $\chi=0$ transforms into the oscillating timelike trajectory $\tanh\pc\chi=-\tanh\zeta\sin \pc t$ with $\pc\phi=\psi$. Of course, ($\psi$, $\zeta$) is the same transformation as ($\psi+\pi$, $-\zeta$). Note also that, for $\tanh\zeta\sin \pc t>0$, we have $\pc\chi<0$. We can thus choose either to allow for negative $\pc\chi$ or change the angle $\pc \phi$ by $\pi$ radians whenever $\tanh\zeta\sin \pc t>0$.\\
\linebreak
Note also that in the unprimed coordinates, the proper time $\tau$ of a stationary particle at the origin $\chi=0$ is given by $\tau=t$, and thus we can use this to express the trajectory of a moving particle in terms of the proper time. By taking the ratio of the first and second equations in \eqref{boosteqs}, after setting $\chi=0$, we obtain
\begin{equation}
\tan\pc t=\tan \tau\cosh\zeta.\label{proptime1}
\end{equation}
From the third equation after setting $\pc\phi=\psi$ we obtain
\begin{equation}
\sinh\pc\chi=-\sin \tau \sinh\zeta.\label{proptime2}
\end{equation}
We can also derive the useful relation
\begin{equation}
\cosh\pc\chi=\frac{\cos \tau }{\cos \pc t}=\sqrt{\cos^2\tau +\sin^2\tau \cosh^2\zeta}.\label{proptime3}
\end{equation}
\linebreak
Massless geodesics can be obtained by taking $\zeta\rightarrow\infty$, and these take the form $\tanh\pc\chi=-\sin \pc t$.

\section{Non-rotating pointlike particles in anti-de Sitter space}\label{ppsec}
\begin{figure}
\begin{center}
\includegraphics[scale=0.7]{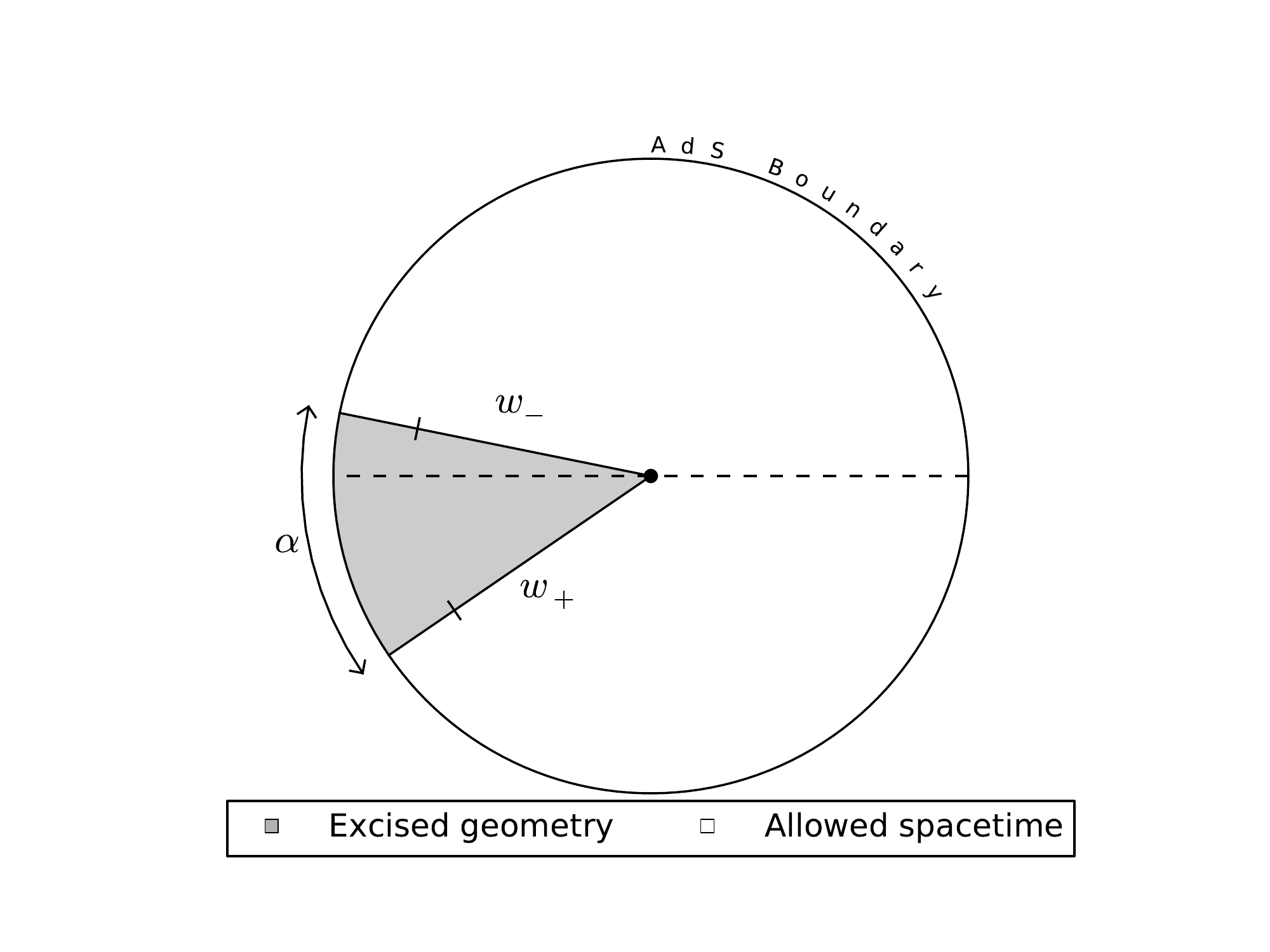}
\caption{\label{staticpp}A static particle in \ads. A wedge has been removed, and the surfaces $w_-$ and $w_+$ are identified by a rotation, resulting in a spacetime with a conical singularity with deficit angle $\alpha$. The wedge is not located symmetrically around $\phi=0$, so a boost in this direction will result in a moving pointlike particle constructed by excising a wedge not located symmetrically around its trajectory (see Figure \ref{1particle}). }
\end{center}
\end{figure}

In three-dimensional anti-de Sitter space, a static pointlike particle will take the form of a conical singularity at the origin of \ads. We can easily construct it by cutting out a piece of geometry (referred to as a {\it wedge}) and identifying the edges of this wedge by a rotation, as in Figure \ref{staticpp} (we would like to remind the reader that all figures in this chapter, unless explicitly stated otherwise, are in the coordinate system with metric \eqref{figurecoord}). The wedge is thus bordered by two surfaces $w_\pm$ at constant angles $\phi_\pm$, such that $\phi_+-\phi_-=2\nu$ is the deficit angle of the conical singularity. These planes are thus mapped to each other by a rotation of angle $2\nu$ which is described by the group element $\omega(\nu)=e^{\gamma_0\nu}=\cos\nu+\gamma_0\sin\nu\in$ \slr. In other words, we have $\omega(\nu)w_-=w_+\omega(\nu)$. The mass of a static particle is proportional to $\nu$ as we explain in Section \ref{ssec_stressenergypp}. More generally, if we have a particle moving along some geodesic, we can excise a piece of geometry that induces a conical singularity along the world line of this particle, and the edges of this piece are then identified by some non-trivial isometry of \ads that has the geodesic as fixed points. We will refer to all such excised pieces of geometry, which induce some conical deficit along a geodesic, as {\it wedges}, and for the special case of a static particle at the origin of \ads we will refer to them as {\it static wedges}. The element in \slr that induces the isometry is called the {\it holonomy} of the particle. In this section, we will construct moving pointlike particles by boosting the static particle, using the isometry \eqref{boosteqs}.
\subsection{Moving particles}
\begin{figure}
\begin{center}
\includegraphics[scale=0.8]{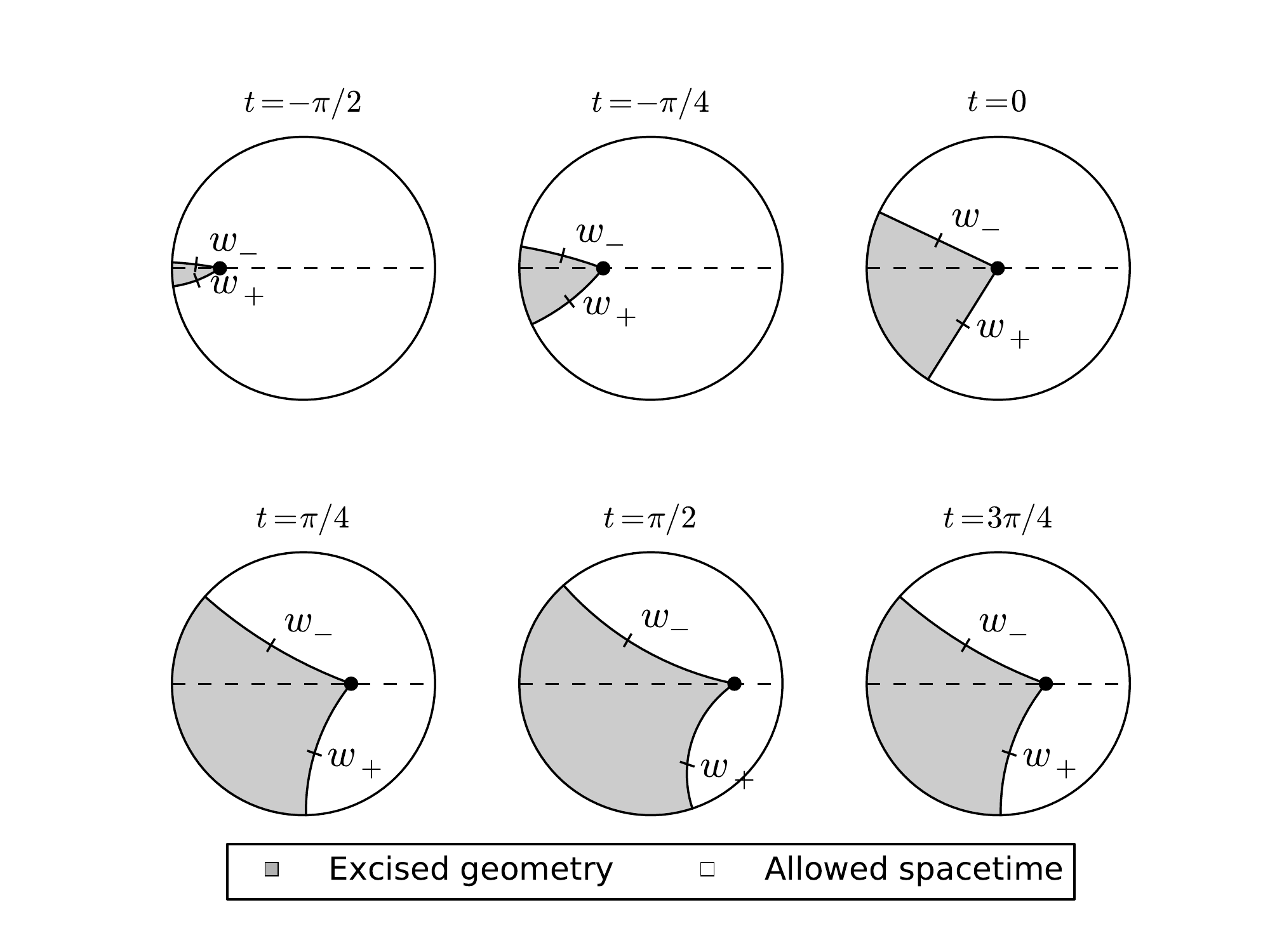}
\caption{\label{1particle} A moving massive particle, obtained by cutting out a wedge from \ads. The wedge in this example is not located symmetrically around the trajectory of the particle, meaning the parameter $p$ in equation \eqref{Gammapm} is non-zero.}
\end{center}
\end{figure}

We will now start with a static particle as in Figure \ref{staticpp}, with a wedge bordered by two surfaces $w_\pm$ at angles $\phi_\pm$, and then boost this spacetime along a direction $\psi$ (dashed line in Figure \ref{staticpp}). The new surfaces bordering the boosted wedge will be denoted by $\pc{w}_\pm$. The most common construction of a moving pointlike particle would be to align the wedge such that the direction $\psi$ is right in between $\phi_+$ and $\phi_-$. We will refer to this parametrization as a {\it symmetric wedge} or {\it symmetric parametrization}. However, this is just an arbitrary choice of coordinates, and it is possible to have many other parametrizations (or in other words, many different types of wedges can be removed which induce the same conical singularity at the trajectory of the particle). In our case, it will be enough to consider a one-parameter family of wedges, which are obtained by boosting a static wedge along an arbitrary direction $\psi$. To this end, we write $\phi_\pm=\psi\pm\nu(1\pm p)$, such that a symmetric wedge is obtained by setting $p=0$ and the deficit angle is given by $2\nu$. When applying the boost, we thus obtain a family of wedges parametrized by a continuous parameter $p$ (note that $p$ has no real physical meaning, and is analogous to a choice of coordinates or choice of gauge, but it will be a crucial ingredient when we consider colliding particles). By using \eqref{boosteqs}, it can be shown that, after applying a boost with boost parameter $\zeta$, the surfaces $\pc w_\pm$ bordering the wedge can be parametrized by
\begin{equation}
\tanh\pc\chi\sin(-\pc\phi +\Gamma_\pm+\psi)=-\tanh\zeta \sin\Gamma_\pm \sin\pc t, \label{c1}
\end{equation}
where
\begin{equation}
\tan\Gamma_\pm =\pm\tan((1 \pm p)\nu) \cosh\zeta.\label{Gammapm}
\end{equation}
These formulas can be derived by taking a linear combination of the last two equations in \eqref{boosteqs} such that it becomes proportional to the second equation in \eqref{boosteqs} (see Appendix \ref{app11} for a detailed derivation). In these formulas it is again clear that $(\psi$, $\zeta)$ is equivalent to $(\psi+\pi$, $-\zeta)$. The ambiguity in $\Gamma_\pm$ is chosen such that $|\Gamma_\pm-(p\pm1)\nu|<\pi$. If we insist on positive radial coordinate $\chi$, the ranges of the parameters for $w_\pm$ are
\begin{equation}
\pc\phi\in(\psi,\arcsin(\tanh\zeta\sin \Gamma_\pm\sin t)+\psi+\Gamma_\pm)
\end{equation}
for $\sin \pc t<0$, and
\begin{equation}
\pc\phi\in(\arcsin(\tanh\zeta\sin \Gamma_\pm\sin t)+\psi+\Gamma_\pm,\psi\pm\pi)
\end{equation}
for $\sin \pc t>0$. For $\sin \pc t=0$  we have $\pc \phi=\psi+\Gamma_\pm$. An illustration of a boosted particle with $p\neq0$ (non-symmetric parametrization), obtained by boosting Figure \ref{staticpp} along $\psi=\pi$ with $\zeta=1.5$, is shown in Figure \ref{1particle}. 
\subsection{Holonomy}
Let us now compute the holonomy of a boosted particle, meaning the group element $\bh$ such that $\pc w_-=\bh^{-1}\pc w_+\bh$. For a static particle the holonomy was given by $\omega(\nu)$ where $\omega$ is defined in \eqref{omegagamma}. In the new coordinate system, the new surfaces $\tilde{w}_\pm$ are given by $\pc{w}_\pm=\bu^{-1}w_\pm\bu$ where $\bu=e^{-\frac{1}{2}\zeta\gamma(\psi-\pi/2)}$. Since $w_-=\omega(\nu)^{-1}w_+\omega(\nu)$, we obtain that $\pc w_-$ and $\pc w_+$ are related by
\begin{equation}
\pc{w}_-=\bu^{-1}\omega(\nu)^{-1}\bu \pc{w}_+\bu^{-1}\omega(\nu)\bu,
\end{equation}
and thus the holonomy is given by
\begin{equation}
\pc{\bh}=\bu^{-1}\omega(\nu)\bu=\cos \nu+\gamma_0\cosh \zeta \sin \nu-\sinh \zeta \sin \nu \gamma(\psi).\label{hol}
\end{equation}
Note moreover that we have the following relation
\begin{equation}
\frac{1}{2}\text{Tr}(\bh)=\cos\nu,\label{pptrace}
\end{equation}
which can be used to obtain the deficit angle (the mass) of a particle moving along any geodesic.

\subsection{The stress-energy tensor}\label{ssec_stressenergypp}
In this section we will compute the stress-energy tensor of a pointlike particle. We will first compute it for a static particle, and then obtain the stress-energy tensor for a moving particle by applying a boost. For a static particle, the stress energy tensor takes the form
\begin{equation}
T^{\tau\tau}=T^{tt}=m\delta^{(2)}(x^\mu).
\end{equation}
Recall that $\tau=t$ is the proper time at the origin of \ads. $\delta^{(2)}$ is the standard two-dimensional delta function at the center of \ads such that the area integral on a constant time slice is equal to one. Assuming we have a conical singularity at the origin with conical deficit $\alpha$, we want to relate $m$ and $\alpha$. The Einstein equations take the form
\begin{equation}
R_{\mu\nu}-\frac{1}{2}Rg_{\mu\nu}+\Lambda g_{\mu\nu}=8\pi G T_{\mu\nu},
\end{equation}
where $G$ is Newton's constant in three dimensions. From this we can derive
\begin{equation}
R=6\Lambda+16\pi G m\delta^{(2)} (x^\mu),
\end{equation}
by taking the trace. Recall that we can choose a gauge where the lapse and shift functions of the metric take any values. By thus choosing a gauge where the metric takes the form $g_{tt}=f(\chi)$ (where $f(\chi)$ is some regular function, for example equal to $-\cosh^2\chi$ to coincide with that of \ads), $g_{ti}=0$ and $g_{ij}=\gamma_{ij}(\chi)$ it can be shown that for the two-dimensional Ricci scalar $\hat{R}[\gamma_{ij}]$ on a constant time slice, we have $\hat R=16\pi G m\delta^{(2)} (x^\mu)+$(finite terms). Now consider a small disc $D$ of radius $\chi$ around the origin. The geodesic curvature around the edge of the disc is $k_g=\coth \chi$. Using the Gauss-Bonnet theorem, as $\chi\rightarrow0$, we obtain
\begin{equation}
2\pi-\alpha=\lim_{\chi\rightarrow0}\int_0^{2\pi-\alpha}k_gds=2\pi\chi(D)-\lim_{\chi\rightarrow0}\frac{1}{2}\int_D {}^{(2)}\!R=2\pi-8\pi G m,
\end{equation}
and thus $\alpha=8\pi G m$, which is the same result as in \cite{Deser:1983tn}. Thus it may be appropriate to refer to the deficit angle $\alpha$ as the mass of the particle. Note that we also used that the Euler characteristic $\chi(D)$ of a disc is equal to one, as can be easily seen from Euler's formula applied to a triangle.\\
\newline
We have seen that a moving pointlike particle can be obtained by applying a boost, and thus we can now obtain the stress-energy tensor by applying the boost to the static stress-energy tensor. The transformation will be given by \eqref{boosteqs} and we will again denote the coordinates in the boosted coordinate system by $(\pc t,\pc\chi,\pc\phi)$ to distinguish them from the static coordinates $(t,\chi,\phi)$. Recall that $t$ coincides with the proper time of the static particle. Let us first figure out what the delta function looks like in the new coordinates, which we will denote by $\delta_{\psi,\zeta}$. The delta function will be taken to be proportional to $\delta(\pc\phi-\psi)\delta(\tanh\pc\chi+\sinh\zeta\sin\pc t)$, such that it singles out the trajectory $\tanh\pc\chi=-\sinh\zeta\sin\pc t$ and $\pc\phi=\psi$. This formula is not well defined for a static particle due to the conical singularity at the origin of \ads due to the polar coordinate nature of the coordinates we are using. Note that since this is a two-dimensional delta function in a three-dimensional space, the normalization must be such that $\int \delta_{\psi,\delta}=\Delta \tau$, where $\Delta \tau$ is the elapse in proper time along the geodesic that is inside the domain of the integral (the domain can be taken as a very thin tube covering a segment of the world line of the particle). By using equations \eqref{proptime1} and \eqref{proptime3} it can be easily shown that
\begin{equation}
\delta_{\psi,\zeta}\equiv\frac{\delta(\pc\phi-\psi)\delta(\tanh\pc\chi+\tanh\zeta\sin \pc t)}{\cosh\pc\chi\sinh\pc\chi\cosh\zeta},
\end{equation}
satisfies
\begin{equation}
\int_{\mathcal{D}}\sqrt{-g} \delta_{\psi,\zeta}\rd\pc\phi \rd\pc\chi \rd\pc t=\Delta \tau,
\end{equation}
where $\mathcal{D}$ is some region that covers $\Delta \tau$ of proper time of the particle's trajectory. The stress-energy tensor is now given by $T^{\mu\nu}=T^{\tau\tau}\dot{x}^\mu\dot{x}^\nu$. For the transformation of the stress-energy tensor, we have the relations $\dot{\pc\chi}=-\sinh\zeta\cos \pc t$ and $\dot{\pc t}=\cosh\zeta/\cosh^2\pc \chi$, which gives us finally
\begin{equation}
\begin{array}{cc}
 T^{\pc t\pc t}&=m\frac{\cosh^2\zeta}{\cosh^4\pc \chi}\delta_{\psi,\zeta},\\
 T^{\pc\chi\pc t}&=-m\frac{\cosh\zeta\sinh\zeta\cos \pc t}{\cosh^2\pc\chi}\delta_{\psi,\zeta},\\
 T^{\pc\chi\pc\chi}&=m\sinh^2\zeta\cos^2\pc t\delta_{\psi,\zeta},\\\label{Tpointparticle}
\end{array}
\end{equation}
while all other components vanish.

\subsection{The massless limit}
We can now use the result for the massive particle to obtain a moving massless particle. The massless limit is obtained by letting $\zeta\rightarrow\infty$ and $\nu\rightarrow0$, such that $\sinh\zeta\tan\nu\rightarrow E$ where $E$ can be interpreted as the energy of the particle. In this limit, the equations \eqref{c1} and \eqref{Gammapm} reduce to
\begin{equation}
\tanh\pc\chi\sin(-\pc\phi +\Gamma_\pm+\psi)=-\sin\Gamma_\pm \sin \pc t, \label{c1_massless}
\end{equation}
where
\begin{equation}
\tan\Gamma_\pm =(p\pm1)E.\label{Gammapm_massless}
\end{equation}
The holonomy is
\begin{equation}
\bh=1+E(\gamma_0-\gamma(\psi)).\label{hol0}
\end{equation}
These formulas are consistent with \cite{Matschull:1998rv}, where the special case of symmetric wedges ($p=0$) was used to study collisions of two massless particles.\\
\lb
Now let's compute the stress-energy tensor for the massless particle. The delta function $\delta_{\psi,\zeta}$ will become $\cosh\zeta\delta_{\psi,\zeta}\rightarrow\delta(\phi-\psi)\delta(v)/\sinh\chi$ where we have defined the infalling time coordinate $v=t+\arcsin(\tanh\chi)$ and used the relation $\cosh\chi=1/\cos t$ on the shell. The delta function was normalized with respect to proper time which is the reason why the normalization diverges in the massless limit. It can now be easily shown that in these infalling coordinates, the only non-zero component of the stress-energy tensor is
\begin{equation}
T^{\chi\chi}=\frac{E}{4\pi G}\frac{\delta(\phi-\psi)\delta(v)}{\cosh^2\chi\sinh\chi}.\label{Tpointparticleml}
\end{equation}

\begin{figure}
\includegraphics[scale=0.7]{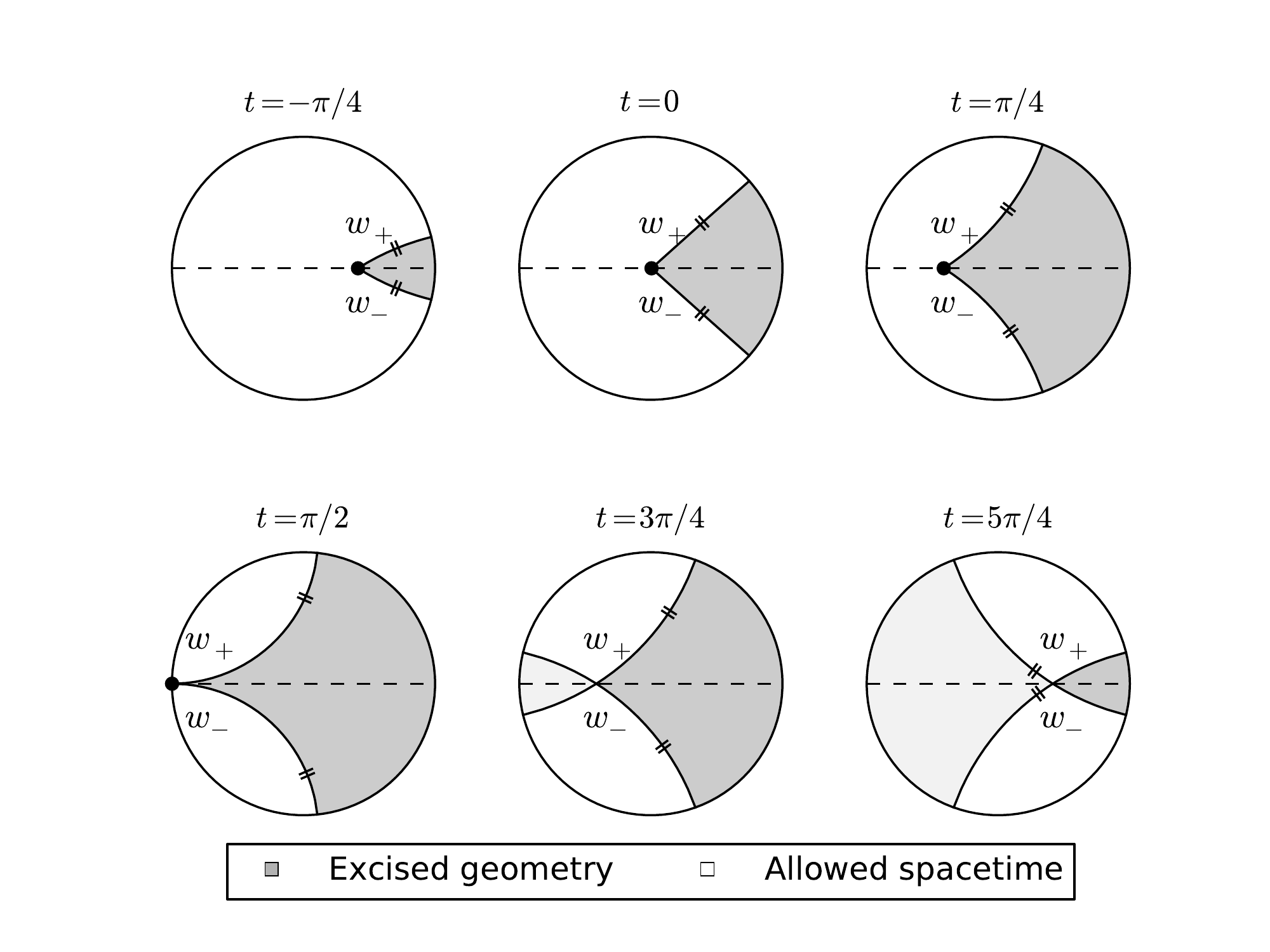}
\caption{\label{1massless}A massless particle moving through \ads, with a symmetric wedge excised behind the particle. The dark grey regions are removed parts of spacetime, the white regions are the allowed spacetime and the light shaded region is just two patches of the allowed geometry overlapping each other. The surfaces $w_+$ and $w_-$ are identified by an isometry of \ads. The particle is annihilated at $t=\pi/2$, and after the annihilation the spacetime is that of empty \ads written in unusual coordinates (the spacetime after the annihilation is covered by two patches that have been boosted in different directions and are glued together along the marked lines).}
\end{figure}

\section{The non-rotating BTZ black hole}\label{bhsec}
\begin{figure}
\includegraphics[scale=0.7]{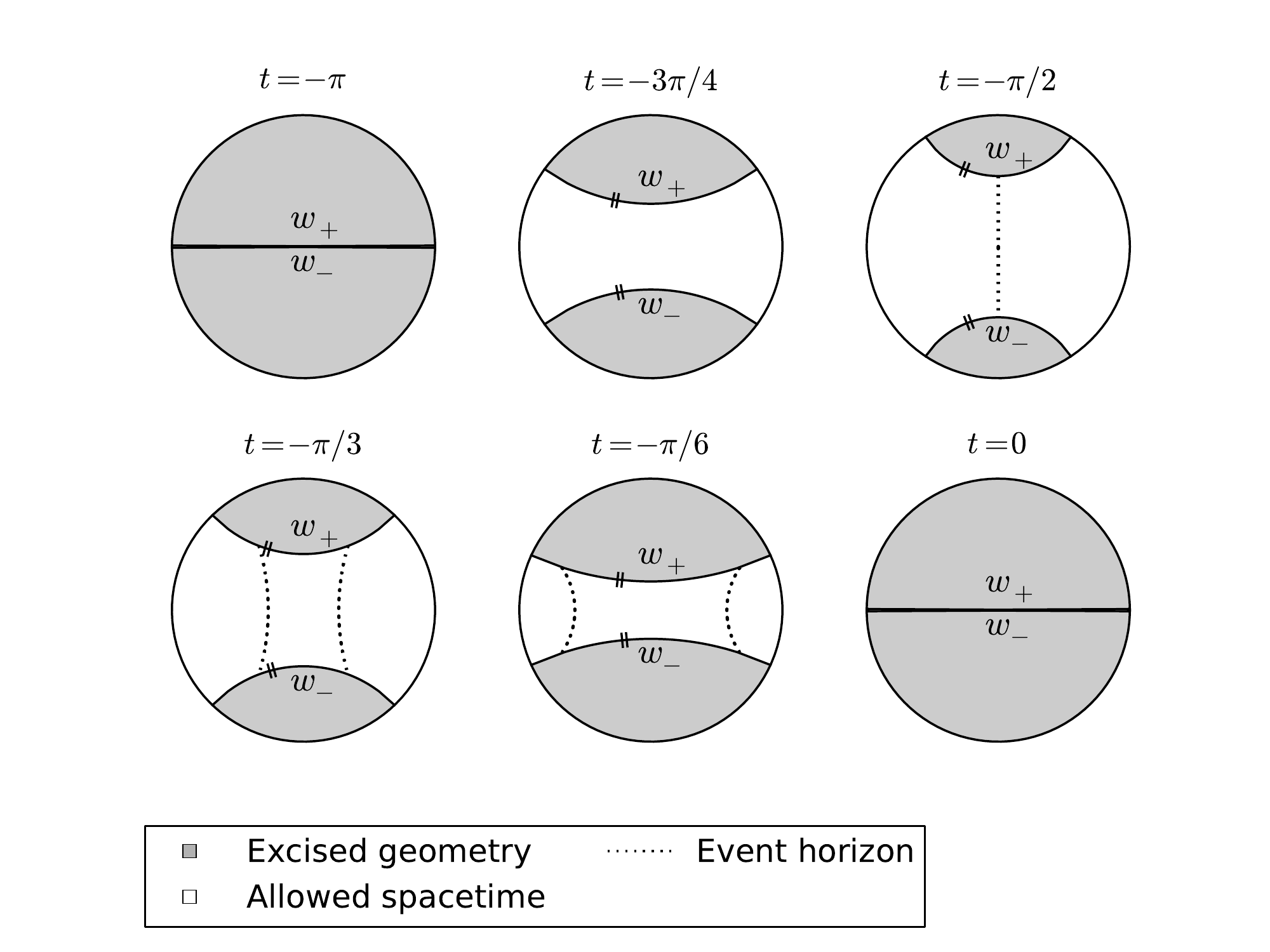}
\caption{\label{btzfig} The construction of the maximally extended BTZ black hole, by excising regions defined by equation \eqref{bhwedge1}. The grey regions are removed parts of the spacetime, and the white region is the allowed part. $w_+$ and $w_-$ are identified, meaning that when we cross $w_\pm$ from the white region, we are mapped to $w_\mp$. In this example we have $\mu=1.2$. The event horizon has been computed using techniques explained in Section \ref{eventsec}, and is defined as the boundary of the region from which lightrays can not reach the AdS boundary. The black hole singularities coincide with the horizontal lines at $t=-\pi$ and $t=0$.}
\end{figure}

\begin{figure}
\includegraphics[scale=0.7]{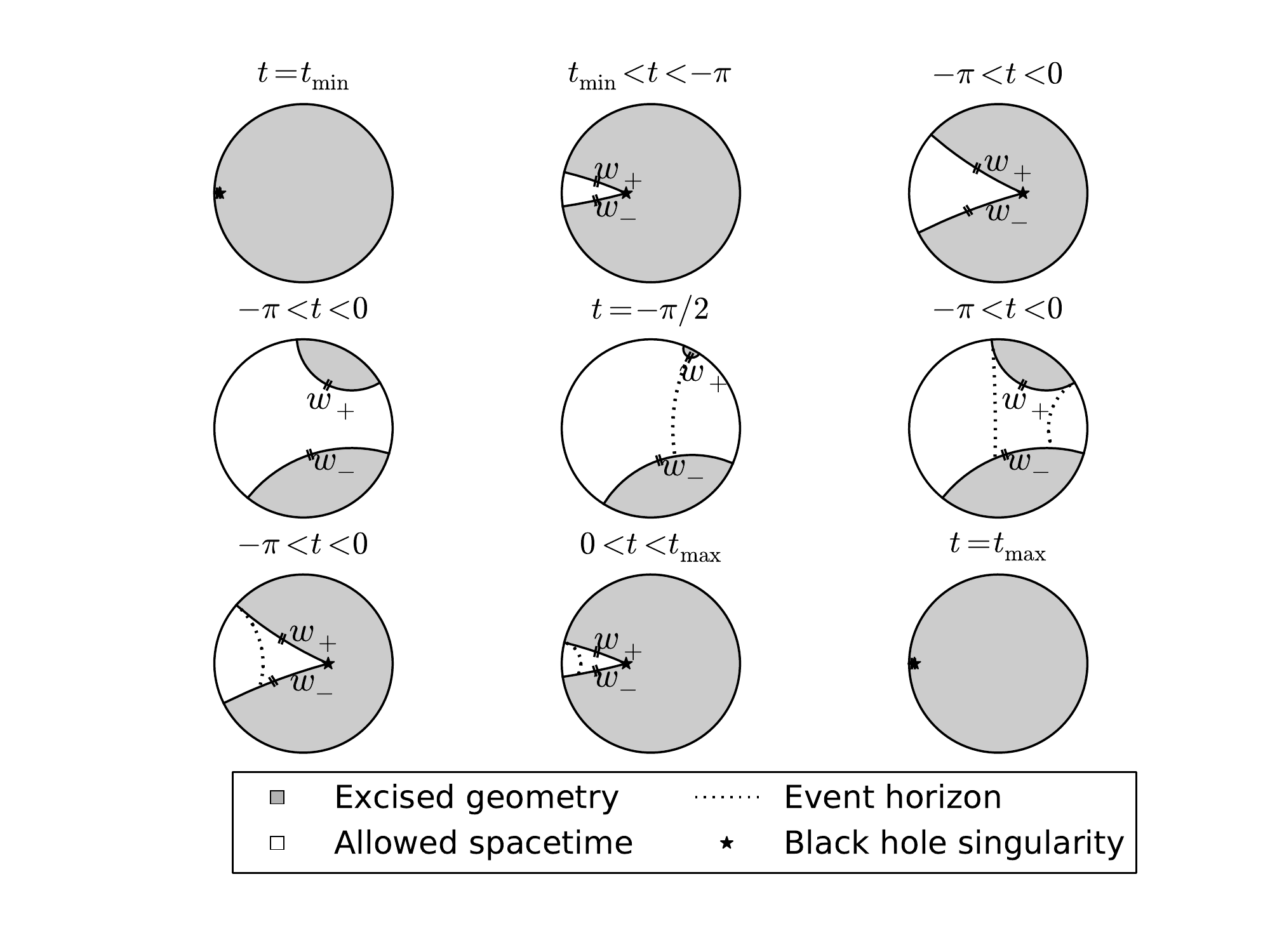}
\caption{\label{btzmodfig} The maximally extended BTZ black hole in the general coordinate system given by excising the regions defined by equation \eqref{bhwedge3}. $w_+$ and $w_-$ are identified, and the grey regions are the removed parts of the spacetime. The spacetime begins at $t=t_{\mathrm{min}}=-\arcsin(\tanh\zeta)-\pi$ and ends at $t=t_{\mathrm{max}}=\arcsin(\tanh\zeta)$. The parameters in this example are $\mu=1.8$, $\zeta=0.5$ and $\xi=1.1$. The event horizon has been computed using techniques explained in Section \ref{eventsec}, and is defined as the boundary of the region from which lightrays can not reach the AdS boundary.}
\end{figure}

In this section we will review the construction of the Banados-Teitelboim-Zanelli (BTZ) black hole \cite{Banados:1992wn,Banados:1992gq} that will be useful for our purposes. We will restrict ourselves to the non-rotating case, but see Chapter \ref{higher} for the rotating black hole. The BTZ black hole can be constructed by doing an identification of points in \ads, similar to that of a conical singularity. The identification will now have a spacelike geodesic as fixed points, compared to when constructing a conical singularity where the set of fixed points is a timelike geodesic. We will refer to this set of fixed points as the {\it singularity} of the black hole. For our purposes, the singularity will be a general radial geodesic passing through the origin of \ads, but we will start by choosing the simple radial geodesic given by $\phi=0$ and $t=0$ and then apply a set of isometries to obtain a singularity given by a more general spacelike geodesic. The isometry we will pick that leaves this geodesic invariant is given by $\bu=e^{\mu\gamma_1}=\cosh\mu+\gamma_1\sinh\mu$. We can now define a region of \ads, bordered by two surfaces $w_\pm$, and then cut out everything outside this region and make sure that these two surfaces are identified by the isometry as $\bu w_-=w_+\bu$. If we write the equations for $w_\pm$ as $w_\pm=\cosh\chi\omega(t)+\sinh\chi\gamma(\phi)$, and we assume that the two surfaces $w_\pm$ are located symmetrically around the singularity, the coordinates must satisfy
\begin{equation}
w_\pm: \tanh\chi\sin\phi=\mp\sin t\tanh\mu.\label{bhwedge1}
\end{equation}
We restrict $t$ to $-\pi\leq t\leq 0$. An illustration of this spacetime is shown in Figure \ref{btzfig}. We will also be interested in surfaces $w_\pm$ that are {\it not} symmetric around the singularity. This can be obtained by acting with an isometry of the form $e^{-\frac{1}{2}\xi\gamma_1}$ for some parameter $\xi$. This group element also has the singularity as fixed points. The surfaces $w_\pm$ transform as $w_\pm\rightarrow e^{\frac{1}{2}\xi\gamma_1}w_\pm e^{-\frac{1}{2}\xi\gamma_1}$ and after applying this isometry, the equations for the surfaces $w_\pm$ are instead
\begin{equation}
\tanh\chi\sin\phi=\mp\sin t\tanh(\mu\pm\xi).\label{bhwedge2}
\end{equation}
Points in this spacetime are still identified under the isometry $\bu=e^{\mu\gamma_1}$. What we have so far is a family of representations of the BTZ black hole where the singularity is given by the geodesic $(\phi=0,t=0)$, parametrized by $\xi$ while $\mu$ is related to the mass of the black hole. Note that $\xi$ has no physical meaning and only specifies our coordinate system, analogous to the parameter $p$ for the pointlike particles. However, we would like to have a general parametrization of the black hole where the singularity is any radial spacelike geodesic (this is analogous to boosting the pointlike particle). This can be obtained by applying the boost $e^{\frac{1}{2}\zeta\gamma_2}$. This is the same type of isometry that was used in Section \ref{ppsec} and will transform the singularity such that it now obeys $\tanh\chi=-\sin t\cosh \zeta$. After also applying a rotation such that the singularity is along an arbitrary angle $\psi$, it can be shown that the surfaces now take the form
\begin{equation}
w_\pm: \tanh \chi \sin(-\phi+\Gamma_\pm+\psi)=-\sin\Gamma_\pm\coth\zeta\sin t,\label{bhwedge3}
\end{equation}
where
\begin{equation}
\tan\Gamma_\pm=\mp\tanh(\mu\pm\xi)\sinh\zeta.\label{Gammapmu}
\end{equation}
This is very similar to the equations for the moving pointlike particle, equations \eqref{c1} and \eqref{Gammapm}, except that we have $\coth\zeta$ instead of $\tanh \zeta$ (which indicates that the geodesic obtained by setting $\phi=\psi$ is spacelike instead of timelike). An illustration of these coordinates is shown in Figure \ref{btzmodfig}. Note also that the parameter $\mu$ can be obtained from the trace of the holonomy via the relation
\begin{equation}
\frac{1}{2}\text{Tr}(\bh)=\cosh\mu,\label{bhtrace}
\end{equation}
which is similar to equation \eqref{pptrace} for a pointlike particle.\\
\linebreak
One way to see that the spacetime we have constructed is really a BTZ black hole is to do a coordinate transformation such that we obtain one of the standard metrics of a black hole. One convenient metric is given by
\begin{equation}
\rd s^2=-\rd t^2 (-M+\rho^2)+\rd \rho^2\frac{1}{-M+\rho^2}+\rho^2 \rd \phi^2\label{BTZ},
\end{equation}
where $M$ is called the mass of the hole. It turns out that we can do a coordinate transformation such that the surfaces $w_\pm$ become simple circle sectors, but defined in a spacetime \eqref{BTZ} with $M=1$. This coordinate transformation can be obtained by a different parametrization of the embedding equation \eqref{embedding_eq}. We start by parametrizing 
\begin{equation}
x_0=\rho\cosh y,\quad\quad x_2=\rho\sinh y,\label{rhoembedding}
\end{equation}
where $\rho\geq0$. It can then be shown that the isometry $\bu$ just acts as a translation by $y\rightarrow y-2\mu$, and thus the surfaces $w_\pm$, given by \eqref{bhwedge1}, are just planes at constant values of $y$, bordering a circle sector with opening angle $2\mu$. Note that this only parametrizes a subset which satisfies $|x_0|>|x_2|$, but this inequality is always obeyed by the spacetime defined by \eqref{bhwedge1}. We also have the restriction $x_0\geq0$ which is consistent with our choice $-\pi\leq t\leq 0$. Our embedding equation \eqref{embedding_eq} has thus turned into
\begin{equation}
 x_3^2-x_1^2=1-\rho^2.
\end{equation}
It is clear that, to continue, we must decide if $\rho$ is larger or smaller than $1$. For $0\leq\rho\leq1$ (corresponding to $|x_3|\geq|x_1|$), we have the parametrization 
\begin{equation}
x_1=\sqrt{1-\rho^2}\sinh (\sigma),\quad x_3=-\sqrt{1-\rho^2}\cosh (\sigma).
\end{equation}
For $\rho\geq1$ ($|x_3|\leq|x_1|$), we can choose the parametrization 
\begin{equation}
x_1=\sqrt{\rho^2-1}\cosh(\sigma),\quad x_3=-\sqrt{\rho^2-1}\sinh(\sigma).
\end{equation}
They both lead to the metric
\begin{equation}
\rd s^2=-(-1+\rho^2)\rd\sigma^2+\frac{\rd\rho^2}{-1+\rho^2}+\rho^2\rd y^2.\label{unitbhmetric}
\end{equation}
which is indeed the BTZ black hole metric \eqref{BTZ} with $M=1$. Note that the two different parametrizations above correspond to inside and outside the event horizon, respectively. Moreover, the parametrization restricts to the case where $x_3\leq0$ inside the horizon and $x_1\geq0$ outside the horizon. By comparing to \eqref{adscoord}, we see that this corresponds to $-\pi/2\leq t\leq0$ inside and $-\pi/2\leq\phi\leq\pi/2$ outside. This means in particular that when we consider the spacetime defined by \eqref{bhwedge1} or \eqref{bhwedge2} in these coordinates, only one of the two disconnected boundaries and only the singularity at $t=0$ are covered by these coordinates (see Figure \ref{btzfig}). This is why the construction in this section is called the maximally extended black hole, since it covers a larger spacetime than the standard coordinates with metric \eqref{BTZ}. Moreover, in this coordinate system the surfaces defined by \eqref{bhwedge2} take the form of static circular sectors with opening angle $2\mu$. By identifying $w_+$ and $w_-$ and then rescaling the coordinate $y$ so that it has the standard range $2\pi$, it is then possible to obtain a relation between $\mu$ and the mass $M$, namely $M=\mu^2/\pi^2$. However, let us consider a more general setup where we instead have many such wedges, defined by $w_\pm^i$, where instead of having $w_-^i$ and $w_+^i$ being identified, we have that $w_-^i$ is identified with $w_+^{i-1}$. It is then clear that the whole spacetime can be transformed to a set of circular sectors with metric \eqref{unitbhmetric}, where each circle sector is linked to the next one. This spacetime then has a total angle of $\alpha=\sum_i2\mu_i$. Rescaling the coordinates by $y\rightarrow\alpha y/(2\pi)$, $\sigma\rightarrow\alpha\sigma/(2\pi)$ and $\rho\rightarrow2\pi\rho/\alpha$ yields the metric\\
\begin{equation}
\rd s^2=-(-M+\rho^2)\rd\sigma^2+\frac{\rd\rho^2}{-M+\rho^2}+\rho^2\rd y^2,\label{blackholemetric}
\end{equation}
which is just the standard metric of a BTZ black hole with mass $M$, and where $M$ is given by $M=\alpha^2/(2\pi)^2=(\sum_i\mu_i)^2/\pi^2$. Now $y$ has the standard range from $0$ to $2\pi$.\\
\linebreak
Another coordinate system that we will use, can be obtained by defining $\cosh\beta=\rho$ in \eqref{unitbhmetric}. This is only defined for $\rho\geq1$ and thus only covers the region outside the event horizon. The metric \eqref{unitbhmetric} then takes the form
\begin{equation}
\rd s^2=-\sinh^2\beta \rd\sigma^2+\rd\beta^2+\cosh^2\beta \rd y^2,\label{betabhcoord}
\end{equation}
which is reminiscent of \eqref{adsmetric}, which is why we will find this coordinate system convenient.

\section{Collision of two particles}
We will now consider the simplest example of colliding particles, namely that of two particles colliding to form a single object. The particles will always be assumed to be colliding head-on in the center of AdS ($t=0,\chi=0$) which implies that the particles are both created at the boundary at time $t=-\pi/2$ and fall along radial geodesics. The first particle will fall in at the angle $\psi_1=0$ and the second particle at the angle $\psi_2=\pi$. We will consider the more general case of two colliding massive particles, and then later specialize to the simpler case of two colliding massless particles which was first considered in \cite{Matschull:1998rv}. The particles will thus be specified by their deficit angles (in their respective rest frames) given by $2\nu_i$ and boost parameters given by $\zeta_i$, where $i=1,2$. It should be pointed out that in the case of massless particles, it is possible to move to a center of momentum frame such that both particles have the same energy. For massive particles however, each particle has two independent parameters (the rest mass and the boost parameter), thus it is not in general possible to move to a frame where $\nu_1=\nu_2$ and $\zeta_1=\zeta_2$. The best we can do is to reduce the number of free parameters to three. We will later use this freedom to simplify the computations, but for now we keep the parameters $\zeta_i$ and $\nu_i$ general.\\
\linebreak
The two particles will be constructed by excising a wedge, as explained in Section \ref{ppsec}. Both wedges will be located behind and symmetrically around each particle's trajectory, meaning that $p_1=p_2=0$ (see equations \eqref{c1} and \eqref{Gammapm}). This is a consequence of the reflection symmetry in the $\phi=0$ plane, and we will see that in the general case for more particles and without any symmetry restrictions, we will need to use general wedges with $p_i\neq0$. The first wedge is thus bordered by two surfaces $w_1^\pm$ given by
\begin{equation}
\tanh\chi\sin(-\phi +\Gamma^1_\pm)=-\tanh\zeta_1 \sin\Gamma^1_\pm \sin t,\label{wpm1}
\end{equation}
where
\begin{equation}
\tan\Gamma^1_\pm =\pm\tan\nu_1 \cosh\zeta_1.
\end{equation}
The wedge of the second particle is bordered by two surfaces given by
\begin{equation}
\tanh\chi\sin(-\phi +\Gamma^2_\pm)=\tanh\zeta_2 \sin\Gamma^2_\pm \sin t,\label{wpm2}
\end{equation}
where
\begin{equation}
\tan\Gamma^2_\pm =\pm\tan\nu_2 \cosh\zeta_2.
\end{equation}
The holonomies of the particles are given by
\begin{align}
\bh_1=&\cos \nu_1+\gamma_0\cosh \zeta_1 \sin \nu_1-\sinh \zeta_1 \sin \nu_1 \gamma_1,\nonumber\\
\bh_2=&\cos \nu_2+\gamma_0\cosh \zeta_2 \sin \nu_2+\sinh \zeta_2 \sin \nu_2 \gamma_1.\label{massiveholonomies}
\end{align}
Past the collision, there is a natural way to continue this spacetime such that the two particles merge and form one joint object, namely by identifying the intersection $I_{1,2}$ between $w_+^1$ and $w_-^2$ and the intersection $I_{2,1}$ between $w_+^2$ and $w_-^1$ as the new joint object, see Figure \ref{2particles}. Note that, due to the identifications of the wedges and the reflection symmetry in the $\phi=0$ plane, these two intersections are really the same spacetime point. Thus the spacetime after the collision is now composed of two separate patches, which are glued to each other via the isometries associated to the particles. Let us call these two wedges of spacetime $c_{1,2}$ and $c_{2,1}$. $c_{1,2}$ is bordered by the surfaces $w_-^{1,2}\equiv w_+^1$ and $w_+^{1,2}\equiv w_-^2$, while $c_{2,1}$ is bordered by $w_-^{2,1}\equiv w_+^2$ and $w_+^{2,1}\equiv w_-^1$. These two wedges can now be identified as being part of either a conical singularity spacetime, or a black hole spacetime, by matching the equations of these wedges to that of either \eqref{c1} and \eqref{Gammapm}, or that of \eqref{bhwedge3} and \eqref{Gammapm_massless}, respectively. The easiest way to see if we have formed a massive pointlike particle (conical singularity) or a black hole, is to see if the resulting radial geodesics, meaning the intersections $I_{1,2}$ and $I_{2,1}$, are timelike or spacelike. Let us first compute the intersection angles $\phi_{1,2}$ and $\phi_{2,1}$. From \eqref{wpm1} and \eqref{wpm2} we easily find that the intersections are given by
\begin{equation}
\tan\phi_{1,2}=-\tan\phi_{2,1}=\tan\nu_1\tan\nu_2\left[\frac{\sinh\zeta_1\cosh\zeta_2+\sinh\zeta_2\cosh\zeta_1}{\sinh\zeta_2\tan\nu_2-\sinh\zeta_1\tan\nu_1}\right],
\end{equation}
with the conventions $0\leq\phi_{1,2}\leq\pi$ and $\pi\leq\phi_{2,1}\leq2\pi$. Now let us focus on the wedge $c_{1,2}$. We can write $w_\pm^{1,2}$ in the following way
\begin{equation}
w_+^{1,2}: \tanh\chi\sin(-\phi+\phi_{1,2}+(\Gamma^2_--\phi_{1,2}))=-\frac{\tanh\zeta_2\sin\Gamma_-^2}{\sin(\phi_{1,2}-\Gamma^2_-)}\sin(\Gamma^2_--\phi_{1,2})\sin t,
\end{equation}
\begin{equation}
w_-^{1,2}: \tanh\chi\sin(-\phi+\phi_{1,2}+(\Gamma^1_+-\phi_{1,2}))=-\frac{\tanh\zeta_1\sin\Gamma_+^1}{\sin(\Gamma^1_+-\phi_{1,2})}\sin(\Gamma^1_+-\phi_{1,2})\sin t.
\end{equation}
By now comparing these equations to \eqref{Gammapm} or \eqref{bhwedge3}, we see that the parameters for the new wedge $c_{1,2}$ can be identified as $\Gamma_+^{1,2}=\Gamma_-^2-\phi_{1,2}$, $\Gamma_-^{1,2}=\Gamma_+^1-\phi_{1,2}$ and
\begin{equation}
\frac{\tanh\zeta_1\sin\Gamma_+^1}{\sin(\Gamma^1_+-\phi_{1,2})}=\frac{\tanh\zeta_2\sin\Gamma_-^2}{\sin(\phi_{1,2}-\Gamma^2_-)}\equiv\Bigg\{\begin{array}{ll}
                                                                                                                                             \tanh\zeta_{1,2},&\text{Pointlike particle}\\
                                                                                                                                             \coth\zeta_{1,2},&\text{Black hole}                                                                                                                                             
                                                                                                                                            \end{array}\label{zetadef2}
\end{equation}
The definition of $\phi_{1,2}$ ensures that the above equality holds and an analogous computation can be done for the wedge $c_{2,1}$. Note that $\zeta_{1,2}$ will be negative, but this is not as issue and could in principle be fixed by rotating the angle $\phi_{1,2}$ by $\pi$. The mass of the resulting object (the parameter $\nu$ for a pointlike particle or the parameter $\mu$ for a black hole) can also be obtained via the relations \eqref{pptrace} or \eqref{bhtrace}, where the holonomy of the resulting object is the product of the incoming particles' holonomies. The trace is given by
\begin{equation}
\frac{1}{2}\text{Tr}(\bh_1\bh_2)=\cos\nu_1\cos\nu_2-\cosh(\zeta_1+\zeta_2)\sin\nu_1\sin\nu_2,
\end{equation}
and black hole formation occurs when $|\text{Tr}(\bh_1\bh_2)|>2$.

\subsection{Collision of massive particles in the center of momentum frame}
We will now use our coordinate freedom to pick a frame such that
\begin{equation}
\tan\nu_1\sinh\zeta_1=\tan\nu_2\sinh\zeta_2\equiv E,\label{restframe}
\end{equation}
which we will interpret as the center of momentum frame of the collision process. This will simplify the computations considerably. We will thus parametrize our setup with $E$, $\zeta_1$ and $\zeta_2$. In this case we have that $\phi_{1,2}=\pi/2$, $\phi_{2,1}=3\pi/2$ and \eqref{zetadef2} can be shown to be equal to $-E$.
\subsubsection{Formation of a conical singularity}
In the case of a formation of a conical singularity, we should use equation \eqref{Gammapm} to compute the deficit angle. For the wedge $c_{1,2}$, it states that
\begin{equation}
\tan\Gamma_\pm^{1,2}=\pm\tan(\nu_{1,2}(1\pm p_{1,2}))\cosh\zeta_{1,2}.
\end{equation}
The deficit angle for $c_{2,1}$ will be the same, so the total deficit angle of the resulting conical singularity in the restframe is then given by $2\pi-2\nu_{1,2}-2\nu_{2,1}=2\pi-4\nu_{1,2}$. It is straightforward to compute $\nu_{1,2}$ as being given by
\begin{equation}
\tan(2\nu_{1,2})=\tan(\nu_{1,2}(1+p_{1,2})+\nu_{1,2}(1-p_{1,2}))=E\sqrt{1-E^2}\left(\frac{\tanh\zeta_1+\tanh\zeta_2}{E^2+(E^2-1)\tanh\zeta_1\tanh\zeta_2}\right).
\end{equation}
For $E>1$ this result is no longer valid, and instead a black hole will form. The collision of two massless particles forming a massive particle is shown in Figure \ref{2particles_pp}.
\subsubsection{Formation of a black hole}
\begin{figure}
\begin{center}
\includegraphics[scale=0.8]{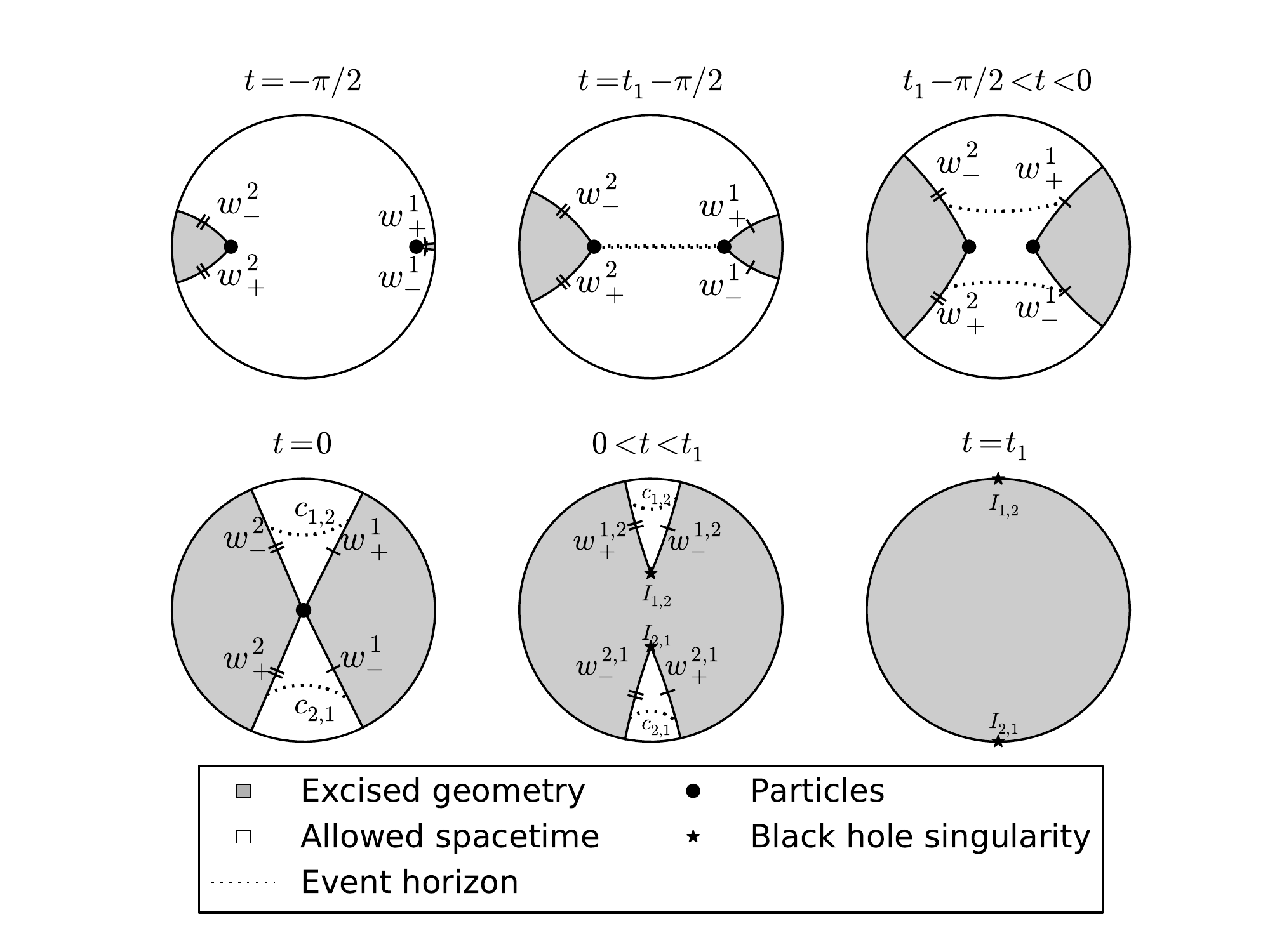}
\caption{\label{2particles} Collision of two particles with different rest masses in the center of momentum frame, forming a black hole. The parameters are $\nu_1=0.3$, $\nu_2=0.9$ and $E=2$. The event horizon has been computed using techniques explained in Section \ref{eventsec}, and is defined as the boundary of the region from which lightrays can not reach the AdS boundary.}
\end{center}
\end{figure}

In the case of a formation of a black hole, we should use equation \eqref{Gammapmu} to compute the mass of the black hole. It states that
\begin{equation}
\tan\Gamma_\pm^{1,2}=\mp\tanh(\mu_{1,2}\pm \xi_{1,2})\sinh\zeta_{1,2}.
\end{equation}
Here $\mu_{1,2}$ will determine the mass of the black hole (see Section \ref{bhsec}) and we then obtain
\begin{equation}
\tanh(2\mu_{1,2})=\tanh(\mu_{1,2}+\xi_{1,2}+\mu_{1,2}-\xi_{1,2})=E\sqrt{E^2-1}\left(\frac{\tanh\zeta_1+\tanh\zeta_2}{E^2+(E^2-1)\tanh\zeta_1\tanh\zeta_2}\right),
\end{equation}
which is only defined for $E>1$. Another way to see that a black hole has really formed is to directly compute the event horizon, which is also marked in Figure \ref{2particles}. It coincides with the backwards lightcone of the last points of the wedges $c_{1,2}$ and $c_{2,1}$ and will be discussed in more detail in Section \ref{eventsec}.

\subsection{Collision of massless particles in the center of momentum frame}
\begin{figure}
\includegraphics[scale=0.7]{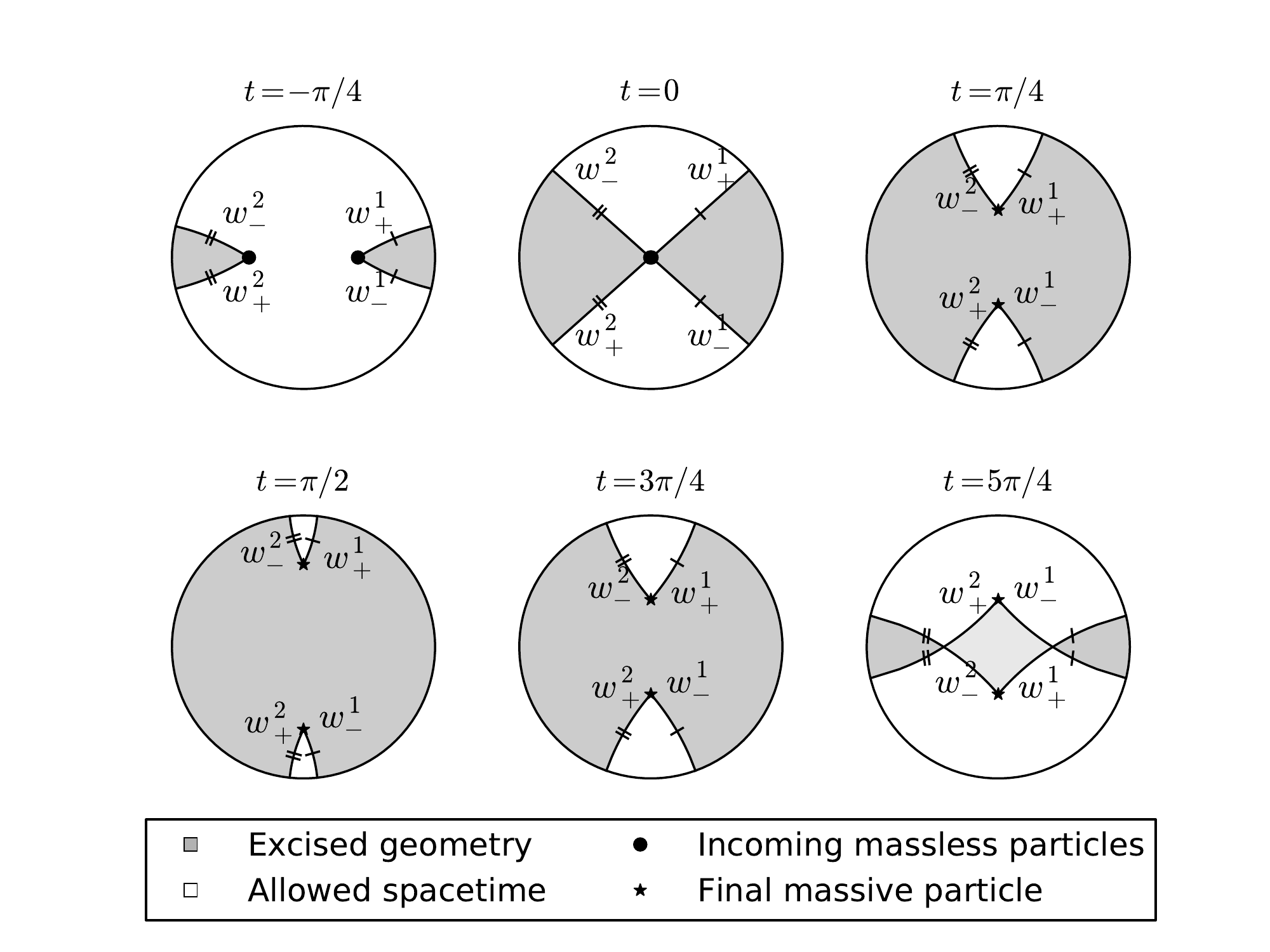}
\caption{\label{2particles_pp} Collision of two massless particles with equal energy forming a massive pointlike particle. The particles fall along radial geodesics and the energies of the particles are given by $E=0.9$. The light grey region in the last panel marks that we have two patches of geometry overlapping each other. The two oscillating geodesics which are identified and represent the resulting massive particle, will keep oscillating forever in this coordinate system.}
\end{figure}

The result for massless particles (with equal energies) is easily obtained from the massive particle result by taking $\zeta_i\rightarrow0$ and $\nu_i\rightarrow0$ while keeping $E$ fixed. This results in the relation $E=\cos\nu_{1,2}$ and $E=\cosh\mu_{1,2}$ for formation of a pointlike particle and a black hole respectively, reproducing the results in \cite{Matschull:1998rv}. The holonomies of the two particles are
\begin{equation}
\bh_1=1+E(\gamma_0- \gamma_1),\quad\bh_2=1+E(\gamma_0+\gamma_1),\label{masslessholonomies}
\end{equation}
and the trace of the product of the two holonomies is
\begin{equation}
\frac{1}{2}\text{Tr}(\bh_1\bh_2)=1-2E^2,
\end{equation}
from which we can also obtain $\nu_{1,2}$ and $\mu_{1,2}$ via \eqref{pptrace} and \eqref{bhtrace}. The threshold for black hole formation is $E=1$. A collision of two massless particles forming a massive pointlike particle is shown in Figure \ref{2particles_pp}. Two massless pointlike particles (with equal energies) colliding to form a black hole is shown in a 3d illustration on the front page of this thesis. Note that at the threshold $E=1$, the trace of the resulting holonomy is minus two instead of plus two. This property was used incorrectly in \cite{Matschull:1998rv} as a way to distinguish the extremal black hole from the massless pointlike particles, but since the isometries act as $x\rightarrow \bh x \bh^{-1}$, two holonomies which only differ by a sign are identical and can not be distinguished by this property.

\subsection{A comment on symmetric wedges}
In this section the wedges corresponding to each particle were located symmetrically behind the particle. Will this also work for more general collision processes? The answer is generically no, and this can be seen already in the two-particle case. Let's assume for example that the two particles are not moving on exactly opposite angles, say $\psi_1=0$ but $\psi_2\neq\pi$. We could then try to play the same game, excising a symmetric wedge behind each particle, and then try to identify the intersections $I_{1,2}$ and $I_{2,1}$ as a final joint object. The problem here is that, even though $w_+^i$ is mapped to $w_-^i$ by $\bh_i$, the intersection $I_{1,2}$ between $w_+^1$ and $w_-^2$ will generically not be mapped to the intersection $I_{2,1}$ between $w_-^1$ and $w_+^2$, and we can not identify this as a joint object (since it is not the same spacetime point) and this coordinate system breaks down after the collision. To solve this problem, one must allow for more general wedges, specified by the parameters $p_i$ according to equation \eqref{c1} and \eqref{Gammapm}, and then adjust these parameters such that the intersections are mapped to each other. For the two-particle case, this is not necessary since we can always make a coordinate transformation where the particles collide head-on and the symmetric wedges are applicable. For many particles this is not the case, and it is crucial to allow for more general wedges to be able to construct consistent solutions for arbitrary initial conditions. We will explore this in more detail in remainder of this chapter.

\section{Collisions of many particles}
\begin{figure}
\includegraphics[scale=0.7]{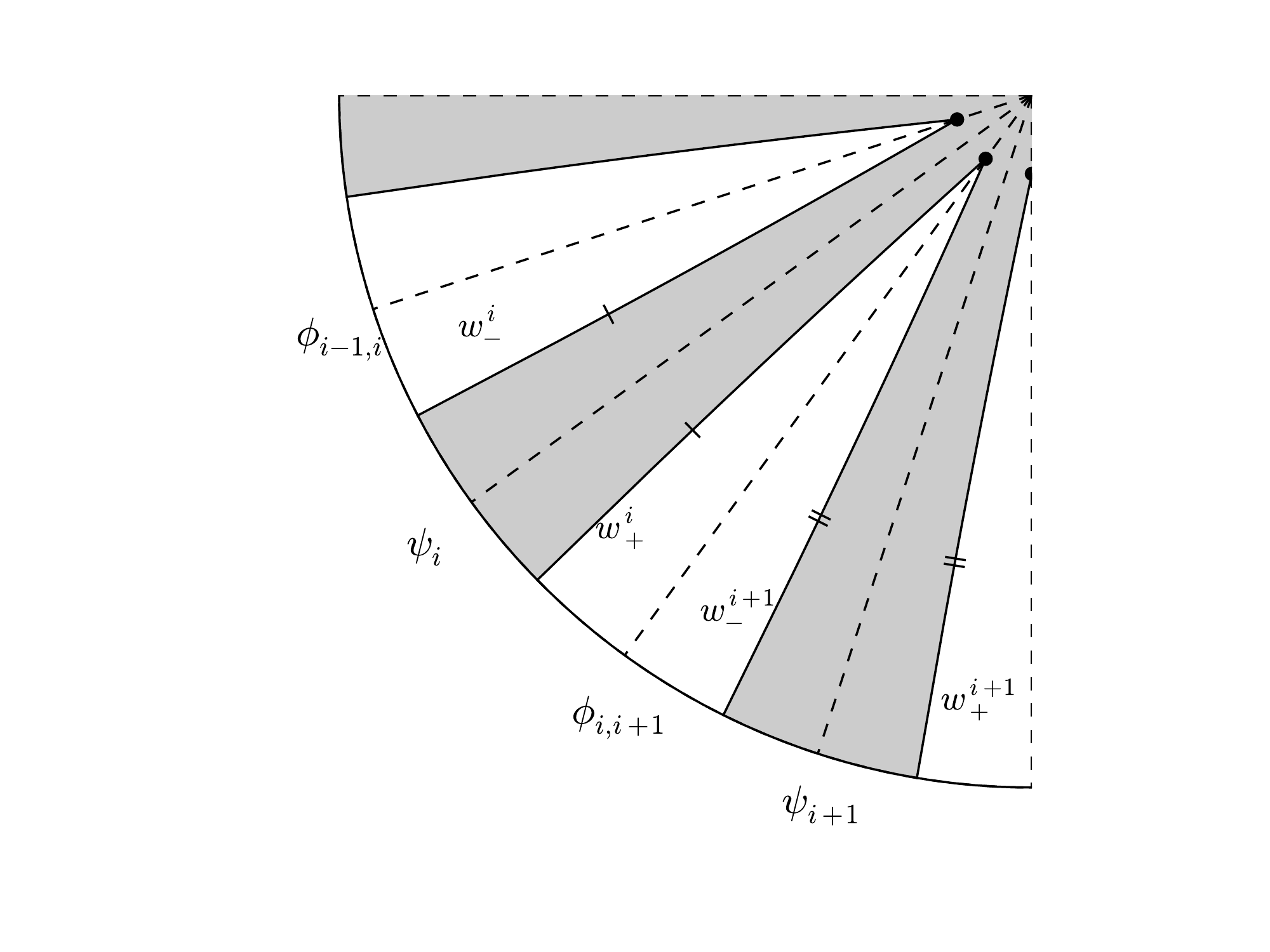}
\caption{\label{illustration} An illustration of the different parameters after the collision has occurred. The grey regions are the removed parts of the spacetime. The particles fall in along the angles $\psi_i$, and the wedge associated to particle $i$ is bounded by $w_\pm^i$. The joint object then moves on angles $\phi_{i,i+1}$, which is the intersection of $w_+^i$ and $w_-^{i+1}$. Parameters with index $i$ (for example $\Gamma_\pm^i$, $p_i$ and $e_i$) are associated to the (grey) wedge at angle $\psi_i$, while parameters with index $i,i+1$ (for example $\Gamma_\pm^{i,i+1}$, $\zeta_{i,i+1}$ and $\nu_{i,i+1}$) are associated to the (white) wedge at angle $\phi_{i,i+1}$. Note also that we will sometimes instead use the notation $w_-^{i,i+1}=w_+^{i}$ and $w_+^{i,i+1}=w_-^{i+1}$.}
\end{figure}

In this section we will consider collisions of $N$ particles, without any symmetry restrictions. This means that each particle can fall in along an arbitrary angle $\psi_i$, and have any mass $\nu_i$ and any boost parameter $\zeta_i$ (for the case of massless particles, this is replaced by an arbitrary energy $E_i$). The particles will be constructed by starting with \ads and then excising a wedge behind each particle as explained in Section \ref{ppsec}. The only restrictions we will impose are that all particles collide at one single point $(t=0,\chi=0)$ and that they join and form a single object. The wedges associated to the particles will be bordered by surfaces denoted by $w_\pm^i$. After the collision, we will identify the intersections $I_{i,i+1}$ of $w_+^i$ and $w_-^{i+1}$ as the final object, and the wedge bordered by these two surfaces will be denoted by $c_{i,i+1}$. The intersection $I_{i,i+1}$ will move on a radial geodesic at angle $\phi_{i,i+1}$. The resulting allowed spacetime after the collision will thus consist of a set of disconnected wedges, linked together via the holonomies of the incoming particles. An illustration of various variables discussed in this section is shown in Figure \ref{illustration}. We will call the wedges of the incoming wedges as {\it initial wedges} and the wedges after the collision as {\it final wedges}. Contrary to the case of two colliding particles, to construct the collision process of $N$ particles we will need to use non-symmetric initial wedges for the particles, so we will thus associate a parameter $p_i$ to each initial wedge quantifying how much it deviates from a symmetric wedge (see equation \eqref{Gammapm} and \eqref{Gammapm_massless}). The reason is that the intersections $I_{i,i+1}$ will not necessarily be mapped to each other via the holonomies of the particles and can thus not always be identified as one and the same point in spacetime. We will thus tune the parameters $p_i$ for the initial wedges such as to make sure this is achieved. Only in the case of discrete rotational symmetry are symmetric wedges enough. We will use this technique in the next sections to study both colliding massive and massless particles. The computations for massive and massless particles are quite similar but there are some important differences, and we will separate them into different sections which are supposed to be as independent as possible. The massless case can of course also be obtained by taking a limit of the massive case. 

\subsection{Massless particles}\label{masslesscollsec}
\begin{figure}
\includegraphics[scale=0.7]{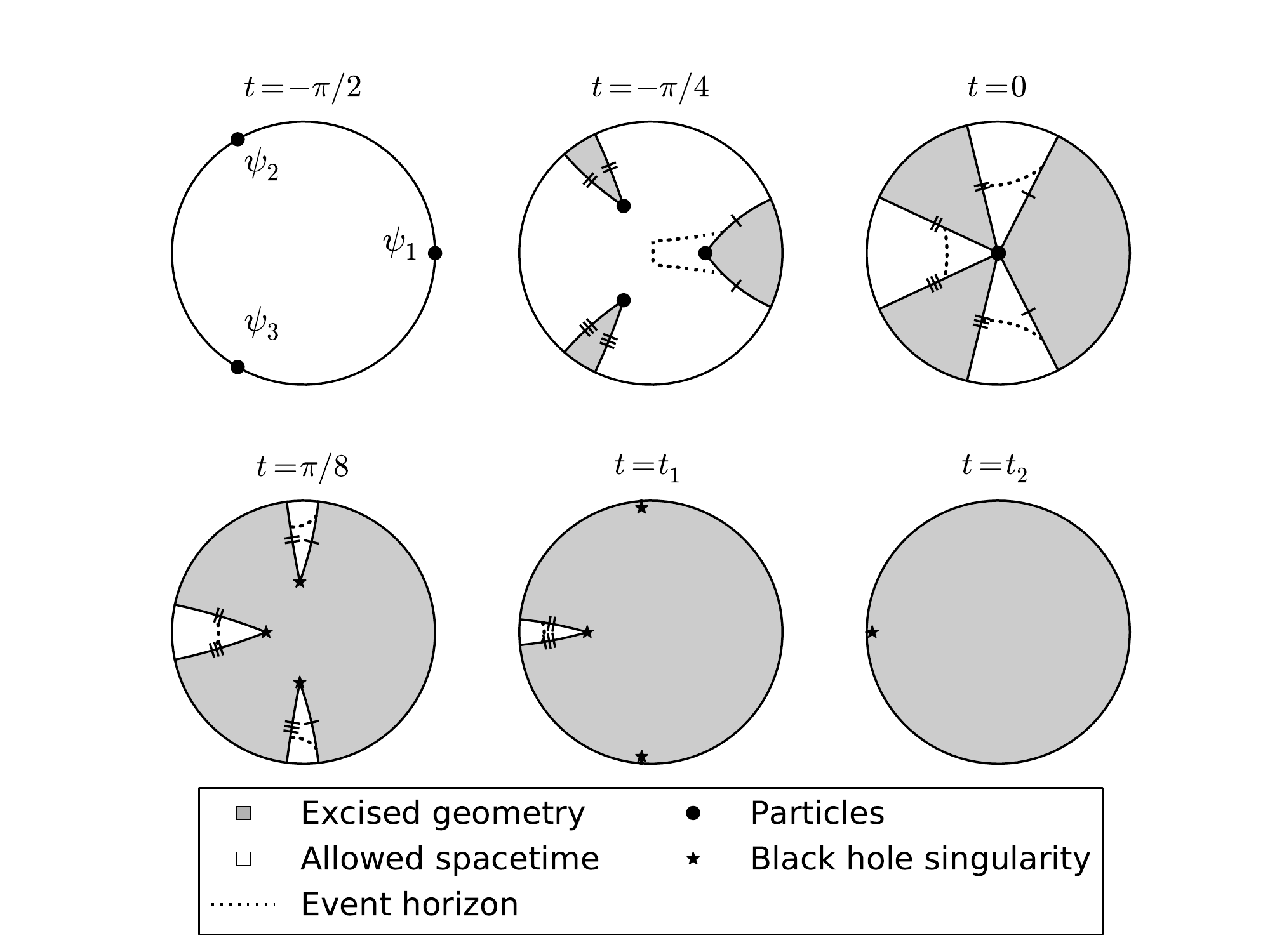}
\caption{\label{3particles} Collision of three massless particles, falling radially on geodesics at angles $\psi_i=(0,2\pi/3,4\pi/3)$ and with $E_i=(2,\frac{1}{2},\frac{1}{2})$. The grey areas constitute removed pieces. The resulting object moves on spatial geodesics (marked by dashed lines), so a black hole has formed instead of a massive particle. The dotted lines are the trajectories of the massless particles, and we see clearly that the wedges for particle 2 and 3 are not located symmetrically behind the particles. Note that lines corresponding to one grey wedge are not nessecarily mapped to each other on the same time slice, which is why it is possible for some of the white wedges in the last panels to disappear first at time $t_1\approx0.61$ (essentially, when jumping from one white area to another when going around one particle, there might be a time shift, which would not be present for wedges that are located symmetrically behind the particle as in the two-particle case). The last wedge disappears at time $t_2\approx0.82$, but these two times are of course mapped to each other by the isometries of the particles. The parameters specifying the wedges, obtained by numerically solving \eqref{peqml} and \eqref{intersectionml}, are $p_i\approx(0, 0.409, -0.409)$, and the angles of the intersections are $\phi_{i,i+1}\approx(1.644,  \pi, -1.644)$.}
\end{figure}

We will here consider the collision of $N$ massless particles. Each particle is described by excising a wedge behind it bordered by two surfaces $w_\pm^i$ governed by \eqref{c1_massless} and \eqref{Gammapm_massless}, with parameters $\psi_i$, $E_i$ and $p_i$, for $i=1,\ldots,N$ and we identify $N+1$ with $1$. After the collision we will identify the intersections $I_{i,i+1}$ of $w_+^i$ and $w_-^{i+1}$ with the resulting joint object. For this to be consistent, the intersection between $w_+^i$ and $w_-^{i+1}$ must be mapped by the holonomy $\bh_{i}$ of particle $i$ to the intersection between $w_+^{i-1}$ and $w_-^{i}$. As we will see by tuning the parameters $p_i$, associated to the wedges, we can accomplish this.\\
\linebreak
The intersection of the surface $I_{i,i+1}$ of $w_+^i$ and $w_-^{i+1}$ will be on a line of constant angle $\phi_{i,i+1}$ (a radial geodesic), and is obtained from \eqref{c1_massless} as the solution of the equation 
\begin{equation}
\frac{\sin(\phi_{i,i+1}-\Gamma_i^+-\psi_i)}{\sin(\phi_{i,i+1}-\Gamma_{i+1}^--\psi_{i+1})}=\frac{\sin\Gamma_i^+}{\sin\Gamma_{i+1}^-}.\label{intersection0}
\end{equation}
After the collision, we identify the intersections of the surfaces bounding two neighbouring wedges as the new joint object formed in the collision. This is only consistent if these intersections are mapped to each other by the isometries associated to the particles. Recall that the intersection between $w_+^i$ and $w_-^{i+1}$ be denoted by $I_{i,i+1}$ and let $\bh_i$ be the holonomy of particle $i$. We then must have the condition
\begin{equation}
I_{i-1,i}=\bh_i^{-1}I_{i,i+1}\bh_i,
\end{equation}
for all $i$. This gives us $N$ conditions to fix the $N$ parameters $p_i$. The holonomies are here given by
\begin{equation}
\bh_i=1+E_i(\gamma_0-\gamma(\psi_i)).
\end{equation}
Assume now that we know the angles $\phi_{i-1,i}$ for intersection $I_{i-1,i}$ and $\phi_{i,i+1}$ for intersection $I_{i,i+1}$, and let us see how we can use this to fix the parameter $p_i$ for particle $i$. Since $I_{i-1,i}$ is the intersection between the plane with constant angle $\phi_{i-1,i}$ and the surface $w_-^i$, they satisfy the equations
\begin{equation}
\tanh\chi\sin(-\phi_{i-1,i}+\Gamma_-^i+\psi_i)=-\sin\Gamma_-^i\sin t,
\end{equation}
where
\begin{equation}
\tan\Gamma_-^i=-(1-p_i)E_i,
\end{equation}
and since $I_{i,i+1}$ is the intersection between $\phi_{i,i+1}$ and $w_+^i$, we have the equations
\begin{equation}
\tanh\chi\sin(-\phi_{i,i+1}+\Gamma_+^i+\psi_i)=-\sin\Gamma_+^i\sin t,
\end{equation}
where
\begin{equation}
\tan\Gamma_+^i=(1+p_i)E_i.
\end{equation}
Our consistency condition now forces us to make sure that these two lines are mapped to each other by the holonomy $\bh_i=1+E_i(\gamma_0-\gamma(\psi_i))$, and by evaluating this (see Appendix \ref{app2}) we obtain
\begin{equation}
p_i=\frac{\tan(\phi_{i,i+1}-\psi_i)+\tan(\phi_{i-1,i}-\psi_i)}{-2E_i\tan(\phi_{i,i+1}-\psi_i)\tan(\phi_{i-1,i}-\psi_i)+\tan(\phi_{i,i+1}-\psi_i)-\tan(\phi_{i-1,i}-\psi_i)}.\label{peqml}
\end{equation}
Equation \eqref{intersection0} can be written as
\begin{equation}
-\sin(\phi_{i,i+1}-\psi_i)\frac{1}{E_i(1+p_i)}+\cos(\phi_{i,i+1}-\psi_i)=\sin(\phi_{i,i+1}-\psi_{i+1})\frac{1}{E_{i+1}(1-p_{i+1})}+\cos(\phi_{i,i+1}-\psi_{i+1})\label{intersectionml}.
\end{equation}
Equation \eqref{peqml} and equation \eqref{intersectionml} constitute $2N$ equations to solve for the $2N$ variables $\phi_{i,i+1}$ and $p_i$, for $i=1,\ldots,N$. Solving this will give a consistent slicing of the spacetime which can be continued after the collision. However, except for the case of discrete rotational symmetry (for which $p_i=0$), it seems difficult to find an analytic solution to these equations. We can solve this numerically and we see as expected that there is a solution that is continuously connected to the rotationally symmetric case. In Section \ref{numericssec} we will explain in detail how one can go about solving these equations in practice for collisions of massive particles from which the massless case can easily be obtained by taking a limit. In Figure \ref{3particles} we illustrate this with the example of three particles, located at angles $\psi_i=(0,2\pi/3,4\pi/3)$ and with energies determined by $E_i=(2,\frac{1}{2},\frac{1}{2})$. Numerically solving \eqref{peqml} and \eqref{intersectionml} we obtain $p_i\approx(0, 0.409, -0.409)$ and $\phi_{i,i+1}\approx(1.644,  \pi, -1.644)$. In this figure it is clear that the wedges are not located symmetrically around the particles. Note also that the surfaces $w_\pm^i$ are not mapped to each other within constant time slices. This explains why it seems like the spacetime disappears earlier in two of the wedges, but this is just a coordinate artefact and the time $t_1$ (when the first two wedges disappear) is mapped to time $t_2$ (when the last wedge disappears) by the holomies of particles two and three.

\subsubsection{Geometry after the collision}\label{geometry}
To investigate the geometry after the collision, we would like to make a coordinate transformation of the late time geometry such that it is either manifestly a massive pointlike particle or a black hole. Just as in the two-particle case, this can be done by mapping the resulting wedges of allowed geometry (for instance the white parts in the last panels in Figure \ref{3particles}), to either the equations describing a boosted conical singularity (as described by equations \eqref{c1} and \eqref{Gammapm}), or a boosted BTZ black hole (as described by equations \eqref{bhwedge3} and \eqref{Gammapmu}). We thus have a set of new wedges of allowed geometry, and we will denote the wedge bounded by $w_+^i$ and $w_-^{i+1}$ by $c_{i,i+1}$, which thus moves along an angle $\phi_{i,i+1}$. We will now rewrite the equations governing $w_+^i\equiv w_-^{i,i+1}$ and $w_-^{i+1}\equiv w_+^{i,i+1}$ to make it manifest that they satisfy either equation \eqref{c1} or \eqref{bhwedge3} for some new parameters $\Gamma_\pm^{i,i+1}$ and $\zeta_{i,i+1}$. For $w_+^i$ we have the equation
\begin{align}
\tanh\chi\sin(-\phi+\psi_i+\Gamma_+^i)&=-\sin \Gamma_+^i\sin t\nonumber
\\
&\Leftrightarrow\nonumber
\\
\tanh\chi\sin(-\pc\phi+(\psi_i+\Gamma_+^i-\phi_{i,i+1})+\phi_{i,i+1})&=-\frac{\sin \Gamma_+^i}{\sin(\psi_i+\Gamma_+^i-\phi_{i,i+1})}\nonumber\\
&\times\sin(\psi_i+\Gamma_+^i-\phi_{i,i+1})\sin t,\label{rec1}
\end{align}
and for $w_-^{i+1}$ we have
\begin{align}
\tanh\chi\sin(-\phi+\psi_{i+1}+\Gamma_-^{i+1})&=-\sin \Gamma_-^{i+1}\sin t\nonumber
\\
&\Leftrightarrow\nonumber
\\
\tanh\chi\sin(-\phi+(\psi_{i+1}+\Gamma_-^{i+1}-\phi_{i,i+1})+\phi_{i,i+1})&=-\frac{\sin \Gamma_-^{i+1}}{\sin(\psi_{i+1}+\Gamma_-^{i+1}-\phi_{i,i+1})}\nonumber\\
&\times\sin(\psi_{i+1}+\Gamma_-^{i+1}-\phi_{i,i+1})\sin t.\label{rec2}
\end{align}
Now the definition of $\phi_{i,i+1}$ as the intersection of $w_+^i$ and $w_-^{i+1}$ requires that
\begin{equation}
\frac{\sin \Gamma_-^{i+1}}{\sin(\psi_{i+1}+\Gamma_-^{i+1}-\phi_{i,i+1})}=\frac{\sin \Gamma_+^i}{\sin(\psi_i+\Gamma_+^i-\phi_{i,i+1})}\equiv \Bigg\{\begin{array}{ll}
                                                                                                                                             \tanh\zeta_{i,i+1},&\text{Pointlike particle}\\
                                                                                                                                             \coth\zeta_{i,i+1},&\text{Black hole}                                                                                                                                             
                                                                                                                                            \end{array}\label{zetadef}
\end{equation}
depending on whether the magnitude of this quantity is smaller or larger than one. Note that this equation is consistent due to the definition of $\phi_{i,i+1}$. We thus see, by comparing to either \eqref{c1} or \eqref{bhwedge3}, that the patch enclosed by these surfaces can indeed be interpreted as a boosted wedge bounded by surfaces $w_\pm^{i,i+1}$ satisfying the equations
\begin{equation}
\tanh\chi\sin(-\phi+\Gamma_\pm^{i,i+1}+\phi_{i,i+1})=-\kappa\sin\Gamma_\pm^{i,i+1}\sin t,
\end{equation}
with 
\begin{align}
\Gamma_-^{i,i+1}&=\psi_i+\Gamma_+^i-\phi_{i,i+1},\nonumber\\
\Gamma_+^{i,i+1}&=\psi_{i+1}+\Gamma_-^{i+1}-\phi_{i,i+1}.\label{Gammaiis}
\end{align}
and
\begin{equation}
\kappa=\Bigg\{\begin{array}{ll}
\tanh\zeta_{i,i+1},&\text{Pointlike particle}\\
\coth\zeta_{i,i+1},&\text{Black hole}                                                                                                                                             
\end{array}
\end{equation}
The mapping from the moving final wedges to the static wedges is illustrated in Figure \ref{3pillustrcs}.\\
\linebreak
For the pointlike particle case, in the coordinate system where this particle is stationary, this patch is a circle segment. From equation \eqref{Gammapm} we can then read off that the parameters $p_{i,i+1}$ and $\nu_{i,i+1}$ associated to this circle segment are given implicitly by the equations
\begin{equation}
\tan \Gamma_\pm^{i,i+1}=\pm\tan\left((1\pm p_{i,i+1})\nu_{i,i+1}\right)\cosh\zeta_{i,i+1},\label{Gammaii}
\end{equation}
such that the angle of this segment is $2\nu_{i,i+1}$. The total deficit angle is then $2\pi-2\sum_i \nu_{i,i+1}$ (recall that these segments now define the {\it allowed} geometry, and not the {\it removed} geometry, which is the reason for why the deficit angle is not just $2\sum_i\nu_{i,i+1}$).\\
\linebreak
For the black hole we want to relate these parameters via equation \eqref{Gammapmu} to the black hole mass and the parameters $\mu_{i,i+1}$ and $\xi_{i,i+1}$, and we find that
\begin{equation}
\tan\Gamma_\pm^{i,i+1}=\mp\tanh(\mu_{i,i+1}\pm\xi_{i,i+1})\sinh\zeta_{i,i+1}.\label{Gammaiibh}
\end{equation}
The total mass of the black hole is then simply
\begin{equation}
M=\frac{1}{\pi^2}\left[\sum_i\mu_{i,i+1}\right]^2.
\end{equation}
Note that under the isometries $\bh_i$ corresponding to the massless particles, spacelike (timelike) geodesics are mapped to spacelike (timelike) geodesics. This is why we know that (the absolute value of) equation \eqref{zetadef} is either larger than one for all $i$, or smaller than one for all $i$ (or equal to one in the extremal case, but we will not consider that in this thesis). This is built in the construction of the consistency equations \eqref{peqml} and \eqref{intersectionml}, although it seems difficult to prove it directly from these equations. \\
\linebreak
Notice that the coordinate transformations to bring the metric to the static metric for a conical singularity or the Schwarschild metric for the black hole, will map the lightlike geodesics that pass through the origin to new lightlike geodesics passing through the origin. Let \LS denote the lightlike surface consisting of all possible lightlike geodesics which end at the origin and originate at $t=-\pi/2$. We can then do the coordinate transformations that bring every moving wedge to a static wedge. These coordinate transformations will only be applied {\it above} the surface \LS. The resulting spacetime will then be the static conical singularity metric, or the Schwarschild AdS black hole metric, glued to empty AdS across the lightlike surface \LS. However, the surface \LS will only have a non-zero stress-energy tensor on the location of the trajectories of the pointlike particles which are inside this surface. We will explore this further in the next section when we investigate the $N\rightarrow\infty$ limit.


\subsubsection{The $N\rightarrow\infty$ limit}\label{secLimit}
\begin{figure}
\begin{center}
\includegraphics[scale=0.8]{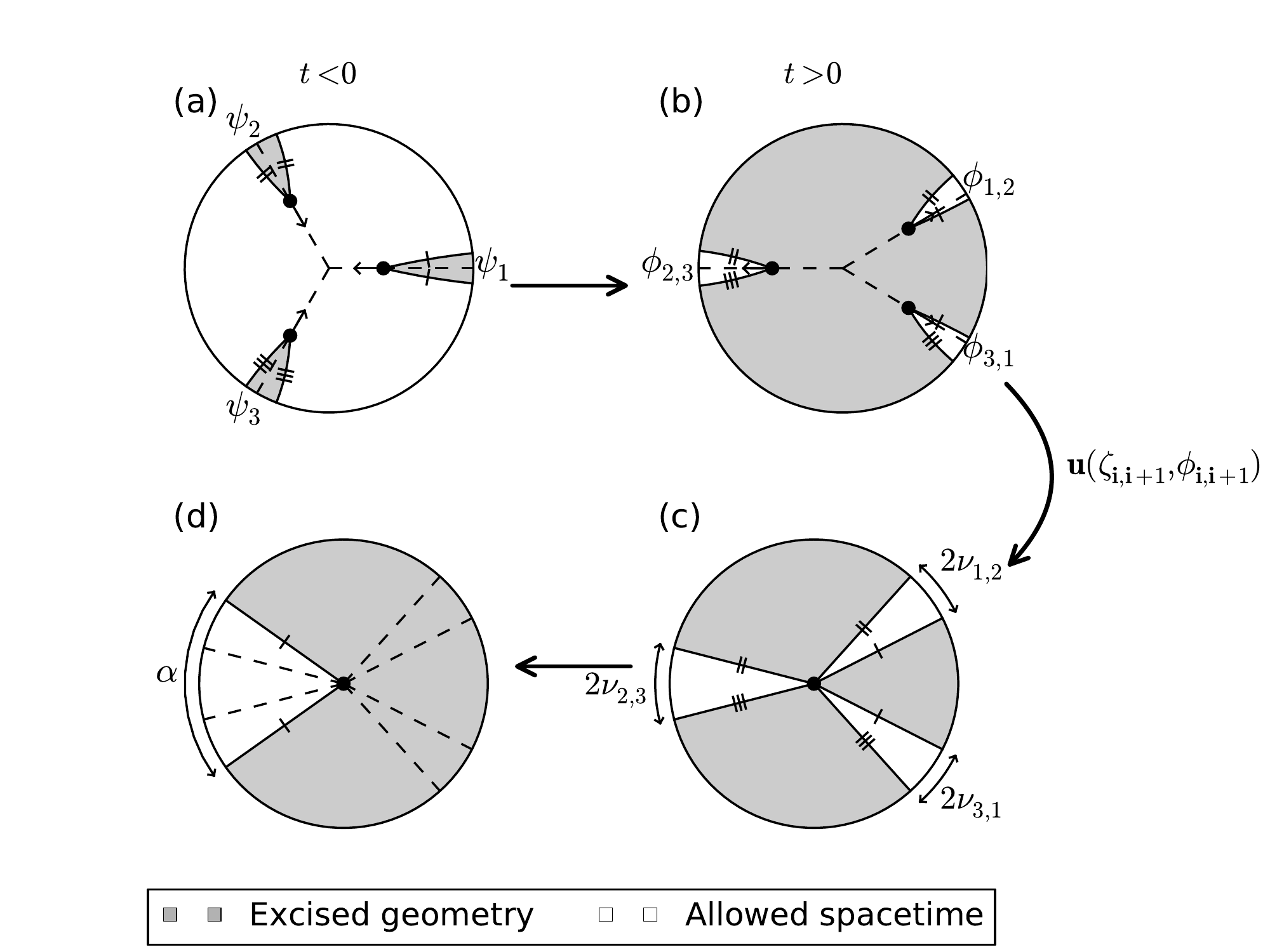}
\caption{\label{3pillustrcs} An illustration of how the transformation of the final geometry to static coordinates is carried out. The parameters are the same as in Figure \ref{3particlessym_cs}. (a): The spacetime before the collision takes place. (b): The spacetime after the collision has taken place, showing three final wedges of allowed geometry. (c): To go from (b) to (c), we transform each wedge with the isometry $\bu_{i,i+1}$ discussed in Section \ref{ppsec}, with parameters $\zeta=\zeta_{i,i+1}$ and $\psi=\phi_{i,i+1}$. This transforms each moving wedge to static wedges with opening angle $2\nu_{i,i+1}$. (d): We finally push the wedges together to form a more common parametrization of a conical singularity, and we can then define a continuous angle $\hat\phi$ covering the whole spacetime which takes values in the range $(0,\alpha)$ where $\alpha=\sum_i2\nu_{i,i+1}$. The coordinates can then be rescaled such that the metric takes the form \eqref{blackholemetric}. Note that we have the same picture when a black hole forms, except that there is an extra coordinate transformation between panel (b) and (c) and in panel (c) and (d) the metric is a BTZ black hole metric with unit mass instead of \ads.}
\end{center}
\end{figure}

In this section we will consider what happens when we take the number of particles to infinity. For $N$ particles, the setup is described by $N$ triples $(E_i,\psi_i,p_i)$. We will assume that in the limit, for every $\phi$ and for every $\delta\phi$, we can find sufficiently large $N$ such that $(\phi,\phi+\delta\phi)$ contains arbitrarily many $\psi_i$. In such a limit, the setup is expected to degenerate into only a continuous energy density function $\theta(\phi)$. For simplicity we will assume that $\psi_i=2\pi i/N$, such that $\psi_{i+1}-\psi_i\equiv d\phi=2\pi/N$, although the result will not depend on the precise way the limit is taken. For $N\gg1$, we expect that we can approximate the $E_i$ and $p_i$ by continuous functions. Thus, to obtain the correct limit for the energy density, we define $2E_i=\theta(\psi_i)d\phi$. It turns out that for large $N$ we have $p_i\sim N$ and that $\phi_{i,i+1}-\psi_i$ approaches a constant (which is generically not zero). We will thus define continuous interpolating functions $P(\phi)$ and $\Phi(\phi)$ such that $E_ip_i=P(\psi_i)$ and $\phi_{i,i+1}-\psi_i=\Phi(\psi_i)$. The equations \eqref{peqml} and \eqref{intersectionml} will in the $N\rightarrow\infty$ limit become differential equations for $P$ and $\Phi$, with $\theta$ as a source. Using $P(\psi_{i+1})=P(\psi_i)+P'(\psi_i)d\phi+O(1/N^2)$ and $\Phi(\psi_{i+1})=\Phi(\psi_i)+\Phi'(\psi_i)d\phi+O(1/N^2)$, it is straightforward to deduce from \eqref{peqml} and \eqref{intersectionml}, that in the limit $N\rightarrow\infty$, $P$ and $\Phi$ must satisfy the differential equations
\begin{subequations}
\begin{align}
P'\tan \Phi&=(\theta-P^2)\tan\Phi-P\label{ode1},\\
P(\tan\Phi)'&=\theta\tan\Phi+(\theta-1)P\tan^2\Phi -P,\label{ode2}
\end{align}
\end{subequations}
with periodic boundary conditions. Note that $'$ denotes derivative with respect to $\phi$. However, when deriving the above equations, we had to use that $|E_i|\ll|p_iE_i|$, which is in general valid when $N\rightarrow\infty$, but it is not valid if $p_i=0$ exactly. Or in other words, the derivation of \eqref{ode1} and \eqref{ode2} is not valid in the rotationally symmetric case, since the limits $N\rightarrow\infty$ and the limit $\theta\rightarrow $ const. do not commute. In the rotationally symmetric case we instead have the solution $P=0$ and $\Phi=0$. This seems to be a solution of \eqref{ode1} and \eqref{ode2} for any $\theta$, but this is also an artifact of the fact that the derivation of these equations is problematic in this case (of course it is not a problem if $P$ or $\Phi$ vanish at isolated points which generically will happen, it is only problematic if $P\equiv\Phi\equiv0$ on a continuous interval which seems to only happen in the case of rotational symmetry). In practice, it turns out to be easier to find the correct solutions of \eqref{ode1} and \eqref{ode2} by directly solving \eqref{intersectionml} and \eqref{peqml} for some large $N$, but one can then verify that these solutions indeed satisfy \eqref{ode1} and \eqref{ode2} up to $O(1/N)$.\\
\linebreak
For future purposes we will need $\Gamma_\pm^{i,i+1}$ expressed in terms of $P$ and $\Phi$. From \eqref{Gammaiis} we obtain
\begin{equation}
\tan\Gamma_-^{i,i+1}=\frac{P-\tan\Phi}{1+\tan\Phi P}+\left[\frac{1}{1+P\tan\Phi}-\frac{P-\tan\Phi}{(1+\tan\Phi P)^2}\tan\Phi\right]\frac{\theta}{2} d\phi+O(\frac{1}{N^2}),
\end{equation}
\begin{align}
\tan\Gamma_+^{i,i+1}&=\frac{P-\tan\Phi}{1+\tan\Phi P}+\Bigg[\frac{-\theta/2+P'+1+\tan^2\Phi}{1+P\tan\Phi}-\nonumber\\
-&\frac{(P-\tan\Phi)((P'-\theta/2)\tan\Phi-P-P\tan^2\Phi)}{(1+\tan\Phi P)^2}\Bigg]d\phi+O(\frac{1}{N^2}),
\end{align}
and thus
\begin{align}
\tan\Gamma_+^{i,i+1}-\tan\Gamma_-^{i,i+1}=\frac{1}{(1+P\tan\Phi)^2}(1+\tan^2\Phi)(-\theta+P'+1+P^2)d\phi.\label{tandiff}
\end{align}
We will now compute the metric of the spacetime explicitly. The spacetime will consist of two patches of geometry separated by a lightlike surface \LS. The spacetime below \LS will be that of \ads while the spacetime above \LS will be either that of a conical singularity or a black hole. Even though the total spacetime is not rotationally symmetric, the spacetime above and below will be written in rotationally symmetric coordinates and there will be a non-trivial angular dependent mapping relating the coordinates below and above the shell. Finding this mapping will be the main goal. We will separate the computations in two different sections, one for the formation of a conical singularity and one for the black hole.\\
\linebreak
{\bf Formation of a conical singularity}\\
In the case of the formation of a conical singularity, the resulting spacetime will in the limit be that of a conical singularity geometry and empty AdS glued together across a lightlike surface, denoted by \LS, which thus consists of all points which satisfy $\tanh\chi=-\sin t$, $-\pi/2\leq t \leq 0$. Above the shell we have the geometry of a massive pointlike particle, but so far it has been written in quite cumbersome coordinates. We would like to describe it by the metric \eqref{adsmetric}, but where $\phi$ has the periodicity $\alpha$, so that $2\pi-\alpha$ is the angular deficit. To go to this coordinate system, we will first apply the coordinate transformation discussed in Section \ref{geometry} to each wedge of allowed geometry (the regions bounded by $w_+^i$ and $w_-^{i+1}$). This brings us to a coordinate system consisting of $N$ static wedges, each denoted by $c^{\text{static}}_{i,i+1}$, with parameters $(\nu_{i,i+1},\phi_{i,i+1},p_{i,i+1})$, with total deficit angle $2\pi-2\sum_i\nu_{i,i+1}$. We will denote the coordinates above the shell by $(\pc \chi,\pc t,\pc\phi)$ and the coordinates below the shell by $(\chi,t,\phi)$. This geometry is valid above the surface \LS, and below \LS we still have empty \ads in the coordinates \eqref{adsmetric} (recall that this coordinate transformation maps \LS into itself). Now, by just pushing the wedges together and defining a continuous angular variable $\hat\phi$, the geometry above the shell becomes the metric \eqref{adsmetric} with periodicity $\alpha=2\sum_i\nu_{i,i+1}$. This is illustrated in Figure \ref{3pillustrcs}. To obtain the relation between $\hat\phi$ and $\phi$ when crossing the shell, note that each time we cross $c^{\text{static}}_{i,i+1}$, $\hat\phi$ will increase by $2\nu_{i,i+1}$. This means that, in the $N\rightarrow\infty$ limit, we can write $\hat\phi=\hat{\phi}_0+\sum_{0\leq j\leq i} 2\nu_{j,j+1}+O(1/N)$ when $\hat\phi\in c^{\text{static}}_{i,i+1}$, where $\hat{\phi}_0$ is an overall angular shift (the approximate value of $\hat\phi$ when $\hat\phi\in c^{\text{static}}_{0,1}$) which can be chosen to vanish.\\
\linebreak
Let us now define continuous interpolating functions $\hat{T}(\psi_i)\equiv p_{i,i+1}\nu_{i,i+1}$ and $\hat{Z}(\psi_i)\equiv\zeta_{i,i+1}$. In the limit we would like to express the functions $\hat Z$ and $\hat T$ in terms of $P$ and $\Phi$. From \eqref{zetadef}, by taking the $N\rightarrow\infty$ limit, we can obtain $\hat Z$ as
\begin{equation}
\tanh \hat Z=\frac{P}{P\cos\Phi-\sin\Phi}.\label{Z}
\end{equation}
From \eqref{Gammaiis} and \eqref{Gammaii} we have in the limit $N\rightarrow\infty$ that
\begin{equation}
\tan \hat T\cosh \hat Z=\frac{P-\tan\Phi}{1+P\tan\Phi},\label{QZ}
\end{equation}
from which we can obtain $\hat T$. Another useful relation (see Appendix \ref{app3}) is
\begin{equation}
\cos\hat T=\cos\Phi+P\sin\Phi.\label{coshatTml}
\end{equation}
To obtain a relation between $\hat\phi$ and $\phi$, we first note that going to the subleading terms in equation \eqref{Gammaii} we can obtain
\begin{equation}
\tan\Gamma_+^{i,i+1}-\tan\Gamma_-^{i,i+1}=2\nu_{i,i+1}\frac{\cosh \hat Z}{\cos^2\hat T}.\label{tandiff0}
\end{equation}
Thus using this together with \eqref{tandiff}, we obtain (see Appendix \ref{app3})
\begin{equation}
\hat\phi=\hat{\phi}_0+\int_0^\phi \frac{1-\cot\Phi P}{\cosh \hat Z}\rd \phi'.\label{anglerel}
\end{equation}
By using \eqref{Z} to obtain $\hat Z$, this now gives us $\hat \phi$ in terms of $P$ and $\Phi$, and thus indirectly in terms of $\theta$, which is the only physical parameter.\\
\linebreak
We are also interested in the relation between the radial coordinates when crossing the surface $\mathcal{L}$, namely we want to compute $\pc\chi=\pc\chi(\phi,\chi)$ on \LS. The mapping to the static coordinate system (before we push the circle sectors together) is given by equations \eqref{boosteqs}. Since lightlike geodesics are mapped to lightlike geodesics, we thus obtain from \eqref{boosteqs} that
\begin{equation}
\sinh\chi=\sinh\pc\chi(\cosh \hat Z+\sinh \hat Z \cos(\phi_{i,i+1}-\pc\phi)).\label{radeqpp}
\end{equation}
Note that any point on \LS at an angle $\phi\in(\psi_i,\psi_{i+1})$ will be mapped to a point with angle $\tilde\phi\in(\phi_{i,i+1}-(1-p_{i,i+1})\nu_{i,i+1},\phi_{i,i+1}+(1+p_{i,i+1})\nu_{i,i+1})$ (by definition of $\nu_{i,i+1}$ and $p_{i,i+1}$). In the limit, we thus have that $\psi_i$ is mapped to $\phi_{i,i+1}+\hat T (\psi_i)$, thus we have from equation \eqref{radeqpp} that the radial coordinates are related on $\mathcal{L}$ as
\begin{equation}
\sinh\pc\chi=\frac{\sinh\chi}{\cosh \hat Z+\sinh \hat Z\cos \hat T}=\frac{\sinh\chi}{\partial_\phi\hat\phi}.\label{radeqpp2}
\end{equation}
See Appendix \ref{app3} for the second equality. The fact that this proportionality factor is the inverse of the function relating the angles in \eqref{anglerel}, turns out to be a requirement for a well defined thin shell spacetime.\\
\linebreak
We would now like to bring the geometry to an \ads-Vaidya type of geometry. Let us first define $A=\frac{1}{\pi}\sum_i\nu_{i,i+1}=\hat\phi(2\pi)$, which thus is the total angle of the conical singularity divided by $2\pi$. We now want to find coordinate transformations such that the metric takes the form
\begin{equation}
\rd s^2=-f(r)\rd v^2+2\rd v\rd r+r^2\rd \phi^2 \label{vaidya}
\end{equation}
below and
\begin{equation}
\rd \bar{s}^2=-\bar{f}(\bar r)\rd \bar{v}^2+2\rd \bar{v}\rd \bar{r}+\bar{r}^2\rd \bar{\phi}^2 \label{vaidyabar}
\end{equation}
above the shell. The lightlike boundary $\mathcal{L}$ between the two patches is given by $v=\bar v=0$. We have $f=1+r^2$ (empty \ads) and $\bar f=A^2+\bar{r}^2$, and we want the periodicity of $\phi$ and $\bar{\phi}$ to be $2\pi$. To do this, we first do the coordinate transformation $\sinh \chi=r$ in the empty \ads part, and for the conical singularity part we have $\sinh\pc\chi=\bar{r}/A$, $\pc t=A \bar{t}$ and $\hat\phi=A\bar{\phi}$. This gives us the metric of the form
\begin{equation}
\rd s^2=-f\rd t^2+\rd r^2/f+r^2\rd \phi^2
\end{equation}
both above and below the shell. From this we obtain the metric \eqref{vaidya} by the standard coordinate transformation to infalling coordinates, given by $\rd v=\rd t+\rd r/f$. The relations between the radial and angular coordinates when crossing the shell can now be written as
\begin{equation}
\bar\phi=H(\phi),
\end{equation}
\begin{equation}
\bar{r}=\frac{r}{H'(\phi)},\label{rpm}
\end{equation}
where
\begin{equation}
\frac{1}{H'(\phi)}=\frac{A}{\cosh \hat Z(\phi)+\sinh \hat Z(\phi)\cos \hat T(\phi)}=\frac{A\cosh \hat Z(\phi)}{1-P(\phi)\cot\Phi(\phi)}.
\end{equation}
The fact that the proportionaly factor relating the radial coordinates turns out to be exactly the inverse of the derivative of the function relating the angular coordinates, is a necessary (and sufficient) condition to have a well defined induced metric on the shell, which is important when analyzing the junction conditions in Section \ref{masslessshellssec}, and can be seen as a non-trivial consistency check. It should be mentioned that these thin shell solutions are probably unphysical, since the thin shell limit of a thick shell composed of some dynamical matter (for example a scalar field) is expected to scatter below the black hole threshold, and not form a conical singularity. Nevertheless, these solutions can be constructed formally.\\
\linebreak
It is instructive to look at the rotationally symmetric case, where $\theta=\mathrm{const.}$, which should reduce to the standard AdS-Vaidya spacetime for a conical singularity. Thus we let $E_i=E=\tan\epsilon$, and we have $\Gamma^i_\pm=\Gamma_\pm=\pm\epsilon$, as well as $\nu_{i,i+1}=\nu=Ad\phi/2$. Defining $2E=\theta d\phi$ we obtain from equation \eqref{zetadef} that
\begin{equation}
\tanh\hat Z=\frac{\theta}{\theta-1},
\end{equation}
and from \eqref{Gammaii} and \eqref{Gammaiis} we obtain
\begin{equation}
1-\theta=A\cosh\hat Z.
\end{equation}
The condition that $-1\leq\tanh\hat Z\leq1$ and $\theta\geq0$ gives us that $\theta\leq1/2$ which thus is the threshhold for black hole formation. We can solve for $A$ in terms of $\theta$ to obtain the simple relation
\begin{equation}
A=\sqrt{1-2\theta}.
\end{equation}
At the threshhold of black hole formation we have $A\rightarrow0$, meaning that the total angle of the conical singularity approaches zero. The relation between $\bar{r}$ and $r$ can be obtained from equation \eqref{rpm} and we obtain that in this case $\bar{r}=r$ when crossing the shell, as expected. Note that the mass $M$, if we write the metric in the form \eqref{BTZ}, is given by $M=-A^2$.\\
\lb
{\bf Formation of a black hole}\\
In the black hole case the situation is a little bit less intuitive than in the formation of a massive particle. In this case, we would like to transform the metric after the shell into the form \eqref{BTZ}. We proceed as in the pointlike particle case and first transform the wedges of allowed geometry to wedges of the form \eqref{bhwedge2}, where the spacelike geodesics take the form of ``straight'' spacelike geodesics in a constant timeslice. The coordinates at this stage will be denoted by ($\pc t,\pc \chi,\pc\phi$). We will then do a coordinate transformation that maps these wedges to static circular sectors $c^{\text{static}}_{i,i+1}$ in a black hole background with unit mass, equation \eqref{unitbhmetric}, with parameters $\mu_{i,i+1}$ and $\xi_{i,i+1}$. The coordinates at this stage will be denoted by ($\sigma,\rho,y)$. The total black hole will then be trivially constructed by gluing together all these static wedges and defining a continuous angular coordinate $\hat y$. Each of these final static wedges will have an opening angle $2\mu_{i,i+1}$ so that $\hat y$ increases by $2\mu_{i,i+1}$ when crossing one such wedge. Thus $\hat{y}=\hat{y}_0+\sum_{0\leq j\leq i}2\mu_{j,j+1}+O(1/N)$ when $\hat{y}\in c^{\text{static}}_{i,i+1}$, where $\hat{y}_0$ is an unimportant overall shift. Note that the spacetime still takes the form \eqref{unitbhmetric}, a black hole metric with unit mass, since $\hat{y}$ takes values in $(0,\alpha)$ for some value $\alpha$. The correct mass is obtained by rescaling the angular coordinate to the standard range $(0,2\pi)$.\\
\lb
Let us now define continuous interpolating functions $\hat X(\psi_i)\equiv \xi_{i,i+1}$ and $\hat Z(\psi_i)\equiv\zeta_{i,i+1}$. In the limit we would like to express the functions $\hat Z$ and $\hat X$ in terms of $P$ and $\Phi$. In the $N\rightarrow\infty$ limit, we obtain from \eqref{zetadef} that 
\begin{equation}
\coth \hat Z=\frac{P}{P\cos\Phi-\sin\Phi}.\label{Zbh}
\end{equation}
From \eqref{Gammaiibh} and \eqref{Gammaiis} we have in the limit $N\rightarrow\infty$ that
\begin{equation}
-\tanh \hat X\sinh \hat Z=\frac{P-\tan\Phi}{1+P\tan\Phi},\label{QX}
\end{equation}
from which we can obtain $\hat X$.  Another useful relation (see Appendix \eqref{app3}) is
\begin{equation}
\cosh\hat X=\cos\Phi+P\sin\Phi.\label{coshhatXml}
\end{equation}
Now from equation \eqref{Gammaiibh} we obtain in the $N\rightarrow\infty$ limit that
\begin{equation}
\tan \Gamma_+^{i,i+1}-\tan \Gamma_-^{i,i+1}=-\frac{2}{\cosh^2\hat X}\mu_{i,i+1}\sinh\hat Z,\label{tandiff0bh}
\end{equation}
and just like in the massive particle case, we can use this result together with \eqref{tandiff}, to obtain the relation between $\hat y$ and $\phi$ when crossing the shell to be (see Appendix \ref{app3})
\begin{equation}
\hat y(\phi)=\hat y_0+\int_0^\phi \frac{P\cot\Phi-1}{\sinh \hat Z}\rd \phi'.\label{anglerelbh}
\end{equation}
Now we want to figure out how the radial coordinate before the shell $\chi$ is related to the radial coordinate $\rho$ (where the metric takes the form \eqref{unitbhmetric}), so consider the wedge bounded by $w_+^i$ and $w_-^{i+1}$. Let $(\pc\chi,\pc t,\pc\phi)$ be the coordinate system after the shell in which a wedge takes the form \eqref{bhwedge2}. From \eqref{bhwedge2} we obtain that lightlike geodesics with equation $\tanh\chi=-\sin t$ at angle $\psi_i$ and $\psi_{i+1}$ are mapped to lightlike geodesics at an angle $\pc\phi$ given by $\sin\pc\phi=\pm\tanh (\mu_{i,i+1}\pm\xi_{i,i+1})$ with the $+$ ($-$) for $\psi_{i+1}$ ($\psi_i$). In the limit, we thus obtain that all lightlike geodesics at an angle in $(\psi_i,\psi_{i+1})$ are mapped to a lightlike geodesic at an angle $\pc\phi$ given by $\sin\pc\phi=\tanh\xi_{i,i+1}=\tanh \hat X$. Now, from \eqref{boosteqs} we obtain that the radial coordinates $\chi$ and $\pc\chi$ are related by
\begin{equation}
\sinh\pc\chi=\frac{\sinh\chi}{\cosh \hat Z+\sinh \hat Z\cos \pc\phi}=\frac{\sinh\chi}{\cosh \hat Z+\frac{\sinh \hat Z}{\cosh \hat X}}.
\end{equation}
Now, it is easy to see that the lightlike geodesics satisfy $x_1^2+x_2^2=x_0^2$, and we can obtain $\rho$ by
\begin{equation}
\rho^2=x_0^2-x_2^2=\sinh^2\chi\cos\pc\phi^2=\frac{\sinh^2\pc\chi}{\cosh^2\hat X}\Rightarrow \rho=\frac{\sinh\chi}{\cosh \hat X\cosh \hat Z+\sinh \hat Z}=\frac{\sinh\chi}{\partial_\phi \hat y},\label{radeqbh}
\end{equation}
where we have used the relations \eqref{rhoembedding}. See Appendix \ref{app3} for the last equality. To make the angular coordinate $\hat y$ have period $2\pi$, we may rescale $\hat y= \bar \phi\sqrt{M}$, $\rho=\bar{r}/\sqrt{M}$ and $t=\bar{t}\sqrt{M}$, where the mass is given by $M=\frac{1}{\pi^2}\left(\sum_i\mu_{i,i+1}\right)^2$, which will give us exactly the metric \eqref{BTZ}. To summarize, our resulting spacetime can be written as
\begin{equation}
\rd s^2=-f(r)\rd v^2+2\rd v\rd r+r^2\rd \phi^2 \label{vaidya2}
\end{equation}
below the shell and 
\begin{equation}
\rd \bar{s}^2=-\bar{f}(\bar{r})\rd \bar{v}^2+2\rd \bar{v}\rd \bar{r}+\bar{r}^2\rd \bar{\phi}^2 \label{vaidya2bar}
\end{equation}
above the shell where $\bar{f}=-M+\bar{r}^2$ and $f=1+r^2$. The coordinates are related on the shell as
\begin{equation}
\bar{\phi}=H(\phi),
\end{equation}
\begin{equation}
\bar{r}=\frac{r}{H'(\phi)},
\end{equation}
where
\begin{equation}
\frac{1}{H'(\phi)}=\frac{\sqrt{M}}{\cosh \hat X\cosh \hat Z+\sinh \hat Z}=\frac{\sqrt{M}\sinh \hat Z}{P\cot\Phi-1}.\label{Gdefbh}
\end{equation}
Let us again consider the case of rotational symmetry. We thus define $2E_i=2E=\theta d\phi$ where $\theta$ is constant, and again we have $\Gamma_\pm^i=\pm\epsilon$. We also have $\mu_{i,i+1}=\mu=\sqrt{M}d\phi/2$, and we obtain from \eqref{zetadef} that
\begin{equation}
 \coth \hat Z=\frac{\theta}{\theta-1},
\end{equation}
and from \eqref{Gammaiis} and \eqref{Gammaiibh} we obtain
\begin{equation}
 \theta-1=\sqrt{M}\sinh \hat Z.
\end{equation}
We can solve for $M$ as
\begin{equation}
M=2\theta-1.
\end{equation}
We also obtain that $\bar{r}=r$ when crossing the shell.

\subsection{Massive particles}\label{massivecollsec}
We will now consider the case of $N$ colliding massive particles. The particles will be located on radial geodesics along angles $\psi_i$, with boost parameters $\zeta_i$ and angular deficits (in the rest frame) of $2\nu_i$, for $i=1,\ldots,N$ and we identify $1=N+1$. Each particle is constructed by excising a wedge bordered by two surfaces $w_\pm^i$ satisfying equations \eqref{c1} and \eqref{Gammapm}. Just as in the massless case, we thus associate a parameter $p_i$ to each wedge which we will then tune to make sure that the intersections $I_{i,i+1}$ of $w_+^i$ and $w_-^{i+1}$ can be identified as the same point in spacetime and can thus be interpreted as the final joint object formed in the collision.\\
\linebreak
The wedges are thus bordered by surfaces $w_\pm^i$, determined by the equations
\begin{equation}
\tanh\chi\sin(-\phi +\Gamma^i_\pm+\psi_i)=-\tanh\zeta_i \sin\Gamma^i_\pm \sin t, \label{c1i}
\end{equation}
where
\begin{equation}
\tan\Gamma_\pm^i =\pm\tan((1 \pm p_i)\nu_i) \cosh\zeta_i.\label{Gammapmi}
\end{equation}
The intersection $I_{i,i+1}$ between $w_+^i$ and $w_-^{i+1}$ is a radial geodesic at an angle $\phi_{i,i+1}$ given by the equation
\begin{equation}
\frac{\tanh\zeta_{i}\sin \Gamma_+^i}{\sin(\psi_i+\Gamma_+^i-\phi_{i,i+1})}=\frac{\tanh\zeta_{i+1}\sin \Gamma_-^{i+1}}{\sin(\psi_{i+1}+\Gamma_-^{i+1}-\phi_{i,i+1})}.\label{intersectionangle}
\end{equation}
The parameters $p_i$ are determined by enforcing that the intersections $I_{i-1,i}$ and $I_{i,i+1}$ are mapped to each other by the isometry $\bh_i$ associated to particle $i$, namely that
\begin{equation}
I_{i-1,i}=\bh_i^{-1}I_{i,i+1}\bh_i,
\end{equation}
for all $i$. This gives us $N$ conditions to fix the $N$ parameters $p_i$. The holonomies are here given by
\begin{equation}
\bh_i=\cos \nu_i+\gamma_0\cosh \zeta_i \sin \nu_i-\sinh \zeta_i \sin \nu_i \gamma(\psi_i).
\end{equation}
This computation is a bit involved, but results in the equation (see Appendix \ref{peqapp})
\begin{equation}
\tan(p_i \nu_i )=\frac{\tan(\phi_{i,i+1}-\psi_i)+\tan(\phi_{i-1,i}-\psi_i)}{\cot\nu_i(\tan(\phi_{i,i+1}-\psi_i)-\tan(\phi_{i-1,i}-\psi_i))-2\cosh \zeta_i\tan(\phi_{i,i+1}-\psi_i)\tan(\phi_{i-1,i}-\psi_i)}.\label{peq}
\end{equation}
We can reformulate \eqref{intersectionangle} in terms of $p_i$ and $\nu_i$ by using \eqref{Gammapmi}, which gives the equation
\begin{align}
\frac{-\sin(\phi_{i,i+1}-\psi_i)}{\sinh\zeta_i\tan((1+p_i)\nu_i)}+\frac{\cos(\phi_{i,i+1}-\psi_i)}{\tanh\zeta_i}=\frac{\sin(\phi_{i,i+1}-\psi_{i+1})}{\sinh\zeta_{i+1}\tan((1-p_{i+1})\nu_{i+1})}+\frac{\cos(\phi_{i,i+1}-\psi_{i+1})}{\tanh\zeta_{i+1}}.\label{nueq}
\end{align}
We now have $2N$ equations for our $2N$ parameters $p_i$ and $\phi_{i,i+1}$, and in practice it seems that this system of equations can always be solved. Except in the rotationally symmetric case, the wedges associated to the particles will be non-symmetric and the parameters are obtained numerically. In Section \ref{numericssec} we will explain in detail how one can go about solving these equations in practice. Several examples are shown in figures \ref{3particlessym_cs}, \ref{3particlesgen} and \ref{4particles}. Note also that when taking the massless limit $\nu_i\rightarrow0$, $\zeta_i\rightarrow\infty$ while $\tan\nu_i\sinh\zeta_i\rightarrow E_i$ goes to a constant, we reproduce the expressions \eqref{peqml} and \eqref{intersectionml}. \\

\subsubsection{Geometry after the collision}
\begin{figure}
\begin{center}
\includegraphics[scale=0.8]{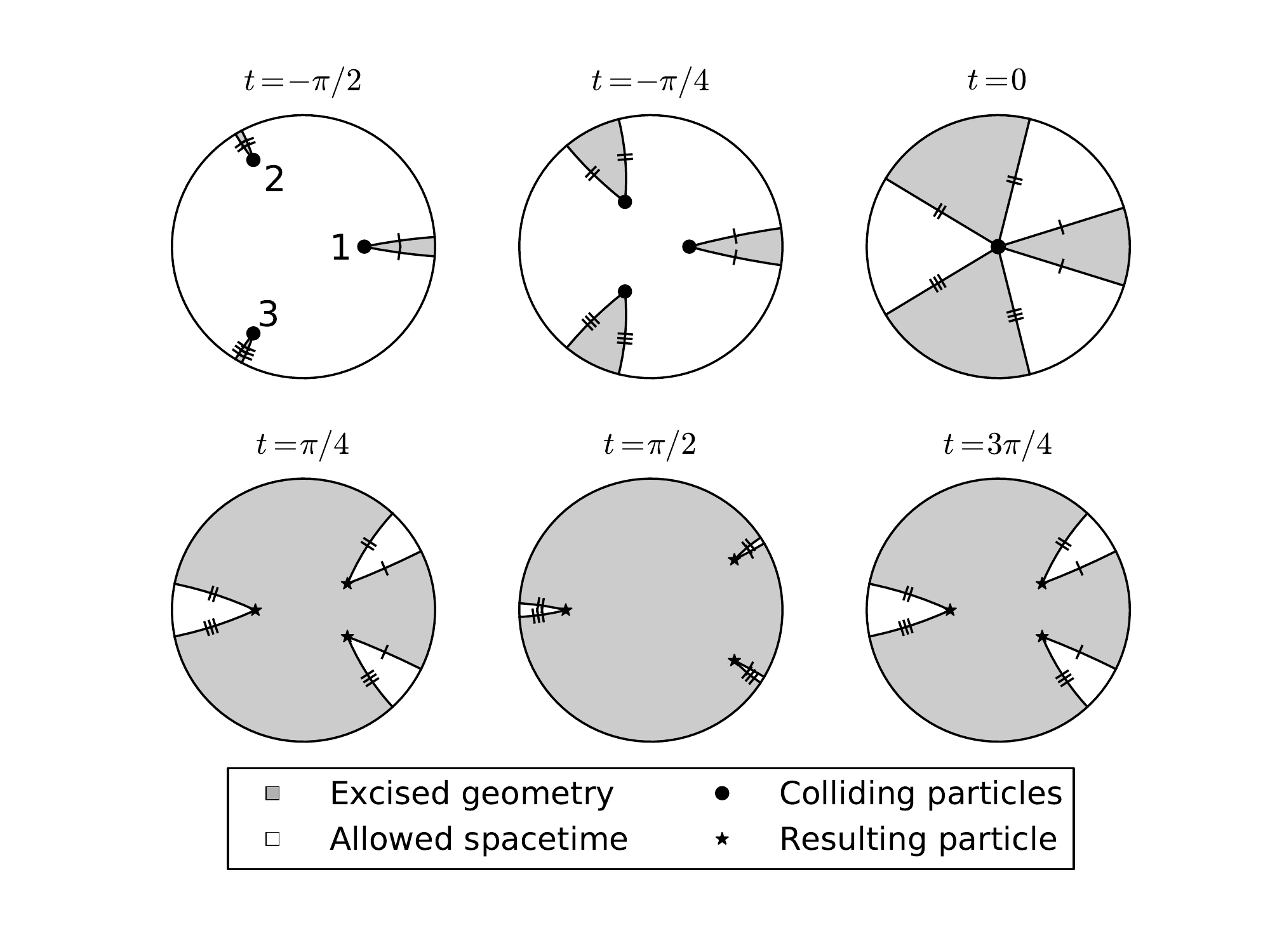}
\caption{\label{3particlessym_cs}Three colliding massive particles forming a new massive particle. The particles have the same rest mass but particle 1 is released from a different radial position compared to the other two (or in other words, it has a different boost parameter). The final wedges after the collision together form a conical singularity spacetime, but where the wedges have been boosted, and they will keep oscillating forever. Note also that, even though we have marked which surfaces are identified via the isometries, there will generically be an additional time shift under the identifications (so curves on the same time slice are generically {\it not} mapped to each other). The parameters $p_i$ which determine the orientation of the wedges have been obtained by solving equations \eqref{peq} and \eqref{nueq}.}
\end{center}
\end{figure}

\begin{figure}
\begin{center}
\includegraphics[scale=0.8]{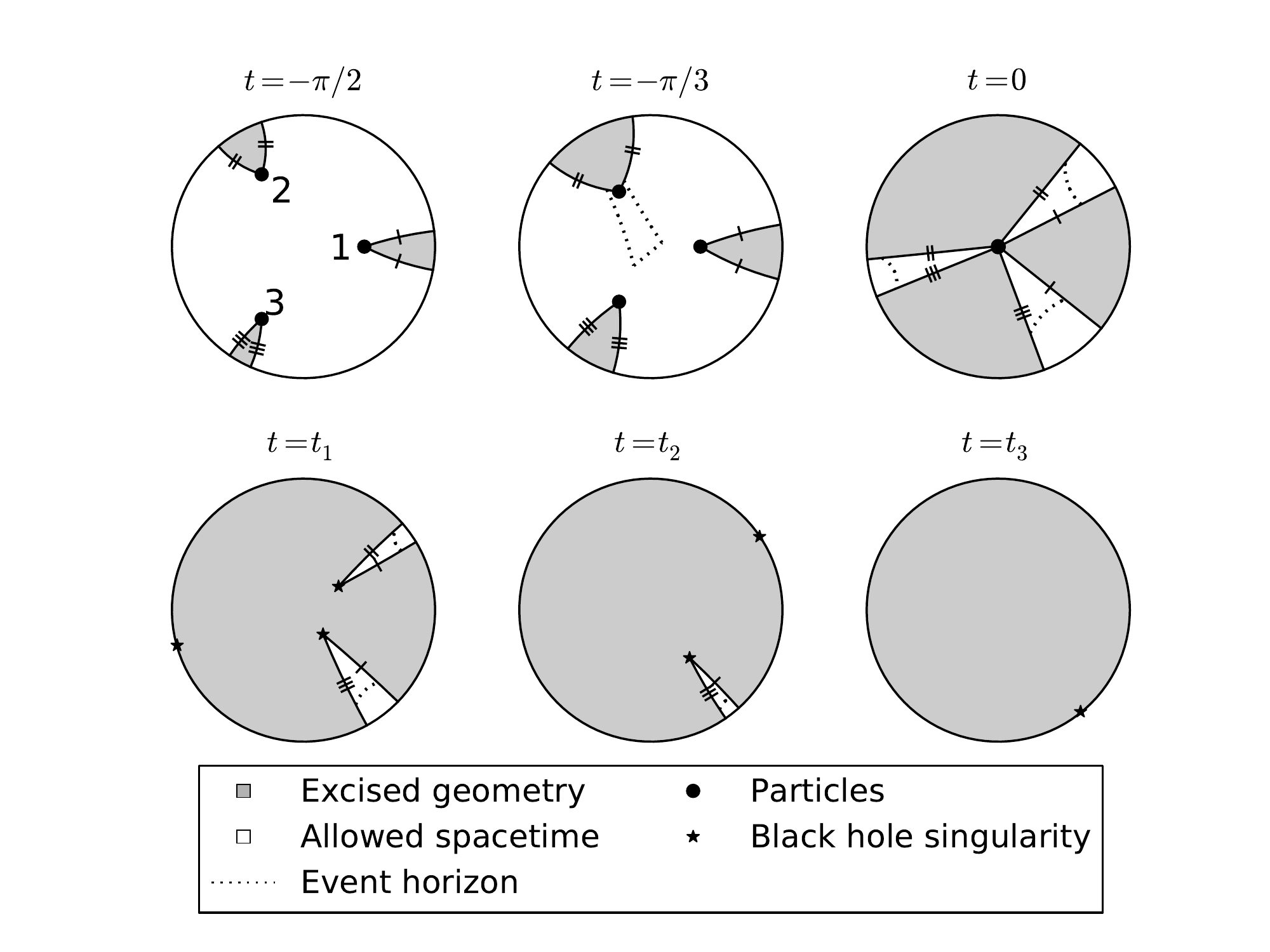}
\caption{\label{3particlesgen}Three colliding massive particles forming a BTZ black hole. In this example, there are no symmetry restrictions, and we have $\zeta_i=(1,1.5,1.5)$ and $\nu_i=(0.4,0.8,0.4)$. The final three wedges after the collision all disappear at different times $t_1$, $t_2$ and $t_3$, but note that all these final points are identified by the holonomies of the particles. The parameters $p_i$ which determine the orientation of the wedges have been obtained by solving equations \eqref{peq} and \eqref{nueq}.}
\end{center}
\end{figure}

\begin{figure}
\begin{center}
\includegraphics[scale=0.8]{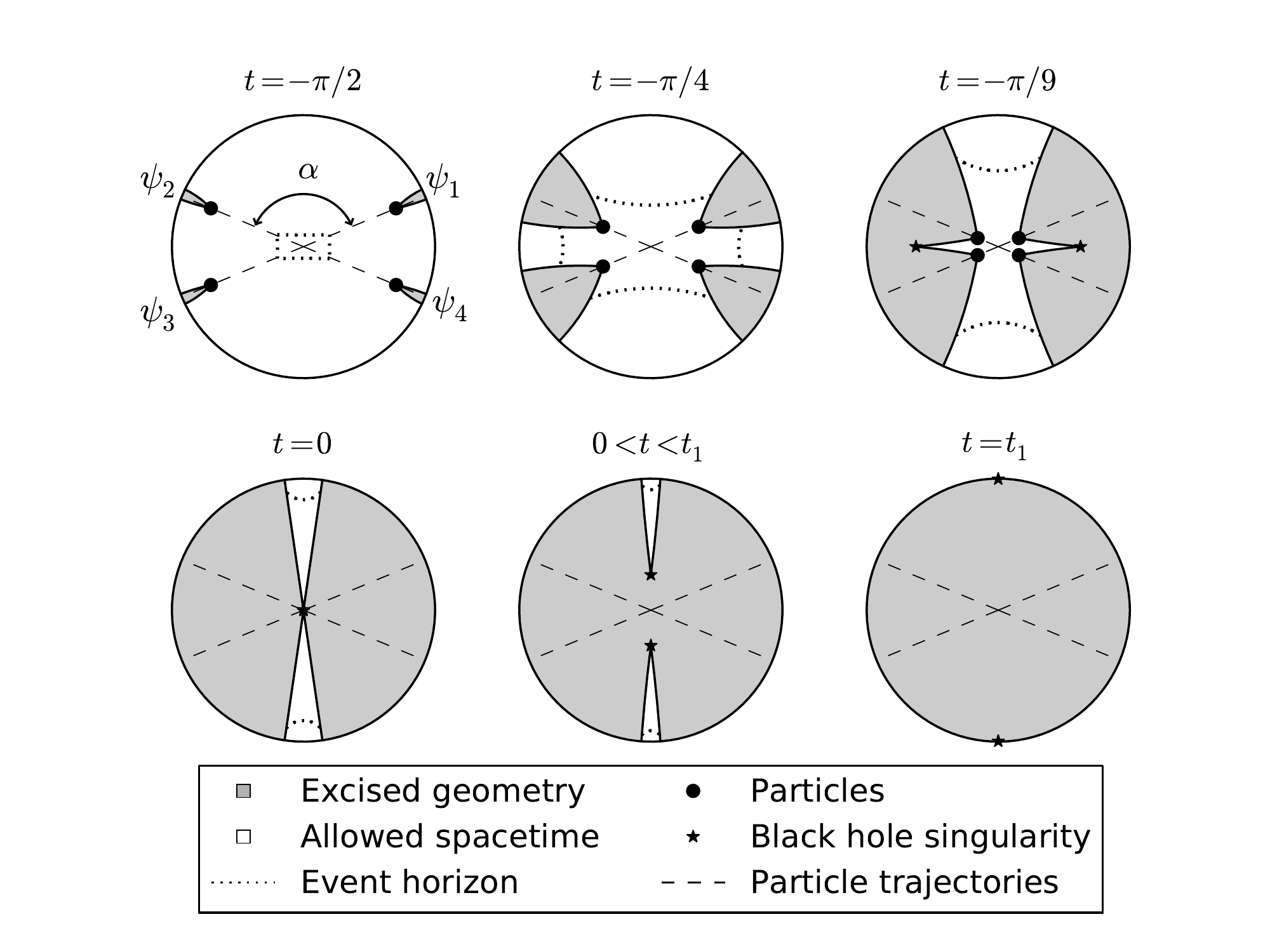}
\caption{\label{4particles} An example of four colliding particles, where they all have the same rest masses and are released from the same radial position, but the rotational symmetry is broken by the angles $\psi_i$. Specifically, we can parametrize the solution in terms of an angle $\alpha\equiv\psi_2-\psi_1=\psi_4-\psi_3$, and then $\psi_3-\psi_2=\psi_1-\psi_4=\pi-\alpha$. In this particular example, we have $\alpha=3\pi/4$. It is clear that the wedges are not symmetric around the trajectories of the particles, and they ``repel'' each other as the angle $\alpha$ increases. Note that in this example, two of the final spacelike geodesics connected to the collision point go ``backwards in time'', but this has no physical significance whatsoever. The parameters $p_i$ which determine the orientation of the wedges have been obtained by solving equations \eqref{peq} and \eqref{nueq}.}
\end{center}
\end{figure}

The geometry after the collision will be either a black hole or a pointlike particle, depending on whether the intersections $I_{i,i+1}$ move on spacelike or timelike geodesics, respectively. We then define parameters $\Gamma_\pm^{i,i+1}$ and $\zeta_{i,i+1}$ such that the wedge $c_{i,i+1}$ can be mapped either to the form \eqref{c1} and \eqref{Gammapm} or to \eqref{bhwedge3} and \eqref{Gammapmu}. By writing
\begin{align}
\tanh\chi\sin(-\phi+\psi_{i+1}+\Gamma_-^{i+1})&=-\tanh\zeta_{i+1}\sin \Gamma_-^{i+1}\sin t\nonumber
\\
&\Leftrightarrow\nonumber
\\
\tanh\chi\sin(-\phi+\phi_{i,i+1}+(\Gamma^{i+1}_-+\psi_{i+1}-\phi_{i,i+1}))&=-\frac{\tanh\zeta_{i+1}\sin\Gamma_-^{i+1}}{\sin(\Gamma^{i+1}_-+\psi_{i+1}-\phi_{i,i+1})}\nonumber\\
&\times\sin(\Gamma^{i+1}_-+\psi_{i+1}-\phi_{i,i+1})\sin t,\nonumber\\
\end{align}
for $w_+^{i,i+1}=w_-^{i+1}$, and 
\begin{align}
\tanh\chi\sin(-\phi+\psi_{i}+\Gamma_+^{i})&=-\tanh\zeta_{i}\sin \Gamma_+^{i}\sin t\nonumber
\\
&\Leftrightarrow\nonumber
\\
\tanh\chi\sin(-\phi+\phi_{i,i+1}+(\Gamma^{i}_++\psi_{i}-\phi_{i,i+1}))&=-\frac{\tanh\zeta_{i}\sin\Gamma_+^{i}}{\sin(\Gamma^{i}_++\psi_i-\phi_{i,i+1})}\nonumber\\
&\times\sin(\Gamma^{i}_++\psi_i-\phi_{i,i+1})\sin t,
\end{align}
for $w_-^{i,i+1}=w_+^i$, we can determine the new parameters $\Gamma_\pm^{i,i+1}$ by
\begin{equation}
\Gamma_-^{i,i+1}=\psi_i+\Gamma_+^i-\phi_{i,i+1},
\end{equation}
\begin{equation}
\Gamma_+^{i,i+1}=\psi_{i+1}+\Gamma_-^{i+1}-\phi_{i,i+1},
\end{equation}
and the parameter $\zeta_{i,i+1}$ is determined by
\begin{equation}
\frac{\tanh\zeta_i\sin\Gamma_+^i}{\sin(\Gamma^i_++\psi_i-\phi_{i,i+1})}=\frac{\tanh\zeta_{i+1}\sin\Gamma_-^{i+1}}{\sin(\Gamma^{i+1}_-+\psi_{i+1}-\phi_{i,i+1})}=\Bigg\{\begin{array}{ll}
                                                                                                                                             \tanh\zeta_{i,i+1},&\text{Pointlike particle}\\
                                                                                                                                             \coth\zeta_{i,i+1},&\text{Black hole}                                                                                                                                             
                                                                                                                                            \end{array},\label{tanhzcothz}
\end{equation}
depending on whether (the absolute value of) the above ratio is smaller or larger than 1. The equality \eqref{tanhzcothz} is consistent due to the definition of $\phi_{i,i+1}$. We are now interested in going to the rest frame of each final wedge $c_{i,i+1}$, such that they take the form of static wedges. In the case of formation of a massive particle, these static wedges will be circle sectors in \ads, while in the case of formation of a black hole the static wedges will be circle sectors in the BTZ black hole background with unit mass. The mapping from the moving final wedges to the static wedges is illustrated in Figure \ref{3pillustrcs}. In the massless case, the trajectories of the particles were again mapped to lightlike geodesics and thus the lightlike surfaces on which they are located is essentially mapped to itself. For the massive particles, this is not the case and there will be a non-trivial transformation of the incoming particles' trajectories which will make the computations later when we take the $N\rightarrow\infty$ limit a bit more complicated.\\
\linebreak
In the case of the formation of a conical singularity, we want to map each final wedge $c_{i,i+1}$ to a static wedge with parameters $\nu_{i,i+1}$ and $p_{i,i+1}$. From \eqref{Gammapm}, we have the relation
\begin{equation}
\tan \Gamma^{i,i+1}_\pm=\pm \tan((1\pm p_{i,i+1})\nu_{i,i+1})\cosh\zeta_{i,i+1},\label{Gammapnu}
\end{equation}
from which we can determine $p_{i,i+1}$ and $\nu_{i,i+1}$. The angle $2\nu_{i,i+1}$ is now the total angle of this static wedge (instead of being the angular deficit). All these $N$ wedges are then glued together by the identifications, forming a conical singularity with total angle $\alpha=\sum_{i}2\nu_{i,i+1}$. By rescaling the coordinates, the spacetime can be written in the form \eqref{blackholemetric}, and we obtain that the parameter $M$ is given by $M=-\alpha^2/(2\pi)^2$. A full solution of three massive particles forming a new massive particle is shown in Figure \ref{3particlessym_cs}.\\
\linebreak
In the case of formation of a black hole, we want to map each final wedge $c_{i,i+1}$ to a circle sector in the black hole geometry, with parameters $\xi_{i,i+1}$ and $\mu_{i,i+1}$. From equation \eqref{Gammapmu}, we have the relation
\begin{equation}
\tan \Gamma^{i,i+1}_\pm=\mp \tanh(\mu_{i,i+1}\pm\xi_{i,i+1})\sinh\zeta_{i,i+1},\label{Gammapmui}
\end{equation}
from which we can determine $\xi_{i,i+1}$ and $\mu_{i,i+1}$. The total angle of such a circle sector is given by $2\mu_{i,i+1}$, but it is a wedge cut out from a black hole spacetime with metric \eqref{unitbhmetric} instead of the \ads metric. These $N$ wedges are now glued together, forming a wedge with total angle $\alpha=\sum_i2\mu_{i,i+1}$. By rescaling the coordinates, we can write the spacetime in the form \eqref{blackholemetric}, and the parameter $M$ is obtained as $M=\alpha^2/(2\pi)^2$. Figures \ref{3particlesgen} and \ref{4particles} show the formation of a black hole from a collision of three and four particles, respectively.


\subsubsection{The $N\rightarrow\infty$ limit}\label{limitsec}
When taking the limit of an infinite number of particles, we will have $\nu_i\rightarrow0$, while $\zeta_i$ and $p_i\nu_i$ will go to constants. It is convenient to introduce continuous interpolating functions $T(\psi_i)=p_i\nu_i$, $\cZ(\psi_i)=\zeta_i$ and also $\Phi(\psi_i)=\phi_{i,i+1}-\psi_i$ which all approach some finite continuous functions in the limit. We will also assume that the density of particles remains constant, namely that $2\nu_i=d \phi \rho(\psi_i)$, where $d\phi=\psi_{i+1}-\psi_i$ (for simplicity we will assume that the angles $\psi_i$ are distributed homogeneously around the boundary so that $d\phi=2\pi/N$). Remember that $2\nu_i$, the deficit angle in the restframe, is equal to the mass of the particle (in units where $8\pi G=1$). We will thus refer to $\rho$ as the rest mass density. Just as in the massless case, the discrete algebraic equations relating the parameters $\nu_i$, $p_i$ and $\zeta_i$ will become differential equations for the functions $T$, $\Phi$ and $Z$. By using the relations $T(\psi_{i+1})=T(\psi_i)+T'(\psi_i)d\phi+O(1/N^2)$, $\Phi(\psi_{i+1})=\Phi(\psi_i)+\Phi'(\psi_i)d\phi+O(1/N^2)$ and $Z(\psi_{i+1})=Z(\psi_i)+Z'(\psi_i)d\phi+O(1/N^2)$ a straightforward calculation shows that the discrete equations \eqref{peq} and \eqref{nueq} then reduce to
\begin{equation}
\tan T (\tan \Phi)'=\rho\tan\Phi+\rho\cosh\cZ \tan^2\Phi\tan T-\tan T\tan^2\Phi-\tan T,\label{odeP}
\end{equation}
\begin{align}
(\tan T)'\sinh Z\tan\Phi&+\tan T\cosh Z Z'\tan\Phi-\tan^2T Z'=\nonumber\\
&\frac{\rho\sinh Z}{\cos^2 T}\tan\Phi-\sinh Z \tan T-\sinh Z\cosh Z\tan^2T\tan\Phi,\label{odeT}
\end{align}
where $'$ denotes derivative with respect to $\phi$. Just as in the massless case, $T=\Phi=0$ seems to be a solution for all $\rho$, but this is just an artefact of the derivation not being valid in the strictly homogeneous case since the two limits $\rho\rightarrow $ const. and $N\rightarrow\infty$ do not commute. Note also that in the massless limit, where $\rho\rightarrow0$ and $Z\rightarrow\infty$, we should let $\tan T\cosh\cZ\rightarrow P$ and $\rho\cosh\cZ\rightarrow\theta$, which reduces the expressions to equations \eqref{ode1} and \eqref{ode2}.\\
\linebreak
We will later be interested in computing $\nu_{i,i+1}$ from \eqref{Gammapnu} or $\mu_{i,i+1}$ from \eqref{Gammapmui}, for large $N$. This will require us to compute the difference $\tan\Gamma_+^{i,i+1}-\tan\Gamma_-^{i,i+1}$ to order $1/N$ (note that it vanishes at zeroth order). A straightforward calculation shows that
\begin{align}
&\tan\Gamma_-^{i,i+1}=\frac{\tan T \cosh \cZ - \tan \Phi}{1+\tan\Phi\tan T\cosh\cZ}+\frac{(1+\tan^2\Phi)\cosh\cZ \rho d\phi}{2\cos^2T(1+\tan\Phi\tan T\cosh\cZ)^2}+O(\frac{1}{N^2}),\nonumber\\
&\nonumber\\
&\tan\Gamma_+^{i,i+1}=\frac{\tan T \cosh \cZ - \tan \Phi}{1+\tan\Phi\tan T\cosh \cZ}+\nonumber\\
&+\frac{(1+\tan^2\Phi)(1+\tan^2T\cosh^2\cZ+\tan T\sinh\cZ\cZ'+\cosh\cZ(1+\tan^2T)(T'-\frac{\rho}{2}))}{(1+\tan T\cosh\cZ\tan\Phi)^2}d\phi\nonumber\\
&+O(\frac{1}{N^2}).\label{gamma}
\end{align}
Thus we obtain
\begin{align}
\tan\Gamma_+^{i,i+1}-\tan\Gamma_-^{i,i+1}=\frac{\cos^2T(\tan T\cosh\cZ)'-\cosh\cZ\rho+\cos^2T+\sin^2T\cosh^2\cZ}{(\cos T\cos\Phi+\sin T\cosh\cZ\sin\Phi)^2}d\phi+O(\frac{1}{N^2}).\label{gammadiff}
\end{align}
\linebreak
We will now compute explicitly the metric in the resulting thin shell spacetimes. The spacetime will consist of two patches, separated by a timelike shell of matter denoted by $\mathcal{L}$. The spacetime outside the shell will either be that of a conical singularity or that of a BTZ black hole, and the inside will just consist of empty \ads. This shell will have a non-trivial embedding in the two patches, specified by the function $\cZ$ and the mass density $\rho$. This is different from the massless shells in Section \ref{masslesscollsec} where there was only one free function (the energy density $\theta$) that specified the properties of the shell. Note also that the shape of the massless shells is trivial to obtain (since all the radial lightlike geodesics essentially look the same) while the shape of the massive shells will be non-trivial (and look different viewed from above or from below the shell). We will write the spacetimes inside and outside $\mathcal{L}$ in rotationally symmetric coordinates, even though the total spacetime is not rotationally symmetric. The coordinate systems will then be discontinuous when crossing $\mathcal{L}$ with a non-trivial and angle dependent mapping, and finding the map from the coordinates inside to the coordinates outside is the main goal of this section, as well as computing the induced metric on $\mathcal{L}$. We will separate the two calculations into that of formation of a conical singularity and that of formation of a black hole. Conceptually these two calculations are different, as they will rely on applying timelike respectively spacelike isometries of \ads, but the end result will essentially be the same but with different signs of the mass. Although the case of formation of a conical singularity is not very physical, the computations are easier to understand and more intuitively visualized, thus it is recommended to understand it first before doing the black hole computation.\\
\lb
{\bf Formation of a conical singularity}\\
In the case of the formation of a conical singularity, the resulting spacetime will in the limit be that of a conical singularity geometry and empty \ads glued together across a timelike shell, denoted by \LS, which (in the coordinates inside the shell) is given by the equation $\tanh\chi=-\tanh\cZ(\phi)\sin t$. For finite $N$, the geometry outside the shell consists of several moving wedges, and before we take the $N\rightarrow\infty$ limit, we will have to transform these wedges to a static geometry. The coordinates after doing this transformation will be denoted by $(\pc t,\pc\chi,\pc\phi)$, and consist of several disconnected (static) wedges, which are glued together via the isometries. The metric is still given by \eqref{adsmetric}. We will then ``push the wedges together'' and define a new continuous angular coordinate $\hat\phi$, see Figure \ref{3pillustrcs} for an illustration. The metric is now still given by \eqref{adsmetric}, but the angular variable takes values in the range $(0,\alpha)$ (so that the angular deficit is $2\pi-\alpha$). In the static coordinates, the wedge labeled by $(i,i+1)$ is a circle sector we will denote $c^{\text{static}}_{i,i+1}$, which has opening angle $2\nu_{i,i+1}$. Thus when passing $c^{\text{static}}_{i,i+1}$, the angle $\hat\phi$ will increase by $2\nu_{i,i+1}$. This means that, in the $N\rightarrow\infty$ limit, we can write $\hat\phi=\hat{\phi}_0+\sum_{0\leq j\leq i} 2\nu_{j,j+1}+O(1/N)$ when $\hat\phi\in c^{\text{static}}_{i,i+1}$, where $\hat{\phi}_0$ is an overall angular shift (the approximate value of $\hat\phi$ when $\hat\phi\in c^{\text{static}}_{0,1}$). The total angle$\alpha$ is of course given by $\alpha=\sum 2\nu_{i,i+1}$. \\
\linebreak
Let us define $\hat \cZ$ and $\hat T$ to be the continuous interpolating functions corresponding to $\zeta_{i,i+1}$ and $p_{i,i+1}\nu_{i,i+1}$. Then, by taking the limit in equation \eqref{tanhzcothz} we obtain
\begin{equation}
\tanh\hat \cZ=\frac{\tan T\sinh\cZ}{\tan T\cosh\cZ\cos\Phi-\sin\Phi},\label{tanhhatZ}
\end{equation}
and from \eqref{Gammapnu} and \eqref{gamma}, we obtain
\begin{equation}
\tan \hat T\cosh \hat \cZ=\frac{\tan T\cosh Z-\tan\Phi}{1+\tan\Phi\tan T\cosh Z}. \label{tanhatTcoshhatZ}
\end{equation}
From this we can also derive the useful relation (see Appendix \ref{usefulrel})
\begin{equation}
\cos\hat T=\cos\Phi\cos T+\sin\Phi\sin T\cosh Z,\label{coshatT}
\end{equation}
and using this relation we can immediately derive from \eqref{tanhhatZ} and \eqref{tanhatTcoshhatZ} that
\begin{equation}
\sin\hat T\cosh\hat Z=\sin T\cosh Z\cos\Phi-\sin\Phi\cos T,\label{sinhatTcoshhatZ}
\end{equation}
\begin{equation}
\sin\hat T\sinh\hat Z=\sin T\sinh Z.\label{sinhatTsinhhatZ}
\end{equation}
From \eqref{Gammapnu}, we can also compute
\begin{equation}
\tan \Gamma_+^{i,i+1}-\tan \Gamma_-^{i,i+1}=\frac{2\cosh\hat Z}{\cos^2\hat T} \nu_{i,i+1}+O(\frac{1}{n^2}).\label{gammadiffcs}
\end{equation}
Now using \eqref{gammadiff}, it can be proven that in the limit we have (see Appendix \ref{anglemap})
\begin{equation}
\hat\phi=\hat{\phi}_0-\int_0^\phi\frac{\sin \hat T}{\sin \Phi}\left(\cos T-\frac{\sin T}{\sinh\cZ}\partial_\phi \cZ\right)\rd\phi.\label{cs_anglemap}
\end{equation}

\begin{figure}
\begin{center}
\includegraphics[scale=0.8]{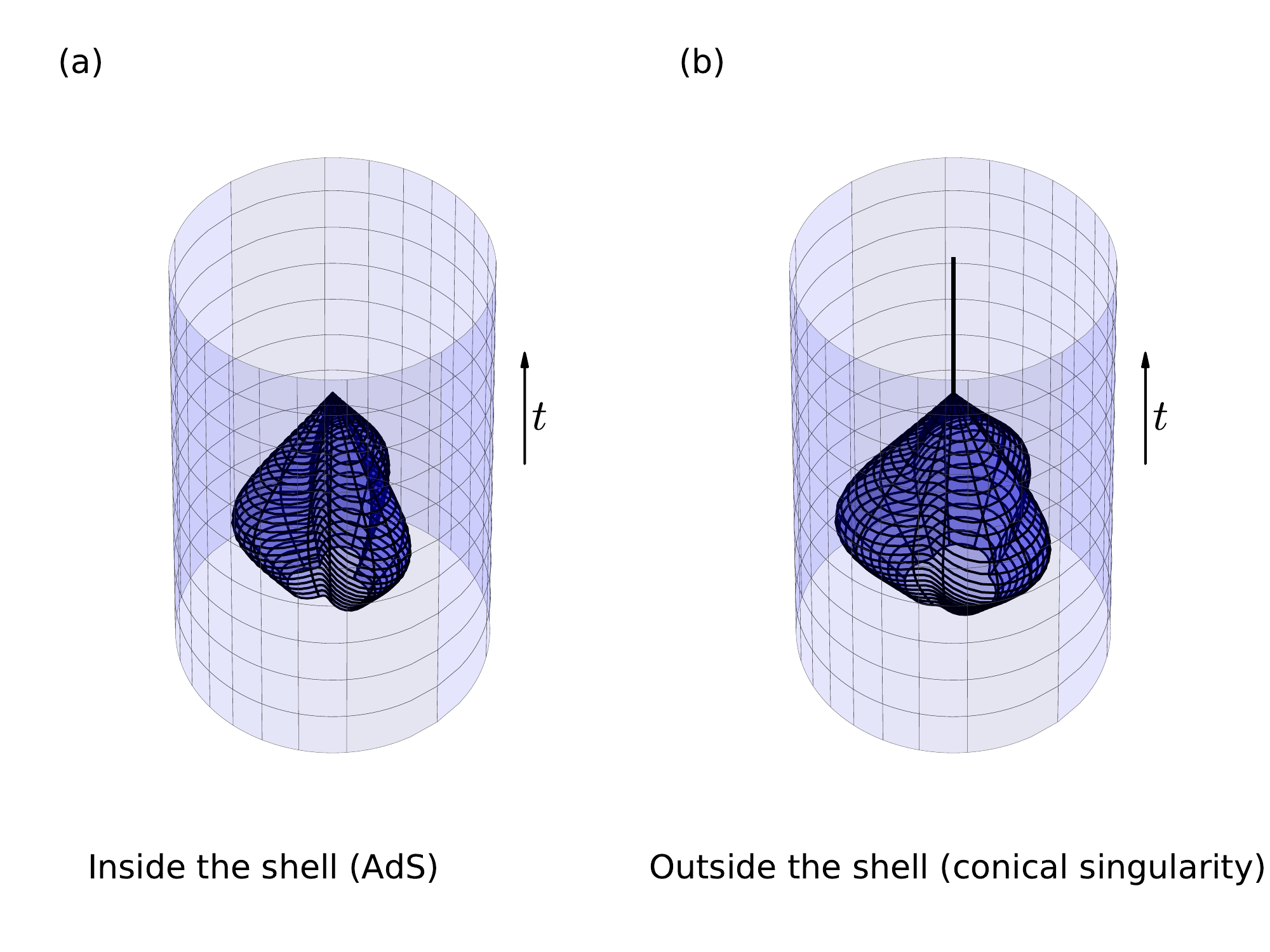}
\caption{\label{shellcsfig} An illustration of a massive thin shell spacetime forming a conical singularity. Panel (a) shows the embedding of the shell in a spacetime with the AdS metric, as seen from the inside part of the shell. The allowed part of spacetime is inside the plotted surface, while everything else should be discarded. The surface in (a) is then glued to the surface in (b) via a non-trivial coordinate transformation. The surface in (b) is embedded in a spacetime with a conical singularity at the origin, which is shown as the thick black line. The allowed part of the spacetime is outside the plotted surface in (b) while the inside part should be discarded. The particular form of the embedding inside the shell has been taken to be $Z(\phi)=1+\frac{1}{2}\cos(3\phi)$. To make the illustration possible we are again using the compactification of the radial coordinate mentioned after equation \eqref{adsmetric}.}
\end{center}
\end{figure}

\noindent We will also be interested in the shape of the shell in the $(\pc\chi,\pc t,\pc\phi)$ coordinates, which will be specified by a function $\pc Z$. Note that $\pc Z$ will depend non-trivially on $Z$ and $\rho$, and this is what makes the computations much more involved compared to the lightlike case (for the lightlike shells, since the lightlike geodesics are invariant under the coordinate transformations that bring us to the static coordinate system, the shape of the shell is essentially the same in both coordinate systems). To determine this, let us first see how generic massive geodesics are mapped under the coordinate transformation that brings us to the static coordinates. It will be useful to work in the static coordinates $(\pc t,\pc\chi,\pc\phi)$ before we push the wedges together. Let us assume that some arbitrary massive geodesic is given by the equation $\tanh\chi=-\tanh\xi\sin t$. The coordinate transformation to go from the static coordinates is given by \eqref{boosteqs} with $\psi=\phi_{i,i+1}$ and $\zeta=-\zeta_{i,i+1}$. This is mapped to a new geodesic given by $\tanh\pc\chi=-\tanh\pc\xi\sin\pc t$ at some angle $\pc\phi$. From \eqref{boosteqs} we obtain then that the radial coordinates are related as
\begin{equation}
\sinh\chi=\sinh\pc\chi\tanh\xi\left(\frac{\cosh\zeta_{i,i+1}}{\tanh\pc\xi}+\sinh\zeta_{i,i+1}\cos(\phi_{i,i+1}-\pc\phi)\right).
\end{equation}
To see how this coordinate transformation acts on the massive geodesics that the particles follow, we can set $\xi=\zeta_i$ and $\pc\xi=\pc\zeta_i$, and in the continuous limit these become $Z$ respectively $\pc Z$. The embedding equation of the shell will then be $\tanh\chi=-\tanh Z\sin t$ inside the shell, and $\tanh\pc\chi=-\tanh \pc Z\sin\pc t$ outside the shell. We also note that in the limit, we have $\pc\phi-\phi_{i,i+1}\rightarrow \hat{T}$, thus
\begin{equation}
\sinh\chi=\sinh\pc\chi\tanh Z\left(\frac{\cosh\hat Z}{\tanh \pc Z}+\sinh\hat Z\cos\hat T\right).
\end{equation}
To determine $\pc Z$, we can use the relations \eqref{proptime2} between the proper time and $\chi$ and $\pc\chi$, which read
\begin{equation}
\begin{array}{ll}
\sinh\chi=-\sinh Z\sin \tau, &\quad \sinh\pc\chi=-\sinh \pc Z\sin \tau.\\\label{ZZpproptime}
\end{array}
\end{equation}
This yields
\begin{equation}
\cosh Z=\cosh \pc Z\cosh\hat Z+\sinh \pc Z\sinh \hat Z\cos\hat T.\label{Zprel1}
\end{equation}
Analogously, it is also possible to obtain a similar relation by looking at the inverse transformation, from which we instead obtain
\begin{equation}
\cosh \pc Z=\cosh Z\cosh\hat Z-\sinh Z\sinh\hat Z\cos\Phi.\label{Zprel2}
\end{equation}
From the above equations, we can eliminate $\cosh \pc Z$, and then after using \eqref{coshatT}, \eqref{sinhatTcoshhatZ} and \eqref{sinhatTsinhhatZ} we can derive the very simple relation
\begin{equation}
\frac{\sinh \pc Z}{\sinh Z}=-\frac{\sin \Phi}{\sin\hat T}.\label{ZpZ}
\end{equation}
This together with \eqref{ZZpproptime}, implies a relation between $\pc\chi$ and $\chi$ when crossing the shell, namely
\begin{equation}
\sinh\chi=-\frac{\sin\hat T}{\sin \Phi}\sinh\pc\chi=\frac{\partial_\phi \hat \phi}{\cos T-\frac{\sin T}{\sinh Z}\partial_\phi Z}\sinh\pc\chi.\label{chimap}
\end{equation}
This equation is useful when comparing with the massless limit. We now know exactly how the coordinates are related when crossing the shell, as well as the shape of the shell in both patches, but it will also be of interest to compute the induced metric. A convenient time coordinate to use for the intrinsic geometry on the shell is the proper time $\tau$ of the pointlike particles. As is shown in Appendix \ref{app_ind_metric}, the induced metric as seen from the patch inside the shell can be simplified to the form

\begin{equation}
\rd s^2=-\rd\tau^2+\sin^2\tau(\sinh^2Z+(\partial_\phi Z)^2)\rd\phi^2.\label{indmetric}
\end{equation}
Analogously, the induced metric outside the shell can be computed as
\begin{equation}
\rd s^2=-\rd\tau^2+\sin^2\tau(\sinh^2\pc Z+(\partial_{\hat\phi} \pc Z)^2)\rd\hat\phi^2.
\end{equation}
Note that $\hat\phi$ still takes values in the range $(0,\alpha)$.\\
\linebreak
We will now move to the final coordinate system that is most suitable for thin shell spacetimes and for applying junction conditions, namely the coordinate system where the metric inside the shell takes the form
\begin{equation}
\rd s^2=-f(r)\rd t^2+\frac{\rd r^2}{f(r)}+r^2\rd\phi^2,\label{rtpmetric}
\end{equation}
while the metric outside the shell is 
\begin{equation}
\rd\bar s^2=-\bar{f}(\bar r)\rd\bar t^2+\frac{\rd\bar r^2}{\bar{f}(\bar r)}+\bar r^2\rd\bar\phi^2,\label{rtpbarmetric}
\end{equation}
where we have defined $f(r)=1+r^2$ and $\bar{f}(\bar r)=-M+\bar{r}^2$. These coordinates are convenient since they take the same form for both the conical singularity case and the black hole case (note that for the conical singularity spacetimes we currently consider, we have $M<0$). Both angular coordinates now take values in $(0,2\pi)$, and the coordinate transformations that brings the metrics to these forms are
\begin{equation}
\begin{array}{ll}
r=\sinh\chi, &\quad \bar r=\sqrt{-M}\sinh\pc\chi,\\
\bar\phi=\frac{\hat \phi}{\sqrt{-M}}, &\quad \bar t=\frac{\pc t}{\sqrt{-M}},\label{bartransf_cs}\\
\end{array}
\end{equation}
while $t$ and $\phi$ remain the same and $M$ is given by
\begin{equation}
M=-\left(\frac{\alpha}{2\pi}\right)^2,
\end{equation}
where we recall that $(0,\alpha)$ is the range of the variable $\hat\phi$. The embedding of the shell is now given by 
\begin{equation}
\frac{r}{\sqrt{f(r)}}=-\frac{R}{\sqrt{f(R)}}\sin t,
\end{equation}
inside, and 
\begin{equation}
\frac{\bar r}{\sqrt{\bar{f}(\bar r)}}=-\frac{\bar R}{\sqrt{\bar{f}(\bar R)}}\sin (\sqrt{-M}\bar t),
\end{equation}
outside, where we have defined $\sinh Z\equiv R$ and $\sqrt{-M}\sinh \pc Z\equiv\bar R$. In terms of the proper time, the embedding of the shell is given by
\begin{equation}
\begin{array}{ll}
r=-R\sin\tau, &\quad \tan t=\sqrt{f(R)}\tan \tau \\
\end{array}
\end{equation}
inside the shell and 
\begin{equation}
\begin{array}{ll}
\bar r=-\bar R\sin\tau, &\quad \tan (\sqrt{-M} \bar t)=\frac{\sqrt{\bar{f}(\bar R)}}{\sqrt{-M}}\tan \tau \\
\end{array}
\end{equation}
outside. The induced metric is now given by
\begin{equation}
\rd s^2=-\rd\tau^2+\sin^2\tau h^2\rd\phi^2=-\rd\tau^2+\sin^2\tau\bar h^2\rd\bar \phi^2,
\end{equation}
where we have defined $h^2\equiv R^2+\frac{(\partial_\phi R)^2}{f(R)}$ and $\bar h^2\equiv \bar{R}^2+\frac{(\partial_{\bar \phi} \bar R)^2}{\bar{f}(\bar R)}$. It can be proven explicitly that the induced metric is the same in both coordinate systems, or in other words that
\begin{equation}
\left(\frac{\rd\bar\phi}{\rd\phi}\right)^2=\frac{h^2}{\bar{h}^2}.\label{indmetric_consistency}
\end{equation}
This is proved in Appendix \ref{app_cons_cond}. This is a necessary condition that must be satisfied to have a consistent geometry. An illustration of a thin shell colliding to form a conical singularity is shown in Figure \ref{shellcsfig}.\\
\linebreak
Defining $R$ and $\bar R$ (or equivalently $h$ and $\bar h$, or $Z$ and $\pc Z$) will determine the whole spacetime. This will turn out to be a useful parametrization when analysing the geometry using the junction conditions where we will take the viewpoint that our thin shell spacetime is {\it defined} by the two free functions $R$ and $\bar R$, from which the embedding of the shell and the energy density can be derived. An interpretation in terms of pointlike particles is not necessary from that point of view.\\
\lb
\linebreak
{\bf Formation of a black hole}\\
When a black hole forms we can no longer map the resulting wedges to circle sectors of \ads, essentially since the tip of a resulting wedge will now be a spacelike geodesic. Instead we must map them to static patches of a BTZ black hole, as specified by equation \eqref{unitbhmetric}. As explained in Section \ref{bhsec}, we do this in two steps. We first go to a coordinate system ($\pc t,\pc\chi,\pc\phi)$ where the final geodesics are mapped to ``straight'' spacelike geodesics in a constant timeslice, such that the wedges take the form \eqref{bhwedge1}. These wedges are still defined in a spacetime with the standard \ads metric \eqref{adsmetric}. We will then do another coordinate transformation to a coordinate system ($\sigma,\rho,y)$ where the wedges are normal circle sectors, but where the metric takes the form \eqref{unitbhmetric}, a BTZ black hole with unit mass. We will then again ``push the wedges together'' and define a continuous angular coordinate $\hat{y}$. Since these final static wedges (denoted by $c^{\text{static}}_{i,i+1}$) will have an opening angle of $2\mu_{i,i+1}$, the coordinate $\hat{y}$ will increase by $2\mu_{i,i+1}$ when crossing one such wedge. Thus $\hat{y}=\hat{y}_0+\sum_{0\leq j\leq i}2\mu_{j,j+1}+O(1/N)$ when $\hat{y}\in c^{\text{static}}_{i,i+1}$, where $\hat{y}_0$ is an unimportant overall shift. Note that the spacetime still takes the form \eqref{unitbhmetric}, a black hole metric with unit mass, since $\hat{y}$ takes values in $(0,\alpha)$ for some value $\alpha$. The correct mass is obtained by rescaling the angular coordinate to the standard range $(0,2\pi)$.\\
\linebreak
Since each final wedge is specified by two parameters $\zeta_{i,i+1}$ and $\xi_{i,i+1}$ (analogous to the $\zeta_{i,i+1}$ and $p_{i,i+1}\nu_{i,i+1}$ in the conical singularity computation), we will define continuous interpolating functions $\hat Z$ and $\hat X$ corresponding to these quantities. By then taking the limit in equation \eqref{tanhzcothz}, we obtain
\begin{equation}
\coth\hat Z=\frac{\tan T\sinh Z}{\tan T\cosh Z\cos\Phi-\sin\Phi},
\end{equation}
and from \eqref{Gammapmui} and \eqref{gamma} we obtain
\begin{equation}
\tanh\hat X\sinh \hat Z=\frac{\tan\Phi-\tan T\cosh Z}{1+\tan T\cosh Z\tan \Phi}.\label{tanhXsinhhatZ}
\end{equation}
From this the following useful relation can be derived (see Appendix \ref{usefulrel}):
\begin{equation}
\cosh\hat X=\cos\Phi\cos T+\sin\Phi\sin T\cosh Z.\label{coshX}
\end{equation}
From the above three relations it then follows that
\begin{equation}
\sinh\hat X\sinh\hat Z=-\sin T\cosh Z\cos\Phi+\sin\Phi\cos T,\label{sinhhatXsinhhatZ}
\end{equation}
\begin{equation}
\sinh\hat X\cosh\hat Z=-\sin T\sinh Z.\label{sinhhatXcoshhatZ}
\end{equation}
From \eqref{Gammapmui} we have
\begin{equation}
\tan\Gamma^{i,i+1}_\pm=\mp\frac{\sinh\hat Z}{\cosh^2\hat X}\mu_{i,i+1}+O(\frac{1}{N^2}).\label{gammadiffbh}
\end{equation}
The angular coordinate is now given by $\hat y=\hat y_0+\sum 2\mu_{i,i+1}+O(1/N)$. By using \eqref{gammadiff}, it can be shown (see Appendix \ref{anglemap}) that in the limit we obtain
\begin{equation}
\hat y=\hat{y}_0-\int_0^\phi\frac{\sinh \hat X}{\sin \Phi}\left(\cos T-\frac{\sin T}{\sinh\cZ}\partial_\phi \cZ\right)\rd\phi.\label{bh_anglemap}
\end{equation}
We will now obtain the relation between the radial coordinates as well as the embedding of the shell. This is a bit more involved than obtaining $\hat{y}$, and requires a more thorough investigation of the two mappings involved to bring us to the static wedges. This is the same set of coordinate transformations that can be induced by going from \eqref{bhwedge3} to \eqref{bhwedge2} and then to a static circle sector in the spacetime with metric \eqref{unitbhmetric} as explained in Section \ref{bhsec}. We will thus first make a coordinate transformation to bring the final wedges to the form \eqref{bhwedge2}, namely given by
\begin{equation}
\sin\pc\phi\tanh\pc\chi=\mp\sin \pc t\tanh(\mu_{i,i+1}\pm\xi_{i,i+1}).\label{muwedge}
\end{equation}
This is done by using a transformation of the form \eqref{boosteqs} with $\zeta=-\zeta_{i,i+1}$ and $\psi=\phi_{i,i+1}$, as well as a convenient rotation to bring the spacelike geodesic to the angle $\pc\phi=0$. We will now bring the wedges to circle sectors in a black hole background with unit mass. However, since we will only be interested in the spacetime outside the horizon, we will use the coordinates $(\sigma,\beta,y)$ with metric \eqref{betabhcoord}. Here,  $\sigma$ is a time coordinate, $\beta$ is the radial coordinate, $y$ is an angular coordinate, and the circle sector has opening angle $2\mu_{i,i+1}$. Note that this only covers the spacetime {\it outside} the horizon which is located at $\beta=0$. As explained in Section \ref{bhsec}, this metric can be obtained by the embedding
\begin{equation}
\begin{array}{cc}
x^0=-\cosh\beta\cosh y, & x^2=\cosh\beta\sinh y,\\
x^1=\sinh \beta \cosh\sigma, & x^3=\sinh \beta \sinh \sigma, \label{bhcoord}
\end{array}
\end{equation}
and we will compute the coordinate transformation by comparing this embedding with \eqref{embedding_eq}.\\
\linebreak
To determine the embedding of the shell in the patch outside the shell, we will have to see how the massive geodesics, where the particles move, transform through the two coordinate transformations. The massive geodesics are given by $\tanh\chi=-\tanh\zeta_i\sin t$, and are first mapped to a geodesic on the form $\tanh\pc\chi=-\tanh\pc\zeta_i\sin\pc t$. Just as in the conical singularity situation, we can obtain a relation between these two from \eqref{boosteqs}. This results in
\begin{equation}
\sinh\pc\chi=\sinh\chi\tanh\pc\zeta_i\left(\frac{\cosh\zeta_{i,i+1}}{\tanh\zeta_i}-\sinh\zeta_{i,i+1}\cos(\phi_{i,i+1}-\psi_i)\right).
\end{equation}
In the limit, where we replace $\zeta_i$ by $Z$, $\pc\zeta_i$ by $\pc Z$ and $\zeta_{i,i+1}$ by $\hat Z$, and by using $\sinh\chi=-\sinh Z\sin \tau$ and $\sinh\pc\chi=-\sinh\pc Z\sin \tau$, this becomes
\begin{equation}
\cosh\pc Z=\cosh Z\cosh\hat Z-\sinh Z\sinh\hat Z\cos\Phi.\label{bhcoshZp}
\end{equation}
We also note that the inverse transformation can be used to obtain
\begin{equation}
\cosh Z=\cosh \pc Z\cosh\hat Z+\sinh\pc Z\sinh\hat Z\cos\pc\Phi,\label{bhcoshZ}
\end{equation}
where $\pc\Phi$ is the continuous version of the angular location of the geodesic in the $(\pc t,\pc\chi,\pc\phi)$ coordinates. By now using \eqref{coshX}, \eqref{sinhhatXsinhhatZ}, \eqref{sinhhatXcoshhatZ} and \eqref{bhcoshZp} we obtain the simple relation
\begin{equation}
\frac{\cosh \pc Z}{\sinh Z}=-\sin\Phi\coth\hat X.\label{bhzzp}
\end{equation}
However, we are interested in the relation between the radial coordinate $\beta$ in the $(\sigma,\beta,y)$ coordinate system, thus we will have to compare the  embedding \eqref{bhcoord} with \eqref{adscoord}. Note first, that equation \eqref{muwedge}, after taking the limit, together with $\tanh\pc\chi=-\tanh \pc Z\sin\pc t$, implies the following relation:
\begin{equation}
\sin\pc\Phi=\frac{\tanh\hat X}{\tanh \pc Z}.
\end{equation}
Now note that the relation $\tanh\pc\chi=-\tanh \pc Z\sin \pc t$, which defines the trajectory of the geodesics, is equivalent to $x_1^2+x_2^2=\tanh^2\pc Z x_0^2$ by comparing with the embedding \eqref{adscoord}. By using $x_0=\cosh\beta\cosh y$ and $x_2=\cosh\beta\sinh y$, as well as $x_1=\sinh\pc\chi\cos\pc\phi$ and $x_2=\sinh\pc\chi\sin\pc\phi$, we obtain for the trajectory of the particles in the $(\sigma,\beta,y)$ coordinates that
\begin{align}
\tanh^2\pc Z\cosh^2\beta&=x_1^2+\frac{x_2^2}{\cosh^2\pc Z}=(\cos^2\pc\Phi+\frac{\sin^2\pc\Phi}{\cosh^2\pc Z})\sinh^2\pc\chi\nonumber\\
&=\frac{\sinh^2\pc\chi}{\cosh^2\hat X}=\frac{\sinh^2\chi\sinh^2\pc Z}{\sinh^2Z\cosh^2\hat X}\nonumber\\
&\Rightarrow \cosh\beta=-\frac{\sin\Phi}{\sinh\hat X}\sinh\chi,\label{betamap}
\end{align}
where we also used \eqref{bhzzp} in the last equality. Thus if we define the ``boost parameter'' $B$ such that $\cosh\beta=-\cosh B\sin\tau$, which specifies the embedding of the shell in the outside patch, we obtain the relation
\begin{equation}
\frac{\cosh B}{\sinh Z}=-\frac{\sin\Phi}{\sinh\hat X},\label{BZ}
\end{equation}
which is similar to what was obtained in the conical singularity case. We also want to know how the time coordinate $\sigma$ is related to $\beta$ on the geodesic. By again using \eqref{bhcoord} in the relation $x_1^2+x_2^2=\tanh^2\pc Z x_0^2$, it is easy to show that
\begin{equation}
\coth\beta=\coth B\cosh\sigma.
\end{equation}
Thus we see for instance, that the maximal distance $\beta=B$ corresponds to $\sigma=0$, and that $\sigma\rightarrow\infty$ when $\beta$ approaches the horizon which is located at $\beta=0$. This is expected to happen; shells collapsing to a black hole, in ``Schwarzschild like" coordinates (which we are using here), will not cross the horizon and these coordinates do not describe the whole spacetime outside the shell (only the part that is outside the horizon). From this we can also obtain the relation between $\sigma$ and the proper time $\tau$, which is
\begin{equation}
\tanh\sigma=-\frac{\cot\tau}{\sinh B},
\end{equation}
from which we also see that $\sigma\rightarrow\infty$ already at some finite value of $\tau<0$, as expected.\\
\linebreak
The induced metric can now also be derived (see Appendix \ref{app_ind_metric}). The induced metric seen from the coordinate patch inside the shell will be the same as computed in the conical singularity case, namely given by
\begin{equation}
\rd s^2=-\rd\tau^2+\sin^2\tau(\sinh^2Z+(\partial_\phi Z)^2)\rd\phi^2.
\end{equation}
Seen from the patch outside the shell, the induced metric is
\begin{equation}
\rd s^2=-\rd\tau^2+\sin^2\tau(\cosh^2B+(\partial_{\hat{y}}B)^2)\rd\hat{y}^2.\label{indmetric_bh}
\end{equation}
\linebreak
We will now, just like in the conical singularity case, move to more conventional metrics decribing a black hole, which are more convenient for using the junction formalism. The metric inside the shell takes the form
\begin{equation}
\rd s^2=-f(r)\rd t^2+\frac{\rd r^2}{f(r)}+r^2\rd\phi^2,
\end{equation}
while the metric outside the shell is
\begin{equation}
\rd\bar s^2=-\bar{f}(\bar r)\rd\bar t^2+\frac{\rd\bar r^2}{\bar{f}(\bar r)}+\bar r^2\rd\bar\phi^2,
\end{equation}
where we now have $M>0$, and we have again defined $f(r)=1+r^2$ and $\bar{f}(\bar r)=-M+\bar r^2$. The coordinate transformation to go to these coordinates is now
\begin{equation}
\begin{array}{ll}
r=\sinh\chi, &\quad \bar r=\sqrt{M}\cosh\beta,\\
\bar\phi=\frac{\hat{y}}{\sqrt{M}}, &\quad \bar t=\frac{\sigma}{\sqrt{M}}.\label{bartransf_bh}\\
\end{array}
\end{equation}
while $t$ and $\phi$ remain the same and $M$ is given by
\begin{equation}
M=\left(\frac{\alpha}{2\pi}\right)^2,\label{Malpha}
\end{equation}
where $(0,\alpha)$ was the range of the variable $\hat y$. The embedding of the shell is given by 
\begin{equation}
\frac{r}{\sqrt{f(r)}}=-\frac{R}{\sqrt{f(R)}}\sin t
\end{equation}
inside, and 
\begin{equation}
\frac{\bar r}{\sqrt{\bar{f}(\bar r)}}=\frac{\bar R}{\sqrt{\bar{f}(\bar R)}} \cosh (\sqrt{M}\bar t)
\end{equation}
outside, where we have defined $\sinh Z\equiv R$ and $\sqrt{M}\cosh B\equiv\bar R$. In terms of the proper time, the embedding inside the shell is again given by
\begin{equation}
\begin{array}{ll}
r=-R\sin\tau, &\quad \tan t=\sqrt{f(R)}\tan \tau\\
\end{array}
\end{equation}
and the embedding outside the shell is given by
\begin{equation}
\begin{array}{ll}
\bar r=-\bar{R} \sin\tau, &\quad \coth (\sqrt{M} \bar t)=-\frac{\sqrt{\bar f(\bar R)}}{\sqrt{M}}\tan \tau \\
\end{array}\label{bh_barcoord}
\end{equation}
Note that the embedding of the $t$ coordinate takes a different form compared to the conical singularity case, and thus we must be careful when applying the junction formalism. The induced metric is now given by
\begin{equation}
\rd s^2=-\rd\tau^2+\sin^2\tau h^2\rd\phi^2=-\rd\tau^2+\sin^2\tau\bar h^2\rd\bar \phi^2,
\end{equation}
where we have defined $h^2\equiv\sinh^2Z+(\partial_\phi Z)^2=R^2+\frac{(\partial_\phi R)^2}{f(R)}$ and $\bar h^2\equiv M\cosh^2B+(\partial_{\bar \phi} B)^2=\bar{R}^2+\frac{(\partial_{\bar\phi} \bar R)^2}{\bar{f}(\bar{R})}$. Again we have the consistency condition $d\bar{\phi}/d\phi=h/\bar{h}$ which says that the induced metric is the same in both patches (see Appendix \ref{app_cons_cond}). In Figure \ref{shellfig} we show an illustration of what a thin timelike shell collapsing to a black hole looks like.
\begin{figure}
\begin{center}
\includegraphics[scale=0.8]{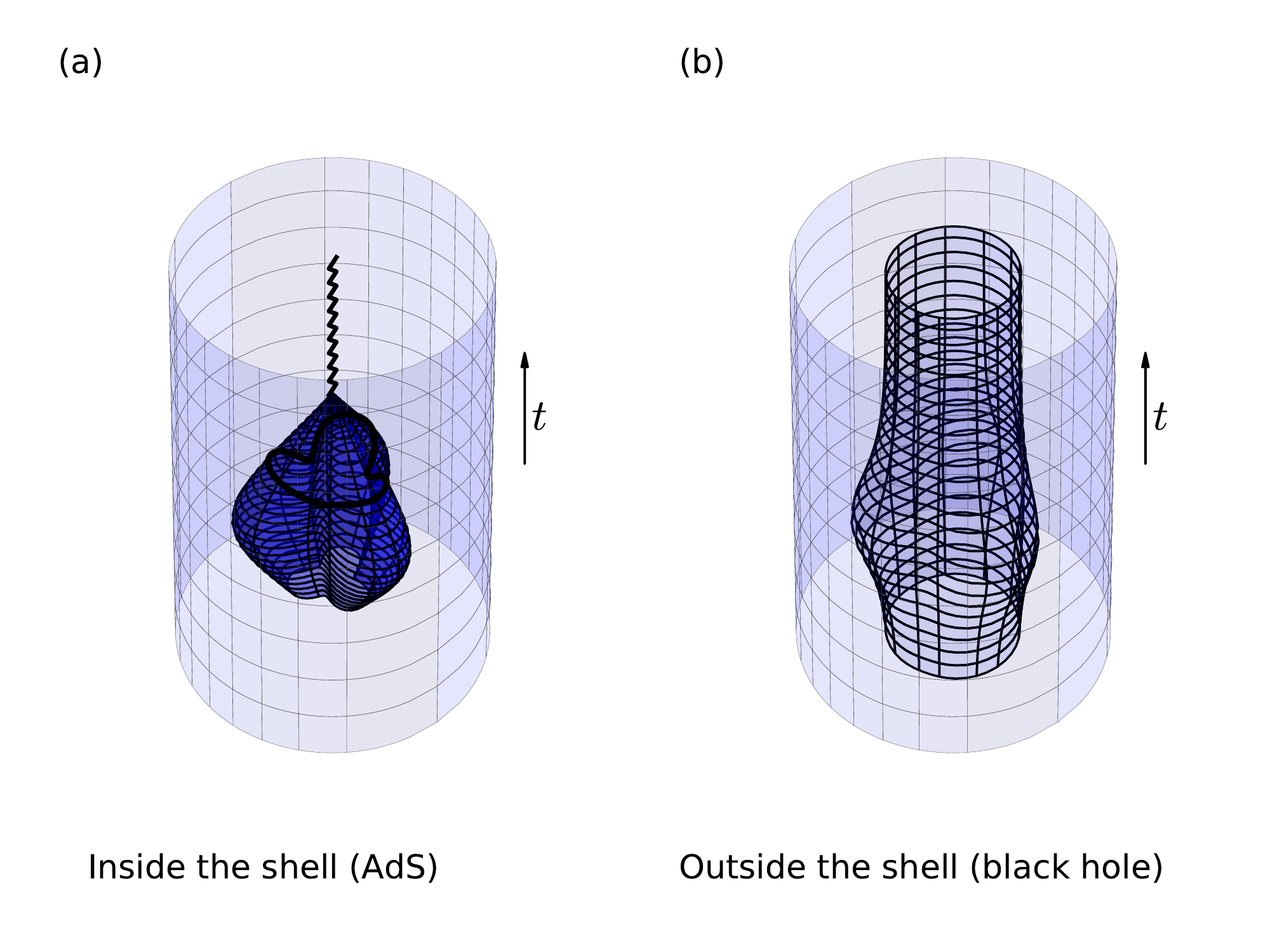}
\caption{\label{shellfig} Illustration of the massive thin shell spacetimes with an angle dependent embedding. Panel (a) shows the embedding of the shell in a spacetime with the AdS metric, as seen from the inside part of the shell. The allowed part of spacetime is inside the plotted surface, while everything else should be removed. The surface in (a) is then glued to the surface in (b) via a non-trivial coordinate transformation. The surface in (b) is embedded in a spacetime with the BTZ black hole metric, and the allowed spacetime is now the part outside the plotted surface, while the inside should be discarded. Note that in (b) we are using coordinates that only cover the region outside the horizon, which is why the shell gets stuck at the horizon for late times. However, in (a) we also cover parts of the spacetime inside the event horizon, and in particular the collision point of the shell when it forms the singularity. The intersection of the horizon with the shell is seen as the thick line in (a), which is mapped to infinite time in the spacetime in (b). When crossing the shell inside the black line one is thus mapped to the inside of the black hole outside the shell, which is not covered by the coordinates we are using. The particular form of the embedding inside the shell has been taken to be $Z(\phi)=1+\frac{1}{2}\cos(3\phi)$. To make this illustration possible, the radial coordinates have been compactified. For panel (a) it is the same compactification used in the rest of the paper (as specified after equation \eqref{adsmetric}), and for the black hole spacetime in panel (b) the radial coordinate $r$ is defined in terms of $\beta$ by $r=\tanh(s/2)$ where $\sinh s=\cosh \beta$.}
\end{center}
\end{figure}

\subsection{The event horizon}\label{eventsec}
We will now discuss how to compute the event horizon when a black hole forms, which is marked in all figures in this chapter that show the formation of a black hole. First let us call the last point, before the spacetime disappears, the {\it last boundary point}, and we will denote the \ads manifold where we are cutting out the wedges by $\mathcal{M}$. Now, in $\mathcal{M}$, this last boundary point is represented by $N$ points, which are identified by the holonomies of the particles. We will label these points by $P_{i,i+1}$, which are the end points of the final wedges $c_{i,i+1}$ of allowed spacetime (and also the endpoints of the intersections $I_{i,i+1}$). Now consider the backward lightcones $\mathcal{L}_{i,i+1}$ in $\mathcal{M}$, of these $N$ last points. We will now show that the restriction of these lightcones to the allowed spacetime can be used to construct the event horizon. It is clear that, in the final wedge of allowed space $c_{i,i+1}$, points outside the lightcone $\mathcal{L}_{i,i+1}$ are causally disconnected from the last point $P_{i,i+1}$. They are also causually disconnected from the whole boundary in wedge $c_{i,i+1}$. One might ask the question if lightrays from these points can somehow cross the bordering surfaces $w_-^{i,i+1}=w_+^i$ and $w_+^{i,i+1}=w_-^{i+1}$ of the wedge $c_{i,i+1}$, enter another wedge, and then reach the boundary, and therefore not be causally disconnected from the whole boundary. We will now show that this is not possible. Let us denote the intersection between the surface $w_\pm^{i,i+1}$ and $\mathcal{L}_{i,i+1}$ by $I_{i,i+1}^\pm$. We will now show that the intersection $I_{i,i+1}^-$ is mapped to the intersection $I_{i-1,i}^+$ via the holonomy $h_i$, and thus when crossing a final wedge's bordering surfaces $w_\pm^{i,i+1}$ from inside (outside) the wedge's lightcone, will always result in again ending up inside (outside) the neighbouring wedge's lightcone. This can be seen from the following three facts:
\begin{itemize}
 \item The lightcone $\mathcal{L}_{i,i+1}$ is, by definition, composed of all lightlike geodesics ending on the point $P_{i,i+1}$.
 \item The intersections $I_{i,i+1}^\pm$ between $\mathcal{L}_{i,i+1}$ and $w_\pm^{i,i+1}$ are geodesics, since all these surfaces are total geodesic surfaces.
 \item Since both the wedge $c_{i,i+1}$, and the lightcone $\mathcal{L}_{i,i+1}$, end on the point $P_{i,i+1}$, the intersections will also end on this point.
\end{itemize}
Thus, since the intersection $I_{i,i+1}^\pm$ is a geodesic, is located on $\mathcal{L}_{i,i+1}$, and end on $P_{i,i+1}$, it must be a lightlike geodesic. Furthermore, since it is located on $w_\pm^{i,i+1}$, it is {\it the} unique lightlike geodesic that lies on this two-dimensional surface and ends on $P_{i,i+1}$. Therefore, since the point $P_{i,i+1}$ is identified with point $P_{i-1,i}$ and $P_{i+1,i+2}$, and the surface $w_-^{i,i+1}$ ($w_+^{i,i+1}$) is identified with $w_+^{i-1,i}$ ($w_-^{i+1,i+2}$), the intersections $I_{i,i+1}^-$ ($I_{i,i+1}^+$), being the unique lightlike geodesics ending on $P_{i,i+1}$ and lying on $w_-^{i,i+1}$ ($w_+^{i,i+1}$), must be mapped to $I_{i-1,i}^+$ ($I_{i+1,i+2}^-$), the unique lightlike geodesics lying on $w_+^{i-1,i}$ ($w_-^{i+1,i+2}$) and ending on $P_{i-1,i}$ ($P_{i+1,i+2}$). This gives a complete characterization of the event horizon in the final wedges of the geometry (which would, if taking the thin shell $N\rightarrow\infty$ limit, correspond to the geometry outside the shell). When the event horizon is outside the final wedges, it will be composed of piecewise geodesic surfaces of $\mathcal{L}_{i,i+1}$ such that the points inside the event horizon are outside {\it all} the $N$ lightcones $\mathcal{L}_{i,i+1}$. It seems difficult to find a nice expression for the horizon in a general spacetime, let alone in the $N\rightarrow\infty$ limit, and we will not pursue that further in this thesis.

\subsection{Numerical algorithm for constructing the solutions}\label{numericssec}
In this section we will explain how to solve the discrete equations \eqref{peq} and \eqref{nueq} in practice. Although we will consider the equations for the massive particles, the procedure outlined here can easily be modified to work with massless particles. Let us denote $\Phi_i\equiv\phi_{i,i+1}-\psi_i$ and $P_i\equiv p_i\nu_i$. The goal is thus to obtain solutions of \eqref{peq} and \eqref{nueq} for $P_i$ and $\Phi_i$ with periodic boundary conditions (meaning $P_{N+1}=P_1$ and $\Phi_{N+1}=\Phi_1$). The equations \eqref{peq} and \eqref{nueq} can be reformulated as
\begin{equation}
P_{i+1}=\nu_{i+1}-\arctan\left(\frac{\sin(\Phi_i+\psi_i-\psi_{i+1})}{\frac{\sinh\zeta_{i+1}\cos\Phi_i}{\tanh\zeta_i}-\frac{\sinh\zeta_{i+1}\sin\Phi_i}{\sinh\zeta_i\tan(\nu_i+P_i)}-\cos(\Phi_i+\psi_i-\psi_{i+1})\cosh\zeta_{i+1}}\right),\label{peqrec}
\end{equation}
\begin{equation}
\Phi_{i+1}=\arctan\left(\frac{\tan(\Phi_i+\psi_i-\psi_{i+1})(\tan\nu_{i+1}+\tan P_{i+1})}{\tan P_{i+1}-2\cosh\zeta_{i+1}\tan(\Phi_i+\psi_i-\psi_{i+1})\tan P_{i+1}\tan\nu_{i+1}-\tan\nu_{i+1}}\right).\label{Phieqrec}
\end{equation}
Note that now the equations are written such that they can be solved recursively: Given $\Phi_i$ and $P_i$ we can determine $P_{i+1}$, and then given $\Phi_i$ and $P_{i+1}$ we can compute $\Phi_{i+1}$. However, we have to make sure that the periodic boundary conditions $\Phi_{N+1}=\Phi_1$ and $P_{N+1}=P_1$ are satisfied. We will start by considering some special cases where the problem has reflection symmetry in some axis.

\subsubsection{Axis of symmetry through one of the particles}
We will first assume that there is a reflection symmetry that passes through particle 1. This implies that the wedge for this particle is symmetric, i.e. $P_1=0$, which will simplify the computation. Assuming then that we know $\Phi_1$, we can from \eqref{peqrec} compute $P_2$. Thereafter we can use \eqref{Phieqrec} to compute $\Phi_2$. The algorithm can then be continued to obtain all values of $\Phi_i$ and $P_i$. When we have moved one lap around the circle and come back to $\psi_1$ however, the periodic boundary conditions will generically not be satisfied for an arbitrary guess of $\Phi_1$. To obtain the correct value of $\Phi_1$ we thus use a Newton-Raphson solver (or other root finding algorithms) to search for the correct value of $\Phi_1$ such that the periodic boundary condition is satisfied. Note that this does not obviously imply that the periodic boundary condition of $P_i$ is satisfied, and indeed there are sometimes spurious solutions where $\Phi_i$ is periodic but not $P_i$. Thus we must now choose the right solution for $\Phi_i$ such that $P_{N+1}=0$ and it seems that there will always exist such a solution. This is the technique used for finding the parameters for Figure \ref{3particlessym_cs}.
\subsubsection{Axis of symmetry between two particles}
The other possible case one can consider that has an axis of symmetry, is when the axis of symmetry lies in between two particles, taken conveniently as particles $N$ and $1$. In this case, the symmetry restriction will then instead determine the value of $\Phi_{0}\equiv\phi_{N,1}-\psi_N$. Thus we start instead by guessing the value of $P_1$. From $\Phi_0$ and $P_1$ we then compute $\Phi_2$, and then we can continue this algorithm and compute all values of $P_i$ and $\Phi_i$ recursively. We can then use a root finder to make sure that $P_i$ is continuous, but again we are not guaranteed that $\Phi_i$ is periodic. Thus we must again choose the correct solution of $P_1$ such that also $\Phi_N=\Phi_0$, and such a solution always seems to exist. This is the method used to obtain the parameters in Figure \ref{4particles}.
\subsubsection{No symmetry restrictions}
In the case without any symmetry restrictions, the procedure is in practice more complicated although conceptually the same. In that case we first have to guess both $P_1$ and $\Phi_1$, recursively construct all other $P_i$ and $\Phi_i$ and then use a two-dimensional root finder to make sure that $P_1=P_{N+1}$ and $\Phi_1=\Phi_{N+1}$. This is the method used when computing the parameters for Figure \ref{3particlesgen}.

\section{Thin shells and the junction formalism}
In this section we will use the junction formalism for general relativity to analyze the thin shell spacetimes that we have constructed by colliding an infinite number of pointlike particles. These spacetimes are defined by the embedding equation of the shell and how the coordinate systems are related across the shell. The junction formalism will then allow us to directly compute the stress-energy tensor supporting this spacetime. This can then be compared with the energy density of the pointlike particles (or more precisely, the stress-energy tensor one obtains by adding up the stress-energy tensor of an infinite number of pointlike particles) and the matching of these two calculations is a very important consistency check. We will again separate the computations for the massless and the massive shells. For massive shells we can use the standard timelike junction formalism \cite{Israel:1966rt,Musgrave:1995ka} of general relativity that is most widely used. For the massless shells, we have to use the junction formalism for lightlike shells which is a bit more involved due to the degenerate induced metric \cite{Barrabes:1991ng,Musgrave:1997sfw}. However, even though the two junction formalisms seem quite different, the final result of the stress-energy tensor for the massless shells can of course be obtained by taking a massless limit of the massive shells.

\subsection{Massless shells}\label{masslessshellssec}

We will now use the thin shell formalism for lightlike shells, as outlined in \cite{Barrabes:1991ng,Musgrave:1997sfw}, to compute the stress-energy tensor on the shell in the thin shell spacetimes that we compute from the collisions of massless particles. Our starting point is thus a spacetime consisting of two patches, $\Rs$ and $\bar{\Rs}$, separated by a null shell. The coordinates below the shell in $\Rs$ will be $(v,r,\phi)$ ($v<0$) and the coordinates above the shell in $\bar\Rs$ are $(v,\bar{r},\bar{\phi})$ ($v>0$). The lightlike shell $\mathcal{L}$ is located at $v=0$, and the metrics are given by
\begin{equation}
\rd s^2=-f\rd v^2+2\rd r\rd v+r^2\rd \phi^2,
\end{equation}
in $\Rs$ and 
\begin{equation}
\rd \bar{s}^2=-\bar{f}\rd v^2+2\rd \bar{r}\rd v+\bar{r}^2\rd \bar{\phi}^2,
\end{equation}
in $\bar\Rs$. We have defined $f\equiv1+r^2$ and $\bar{f}\equiv-M+r^2$ ($M<0$ for the creation of a conical singularity and $M>0$ for the creation of a black hole). The normal vectors $n$ and $\bar{n}$ of the shell, which satisfy $n\cdot n=\bar{n}\cdot\bar{n}=0$ and are orthogonal to the tangent space of the shell are given by
\begin{equation}
n^\mu=a(0,-1,0),\quad \bar{n}^\mu=\bar{a}(0,-1,0),
\end{equation}
for some arbitrary $a$ and $\bar{a}$. These are determined solely in terms of the embedding of the null surface, and do not say anything about the matter content of the shell (unlike the case with timelike shells \cite{Israel:1966rt,Musgrave:1995ka}). Note that if all coordinates are continuous across the shell, this is just the standard AdS$_3$ Vaidya spacetime described in Section \ref{ssec_vaidya}. However, for our thin shell spacetime there is a non-trivial relation between the coordinates when crossing the shell, which reads $\bar\phi=H(\phi)$ and $\bar{r}=r/H'(\phi)$ for some function $H(\phi)$, and as we will see, to be able to consistently define an induced metric on the shell any of these two relations will imply the other. Note that these relations are exactly what we obtained from the pointlike particle construction. We will fix $\bar a$ in terms of $a$ by requiring that $n$ and $\bar n$, when projected onto the shell, is the same from each side of the shell. We could then fix $a$ to be say equal to one, but we will keep it through the whole calculation and we will explicitly see in the end that the final result is independent of $a$.\\
\lb
The shell will be parametrized by two coordinates, and for convenience we will choose the coordinates $\phi$ and $r$, such that the embedding in $\Rs$ is simple (but it will be more complicated in $\bar\Rs$). The basis vectors of the tangent space on the shell are given by
\begin{equation}
e_{r}^\mu=\frac{\partial x^\mu}{\partial r}=(0,1,0),
\end{equation}
\begin{equation}
e_{\phi}^\mu=\frac{\partial x^\mu}{\partial \phi}=(0,0,1),
\end{equation}
in $\Rs$, and in $\bar\Rs$ we have
\begin{equation}
\bar{e}_{r}^\mu=\frac{\partial \bar{x}^\alpha}{\partial r}=(0,1/H'(\phi),0),
\end{equation}
\begin{equation}
\bar{e}_{\phi}^\mu=\frac{\partial \bar{x}^\alpha}{\partial \phi}=(0,-H''(\phi)r/(H'(\phi))^2,H'(\phi)).
\end{equation}
The (degenerate) metric $g_{ab}=e^{\alpha}_{a}e^{\beta}_{b}g_{\alpha\beta}=\bar{e}^{\alpha}_{a}\bar{e}^{\beta}_{b}\bar{g}_{\alpha\beta}$ on the shell is then
\begin{equation}
\rd s^2=r^2\rd \phi^2.
\end{equation}
The induced metric is the same from both sides of the shell since $\frac{\rd\bar{\phi}}{\rd\phi}=\frac{r}{\bar{r}}$, which is the consistency condition we mentioned earlier that must be satisfied for the spacetime to be well defined (such that the junction formalism is applicable), which came out naturally from the pointlike particle construction. Note that since the normal vector $n$ is proportional to $e_r$, we will now fix $\bar a=a/ H'(\phi)$ such that the proportionality factor is the same and $n$ and $\bar n$ are the ``same'' vector on each side of the shell.\\
\lb
To determine the shell's stress-energy tensor, we will follow the procedure in \cite{Barrabes:1991ng,Musgrave:1997sfw}. Since the normal vectors for lightlike shells do not give any information about the gluing, the null shell formalism requires that we define two transverse vectors $N$ and $\bar{N}$ given by the conditions $N_\alpha e^\alpha_a=\bar{N}_\alpha \bar{e}^\alpha_a$ and $N\cdot N=\bar{N}\cdot\bar{N}$, and such that $N\cdot n=\bar{N}\cdot \bar{n}\neq 0$. For convenience we choose $N\cdot n=\bar{N}\cdot \bar{n}=-1$, and this completely determines $N$ and $\bar{N}$ up to shifts of tangent vectors. We will thus pick
\begin{equation}
N^\mu=\frac{1}{a}(1,0,0),
\end{equation}
where we have used the freedom of tangential shifts to set the $\phi$ and $r$ components of $N$ to zero (by convention $N$ points away from $\Rs$ and into $\bar\Rs$, which would require $a>0$). It then follows that
\begin{equation}
\bar{N}^\mu=\left(\frac{H'}{a},\frac{H'}{2a}\bar{f}-\frac{1}{2aH'}f-\frac{(H'')^2}{2a(H')^3},\frac{H''}{ra}\right).
\end{equation}
Due to the degeneracy of the metric, the junction formalism relies on the so called generalized extrinsic curvatures, which follow the same definition as the standard extrinsic curvatures but with the normal vector replaced by an arbitrary vector. Thus the generalized extrinsic curvature corresponding to an arbitrary vector $v$ is defined by
\begin{equation}
K_{ab}\equiv-v_\mu e^\nu_{(a)}\nabla_\nu e^\mu_{(b)}=-v_\mu\left(\frac{\partial^2x^\mu}{\partial \xi^{(a)}\partial \xi^{(b)}}+\Gamma^\mu_{\alpha\beta}e^\alpha_{(a)}e^\beta_{(b)}\right),
\end{equation}
with a similar definition in $\bar\Rs$. To compute these we need the Christoffel symbols, which in $\Rs$ are given by
\begin{align}
\begin{array}{lll}
 \Gamma^v_{vv}=\frac{f'}{2},&\Gamma^v_{\phi\phi}=-r,&\Gamma^r_{vv}=\frac{f f'}{2},\\
 \Gamma^r_{vr}=-\frac{f'}{2},&\Gamma^r_{\phi\phi}=-r f,&\Gamma^\phi_{r\phi}=\frac{1}{r},
\end{array}
\end{align}
with analogous formulas in $\bar\Rs$. The only non-zero extrinsic curvature component turns out to be
\begin{align}
\bar{K}_{\phi\phi}=&\frac{r\bar{f}}{2a(H')^2}-\frac{rf}{2a}+\frac{rH'''}{aH'}-\frac{3}{2}\frac{r(H'')^2}{a(H')^2}\nonumber\\
=&-\frac{r}{2a}\left(M(H')^2+1\right)+\frac{rH'''}{aH'}-\frac{3}{2}\frac{r(H'')^2}{a(H')^2}.\label{Keq}
\end{align}
To be consistent with the notation in \cite{Musgrave:1997sfw}, we now define $\gamma_{ab}\equiv2(\bar{K}_{ab}-K_{ab})$, which thus has the only non-zero component $\gamma_{\phi\phi}=2\bar{K}_{\phi\phi}$. This object is the main ingredient when computing the stress-energy tensor.

\subsubsection{Intrinsic stress-energy tensor}
The intrinsic stress-energy tensor $S_{ab}$ of the shell is uniquely determined by $\gamma_{ab}$. However, to get there, we will first define a method for raising indices in the geometry on the shell. Since the metric is degenerate, we can not use it for this purpose, and instead indices are raised by a different tensor. Following \cite{Musgrave:1997sfw}, to construct this tensor, we first decompose the normal vector in terms of the basis $(N,e)$ as
\begin{equation}
n^\alpha=\ell^a e_a^\alpha,
\end{equation}
from which we immediately see that
\begin{equation}
\ell^r=-a,\quad\ell^\phi=0.
\end{equation}
Indices are now raised by the quantity $g_*^{ab}$ defined by
\begin{equation}
g_*^{ac}g_{cb}=\delta_b^a+\ell^a e^\alpha_{(b)} N_\alpha=\left(\begin{array}{cc} 0 & 0 \\ 0 & 1 \\ \end{array}\right).
\end{equation}
$g_*$ is not uniquely defined, but we can choose
\begin{equation}
g^{ab}_*=\left(\begin{array}{cc} 0 & 0 \\ 0 & \frac{1}{r^2} \\ \end{array}\right).
\end{equation}
The instrinsic stress-energy tensor of the shell is now given by
\begin{equation}
-16\pi G S^{ab}=\left(g_*^{ac}\ell^b\ell^d+\ell^a\ell^cg_*^{bd}-g_*^{ab}\ell^c\ell^d-\ell^a\ell^bg_*^{cd}\right)\gamma_{cd},
\end{equation}
from which we obtain the only non-zero component of the stress-energy tensor as
\begin{equation}
16\pi G S^{rr}=\frac{a^2}{r^2}\gamma_{\phi\phi}.
\end{equation}
Note that the result depends on the arbitrary normalization factor $a$ (which is not even assumed to be a constant). To obtain a result independent of this factor, we must compute the extrinsic stress-energy tensor as it appears in the Einstein equations.

\subsubsection{Extrinsic stress-energy tensor}
The extrinsic stress-energy tensor of the shell, $S^{\mu\nu}$, is related to the intrinsic one by $S^{\mu\nu}=S^{ij}e_i^\mu e_j^\nu$, but we will compute it using the algorithm outlined in \cite{Barrabes:1991ng,Musgrave:1997sfw}. We will for concreteness only consider the form of the stress-energy tensor in the coordinates below the shell. We first need to define a tensor $\gamma_{\mu\nu}$, whose projection onto the surface $\mathcal{L}$ is $\gamma_{ij}$. It is easy to see that $\gamma_{\mu\nu}$ is given by $\gamma_{\mu\nu}|_{\mu=\phi,\nu=\phi}=\gamma_{ij}|_{i=\phi,j=\phi}$ while all other components vanish. The extrinsic stress-energy tensor is then given by
\begin{equation}
-16 \pi G S^{\mu\nu}=2\gamma^{(\mu}n^{\nu)}-\gamma n^\mu n^\nu-\tilde\gamma g^{\mu\nu},
\end{equation}
where $\gamma^\alpha=\gamma^{\alpha\beta}n_\beta$, $\tilde{\gamma}=\gamma^\alpha n_\alpha$ and $\gamma=\gamma_{\alpha\beta}g^{\alpha\beta}$. We see that $\gamma^\alpha=0$ and $\tilde{\gamma}=0$ while $\gamma=\gamma_{\phi\phi}r^{-2}$. Thus we obtain that the only non-zero component of $S^{\alpha\beta}$ is
\begin{equation}
16\pi G S^{rr}=\frac{a^2}{r^2}\gamma_{\phi\phi}.
\end{equation}
and we see indeed that we have $S^{\mu\nu}=S^{ij}e_i^\mu e_j^\nu$.\\
\lb
The full stress-energy tensor, as it shows up in the right hand side in the Einstein equations, is now given by
\begin{equation}
T^{\mu\nu}=\alpha S^{\mu\nu}\delta(F),
\end{equation}
where $F=0$ defines the shell and such that the normal vector is given by $n_\mu=\frac{1}{\alpha}\partial_\mu F$ (which defines $\alpha$). Choosing $F=v$ (and thus $\alpha=-1/a$) we obtain the only non-zero component as
\begin{align}
T^{rr}&=\frac{1}{16\pi G r}\left[1+(H')^2M-\frac{2H'''}{H'}+3\frac{(H'')^2}{(H')^2}\right]\delta(v)\nonumber\\
&=\frac{1}{16\pi G r}\left[1+(H')^2M-2\{H,\phi\}\right],\label{masslessTrr}
\end{align}
where $\{F(x),x\}$ is the Schwarzian derivative. This result is independent of the arbitrary normalization $a$, as expected. The dependence on the normalization $a$ in $S^{\mu\nu}$ can thus be understood as an ambiguity in the delta function for the stress-energy tensor as it appears in the right hand side of Einstein's equations. 

\subsubsection{Connection with colliding pointlike particles}
So far we have defined our thin shell spacetime by the function $H(\phi)$. We will now assume that this function is the result of a limit of solutions of colliding pointlike particles. In this case (see Appendix \ref{app4}) we obtain the simple result
\begin{equation}
T^{rr}=\frac{\theta(\phi)}{8\pi G r}\delta(v).\label{awesome}
\end{equation}
Recall that $\theta$ is the continuous distribution of the pointlike particles, namely such that $2E_i=\theta(\psi_i)d\phi$ for large finite $N$ in the limit. Thus the energy density computed by the thin shell formalism is proportional to the energy distribution of the pointlike particles, which is expected but is still a very non-trivial consistency check. We can also be more rigorous and compare this result with adding up an infinite number of stress-energy tensor contributions for a massless particle as given by \eqref{Tpointparticleml}. The shell is thus build by a large $N$ number of particles with energies $2E_i=\theta(\psi_i)d\phi$ at angles $\psi_i$, each having a stress-energy tensor given by \eqref{Tpointparticleml}. Letting $N\rightarrow\infty$ and replacing the sum by an integral over $\phi$, and using the relation $r=\sinh\chi$, it is easy to see that we again obtain exactly \eqref{awesome}.

\subsection{Massive shells}\label{massiveshellssec}

In this section we will use the Israel junction formalism for timelike shells, as outlined in \cite{Israel:1966rt,Musgrave:1995ka}, to compute the stress-energy tensor of the timelike thin shell spacetimes that were constructed from an infinite number of colliding massive particles. Conceptually this procedure is simpler than for the lightlike shells since the induced metric is non-degenerate, but the calculations will be more tedious.\\
\linebreak
The starting point is now the two metrics
\begin{equation}
\rd s^2=-f(r)\rd t^2+\frac{\rd r^2}{f(r)}+r^2\rd\phi^2,\label{metricads}
\end{equation}
and
\begin{equation}
\rd\bar s^2=-\bar f(\bar r)\rd\bar t^2+\frac{\rd\bar r^2}{\bar f(\bar r)}+\bar r^2\rd\bar\phi^2,\label{metricbh}
\end{equation}
where the barred quantities are in $\bar\Rs$ (outside the shell) and non-barred quantities are in $\Rs$ (inside the shell). We have also defined $f=1+r^2$ and $\bar f=-M+\bar r^2$. For the formation of a conical singularity (black hole), we have $M<0$ ($M>0$). The embedding of the shell, as was obtained in the pointlike particle construction or can be obtained by just assuming that each point on the shell follows a radial geodesic, is given inside the shell by
\begin{equation}
\frac{r}{\sqrt{f(r)}}=-\frac{R}{\sqrt{f(R)}}\sin t,\label{trajinside}
\end{equation}
or in terms of the proper time $\tau$ by
\begin{equation}
\begin{array}{ll}
r=-R\sin\tau, &\quad \tan t=\sqrt{f(R)}\tan \tau.\\
\end{array}\label{proptimeinside}
\end{equation}
Outside the shell it is given by
\begin{equation}
\frac{\bar r}{\sqrt{\bar{f}(\bar r)}}=\frac{\bar R}{\sqrt{\bar{f}(\bar R)}} \times\Big\{\begin{array}{cc} \cosh (\sqrt{M}\bar t),& M>0 \\ \sin(\sqrt{-M}\bar t), & M<0\\ \end{array}\label{trajoutside}
\end{equation}
or in terms of the proper time $\tau$ by
\begin{equation}
\begin{array}{ll}
\bar r=-\bar{R} \sin\tau, &\quad   \Bigg\{\begin{array}{cc} \coth (\sqrt{M} \bar t)=-\frac{\sqrt{\bar{f}(\bar R)}}{\sqrt{M}}\tan \tau,& M>0 \\ \tan (\sqrt{-M} \bar t)=\frac{\sqrt{\bar{f}(\bar R)}}{\sqrt{-M}}\tan \tau, & M<0\\ \end{array}\\
\end{array}\label{proptimeoutside}
\end{equation}
for two arbitrary functions $R$ and $\bar R$, which we will here use to {\it define} these spacetimes (so at this point, no reference to the other quantities that were defined or computed in the pointlike particle construction is needed). These two functions, together with how the coordinates are related when crossing the shell, completely specify the spacetime. The induced metric is
\begin{equation}
\rd s^2=-\rd\tau^2+\sin^2\tau h^2\rd\phi^2=-\rd\tau^2+\sin^2\tau\bar h^2\rd\bar \phi^2,\label{indmetricjunction}
\end{equation}
where $h^2=R^2+\frac{(\partial_{\phi} R)^2}{f(R)}$ and $\bar h^2=\bar{R}^2+\frac{(\partial_{\bar\phi} \bar R)^2}{\bar{f}(\bar{R})}$, and the angular coordinates when crossing the shell are related by
\begin{equation}
 \frac{\rd\bar\phi}{\rd\phi}=\frac{h}{\bar h},
\end{equation}
which must hold to make sure that the induced metric is the same from both sides (this is a necessary condition that must be satisfied for the junction formalism to be applicable, and we also showed that this follows in the pointlike particle construction). It might be convenient to specify $\bar R$ as a function of $\phi$ instead as of as a function of $\bar \phi$. In that case we can obtain $\bar h$ in the following way:
\begin{align}
\bar h^2=\bar{R}^2+\frac{(\partial_{\bar\phi} \bar R)^2}{\bar{f}(\bar{R})}=\bar{R}^2+\frac{(\partial_\phi \bar R)^2}{\bar{f}(\bar{R})}\frac{\bar h^2}{h^2}\Rightarrow \bar h=\frac{\sqrt{\bar{f}(\bar{R})}h\bar R}{\sqrt{h^2\bar{f}(\bar R)-(\partial_{\phi} \bar R)^2}}
\end{align}
We will show later in this section that the stress-energy tensor, as computed from the junction formalism, has no pressure, and that when computing $R$ and $\bar R$ from the pointlike particle construction, the energy density coincides with the density of the pointlike particles, as expected.\\
\linebreak
By using $\phi$ as the angular coordinate in the intrinsic geometry of the shell, a basis for the tangent vectors of the shell is given by
\begin{equation}
\begin{array}{ll}
e^\mu_\tau=u^\mu\equiv \dot{x}^\mu=(\dot t,\dot r,0), &\quad\quad \bar e^\mu_\tau=\bar u^\mu\equiv \dot{\bar{x}}^\mu=(\dot{ \bar{ t} },\dot{ \bar{ r}},0),\\
\end{array}
\end{equation}
\begin{equation}
\begin{array}{ll}
e^\mu_\phi\equiv\frac{\partial x^\mu}{\partial \phi}=(\partial_\phi t,\partial_\phi r,1), &\quad\quad \bar e^\mu_{\phi}\equiv\frac{\partial \bar{x}^\mu}{\partial \phi}=(\partial_{\phi} \bar t,\partial_{\phi}\bar r,\frac{h}{\bar h}),\\
\end{array}
\end{equation}
where $\dot{\color{white}{t}}$ means derivative with respect to the proper time. Note that $e_\tau$ and $\bar{e}_\tau$ also coincide with the velocities of the shell. We can also use $\bar \phi$ as the angular coordinate on the shell, and in that case we have
\begin{equation}
\begin{array}{ll}
e^\mu_{\bar \phi}=(\partial_{\bar \phi} t,\partial_{\bar \phi} r,\frac{\bar h}{h})=\frac{\bar h}{h} e^\mu_{\phi}, &\quad\quad \bar e^\mu_{\bar \phi}=(\partial_{\bar \phi} \bar t,\partial_{\bar \phi} \bar r,1)=\frac{\bar h}{h} \bar e^\mu_{\phi}.\\
\end{array}
\end{equation}
From now on we will let $'$ denote derivative with respect to $\phi$. For completeness, we will list all first and second derivatives of $r$, $t$, $\bar r$ and $\bar t$ with respect to $\phi$ and $\tau$. These are obtained from equations \eqref{trajinside}-\eqref{proptimeoutside} and are as follows: 
\begin{subequations}
\begin{align}
\dot{r}=-R\cos\tau,&\quad\quad \dot{\bar{r}}=\bar{R}\cos\tau,\label{dotrdotbarr}\\
r'=-R'\sin\tau,&\quad\quad \bar{r}'=-\bar{R}'\sin\tau,\\
\dot{t}=\frac{\sqrt{f(R)}}{f(r)},&\quad\quad \dot{\bar{t}}=\frac{\sqrt{\bar{f}(\bar R)}}{\bar{f}(\bar r)},\label{dottdotbart}\\
t'=\frac{\sin\tau\cos\tau RR'}{f(r)\sqrt{f(R)}},&\quad\quad \bar{t}'=\frac{\sin\tau\cos\tau \bar{R}\bar{R}'}{\bar{f}(\bar r)\sqrt{\bar{f}(\bar{R})}},\\
\ddot{r}=R\sin\tau,&\quad\quad \ddot{\bar{r}}=\bar{R}\sin\tau,\\
r''=-R''\sin\tau,&\quad\quad \bar{r}''=-\bar{R}''\sin\tau,\\
\ddot{t}=-\frac{\sqrt{f(R)}R^2\sin(2\tau)}{f(r)^2},&\quad\quad \ddot{\bar{t}}=-\frac{\sqrt{\bar{f}(\bar{R})}\bar{R}^2\sin(2\tau)}{\bar{f}(\bar{r})^2},\\
\dot{r}'=-R'\cos\tau,&\quad\quad t''=\frac{\sin\tau\cos\tau(-(RR')^2+(RR''+(R')^2)f(R))}{f(r)\sqrt{f(R)}^3},\\
\dot{\bar{r}}'=-\bar{R}'\cos\tau,&\quad\quad \bar{t}''=\frac{\sin\tau\cos\tau(-(\bar{R}\bar{R}')^2+(\bar{R}\bar{R}''+(\bar{R}')^2)\bar{f}(\bar{R}))}{\bar{f}(\bar{r})\sqrt{\bar{f}(\bar{R})}^3},\\
\dot{t}'=\frac{RR'}{f(r)\sqrt{f(R)}},&\quad\quad \dot{\bar{t}}'=\frac{\bar{R}\bar{R}'}{\bar{f}(\bar{r})\sqrt{\bar{f}(\bar{R})}},
\end{align}
\end{subequations}
Note that these equations are only defined if $\bar{R}^2\geq M$. This will always be the case when a conical singularity forms. When a black hole forms, it only holds if $\bar{R}$, which is the maximal radial position of the shell, is outside the horizon (which we will always assume). The normal vectors $n$ and $\bar{n}$, which will be spacelike, are defined by the normalization $n\cdot n=\bar n\cdot \bar n=1$ and that they are orthogonal to the tangent vectors of the shell. We will also adopt the standard convention that the normal vectors point from the inside (the \ads part) to the outside.  These conditions result in the normal vectors
\begin{equation}
n_\mu=\frac{R}{\sqrt{R^2+\frac{(R')^2}{f(R)}}}\left(R\cos \tau,\frac{\sqrt{f(R)}}{f(r)},\frac{R'}{\sqrt{f(R)}}\sin\tau\right)
\end{equation}
and
\begin{equation}
\bar{n}_\mu=\sqrt{\frac{R^2-\frac{(\bar{R}')^2}{\bar{f}(\bar{R})}+\frac{(R')^2}{f(R)}}{R^2+\frac{(R')^2}{f(R)}}}\left(\bar{R}\cos \tau,\frac{\sqrt{\bar{f}(\bar{R})}}{\bar{f}(\bar{r})},\frac{\bar{R}'}{\sqrt{\bar{f}(\bar{R})}}\sin\tau\right).
\end{equation}
We will also need the extrinsic curvatures, which are defined by
\begin{equation}
K_{ij}\equiv - n_\mu\left(\frac{\partial^2x^\mu}{\partial\xi^i\partial\xi^j}+\Gamma^\mu_{\nu\rho}\frac{\partial x^\nu}{\partial \xi^i}\frac{\partial x^\rho}{\partial \xi^j}\right).\label{Kij}
\end{equation}
The intrinsic stress-energy tensor $S_{ij}$ of the shell can now be computed from the junction conditions \cite{Misner:1974qy,Israel:1966rt,Musgrave:1995ka} as
\begin{equation}
[K_{ij}-\gamma_{ij}K]=-8\pi G S_{ij},\label{junctioncond}
\end{equation}
where $[X]=\bar{X}-X$, $\gamma_{ij}$ is the induced metric and $K$ is the trace of the extrinsic curvature. For simplicity we will use the coordinate $\phi$ for describing the intrinsic geometry on the shell, but it is possible to translate the results as a function of $\bar{\phi}$ by using the relation between $\bar\phi$ and $\phi$. Using \eqref{junctioncond} and our explicit form of the induced metric, \eqref{indmetricjunction}, the stress-energy tensor is then expressed in terms of the extrinsic curvatures as
\begin{align}
8\pi G S_{\tau\tau}&=-\frac{\bar{K}_{\phi\phi}-K_{\phi\phi}}{\sin^2\tau(R^2+\frac{(R')^2}{f(R)})},\label{Stautau}\\
8\pi G S_{\tau\phi}&=8\pi S_{\phi\tau}=-\bar{K}_{\tau\phi}+K_{\tau\phi},\label{Stauphi}\\
8\pi G S_{\phi\phi}&=-\sin^2\tau(R^2+\frac{(R')^2}{f(R)})(\bar{K}_{\tau\tau}-K_{\tau\tau}).\label{Sphiphi}
\end{align}
\linebreak
The Christoffel symbols computed from a metric of the form \eqref{metricads} are
\begin{align}
\begin{array}{lll}
 \Gamma^t_{tr}=\frac{f'(r)}{2f(r)},&\Gamma^r_{tt}=\frac{f'(r)f(r)}{2},&\Gamma^r_{rr}=-\frac{f'(r)}{2f(r)},\\
 \Gamma^r_{\phi\phi}=-rf(r),&\Gamma^\phi_{r\phi}=\frac{1}{r},&
\end{array}
\end{align}
with analogous expressions outside the shell. From this it is straightforward to compute the extrinsic curvatures, and we obtain that
\begin{equation}
K_{\tau\phi}=\bar{K}_{\tau\phi}=K_{\tau\tau}=\bar{K}_{\tau\tau}=0,
\end{equation}
which implies that the only non-zero component of the stress-energy tensor is $S_{\tau\tau}$ (the energy density). In other words, the stress-energy tensor is diagonal (no momentum flux in the angular direction) and has no pressure. This is consistent with the interpretation of the shell as composed of pointlike particles falling in radially. The only non-trivial extrinsic curvatures are $K_{\phi\phi}$ and $\bar{K}_{\phi\phi}$, which are given by
\begin{equation}
\frac{K_{\phi\phi}}{\sin \tau}=-\frac{R^2\sqrt{f}-R \left(\frac{R''}{\sqrt{f}}-\frac{R}{\sqrt{f}^3}(R')^2\right)+2\frac{(R')^2}{\sqrt{f}}}{(R^2+\frac{(R')^2}{f})^\frac{1}{2}}\label{Kphiphi}
\end{equation}
and 
\begin{align}
\frac{\bar{K}_{\phi\phi}}{\sin\tau}=&-\Bigg[\left(R^2+\frac{(R')^2}{f}\right)^2\frac{\sqrt{\bar{f}}}{\bar R}-\left(R^2+\frac{(R')^2}{f}\right)\frac{(\bar R')^2}{\sqrt{\bar f}\bar{R}}+RR' \frac{\bar R'}{\sqrt{\bar f}}\nonumber\\
&-(R^2+\frac{(R')^2}{f})\left(\frac{\bar R''}{\sqrt{\bar f}}-\frac{\bar R}{\sqrt{\bar f}^3}(\bar R')^2\right)+\left(\frac{R''}{\sqrt{f}}-\frac{R}{\sqrt{f}^3}(R')^2\right)\frac{R'}{\sqrt{f}}\frac{\bar R'}{\sqrt{\bar f}}\Bigg]\nonumber\\
&\times \frac{1}{(R^2+\frac{(R')^2}{f})^\frac{1}{2}(R^2+\frac{(R')^2}{f}-\frac{(\bar R')^2}{\bar f})^{\frac{1}{2}}}.\label{barKphiphi}
\end{align}
Here we are using a shorthand notation where $f=f(R)$ and $\bar{f}=\bar{f}(\bar R)$. We have written them in terms of $\bar R$ and $R$ to be consistent with the rest of the notation in this section, but it should be pointed out that the expressions take a simpler form when written in terms of the quantities $Z$ and $\bar Z$ which are used in Section \ref{limitsec}. An explicit expression where we have made the substitution $R=\sinh Z$ can be found in Appendix \ref{app_shellT}. The calculation of these curvatures are a bit lengthy but in principle straightforward, and can be done using symbolic manipulation softwares. The only non-zero component of the stress-energy tensor, $S_{\tau\tau}$, is then obtained from \eqref{Stautau}.\\
\linebreak
%
%
So far we have not assumed that this spacetime is composed of pointlike particles, but instead we took the metrics and the embedding of the shell, given by the functions $R$ and $\bar R$, as the definition. If we thus want a spacetime with a particular energy density and particular starting position of the shell (which are the two physically reasonable quantities to parametrize such a spacetime with), we would have to tune $\bar{R}$ such that $S_{\tau\tau}$ equals a particular desired profile as a function of $\phi$, which would requiring solving a relatively complicated second order ordinary differential equation with periodic boundary conditions. However, if we assume that the spacetime arises due to the pointlike particle construction developed in this paper, $\bar{R}$ can in principle be computed in terms of $\rho$. It can be proven that (see Appendix \ref{app_shellT})
\begin{equation}
\frac{\sqrt{\bar{f}(\bar{R})}}{\bar R}=\frac{1}{R}\left(\cosh Z\cos T-\cot \Phi\sin T\right),\label{barRrel2}
\end{equation}
where $T$ and $\Phi$ are the quantities introduced in Section \ref{limitsec}. Thus to obtain $\bar R$ in terms of $\rho$, we first solve the differential equations \eqref{odeP} and \eqref{odeT} with periodic boundary conditions to obtain $T$ and $\Phi$ in terms of $\rho$, and then use equation \eqref{barRrel2}. This looks quite involved and nothing indicates that it is possible to express $\bar R$ analytically in terms of $\rho$, but it can be proven (see Appendix \ref{app_shellT}) that the stress-energy tensor simplifies to the form
\begin{equation}
8\pi G S_{\tau\tau}^{\mathrm{point-particles}}=-\frac{\rho}{\sin\tau\sqrt{R^2+\frac{(R')^2}{f(R)}}}.\label{Stautaupp}
\end{equation}
This result makes sense intuitively, since the denominator is just the length element on the shell, and this thus represents the rest mass per unit length on the shell. However, we need to be careful with such naive interpretations, since to really know what we are talking about we must understand how $S_{ij}$ is related to the stress-energy tensor that shows up in the right hand side in Einsteins equations. We can then compare the result to the stress-energy tensor of a pointlike particle, equation \eqref{Tpointparticle}, which we will do later in this section. Note also that the result is positive, since $\tau<0$. This pointlike particle limit can be used as a useful tool for finding thin shell solutions with a particular energy profile, which seems to be easier than solving the differential equation for $\bar R$ arising from  \eqref{Stautau} directly. We also want to stress that even if $\rho$ is more relevant when specifying initial data, knowing $\bar R$ is still crucial if we want to understand how the spacetime after the shell is linked to the spacetime before the shell and exactly how the shell is embedded.\\
\linebreak
Now the Einstein equations look like
\begin{equation}
R^{\mu\nu}-\frac{1}{2}Rg^{\mu\nu}+\Lambda g^{\mu\nu}=8\pi G T^{\mu\nu}=8\pi G\alpha \delta(F(x^\mu))S^{\mu\nu},
\end{equation}
where $F=0$ determines the embedding of the shell and $\alpha$ is a function to be determined. The tensor $S^{\mu\nu}$ is defined by the relation
\begin{equation}
S^{\mu\nu}=e_i^\mu e_j^\nu S^{ij}.
\end{equation}
This ensures that $S_{ij}$ is the induced tensor of $S_{\mu\nu}$ on the shell and that $S_{\mu\nu}n^\mu=0$. The parameter $\alpha$ should be determined such that $\alpha\int \delta(F)\rd n=1$, meaning that $\alpha\delta(F)$ has the standard delta-function normalization when we integrate along the direction which is normal to the surface. By expanding $F$ in the normal direction around the shell, we thus see that $\alpha=|\partial_\mu F n^\mu|$. $F$ will be chosen as
\begin{equation}
F=t+\arcsin\left(\frac{r\sqrt{1+R^2}}{R\sqrt{1+r^2}}\right),\label{Feq1}
\end{equation}
by virtue of \eqref{trajinside}. This has the advantage that, in the lightlike limit where $R\rightarrow\infty$, we have $F\rightarrow t+\arctan(r)=v$, where $v$ is the standard infalling coordinate that is commonly used when describing \ads-Vaidya type spacetimes. Note that $\partial_\mu F\propto n_\mu$, and a straightforward computation shows that 
\begin{equation}
\alpha=\frac{\sqrt{R^2+\frac{(R')^2}{f(R)}}}{R^2\cos \tau}.\label{alphaeq}
\end{equation}
Since $S^{\tau\tau}$ is the only non-zero component, the stress-energy tensor in the $(t,r,\phi)$ coordinates is given by $S^{\mu\nu}=e_\tau^\mu e_\tau^\nu S^{\tau\tau}$. By using the relations \eqref{proptimeinside}, \eqref{dotrdotbarr} and \eqref{dottdotbart} we obtain
\begin{equation}
\begin{array}{cc}
 8\pi GT^{tt}=&\alpha \frac{f(R)}{f(r)^2}\delta(F(x^\mu))S^{\tau\tau},\\
8\pi G T^{rt}=&-\alpha \frac{R\cos\tau\sqrt{f(R)}}{f(r)} \delta(F(x^\mu))S^{\tau\tau},\\
 8\pi GT^{rr}=&\alpha R^2\cos^2\tau\delta(F(x^\mu))S^{\tau\tau}.\\\label{Tthinshell}
\end{array}
\end{equation}
If we use the relation \eqref{Stautaupp} to relate $S^{\tau\tau}$ to $\rho$, we obtain
\begin{equation}
\begin{array}{cc}
 8\pi GT^{tt}_{\mathrm{point-particles}}=&\rho\frac{f(R)}{\sqrt{f(r)}^5R r\cos t}\delta(F(x^\mu)),\\
 8\pi GT^{rt}_{\mathrm{point-particles}}=&-\rho \frac{\sqrt{f(R)}}{f(r)r} \delta(F(x^\mu)),\\
 8\pi GT^{rr}_{\mathrm{point-particles}}=&\rho \frac{\cos t\sqrt{f(r)} R}{r}\delta(F(x^\mu)),\\\label{Tthinshell2}
\end{array}
\end{equation}
where we also used the relations $\cos\tau=\sqrt{f(r)}\cos t$ and $r=-R\sin\tau$. \\
\subsubsection{Comparing to the stress-energy tensor of a pointlike particle}
We could also compute this stress-energy tensor by adding up the stress-energy tensor contributions of an infinite number of massive pointlike particles, by using the expression \eqref{Tpointparticle} (just as we did when we looked at massless shells). To build a thin shell, we place $N$ particles with such a stress-energy tensor at angles $\psi_i=2\pi i/N$, with masses and boost parameter given by $8\pi G m=\rho(\psi_i) d\phi$ and $\zeta_i=Z(\psi_i)$, where $d\phi=2\pi/N$. To translate to the quantities used in this section, we use the relations $\sinh\chi=r$, $\cosh\chi=\sqrt{f(r)}$, $\sinh Z=R$ and $\cosh Z=\sqrt{f(R)}$ to eliminate $Z$ and $\chi$. To transform the delta function in \eqref{Tpointparticle} to the one used here, we use
\begin{equation}
\delta(\tanh\chi+\tanh \zeta_i\sin t)=\frac{\sqrt{1+R^2}\delta(F(x^\mu))}{R\cos t}.
\end{equation}
We thus obtain that such a particle at angle $\phi=\psi_i$, with mass $m=\rho d\phi/8\pi G$, has stress-energy tensor
\begin{equation}
\begin{array}{cc}
8\pi G T^{tt}_i=&\rho d\phi\delta(\phi-\psi_i)\delta(F(x^\mu))\frac{f(R)}{\sqrt{f(r)}^5R r\cos t},\\
 8\pi GT^{\chi t}_i=&-\rho d\phi\delta(\phi-\psi_i) \delta(F(x^\mu))\frac{\sqrt{f(R)}}{\sqrt{f(r)}^3r},\\
 8\pi GT^{\chi\chi}_i=&\rho d\phi\delta(\phi-\psi_i)\delta(F(x^\mu))\frac{\cos t R}{r\sqrt{f(r)}},\\\label{Tthinshellpp}
\end{array}
\end{equation}
When taking the limit, we have $\sum_id\phi\delta(\psi_i-\phi)\rightarrow1$, and after transforming the tensor indices by using $\partial r/\partial\chi=\sqrt{f(r)}$, we again obtain \eqref{Tthinshell2}. This is a very non-trivial consistency check on the computations carried out in this chapter.

%

\subsubsection{The massless limit}
We will now take the massless limit, recovering the results in Section \ref{masslessshellssec}. Note that in the limit of the massive shells, we really obtain a shell that bounces at the boundary instead of being created at the boundary. However, from a physical perspective, the lightlike shells we studied in Section \ref{masslesscollsec} are more interesting and we will thus only consider the part of the shell at $t>0$. From an AdS/CFT point of view, the massless shells that are created at the boundary are more interesting since they correspond to an instantaneous quench in the dual field theory. The interpretation of the massive shells is less clear. It can be easily shown that in the limit where $R,\bar{R}\rightarrow\infty$, the spacetime takes the form
\begin{align}
\rd s^2=-f\rd v^2+2\rd v\rd r+r^2\rd\phi^2,&\hspace{10pt} v<0,\nonumber\\
\rd s^2=-\bar{f}\rd v^2+2\rd v\rd\bar{r}+\bar{r}^2\rd\bar{\phi}^2,&\hspace{10pt} v>0,
\end{align}
where again $f=1+r^2$ and $\bar{f}=-M+\bar{r}^2$. The relation between the coordinates when crossing the shell follows from \eqref{cs_anglemap} and \eqref{chimap} in the conical singularity derivation, and \eqref{bh_anglemap} and \eqref{betamap} in the black hole derivation, and takes the form
\begin{equation}
\frac{R}{\bar R}=\frac{r}{\bar r}= H'(\phi),\quad \bar{\phi}(\phi)= H(\phi),\label{RoverbarR}
\end{equation}
for some function $H(\phi)$, and we use $'$ to denote derivative with respect to $\phi$. This is the same relation that was obtained in \eqref{masslesscollsec} and ensures that the induced metric is well defined.\\
\lb
We can also reproduce the stress-energy for the massless shell from the result of the massive shell. As we explained in Section \ref{massiveshellssec}, the stress-energy tensor that shows up in the Einstein equations takes the form
\begin{equation}
16\pi G T^{\mu\nu}=16\pi G\alpha \delta(F(x^\mu))e^\mu_i e^\nu_j S^{ij}.\label{Tmunu}
\end{equation}
From \eqref{alphaeq} and \eqref{Feq1} it is clear that in the limit $R\rightarrow\infty$, we have $\alpha=1/R+O(1/R^2)$ and $F\rightarrow v$. Since the only non-zero component of $S$ is $S^{\tau\tau}$, and $\dot v\rightarrow0$, we obtain in the limit that the only component of the stress-energy tensor $T$ is
\begin{equation}
T^{rr}=\delta(v)\lim_{R\rightarrow\infty}\frac{\dot{r}^2 S^{\tau\tau}}{R},
\end{equation}
where $S^{\tau\tau}$ is given by
\begin{equation}
16\pi G S^{\tau\tau}=2\frac{\bar{K}_{\phi\phi}-K_{\phi\phi}}{\sin^2\tau(R^2+\frac{(R')^2}{f(R)})}.
\end{equation}
The extrinsic curvatures $K_{\phi\phi}$ and $\bar{K}_{\phi\phi}$ are given by \eqref{Kphiphi} and \eqref{barKphiphi}
Now we want to extract the leading behaviour of $S^{\tau\tau}$ when $R\rightarrow\infty$. Starting with the expressions for the extrinsic curvatures given by \eqref{Kphiphi} and \eqref{barKphiphi}, we eliminate $\tau$ via the relation $\sin\tau=-r/R$ (see equation \eqref{proptimeinside}) as well as express $\bar{R}$ and its derivatives in terms of $R$ and $H'$ using \eqref{RoverbarR}. A straightforward calculation results in
\begin{equation}
16\pi GS^{\tau\tau}=\frac{1}{Rr}\left((1+M(H')^2)+3\left(\frac{H''}{H'}\right)^2-2\frac{H'''}{H'}\right)+O(\frac{1}{R^2}).
\end{equation}
Note that the leading behaviour in $\bar{K}_{\phi\phi}$ and $K_{\phi\phi}$ will cancel out, and thus it is the first subleading quantities that are relevant. Now, since $\dot{r}=-R\cos\tau$, the stress-energy tensor in the Einstein equations, given by \eqref{Tmunu}, reads
\begin{align}
16\pi GT^{rr}=&\delta(v)\frac{1}{r}\left((1+M(H')^2)+3\left(\frac{H''}{H'}\right)^2-2\frac{H'''}{H'}\right)\nonumber\\
=&\delta(v)\frac{1}{r}\left(1+M(H')^2-2\{H,\phi\}\right).\label{Trrmassless}
\end{align}
where $\{F(x),x\}$ denotes the Schwarzian derivative. This result agrees with \eqref{masslessTrr} obtained using the junction formalism for massless shells directly.\\

\subsection{The stress-energy tensor of the dual CFT}\label{contcoord}
In the formulation of the thin shell spacetimes that we have presented so far, where the coordinates are discontinuous across the shell, the lightlike shells will have discontinuous boundary coordinates. This makes it difficult to make a connection with observables in the dual CFT. Inspired by this, we will construct a new coordinate system that is continuous at the boundary. In other words, the spacetime we have is not asymptotically \ads (since the boundary coordinates are not continuous) and to make connection to dual CFT observables we should make a coordinate transformation that brings the metric to the (asymptotically \ads) Fefferman-Graham gauge, which in particular will make the coordinates at the boundary continuous. This is achieved by applying a {\it large} diffeomorphism on the part of the spacetime which is after the shell. A large diffeomorphism (if applied to the whole spacetime) will typically change the physical state, and this is why the final spacetime after the shell should not really be interpreted as a pure BTZ black hole but rather a solution which has non-trivial Virasoro charges turned on (boundary gravitons). Thus the spacetime after the shell will now manifestly break rotational symmetry, and the breaking of rotational symmetry is no longer encoded in discontinuities across the shell. We will only compute the first few terms of this diffeomorphism, in an expansion close to the boundary, since obtaining the full coordinate transformation is quite difficult. This is enough to compute the energy density modes (that are dual to the stress-energy tensor in the dual CFT) in the resulting spacetime. Such a formulation is more natural from an AdS/CFT point of view and much more suitable when computing AdS/CFT observables. It is also very likely that such ``natural coordinates'' also exist for the massive shells. This is however expected to be more difficult since we do not have the discontinuity of the coordinates at the boundary to guide us, and we will not pursue this further in this thesis.\\
\linebreak
To be able to read off the stress-energy tensor modes easily, we will try to find a coordinate transformation that makes the coordinates continuous at the boundary as well as brings the metric for $v>0$ to the form
\begin{equation}
\rd s^2=\rd\rho^2+T_+(y_+)\rd y_{+}^2+T_-(y_-)\rd y_{-}^2-(e^{2\rho}+T_+(y_+)T_-(y_-)e^{-2\rho})\rd y_-\rd y_+.\label{TTmetric}
\end{equation}
As was pointed out in e.g. \cite{Banados:1998gg}, all metrics of the form \eqref{TTmetric} solve Einstein's equations for arbitrary $T_{\pm}$. According to the standard $AdS_3/CFT_2$ correspondence, the functions $T_\pm/8\pi G$ are identified with the stress-energy tensor modes of the dual CFT (see for instance \cite{Kraus:2006wn}).
Let us first try to get some intuition of what we expect from such a coordinate transformation. The relations between the radial coordinates and angular coordinates when crossing the shell are $\bar r=r/H'(\phi)$ and $\bar\phi=H(\phi)$, so this must be the boundary condition of the coordinate transformation on the shell ($v=0$) when $\bar r\rightarrow\infty$. In other words, if we try to define new coordinates $\tilde r$ and $\tilde \phi$ that make the coordinates continuous at the boundary, they must satisfy $\bar\phi=H(\tilde\phi)$ and $\bar r=\tilde r/H'(\tilde\phi)$ when $\bar r\rightarrow\infty$ and $v=0$. Now if we require a flat boundary metric for all times $v>0$, it can be easily shown that this will imply that the coordinate transformation takes the form
\begin{equation}
\bar\phi\pm v=H(\tilde\phi\pm v),\label{phibc}
\end{equation}
\begin{equation}
 \bar r=\tilde r/\sqrt{H'(v+\tilde{\phi})H'(v-\tilde{\phi})}, \label{rbc}
\end{equation}
when $\bar r\rightarrow\infty$. This will thus be our boundary conditions on our coordinate transformation for all times $v>0$.\\
\linebreak
To construct the full coordinate transformation that brings the metric to the form \eqref{TTmetric}, we will first do a change of coordinates such that the metric for $v>0$ takes the form
\begin{equation}
\rd s^2=\rd\rho_1^2+\frac{M}{4}(\rd x_{+}^2+\rd x_{-}^2)-(e^{2\rho_1}+\frac{M^2}{16}e^{-2\rho_1})\rd x_-\rd x_+.
\end{equation}
As can be easily verified, this is achieved by the following coordinate transformation.
\begin{equation}
\bar t=v+\frac{1}{\sqrt{M}}\text{arccoth}\left(\frac{1}{\sqrt{M}}e^{\rho_1}+\frac{\sqrt{M}}{4}e^{-\rho_1}\right),\hspace{10pt}\bar{r}=e^{\rho_1}+\frac{M}{4}e^{-\rho_1},
\end{equation}
and where $x_\pm=\bar t\pm\bar{\phi}$. For the region inside the horizon, arccoth is replaced by arctanh, and for the case of formation of a conical singularity, we replace it by arctan and $\sqrt{M}$ by $\sqrt{-M}$. This step is partially just going from our infalling coordinates back to our global time coordinates. Now we will construct the rest of the coordinate transformation enforcing the boundary conditions \eqref{phibc} and \eqref{rbc} that bring the metric to the form \eqref{TTmetric}. By expanding the coordinate transformation in an expansion close to the boundary, we obtain
\begin{align}
x_\pm=&H(y_\pm)+e^{-2\rho}\frac{H''(y_\mp)H'(y_\pm)}{2H'(y_\mp)}+e^{-4\rho}\left[\frac{(H''(y_\mp))^2H''(y_\pm)}{8(H'(y_\mp))^2}+M\frac{(H'(y_\mp))^2H''(y_\pm)}{8}\right]+O(e^{-6\rho}),\nonumber\\
e^{\rho_1}=&\frac{e^\rho}{\sqrt{H'(y_-)H'(y_+)}}-e^{-\rho}\frac{H''(y_-)H''(y_+)}{4\sqrt{H'(y_-)H'(y_+)}^3}+O(e^{-3\rho}),\label{coordtransf}
\end{align}
which is such that the metric takes the form
\begin{equation}
\rd s^2=\rd\rho^2+T(y_+)\rd y_{+}^2+T(y_-)\rd y_{-}^2-(e^{2\rho}+T(y_+)T(y_-)e^{-2\rho})\rd y_-\rd y_+,\label{TTmetric2}
\end{equation}
which is a special case of \eqref{TTmetric} where $T_+(x)=T_-(x)\equiv T(x)$. Of course, we only obtain \eqref{TTmetric2} up to higher orders in $e^{-\rho}$, but since we know that \eqref{TTmetric2} is a solution of Einstein's equations, we also know that the full coordinate transformation exists and can in principle be computed to any order. Finite forms of such coordinate transformations have been obtained in the literature (see for instance \cite{Barnich:2016lyg} and references therein), but for our purposes this is enough, since we are mainly interested in determining $T_\pm$ (however, to determine the full shape of the shell in the new coordinates as well as the map between the coordinates when crossing the shell, knowledge of the full transformation is required). The coordinate transformation \eqref{coordtransf} now implies that
\begin{equation}
T(x)=\frac{M}{4}(H'(x))^2-\frac{H'''(x)}{2H'(x)}+\frac{3}{4}\frac{(H''(x))^2}{(H'(x))^2}=\frac{M}{4}(H'(x))^2-\frac{1}{2}\{H,x\},
\end{equation}
which looks very similar to the result we obtained for the stress-energy tensor of the massless shell, equation \eqref{masslessTrr}. To be more precise, let us denote the stress-energy tensor modes in the dual CFT by $8\pi GT^{CFT}_\pm(y_\pm)=T(y_\pm)$ and write $8\pi G T^{rr}_{\text{shell}}=\delta(v)\rho_{\text{shell}}(\phi)/r$. By comparing with \eqref{masslessTrr}, we thus obtain
\begin{equation}
T_\pm^{CFT}(\phi)=\frac{1}{2}\left(\rho_\text{shell}(\phi)+M_{AdS}\right).
\end{equation}
The interpretation is thus as follows: The shell has a certain angular dependent energy density, directly determined by the angular dependence of the energy in the instantaneous quench in the dual CFT that sourced the shell. This energy density then directly determines the stress-energy tensor modes of the CFT after the quench such that the energy density of the CFT directly after the instantaneous quench is equal to the energy distribution of the shell plus $M_{AdS}$, the energy density of \ads. The energy density in the CFT will then attain a time dependence by splitting up into the right- and left-moving modes $T^{CFT}_+(y_+)$ and $T^{CFT}_-(y_-)$. Note also that the fact that the energy density just splits up into left- and right-moving modes is expected from the conformal symmetry of the boundary theory. As we discussed in \ref{ssec_weak} (and see in particular Figure \ref{Ttt_weak_inhom}), similar behaviour is expected to occur in higher dimensions, but the two left- and right-moving modes should then also be subject to dissipative effects. It is not clear how this analysis would work for the massive shells, and we will leave that as an interesting open problem.
\pagebreak
\section{Conclusions and outlook}
In this chapter we have studied particle collisions in three-dimensional AdS. We constructed solutions corresponding to an arbitrary number of pointlike particles (massive or massless) that all collide at a single point in \ads and merge into a single object. The pointlike particles can be constructed by excising a wedge (a piece of geometry) from \ads. For setups with (discrete) rotational symmetry these wedges can take the standard symmetric form, but for general setups they are not symmetric and must be chosen in a very specific way for the spacetime to be consistent. We also showed how to take the limit of an infinite number of particles. This resulted in new solutions corresponding to a thin shell which collapses to form a black hole (or a pointlike particle), where the shell in general is not rotationally symmetric. We also analyzed these thin shell spacetimes using the well-known junction formalism of general relativity and found that the stress-energy tensor computed in this way agrees with the stress-energy tensor of the pointlike particles. We also computed the boundary stress-energy tensor for the massless shells, which are dual to inhomogeneous energy injection in the boundary CFT. The solutions presented in this chapter add several more examples of analytical models of the formation of a black hole, which are in general quite hard to construct in general relativity. Moreover, the solutions here are the first examples of solutions with thin shells collapsing to black holes that break rotational symmetry.\\
\lb
The solutions we found in this chapter illustrate the power of the cutting-and-gluing techniques in three-dimensional gravity that we used, not only to construct solutions with pointlike objects but also to construct solutions with continuous matter. Further investigating such methods for more complicated setups is an interesting direction for future research. For example, one could consider particles that do not move along radial geodesics but along some arbitrary geodesic in \ads. Using such techniques it should be possible to construct thin shell spacetimes sourced by a boundary source that not only has an inhomogeneous energy profile, but also where the source turns on at different times along the boundary. Black holes forming from pointlike particles colliding in many stages (not only one collision point) can also be constructed if we allow the particles to move along non-radial geodesics. Another interesting extension, which would also require particles on non-radial geodesics, would be to construct ``thick'' shells, namely solutions generalizing the general \ads Vaidya spacetime (see equation \eqref{globalvaidya}). Such solutions can probably also be obtained by adding an infinite number of (inhomogeneous) thin shells. Other spacetimes that could be constructed by allowing non-radial geodesics include shells with pressure, and for the rotationally symmetric case such thin shell spacetimes exist (see for instance \cite{Erdmenger:2012xu,Keranen:2015fqa}). All thin shell spacetimes in this chapter are pressureless, and one could possibly add pressure by allowing non-radial movement of the particles inside the shell.\\
\lb
Obtaining a deeper understanding of the CFT dual of the spacetimes we constructed in this thesis is also a question of great importance, since it is not clear how physical the spacetimes we have constructed are. For example, one could look at non-local quantities, such as entanglement entropies. According to the Ryu-Takayanagi formula \cite{Ryu:2006bv} and its time dependent generalization \cite{Hubeny:2007xt}, entanglement entropies in the dual CFT can be computed by extremizing surfaces in the bulk. In \ads it reduces to studying spacelike geodesics and computing their lengths. This has been done for the simpler case of a rotationally symmetric \ads Vaidya spacetime \cite{Ziogas:2015aja}, as well as for Poincar\'e patch AdS Vaidya \cite{Balasubramanian:2011ur}, and we expect that similar techniques will also work for the non-rotationally symmetric thin shell spacetimes exhibited in this thesis (although the final results may have to be computed numerically). The geodesics would be constructed by patching together geodesics in empty \ads with geodesics in a black hole background and the gluing procedure will depend on the map between the coordinates across the shell. To obtain the final value of the entropy, we would have to go to the continuous coordinate system constructed in \ref{contcoord} to correctly regularize the geodesic length. Computing the thermalization of enganglement entropies can also be done using CFT techniques, as was done for the rotationally symmetric thermalization process (the dual of the \ads Vaidya spacetime) in \cite{Anous:2016kss} with perfect agreement with the gravity computation. Generalizing this computation to the non-rotationally symmetric case and comparing with gravity calculations using backgrounds obtained in this chapter would constitute another non-trivial check on the \ads/CFT$_2$ correspondence.\\
\lb
Studying collisions of charged particles is also an obvious extension. However, charging the particles under a normal $U(1)$ gauge field will spoil the topological character of the theory, and the cutting-and-gluing procedures will then not work. In Chapter \ref{higher}, although mostly focused on higher spin gravity, we will show how one can use the techniques in that chapter to construct a solution with charged colliding particles, where the particles are charged under a $U(1)$ Chern-Simons term (such that the theory still has no propagating degrees of freedom). Extending the techniques in Chapter \ref{higher} to construct solutions with colliding higher spin pointlike particles will be the focus of future research projects and is indeed the main motivation for developing the methods in that chapter.




\chapter{Black holes in higher spin gravity}\label{higher}


In this chapter we are going to step away from the main topic of this thesis, which has so far been focused on dynamical solutions and formation of black holes. Instead, we will now consider a theory called {\it higher spin gravity}, and we will focus on new ways to construct solutions in this theory (in particular generalizations of black holes). Although dynamics will not be the main focus in this chapter, we hope that the constructions we develop will be useful for such applications and will hopefully be the focus of future research. The idea is then that dynamical solutions in three-dimensional higher spin theories can be useful toy models for studying dynamics in field theories dual to such higher spin theories. We provide some hints to this by considering a simpler setup with gravity coupled to a $U(1)$ Chern-Simons field where it is possible to construct solutions corresponding to the formation of a charged black hole from collisions of charged pointlike particles.\\
\linebreak
Higher spin gauge theories are theories containing at least one massless field that has spin $5/2$ or higher. These are interesting from a theoretical point of view, but also because they are exepected to arise in the tensionless limit of string theory \cite{Gross:1988ue}. Higher spin theories in anti-de Sitter space are also interesting from the point of view of the AdS/CFT correspondence and several dualities between field theories and higher spin (gravity) theories have been proposed \cite{Klebanov:2002ja,Sezgin:2002rt,Giombi:2009wh,Giombi:2010vg,Bekaert:2012ux}. In higher dimensions (four or more), higher spin theories are very non-trivial, and there are many no-go theorems restricting what possible theories might exist (see for instance \cite{Bekaert:2010hw}). One of the biggest obstacles is that a consistent interacting higher spin theory must typically include an infinite number of higher spin fields. Currently only one complete description of such a higher spin theory is known, called Vasiliev theory after its discoverer \cite{Vasiliev:1990en,Vasiliev:2003ev}, and it must be defined in anti-de Sitter space. Note furthermore that higher spin gauge theories enjoy a very large set of gauge transformations which mix the higher spin fields and the metric essentially making standard geometrical quantities (defined using the metric) gauge dependent. \\
\linebreak
An exception to the above rule, that a consistent higher spin theory interacting with gravity must include an infinite tower of higher spin fields, occurs for certain topological theories in three dimensions. These theories are natural generalizations of three-dimensional gravity and can be formulated as a Chern-Simons theory with gauge group \sln{N}$\times$\sln{N} and with some particular boundary conditions. Such a theory will generically include higher spin fields with spin up to $N$ coupled to gravity, although the exact content of the theory depends on the boundary conditions we impose. These theories also support black hole solutions which are natural generalizations of the BTZ black hole but which carry extra charges \cite{Gutperle:2011kf,Ammon:2012wc,Perez:2014pya,Bunster:2014mua}. Note that due to the higher spin gauge transformations, the existence of a horizon is a gauge dependent statement and thus higher spin black holes are defined by other means. In this thesis we will only consider such three-dimensional topological theories, focusing on the case where $N=3$, and our main goal will be to explore a new method to define higher spin black holes that naturally generalizes the method of identification of points which we used in Chapter \ref{threed} to construct black hole solutions and conical singularities in three-dimensional gravity. Although we here only use these methods to construct the most simple static solutions, the hope is that such methods can be used in future research to construct more complicated dynamical situations such as the pointparticle collisions we studied in Chapter \ref{threed}. To this end, we will first review the Chern-Simons formulation of three-dimensional gravity as well as the method of identifying points in \ads (extending the method in Chapter \ref{threed} to rotating black holes and rotating conical singularities). Thereafter we will discuss higher spin gravity in three dimensions, focusing on the case where we only have a spin-2 and a spin-3 field. We will then discuss the standard construction of black hole solutions in these theories. We will then proceed to our new results (based on unpublished work together with Andrea Campoleoni and Xavier Bekaert), and show how these solutions can be obtained as orbifolds of a manifold extending \ads (meaning that \ads is a submanifold of the extended manifold) generalizing the construction of the BTZ black hole and conical singularities in three-dimensional gravity.
\pagebreak



\section{Three-dimensional gravity and its Chern-Simons formulation}
Certain aspects of three-dimensional gravity were introduced in Chapter \ref{threed} where we studied black hole formation from collisions of pointlike particles. That construction was possible due to the fact that gravity in three dimensions does not have propagating degrees of freedom (no gravitational waves). We will now explore a reformulation of three-dimensional gravity which makes this fact manifest. It turns out that three-dimensional gravity with negative cosmological constant\footnote{Gravity with vanishing or positive cosmological constant also have Chern-Simons formulations but with a different gauge group.} can be formulated as a Chern-Simons theory with gauge group $SO(2,2)$ \cite{Achucarro:1987vz,Witten:1988hc}. Since $SO(2,2)\cong$\sln{2}$\times$\sln{2}, we can also formulate it using two \sln{2} Chern-Simons connections. We will start with the $SO(2,2)$ theory, given by the action
\begin{equation}
I_{CS}[A]=\frac{k}{2\pi}\int \rd^3x\tr\left(A\wedge dA+\frac{2}{3}A\wedge A\wedge A\right).\label{so22CSaction}
\end{equation}
Here $k$ is called the Chern-Simons level and is related to the Newton constant in three dimensions by $k=L/4G$ where $L$ is the AdS radius. We will henceforth set $L=1$. Note also that $\tr$ just stands for a non-degenerate invariant bilinear form on the Lie algebra and is in general not unique. The Chern-Simons level $k$ is important for the quantum theory, and has to be equal to an integer to have a well defined path integral formulation, but for the classical theory (and our purposes) it plays no role. The equations of motion are
\begin{equation}
\rd A+A\wedge A=0.
\end{equation}
These equations of motion are equivalent to Einstein's field equations, and make it manifest that three-dimensional gravity has no propagating degrees of freedom. The connection to the standard variables in three-dimensional gravity is given by
\begin{equation}
A=\omega^aL_a+e^aP_a,
\end{equation}
where $L_a$ and $P_a$ are generators for $SO(2,2)$ with the commutation relations
\begin{equation}
[P_a,P_b]=\epsilon_{ab}^{\phantom{ab}c}L_c,
\end{equation}
\begin{equation}
[L_a,P_b]=\epsilon_{ab}^{\phantom{ab}c}P_c,
\end{equation}
\begin{equation}
[L_a,L_b]=\epsilon_{ab}^{\phantom{ab}c}L_c.
\end{equation}
Here $e^a=e^a_\mu \rd x^\mu$ is the vielbein and $\omega^a=\omega^a_\mu \rd x^\mu$ is the dualized spin connection defined in terms of the standard spin connection $\omega^{ab}=\omega^{ab}_\mu \rd x^\mu$ by $\omega^c=\frac{1}{2}\epsilon^{cab}\omega_{ab}$. We can now decompose this algebra into two copies of $\mathfrak{sl}(2,\R)$ as $Y^{\pm}_a=\frac{1}{2}(L_a\pm P_a)$. The commutation relations are then
\begin{equation}
[Y^+_a,Y^-_b]=0,\quad [Y_a^\pm,Y_b^\pm]=\epsilon_{ab}^{\phantom{ab}c}Y^\pm_c.
\end{equation}
We now write the $Y_i^\pm$ in block diagonal form and use the same generators for the two copies as
\begin{equation}
Y_a^+=\left(\begin{array}{cc}
           T_a&0\\
	   0&0\\
          \end{array}\right),\quad\quad Y_a^-=\left(\begin{array}{cc}
           0&0\\
	   0&T_a\\
          \end{array}
\right).
\end{equation}
Note that we use the convention $\epsilon_{012}=1$. In terms of these new generators, the Chern-Simons theory is reformulated using two \sln{2} connections $A_\pm$, and the action is
\begin{equation}
I=I_{CS}[A_+]-I_{CS}[A_-],\label{CSdiffaction}
\end{equation}
where the $\tr$ will now just mean the standard trace on the \sln{2} generators in some convenient representation to be specified below. Note that the relative factor between the two terms $I_{CS}[A_+]$ and $I_{CS}[A_-]$ depends on the choice of the non-degenerate bilinear form in \eqref{so22CSaction}. The (unintuitive) choice taken here has a relative sign between the two \sln{2} sectors such that the final action coincides with the Einstein-Hilbert action. The equations of motion are just the flatness condition of both connections, namely
\begin{equation}
\rd A_\pm+A_\pm\wedge A_\pm=0.\label{pmflatness}
\end{equation}
The map between the gauge connections $A_\pm$ and the metric variables are now given by
\begin{equation}
A_\pm^a=\omega^a\pm e^a.\label{Apmomegae}
\end{equation}
Note that the metric is obtained by $g_{\mu\nu}=2\textrm{tr}(e_\mu e_\nu)=e_\mu^a e_\nu^b 2\textrm{tr}(T_aT_b)$, where the $T_a$ is the basis we have chosen for the Lie algebra. This means that $e^a_\mu$ and $\omega_\mu^a$ can only be identified with the standard vielbein and spin connection in the gravity theory if $2\textrm{tr}(T_aT_b)=\eta_{ab}=\mathrm{diag}(-1,1,1)$. This is true for the basis
\begin{equation}
T_0=\left(\begin{array}{cc}
 0&-\frac{1}{2}\\
 \frac{1}{2}&0\\
\end{array}\right),\quad
T_1=\left(\begin{array}{cc}
 \frac{1}{2}&0\\
 0&-\frac{1}{2}\\
\end{array}\right),\quad
T_2=\left(\begin{array}{cc}
 0&\frac{1}{2}\\
 \frac{1}{2}&0\\
\end{array}\right),\label{theTbasis}
\end{equation}
which indeed also satisfies the commutation relations $[T_a,T_b]=\epsilon_{ab}^{\phantom{ab}c}T_c$.\\
\linebreak
Another common basis for the Lie algebra $\mathfrak{sl}(2,\R)$ is given by
\begin{equation}
L_{-1}=\left(\begin{array}{cc}
 0&0\\
 1&0\\
\end{array}\right),\hspace{20pt}
L_{0}=\left(\begin{array}{cc}
 -\frac{1}{2}&0\\
 0&\frac{1}{2}\\
\end{array}\right),\hspace{20pt}
L_{1}=\left(\begin{array}{cc}
 0&-1\\
 0&0\\
\end{array}\right),\hspace{20pt}\label{theLbasis}
\end{equation}
which satisfies the commutation relations $[L_i,L_j]=(i-j)L_{i+j}$. The relations between the two bases are
\begin{equation}
T_0=(L_{-1}+L_{1})/2,\quad T_1=-L_0,\quad T_2=(L_{-1}-L_1)/2.\label{TLrelations}
\end{equation}
\lb
We will now prove that the Chern-Simons description in this section is equivalent to three-dimensional gravity with a negative cosmological constant, by directly proving that the equations \eqref{pmflatness} are equivalent to the Einstein equations. By using equation \eqref{Apmomegae}, and the commutation relations for the basis $T_i$ we obtain that \eqref{pmflatness} reduces to the two equations
\begin{align}
\partial_\mu\omega_\nu^a-\partial_\nu\omega_\mu^a+(\omega^b_\mu\omega^c_\nu+e^b_\mu e^c_\nu)\epsilon^a_{\phantom{a}bc}&=0,\label{EEomega}\\
\partial_\mu e_\nu^a-\partial_\nu e^a_\mu+(\omega^b_\mu e^c_\nu+e^b_\mu \omega^c_\nu)\epsilon^a_{\phantom{a}bc}&=0.\label{EEtorsion}
\end{align}
We will now show that the second equation is equivalent to the zero torsion condition (which can be used to algebraically solve $\omega$ in terms of $e$) and then the first equation is equivalent to the Einstein equations with negative cosmological constant. By substituting $\omega^a$ for $\omega^{ab}$, the second equation can be written as
\begin{equation}
(\partial_{[\mu}e_{\nu]a}+e_{[\nu}^d\omega_{\mu]ad})e^\mu_be^\nu_c=0.
\end{equation}
Now by adding two of the cyclic permutations of $(a,b,c)$, and subtracting the other, we can then solve for $\omega^{ab}$ as
\begin{align}
2e^\mu_b\omega_{\mu a c}&+\partial_{[\mu}e_{\nu]a}e^\mu_be^\nu_c+\partial_{[\mu}e_{\nu]c}e^\mu_ae^\nu_b-\partial_{[\mu}e_{\nu]b}e^\mu_ce^\nu_a=0\nonumber\\
&\Rightarrow \omega_\nu^{ab}=e^{\nu b}e^a_\rho \Gamma_{\nu\mu}^\rho-e^{\nu b}\partial_\mu e^a_\nu.\label{omegaGamma}
\end{align}
This is the so called tetrad postulate (which states that the vielbein is covariantly conserved), but note that in the Palatini formalism where $\omega$ is treated as an independent field this is not postulated but instead follows from the action.\\
\lb
Now let us consider the first equation. We can define a curvature tensor in terms of the spin connection by
\begin{equation}
R_{\mu\nu}^{\phantom{\mu\nu}ab}\equiv\partial_\mu\omega_\nu^{\phantom{a}ab}-\partial_\nu\omega_\mu^{\phantom{a}ab}+\omega_{\mu}^{\phantom{a}ac}\omega_{\nu c}^{\phantom{\nu c}b}-\omega_{\nu}^{\phantom{\nu}ac}\omega_{\mu c}^{\phantom{\mu c}b}.
\end{equation}
This tensor, which is a function of $\omega$, is related to the standard Riemann tensor as a function of the Levi-Civita connection $\Gamma$, by
\begin{equation}
R_{\mu\nu a b}(\omega)=R^\alpha_{\phantom{\alpha}\tau\mu\nu}(\Gamma)e_{a\alpha}e^\tau_b,\label{RRrelation}
\end{equation}
if we assume the relation \eqref{omegaGamma} between $\omega$ and $\Gamma$ (see for instance \cite{van1981supergravity}). Equation \eqref{EEomega} can be written as
\begin{equation}
\partial_\mu\omega_\nu^{ab}-\partial_\nu\omega_\mu^{ab}+\omega^a_\mu\omega^b_\nu-\omega^b_\mu\omega^a_\nu+e^a_\mu e^b_\nu-e^b_\mu e^a_\nu=0.
\end{equation}
It can now be shown, by using $\omega^c=\frac{1}{2}\epsilon^{cab}\omega_{ab}$, that $\omega^a_\mu\omega^b_\nu-\omega^b_\mu\omega^a_\nu=\omega_\mu^{\phantom{\mu}ac}\omega_{\nu c}^{\phantom{\nu c}b}-\omega_\nu^{\phantom{\nu}ac}\omega_{\mu c}^{\phantom{\mu c}b}$, and we obtain
\begin{equation}
R_{\mu\nu}^{\phantom{\mu\nu}ab}+e^a_\mu e^b_\nu-e^b_\mu e^a_\nu=0.\label{RiemannEE}
\end{equation}
Now we use \eqref{RRrelation} and contract this equation with $e_{b\sigma}e_a^\mu$ to obtain
\begin{equation}
R^\mu_{\phantom{\mu}\sigma\mu\nu}+2g_{\sigma\nu}=0,\label{EEcontracted}
\end{equation}
which is the form of the Einstein equations with a cosmological constant $\Lambda=-1$, after the Ricci scalar has been eliminated by taking a trace (known as the ``trace reversed'' form of the Einstein equations). It should be pointed out that, at first sight, equation \eqref{RiemannEE} for the full Riemann tensor looks much stronger than the contracted equation \eqref{EEcontracted} for the Ricci tensor and one may wonder if some information is lost in this step. However, in three dimensions, the Riemann tensor is completely determined in terms of the Ricci tensor, and the contraction from \eqref{RiemannEE} to \eqref{EEcontracted} is thus actually invertible and these two equations are equivalent (but this would not be the case in higher dimensions). It is also possible to directly prove that the action \eqref{CSdiffaction} is equivalent to the Einstein-Hilbert action and it yields the relation
\begin{equation}
k=\frac{L}{4 G},\label{kGrelation}
\end{equation}
where $G$ is Newton's constant in three dimensions and we have restored the AdS radius $L$.

\subsection{Boundary conditions and asymptotic symmetry algebra}\label{ads3asa}

To specify the boundary conditions on the gauge fields $A^\pm$, we will factor out the radial dependence by writing $A^\pm=b_\pm^{-1}(\chi)a^\pm b_\pm(\chi)$ where $b_\pm=e^{\pm\chi L_0}$. The boundary conditions can then be specified in terms of $a^\pm$ as
\begin{align}
a_\phi^\pm&=L_{\pm1}-(\mathcal{L}_\pm+O(e^{-\chi}))L_{\mp1}+O(e^{-\chi}),\nonumber\\
a_\chi^\pm&=O(1).\label{HWgaugegrav}
\end{align}
at the boundary $\chi\rightarrow\infty$ and $\mathcal{L}_\pm$ is a function of $\phi$ and $t$. As was shown in \cite{Coussaert:1995zp}, this is equivalent to the standard Brown-Henneaux boundary conditions formulated using the metric formulation in \cite{Brown:1986nw}. Actually, most of the solutions we will consider take this form exactly (not only at the boundary) and with constant $\L(\phi)$. Given these boundary conditions, solutions with different $\mathcal{L}_\pm$ will be interpreted as different physical states, and only gauge transformations that leave these invariant are really true gauge transformations. Gauge transformations that change $\L_\pm$ are called {\it large} or {\it improper} gauge transformations and will change the physical state. The asymptotic symmetry algebra is defined as the Poisson algebra obeyed by the charges which generate the large gauge transformations (but leave the form of the boundary conditions \eqref{HWgaugegrav} invariant).
%
%
%
Starting with $a^\pm_\phi$, we thus want to apply a gauge transformation $\delta a^\pm_\phi=\partial_\phi \Lambda^\pm+[a_\phi^\pm,\Lambda^\pm]$, and then determine the parameters $\Lambda^\pm$ that leave the asymptotic form \eqref{HWgaugegrav} invariant. We start with the ansatz $\Lambda^\pm=\alpha_- L_{-1}+\alpha_0 L_0+\alpha_+ L_1$, giving
\begin{equation}
\delta a^\pm_\phi=\alpha_-' L_{-1}+\alpha_0' L_0+\alpha_+' L_1\pm\left(\alpha_\mp 2L_0+L_{\pm1} \alpha_0+\L_\pm\left( \alpha_\pm 2L_0+L_{\mp1} \alpha_0\right)\right),
\end{equation}
where $'$ is derivative with respect to $\phi$. We want to absorb this into only a variation of $\L_\pm$, which gives the equations 
$$\alpha_0'\pm2\alpha_\mp\pm2\L_\pm\alpha_\pm=0,$$
$$\alpha'_\pm\pm\alpha_0=0,$$
$$\alpha'_\mp\pm\L_\pm\alpha_0\equiv-\delta \L_\pm,$$
We can now let $\alpha_\pm=\epsilon_\pm$ to obtain the solution $\alpha_0=\mp\epsilon_\pm'$, $2\alpha_\mp=\epsilon_\pm''-2\L_\pm\epsilon_\pm$ and then read of that $\L_\pm$ will change as
\begin{equation}
\delta \L_\pm=2\epsilon_\pm' \L_\pm-\frac{1}{2}\epsilon_\pm'''+ \L_\pm'\epsilon_\pm,\label{Lpmvar}
\end{equation}
and the gauge parameter is
\begin{equation}
\Lambda^\pm=\epsilon_\pm L_\pm\mp\epsilon_\pm'L_0+\frac{1}{2}\left(\epsilon_\pm''-2\L_\pm\epsilon_\pm\right)L_\mp,\label{gaugeparamrel}
\end{equation}
Also requiring that the $a_t^\pm$ component is mapped into itself will require that $\mathcal{L}$ and $\epsilon_\pm$, namely that they are only a function of $t\pm\phi$. The transformation of $\L_\pm$ is the same as obtained for the transformation of the stress energy tensor under a conformal transformation in a CFT, and the boundary stress-energy tensor is given by $T_\pm=\frac{k}{2\pi}\L_\pm$\ \cite{Kraus:2006wn}. We will now outline how we can obtain the Virasoro algebra as a Poisson bracket algebra of the charges that generate the gauge transformation that leaves the asymptotic boundary conditions invariant, by using the Regge-Teitelboim method \cite{Regge:1974zd} (see also \cite{Banados:1994tn,Perez:2014pya,Benguria:1976in}). For simplicity we will now drop the $\pm$ index. In Hamiltonian form, the Chern-Simons action can be written as
\begin{equation}
I_H=\frac{k}{4\pi}\int \rd t\int \rd^2x\epsilon^{ij}g_{ab}(\dot{A}_i^aA_j^b+A_t^aF_{ij}^b),
\end{equation}
where $F_{ij}=\partial_iA_j-\partial_jA_i+[A_i,A_j]$ and these constraints generate the gauge transformations of $A_i$. To be more specific, we can define $G=\frac{k}{4\pi}\epsilon^{ij}F_{ij}$ and then the smeared gauge generators are given by
\begin{equation}
G(\Lambda)=\int\rd^2x\tr(\Lambda G)+Q(\Lambda),
\end{equation}
which generate gauge transformations via the Poisson bracket $\delta A_i=\{A_i,G(\Lambda)\}=-\partial_i\Lambda-[A_i,\Lambda]$ and where $Q$ is a boundary term to make the functional derivatives of $G$ well defined. The charges $Q$, which are constants of motion, can be found to be
\begin{equation}
Q(\Lambda)=-\frac{k}{2\pi}\int\rd\phi\tr(A_\phi\Lambda)=-\frac{k}{2\pi}\int\rd\phi\epsilon\mathcal{L},
\end{equation}
where we used the relations \eqref{HWgaugegrav} and \eqref{gaugeparamrel}. We still require that $\Lambda$ satisfies \eqref{gaugeparamrel} such that the boundary conditions are satisfied, meaning that even if $\L$ may change (in which case the gauge transformation is improper and the physical state changes) the new solution must still belong to the same space of functions. The charges satisfy $\delta_{\Lambda_1}Q(\Lambda_2)=\{Q(\Lambda_2),Q(\Lambda_1)\}$ and this can be used to evaluate the algebra of the gauge transformations. By expanding $\mathcal{L}$ in Fourier modes as
\begin{equation}
\mathcal{L}=\frac{1}{k}\sum_n \mathcal{L}_ne^{in\phi},
\end{equation}
and letting $\epsilon_1=e^{-ni\phi}$ and $\epsilon_2=e^{-mi\phi}$, we obtain by using \eqref{Lpmvar} that
\begin{equation}
i\{\mathcal{L}_m,\mathcal{L}_n\}=\mathcal{L}_{m+n}(m-n)+\frac{k}{2}\delta_{m+n,0}m^3.
\end{equation}
We thus conclude that the asymptotic symmetry algebra of the boundary conditions \eqref{HWgaugegrav} are given by two copies of a Virasoro algebra with central charge $c=6k=3L/2G$ (where we restored the AdS radius $L$ and used the relation \eqref{kGrelation}). This result was first obtained in \cite{Brown:1986nw} by using the metric formulation.\\

\subsection{Anti-de Sitter space in the Chern-Simons formulation}
Recall that \ads can be defined by the metric
\begin{equation}
\rd s^2=\rd \chi^2+\sinh^2\chi\rd \phi^2-\cosh^2\chi\rd t^2.\label{adschimetric}
\end{equation}
Let us directly compute the gauge connections $A^\pm$. We may factor \eqref{adschimetric} as
\begin{equation}
g_{\mu\nu}=e^a_\mu e^b_\nu \eta_{ab},
\end{equation}
where
\begin{equation}
e^a_\mu=\mathrm{diag}(\cosh\chi,-1,-\sinh\chi).
\end{equation}
Different signs in the components of the vielbein correspond to reversing the coordinates by $x^\mu\rightarrow-x^\mu$, and they are chosen such that the gauge connections we will later derive agree with conventions in the previous literature. The spin connection can now be computed as
\begin{equation}
\omega^{ab}_\mu=\frac{1}{2}e^{\nu[a}\left(e^{b]}_{\nu,\mu}-e^{b]}_{\mu,\nu}+e^{b]\sigma}e^c_\mu e_{\nu c,\sigma}\right).
\end{equation}
We can then compute the dualized spin connection $\omega^c=\frac{1}{2}\epsilon^c_{ab}\omega^{ab}$ which turns out to be given by
\begin{equation}
\omega=\left(\begin{array}{ccc}
              0&0&\cosh\chi\\
              0&0&0\\
              -\sinh\chi&0&0\\
             \end{array}\right),
\end{equation}
and thus the Chern-Simons connections $A^\pm=\omega\pm e$ are
\begin{equation}
A^\pm=\left(\begin{array}{ccc}
              \pm\cosh\chi&0&\cosh\chi\\
              0&\mp1&0\\
              -\sinh\chi&0&\mp\sinh\chi\\
             \end{array}\right).
\end{equation}
In the matrix notation above, the vertical indices are the local Lorentz indices (lower case Latin letters) or equivalently the Lie algebra indices corresponding to the basis $T_a$ given by \eqref{theTbasis}, and the horizontal are the curved indices (Greek letters). The order of the coordinates is $x^\mu=(t,\chi,\phi)$. In terms of the basis $L_a$ (as given by \eqref{theLbasis} or \eqref{TLrelations}), the gauge connections for \ads take the form
\begin{align}
A_\phi^\pm&=\frac{1}{2}(e^{\chi}L_{\pm1}+e^{-\chi}L_{\mp1}),\nonumber\\
A_t^\pm&=\pm\frac{1}{2}(e^{\chi}L_{\pm1}+e^{-\chi}L_{\mp1}),\nonumber\\
A_\chi^\pm&=\pm L_0.\label{chernsimonsads}
\end{align}
The metric can be restored by evaluating
\begin{equation}
g_{\mu\nu}=\frac{1}{2}\tr((A^+_{\mu}-A^-_{\mu})(A^+_{\nu}-A^-_{\nu})).
\end{equation}
Often in the literature, a different gauge is used where
\begin{align}
A_\phi^\pm&=e^{\chi}L_{\pm1}-\mathcal{L}_\pm e^{-\chi}L_{\mp1},\nonumber\\
A_t^\pm&=\pm (e^{\chi}L_{\pm1}-\mathcal{L}_\pm e^{-\chi}L_{\mp1}),\nonumber\\
A_\chi^\pm&=\pm L_0,\label{Lgauge}
\end{align}
which also makes it manifest that \ads satisfies the boundary conditions \eqref{HWgaugegrav} (it is actually equal to \eqref{HWgaugegrav} not only when $\chi\rightarrow\infty$, but for all $\chi$. This will actually be true for all solutions that we consider). To go to this gauge, one applies a gauge transformation of the form $A_\pm\rightarrow e^{\mp L_0\log 2}A_\pm e^{\pm L_0\log 2}$, and we obtain that for \ads we have $\mathcal{L}_\pm=-1/4$.\\

\subsection{Relation with \ads as the group manifold \sln{2}}
An interesting question to ask is if there is a connection between the fact that three-dimensional gravity with negative cosmological constant can be formulated as an \sln{2}$\times$\sln{2} Chern-Simons theory and the fact that \ads is isomomorphic to the group manifold \sln{2} which we relied heavily upon in Chapter \ref{threed}. We will now explicitly establish a connection between these two pictures. Let us first write the gauge connections in terms of two group elements $g_\pm$, such that
\begin{equation}
A^\pm_\mu=g_\pm^{-1}\partial_\mu g_\pm.
\end{equation}
To find $g_\pm$, consider the following matrix
\begin{equation}
F=e^{yT_0}e^{x T_1}.
\end{equation}
We have that $F^{-1}\partial_x F=T_1$. Now consider $G\equiv F^{-1}\partial_yF=e^{-x T_1}T_0e^{x T_1}$. We have
\begin{equation}
\partial_x G=e^{-x T_1}[T_0,T_1]e^{x T_1}=e^{-x T_1}T_2e^{x T_1}
\end{equation}
\begin{equation}
\partial^2_x G=e^{-x T_1}[T_2,T_1]e^{x T_1}=e^{-x T_1}T_0e^{x T_1}=G
\end{equation}
This differential equation has the solution $G=A\cosh x+B\sinh x$ and by setting $x=0$ to determine the integration constants we obtain 
\begin{equation}
G=T_0\cosh x+T_2\sinh x=\frac{1}{2}(e^x L_{-1}+e^{-x}L_1)
\end{equation}
From this we can deduce that 
\begin{equation}
g_\pm=e^{(\pm t+\phi)T_0}e^{\mp\chi T_1}.\label{gpmads}
\end{equation}
Now to establish a connection to \ads viewed as the group manifold \sln{2} we simply consider the matrix $G\equiv g_+g_-^{-1}$. This group element takes the form
\begin{equation}
G=e^{(t+\phi)T_0}e^{-2\chi T_1}e^{(t-\phi)T_0},\label{Geq}
\end{equation}
which parametrizes a full \sln{2} manifold (and is similar in structure to the parametrization in Chapter \ref{threed} and \cite{Matschull:1998rv}). Now consider $E_\mu$ defined by
\begin{equation}
E_\mu=\frac{1}{2}G^{-1}\partial_\mu G.
\end{equation}
In terms of $g_\pm$ we obtain
\begin{equation}
E_\mu=\frac{1}{2}(G^{-1} \partial_\mu g_+ g_-^{-1}-G^{-1} g_- g_-^{-1}\partial_\mu g_-g_-^{-1})=\frac{1}{2}g_-(A^+_\mu-A^-_\mu)g_-^{-1}
\end{equation}
Thus we see that $E_\mu$ is not the same as $e_\mu$, but they are conjugate and the extra $g_-$ will cancel when taking the trace of any product. An equivalent way of phrasing this is that $E_\mu$ and $e_\mu$ are related by a Lorentz transformations, since Lorentz transformations act as $e_\mu\rightarrow he_\mu h^{-1}$. Thus in particular we have for the Killing metric $K_{\mu\nu}$ that
\begin{equation}
K_{\mu\nu}\equiv2Tr(E_\mu E_\nu )=2Tr(e_\mu e_\nu )=g_{\mu\nu}.\label{killingmetric}
\end{equation}
Thus we establish the connection between the Chern-Simons formulation and the \sln{2} formulation in the sense that the metric computed in the Chern-Simons formulation is the same as the Killing metric on the \sln{2} manifold. Of course this statement is true regardless of what coordinates we choose, although other coordinates might only parametrize a patch of \sln{2}. Moreover, the matrix $G$ can be expanded as
\begin{equation}
G=\cos t\cosh\chi-2\cos\phi\sinh\chi T_1+2\cosh\chi\sin t T_0-2\sin\phi\sinh\chi T_2,
\end{equation}
and thus corresponds to the embedding
\begin{equation}
\begin{array}{cc}
x^3=\cosh \chi \cos t, & x^0=\cosh \chi \sin t,\\
x^1=-\sinh \chi \cos \phi, & x^2=-\sinh \chi \sin \phi, \label{adscoordhigher}
\end{array}
\end{equation}
This should be contrasted with the parametrization \eqref{matrixembads} from which it differs only with a few signs (however note that a different basis of 2$\times$2 matrices is used in Chapter \ref{threed}). Note also that isometries of \ads, when described as the group manifold \sln{2}, are implemented by left and right multiplications as $G\rightarrow h_+Gh_-^{-1}$. These transformations are precisely equivalent to left multiplications of $g_\pm$ as $g_\pm\rightarrow h_\pm g_\pm$, which in the Chern-Simons formulation are just interpreted as redefinitions of $g_\pm$ which do not affect $A_\pm$. 

\subsection{Identification of points in AdS}\label{sec_ident}
In this section we will describe how to create spinning pointlike particles and rotating BTZ black holes by making identifications of points in \ads. This will extend some of the results in Chapter \ref{threed}, but we will now add angular momentum and emphasize the form of the solutions in terms of the Chern-Simons connections.
\subsubsection{Spinning conical singularity}\label{sec_scs}
A spinning pointlike particle can be obtained by identifying points with an isometry that, on top of the angular shift, also includes a shift in time. Let us construct this by excising a wedge between two planes $w_-$ and $w_+$ such that $w_-$ is mapped to $w_+$ by a rotation clockwise of angle $\alpha$. This means that when crossing $w_-$, our angular coordinate is mapped as $\phi\rightarrow\phi-\alpha$. We will also allow for a time shift $t\rightarrow t+\tau$ and let us define $x=\tau/\alpha$. By defining a new time coordinate $t'= t+x\phi$, the isometry acts trivially as $\phi\rightarrow\phi-\alpha$. The metric then turns into
\begin{align}
\rd s^2&=r^2\rd\phi^2+\frac{\rd r^2}{1+r^2}-(1+r^2)\rd t^2\nonumber\\
&= r^2\rd\phi^2+\frac{\rd r^2}{1+r^2}-(1+r^2)\rd(t'-x\phi)^2\nonumber\\
&=\left(r^2-(1+r^2)x^2\right)\rd\phi^2+2(1+r^2)x\rd\phi \rd t'-(1+r^2)\rd t'^2+\frac{\rd r^2}{1+r^2}.
\end{align}
This motivates us to introduce the new radial coordinate $r^2-(1+r^2)x^2=\rho^2\Rightarrow r^2=\frac{\rho^2+x^2}{1-x^2}$ and the $\rd r^2$ term becomes
\begin{equation}
\frac{\rd r^2}{1+r^2}=\frac{\rd\rho^2}{\rho^2+1+x^2+\frac{x^2}{\rho^2}}.
\end{equation}
The metric is now
\begin{align}
\rd s^2&=\rho^2\left(\rd\phi+\frac{x(1+\rho^2)}{(1-x^2)\rho^2}\rd t'\right)^2-\left(\rho^2+1+x^2+\frac{x^2}{\rho^2}\right)\frac{\rd t'^2}{(1-x^2)^2}+\frac{\rd\rho^2}{\rho^2+1+x^2+\frac{x^2}{\rho^2}}
\end{align}
Defining a new angular coordinate by $\phi'=\phi+\frac{x}{1-x^2}t'$ and then rescaling our time coordinate as $t'\rightarrow t'(1-x^2)$ we have

\begin{align}
\rd s^2&=\rho^2\left(\rd\phi'+\frac{x}{\rho^2}\rd t'\right)^2-\left(\rho^2+1+x^2+\frac{x^2}{\rho^2}\right)\rd t'^2+\frac{\rd\rho^2}{\rho^2+1+x^2+\frac{x^2}{\rho^2}}
\end{align}
The range of $\phi'$ is now $0\leq\phi'\leq\alpha$. To make the range again from 0 to $2\pi$ we first define $\beta=\alpha/2\pi$ and then rescale our coordinates as $\phi'\rightarrow\beta\phi'$, $\rho\rightarrow\rho/\beta$, $t'\rightarrow\beta t'$ to obtain (dropping the primes)

\begin{align}
\rd s^2&=\rho^2\left(\rd\phi+\frac{x\beta^2}{\rho^2}\rd t\right)^2-\left(\rho^2+(1+x^2)\beta^2+\frac{x^2\beta^4}{\rho^2}\right)\rd t^2+\frac{\rd\rho^2}{\rho^2+(1+x^2)\beta^2+\frac{x^2\beta^4}{\rho^2}}.\label{metricspinpp}
\end{align}
This is identical to a spinning BTZ black hole with negative mass if we identify 
\begin{equation}
M=-(1+x^2)\beta^2=-(\alpha^2+\tau^2)/(2\pi)^2,
\end{equation}
\begin{equation}
J=2x\beta^2=2\tau\alpha/(2\pi)^2,
\end{equation}
which thus is interpreted as a spinning conical singularity.\\
\linebreak
This identification can also be described in terms of group actions on the \sln{2} manifold. In terms of the group element $G$ parametrizing \sln{2} (see equation \eqref{Geq}), the map $\phi\rightarrow\phi-\alpha$, $t\rightarrow t+\tau$ can be written as
\begin{equation}
G\rightarrow h_+ G h_-^{-1},
\end{equation}
where $h_\pm=e^{(-\alpha\pm\tau)T_0}$. This can also be formulated in terms of right multiplications of $g_\pm$ as $g_\pm\rightarrow h_\pm g_\pm = e^{(-\alpha\pm\tau)T_0}g_\pm$.\\
\linebreak
The gauge connections corresponding to the metric \eqref{metricspinpp} are
\begin{align}
A_{\phi}^\pm&=\frac{(\alpha\mp \tau)}{4\pi}(e^{\chi}L_{\pm 1}+L_{\mp 1}e^{-\chi}),\nonumber\\
A_{t}^\pm&=\pm\frac{(\alpha\mp \tau)}{4\pi}(e^{\chi}L_{\pm1}+L_{\mp1}e^{-\chi}),\nonumber\\
A_\chi^\pm&=\pm L_0,
\end{align}
which can be obtained by just carrying out the coordinate transformations described above in the metric formulation. To compare with previous results in the literature, we would like to do a gauge transformation such that the leading first term becomes $\sim e^{\chi}L_{\pm1}$. This can be done by either shifting $\chi$ or by doing the following gauge transformation
\begin{align}
A^\pm_\phi\rightarrow&\lambda_\pm^{-1}A^\pm_\phi \lambda_\pm=e^{\chi}L_{\pm 1}+L_{\mp 1}e^{-\chi}[\frac{\alpha\mp \tau}{4\pi}]^2\nonumber\\
=&e^{\chi}L_{\pm 1}-L_{\mp 1}e^{-\chi}\frac{M\pm J}{4}\nonumber\\
=&e^{\chi}L_{\pm 1}-L_{\mp 1}e^{-\chi}\mathcal{L}_\pm,
\end{align}
where $\lambda=e^{\mp \log[\frac{\alpha\mp \tau)}{4\pi}]L_0}$. The spacetime now manifestly obeys the boundary condition \eqref{HWgaugegrav}.\\
\lb
Note that the coordinate/gauge transformations in this section do now work for the special case where $\alpha=\pm\tau$ which corresponds to extremal pointlike particles with $|J|=-M$.

\subsubsection{BTZ black hole}\label{sec_btz}
The BTZ black hole can also be obtained by identifying points under an isometry of AdS$_3$. The easiest way to do this is to first introduce a different set of coordinates for \ads such that the isometry becomes manifest. Recall that \ads is described by the embedding \eqref{embedding_eq}. Instead of the parametrization \eqref{adscoordhigher}, we can instead use
\begin{equation}
\begin{array}{cc}
x^2=r \sinh \phi, & x^3=-r \cosh \phi,\label{bhcoords1}\\
\end{array}
\end{equation}
for $x^2$ and $x^3$.
The other coordinates $x^1$ and $x^3$ can then, assuming $|x^1|\geq|x^0|$, be parametrized as
\begin{equation}
\begin{array}{cc}
x^1=-\sqrt{r^2-1} \cosh t, & x^0=\sqrt{r^2-1} \sinh t,\\
\end{array}
\end{equation}
which covers the region where $r>1$. For $r<1$, we can instead choose
\begin{equation}
\begin{array}{cc}
x^1=-\sqrt{1-r^2} \sinh t, & x^0=\sqrt{1-r^2} \cosh t,\label{bhcoords3}\\
\end{array}
\end{equation}
by assuming that $|x^1|\leq|x^0|$. The metric takes the form 
\begin{equation}
\rd s^2=\frac{1}{r^2-1}\rd r^2-(r^2-1)\rd t^2+\rd\phi^2r^2,
\end{equation}
which is the metric of a BTZ black hole. For $r>1$, it is more convenient to work with the radial coordinate $\chi$ defined by $\cosh\chi=r$. Note that this parametrization is not exactly the same as was used in Section \ref{bhsec}, but corresponds essentially to swapping $x^3$ and $x^0$ (or more precisely, doing a shift in $t$ by $\pi/2$ in the coordinates given by \eqref{adscoord}). This is another unfortunate (but trivial) difference in conventions arising from different conventions in the literature of the different research areas.\\
\linebreak
We can now identify points under the isometry that acts as a shift $\phi\rightarrow \phi-\alpha$, $t\rightarrow t+\tau$. By again introducing $t'=t+x\phi$ where $x=\tau/\alpha$ we obtain the metric
\begin{equation}
\rd s^2=\left(r^2-(r^2-1)x^2\right)\rd\phi^2+2(r^2-1)x\rd\phi \rd t'-(r^2-1)\rd t'^2+\frac{\rd r^2}{r^2-1}.
\end{equation}
We can now introduce the radial coordinate $\rho$ by $r^2-(r^2-1)x^2=\rho^2\Rightarrow r^2=\frac{\rho^2-x^2}{1-x^2}$ and the $\rd r^2$ term becomes
\begin{equation}
\frac{\rd r^2}{-1+r^2}=\frac{\rd\rho^2}{\rho^2-1-x^2+\frac{x^2}{\rho^2}}.
\end{equation}
The metric is now
\begin{align}
\rd s^2&=\rho^2\left(\rd\phi+\frac{x(\rho^2-1)}{(1-x^2)\rho^2}\rd t'\right)^2-\left(\rho^2-1-x^2+\frac{x^2}{\rho^2}\right)\frac{\rd t'^2}{(1-x^2)^2}+\frac{\rd\rho^2}{\rho^2-1-x^2+\frac{x^2}{\rho^2}}
\end{align}
Defining a new angular coordinate by $\phi'=\phi+\frac{x}{1-x^2}t'$ and then rescaling our time coordinate as $t'\rightarrow t'(1-x^2)$ we have
\begin{align}
\rd s^2&=\rho^2\left(\rd\phi'-\frac{x}{\rho^2}\rd t'\right)^2-\left(\rho^2-1-x^2+\frac{x^2}{\rho^2}\right)\rd t'^2+\frac{\rd\rho^2}{\rho^2-1-x^2+\frac{x^2}{\rho^2}}
\end{align}
To make the range of $\phi'$ again from 0 to $2\pi$ we first define $\beta=\alpha/2\pi$ and then rescale our coordinates as $\phi'\rightarrow\beta\phi'$, $\rho\rightarrow\rho/\beta$, $t'\rightarrow\beta t'$ to obtain (dropping the primes)

\begin{align}
\rd s^2&=\rho^2\left(\rd\phi-\frac{x\beta^2}{\rho^2}\rd t\right)^2-\left(\rho^2-(1+x^2)\beta^2+\frac{x^2\beta^4}{\rho^2}\right)\rd t^2+\frac{\rd\rho^2}{\rho^2-(1+x^2)\beta^2+\frac{x^2\beta^4}{\rho^2}}.\label{metricspinbh}
\end{align}
We can now identify the mass and angular momentum of the black hole as
\begin{equation}
M=(1+x^2)\beta^2=(\alpha^2+\tau^2)/(2\pi)^2
\end{equation}
\begin{equation}
J=-2x\beta^2=-2\tau\alpha/(2\pi)^2.
\end{equation}
The gauge connections in the ``black hole'' coordinates \eqref{bhcoords1}-\eqref{bhcoords3} take the form
\begin{align}
A^\pm_t=&\pm\frac{1}{2}(e^{\chi}L_{\pm1}-e^{-\chi}L_{\mp1}),\nonumber\\
A^\pm_\phi=&\frac{1}{2}(e^{\chi}L_{\pm1}-e^{-\chi}L_{\mp1}),\nonumber\\
A^\pm_\chi=&\pm L_0.
\end{align}
These can be written as $A_\mu^\pm=g_\pm^{-1}\partial_\mu g_\pm$ where 
\begin{equation}
g_\pm=e^{-(t\pm\phi)T_2}e^{\mp \chi T_1}.\label{gpmbh}
\end{equation}
The identifications then take the form $g_\pm\rightarrow h_\pm g_\pm = e^{(\pm\alpha-\tau)T_2}g_\pm$ and the group element $G=g_+g_-^{-1}$ is
\begin{equation}
G=e^{-(t+\phi)T_2}e^{-2\chi T_1}e^{(t-\phi)T_2}.\label{Geqbh}
\end{equation}
Furthermore, by doing the above identifications and coordinate transformations, the gauge connection for the BTZ black hole takes the form
\begin{align}
A^\pm_t=&\pm\frac{\alpha\mp\tau}{4\pi}(e^{\chi}L_{\pm1}-e^{-\chi}L_{\mp1}),\nonumber\\
A^\pm_\phi=&\frac{\alpha\mp\tau}{4\pi}(e^{\chi}L_{\pm1}-e^{-\chi}L_{\mp1}),\nonumber\\
A^\pm_\chi=&\pm L_0.\label{BTZinchernsimons}
\end{align}
Just as in the case of a spinning pointlike particle, we can now do a gauge transformation to make the coefficient of the leading term equal to one
\begin{align}
A^\pm_\phi\rightarrow&\lambda_\pm^{-1}A^\pm_\phi \lambda_\pm=e^{\chi}L_{\pm 1}-L_{\mp 1}e^{-\chi}[\frac{\alpha\mp \tau}{4\pi}]^2\nonumber\\
=&e^{\chi}L_{\pm 1}-L_{\mp 1}e^{-\chi}\frac{M\pm J}{4}\nonumber\\
=&e^{\chi}L_{\pm 1}-L_{\mp 1}e^{-\chi}\mathcal{L}_\pm,\label{btzgaugetransform}
\end{align}
where $\lambda=e^{\mp \log[\frac{\alpha\mp \tau)}{4\pi}]L_0}$. The spacetime now manifestly obeys the boundary condition \eqref{HWgaugegrav}.\\
\lb
Note that the coordinate/gauge transformations in this section do now work for the special case where $\alpha=\pm\tau$ which corresponds to extremal black holes with $|J|=M$.

\subsubsection{Eternal BTZ black hole}
\begin{center}
\begin{figure}
 \includegraphics[scale=.8]{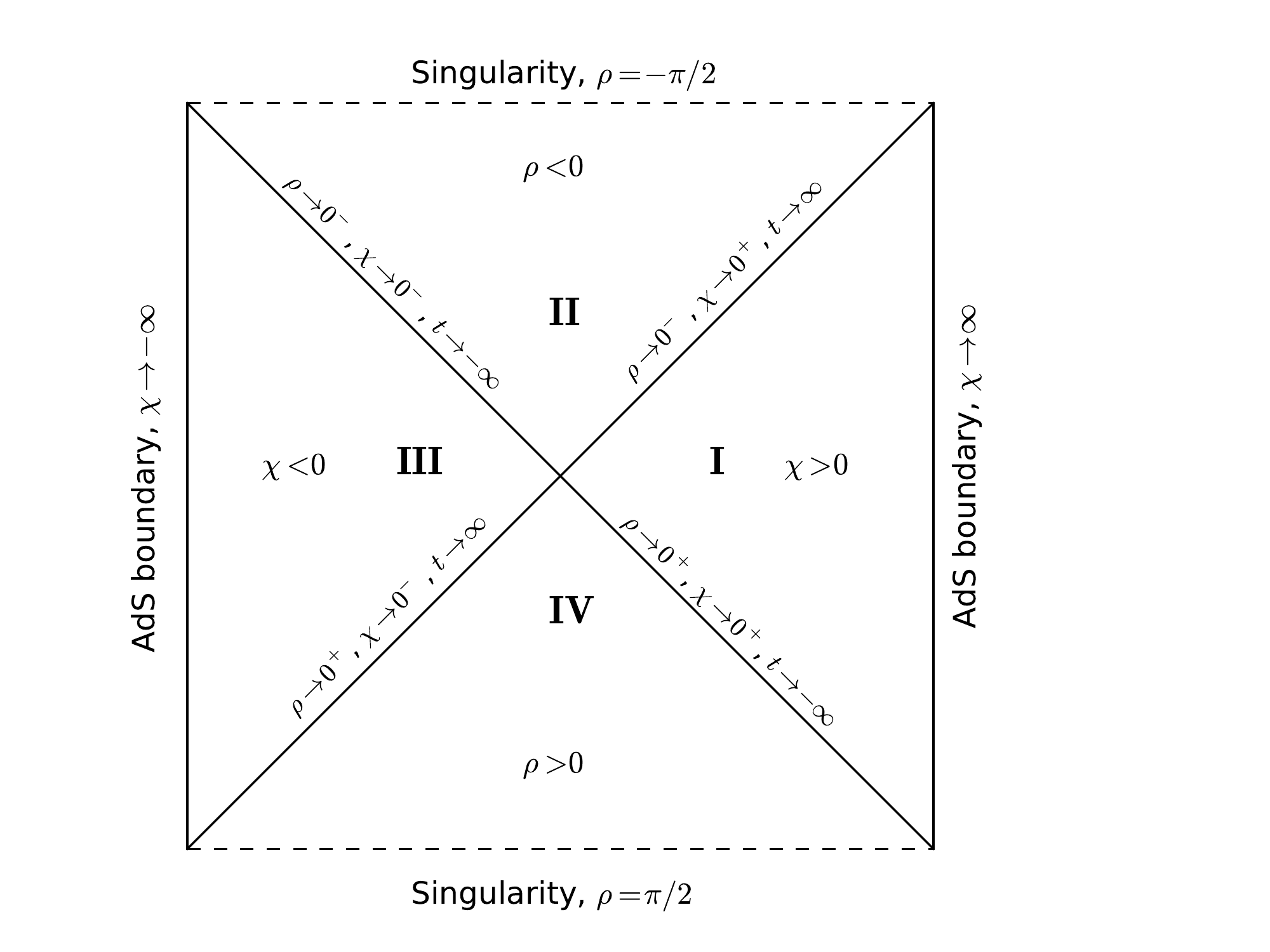}
 \caption{\label{penroseBTZ} Penrose diagram for the eternal static BTZ black hole.}
\end{figure}
\end{center}

In the previous subsection we parametrized (a submanifold of) \sln{2} by the group element \eqref{Geqbh}, however after doing the identification we only obtain the part of the BTZ black hole that is outside the horizon. Note also that this group element parametrizes two disconnected patches for $\chi\geq0$ and $\chi\leq0$, which will correspond to the two asymptotic regions of the eternal BTZ black hole. The two regions are connected at $\chi=0$, but the manifold here degenerates to a one-dimensional submanifold parametrized by $\phi$ only. However, the full boundary of the region is much larger than this one-dimensional manifold. To obtain the full boundary of the submanifold parametrized by \eqref{Geqbh}, we can take the limit $\chi\rightarrow0$ and $t\rightarrow\pm\infty$ while keeping $\chi e^t$ fixed. When $\chi\rightarrow0$ in this way, we do not obtain a one-dimensional submanifold, but rather obtain a two-dimensional manifold which will conincide with the event horizon after the identification. If we define a new coordinate by the limit $\chi \cosh t \rightarrow\lambda$, the group element \eqref{Geqbh} takes the form
\begin{equation}
G=e^{-\phi T_2}e^{-2\lambda (T_1\mp T_0)}e^{-\phi T_2},\label{grplmntbhhor}
\end{equation}
when $t\rightarrow\pm\infty$. We can now define another set of coordinates that connects to this two-dimensional submanifold. Consider for example
\begin{equation}
G=e^{-(t+\phi)T_2}e^{-2\rho T_0}e^{(t-\phi)T_2}.\label{grplmntbhnsd}
\end{equation}
By taking the limit $\rho\rightarrow0$ and $t\rightarrow\pm\infty$ while $\rho \sinh t \rightarrow -\lambda$, we again obtain \eqref{grplmntbhhor} and the coordinates \eqref{grplmntbhnsd} are thus connected to \eqref{Geqbh} via the horizon given by \eqref{grplmntbhhor}. This can be illustrated in the penrose diagram (for the static black hole) in Figure \ref{penroseBTZ} (for the rotating black hole, the singularity is replaced by the inner horizon and the spacetime can be extended further). We can also define a natural coordinate system that covers all regions. This coordinate system is similar to the well known Kruskal coordinate system. Note that the parametrizations \eqref{Geqbh} and \eqref{grplmntbhnsd} can be rewritten using the formulas
\begin{equation}
e^{-t T_2}e^{-2\rho T_0}e^{t T_2}=e^{-2\rho(\cosh t T_0-\sinh t T_1)},\quad e^{-t T_2}e^{-2\chi T_1}e^{t T_2}=e^{-2\chi(\cosh t T_1-\sinh t T_0)}.
\end{equation}
This motivates us to instead use the parametrization
\begin{equation}
G=e^{-\phi T_2}e^{-2(\lambda_1T_1+\lambda_0T_0)}e^{-\phi T_2},\label{grplmntbhlambda}
\end{equation}
which then covers all regions in \ref{penroseBTZ}. We see that region I corresponds to $\lambda_1>|\lambda_0|$, region II corresponds to $-\lambda_0>|\lambda_1|$, region III corresponds to $-\lambda_1>|\lambda_0|$ and region IV corresponds to $\lambda_0>|\lambda_1|$. Region II and IV also satisfy the constraint $\lambda_0^2-\lambda_1^2<\pi^2/4$. All regions are separated by null surfaces. 

\subsubsection{Chemical potentials and Euclidean formulation}

Black holes can also be defined in Euclidean signature where time has been Wick rotated by $t=-i\tau$. They are defined as spacetimes which have the topology of a solid torus, where the horizon is located at the center of the torus. To have a smooth geometry (and a well defined variational principle), one imposes that the torus has no conical singularity at the center (which is the horizon in Lorentzian signature). This results in a constraint on the periodicity in Euclidean time. In the Chern-Simons formulation, the gauge connections $A^\pm$ are replaced by one complex connection $A$ and the gauge group is replaced by $SL(2,\C)$. The regularity condition is phrased as the condition that the holonomy in Euclidean time should be trivial (namely that the integral of the gauge connection along the closed loop in the Euclidean time direction should have trivial eigenvalues). The time component of the gauge connection in Euclidean signature takes the form $A_\tau=-i\xi A_\phi$, and one can either choose to tune the periodicity of $\tau$, or to fix the periodicity and instead tune $\xi$ \cite{Bunster:2014mua}. The parameter $\xi$ is the chemical potential and is in general complex. To absorb $\xi$ into the periodicity of $\tau$, we start by assuming that the two coordinates $\phi$ and $\tau$ are periodic with periodicity $\tau\sim\tau+2\pi$ and $\phi\sim\phi+2\pi$. If we define new coordinates $\tau'=\tau \alpha/2\pi$ and $\phi'=\phi+\tau\ell/2\pi$, the coordinates are now instead identified under the identifications $(\tau',\phi')\sim(\tau',\phi'+2\pi)$ and $(\tau',\phi')\sim(\tau'+\alpha,\phi'+\ell)$. In terms of a single complex coordinate $z=\phi'+i \tau'$, the identification takes the form $z\sim z+2\pi\sim z+w$, where $w=\ell+i\alpha$. This can be interpreted as changing the modular parameter of the torus. It is easy to see that by choosing $\alpha=2\pi\mathrm{Re}\xi$ and $\ell=2\pi\mathrm{Im} \xi$, we can absorb the chemical potential $\xi$.\\
\linebreak
If we now Wick rotate back to Lorentzian signature, the two gauge connections $A_t^\pm$ take the form \cite{Bunster:2014mua}
\begin{equation}
A_t^\pm=\pm\xi_\pm A_\phi^\pm,\label{Atchempot}
\end{equation}
for two real chemical potentials $\xi_\pm$. This expression differs from \eqref{BTZinchernsimons} only with an overall factor. However, in Lorentzian signature, we do not interpret the time coordinate as periodic, thus we can now absorb $\xi_\pm$ completely without the cost of it showing up elsewhere in the periodicities of the coordinates. This means that in Lorentzian signature the chemical potentials have no physical meaning and can be gauged away. The specific coordinate transformation that absorbs $\xi_\pm$ is given by $t'=(\xi_++\xi_-)t/2$ and $\phi'=\phi+(\xi_+-\xi_-)t/2$. \\
\linebreak
In our formulation of creating the black hole as an identification of points of AdS$_3$, there is actually a much easier way to obtain the periodicity of the Euclidean time coordinate. The Euclidean version of \eqref{Geqbh} reads
\begin{equation}
G=e^{-(-i\tau+\phi)T_2}e^{-2\chi T_1}e^{(-i\tau-\phi)T_2}=\cosh \phi\cosh \chi-\sin \tau \sinh \chi 2 iT_0-\cos \tau \sinh \chi 2 T_1-\sinh \phi \cosh \chi 2 T_2.\label{Ghsbheucl}
\end{equation}
In this formulation the Euclidean time coordinate $\tau$ is already automatically periodic and we do not need to impose any additional conditions on the connections.


\subsubsection{Classification of solutions}

It is possible to classify what kind of different solutions can be obtained by identifying points in \ads. This is done by classifying the generators $\xi_\pm$ in $\mathfrak{sl}(2,\R)$, up to conjugation of matrices in \sln{2}, that generate the isometries used to do the identification. It can be shown that the black hole solutions and the conical singularity solutions belong to different classes, where the former have $\xi_\pm$ with real eigenvalues and the latter have $\xi_\pm$ with imaginary eigenvalues. Identifications where for example $\xi_+$ has real eigenvalues and $\xi_-$ has imaginary eigenvalues (or vice versa) will correspond to unphysical solutions where $|J|>|M|$. Degeneration of eigenvalues corrrespond to extremal solutions. The details of this procedure can be found in Appendix \ref{sl2class}. This classification was done previously in \cite{Banados:1992gq} by instead classifying generators in $SO(2,2)$ directly.

\subsection{Summary}
In this section we have considered some aspects of three-dimensional gravity in the Chern-Simons formulation, paving the way for generalizations including higher spin fields. We also provided a link between the Chern-Simons formulation and \ads viewed as the group manifold \sln{2}. We then constructed rotating black holes and rotating conical singularities by identifying points in \ads and showed how this process can be interpreted using the Chern-Simons language. Our goal of the remainder of this chapter is to generalize the concepts in this section to gravity coupled to higher spin fields.
\pagebreak

\section{Gravity coupled to a $U(1)$ charge}\label{sec_U1}
Before we look at the more complicated setup of gravity coupled to higher spin fields, we will first consider gravity coupled to a $U(1)$ Chern-Simons term and extend the orbifold techniques of pure gravity to this case. We will show that it is possible to construct black holes and pointlike particles charged under a $U(1)$ Chern-Simons term $a_\mu$ by doing identifications of points in an extended manifold, namely $\mathcal{M}=\mathcal{N}\times \R$ where $\mathcal{N}$ is a submanifold of \ads (possible equal to \ads itself). This example will work as a toy model before tackling the more non-trivial case of gravity coupled to a spin-3 field.\\
\linebreak
Let us define $g_0(t,\chi,\phi)$ to be a group element in \sln{2}$\times$\sln{2}, such that $A^0_\mu=g_0^{-1}\partial_\mu g_0$ is a gauge connection for (a patch of) \ads. $\partial_t$ and $\partial_\phi$ are two commuting Killing vectors, meaning essentially that $A_\mu$ is independent of $t$ and $\phi$. Typically we have in mind that $g_0=g_+\oplus g_-$, where $g_\pm$ is either given by \eqref{gpmads} and $(t,\chi,\phi)$ parametrizes all of \ads, or equal to \eqref{gpmbh} and where $(t,\chi,\phi)$ parametrizes a part of \ads which is natural for constructing the BTZ black hole. The manifold covered by the coordinates $(t,\chi,\phi)$ will be denoted by $\mathcal{N}$. Thus doing identifications along an isometry spanned by $\partial_t$ and $\partial_\phi$ will result in pointlike particle solutions or in black hole solutions. Note that $g_0$ can be written as a product of exponentials of generators of \sln{2}$\times$\sln{2}$\simeq SO(2,2)$, which we will collectively denote by $Q_i$, and $A_\mu=A_\mu^i Q_i$ is a linear combination of these generators. To incorporate $a_\mu$ on the same footing, we will define a new generator $K$ that satisfies $[Q_i,K]$. We then define a group element $g\in$\sln{2}$\times$\sln{2}$\times \R$ that satisfies
\begin{equation}
g^{-1}\partial_\mu g=A_\mu^iQ_i+a_\mu K.
\end{equation}
For $g=g_0$ we obviously have $a_\mu=0$. With the aim to carry out identifications on an extended manifold, we now extend our manifold with one extra coordinate $\theta$, and define $g$ as
\begin{equation}
g=e^{\theta K} g_0.
\end{equation}
For $\mu\in(t,\chi,\phi)$ nothing has changed and $a_\mu=0$ and $A_\mu=A^0_\mu$, but if we also consider $\mu=\theta$ we see that $A_\theta=0$ but $a_\theta=1$. This four dimensional manifold $\mathcal{M}=\mathcal{N}\times \R$ which is covered by the coordinates $(t,\phi,\chi,\theta)$, together with $A_\mu$ and $a_\mu$, will be our starting point for carrying out identifications that can be interpreted as adding non-trivial $U(1)$ charge.\\
\linebreak
As explained in Section \ref{sec_ident}, we can obtain new solutions (pointlike particles or black holes) by doing identifications of points under an isometry spanned by the Killing vectors $\partial_t$ and $\partial_\phi$, which amounts to a coordinate shift by for instance $\phi\rightarrow \phi-\alpha, t\rightarrow t+\tau$. Now let us add to this an extra shift also in the extra $\theta$ coordinate, such that the total Killing vector is some linear combination of $\partial_t$, $\partial_\phi$ and $\partial_\theta$. The map under which we do the identification, denoted by $I$, can thus be described as
\begin{equation}
I:\begin{array}{c}
   \phi\rightarrow\phi-\alpha\\
   t\rightarrow t+\tau\\
   \theta\rightarrow\theta+q\\
  \end{array}
\end{equation}
The total manifold $\mathcal{M}/I$ is now no longer of the simple form $\mathcal{N}\times \R$, and it is no longer possible to project down to the physical $\mathcal{N}\subset$\ads part by just ignoring the $\R$ part. However, the new manifold can still be written as a product manifold. The coordinate transformation for which this becomes manifest is very similar to the coordinate transformations used in Section \ref{sec_ident}. We thus define new coordinates $\theta'$ and $t'$ such that the map $I$ only acts in the $\phi$ coordinate. This can be done by for instance considering
\begin{equation}
\begin{array}{c}
   t'=t+\frac{\tau}{\alpha}\phi\\
   \theta'=\theta+\frac{q}{\alpha}\phi\\
      \phi'=\phi\\
      \chi'=\chi\\
      
  \end{array}
\end{equation}
Now the map $I$ acts trivially as just $\phi\rightarrow\phi-\alpha$. The manifold can now be written as $\mathcal{M}=\mathcal{N}'\times \R$, where $\mathcal{N}'$ is covered by the coordinates $(t',\phi',\chi')$ (with the identification $\phi'\sim\phi'-\alpha$) and $\R$ by $\theta'$. We can now again ignore the $\R$ part and consider $A_{\mu'}(t',\phi',\chi')$ and $a_{\mu'}(t',\phi',\chi')$ to be a solution of three-dimensional gravity coupled to a $U(1)$ Chern-Simons field (it is indeed easy to see that the resulting connections satisfy the flatness conditions). However, now we will have $a_{\mu'}\neq0$ for $\mu'\in(t',\phi',\chi')$, and we have
\begin{equation}
a_{\phi'}=a_\theta\frac{\partial \theta}{\partial\phi'}=-\frac{q}{\alpha}.
\end{equation}
$A_\mu$ can be computed using the exactly the same methods as in Section \ref{sec_ident}, and changing $\alpha$ and $\tau$ will result in different values of the mass and angular momentum of the resulting conical singularity or black hole. The difference in the present case is that the pointlike particle or black hole solution have also aquired a $U(1)$ charge given by
\begin{equation}
Q=\int_0^\alpha a_{\phi'}d\phi'=-q.
\end{equation}
Thus we see that this extra shift in the $\theta$ coordinate when identifying points in the extended manifold can be interpreted as charging the solution under a $U(1)$ Chern-Simons gauge field. In Section \ref{higher} we will extend these methods to higher spin gravity, where the extended manifold will be more non-trivial since in that case the extra generators will not commute with all generators of \ads.
\subsection{Collisions of charged particles}
The observation that adding a $U(1)$ Chern-Simons charge can be interpreted as an identification in an extended manifold can be very useful since some solutions that might not have managable analytical expressions in terms of gauge connections or metrics can be easily formulated in terms of identifications in the extended manifold. Examples of this include collisions of particles which we explored in Chapter \ref{threed}, and with the tools developed here such dynamical processes can be easily generalized to include charged particles. Since isometries of \ads only act within the \ads part, and do not touch the extended part of the manifold parametrized by the extra coordinate $\theta$, it is trivial to apply a boost to a static particle and obtain a moving one, and the charge will remain the same. The case of two colliding particles is also easy to obtain. Note that charge conservation is implemented by the fact that the total $U(1)$ holonomy must be preserved: If we collide two particles which have shifts in the extra coordinate of $q_1$ and $q_2$ when moving around their respective world lines, there will be a shift $q_3=q_1+q_2$ in $\theta$ when moving around the resulting object's world line. In general, it is trivial to add charge to any solution of three-dimensional gravity with any number of pointlike particles, by just adding a shift in the extra coordinate $\theta$ when moving around each particle's world line such that charge conservation holds at every intersection. Of course, formulating such solutions using metric variables and the three-dimensionsal manifold would be very difficult, but this is already the case for the collisions without charge that we constructed in Chapter \ref{threed}.

\pagebreak
\section{Higher spin gravity in three dimensions}
As we mentioned in the introduction, there is a class of three-dimensional higher spin theories that contain a finite number of fields. These theories are natural generalizations of the Chern-Simons description of three-dimensional gravity, and also do not contain any propagating degrees of freedom. They are obtained by generalizing the gauge group in the Chern-Simons formulation of three-dimensional gravity as $\sln{2}\times\sln{2}\rightarrow \sln{N}\times\sln{N}$ for some integer $N\geq 3$. Typically, the field content will be that of a metric (spin-2 field) coupled to a number of fields with varying spins. The spins of the extra fields will depend on how we embed the \sln{2} into \sln{N} and essentially corresponds to different ways of identifying the metric and the gravity sector. Generically, the extra fields can have a spin greater or equal to 5/2 (which is the definition of a higher spin field) but there are examples where the extra field content only has fields with lower spin. We will only look at the special case where $N=3$. In this case, there are two inequivalent embeddings of \sln{2} into \sln{3}, called the {\it principal} embedding and the {\it diagonal} embedding, and the choice of embedding corresponds to choosing different boundary conditions at the AdS boundary. The principal embedding can be interpreted as a theory with a spin-3 field non-minimally coupled to a spin-2 field (metric), and we will focus almost exclusively on this embedding. The diagonal embedding consists of a spin-2 field coupled to a scalar and two (bosonic) spin-3/2 fields.\\
\linebreak
The theory is specified by the action 
\begin{equation}
I[A_+,A_-]=I_{CS}[A_+]-I_{CS}[A_-],
\end{equation}
where
\begin{equation}
I_{CS}[A]=\frac{k}{2\pi}\int \rd^3x\tr\left(A\wedge dA+\frac{2}{3}A\wedge A\wedge A\right),
\end{equation}
but now the gauge connections take values in \sln{3}. The equations of motion are again flatness of the connections
\begin{equation}
\rd A_\pm+A_\pm\wedge A_\pm=0.
\end{equation}
The Lie algebra $\mathfrak{sl}(3,\R)$ can then be described using 8 generators $L_{\pm1}$, $L_0$, $W_{\pm2}$, $W_{\pm1}$ and $W_0$ which can be chosen as
\begin{align}
L_{-1}&=\left(\begin{array}{ccc}
         0&-\sqrt{2}&0\\
         0&0&-\sqrt{2}\\
         0&0&0\\
        \end{array}\right),\hspace{10pt}L_{0}=\left(\begin{array}{ccc}
         1&0&0\\
         0&0&0\\
         0&0&-1\\
        \end{array}\right),\hspace{10pt}L_{1}=\left(\begin{array}{ccc}
         0&0&0\\
         \sqrt{2}&0&0\\
         0&\sqrt{2}&0\\
        \end{array}\right),\nonumber\\
W_{-1}&=\left(\begin{array}{ccc}
         0&-\sqrt{2}&0\\
         0&0&\sqrt{2}\\
         0&0&0\\
        \end{array}\right),\hspace{10pt}W_{0}=\frac{2}{3}\left(\begin{array}{ccc}
         1&0&0\\
         0&-2&0\\
         0&0&1\\
        \end{array}\right),\hspace{10pt}W_{1}=\left(\begin{array}{ccc}
         0&0&0\\
         \sqrt{2}&0&0\\
         0&-\sqrt{2}&0\\
        \end{array}\right),\nonumber\\
        &W_{-2}=\left(\begin{array}{ccc}
         0&0&4\\
         0&0&0\\
         0&0&0\\
        \end{array}\right),\hspace{25pt}W_{2}=\left(\begin{array}{ccc}
         0&0&0\\
         0&0&0\\
         4&0&0\\
        \end{array}\right).\label{highspingen}
\end{align}
The commutation relations are
\begin{equation}
[L_i,L_j]=(i-j)L_{i+j},
\end{equation}
\begin{equation}
[L_i,W_m]=(2i-m)W_{i+m},
\end{equation}
\begin{equation}
[W_m,W_n]=-\frac{1}{3}(m-n)(2m^2+2n^2-mn-8)L_{m+n}.
\end{equation}
The $L_i$ generate the $\mathfrak{sl}(2,\R)$ algebra, and this choice corresponds to the principal embedding of \sln{2} in \sln{3}. If we had chosen the diagonal embedding we would have identified a different set of three matrices to generate the group \sln{2}.\\
\linebreak
This theory is invariant under a much larger gauge group than in the pure gravity case, corresponding to the gauge invariance of the two \sln{3} Chern-Simons actions. These {\it higher spin gauge transformations} mix the metric and the higher spin field, and thus geometrical quantities (such as distance, area and causality properties) defined using only the metric are no longer gauge invariant, and can thus not be considered sensible physical observables. The only really gauge invariant observables are holonomies obtained by integrating the gauge connection along a closed curve. There are also the surface charges at the boundary, which depend on the boundary conditions we choose. These charges will be invariant (by definition) under {\it proper} gauge transformations, and thus if we only allow for such gauge transformations the surface charges also become true physical observables.

\subsection{Boundary conditions and asymptotic symmetries}
There is a natural boundary condition which generalizes the Brown-Henneaux boundary condition \eqref{HWgaugegrav}. Assuming the boundary is located at $\chi\rightarrow\infty$, the boundary conditions can be specified as
\begin{align}
a_\phi^\pm&=L_{\pm1}-(\mathcal{L}_\pm+O(e^{-\chi}))L_{\mp1}-(\mathcal{W}_\pm+O(e^{-\chi})) W_{\mp2},\nonumber\\
a_\chi^\pm&=O(1),\label{HWgaugegravhs}
\end{align}
where $\mathcal{L}_\pm$ and $\mathcal{W}_\pm$ depend on $\phi$ and $t$. $A^\pm$ is then given in terms of $a^\pm$ as $A^\pm=b_\pm^{-1}(\chi)(\rd+a^\pm )b_\pm(\chi)$ with $b_\pm=e^{\pm\chi L_0}$. We will refer to this procedure as ``gauging away the radial dependence'', and will often just use $A^\pm$  on the form \eqref{HWgaugegravhs} instead of $a^\pm$. Here $\mathcal{L}_\pm$ and $\mathcal{W}_\pm$ are the asymptotic charges. The asymptotic symmetry algebra is then obtained by considering gauge transformations that leave the form of the boundary conditions invariant (although they might change the form of $\mathcal{L}_\pm$ and $\mathcal{W}_\pm$). Gauge transformations that change $\L_\pm$ and $\W_\pm$ are again called {\it improper} gauge transformations while {\it proper} gauge transformations leave these charges invariant. These boundary conditions are interesting since the asymptotic symmetry algebra of the charges satisfies the $W_3$ algebra with the same central charge as in gravity, and thus naturally generalizes the gravity result \cite{Campoleoni:2010zq,Henneaux:2010xg}. Specifically, one defines
\begin{equation}
\mathcal{L}=\frac{1}{k}\sum_n \mathcal{L}_ne^{in\phi},\quad \mathcal{W}=\frac{1}{4k}\sum_n \mathcal{W}_ne^{in\phi},
\end{equation}
one finds that the $\L_n$ and $\W_n$ satisfy the Poisson bracket algebra
\begin{align}
i\{\mathcal{L}_m,\mathcal{L}_n\}&=(m-n)\mathcal{L}_{m+n}+\frac{k}{2}\delta_{m+n,0}m^3,\nonumber\\
i\{\mathcal{L}_m,\mathcal{W}_n\}&=(2m-n)\mathcal{W}_{m+n}+\frac{k}{2}\delta_{m+n,0}m^3,\nonumber\\
i\{\mathcal{W}_m,\mathcal{W}_n\}&=\frac{1}{3}(m-n)(2m^2-mn+2n^2)\mathcal{L}_{m+n}+\frac{16}{3k}(m-n)\Lambda_{m+n}+\frac{k}{6}m^5\delta_{m+n,0}m^3,
\end{align}
where $\Lambda_n=\sum_m \mathcal{L}_{n-m}\mathcal{L}_m$.

\subsection{Higher spin black holes}\label{HSblackholes}
Generalizations of the BTZ black hole solution in three-dimensional gravity have been constructed in the literature \cite{Gutperle:2011kf,Ammon:2012wc,Perez:2014pya,Bunster:2014mua,Banados:2015tft} and we will mostly refer to the results in \cite{Bunster:2014mua}. The black hole solutions are usually defined in Euclidean signature, where we replace the two gauge connections $A^\pm$ by a single complex connection $A$ which now takes values in $SL(3,\C)$. The analytic continuation rules are given by \cite{Bunster:2014mua}
\begin{equation}
A^+=A,\quad A^-=-A^\dagger.
\end{equation}
The black hole solutions satisfy the boundary conditions \eqref{HWgaugegravhs} with constant $\L_\pm$ and $\W_\pm$ which in Euclidean signature are replaced by two complex $\L$ and \W. The gauge connections are written in Euclidean signature as
\begin{align}
A_\phi&=L_{1}-\mathcal{L}L_{-1}-\mathcal{W}W_{-2},\nonumber\\
A_\tau&=-i\left[\xi A_\phi +\eta \left(W_{2}+8\mathcal{W} L_{-1}+\mathcal{L}^2W_{-2}-2\mathcal{L} W_0\right)\right].\nonumber\\
A_\chi&= L_0.\label{HSbhEucl}
\end{align}
where we have gauged away the radial $\chi$ dependence. The parameters $\eta$ and $\xi$ are the chemical potentials, and are defined such that the the holonomy $\int_0^{2\pi}A_\tau\rd \tau$ in the compact Euclidean time coordinate is trivial. This results in two cubic equations relating $\L$ and $\W$ to $\xi$ and $\eta$. The precise relation will not be relevant to us and we refer to \cite{Bunster:2014mua} for more details. In Lorentzian signature, the black hole solution is
\begin{align}
A_\phi^\pm&=L_{\pm1}-\mathcal{L}_\pm L_{\mp1}-\mathcal{W}_\pm W_{\mp2},\nonumber\\
A_t^{\pm}&=\pm\left[\xi_\pm A_\phi^{\pm} +\eta_\pm \left(W_{\pm2}+8\mathcal{W}_\pm L_{\mp1}+(\mathcal{L}_\pm)^2W_{\mp2}-2\mathcal{L}_\pm W_0\right)\right].\nonumber\\
A_\chi^\pm&= L_0.\label{HSbhLor}
\end{align}
Recall that in pure gravity it was possible to absorb the chemical potential $\xi_\pm$ by a coordinate transformation involving $t$ and $\phi$ and they thus played no role in Lorentzian signature. However, it is not immediately clear whether or not we can get rid of $\eta_\pm$, but we will show later in Section \ref{sec_chempothigh} that we can indeed get rid of $\eta_\pm$ by applying a higher spin gauge transformation that can be interpreted as a coordinate transformation in an extended manifold. This exact question was asked in \cite{Henneaux:2013dra}.\\
\linebreak
The black hole solutions also satisfy the following inequality:
\begin{equation}
\mathcal{W}_\pm^2\leq \frac{4}{27}\mathcal{L}_\pm^3,\label{entropy_ineq}
\end{equation}
which in particular implies that $\mathcal{L}_\pm\geq0$. This inequality can be derived by imposing that the entropy is real and we refer to \cite{Bunster:2014mua} for a derivation along these lines. We will later prove this inequality when we define the higher spin black holes by identifying points in \sln{3}, and is a direct result of the particular conjugacy class to which the isometry we use to identify points belong to (and this conjugacy class naturally generalizes the conjugacy class in gravity used to construct the BTZ black hole). Note also that \eqref{entropy_ineq} generalizes the one for gravity where $\L_\pm\geq0$, which could be reformulated in terms of the mass $M$ and angular momentum $J$ as $M\geq|J|$. The conical singularity solutions instead satisfy $\L_\pm\leq0$, which could be reformulated as $M\leq-|J|$. Solutions which had $\L_+\geq0$ and $\L_-\leq0$ (or vice versa) were interpreted as unphysical solutions. By using the method of identifying points in \sln{3} to construct solutions of higher spin gravity, we will see that there are also a class of solutions which can be interpreted as conical singularities with higher spin charge. These solutions are natural generalizations of the normal conical singularities in gravity and always violate inequality \eqref{entropy_ineq}.
\pagebreak
\section{Higher spin gravity solutions by identifying points in \sln{3}}
The black hole solutions described in Section \ref{HSblackholes} are in many ways natural generalizations of the black hole solutions in three-dimensional gravity to three-dimensional higher spin gravity. However, the interpretation of them as black hole solutions relies on the formulation of gravity in the Chern-Simons formulation, and thus it is not clear how to identify black hole solutions in the metric formulation. In particular, we have seen in Section \ref{sec_ident} that in three-dimensional gravity we can construct black hole solutions (and pointlike particle solutions) by identifying points in \ads. We have also seen in Section \ref{sec_U1} that these techniques can be extended to gravity coupled to a $U(1)$ Chern-Simons field by considering identifications in an extended manifold. We will in this section show that similar techniques exist in higher spin gravity, where the identification of points also take place in a manifold extending \ads. The extended manifold will by definition be a submanifold of \sln{3} and the identifications will be natural extensions of the identifications done in \ads. We will first describe the general proposal and then illustrate the technique by explicitly constructing higher spin black hole solutions and solutions that we will interpret as pointlike particles (or conical singularities) with higher spin charge.
\subsection{The general prescription}\label{sec_cov}
Let us now describe the method we will employ to construct our extended manifold, carry out our identifications of points and then relate the resulting manifolds to solutions of three-dimensional higher spin gravity. It will naturally extend the procedure for the case of gravity coupled to a $U(1)$ field. We will first describe how one proceeds to interpret an identification of points of \sln{N}, taking $N=3$ for simplicity, as a solution to a Chern-Simons theory, as well as comment on some important subtleties, and then we will give a step by step procedure that can be used in practice.\\
\linebreak
The starting point is to pick an isometry $I$ of \sln{3} and form the quotient \sln{3}$/I$. To interpret this as a solution of a three-dimensional Chern-Simons theory we then pick out a three-dimensional slice of this eight-dimensional manifold and construct the Chern-Simons connections. Note that we are guaranteed that the flatness conditions are satisfied and different slices will correspond to different gauges of the higher spin gauge transformations, but we are not guaranteed that the Chern-Simons connections we then obtain satisfy the correct boundary conditions. In this picture, gauge transformations correspond to coordinate transformations (and frame rotations) on the full eight-dimensional manifold. However, a priori there is no way to distinguish between permissible and non-permissible gauge transformations at this point (in the sense that they preserve a given set of boundary conditions on the Chern-Simons connections). Distinguishing between permissible and non-permissible gauge transformations would force us to restrict our possible coordinates we are allowed to use to parametrize our \sln{3} manifold, or in other words imposing some criteria on the behaviour of the group element $G$ parametrizing the manifold. Note that the same issue exists for standard three-dimensional gravity defined as a Chern-Simons theory. In three-dimensional gravity, we can define the BTZ black hole without referring to boundary conditions or equations of motion, since we can just define it as a manifold obtained via a specific quotient of \sln{2}. Indeed, the existence of an event horizon is obviously independent of the equations of motion or what boundary condition we would like to impose in the Chern-Simons formulation. The fact that the BTZ black hole satisfies the Einstein's equations can be seen as an unexpected bonus that we might deem important.
\linebreak
Let us now give a step-by-step procedure to obtain the higher spin generalization of a solution in three-dimensional gravity. This procedure can be divided into the following steps:
\begin{enumerate}
 \item Choose an embedding of \sln{2} in \sln{N}, as well as an isometry $\partial_s$ of \sln{2}. Identifications of points in \sln{2} under this isometry should result in some desired solution of three-dimensional gravity for which we want to construct our higher spin gravity analog. 
 We will focus on \sln{3} where \sln{2} is principally embedded.
 \item Now choose $n$ number of extra isometries $\partial_{\theta_i}$ such that $[\partial_{\theta_i},\partial_s]=0$. We thus have that $n+2$ can be at most equal to twice the rank of \sln{N}, and will correspond to the number of non-trivial eigenvalues of the holonomy we will induce (in the principal embedding it is also equal to the number of charges for the appropriate boundary conditions but this is not necessarily the case for other embeddings). The criteria that the isometries commute with $\partial_s$ implies that any linear combination is also an isometry. We will denote the resulting manifold spanned by \sln{2} and  $(\theta_1,\ldots,\theta_n)$ by $\mathcal{Y}$. For \sln{3} we can only find two such extra isometries and in this case we will denote them by $\partial_\theta$ and $\partial_\psi$.
 \item Now form the quotient $\mathcal{Y}/I$, by identifying points under the isometry $I$ which is a linear combination of $\partial_s$ and the $\partial_{\theta_i}$. What we mean by quotient here is that we pick out a cell, a region spanned by surfaces that are mapped to each other by $I$, and then identify these surfaces by $I$ and ignore the rest of the manifold.
 \item Now we want to project this manifold on a three-dimensional manifold without the extra higher spin coordinates. This is easiest done by parametrizing our extended manifold by coordinates along the extra isometries $\theta_i$. By a linear coordinate transformation we can obtain a coordinate along our chosen isometry, and thus the isometry $I$ just acts trivially along this coordinate. It is then trivial to perform the identification and then dropping dependence on the other coordinates.
%
%
 \item Now reconstruct $g_\pm$ and $A_\pm$ from the relation $G=g_+g_-^{-1}$, where $G$ is the group element parametrizing the resulting manifold $\mathcal{M}_1/I$. Now $A_\pm$ will be gauge connections solving the flatness condition of three-dimensional higher spin gravity endowed with some non-trivial higher spin charge. However, it will still not necessarily satisfy the appropriate boundary conditions and we will typically have to apply another gauge transformation (that would be non-permissible from the Chern-Simons theory point of view). This just corresponds to moving our three-dimensional slice on which our solution is defined, and could have been done simultaneously with any of the above steps (in particular at the stage where we choose our coordinates on the manifold). However, it is not until we interpret our manifold as a solution of a Chern-Simons theory with prescribed boundary conditions that we know what coordinate system we should have picked and it thus makes more logical sense to do this step in the end.
\end{enumerate}
We will illustrate this technique in detail for constructing a higher spin black hole and a higher spin conical singularity in the \sln{3}$\times$\sln{3} Chern-Simons theory. However, note that it is not clear that at this point that the choice of embedding in this prescription has no meaning, and since we allow for very general gauge transformations in the end one may be able to make this choice redundant and map a solution obtained from a principally embedded gravity solution to a diagonally embedded one. 

\subsubsection{Example: Higher spin pointlike particle}\label{sec_HSPP}
In this section, we will show how to do identifications of points in a manifold extending \ads following the prescription in \ref{sec_cov}, such as to create pointlike particles with higher spin charge. We will focus on the principal embedding, thus the generators $L_0$, $L_1$ and $L_{-1}$ in \eqref{highspingen} will correspond to the \ads part embedded in the larger manifold. The group element spanning the \sln{2} submanifold is again given by \eqref{Geq}, namely 
\begin{equation}
G_{0}=e^{(t+\phi)T_0}e^{-2\chi T_1}e^{(t-\phi)T_0}.
\end{equation}
The gauge connections are given by $A_\mu^{\pm,0}=g_{\pm,0}^{-1}\partial_\mu g_{\pm,0}$, where
\begin{equation}
g_{\pm,0}=e^{(\pm t+\phi)T_0}e^{\mp\chi T_1}.
\end{equation}
We will now extend our manifold to more coordinates, where the extra coordinates should correspond to isometries commuting with $\partial_\phi$ and $\partial_t$. We thus multiply $g_\pm$ from the left, namely we define
\begin{equation}
g_\pm\equiv e^{-(\theta\pm\psi)\hat W}e^{(\pm t+\phi)T_0}e^{\mp \chi T_1}.
\end{equation}
So far we could interpret $\theta$ and $\psi$ as just some parameters parametrizing different forms of $g_\pm$, which do not affect $A_t$, $A_\phi$ and $A_\chi$. However, if we treat $\theta$ and $\psi$ as new coordinates and define $A_\theta^\pm\equiv g_\pm^{-1}\partial_\theta g_\pm$ and $A_\psi^\pm\equiv g_\pm^{-1}\partial_\psi g_\pm$, we obtain
\begin{equation}
A^\pm_\theta=-e^{\pm \chi T_1}e^{-(\pm t+\phi)T_0}\hat We^{(\pm t+\phi)T_0}e^{\mp \chi T_1},
\end{equation}
\begin{equation}
A^\pm_\psi=\pm A^\pm_\theta.
\end{equation}
For the extra coordinate directions to be isometries commuting with $\partial_\phi$ and $\partial_t$, it is enough that $\hat W$ commutes with $T_0$, and it can be seen that the only non-trivial matrix (up to an overall constant) commuting with $T_0$ is
\begin{equation}
\hat W\equiv 2W_0+W_2+W_{-2},\label{What}
\end{equation}
and from now on $\hat W$ will always refer to this matrix. Thus we obtain
\begin{equation}
A^\pm_\theta=-2W_0-e^{\pm \chi T_1}(W_2+W_{-2})e^{\mp \chi T_1}.
\end{equation}
To simplify this expression, we use the same trick as when we computed $G$ in the AdS case, namely by computing the derivatives and solving the resulting ODE. We find that
\begin{equation}
\partial_\chi A^\pm_\theta=\mp e^{\pm \chi T_1}([T_1,W_2]+[T_1,W_{-2}])e^{\mp \chi T_1}=\mp 2e^{\pm \chi T_1}(W_2-W_{-2})e^{\mp \chi T_1},
\end{equation}
and
\begin{equation}
\partial^2_\chi A^\pm_\theta=-4e^{\pm \chi T_1}(W_2+W_{-2})e^{\mp \chi T_1}=4A^\pm_\theta+8W_0.
\end{equation}
This has the solution 
\begin{align}
A^\pm_\theta&=-2W_0-(W_2+W_{-2})\cosh2\chi\mp(W_2-W_{-2})\sinh2\chi\nonumber\\
&=-2W_0-W_2e^{\pm2\chi}-W_{-2}e^{\mp2\chi},
\end{align}
and then $A^\pm_\psi$ is given by $A^\pm_\psi=\pm A^\pm_\theta$. Note also that the $G=g_+g_-^{-1}$, which parametrizes the extended manifold, is given by
\begin{equation}
G=e^{-(\theta+\psi)\hat{W}}e^{(t+\phi)T_0}e^{-2\chi T_1}e^{(t-\phi)T_0}e^{(\theta-\psi)\hat{W}},\label{sl3parametr}
\end{equation}
and is clearly some submanifold of \sln{3}.\\
\linebreak
We will now use this extended manifold to construct a pointlike particle with higher spin charge by identifying points under an isometry that is a linear combination of $\partial_\phi$, $\partial_t$, $\partial_\theta$ and $\partial_\psi$. Equivalently, just like the spinning pointlike particle can be constructed by excising a wedge and identifying the sides by a rotation and a shift in the time coordinate, we can do the same here but also allowing for a shift in the $\theta$ and $\psi$ coordinates when crossing the wedge. The identification, denoted by $I$, will thus be taken as
\begin{equation}
I:\begin{array}{c}
   \phi\rightarrow\phi-\alpha,\\
   t\rightarrow t+\tau,\\
   \theta\rightarrow\theta+q,\\
   \psi\rightarrow\psi+w.\\
  \end{array}\label{coordtrpp}
\end{equation}
This should be interpreted as $\phi$ having periodicity $\alpha$, but when going around in the $\phi$ coordinate there is a ``twist'' in the other coordinates. Just like in the $U(1)$ example, we now introduce new coordinates such that the isometry acts trivially and the manifold can be written as a product manifold. We thus do a change of coordinates such that
\begin{equation}
\begin{array}{c}
   t'=t+\frac{\tau}{\alpha}\phi,\\
   \theta'=\theta+\frac{q}{\alpha}\phi,\\
   \psi'=\psi+\frac{w}{\alpha}\phi,\\
      \phi'=\frac{2\pi}{\alpha}\phi,\\
      \chi'=\chi.\\
      
  \end{array}
\end{equation}
In these new coordinates the identification just acts as $\phi'\rightarrow\phi'-2\pi$. Doing this coordinate transformation, the $A^\pm_{\phi'}$ will become
\begin{align}
2\pi A^\pm_{\phi'}=&\alpha A^\pm_\phi-(q\pm w)A^\pm_\theta-\tau A_t^\pm \nonumber\\
=&\frac{\alpha\mp\tau}{2}(e^{\pm\chi}L_1+L_{-1}e^{\mp\chi})+(q\pm w)(2W_0+W_2e^{\pm2\chi}+W_{-2}e^{\mp2\chi}).\label{Aphipp}
\end{align}
We also have
\begin{equation}
2\pi A^\pm_{t'}=\pm\frac{\alpha\mp\tau}{2}(e^{\pm\chi}L_1+L_{-1}e^{\mp\chi}).
\end{equation}
Note the asymmetry between $A_{\phi'}^\pm$ and $A_{t'}^\pm$ which is not present in the pure gravity case. However, note also that the coordinate transformation \eqref{coordtrpp} is not unique and other choices will change $A_{t'}^\pm$.\\
\linebreak
We can now compute the holonomy of the charged conical singularity, which is given by
\begin{equation}
H^\pm=\int_0^{2\pi}A_{\phi'}^\pm \rd\phi'.
\end{equation}
The eigenvalues of $H^\pm$ are 
\begin{equation}
\lambda_i^\pm\in \left\{\frac{16}{3}(q\pm w),-\frac{8}{3}(q\pm w)-i(\alpha\mp\tau),-\frac{8}{3}(q\pm w)+i(\alpha\mp\tau)\right\}.
\end{equation}
We see that in general we have induced a non-trivial holonomy. The special case of AdS where the holonomy is trivial is given by $q=w=\tau=0$ and $\alpha=2\pi$.

\subsubsection{Example: Higher spin black hole}\label{sec_HSBH}
In this section we will apply the same technique to construct charged higher spin black holes. This is done by identifying points under a different isometry that extends the spacelike isometry used when constructing the standard BTZ black hole. Our starting point is now the matrix
\begin{equation}
G_0=e^{-(t+\phi)T_2}e^{-2\chi T_1}e^{(t-\phi)T_2},
\end{equation}
which parametrizes a submanifold of \sln{2}. The $\phi$ and $t$ directions are isometries and were used in \ref{sec_btz} to construct the BTZ black hole. The gauge connections are given by $A_\mu^{\pm,0}=g_{\pm,0}^{-1}\partial_\mu g_{\pm,0}$ where 
\begin{equation}
g_{\pm,0}=e^{-(t\pm\phi)T_2}e^{\mp \chi T_1}.
\end{equation}
Just like as in the case of the conical singularity, we now want to extend the manifold to two new coordinates $\theta$ and $\psi$ where $\partial_\theta$ and $\partial_\psi$ commute with $\partial_\phi$ and $\partial_t$. We do this by multiplying $g_{\pm,0}$ from the left and define
\begin{equation}
g_\pm=e^{-(\pm\psi+\theta)\bar W}e^{-(t\pm\phi)T_2}e^{\mp \chi T_1},
\end{equation}
with $\bar{W}$ commuting with $T_2$. The matrix $G=g_+g_-^{-1}$ parametrizing the extended manifold takes the form
\begin{equation}
G=e^{-(\psi+\theta)\bar W}e^{-(t+\phi)T_2}e^{-2 \chi T_1}e^{(t-\phi)T_2}e^{(-\psi+\theta)\bar W},\label{Ghsbh}
\end{equation}
which is a submanifold of \sln{3}. This does not have any effect on $A_t$, $A_\phi$, $A_\chi$, but if we treat $\theta$ and $\psi$ as new coordinates and define $A_\theta^\pm\equiv g_\pm^{-1}\partial_\theta g_\pm$ and $A_\psi^\pm\equiv g_\pm^{-1}\partial_\psi g_\pm$, we obtain
\begin{equation}
A^\pm_\theta=- e^{\pm \chi T_1}\bar We^{\mp \chi T_1},
\end{equation}
\begin{equation}
A^\pm_\psi=\pm A^\pm_\theta.
\end{equation}
It can be seen that there is only one non-trivial matrix $\bar W$ which commutes with $T_2$, and up to an overall constant it is given by
\begin{equation}
\bar W\equiv 2W_0-W_2-W_{-2}.\label{Wbar}
\end{equation}
Thus we obtain
\begin{equation}
A^\pm_\theta=-2W_0+ e^{\pm \chi T_1}(W_2+W_{-2})e^{\mp \chi T_1}.
\end{equation}
Now let us simplify this expression. Note that
\begin{equation}
\partial_\chi A^\pm_\theta=\pm e^{\pm \chi T_1}([T_1,W_2]+[T_1,W_{-2}])e^{\mp \chi T_1}=\pm 2e^{\pm \chi T_1}(W_2-W_{-2})e^{\mp \chi T_1},
\end{equation}
and that
\begin{equation}
\partial^2_\chi A^\pm_\theta= 4e^{\pm \chi T_1}(W_2+W_{-2})e^{\mp \chi T_1}=4A^\pm_\theta+8W_0.
\end{equation}
This differential equation has the solution
\begin{align}
A^\pm_\theta&=-2W_0+(W_2+W_{-2})\cosh2\chi\pm(W_2-W_{-2})\sinh2\chi\nonumber\\
&=- 2W_0+ W_2e^{\pm2\chi}+ W_{-2}e^{\mp2\chi}.
\end{align}
We will now do identifications of points in this extended manifold to construct a higher spin black hole, extending the procedure for the BTZ black hole. The identification takes the same form as in the conical singularity case (although the coordinates now have a different meaning), namely we consider the map $I$ defined by
\begin{equation}
I:\begin{array}{c}
   \phi\rightarrow\phi-\alpha,\\
   t\rightarrow t+\tau,\\
   \theta\rightarrow\theta+q,\\
   \psi\rightarrow\psi+w,\\
  \end{array}
\end{equation}
and just as in the conical singularity case, we introduce the new coordinates
\begin{equation}
\begin{array}{c}
   t'=t+\frac{\tau}{\alpha}\phi,\\
   \theta'=\theta+\frac{q}{\alpha}\phi,\\
   \psi'=\psi+\frac{w}{\alpha}\phi,\\
      \phi'=\frac{2\pi}{\alpha}\phi,\\
      \chi'=\chi,\\
      
  \end{array}\label{coordtrbh}
\end{equation}
such that the action of $I$ is purely in the $\phi'$ coordinate as $I: \phi'\rightarrow\phi'-2\pi$. We then obtain that
\begin{align}
2\pi A^\pm_{\phi'}=\frac{\alpha\mp\tau}{2}(e^{\chi}L_{\pm1}-L_{\mp1}e^{-\chi})+(q\pm w)(2W_0-W_2e^{\pm2\chi}-W_{-2}e^{\mp2\chi}).\label{Aphibh}
\end{align}
We also have
\begin{equation}
2\pi A^\pm_{t'}=\pm\frac{\alpha\mp\tau}{2}(e^{\chi}L_{\pm1}-L_{\mp1}e^{-\chi}).
\end{equation}
Again note that the coordinate transformation \eqref{coordtrbh} is not unique, and other linear coordinate transformations can result in different results for $A_t^\pm$. \\
\linebreak
We can now compute the holonomy around the $\phi'$ circle. It is given by
\begin{equation}
H^{\pm}=\int_0^{2\pi}A_{\phi'}^\pm \rd\phi'.
\end{equation}
The eigenvalues of this matrix are given by
\begin{equation}
\lambda_i^\pm\in \left\{\frac{16}{3}(q\pm w),-\frac{8}{3}(q\pm w)-(\alpha\mp\tau),-\frac{8}{3}(q\pm w)+(\alpha\mp\tau)\right\}.
\end{equation}
Note that compared to the conical singularity case, all eigenvalues are now real and the holonomy is thus always non-trivial. In Section \ref{sec_chempothigh} we will relate the solutions obtained here to the standard higher spin black hole solutions in the literature by writing them on the form \eqref{HWgaugegravhs}.

\subsubsection{Classification of solutions}
The different solutions that can be constructed using this procedure can be classified by classifying generators $\xi_\pm$ in $\mathfrak{sl}(3,\R)$ up to conjugations by matrices in \sln{3}. This is analogous to the classifications of generators in $\mathfrak{sl}(2,\R)$ that we did for the gravity case. It can be shown that the black hole solutions correspond to generators where $\xi_\pm$ has all real eigenvalues, while for the conical singularity $\xi_\pm$ has two complex (conjugate) eigenvalues and one real. The inequality \eqref{entropy_ineq} can then be derived by just using the properties of the eigenvalues, namely that they are all real and that their sum is zero (the generators are traceless). Analogously, for the conical singularities we can show that \eqref{entropy_ineq} is always violated by using that two eigenvalues are complex conjugates and that the sum of all eigenvalues is zero. Solutions where for instance $\xi_+$ has real eigenvalues and $\xi_-$ has two complex eigenvalues will be interpreted as unphysical. The details of this classification can be found in Appendix \ref{sl3class}.

\subsection{Some further comments}
\subsubsection{Gauge transformation to highest weight gauge}\label{sec_gaugetr}

We will not be interested in performing a gauge transformation such that the connections take the form
\begin{equation}
A_{\phi}^{'\pm}=L_{\pm1}e^{\chi}-\mathcal{L}_\pm L_{\mp1}e^{-\chi}-\mathcal{W}_\pm W_{\mp2}e^{-2\chi}.\label{finalgauge}
\end{equation}
This is the standard highest weight gauge used in the literature and manifestly obeys the boundary condition \eqref{HWgaugegravhs}. We thus want to map $A_\phi^\pm$, given by either \eqref{Aphipp} or \eqref{Aphibh}, to $A_{\phi}^{'\pm}$ given by \eqref{finalgauge}. To do this we will first factor out the radial dependence by writing 
\begin{equation}
A^\pm_\mu=b_\pm^{-1}(\chi)a^\pm_\mu b(\chi)+b(\chi)^{-1}\partial_\mu b(\chi),
\end{equation}
and the same for the primed connection, and the gauge transformation can be formulated as a transformation acting on $a_\mu^\pm$ as
\begin{equation}
a_\phi^{'\pm}= P_\pm a_\phi^{\pm} P_\pm^{-1}=L_{\pm1}-\mathcal{L}_\pm L_{\mp1}-\mathcal{W}_\pm W_{\mp2}.\label{gaugetr}
\end{equation}
The precise form of $P_\pm$ depends on whether we are dealing with a conical singularity or a black hole. In the case of the black hole, we have 
\begin{equation}
a_\phi^\pm=\frac{\alpha\mp\tau}{4\pi}(L_\pm-L_\mp)+\frac{q\pm w}{2\pi}(2W_0-W_2-W_{-2}).
\end{equation}
For simplicity we let $\sigma_\pm=\frac{q\pm w}{\alpha\mp\tau}$ and $\kappa_\pm=\frac{\alpha\mp\tau}{4\pi}$. It can then be shown that the gauge transformation is given by $P_\pm=e^{\mp\log \kappa_\pm T_1} S_\pm(\sigma_\pm)$, where 
\begin{equation}
S_+(\sigma)=\left(\frac{1}{1-64\sigma^2}\right)^{2/3}\left[
\begin{array}{ccc}
\frac{1}{3}(3+64\sigma^2)&\frac{4}{9}\sqrt{2}\sigma(-21+64\sigma^2)&\frac{512}{9}\sigma^2\\
-8\sqrt{2}\sigma&\frac{1}{3}(3+64\sigma^2)&-\frac{8}{3}\sqrt{2}\sigma\\
0&-4\sqrt{2}\sigma&1\\
\end{array}
\right],\label{Splusbh}
\end{equation}
and
\begin{equation}
S_-(\sigma)=\left(\frac{1}{1-64\sigma^2}\right)^{2/3}\left[
\begin{array}{ccc}
1&4\sqrt{2}\sigma&0\\
\frac{8}{3}\sqrt{2}\sigma&\frac{1}{3}(3+64\sigma^2)&8\sqrt{2}\sigma\\
\frac{512}{9}\sigma^2&\frac{4}{9}\sqrt{2}\sigma(-21+64\sigma^2)&\frac{1}{3}(3+64\sigma^2)\\
\end{array}
\right].\label{Sminusbh}
\end{equation}
In the case of the conical singularity, we have 
\begin{equation}
a_\phi^\pm=\frac{\alpha\mp\tau}{4\pi}(L_\pm+L_\mp)+\frac{q\pm w}{2\pi}(2W_0+W_2+W_{-2}).
\end{equation}
Again we define $\sigma=\frac{q\pm w}{\alpha\mp\tau}$. The gauge transformation is again written on the form $P_\pm=e^{\mp\log \kappa_\pm T_1} S_\pm(\sigma_\pm)$, where now
\begin{equation}
S_+(\sigma)=\left(\frac{1}{1+64\sigma^2}\right)^{2/3}\left[
\begin{array}{ccc}
\frac{1}{3}(3-64\sigma^2)&-\frac{4}{9}\sqrt{2}\sigma(21+64\sigma^2)&\frac{512}{9}\sigma^2\\
8\sqrt{2}\sigma&\frac{1}{3}(3-64\sigma^2)&-\frac{8}{3}\sqrt{2}\sigma\\
0&4\sqrt{2}\sigma&1\\
\end{array}
\right],\label{Spluscs}
\end{equation}
and
\begin{equation}
S_-(\sigma)=\left(\frac{1}{1+64\sigma^2}\right)^{2/3}\left[
\begin{array}{ccc}
1&-4\sqrt{2}\sigma&0\\
\frac{8}{3}\sqrt{2}\sigma&\frac{1}{3}(3-64\sigma^2)&-8\sqrt{2}\sigma\\
\frac{512}{9}\sigma^2&\frac{4}{9}\sqrt{2}\sigma(21+64\sigma^2)&\frac{1}{3}(3-64\sigma^2)\\
\end{array}
\right].\label{Sminuscs}
\end{equation}
%
\subsubsection{Computing $\mathcal{L}_\pm$ and $\mathcal{W}_\pm$}\label{LWcomp}
Although the charges $\mathcal{L}_\pm$ and $\mathcal{W}_\pm$ can be read off after we performed the gauge transformations in the previous subsection, it is easier to compute them by comparing invariants. We have $\tr (A_\phi^\pm)^2=8\mathcal{L}_\pm$ and $\tr (A_\phi^\pm)^3=-24\mathcal{W}_\pm$. By comparing with traces of $A_\phi^\pm$ given by \eqref{Aphipp} and \eqref{Aphibh}, we obtain
\begin{equation}
\mathcal{L}_\pm^{\textrm{CS}}=\frac{1}{16\pi^2}\left(-(\alpha\mp\tau)^2+\frac{64}{3}(q\pm w)^2\right),\label{Lpp}
\end{equation}
\begin{equation}
\mathcal{W}_\pm^{\textrm{CS}}=-\frac{1}{108\pi^3}(q\pm w)\left(9(\alpha\mp\tau)^2+64(q\pm w)^2\right),\label{Wpp}
\end{equation}
for the conical singularity and
\begin{equation}
\mathcal{L}_\pm^{\textrm{BH}}=\frac{1}{16\pi^2}\left((\alpha\mp\tau)^2+\frac{64}{3}(q\pm w)^2\right),\label{Lbh}
\end{equation}
\begin{equation}
\mathcal{W}_\pm^{\textrm{BH}}=\frac{1}{108\pi^3}(q\pm w)\left(9(\alpha\mp\tau)^2-64(q\pm w)^2\right),\label{Wbh}
\end{equation}
for the black hole.

\subsubsection{Interpretation of gauge transformations as shifts in coordinates}

In the pure gravity case, the gauge transformation one has to do to bring the connections to the highest weight gauge can be interpreted as just being shifts in the radial coordinate $\chi$. In the higher spin case, the gauge transformations can be interpreted as shifts in coordinates that parametrize the additional regions of the \sln{3} manifold. Recall that the connection to the \sln{3} manifold is via the group element $G=g_+g_-^{-1}$. Here $g_\pm$ determines the gauge connections $A_\mu^\pm$ via $A_\mu^\pm=g_\pm^{-1}\partial_\mu g_\pm$. Let us write $g_\pm=h_\pm(t,\phi,\theta,\psi) b_\pm(\chi)$ such that $a_\mu^\pm=h_\pm^{-1}\partial_\mu h_\pm$. The group element $G$, which parametrizes a submanifold of \sln{3}, is then given by
\begin{equation}
G=h_+(t,\phi,\theta,\psi)b_+(\chi)b_-^{-1}(\chi)h_-^{-1}(t,\phi,\theta,\psi).
\end{equation}
We can now extend this manifold to cover some larger submanifold of \sln{3} by adding more matrices in the middle. We can thus for example consider
\begin{equation}
G=h_+(t,\phi,\theta,\psi)c_+^{-1}(\rho)b_+(\chi)b_-^{-1}(\chi)c_-(\rho)h_-^{-1}(t,\phi,\theta,\psi),\label{gaugetr_as_coord}
\end{equation}
where now the coordinate $\rho$ extends our manifold further. We can assume that $c_+(0)=c_-(0)=1$ and thus $\rho=0$ reduces to our previos manifold spanned by the coordinates $t,\phi,\theta,\psi,\chi$. Note that it is important that the $h_+$ and $h_-$ remain on the sides since we still want the coordinates $t$, $\phi$, $\theta$ and $\psi$ to be isometries. Now it is clear that changing $\rho$ corresponds to right multiplications of $h_\pm$ by $c_\pm^{-1}$, and thus corresponds to gauge transformations of $a_\mu^\pm$. So the gauge transformations to go to the highest weight gauge can be obtained by replacing $c_+$ ($c_-$) by either \eqref{Splusbh} (\eqref{Sminusbh}) or \eqref{Spluscs} (\eqref{Sminuscs}), as well as doing a shift in $\chi$. Thus we conclude that all gauge transformations on $a_\mu^\pm$ that we have discussed can be interpreted as shifts in some coordinates on the \sln{3} manifold, naturally generalizing the case of pure three-dimensional gravity.

\subsubsection{Chemical potentials for higher spin black holes}\label{sec_chempothigh}
In Euclidean signature, where we Wick rotate $t=-i\tau$, the two gauge connections $A^\pm$ are replaced by a single complex gauge connection $A$ taking values in $\mathfrak{sl}(3,\C)$. When defining a higher spin black hole in Euclidean signature, we would also impose trivial holonomy in Euclidean time, meaning that the integral of $A_{\tau}$ over the compact Euclidean time coordinate has a trivial holonomy matrix. This is the natural generalization of the same condition one imposes in gravity and now forces us to introduce two chemical potentials $\xi$ and $\eta$ in $A_{\tau}$. After Wick rotating back to Lorentzian signature, the expression for $A_t^\pm$ analogous to \eqref{Atchempot} includes the Lorentzian chemical potentials $\xi_\pm$ and $\eta_\pm$ and takes the form \cite{Bunster:2014mua}
\begin{equation}
A_t^{'\pm}=\pm\left[\xi_\pm A_\phi^{'\pm} +\eta_\pm \left(W_{\pm2}+8\mathcal{W}_\pm L_{\mp1}+(\mathcal{L}_\pm)^2W_{\mp2}-2\mathcal{L}_\pm W_0\right)\right].\label{Atchempoths}
\end{equation}
In the pure gravity case where we only have $\xi_\pm$, this chemical potential can be absorbed into a reparametrization of $t$ and $\phi$ and thus has no physical meaning in Lorentzian signature. It is clear that in the higher spin case, the chemical potentials $\xi_\pm$ and $\eta_\pm$ can not just be absorb by a reparametrization of the coordinates $t$ and $\phi$. However, we will now show that they can indeed be absorbed via a reparametrization including the extra coordinates $\theta$ and $\psi$. Recall the coordinate transformation \eqref{coordtrbh} that we did in order to render the identification of points trivial (such that it only acted in the angular direction). Since the identification left the extra coordinates $\theta$ and $\psi$ untouched, it is a consistent procedure to truncate our manifold and only consider the coordinates $t$, $\phi$ and $\chi$. However, note that the coordinate transformation \eqref{coordtrbh} is not unique, and is just one arbitrary choice since we can add to the right hand side any combination of the coordinates that is invariant under the identification $I$. One way to parametrize (a subset of) the possible (linear) coordinate transformations, is to perform a new coordinate transformation given by
\begin{align}
t''&=\frac{1}{c_t}t',\quad\theta''=\theta'-\frac{c_\theta}{c_t} t',\nonumber\\
\psi''&=\psi'-\frac{c_\psi}{c_t} t',\quad\phi''=\phi'-\frac{c_\phi}{c_t} t',\nonumber\\
\chi''&=\chi'.\label{coordtrprimeprime}
\end{align}
Note that the identification $I$ would still act trivially as $\phi''\rightarrow\phi''-\alpha$ and thus all these coordinate transformations are equally justified. This coordinate transformation has the effect that the gauge connection for the new time coordinate $t''$ takes the form
\begin{equation}
A_{t''}=c_tA_{t'}+c_\phi A_{\phi'}+c_\theta A_{\theta'}+c_\psi A_{\psi'},
\end{equation}
while the other gauge connections are left invariant. This gives us four free parameters in the gauge connection $A_{t''}$ that can be used to tune the four chemical potentials $\xi_\pm$ and $\eta_\pm$. Of course we could consider more general gauge transformations that also change $A_{\theta'}^\pm$ and $A_{\psi'}^\pm$ but since we are not interested in the form of these connections there is no reason to consider more general coordinate transformations. For completeness we will write out the complete relation between the $c_i$ and the chemical potentials. We will henceforth drop the $''$. After the identification of points, the $A_t^\pm$ can thus be written as
\begin{equation}
A_t^\pm=(c_\phi\pm c_t)\frac{\alpha\mp\tau}{2}(L_{\pm1}-L_{\mp1})+c_\phi(q\pm w)\bar{W}-(c_\theta\pm c_\psi)\bar{W}.\label{Atc}
\end{equation}
Now we would like to map this to the expression \eqref{Atchempoths} including the chemical potentials $\xi_\pm$ and $\eta_\pm$ to show that the chemical potentials can be obtained from a reparametrization of the coordinates including the extended manifold. However, \eqref{Atchempoths} is the expression for $A_t^{'\pm}$. We are interested in the expression $A_t^\pm$, which can be obtained by applying the (inverse of the) gauge transformation \eqref{gaugetr}. Evaluating the gauge transformation \eqref{gaugetr} on the expression \eqref{Atchempoths} results in
\begin{align}
A_t^\pm=\pm &\xi_\pm \frac{\alpha\mp\tau}{2}(L_{\pm1}-L_{\mp1})\pm\xi_\pm(q\pm w)\bar{W}\nonumber\\
&\pm\eta_\pm\left(-\frac{32(\alpha\mp\tau)(q\pm w)}{6}(L_{\pm1}-L_{\mp1})+(\frac{16(q\pm w)^2}{3}-\frac{1}{4})\bar{W}\right).
\end{align}
To identify what reparametrization of the extra coordinates this corresponds to, we compare this to \eqref{Atc}. The coefficients $c_i$ are therefore related to the chemical potentials by
\begin{align}
c_\phi(q\pm w)-(c_\theta\pm c_\psi)&=\pm\xi_\pm(q\pm w)\pm\eta_\pm(\frac{16}{3}(q\pm w)^2-\frac{1}{4}),\nonumber\\
c_\phi\pm c_t&=\pm\xi_\pm\mp\eta_\pm\frac{32}{3}(q\pm w).
\end{align}
Another way to phrase this is that if we do the inverse of the transformation \eqref{coordtrprimeprime}, we can absorb the chemical potentials $\xi_\pm$ and $\eta_\pm$ into redefinitions of the coordinates on the extended manifold, thus generalizing the result for pure gravity. This question, if the higher spin chemical potentials could be absorbed in a coordinate transformation in some higher spin geometry, was posed as an open problem in \cite{Henneaux:2013dra}. Note that the special case of gravity, where $\eta_\pm=q=w=0$, we have $c_\psi=c_\theta=0$, $2c_\phi=\xi_+-\xi_-$ and $2c_t=\xi_++\xi_-$ reproducing our previous result. \\
\linebreak
Let us now briefly discuss the same procedure in Euclidean signature. In this case, we can also absorb the chemical potentials into the coordinates, but they will now show up by changing the periodicity of the manifold. The gauge connection in Euclidean time reads \cite{Bunster:2014mua}
\begin{equation}
A_\tau^{'\pm}=-i\left[\xi \left(L_{1}-\mathcal{L}L_{-1}-\mathcal{W}W_{-2}\right) +\eta \left(W_{2}+8\mathcal{W} L_{-1}+(\mathcal{L})^2W_{-2}-2\mathcal{L} W_0\right)\right],
\end{equation}
obtained by setting $t=-i\tau$ and continuing $A=A^+$ and $A^\dagger=-A^-$. Recall that points on the extended manifold are given by $G$ in \eqref{Ghsbh} and we can compute the metric on this extended manifold by
\begin{align}
\rd s^2=\frac{1}{8}\tr\left(G^{-1}\rd G G^{-1}\rd G\right)=&-\sinh^2\chi\rd t^2+\rd\chi^2+\cosh^2\chi\rd\phi^2\nonumber\\
&+\left(\frac{64}{3}+16\sinh^22\chi\right)\rd\psi^2-16\sinh^22\chi\rd\theta^2.
\end{align}
Note that both $t$ and $\theta$ are timelike directions and we will use this as a motivation to also Wick rotate $\theta=-i\nu$ (note that this is also consistent with the rule that we should identify $A=A^+$ and $A^\dagger=-A^-$). The Euclidean connections for the extra coordinates thus read (after we gauge away the $\chi$ dependence)
\begin{equation}
A_\theta^\pm=i(2W_0-W_{-2}-W_2),\quad A_\psi=-2W_0+W_{-2}+W_2.
\end{equation}
The fact that we have one complex and one real connection in the extra coordinate is crucial if we want to absorb the complex chemical potentials $\xi$ and $\eta$. Let us now assume that the chemical potentials $\xi$ and $\eta$ are given such that the holonomy in the coordinate $\tau$ is trivial and that the period of $\tau$ is $2\pi$ (or in other words, the manifold is identified by $\tau\sim\tau+2\pi$ as well as $\phi\sim\phi+2\pi$). We can now perform a coordinate transformation of the form
\begin{align}
\tau'&=\frac{\alpha_\tau}{2\pi}\tau,\quad\nu'=\nu+\frac{\alpha_\nu}{2\pi} \tau,\nonumber\\
\psi'&=\psi+\frac{\alpha_\psi}{2\pi} \tau,\quad\phi'=\phi+\frac{\alpha_\phi}{2\pi} \tau.\label{coordtrprimeprimeEucl}
\end{align}
This will transform $A_\tau$ and it is clear that the coefficients $\alpha_i$, just as in the Lorentzian case, can be chosen such that we absorb the chemical potentials $\xi$ and $\eta$. The identification on the manifold, which originally read $\tau\sim\tau+\alpha$, now changes to $(\phi',\tau',\nu',\psi')\sim(\phi'+\alpha_\phi,\tau'+\alpha_\tau,\nu'+\alpha_\nu,\psi'+\alpha_\psi)$. We see that just as in the gravity case, the chemical potentials can be absorbed also in Euclidean signature but at the price of changing the identification of the coordinates.\\
\linebreak
Just like in the BTZ black hole case, the smoothness conditions can also be seen to follow directly from the identification of points procedure. We can use the Euclidean version of \eqref{Ghsbh}, which reads
\begin{equation}
G=e^{-(\psi-i\nu)\bar{W}}e^{-(-i\tau+\phi)T_2}e^{-2\chi T_1}e^{(-i\tau-\phi)T_2}e^{-(\psi+i\nu)\bar{W}}.
\end{equation}
The periodicity here of the coordinates $\tau$ and $\nu$ would automatically result in the smoothness conditions obtained via the Chern-Simons formulation. The simplicity of this argument, of which the details are left as an exercise for the reader, is another indicition that using the interpretation of the higher spin black holes as an identification of points of the group manifold \sln{3} or $SL(3,\C)$ can be very useful.

\subsubsection{Solution space}
\begin{figure}
\includegraphics[scale=0.7]{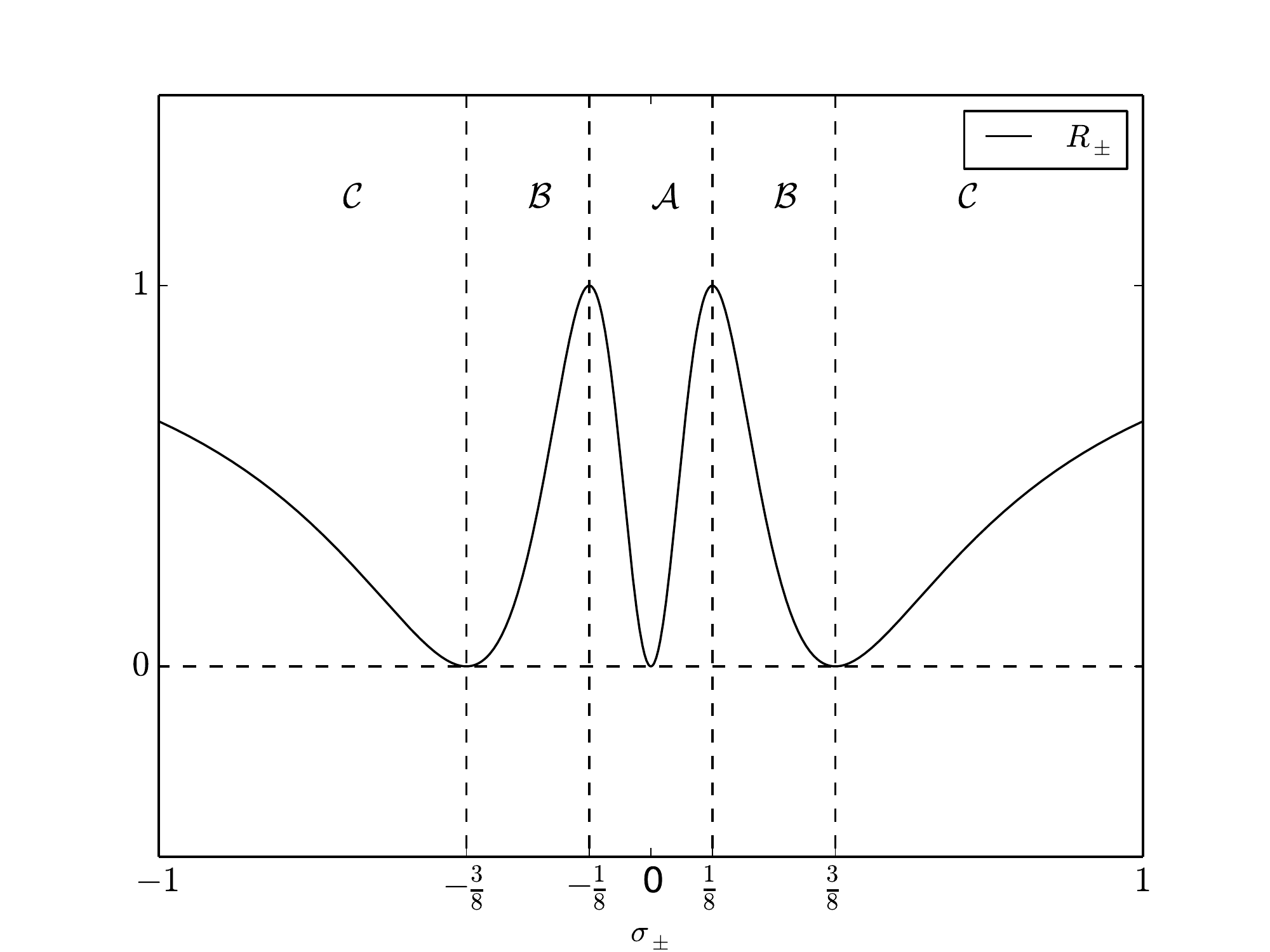}
\caption{\label{bhRfig} Plot of $R_\pm$ for the black hole case as a function of $\sigma_\pm$, as given by \eqref{Rbhdef}. All values of $R_\pm$ with $0\leq R_\pm\leq 1$ and with both positive and negative $\sigma_\pm$ can be obtained in each of the regions $\mathcal{A}$, $\mathcal{B}$ and $\mathcal{C}$ thus covering all higher spin black holes three times (not six times since we assume that $\alpha\pm\tau\geq0$). For large $|\sigma_\pm|$, $R_\pm$ approaches one.}
\end{figure}

From the condition that the entropy of a higher spin black hole is real, the inequality \eqref{entropy_ineq} on $\mathcal{W}_\pm$ and $\mathcal{L}_\pm$ can be obtained which in particular implies that
$\mathcal{L}_\pm\geq0$. Interestingly, in the black hole case, this inequality is always satisfied, showing that these solutions really correspond to (the Lorentzian version of) the higher spin black hole solutions constructed in previous literature. Moreover, if we define
\begin{equation}
R_\pm\equiv\frac{27\mathcal{W}^2_\pm}{4\mathcal{L}_\pm^3}=\frac{256\sigma_\pm^2(9-64\sigma_\pm^2)^2}{27(1+64\sigma_\pm^2)^3},\label{Rbhdef}
\end{equation}
it can be seen that every $R_\pm$ with $0\leq R_\pm\leq 1$ are obtained for $-\frac{1}{8}\leq\sigma_\pm\leq\frac{1}{8}$ (which we will call region $\mathcal{A}$), and all these solutions can be brought to the form \eqref{finalgauge} using the gauge transformation discussed in Section \ref{sec_gaugetr}. From this it is then clear from \eqref{Lbh} and \eqref{Wbh} that we can obtain any values of $\mathcal{L}_\pm$ and $\mathcal{W}_\pm$ which satisfy these inequalities, and thus all possible higher spin black holes can be constructed using this technique. However, there are two more regions where we can obtain all possible values of $R_\pm$, namely $\frac{1}{8}\leq|\sigma_\pm|\leq\frac{3}{8}$ (region $\mathcal{B}$) and $|\sigma_\pm|\geq\frac{3}{8}$ (region $\mathcal{C}$), see Figure \ref{bhRfig}. These regions are also gauge equivalent to the higher spin black hole solutions and corresponds to different permutations of the eigenvalues of $A_\phi^\pm$. This is made more precise in Appendix \ref{perm_app}. Thus the identifications of points will cover all higher spin black holes three times over.\\
\linebreak
\begin{figure}
\includegraphics[scale=0.7]{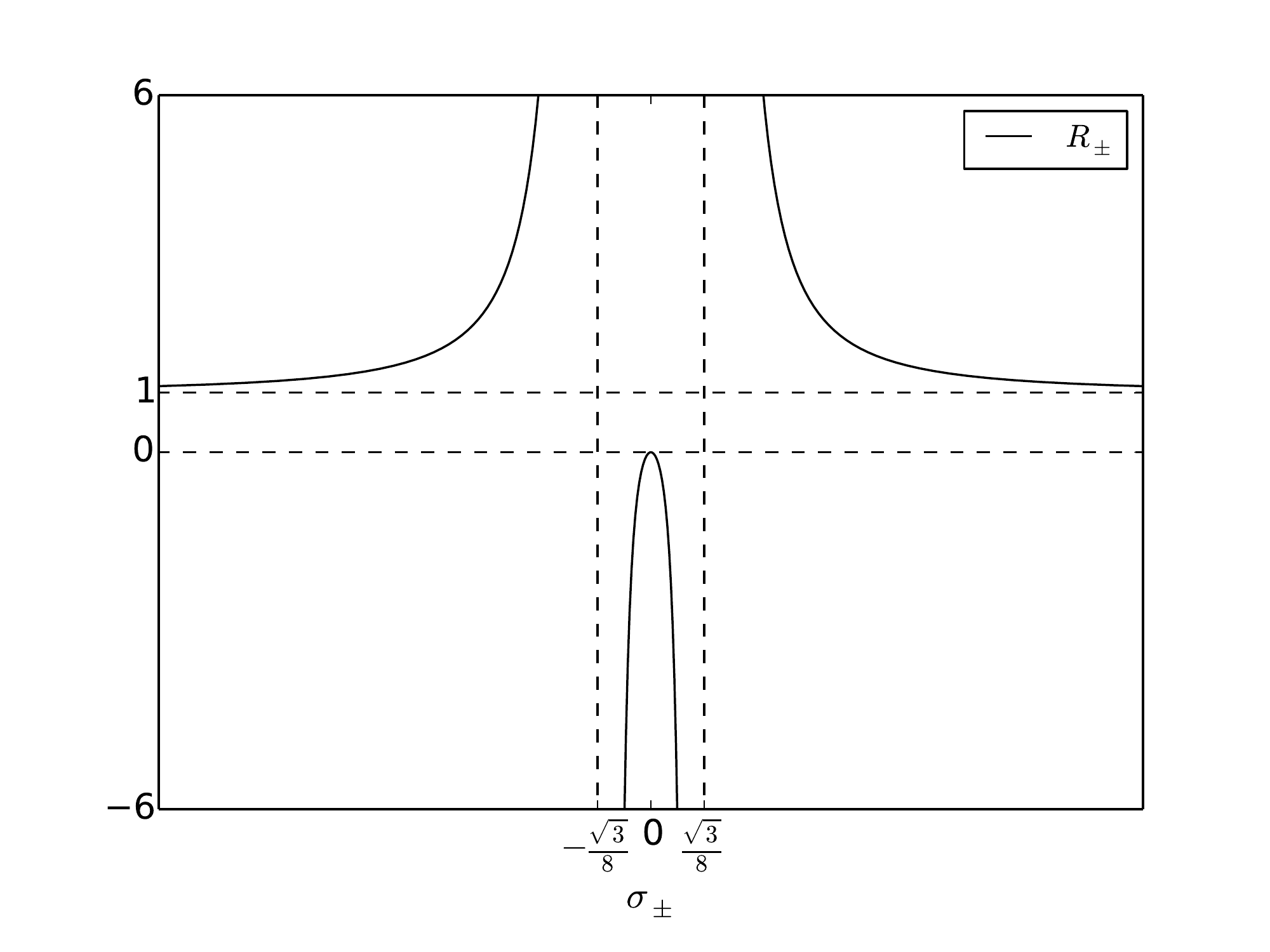}
\caption{\label{Rfig} Plot of $R_\pm$ as a function of $\sigma_\pm$ in the conical singularity case, as given by \eqref{Rdef}. Note in particular that all possible values of $R_\pm\leq0$ can be obtained when $\mathcal{L}_\pm\leq0$ (in the range $-\sqrt{3}/8\leq\sigma_\pm\leq\sqrt{3}/8$), and for $\mathcal{L}_\pm\geq0$ we can obtain all values of $R_\pm$ larger than one. Thus these solutions cover exactly the solution space of $\mathcal{L}_\pm$ and $\mathcal{W}_\pm$ that is {\it not} covered by the higher spin black hole solutions (except for solutions where the $+$ quantities satisfy the inequality \eqref{entropy_ineq} while the $-$ quantities violate it, or vice versa).}
\end{figure}
\linebreak
In the construction of the conical defects, we have instead
\begin{equation}
R_\pm\equiv\frac{27\mathcal{W}^2_\pm}{4\mathcal{L}_\pm^3}=\frac{256\sigma_\pm^2(9+64\sigma_\pm^2)^2}{27(-1+64\sigma_\pm^2)^3}.\label{Rdef}
\end{equation}
It can now be shown that even though we can obtain solutions which satisfy $\mathcal{L}_\pm\geq0$, they will violate the inequality \eqref{entropy_ineq} and thus do not correspond to higher spin black hole solutions. In fact, it can be shown that the conical defect solutions cover the ranges of $\mathcal{L}_\pm$ and $\mathcal{W}_\pm$ that are not covered by the higher spin black hole solutions\footnote{Except the solutions where for instance the $+$ quantities satisfy \eqref{entropy_ineq} while the $-$ quantities do not, or vice versa.}; any sign of $\mathcal{L}_\pm$ with any negative value of $R_\pm$, as well as positive $\mathcal{L}_\pm$ together with $R_\pm>1$. This is illustrated in Figure \ref{Rfig}. Note also that in the conical singularity case we have two complex eigenvalues and one real, thus we do not have the same ambiguity when determining the matrix parameters in terms of the eigenvalues as we had in the black hole case with three real eigenvalues and therefore each identification of points gives a unique solution.

\subsubsection{Global structure of the higher spin black hole}
One of the more interesting applications of these techniques is that one can immediately obtain the global structure of the solution, in particular go behind the horizon in the black hole case. This is quite difficult to do in the Chern-Simons formulation where the focus is on the asymptotic properties of the connections. In \cite{Castro:2016ehj} a prescription for constructing Kruskal coordinates for the higher spin black holes in the Chern-Simons formulation was proposed. However, their method has a few drawbacks, for instance it relies on imposing regularity in Euclidean signature and then analytically continue to Lorentzian signature. In the method of identification of points, the global structure immediately follows from the global structure inherited from the \sln{3} manifold. In particular, instead of using the parametrization \eqref{sl3parametr} for the submanifold of \sln{3}, we can instead use one that naturally extends the paramtrization \eqref{grplmntbhlambda} for the eternal BTZ black hole, namely one of the form
\begin{equation}
G=e^{-(\theta+\psi)\hat{W}}e^{-\phi T_2}e^{-2(\lambda_1T_1+\lambda_0T_0)}e^{-\phi T_2}e^{(\theta-\psi)\hat{W}},
\end{equation}
Such a parametrization then naturally generalizes the eternal BTZ black hole to the higher spin case and we have obtained a very natural definition of the global structure of the higher spin black hole, including the region inside the horizon and the two asymptotic boundaries. We will leave a more detailed investigation of these ideas to future research.

\subsection{Summary}
In this section we have considered higher spin gravity in three dimensions, described as a Chern-Simons theory with gauge group \sln{N}. We have proposed a new geometric method to define higher spin black holes where the solutions are constructed by identifying points in a submanifold of \sln{3} that naturally extends AdS$_3$. Applying this procedure to the \sln{3} case, we reproduced the higher spin black holes found in the literature and found solutions that are interpreted as higher spin conical singularities. The solutions can be classified according to the structure of the eigenvalues of the generators used to do the identification, just like one classifies conical singularity and black holes in three-dimensional gravity. The inequality \eqref{entropy_ineq}, which is satisfied for the higher spin black holes and was originally proven by requiring that the entropy (computed in the Euclidean formulation) is real, now follows naturally only from properties of the eigenvalues of the generators. We also showed that the chemical potentials, which are necessary in Euclidean signature to make the holonomy in Euclidean time trivial, can be absorbed completely in reparametrizations of the coordinates in the extended manifold generalizing a similar procedure for the BTZ black hole.

\pagebreak
\section{Conclusions and outlook}
In this chapter we have reviewed three-dimensional gravity in the Chern-Simons formulation and we gave an explicit connection between the Chern-Simons formulation and the formulation of \ads as the group manifold \sln{2}. We then showed how (spinning) black holes and pointlike particles can be constructed by identifying points in \ads, and showed that the black holes and pointlike particles belong to different classes of isometries which can be characterized by the eigenvalues of their generators. We then showed that these ideas can be extended to gravity coupled to a $U(1)$ Chern-Simons field and outlined some ideas on how it can also be extended to gravity coupled to higher spin fields provided that we extend our base manifold on which we carry out the identification procedure. We obtained agreement with previous literature on higher spin black holes as constructed in the Chern-Simons formulation in Euclidean signature and showed, among other things, that an inequality arising from requiring that the black hole entropy is real now arises naturally from the identification method. However, more work certainly needs to be done to understand the global structure of these manifolds. \\
\lb
The investigation in this chapter is only the beginning of understanding this ``higher spin geometry'' and how it can be used to describe solutions of higher spin gravity. In particular, studying the geometry of the extended manifold is a particularly interesting direction for future research. As we have alluded to already, these methods provide a very natural tool to study global properties of solutions to higher spin gravity which is not trivial to do in the Chern-Simons formulation. Moreover, one can try to understand the causal structure of the extended manifold and then perhaps it would be possible to define a generalization of an event horizon  which would be another way to characterize a higher spin black hole. Whether or not there is a generalization of the Bekenstein-Hawking formula for computing the entropy as the area of some horizon is also an intriguing problem. \\
\lb
We also saw in this chapter that these methods provide a natural way to construct collisions of particles that are charged under a $U(1)$ Chern-Simons field. It would be interesting to see if this can be generalized to the higher spin case. This is expected to be much more complicated and would most likely involve extending the extended manifold even further to include the whole \sln{3} manifold.











\chapter{Conclusions}\label{conclusions}

In this thesis we have investigated some aspects of the AdS/CFT duality, with particular focus on time-dependent effects and thermalization processes in quantum field theories. According to the AdS/CFT correspondence, these processes can be modelled by time-dependent solutions of gravity and formation of black holes in anti-de Sitter space. Such solutions can be obtained in different ways, and we have used both numerical and analytical methods to find new solutions of Einstein's equations with negative cosmological constant corresponding to the dynamical formation of a black hole.\\
%
%
%
\lb
In Chapter \ref{confinement} we studied time-dependent processes in confining holographic models. These are holographic theories dual to confining quantum field theories, and can have applications to dynamics in QCD (which is a confining theory). We studied in detail two holographic models, the hard wall model and the AdS Soliton. We found that there are dynamical solutions that never equilibrate (never result in black hole formation) and that observables in the dual field theory then undergo eternal oscillations in time. Moreover, in both models we found that there are sometimes long-time amplitude modulations of the oscillating field theory observables (although resulting from different mechanisms in the two models). It is interesting that such a simple model as the hard wall model can reproduce many qualitative features of top-down models such as the AdS Soliton and one may ask what features are universal for all confining models. We conclude that confining holographic models can result in interesting predictions of dynamical behaviour of field theory quantities, and it would be interesting to understand how to explain this behaviour from a field theory point of view and if similar behaviour can be seen in experiments. We hope that future studies of dynamics in confining holographic models can shed some more light on these issues.\\
\lb
In Chapter \ref{threed} we studied black hole formation from a very different mechanism, namely that of collisions of pointlike particles in three-dimensional gravity. We showed how to construct solutions with an arbitrary number of pointlike particles colliding to form a single object (either a new particle or a black hole). We also showed that we can take the limit of an infinite number of particles which resulted in solutions with a thin shell of matter collapsing to form a black hole (or a conical singularity). In particular, the lightlike thin shell solutions (resulting from a limit of colliding massless particles) are very relevant for the AdS/CFT correspondence since they correspond to the time-dependent setup where a field theory is perturbed by an inhomogeneous injection of energy and opens up possibilities for very non-trivial tests of the AdS/CFT correspondence in time-dependent situations. The same constructions work for massive particles and timelike shells, although in this case the interpretation using the AdS/CFT corresondence is less clear. We conclude that these cutting-and-gluing techniques that we have used, where solutions of three-dimensional gravity are constructed by excising regions in \ads and identifying the sides of these regions, can be very useful for contructing solutions of gravity relevant for the AdS/CFT correspondence (and in particular for time-dependent situations) and we hope to expand on these methods in future work.\\
\lb
In Chapter \ref{higher} we explored new methods for constructing solutions (black holes and conical singularities) in three-dimensional higher spin gravity, an extension of gravity including higher spin fields. Focusing on the case of gravity coupled to a spin-3 field, we showed that we can construct solutions  by identifying points under an isometry in the group manifold \sln{3}. We showed that some of these solutions can be mapped to so-called higher spin black holes, which are black holes with a higher spin charge that have been studied previously in the literature. We also identified solutions which can be interpreted as conical singularities with higher spin charge. These techniques naturally generalize the analogous methods used in three-dimensional gravity to construct standard black holes and conical singularities. We also commented on collisions of particles charged under a $U(1)$ Chern-Simons field, and we believe that the techniques explored in this chapter can be used to study collisions of higher spin particles and formation of higher spin black holes (and perhaps other non-trivial time-dependent solutions in higher spin gravity). We leave this interesting question for future research. We have only scratched the surface of these new methods of constructing higher spin black holes and higher spin conical defects and we hope that we can explore these techniques further in future research projects, in particular using these technniques to better understand the global structure of higher spin black holes which is typically difficult in the Chern-Simons formulation. Generalizing the results to gravity coupled to fields with arbitrary high spin as well as to other topological theories in three dimensions is also an interesting future direction.\\
\lb
In conclusion, holographic methods can be very useful for studying time-dependent effects in strongly coupled field theories and even though many situations require numerical methods, analytical tools can still be used to find new non-trivial solutions in gravity that can be relevant for the AdS/CFT correspondence. The techniques explored in this thesis can be used to study many different time-dependent setups and we hope that they can in the future provide new non-trivial tests of the AdS/CFT correspondence and hopefully even say something about phenomena in real world systems.


\appendix
\chapter{Computations related to Chapter 3}\label{appendix}
\section{Shape of boosted wedges}\label{app11}

In this appendix we derive equations \eqref{c1} and \eqref{Gammapm}, from equations \eqref{boosteqs}. The last three equations in \eqref{boosteqs} are
\begin{equation}
\sin \pc t\cosh \pc\chi=\cosh\chi\sin t \cosh \zeta - \sinh\chi\sinh\zeta\cos(\phi-\psi),\label{appeq1} 
\end{equation}
\begin{equation}
\sinh\pc\chi\cos\pc\phi=-\cosh\chi\sinh\zeta\sin t\cos\psi+\sinh\chi(-\sin\psi\sin(\phi-\psi)+\cosh\zeta\cos\psi\cos(\phi-\psi)),\label{appeq2}
\end{equation}
\begin{equation}
\sinh\pc\chi\sin\pc\phi=-\cosh\chi\sinh\zeta\sin t\sin\psi+\sinh\chi(\cos\psi\sin(\phi-\psi)+\cosh\zeta\sin\psi\cos(\phi-\psi)).\label{appeq3}
\end{equation}
We would like to take linear combinations of the two last equations and make them proportional to the first one, thus we write $\sin A$\eqref{appeq2}+$\cos A$\eqref{appeq3}$\equiv K$\eqref{appeq1} for some parameters $A$ and $K$ to be determined. Evaluating this we obtain
\begin{align}
\sinh\pc\chi\sin(\pc\phi+A)=&-\cosh\chi\sin t\sinh\zeta\sin(\psi+A)+\sinh\chi\cos(\psi+A)\sin(\phi-\psi)+\nonumber\\
 &+\sinh\chi\cosh\zeta\sin(\psi+A)\cos(\phi-\psi)\nonumber\\
 \equiv&K\left(\cosh\chi\sin t\cosh\zeta-\sinh\chi\sinh\zeta\cos(\phi-\psi)\right)\nonumber\\
  =&K\sin \pc t \cosh \pc \chi\label{appeq4}
\end{align}
By comparing the coefficients of $\cosh\chi\sin t$ and $\sinh\chi$ we thus obtain the equation
\begin{align}
\cos(\phi-\psi)\tanh\zeta&=\sinh^{-1}\zeta\cot(\psi+A)\sin(\phi-\psi)+\coth\zeta\cos(\phi-\psi)\nonumber\\
&\Rightarrow \tan(\psi+A)=\tan(\psi-\phi)\cosh\zeta,
\end{align}
and $K$ is then given as
\begin{equation}
K=-\tanh\zeta\sin(\psi+A).
\end{equation}
Now we let $\phi=\phi_\pm=\psi\pm(1\pm p)\nu$ and $\Gamma_\pm\equiv -\psi-A$, which gives
\begin{equation}
K=\tanh\zeta\sin\Gamma_\pm,
\end{equation}
and
\begin{equation}
\tan\Gamma_\pm=-\cosh\zeta\tan(\psi-\phi_\pm)=\pm\cosh\zeta\tan((1\pm p)\nu),
\end{equation}
which is equation \eqref{Gammapm}. Equation \eqref{appeq4} then reduces exactly to \eqref{c1}.

\section{Equations relating to collisions of massless particles}

\subsection{Derivation of equation \eqref{peqml}}\label{app2}
In this section we will outline how to derive equation \eqref{peqml}, which is a consistency condition imposed on the parameters of the wedges such that the geometry after the collision (of massless particles) is consistent. The massive case is dealt with in Appendix \ref{peqapp}. The goal is to find a constraint that will make sure that the intersection $I_{i-1,i}$ is mapped to $I_{i,i+1}$ via the holonomy $\bh_i$ of particle $i$. Recall that $I_{i,i+1}$ is the intersection between the surface $w_+^i$ and $w_-^{i+1}$, and is a radial geodesic at angle $\phi_{i,i+1}$. We will assume that $\phi_{i-1,i}$ and $\phi_{i,i+1}$ are known, and then determine $p_i$ in terms of these angles. The holonomy $\bh_i$ is given by
\begin{equation}
\bh_i=\mathbf{1}+E_i(\gamma_0-\cos\psi_i\gamma_1-\sin\psi_i\gamma_2).
\end{equation}
We parametrize the intersection $I_{i-1,i}$ between $w_-^i$ and $w_+^{i-1}$ as
\begin{equation}
I_{i-1,i}=\cosh\pc\chi\sin \pc t\gamma_0+\sinh\pc\chi\cos\phi_{i-1,i}\gamma_1+\sinh\pc\chi\sin\phi_{i-1,i}\gamma_2+\cosh\pc\chi\cos \pc t,
\end{equation}
and the intersection $I_{i,i+1}$ between $w_+^i$ and $w_-^{i+1}$ as
\begin{equation}
I_{i,i+1}=\cosh\chi\sin t\gamma_0+\sinh\chi\cos\phi_{i,i+1}\gamma_1+\sinh\chi\sin\phi_{i,i+1}\gamma_2+\cosh\chi\cos t.
\end{equation}
Computing $\bh_i^{-1}I_{i,i+1}\bh_i=I_{i-1,i}$ gives us the equations
\begin{subequations}
\begin{align}
\sin \pc t\cosh\pc\chi=&2E_i^2\sin t\cosh\chi+2E_i^2\cos(\phi_{i,i+1}-\psi_i)\sinh\chi\nonumber\\
&-2E_i\sin(\phi_{i,i+1}-\psi_i)\sinh\chi+\sin t\cosh\chi,\label{weqsa}\\
\cosh\phi_{i-1,i}\sinh\pc\chi=&-2E_i^2\sin\psi_i\sin(\phi_{i,i+1}-\psi_i)\sinh\chi-2E_i^2\sin t\cos\psi_i\cosh\chi\nonumber\\
&-2E_i^2\cos\phi_{i,i+1}\sinh\chi+2E_i\sin\phi_{i,i+1}\sinh\chi\nonumber\\
&+2E_i\sin\psi_i\sin t\cosh \chi+\cos\phi_{i,i+1}\sinh\chi\label{weqsb},\\
\sin\phi_{i-1,i}\sinh\pc\chi=&-2E_i^2\sin\psi_i\sin t\cosh\chi-2E_i^2\sin\psi_i\cos(\phi_{i,i+1}-\psi_i)\sinh\chi\nonumber\\
&-2E_i\sin t\cos\psi_i\cosh\chi-2E_i\cos\phi_{i,i+1}\sinh\chi+\sin\phi_{i,i+1}\sinh\chi,\label{weqsc}\\
\cos t\cosh\chi=&\cos \pc t\cosh\pc\chi.\label{weqsd}
\end{align}
\end{subequations}
The above equations correspond to equating the coefficients of $\gamma_0$, $\gamma_1$, $\gamma_2$ and $\mathbf{1}$, respectively. We also have the following equations that must be satisfied
\begin{subequations}
\begin{align}
\tanh\pc\chi\sin(-\phi_{i-1,i}+\Gamma_-^i+\psi_i)=&-\sin\Gamma_-^i\sin \pc t\label{weqconsta},\\
\tanh\chi\sin(-\phi_{i,i+1}+\Gamma_+^i+\psi_i)=&-\sin\Gamma_+^i\sin t\label{weqconstb},
\end{align}
\end{subequations}
where
\begin{equation}
\tan\Gamma_{\pm}^i=\pm(1\pm p_i)E_i.
\end{equation}
The goal is now to find $p_i$ such that all these equations are satisfied. First, we can eliminate $\pc t$ and $t$ in terms of $\pc\chi$ and $\chi$, respectively, which only leaves us the coordinates $\chi$ and $\pc\chi$. Then we must specify $p_i$, such that, for every $\chi$\footnote{That is, $\chi$ is a free variable, but there might be certain bounds for $\chi$.}, we can find a $\pc\chi$ such that \eqref{weqsa}-\eqref{weqsd} are satisfied. Now we use \eqref{weqconsta} and \eqref{weqconstb} to eliminate $\cosh\chi\sin t$ in \eqref{weqsc}, to obtain
\begin{equation}
\sinh\pc\chi=\frac{-(E_ip_i\cos\phi_{i,i+1}-E_ip_i\cos(\phi_{i,i+1}-2\psi_i)-p_i\sin\phi_{i,i+1}+\sin(\phi_{i,i+1}-2\psi_i))\sinh\chi}{(p_i+1)\sin\phi_{i-1,i}}.
\end{equation}
Now it is convenient to substitute this into the linear combination \eqref{weqsb}$-E_i$\eqref{weqsc}, from which we can then solve for $p_i$ as
\begin{equation}
p_i=\frac{\sin(\phi_{i,i+1}+\phi_{i-1,i}-2\psi_i)}{-E_i\cos(\phi_{i,i+1}-\phi_{i-1,i})+E_i\cos(\phi_{i,i+1}+\phi_{i-1,i}-2\psi_i)+\sin(\phi_{i,i+1}-\phi_{i-1,i})},
\end{equation}
which is equivalent to \eqref{peq}. To obtain this result we only used equations \eqref{weqsb} and \eqref{weqsc}. It is now possible to explicitly check that equation \eqref{weqsa} is satisfied, although the computations are quite tedious. We then know that equation \eqref{weqsd} must be satisfied, because the equations are not independent (or to be more precise, equations \eqref{weqsa}-\eqref{weqsc} imply via the embedding equation (or condition of unit determinant) that $\cos t\cosh\chi=\pm\cos \pc t\cosh\pc\chi$, and the condition to have the plus sign just tells us in what range we have to pick $\pc t$). 

%

\subsection{Some more derivations}\label{app3}
In this section we derive equations \eqref{coshatTml}, \eqref{anglerel}, \eqref{radeqpp2}, \eqref{coshhatXml}, \eqref{anglerelbh} and \eqref{radeqbh}. Starting with
\begin{equation}
\hat\phi'(\phi)=\frac{\cos^2\hat T(1+\tan^2\Phi)}{\cosh \hat Z(1+P\tan\Phi)^2}\Big(-2\rho+P'+1+P^2\Big),
\end{equation}
which is easily obtained from \eqref{tandiff} and \eqref{tandiff0}, we first use equation \eqref{ode1} to write this as
\begin{align}
\hat\phi'(\phi)=\frac{\cos^2\hat T(1+\tan^2\Phi)}{\cosh \hat Z(1+P\tan\Phi)^2}(1-\cot\Phi P),
\end{align}
From equation \eqref{Z} we have
\begin{equation}
\frac{1}{\cosh^2 \hat Z}=1-\tanh^2\hat Z=1-\frac{P^2}{(P\cos\Phi-\sin\Phi)^2},
\end{equation}
and we then obtain from \eqref{QZ} that
\begin{align}
\frac{1}{\cos^2\hat T}=1+\tan^2\hat T=&1+\frac{(P-\tan\Phi)^2}{(1+P\tan\Phi)^2}\frac{1}{\cosh^2\hat Z}\nonumber\\
=&1+\frac{-P^2+(P\cos\Phi-\sin\Phi)^2}{(\cos\Phi+P\sin\Phi)^2}\nonumber\\
=&\frac{1}{(\cos\Phi+P\sin\Phi)^2}.
\end{align}
From this we immediately obtain
\begin{align}
\hat\phi(\phi)=\frac{1-\cot\Phi P}{\cosh \hat Z}.
\end{align}
We would now like to derive the second equality in equation \eqref{radeqpp2}. Since $\tanh \hat Z=\frac{P}{P\cos\Phi-\sin\Phi}$, we have $\sinh \hat Z\cosh \hat Z=\frac{P(P\cos\Phi-\sin\Phi)}{-P^2+(P\cos\Phi-\sin\Phi)^2}$ and $\cosh^2 \hat Z=\frac{(P\cos\Phi-\sin\Phi)^2}{-P^2+(P\cos\Phi-\sin\Phi)^2}$, and we obtain that
\begin{align}
\frac{1}{\cosh \hat Z+\sinh \hat Z\cos \hat T}=&\cosh \hat Z\frac{-P^2+(P\cos\Phi-\sin\Phi)^2}{(P\cos\Phi-\sin\Phi)^2+P(P\cos\Phi-\sin\Phi)(\cos\Phi+P\sin\Phi)}\nonumber\\
=&\frac{\cosh \hat Z}{1-P\cot\Phi}.
\end{align}
which proves \eqref{radeqpp2}, and where we also used $\cos Q=\cos\Phi+P\sin\Phi$ since it can be seen that $\cos\Phi+P\sin\Phi>0$.\\
\lb
For the black hole case, the situation is very similar. We then start with
\begin{equation}
\hat y(\phi)=\hat y_0-\int_0^{\phi}\frac{\cosh^2X(1+\tan^2\Phi)}{\sinh Z(1+P\tan\Phi)^2}\Big(-\theta+P'+1+P^2\Big)\rd \phi',
\end{equation}
which is easily obtained from \eqref{tandiff} and \eqref{tandiff0bh}. This can now be simplified to
\begin{equation}
\hat y'(\phi)=-\frac{\cosh^2\hat X(1+\tan^2\Phi)}{\sinh \hat Z(1+P\tan\Phi)^2}(1-\cot\Phi P),\label{Npbh}
\end{equation}
by again using \eqref{ode1}. From equation \eqref{Zbh} we have
\begin{equation}
\frac{1}{\sinh^2 \hat Z}=\coth^2\hat Z-1=\frac{P^2}{(P\cos\Phi-\sin\Phi)^2}-1
\end{equation}
and we then obtain from \eqref{QX} that
\begin{align}
\frac{1}{\cosh^2\hat X}=1-\tanh^2\hat X=&1-\frac{(P-\tan\Phi)^2}{(1+P\tan\Phi)^2}\frac{1}{\sinh^2\hat Z}\nonumber\\
=&1-\frac{P^2-(P\cos\Phi-\sin\Phi)^2}{(\cos\Phi+P\sin\Phi)^2}\nonumber\\
=&\frac{1}{(\cos\Phi+P\sin\Phi)^2}.
\end{align}
Equation \eqref{Npbh} then becomes
\begin{equation}
\hat y'(\phi)=\frac{\cot\Phi P-1}{\sinh \hat Z}.
\end{equation}
We now want to derive expression \eqref{radeqbh}. Since $\coth\hat Z=\frac{P}{P\cos\Phi-\sin\Phi}$, we have $\sinh \hat Z\cosh \hat Z=\frac{P(P\cos\Phi-\sin\Phi)}{P^2-(P\cos\Phi-\sin\Phi)^2}$ and $\sinh^2 \hat Z=\frac{(P\cos\Phi-\sin\Phi)^2}{P^2-(P\cos\Phi-\sin\Phi)^2}$, and we obtain that
\begin{align}
\frac{1}{\cosh \hat X \cosh \hat Z+\sinh \hat Z}&=\sinh \hat Z \frac{P^2-(P\cos\Phi-\sin\Phi)^2}{P(P\cos\Phi-\sin\Phi)(\cos\Phi+P\sin\Phi)+(P\cos\Phi-\sin\Phi)^2}\nonumber\\
&=\frac{\sinh \hat Z}{P\cot\Phi-1},
\end{align}
which proves \eqref{radeqbh}, and where we have used that $\cosh \hat X=P\cos\Phi+\sin\Phi$ since it can be seen that $P\cos\Phi+\sin\Phi>0$.\\
\lb
For an explanation of the inequalities $P\cos\Phi+\sin\Phi>0$ and $\cos\Phi+P\sin\Phi>0$, see the derivation for the analogous inequalities for the massive case in Appendix \ref{usefulrel}.

\subsection{The stress-energy tensor for the shell}\label{app4}
Let us for concreteness assume that a black hole has formed. The formation of a conical singularity works similarly. We start by proving the following identity
\begin{equation}
\frac{G'}{G}=P.\label{GPGP}
\end{equation}
From \eqref{Gdefbh} and \eqref{Zbh} it is possible to derive that
\begin{equation}
G=\frac{\sqrt{M}|\sin\Phi|}{\sqrt{P^2\sin^2\Phi+2P\cos\Phi\sin\Phi-\sin^2\Phi}}.\label{Geqapp}
\end{equation}
We now have
\begin{equation}
\frac{G'}{G}=\frac{P\cos\Phi(\sin\Phi)'-PP'\sin^2\Phi-P'\cos\Phi\sin\Phi-P(\cos\Phi)'\sin\Phi}{P^2\sin^2\Phi+2P\cos\Phi\sin\Phi-\sin^2\Phi}
\end{equation}
After using $(\sin\Phi)'\cos\Phi-(\cos\Phi)'\sin\Phi=\cos^2\Phi(\tan\Phi)'$, we can use equations \eqref{ode1} and \eqref{ode2} to eliminate the derivatives, and this then reduces precisely to \eqref{GPGP}.\\
\lb
Now it is easy to prove \eqref{awesome}. Using \eqref{Geqapp}, \eqref{GPGP}, $(G'/G)'=G''/G-(G')^2/G^2$ and the differential equation \eqref{ode1} we obtain from \eqref{Keq} that
\begin{align}
16\pi G T^{rr}=&\frac{1}{r}\left[1+\frac{M}{G^2}+2\left(\frac{G'}{G}\right)'+\frac{(G')^2}{G^2}\right]\delta(v)\nonumber\\
=&\frac{1}{r}\left[1+P^2+2P\cot\Phi-1+2P'+P^2\right]\delta(v)\nonumber\\
=&\frac{1}{r}\left[1+P^2+2P\cot\Phi-1+2(-P\cot\Phi+2\rho-P^2)+P^2\right]\nonumber\\
=&\frac{4\rho}{r}.
\end{align}

\section{Equations relating to collisions of massive particles}

\subsection{Derivation of equation \eqref{peq}}\label{peqapp}
In this section we will outline how to derive equation \eqref{peq}, which is a consistency condition imposed on the parameters of the wedges such that the geometry after the collision is consistent. The derivation here is for the massive particles, and is very similar to that of massless particles done in Appendix \ref{app2}. The goal is to find a constraint that will make sure that the intersection $I_{i-1,i}$ is mapped to $I_{i,i+1}$ via the holonomy of particle $i$. Recall that $I_{i,i+1}$ is the intersection between the surface $w_+^i$ and $w_-^{i+1}$, and is a radial geodesic at angle $\phi_{i,i+1}$. We will assume that $\phi_{i-1,i}$ and $\phi_{i,i+1}$ are known, and then find a constraint on the value of $p_i$ in terms of these angles. The holonomy of particle $i$ is given by
\begin{equation}
\bh_i=\cos\nu_i+\sin\nu_i\cosh\zeta_i\gamma_0-\sin\nu_i\sinh\zeta_i\cos\psi_i\gamma_1-\sin\nu_i\sinh\zeta_i\sin\psi_i\gamma_2.
\end{equation}
$I_{i-1,i}$ will be parametrized as
\begin{equation}
I_{i-1,i}=\cosh\pc\chi\cos\pc t+\cosh\pc\chi\sin\pc t\gamma_0+\sinh\pc\chi\cos\phi_{i-1,i}\gamma_1+\sinh\pc\chi\sin\phi_{i-1,i}\gamma_2,
\end{equation}
and $I_{i,i+1}$ as
\begin{equation}
I_{i,i+1}=\cosh\chi\cos t+\cosh\chi\sin t\gamma_0+\sinh\chi\cos\phi_{i,i+1}\gamma_1+\sinh\chi\sin\phi_{i,i+1}\gamma_2.
\end{equation}
Now computing $\bh_i^{-1}I_{i,i+1}\bh_i=I_{i-1,i}$, and extracting the coefficients of $\bm{1}$ and $\gamma_i$, gives the equations
\begin{subequations}
\begin{align}
\cos \pc t\cosh\pc\chi=&\cos t\cosh\chi,\label{eq1}\\
\sin \pc t\cosh\pc\chi=&2\sin^2\nu_i\cosh\zeta_i\sinh\zeta_i\cos(\phi_{i,i+1}-\psi_i)\sinh\chi\nonumber\\
 &-2\sin\nu_i\cos\nu_i\sinh\zeta_i\sin(\phi_{i,i+1}-\psi_i)\sinh\chi\nonumber\\
 &+2\sin^2\nu_i\sin t\sinh^2\zeta\cosh\chi+\sin t\cosh\chi,\label{eq2}\\
\cos\phi_{i-1,i}\sinh\pc\chi=&-2\sin^2\nu_i\sin\psi_i\sinh^2\zeta_i\sin(\phi_{i,i+1}-\psi_i)\sinh\chi\nonumber\\
&-2\sin^2\nu_i\cos\psi_i\cosh\zeta_i\sinh\zeta_i\sin t\cosh\chi\nonumber\\
&-2\sin^2\nu_i\cos\phi_{i,i+1}\cosh^2\zeta_i\sinh\chi\nonumber\\
&+2\sin\nu_i\cos\nu_i\cosh\zeta_i\sin\phi_{i,i+1}\sinh\chi\nonumber\\
&+2\sin\nu_i\cos\nu_i\sin\psi_i\sinh\zeta_i\sin t\cosh\chi+\cos\phi_{i,i+1}\sinh\chi,\label{eq3}\\
\sin\phi_{i-1,i}\sinh\pc\chi=&-2\sin^2\nu_i\sin\psi_i\sinh^2\zeta_i\cos(\phi_{i,i+1}-\psi_i)\sinh\chi\nonumber\\
&-2\sin^2\nu_i\sin\psi_i\sinh\zeta_i\cosh\zeta_i\sin t\cosh\chi\nonumber\\
&-2\sin\nu_i\cos\nu_i\sinh\zeta_i\cos\psi_i\sin t\cosh\chi\nonumber\\
&-2\sin\nu_i\cos\nu_i\cos\phi_{i,i+1}\sinh\chi\cosh\zeta_i\nonumber\\
&-2\sin^2\nu_i\sin\phi_{i,i+1}\sinh\chi+\sin\phi_{i,i+1}\sinh\chi.\label{eq4}
\end{align}
\end{subequations}
Since the intersections $I_{i-1,i}$ and $I_{i,i+1}$ are located on $w_-^{i}$ and $w_+^i$ respectively, we also have the equations
\begin{subequations}
\begin{align}
\tanh\pc\chi\sin(-\phi_{i-1,i} +\Gamma_-^i+\psi_i)&=-\tanh\zeta_i \sin\Gamma_-^i \sin \pc t,\nonumber\\
\tanh\chi\sin(-\phi_{i,i+1} +\Gamma_+^i+\psi_i)&=-\tanh\zeta_i \sin\Gamma_+^i \sin t,\label{weqapp}
\end{align}
\end{subequations}
where
\begin{equation}
\tan\Gamma_\pm^i =\pm\tan((1 \pm p_i)\nu_i) \cosh\zeta_i.
\end{equation}
Now we will find a value of $p_i$ such that the above equations are solved. This can be done by first solving for $\sinh\pc\chi$ in \eqref{eq4}, and then substituting that into \eqref{eq3}. By also using \eqref{weqapp} to eliminate $\cosh\chi$ and $\cosh\pc\chi$, we can solve for $p_i$ as
\begin{align}
  &\tan(p_i\nu_i)=\nonumber\\
  &\frac{\sin(\phi_{i-1,i}+\phi_{i,i+1}-2\psi_i)\sin\nu_i}{\sin\nu_i\cosh\zeta_i\cos(\phi_{i-1,i}+\phi_{i,i+1}-2\psi_i)+\cos\nu_i\sin(\phi_{i,i+1}-\phi_{i-1,i})-\cosh\zeta_i\sin\nu_i\cos(\phi_{i,i+1}-\phi_{i-1,i})},
\end{align}
which is equivalent to \eqref{peq}. It can now be checked explicitly that also \eqref{eq2} is satisfied, and then we know from the embedding equation \eqref{embedding_eq} that \eqref{eq1} must be satisfied as well (to be more precise, from \eqref{embedding_eq} we obtain that \eqref{eq1} is satisfied up to a sign, and this sign determines in what interval we choose $\pc t$).


\subsection{Derivation of equations \eqref{coshatT} and \eqref{coshX}}\label{usefulrel}
We will start by deriving the relation 
\begin{equation}
\cos\hat T=\cosh\hat X=\cos\Phi\cos T+\sin\Phi\sin T\cosh Z,\label{coshatTX}
\end{equation}
where it is implicit that $\hat T$ and $\hat X$ are only defined in the case of a formation of a conical singularity or a black hole, respectively. This follows from \eqref{tanhhatZ} and \eqref{tanhatTcoshhatZ}, since we have
\begin{align}
\frac{1}{\cos^2\hat T}&=1+\tan^2\hat T=1+\frac{(\tan T\cosh Z-\tan\Phi)^2}{(1+\tan\Phi\tan T\cosh Z)^2}\frac{1}{\cosh^2 \hat Z}\\
&=1+\frac{(\tan T\cosh Z\cos\Phi-\sin\Phi)^2-\tan^2T\sinh^2Z}{(\cos\Phi+\sin\Phi\tan T\cosh Z)^2}\\
&=\frac{1+\tan^2 T}{(\cos\Phi+\sin\Phi\tan T\cosh Z)^2}\\
&=\frac{1}{(\cos\Phi\cos T+\sin\Phi\sin T\cosh Z)^2}.\\
\end{align}
Thus we can conclude that 
\begin{equation}
\cos\hat T=\pm(\cos\Phi\cos T+\sin\Phi\sin T\cosh Z).
\end{equation}
A very similar computation shows that in the case of a formation of a black hole, we have
\begin{equation}
\cosh\hat X=\pm(\cos\Phi\cos T+\sin\Phi\sin T\cosh Z).
\end{equation}
The difficult part when deriving this relation is  to determine the sign, and we will argue that we should always pick the plus sign although we will not be completely rigorous. First of all, in the homogeneous case where $\Phi=T=\hat T=\hat X=0$, we must have $\cos\hat T=\cos\hat X=1$. In the inhomogeneous case, we will only be interested in solutions that can be continuously deformed to the homogeneous case, and it is thus reasonable to expect that we should always choose the plus sign. To make this argument a bit more solid, we will now argue that $\sin\Phi$ and $\sin T$ will always have the same sign. This would mean that the right hand side of \eqref{coshatTX} is always (strictly) positive, and when continuously deforming a homogeneous solution to a non-homogeneous one, the plus sign must be kept. The only caveat that remains is then for solutions where $\Phi$ or $T$ would be very large such that $\cos \Phi$ or $\cos T$ can change sign, but we have plenty of numerical evidence that $\Phi$ or $T$ are never that large.\\
\linebreak
To see that $\Phi$ and $T$ always have the same sign, we will start by looking at the differential equations \eqref{odeP} and \eqref{odeT}. It is easy to see that we have $T(\phi_0)=0$ if and only if $\Phi(\phi_0)=0$, at some point $\phi_0$. Moreover, it can be shown that generically $T'$ and $\Phi'$ are non-zero at these points, and that we can only have $T'(\phi_0)=0$ if and only if $\Phi'(\phi_0)=0$. This indicates that $T$ and $\Phi$ flip signs at the same points. Now consider solutions that are small deviations from a homogeneous solution (so that we only keep terms linear in $\tan \Phi$ and $\tan T$). It can be easily shown that these satisfy $\tan T/\tan \Phi=\rho>0$. Thus for perturbative solutions, $\tan \Phi$ and $\tan T$ always have the same sign, and together with the fact that for non-perturbative solutions $\tan\Phi$ and $\tan T$ flip sign at the same points, one can easily convince oneself that $\Phi$ and $T$ always have the same sign.

\subsection{Derivation of equations \eqref{cs_anglemap} and \eqref{bh_anglemap}}\label{anglemap}
%
In this section we will prove the mapping between the angular coordinates across the shell, equations \eqref{cs_anglemap} and \eqref{bh_anglemap}. These are computed as the sum of the opening angles of the static wedges. Thus in the conical singularity case, we have
\begin{equation}
\hat\phi=\hat{\phi}_0+\sum_{0\leq i<j} 2\nu_{i,i+1}+O(1/N),
\end{equation}
when $\hat\phi$ is inside wedge $c_{j,j+1}^\text{static}$, while in the black hole case\footnote{To be able to do the conical singularity and the black hole case simultaneously, we use $\hat\phi$ to also denote the quantity that was called $\hat y$ in Section \ref{massivecollsec}}
\begin{equation}
\hat\phi=\hat{\phi}_0+\sum_{0\leq i<j} 2\mu_{i,i+1},
\end{equation}
when $\hat\phi$ is inside wedge $c_{j,j+1}^\text{static}$. Let us define (see equation \eqref{gammadiff})
\begin{align}
\Delta \Gamma&\equiv\lim_{n\rightarrow\infty} \frac{1}{d\phi}\left[\tan\Gamma_+^{i,i+1}-\tan\Gamma_-^{i,i+1} \right]\nonumber\\
&=\frac{\cos^2T(\tan T\cosh\cZ)'-2\cosh\cZ\rho+\cos^2T+\sin^2T\cosh^2\cZ)}{(\cos T\cos\Phi+\sin T\cosh\cZ\sin\Phi)^2},
\end{align}
where $'$ is derivative with respect to $\phi$. By equating equation \eqref{gammadiff} with either \eqref{gammadiffcs} or \eqref{gammadiffbh} we obtain in the limit that
\begin{equation}
\hat\phi=\hat{\phi}_0+\int_0^\phi \Delta \Gamma d\phi_1\times\bigg\{ \begin{array}{ll} \frac{\cos^2\hat T}{\cosh \hat Z},&\text{Pointlike particle}\\-\frac{\cosh^2\hat X}{\sinh \hat Z},&\text{Black hole}\\\end{array}
\end{equation}
After using \eqref{odeT} to substitute $T'$ we obtain
\begin{equation}
\Delta \Gamma=\frac{(\cos T-\frac{\sin T}{\sinh\cZ}\cZ')(\cos T-\cosh Z \sin T \cot\Phi)}{(\cos T\cos\Phi+\sin T\cosh\cZ\sin\Phi)^2\cosh \hat Z}
\end{equation}
%
%
Now using equations \eqref{coshatTX} we obtain
\begin{equation}
\hat\phi=\hat{\phi}_0+\int_0^\phi\frac{(\cos T-\frac{\sin T}{\sinh\cZ}\cZ')(\cos T-\cosh Z \sin T \cot\Phi)}{\cosh \hat Z}d\phi,
\end{equation}
in the case of formation of a conical singularity, and 
\begin{equation}
\hat\phi=\hat{\phi}_0-\int_0^\phi\frac{(\cos T-\frac{\sin T}{\sinh\cZ}\cZ')(\cos T-\cosh Z \sin T \cot\Phi)}{\sinh \hat Z}d\phi,
\end{equation}
in the case of formation of a black hole. $\hat\phi$ is still increasing despite the suspicious sign, since $\hat Z<0$. We can simplify this even further by using equations \eqref{tanhatTcoshhatZ} and \eqref{tanhXsinhhatZ} to get rid of $\hat Z$. This will result in the equations \eqref{cs_anglemap} and \eqref{bh_anglemap}.

\subsection{Derivation of the induced metric}\label{app_ind_metric}
In this appendix we will prove equations \eqref{indmetric} and \eqref{indmetric_bh}, namely the form of the induced metric on the shell for the collision process of massive particles, and we will start with the conical singularity case. We will focus on the inside patch, the outside is completely analogous. Let us restate the problem: We want to compute the induced metric on a surface given by the equation
\begin{equation}
\tanh\chi=-\tanh Z(\phi)\sin t,
\end{equation}
in the spacetime with metric
\begin{equation}
ds^2=-\cosh^2\chi dt^2+d\chi^2+\sinh^2\chi d\phi^2.
\end{equation}
We will use the proper time $\tau$ and the angular coordinate $\phi$ to parametrize the geometry on this surface. We will restate some useful relations involving the proper time which are also found in Section \ref{geodesicssec}. They are
\begin{equation}
\sinh\chi=-\sinh Z\sin \tau,\label{ur1}
\end{equation}
\begin{equation}
\cosh\chi=\frac{\cos \tau}{\cos t}=\sqrt{\cos^2\tau+\sin^2\tau\cosh^2Z},\label{ur2}
\end{equation}
\begin{equation}
\tan t=\cosh Z\tan \tau.\label{ur3}
\end{equation}
The induced metric is computed as
\begin{equation}
ds^2=\gamma_{\tau\tau}d\tau^2+2\gamma_{\tau\phi}d\tau d\phi+\gamma_{\phi\phi}d\phi^2,
\end{equation}
where
\begin{equation}
\begin{array}{cc}
 \gamma_{\tau\tau}=&-\cosh^2\chi\dot{t}^2+\dot{\chi}^2\\
 \gamma_{\tau \phi}=&-\cosh^2\chi\dot{t}t'+\dot{\chi}\chi'\\
 \gamma_{\phi\phi}=&-\cosh^2\chi(t')^2+(\chi')^2+\sinh^2\chi\\
\end{array}
\end{equation}
Here $\cdot$ is derivative with respect to $\tau$ and $'$ is derivative with respect to $\phi$. From \eqref{ur1}, \eqref{ur2} and \eqref{ur3} we can derive that
\begin{equation}
\dot\chi=-\sinh Z\cos t,\label{chidot}
\end{equation}
\begin{equation}
\dot t=\frac{\cosh Z}{\cosh^2\chi},\label{tdot}
\end{equation}
\begin{equation}
\chi'=-\frac{\cosh Z\sin \tau}{\cosh\chi}Z',\label{chiprime}
\end{equation}
\begin{equation}
t'=\frac{\sinh Z\sin \tau\cos t}{\cosh\chi}Z'.\label{tprime}
\end{equation}
What we want to show is thus $\gamma_{\tau\tau}=-1$, $\gamma_{\tau\phi}=0$ and $\gamma_{\phi\phi}=\sin^2\tau(\sinh^2Z+(\partial_\phi Z)^2)$. The outside patch works analogously.\\

{\bf $\mathbf{\gamma_{\tau\tau}}$:} We already know that $\gamma_{\tau\tau}=-1$ must hold, since we know that $\tau$ is the proper time, but for completeness we will prove this explicitly here anyway. By using the relations \eqref{ur2}, \eqref{chidot} and \eqref{tdot} we obtain
\begin{align}
\gamma_{\tau\tau}&= -\cosh^2\chi\dot{t}^2+\dot{\chi}^2=-\frac{\cosh^2Z}{\cosh^2\chi}+\sinh^2Z\cos^2t\nonumber\\
&=\frac{1}{\cosh^2\chi}\left(\underbrace{-\cosh^2Z+\sinh^2Z\cos^2\tau}_{=-\cosh^2Z\sin^2\tau-\cos^2\tau}\right)\nonumber\\
&=-1.
\end{align}

{\bf $\mathbf{\gamma_{\phi\phi}}$:} It is also straightforward to compute $\gamma_{\phi\phi}$. We obtain by using \eqref{ur1}, \eqref{ur2}, \eqref{chiprime} and \eqref{tprime} that
\begin{align}
\gamma_{\phi\phi}&=-\cosh^2\chi(t')^2+(\chi')^2+\sinh^2\chi\nonumber\\
&=\sin^2\tau\left(-\sinh^2Z\cos^2t(Z')^2+\frac{\cosh^2Z(Z')^2}{\cosh^2\chi}\right)+\sinh^2\chi\nonumber\\
&=\frac{\sin^2\tau}{\cosh^2\chi}\left(-\sinh^2Z\cos^2\tau+\cosh^2Z\right)(Z')^2+\sin^2\tau\sinh^2Z\nonumber\\
&=\sin^2\tau(\sinh^2Z+(Z')^2).
\end{align}

{\bf $\mathbf{\gamma_{\tau\phi}}$:} From \eqref{chidot}-\eqref{tprime} it directly follows that $\gamma_{\tau\phi}=0$.\\
\linebreak
Now let us consider the black hole case. We thus consider a metric on the form
\begin{equation}
ds^2=-\sinh^2\beta d\sigma^2+d\beta^2+\cosh^2\beta dy^2.
\end{equation}
In this spacetime our shell is specified by a function $B(y)$ by the equation $\coth\beta=\coth B\cosh\sigma$ and we want to compute the induced metric on this shell. We use $y$ and the proper time $\tau$ as our intrinsic coordinates on the shell. The induced metric is computed as
\begin{equation}
ds^2=\gamma_{\tau\tau}d\tau^2+2\gamma_{\tau\phi}d\tau dy+\gamma_{\phi\phi}dy^2.
\end{equation}
where
\begin{equation}
\begin{array}{cc}
 \gamma_{\tau\tau}=&-\sinh^2\beta\dot{\sigma}^2+\dot{\beta}^2,\\
 \gamma_{\tau\phi}=&-\sinh^2\beta\dot{\sigma}\sigma'+\dot{\beta}\beta',\\
 \gamma_{\phi\phi}=&-\sinh^2\beta(\sigma')^2+(\beta')^2+\cosh^2\beta.
\end{array}
\end{equation}
Now $'$ means derivative with respect to $y$. The equation for the shell can be described in terms of the proper time by
\begin{equation}
\cosh\beta=-\cosh B\sin\tau,\label{ur4}
\end{equation}
\begin{equation}
\sinh \beta=\frac{\cos\tau}{\sinh\sigma}=\sqrt{\sin^2\tau\sinh^2B-\cos^2\tau},\label{ur5}
\end{equation}
\begin{equation}
\coth\sigma=-\sinh B\tan\tau.\label{ur6}
\end{equation}
The derivatives are
\begin{equation}
\dot\beta=-\cosh B\sinh \sigma,\label{betadot}
\end{equation}
\begin{equation}
\dot \sigma=\frac{\sinh B}{\sinh^2\beta},\label{sigmadot}
\end{equation}
\begin{equation}
\beta'=-\frac{\sinh B\sin \tau}{\sinh\chi}B',\label{betaprime}
\end{equation}
\begin{equation}
\sigma'=\frac{\cosh B\sin \tau\sinh \sigma}{\sinh\beta}B'.\label{sigmaprime}
\end{equation}
The computations are very similar to the conical singularity case.\\

{\bf $\mathbf{\gamma_{\tau\tau}}$:} By using the relations \eqref{ur5}, \eqref{betadot} and \eqref{sigmadot} we obtain
\begin{align}
\gamma_{\tau\tau}&= -\sinh^2\beta\dot{\sigma}^2+\dot{\beta}^2=-\frac{\sinh^2B}{\sinh^2\beta}+\cosh^2B\sinh^2\sigma\nonumber\\
&=\frac{1}{\sinh^2\beta}\left(\underbrace{-\sinh^2B+\cosh^2B\cos^2\tau}_{=-\sinh^2B\sin^2\tau+\cos^2\tau}\right)\nonumber\\
&=-1.
\end{align}

{\bf $\mathbf{\gamma_{yy}}$:} It is also straightforward to compute $\gamma_{\phi\phi}$. We obtain by using \eqref{ur4}, \eqref{ur5}, \eqref{betaprime} and \eqref{sigmaprime} that
\begin{align}
\gamma_{yy}&=-\sinh^2\beta(\sigma')^2+(\beta')^2+\cosh^2\beta\nonumber\\
&=\sin^2\tau\left(-\cosh^2B\sinh^2\sigma(B')^2+\frac{\sinh^2B(B')^2}{\sinh^2\beta}\right)+\cosh^2\beta\nonumber\\
&=\frac{\sin^2\tau}{\sinh^2\beta}\left(-\cosh^2B\cos^2\tau+\sinh^2B\right)(B')^2+\sin^2\tau\cosh^2B\nonumber\\
&=\sin^2\tau(\cosh^2B+(B')^2).
\end{align}

{\bf $\mathbf{\gamma_{\tau y}}$:} From \eqref{betadot}-\eqref{sigmaprime} it directly follows that $\gamma_{\tau y}=0$.

\subsection{Consistency condition on the induced metric}\label{app_cons_cond}
In this section we will prove equation \eqref{indmetric_consistency}. This equation states that the relation between $R$ and $\bar{R}$ (or equivalently between $Z$ and $\pc Z$ in the conical singularity case, and between $Z$ and $B$ in the black hole case) is consistent with how $\phi$ is related to $\bar \phi$ when crossing the shell, in the sense that the induced metric are the same in both patches. The ($d\bar \phi^2$ part of the) metric outside the shell can be written as
\begin{equation}
\bar{\gamma}_{\bar\phi\bar\phi}d\bar\phi^2=(\bar{R}^2+\frac{(\partial_{\bar\phi} R)^2}{\bar{f}(\bar R)})d\bar\phi^2=\left(\bar{R}^2\left(\frac{d\bar\phi}{d\phi}\right)^2+\frac{(\partial_{\phi} R)^2}{\bar{f}(\bar R)}\right)d\phi^2\equiv\bar{\gamma}_{\phi\phi}d\phi^2.
\end{equation}
From the inside of the shell the ($d \phi^2$ part of the) induced metric is $\gamma_{\phi\phi}d\phi^2=(\sinh^2Z+(\partial_{\phi} Z)^2)d\phi^2$, and thus what we want to show is that $\bar{\gamma}_{\phi\phi}=\gamma_{\phi\phi}$.\\
\linebreak
We will start by proving the useful relation \eqref{barRrel2}, namely that
\begin{equation}
\frac{\sqrt{\bar{f}(\bar{R})}}{\bar R}=\frac{1}{\sinh Z}\left(\cosh Z\cos T-\cot \Phi\sin T\right).\label{barRrel}
\end{equation}
Note also that the above left hand side is equal to $\coth \pc Z$ in the conical singularity case, and equal to $\tanh B$ in the black hole case. In the conical singularity case, it can be obtained by starting with equations \eqref{Zprel1} and \eqref{Zprel2}. By solving for $\coth \pc Z$ we immediately obtain
\begin{equation}
\frac{\coth \pc Z}{\cos \hat{T}}=\frac{\cosh Z\cosh\hat Z-\sinh Z\sinh\hat Z\cos \Phi}{\sinh Z\cosh\hat Z\cos\Phi-\sinh\hat Z\cosh Z}.
\end{equation}
Now we can use \eqref{sinhatTcoshhatZ} and \eqref{sinhatTsinhhatZ} to eliminate $\hat Z$, and then using \eqref{coshatT} to eliminate $\cos\hat T$, the result \eqref{barRrel} follows. In the case of a formation of a black hole the same relation can be proved. By comparing the embedding equations \eqref{adscoord} and \eqref{bhcoord} at the point $\tau=-\pi/2$ we can obtain $\sinh B=\sinh \pc Z\cos\pc\Phi$ and $\cosh B\cosh\hat X=\cosh \pc Z$. Now By using \eqref{bhcoshZp} and \eqref{bhcoshZ} we obtain
\begin{equation}
\frac{\tanh B}{\cosh \hat X}=\tanh \pc Z\cos\pc\Phi=\frac{\cosh\hat Z\sinh Z\cos\Phi-\sinh\hat Z\cosh Z}{\cosh Z\cosh\hat Z-\sinh Z\sinh\hat Z\cos\Phi}.
\end{equation}
We then use \eqref{coshX}, \eqref{sinhhatXsinhhatZ} and \eqref{sinhhatXcoshhatZ} to eliminate $\hat Z$ and $\hat X$, and then equation \eqref{barRrel} follows.\\
\linebreak
The next step is to obtain an expression for $\partial_\phi R$. This can be obtained by taking the derivative of \eqref{barRrel} with respect to $\phi$. This yields
\begin{align}
\frac{\partial_\phi \bar{R}}{\sqrt{\bar f(\bar R)}}\left(1-\frac{\bar f(\bar R)}{\bar{R}^2}\right)=&\partial_\phi T\frac{1}{\sinh Z}\left(\cosh Z\sin T+\cos T\cot \Phi\right)+\partial_\phi\Phi\frac{\sin T}{\sinh Z\sin^2\Phi}\nonumber\\
&+\partial_\phi Z\frac{1}{\sinh^2Z}(-\cos T+\cosh Z\sin T\cot \Phi).
\end{align}
Now we can use \eqref{odeP} and \eqref{odeT} to eliminate $\partial_\phi \Phi$ and $\partial_\phi T$, and \eqref{barRrel} to substitute $\bar{f}/\bar{R}^2$. After simplifying the result, we end up with the simple expression
\begin{equation}
\frac{\partial_\phi \bar{R}}{\sqrt{\bar f(\bar R)}}=\partial_\phi Z\cos T+\sinh Z\sin T.\label{partialR}
\end{equation}
Now, equations \eqref{cs_anglemap} and \eqref{ZpZ}, or \eqref{bh_anglemap} and \eqref{BZ}, both imply that
\begin{equation}
\frac{d\bar{\phi}}{d\phi}=\frac{1}{\bar{R}}\left(\sinh Z\cos T-\sin T\partial_\phi Z\right).
\end{equation}
The remainder of the proof is now straightforward. Starting with $\bar{\gamma}_{\phi\phi}$, we obtain
\begin{align}
\bar{\gamma}_{\phi\phi}&=\bar{R}^2\left(\frac{d\bar\phi}{d\phi}\right)^2+\frac{(\partial_{\phi} R)^2}{\bar{f}(\bar R)}\nonumber\\
&=\left(\sinh Z\cos T-\sin T\partial_\phi Z\right)^2+(\partial_\phi Z\cos T+\sinh Z\sin T)^2\nonumber\\
&=\sinh^2Z+\left(\partial_\phi Z\right)^2\nonumber\\
&=\gamma_{\phi\phi}.\label{finalgammagamma}
\end{align}
This completes the proof that the induced metric is the same from both sides of the shell.

\subsection{The stress-energy tensor for the shell}\label{app_shellT}
In this appendix we will outline how to prove equation \eqref{Stautaupp}, namely we will evaluate the $\tau\tau$ component of the stress-energy tensor of the shell in the case where the shell is obtained as a limit of pointlike particles. Just as in Appendix \ref{app_cons_cond} we put the black hole and conical singularity on the same footing by using $\bar{R}(\phi)$ to characterize the shell in the outside patch, but we will now use $Z(\phi)$ inside the shell to simplify some of the expressions. We will use $'$ to denote derivatives with respect to $\phi$.\\
\linebreak
The energy density is then given by
\begin{equation}
8\pi G S_{\tau\tau}=-\frac{\bar{K}_{\phi\phi}-K_{\phi\phi}}{\sin^2\tau(R^2+\frac{(R')^2}{f(R)})},\label{SmS}
\end{equation}
The extrinsic curvatures $K_{\phi\phi}$ and $\bar{K}_{\phi\phi}$ inside respectively outside the shell are defined by \eqref{Kij}, and can be computed using some symbolic manipulation software. The result is
\begin{equation}
\frac{K_{\phi\phi}}{\sin \tau}=-\frac{\sinh^2Z\cosh Z-\sinh Z Z''+2\cosh Z(Z')^2}{(\sinh^2Z+(Z')^2)^\frac{1}{2}},
\end{equation}
and 
\begin{align}
\frac{\bar{K}_{\phi\phi}}{\sin\tau}=&-\Bigg[(\sinh^2Z+(Z')^2)^2\frac{\sqrt{\bar{f}}}{\bar R}-(\sinh^2Z+(Z')^2)\frac{(\bar R')^2}{\sqrt{\bar f}\bar{R}}+\sinh Z\cosh Z Z'\frac{\bar R'}{\sqrt{\bar f}}\\
&-(\sinh^2Z+(Z')^2)\left(\frac{\bar R''}{\sqrt{\bar f}}-\frac{\bar R}{\sqrt{\bar f}^3}(\bar R')^2\right)+Z'Z''\frac{\bar R'}{\sqrt{\bar f}}\Bigg]\\
&\times \frac{1}{(\sinh^2Z+(Z')^2)^\frac{1}{2}(\sinh^2Z+(Z')^2-\frac{(\bar R')^2}{\bar f})^{\frac{1}{2}}},
\end{align}
The goal is to write this in terms of $\rho$, by using the differential equations \eqref{odeP} and \eqref{odeT} to eliminate the derivatives. From \eqref{partialR} we obtain that
\begin{equation}
\frac{\bar R''}{\sqrt{\bar f}}-\frac{\bar R}{\sqrt{\bar f}^3}(\bar R')^2=Z''\cos T+Z'\cosh Z\sin T+T'(\sinh Z\cos T-Z'\sin T).\label{partial2R}
\end{equation}
$T'$ can be eliminated using \eqref{odeT}, which states that
\begin{equation}
T'=\rho-\sin T\cos T\cot \Phi-\cosh Z\sin^2T-\frac{Z'\sin T}{\sinh Z}(\cosh Z\cos T-\sin T\cot\Phi).\label{eqTp}
\end{equation}
Note also that $\sqrt{\sinh^2Z+(Z')^2-\frac{(\bar R')^2}{\bar f}}=\sinh Z\cos T-Z'\sin T$ (see equation \eqref{finalgammagamma}). By using \eqref{barRrel}, \eqref{partialR}, \eqref{partial2R} and \eqref{eqTp} and then simplifying the expression using for example Mathematica, we end up with the remarkably simple result
\begin{equation}
8\pi G S_{\tau\tau}\sin\tau=-\frac{\rho}{\sqrt{\sinh^2Z+(Z')^2}},
\end{equation}
which is the same as \eqref{Stautaupp} after the substitution $\sinh Z=R$.

\chapter{Computations related to Chapter 4}\label{appendix2}

\section{Classification of solutions for gravity}\label{sl2class}
In this section we will classify all one-parameter subgroups of \sln{2}$\times$\sln{2} up to conjugation. This will in particular allow us to determine to which family of isometries pointlike particles and BTZ black holes belong. This was also done in \cite{Banados:1992gq} by instead looking at the $SO(2,2)$. 
\subsection{Relations between $SO(2,2)$ and $\sln{2}\times \sln{2}$}
We will first restate in more detail the relations between $SO(2,2)$ and $\sln{2}\times \sln{2}$. We define the generators of $SO(2,2)$ as
\begin{equation}
J_{\mu\nu}=x_\nu\partial_\mu-x_\mu\partial_\nu.
\end{equation}
The commutation relations are
\begin{equation}
[J_{\mu\nu},J_{\rho\sigma}]=\eta_{\nu\rho}J_{\sigma\mu}+\eta_{\mu\sigma}J_{\rho\nu}+\eta_{\mu\rho}J_{\nu\sigma}+\eta_{\nu\sigma}J_{\mu\rho}.
\end{equation}
The metric is $\eta_{\mu\nu}=\mathrm{diag}(-,-,+,+)$. For further convenience we label these indices as $3,0,1,2$ and $i,j,k$ now runs over only $0,1,2$. Now we define $P_i=J_{3i}$, $L_i=\frac{1}{2}\epsilon_i^{jk}J_{jk}$. Using $\epsilon_{ij}^k\epsilon_{kmn}=\eta_{in}\eta_{jm}-\eta_{im}\eta_{jn}$ gives
\begin{equation}
[P_i,P_j]=\epsilon_{ij}^kL_k,
\end{equation}
\begin{equation}
[L_i,P_j]=\epsilon_{ij}^kP_k,
\end{equation}
\begin{equation}
[L_i,L_j]=\epsilon_{ij}^kL_k,
\end{equation}
\begin{equation}
J_{ij}=-\epsilon_{ij}^kL_k.
\end{equation}
Now we decompose this algebra into two copies of $\mathfrak{sl}(2,\R)$ as $Y^{\pm}_i=\frac{1}{2}(L_i\pm P_i)$. The commutation relations are now
\begin{equation}
[Y^+_i,Y^-_j]=0 ,
\end{equation}
\begin{equation}
 [Y_i^\pm,Y_j^\pm]=\epsilon_{ij}^kY^\pm_k.
\end{equation}
Now considering a general generator on the form $\omega_{\mu\nu}J^{\mu\nu}$ we would like to write it on the form $\omega^i_+Y^+_i+\omega^i_-Y^-_i$. We have
\begin{equation}
\omega_{\mu\nu}J^{\mu\nu}=-2\omega_{3i}P^i+\omega_{ij}(-\epsilon^{ijk}L_k)=-2\omega_{3i}(Y_+^i-Y_-^i)+\omega_{ij}(-\epsilon^{ijk}(Y^+_k+Y^-_k))
\end{equation}
which gives
\begin{equation}
\omega^\pm_k=\mp2\omega_{3k}-\omega_{ij}\epsilon^{ij}_k
\end{equation}
Computing the inner products of these vectors gives
\begin{equation}
\omega_\pm^k\omega_k^\pm=-4\omega_{3k}\omega^{3k}-2\omega_{ij}\omega^{ij}\mp 4\omega_{ij}\epsilon^{ij}_k\omega^{3k}.
\end{equation}
The Casimir invariants are extracted as
\begin{align}
I_1\equiv&\omega^{\mu\nu}\omega_{\mu\nu}=-\frac{1}{4}(\omega_+^k\omega_k^++\omega_-^k\omega_k^-)\nonumber\\
=&-\frac{1}{2}(Tr(\omega_+^kT_k\omega_+^lT_l)+Tr(\omega_-^kT_k\omega_-^lT_l)),
\end{align}
\begin{align}
I_2\equiv&\frac{1}{2}\epsilon^{\mu\nu\rho\sigma}\omega_{\mu\nu}\omega_{\rho\sigma}=\frac{1}{4}(-\omega_+^k\omega_k^++\omega_-^k\omega_k^-)\nonumber\\
=&\frac{1}{2}(-Tr(\omega_+^kT_k\omega_+^lT_l)+Tr(\omega_-^kT_k\omega_-^lT_l)).
\end{align}
where we have chosen a basis $T_i$ such that the $Y_i^\pm$ can be written in block diagonal form as
\begin{equation}
Y_i^+=\left(\begin{array}{cc}
           T_i&0\\
	   0&0\\
          \end{array}\right),\quad\quad Y_i^-=\left(\begin{array}{cc}
           0&0\\
	   0&T_i\\
          \end{array}
\right).
\end{equation}
Note that we use the convention $\epsilon_{3012}=\epsilon_{012}=1$.

\subsection{One-parameter subgroups in \sln{2}$\times$\sln{2}}

We will now interested in classifying one-parameter subgroups $g(t)\in \sln{2}\times \sln{2}$ under conjugation, namely under the transformation $g(t)\rightarrow kg(t)k^{-1}$. This is equivalent to classifying Lie algebra elements $e\in \mathfrak{sl}(2,R)$ under conjugation in \sln{2}. This can either be done by looking at the eigenvalues of $e$, or by looking at the vectors $\omega_i$ where $e=\omega^i T_i$. The eigenvalues of a real $2\times2$ traceless matrix satisfies the requirement that if $\lambda$ is an eigenvalue, then both $-\lambda$ and $\lambda^*$ are eigenvalues. Thus given two copies of $\mathfrak{sl}(2,R)$ with two generators $e_\pm$, we have the following cases:
\begin{enumerate}
 \item The eigenvalues $\lambda_+$, $-\lambda_+$, $\lambda_-$ and $-\lambda_-$ are all real.
 \item The eigenvalues $\lambda_+$, $-\lambda_+$, $\lambda_-$ and $-\lambda_-$ are all imaginary.
 \item The eigenvalues $\lambda_+$, $-\lambda_+$ are imaginary, $\lambda_-$ and $-\lambda_-$ are real.
 \item The eigenvalues $\lambda_+$, $-\lambda_+$ are real, $\lambda_-$ and $-\lambda_-$ are imaginary.
\end{enumerate}
We also have to be careful with the special cases when some eigenvalues are equal to zero. Note also that imaginary eigenvalues would correspond to a timelike vector $\omega_i$ and real eigenvalues correspond to a spacelike $\omega_i$. In the cases where the eigenvalues are non-zero, the matrix can be diagonalized, however it will be useful to write the matrix in a non-diagonal basis. If the eigenvalues of a matrix $e_\pm$ are real, we will perform a conjugation (Lorentz transformation) such that it is proportional to $T_2$, and if the eigenvalues are imaginary we perform a conjugation (in $SL(2,\C)$) such that it is proportional to $T_0$. The basis $T_i$ is again given by
\begin{equation}
T_0=\left(\begin{array}{cc}
 0&-\frac{1}{2}\\
 \frac{1}{2}&0\\
\end{array}\right),\quad
T_1=\left(\begin{array}{cc}
 \frac{1}{2}&0\\
 0&-\frac{1}{2}\\
\end{array}\right),\quad
T_2=\left(\begin{array}{cc}
 0&\frac{1}{2}\\
 \frac{1}{2}&0\\
\end{array}\right).
\end{equation}
Note that it is always possible to move to this basis, and two matrices that are both proportional to either $T_0$ or $T_2$ will be conjugate to each other in \sln{2} only if they are the same matrix. Thus this will cover all diagonalizable matrices in \sln{2} with non-vanishing eigenvalues.\\
\linebreak
{\bf Case 1:}\\
In this case the generators can thus be written as $e_\pm=\lambda_\pm T_2$, and thus $\omega_i^\pm=(0,0,\lambda_\pm)$. We then obtain
\begin{equation}
\omega_{10}=-\frac{\lambda_++\lambda_-}{4}\equiv\lambda_1,\quad\quad\omega_{23}=\frac{\lambda_+-\lambda_-}{4}\equiv\lambda_2.
\end{equation}
This gives the killing vector $J=2\lambda_1 J_{01}+2\lambda_2J_{32}$, and the Casimir invariants
\begin{equation}
I_1=-\frac{1}{4}(\lambda_+^2+\lambda_-^2)=-2(\lambda_1^2+\lambda_2^2),\quad\quad I_2=\frac{1}{4}(\lambda_-^2-\lambda_+^2)=4\lambda_1\lambda_2.
\end{equation}
By comparing with \cite{Banados:1992gq}, we see that this is what they call type I$_b$ and corresponds to a standard BTZ black hole (note that to do the comparison, we first have to do the relabeling $(3,0,1,2)\rightarrow(0,1,2,3)$). The notation using $\lambda_1$ and $\lambda_2$ are used to reproduce \cite{Banados:1992gq}. By comparing with Section \ref{sec_btz}, we see that we can identify $\lambda_1=\tau/2$ and $\lambda_2=\alpha/2$ and thus
\begin{equation}
I_1=-2\pi^2M,\quad\quad I_2=-2\pi^2J.\label{I1I2MJ}
\end{equation}
These solutions satisfy $M>|J|$. This does not include the extremal case where $M=|J|$, because it can not be brought to the standard form \eqref{metricspinbh} (metric formulation) or \eqref{btzgaugetransform} (Chern-Simons formulation) since the coordinate/gauge transformations break down.\\
\linebreak
{\bf Case 2:}\\
Here we go to a frame where $e_\pm=\lambda_\pm T_0$, and thus $\omega^\pm_i=(\lambda_\pm,0,0)$. We obtain
\begin{equation}
\omega_{30}=\frac{\lambda_--\lambda_+}{4}\equiv b_1,\quad\quad\omega_{12}=-\frac{\lambda_++\lambda_-}{4}\equiv b_2.
\end{equation}
The killing vector is $J=b_1 J_{30}+b_2J_{12}$ and the Casimir invariants become
\begin{equation}
I_1=\frac{1}{4}(\lambda_+^2+\lambda_-^2)=2(b_1^2+b_2^2),\quad\quad I_2=\frac{1}{4}(\lambda_+^2-\lambda_-^2)=4b_1b_2.
\end{equation}
This is type I$_c$ in \cite{Banados:1992gq} and corresponds to a (spinning) pointlike particle. The notation using $b_1$ and $b_2$ are used to reproduce \cite{Banados:1992gq}. By comparing to the results in Section \ref{sec_scs} we obtain $b_2=\alpha/2$ and $b_1=-\tau/2$ and again we have the relations $I_1=-2\pi^2M$, $I_2=-2\pi^2J$. These solutions satisfy $-M>|J|$ and also in this case the extremal solutions $-M=|J|$ is not included since the coordinate/gauge transformations in Section \ref{sec_scs} break down.\\
\linebreak
{\bf Case 3:}\\
We will now assume that $\lambda_+$ is imaginary and $\lambda_-$ is real. In that case we write $\omega_i^+=(\lambda_+,0,0)$ and $\omega_i^-=(0,0,\lambda_-)$. We obtain
\begin{equation}
\omega_{30}=-\frac{\lambda_+}{4}\equiv b,\quad\omega_{12}=-\frac{\lambda_+}{4}\equiv b,\quad\omega_{32}=\frac{\lambda_-}{4}\equiv a,\quad\omega_{01}=\frac{\lambda_-}{4}\equiv a.
\end{equation}
The killing vector is then $J=2b(J_{12}+J_{30})-2a(J_{01}+J_{32})$, and the Casimir invariants are
\begin{equation}
I_1=\frac{1}{4}\lambda_+^2-\frac{1}{4}\lambda_-^2=4(b^2-a^2),\quad\quad I_2=\frac{1}{4}\lambda_+^2+\frac{1}{4}\lambda_-^2=4(b^2+a^2)
\end{equation}
This is type I$_a$ in \cite{Banados:1992gq}. By assuming that the relation \eqref{I1I2MJ} also holds in this case, we see that these solutions obey $-J>|M|$  and thus correspond to unphysical solutions that should be discarded.\\
\linebreak
{\bf Case 4:}\\
Here we instead have $\omega_i^+=(0,0,\lambda_+)$ and $\omega_i^-=(\lambda_-,0,0)$ and
\begin{equation}
\omega_{30}=\frac{\lambda_-}{4}\equiv b,\quad\omega_{12}=-\frac{\lambda_-}{4}\equiv -b,\quad\omega_{32}=-\frac{\lambda_+}{4}\equiv -a,\quad\omega_{01}=\frac{\lambda_+}{4}\equiv a.
\end{equation}
The Killing vector is then $J=2b(-J_{12}+J_{30})-2a(J_{01}-J_{32})$, and the Casimir invariants are
\begin{equation}
I_1=-\frac{1}{4}\lambda_+^2+\frac{1}{4}\lambda_-^2=4(b^2-a^2),\quad\quad I_2=-\frac{1}{4}\lambda_+^2-\frac{1}{4}\lambda_-^2=-4(a^2+b^2)
\end{equation}
This is also of type I$_a$, but was not listed explicitly in \cite{Banados:1992gq} since they actually classify subgroups of $O(2,2)$ for which both type I$_a$ are equivalent. Again by assuming that the relation \eqref{I1I2MJ} holds, this solution satisfies $J>|M|$ and should be discarded.\\
\linebreak
{\bf Two vanishing eigenvalues:}\\
Now we look at the case where either the eigenvalues of $e_+$ or the eigenvalues of $e_-$ vanish (but not both). In that case, the matrix will not be diagonalizible and instead be nilpotent of order two. The corresponding vector $\omega_i^\pm$ will be a null vector. In two dimensions there are two inequivalent nilpotent matrices, which in an approprate basis can be chosen as
\begin{equation}
\frac{\mu}{2}\left(\begin{array}{cc}
 1&1\\
 -1&-1\\
\end{array}\right),
\end{equation}
where $\mu=\pm 1$. This is not the standard off-diagonal form that is usually used for nilpotent matrices, but we will choose this basis such as to reproduce results in \cite{Banados:1992gq}. We will start by considering when $e_+$ is nilpotent and when $e_-$ has real eigenvalues, which is equivalent to $\omega_i^+=(\mu,\mu,0)$ and $\omega_i^-=(0,0,\lambda)$. A straightforward calculation gives us
\begin{equation}
\omega_{30}=-\frac{\mu}{4},\quad\omega_{12}=-\frac{\mu}{4},\quad\omega_{20}=\frac{\mu}{4},\quad\omega_{31}=-\frac{\mu}{4},\quad\omega_{32}=\frac{\lambda}{4},\quad\omega_{01}=\frac{\lambda}{4},
\end{equation}
which essentially reproduces type II$_a$ in \cite{Banados:1992gq}. The Casimir invariants are
\begin{equation}
I_1=-\frac{\lambda^2}{4},\quad\quad I_2=\frac{\lambda^2}{4}.
\end{equation}
This corresponds to an extremal black hole with $M=-J$. The other possibility where $\omega_i^-=(\mu,\mu,0)$ and $\omega_i^+=(0,0,\lambda)$ yields
\begin{equation}
\omega_{30}=\frac{\mu}{4},\quad\omega_{12}=-\frac{\mu}{4},\quad\omega_{20}=\frac{\mu}{4},\quad\omega_{31}=\frac{\mu}{4},\quad\omega_{32}=-\frac{\lambda}{4},\quad\omega_{01}=\frac{\lambda}{4},
\end{equation}
with Casimir invariants
\begin{equation}
I_1=-\frac{\lambda^2}{4},\quad\quad I_2=-\frac{\lambda^2}{4}.
\end{equation}
This corresponds to an extremal black hole with $M=J$ which also belongs to type II$_a$.\\
\linebreak
Now we consider the case where we have imaginary eigenvalues instead of real ones, starting with $\omega_i^-=(\mu,\mu,0)$ and $\omega_i^+=(\lambda,0,0)$. We obtain
\begin{equation}
\omega_{20}=\frac{\mu}{4},\quad\omega_{31}=\frac{\mu}{4},\quad\omega_{30}=\frac{\mu-\lambda}{4},\quad\omega_{12}=-\frac{\lambda+\mu}{4},
\end{equation}
with Casimir invariants 
\begin{equation}
I_1=\frac{\lambda^2}{4},\quad\quad I_2=\frac{\lambda^2}{4},
\end{equation}
corresponding to a spinning pointlike particle with $M=J$. This is type II$_b$ in \cite{Banados:1992gq}. Choosing instead $\omega_i^+=(\mu,\mu,0)$ and $\omega_i^-=(\lambda,0,0)$ results in
\begin{equation}
\omega_{20}=\frac{\mu}{4},\quad\omega_{31}=-\frac{\mu}{4},\quad\omega_{30}=\frac{\lambda-\mu}{4},\quad\omega_{12}=-\frac{\lambda+\mu}{4}.
\end{equation}
The Casimir invariants are now
\begin{equation}
I_1=\frac{\lambda^2}{4},\quad\quad I_2=-\frac{\lambda^2}{4},
\end{equation}
corresponding to a spinning pointlike particle with $M=-J$. This is also type II$_b$ in \cite{Banados:1992gq}.\\
\linebreak
{\bf Four vanishing eigenvalues:}\\
Having all eigenvalues vanish corresponds to both $\omega_i^+$ and $\omega_i^-$ being null vectors. We thus choose $\omega_i^-=(\mu_-,\mu_-,0)$ and $\omega_i^+=(\mu_+,\mu_+,0)$ where only the signs of $\mu_\pm$ are relevant. This yields
\begin{equation}
\omega_{31}=\frac{\mu_--\mu_+}{4},\quad\omega_{30}=\frac{\mu_--\mu_+}{4},\quad\omega_{20}=\frac{\mu_-+\mu_+}{4},\quad\omega_{21}=\frac{\mu_-+\mu_+}{4}.
\end{equation}
This gives four inequivalent classes corresponding to all possible signs of $\mu_\pm$ and includes type III in \cite{Banados:1992gq}. The Casimir invariants both vanish and these solutions include extremal black holes with zero mass and massless pointlike particles.\\
\linebreak
Note that obtaining the metric for the extremal cases would require parametrizing \ads with a different coordinate system than what was used for the non-extremal cases, and we will not explore that further here.

\section{Classification of solutions for \sln{3}$\times$\sln{3} higher spin gravity}\label{sl3class}
We will now classify what kind of different solutions can be obtained by the technique of identifying points discussed in this section. This is done by classifying one-parameter subgroups of $\sln{3}\times \sln{3}$ which works similarly to that of $\sln{2}\times \sln{2}$ discussed in Section \ref{sl2class}. A matrix $e\in \mathfrak{sl}(3,\R)$ has three eigenvalues $\lambda_i$, that satisfy $\sum\lambda_i=0$ and if $\lambda$ is an eigenvalue, so is $\lambda^*$. One possibility is that the eigenvalues are all real. If we instead assume that one of them, say $\lambda_1$, is not real, then we must have $\lambda_2=\lambda_1^*$ and then $\lambda_3=-2\mathrm{Re}(\lambda_1)$. The basis $T_i$, that generates the (principally embedded) \sln{2} subgroup of \sln{3}, takes the form
\begin{equation}
T_0=\frac{1}{2}\left(\begin{array}{ccc}
         0&-\sqrt{2}&0\\
         \sqrt{2}&0&-\sqrt{2}\\
         0&\sqrt{2}&0\\
        \end{array}\right),\quad T_{1}=\left(\begin{array}{ccc}
         -1&0&0\\
         0&0&0\\
         0&0&1\\
        \end{array}\right),\quad T_{2}=\frac{1}{2}\left(\begin{array}{ccc}
         0&\sqrt{2}&\\
         \sqrt{2}&0&\sqrt 2\\
         0&\sqrt{2}&0\\
        \end{array}\right)\\
\end{equation}
We will again use the matrices $\hat W$ and $\bar W$, given by
\begin{align}
\hat{W}&=2W_0+W_2+W_{-2}=\left(\begin{array}{ccc}
         \frac{4}{3}&0&4\\
         0&-\frac{8}{3}&0\\
         4&0&\frac{4}{3}\\
        \end{array}\right),\\
\bar{W}&=2W_0-W_2-W_{-2}=\left(\begin{array}{ccc}
         \frac{4}{3}&0&-4\\
         0&-\frac{8}{3}&0\\
         -4&0&\frac{4}{3}\\
        \end{array}\right).\\
\end{align}
A traceless matrix with three real eigenvalues can then be brought to the form $aT_2+b\frac{3}{8}\bar W$ where $\lambda_1=a-b,\lambda_2=-a-b,\lambda_3=2b$. Similarly, a matrix with two complex eigenvalues and one real eigenvalue can be brought to $aT_0+b\frac{3}{8}\hat W$ where $\lambda_1=-b-ai,\lambda_2=-b+ai,\lambda_3=2b$. Moreover, since these expressions can cover any set of three eigenvalues by tuning the parameters $a$ and $b$, they will parametrize all possible diagonalizable matrices in $\mathfrak{sl}(3,\R)$ up to conjugation. For two generators $e_\pm$ we thus have the following cases:
\begin{enumerate}
 \item The eigenvalues $\lambda^\pm_i$, are all real.
 \item The eigenvalues $\lambda^\pm_1$ and $\lambda^\pm_2$ are complex with $\lambda^\pm_1=\lambda^{\pm*}_2$, $\lambda^\pm_3=-2\mathrm{Re}(\lambda^\pm_1)$ is real.
 \item The eigenvalues $\lambda^+_1$ and $\lambda^+_2$ are complex with $\lambda^+_1=\lambda^{+*}_2$, $\lambda^+_3=-2\mathrm{Re}(\lambda^+_1)$ and $\lambda^-_i$ are all real.
\item The eigenvalues $\lambda^-_1$ and $\lambda^-_2$ are complex with $\lambda^-_1=\lambda^{-*}_2$, $\lambda^-_3=-2\mathrm{Re}(\lambda^-_1)$ and $\lambda^+_i$ are all real.
\end{enumerate}
Again we must take extra care when two eigenvalues take the same value, in which case the matrix may not be diagonalizable. Recall that a higher spin black hole is a solution that satisfies the inequality
\begin{equation}
\mathcal{W_\pm}^2\leq \frac{4}{27}\mathcal{L_\pm}^3.\label{WLineq}
\end{equation}
$\mathcal{W}_\pm$ and $\mathcal{L}_\pm$ are defined from equation \eqref{HWgaugegravhs}, after applying a gauge transformation to bring it to that gauge. They can be computed in terms of the generators $e_\pm$ by $\mathrm{Tr}(e_\pm^2)=8\mathcal{L}_\pm$ and $\mathrm{Tr}(e_\pm^3)=-24\mathcal{W}_\pm$.\\
\linebreak
{\bf Case 1:}\\
For this case we write $e_\pm=a_\pm T_2+b_\pm\frac{3}{8}\bar W$ and thus this corresponds to the black hole solutions constructed in \ref{sec_HSBH}. We also know from \ref{sec_HSBH} that $\mathcal{L}_\pm$ and $\mathcal{W}_\pm$ must satisfy the inequality \eqref{WLineq}. However, the inequality \eqref{WLineq} also follows naturally from the property that the eigenvalues are real and that the matrices are traceless.
For a traceless matrix with eigenvalues $\lambda_1,\lambda_2,\lambda_3$ it can be shown that
\begin{equation}
(\lambda_1^2+\lambda_2^2+\lambda_3^2)^3-6(\lambda_1^3+\lambda_2^3+\lambda_3^3)^2\geq 0,\label{lambdasineq}
\end{equation}
Since $\mathrm{Tr}(e_\pm^2)=\lambda_{1\pm}^2+\lambda_{2\pm}^2+\lambda_{3\pm}^2=8\mathcal{L}_\pm$ and $\mathrm{Tr}(e_\pm^3)=\lambda_{1\pm}^3+\lambda_{2\pm}^3+\lambda_{3\pm}^3=-24\mathcal{W}_\pm$ we can reformulate the inequality \eqref{lambdasineq} in terms of $\mathcal{W}_\pm$ and $\mathcal{L}_\pm$ and the inequality \eqref{WLineq} follows. Since we also have $\mathcal{L}_\pm\geq0$ these solutions correspond to the higher spin black holes. The inequality \eqref{lambdasineq} can be shown by substituting $\lambda_3=-\lambda_1-\lambda_2$ after which the expression \eqref{lambdasineq} can be simplified to
\begin{equation}
2(\lambda_1-\lambda_2)^2(\lambda_1+2\lambda_2)^2(2\lambda_1+\lambda_2)^2\geq 0.
\end{equation}
Equality thus holds only when either $\lambda_1=\lambda_2$, $\lambda_3=\lambda_1$ or when $\lambda_3=\lambda_2$. Note that such solutions do not correspond to extremal black holes since these solutions can not be brought to the highest weight gauge \eqref{HWgaugegravhs} since the gauge transformations break down. Note that setting for instance $\lambda_{3}=0$ will give us the standard BTZ black hole in three-dimensional gravity but embedded in the higher spin theory. \\
\linebreak
{\bf Case 2:}\\
In this case we can write $e_\pm=aT_0+b\frac{3}{8}\hat W$, and would correspond to the pointlike particle solutions constructed in \ref{sec_HSPP}. The eigenvalues will be $\lambda_{3\pm}^*=\lambda_{2\pm}=a_\pm+b_\pm i$ and $\lambda_{1\pm}=-2a_\pm$. Now it can be shown that
\begin{equation}
(\lambda_{1\pm}^2+\lambda_{2\pm}^2+\lambda_{3\pm}^2)^3-6(\lambda_{1\pm}^3+\lambda_{2\pm}^3+\lambda_{3\pm}^3)^2\leq 0,\label{lambdasineq2}
\end{equation}
as can be shown by substituting the eigenvalues $\lambda_{i\pm}$ for $a_\pm$ and $b_\pm$, for which the expression takes the form
\begin{equation}
-8b_\pm^2(9a_\pm^2+b_\pm^2)^2\leq0.
\end{equation}
Again by using $\mathrm{Tr}(e_\pm^2)=\lambda_{1\pm}^2+\lambda_{2\pm}^2+\lambda_{3\pm}^2=8\mathcal{L}_\pm$ and $\mathrm{Tr}(e_\pm^3)=\lambda_{1\pm}^3+\lambda_{2\pm}^3+\lambda_{3\pm}^3=-24\mathcal{W}_\pm$ we see that all these solutions violate \eqref{WLineq}. Since $8\mathcal{L}_\pm=-6a_\pm^2+2b_\pm^2$, there is no restriction on the sign of this quantity. Note that setting $\lambda_{1\pm}=0$ will give us the conical singularity solutions in three-dimensional gravity but embedded in the higher spin theory.\\
\linebreak
{\bf Case 3 and 4:}\\
These cases are the analogs of the unphysical BTZ black holes where $|J|>|M|$. They could for instance have $\mathcal{W}_+$ and $\mathcal{L}_+$  satisfying the bounds \eqref{WLineq} and $\mathcal{L}_\pm\leq0$, while $\mathcal{W}_-$ and $\mathcal{L}_-$ could violate them, or vice versa.\\
\linebreak
{\bf Degenerate eigenvalues:}\\
It is clear that if two eigenvalues are equal, all eigenvalues must be real. Let us assume that the eigenvalues of $e_+$ degenerate, such that $\lambda_2^+=\lambda_3^+$ and $\lambda_1^+=-2\lambda_2^+$. In this case, the matrix can not necessarily be diagonalized, but must also include a nilpotent matrix of either order 1 or 2. Thus we will have $e_+=d+N$ where $N$ is a nilpotent matrix (of order 1 or 2) and $d$ is a diagonal matrix with two eigenvalues equal, while $e_-$ is diagonalizable. This will correspond to $\mathcal{L}_+$ and $\mathcal{W}_+$ saturating the bound \eqref{WLineq} and is identified with an ``extremal'' solution analogous to the extremal BTZ black holes (or conical singularities) in three-dimensional gravity where $|M|=|J|$. An analogous extremal solution of course exists where we swap $+$ and $-$ and where instead $\mathcal{L}_-$ and $\mathcal{W}_-$ saturate the bound \eqref{WLineq}. If both inequalities are saturated, we obtain an extremal solution which is analogous to the extremal BTZ black hole with zero mass.\\
\linebreak
Extremal higher spin black holes were constructed in \cite{Banados:2015tft} by looking directly at the gauge connections and it was argued that extremal higher spin black holes correspond to gauge connections that can not be diagonalized (and thus include a nilpotent part). In fact, it is easy to see that a classification in terms of the gauge connection would be equivalent to the classification of the generators of isometries that we have used here since they both in essence classify the holonomy matrix along the angular coordinate. For an isometry $\xi_\pm$ along a coordinate $\lambda$ we have
\begin{equation}
A^\pm_\lambda=g_\pm(0)^{-1}\xi_\pm g_\pm(0).
\end{equation}
It is clear from this equation that $A_\lambda^\pm$ and $\xi_\pm$ will be of the same Jordan normal form and thus classifications based on $A_\lambda^\pm$ or on $\xi_\pm$ will be equivalent.

\section{Comments on permutation of eigenvalues of $A_\phi$}\label{perm_app}
Here we will elaborate on the degeneracy in the higher spin black hole case, namely the fact that the identification of points covers all higher spin black hole solutions three times. Thus let us consider a matrix $U$ with eigenvalues $\{\lambda_1,\lambda_2,\lambda_3\}$, that we want to write on the form
\begin{equation}
U=\kappa\left(\frac{1}{2}(L_{\pm1}-L_{\mp1})+\sigma(2W_0-W_{-2}-W_{2})\right),\label{Umatrix}
\end{equation}
where we can assume that $\kappa>0$. This is the form on which $A_\phi^\pm$ is written after the identification of points has been carried out. By computing the eigenvalues of a matrix on this form we obtain that $\{\lambda_1,\lambda_2,\lambda_3\}=\{-\frac{8}{3}\sigma+\kappa,-\frac{8}{3}\sigma-\kappa,\frac{16}{3}\sigma\}$ as an equivalence between two sets. There are thus three ways to obtain $\sigma$, namely $\sigma_3=3\lambda_3/8|\lambda_1-\lambda_2|$, $\sigma_2=3\lambda_2/8|\lambda_3-\lambda_1|$ or $\sigma_1=3\lambda_1/8|\lambda_2-\lambda_3|$ and each choice will give a $\sigma$ in one of the three regions $\mathcal{A}=\{|\sigma|<1/8\}$, $\mathcal{B}=\{1/8<|\sigma|<3/8\}$ or $\mathcal{C}=\{3/8<|\sigma|\}$ shown in Figure \ref{bhRfig}. We will now show explicitly that these three choices correspond to the three regions. Let us assume for instance that $\sigma_3\in \mathcal{A}$, namely that $|\lambda_3|/|\lambda_1-\lambda_2|=|\lambda_1+\lambda_2|/|\lambda_1-\lambda_2|<1/3$, which is equivalent to the inequality
\begin{equation}
2\lambda_1^2+2\lambda_2^2+5\lambda_1\lambda_2<0.\label{firstlambdaineq}
\end{equation}
We then want to first prove that $\sigma_2,\sigma_1 \notin \mathcal{A}$, namely that $|\lambda_1|/|\lambda_3-\lambda_2|=|\lambda_1|/|2\lambda_2+\lambda_1|>1/3$ (the case where we swap $\lambda_1$ and $\lambda_2$ follows analogously). This inequality is equivalent to $9\lambda_1^2>4\lambda_2^2+4\lambda_1\lambda_2+\lambda_1^2$ which follows from
\begin{equation}
2\lambda_1^2-\lambda_2^2-\lambda_1\lambda_2>2\lambda_1^2-\lambda_2^2-\lambda_1\lambda_2+(2\lambda_1^2+2\lambda_2^2+5\lambda_1\lambda_2)=(2\lambda_1+\lambda_2)^2>0.
\end{equation}
Next we want to show that for instance if $\sigma_1\in\mathcal{B}$, then $\sigma_2\in\mathcal{C}$. This is equivalent to first assuming that $|\lambda_1|/|2\lambda_2+\lambda_1|<1$, which is equivalent to the inequality $\lambda_2^2+\lambda_1\lambda_2>0$, and then proving that $\lambda_1^2+\lambda_1\lambda_2<0$. This is easily proven by assuming that either $\lambda_1<0, \lambda_2>0$ or $\lambda_2<0, \lambda_1>0$ (since we know that $\lambda_1\lambda_2<0$ from \eqref{firstlambdaineq}).\\
\linebreak
We can also explicitly compute the gauge transformation that changes $\kappa$ and $\sigma$ but leaves the eigenvalues invariant. For instance, to transform $U$ given by \eqref{Umatrix} to a new matrix $U'=bUb^{-1}$ with parameters $\kappa'$ and $\sigma'$ but with the same eigenvalues, we can for instance pick
\begin{equation}
b=\pm\frac{17}{16}T_0+\frac{17}{16}T_1+\frac{3}{4}W_0\pm\frac{15}{32}(W_{1}-W_{-1})-\frac{7}{64}W_2+\frac{23}{64}W_{-2}.
\end{equation}
These two transformations will move the parameter $\sigma$ from one of the regions to another (depending on the above signs), and the new parameters are given by $\kappa'=\mp4\kappa\sigma-\kappa/2$ and $\kappa'\sigma'=\pm4\kappa/3-\sigma\kappa/2$ (there are two more possible $b$ that will give a different sign of $\kappa'$ but we will not list them here).\\

\section{Relations with previous conventions}
In \cite{Bunster:2014mua} and other previous work, a different convention for the $\mathcal{W}_\pm$ and $\mathcal{L}_\pm$ was used. Let us denote these quantities in \cite{Bunster:2014mua} by $\mathcal{\tilde W}_\pm$ and $\mathcal{\tilde L}_\pm$. Then the relation to the quantities in this paper are
\begin{equation}
\mathcal{\tilde L}_\pm=\frac{k}{2\pi}\mathcal{L}_\pm,\quad \mathcal{\tilde W}_\pm=\frac{2k}{\pi}\mathcal{W}_\pm,
\end{equation}
where $k$ is the Chern-Simons level.

\bibliography{mybib}

\begin{thebibliography}{100}

\bibitem{Maldacena:1997re}
Juan~Martin Maldacena.
\newblock {The Large N limit of superconformal field theories and
  supergravity}.
\newblock {\em Int. J. Theor. Phys.}, 38:1113--1133, 1999.
\newblock [Adv. Theor. Math. Phys.2,231(1998)].

\bibitem{Craps:2014eba}
Ben Craps, E.~J. Lindgren, Anastasios Taliotis, Joris Vanhoof, and Hong-bao
  Zhang.
\newblock {Holographic gravitational infall in the hard wall model}.
\newblock {\em Phys. Rev.}, D90(8):086004, 2014.

\bibitem{Craps:2015upq}
Ben Craps, Erik~Jonathan Lindgren, and Anastasios Taliotis.
\newblock {Holographic thermalization in a top-down confining model}.
\newblock {\em JHEP}, 12:116, 2015.

\bibitem{Lindgren:2015fum}
E.~J. Lindgren.
\newblock {Black hole formation from point-like particles in three-dimensional
  anti-de Sitter space}.
\newblock {\em Class. Quant. Grav.}, 33(14):145009, 2016.

\bibitem{Lindgren:2016wtw}
Jonathan Lindgren.
\newblock {Collisions of massive particles, timelike thin shells and formation
  of black holes in three dimensions}.
\newblock {\em JHEP}, 12:048, 2016.

\bibitem{Hawking:1974sw}
S.~W. Hawking.
\newblock {Particle Creation by Black Holes}.
\newblock {\em Commun. Math. Phys.}, 43:199--220, 1975.
\newblock [,167(1975)].

\bibitem{Gregory:1993vy}
R.~Gregory and R.~Laflamme.
\newblock {Black strings and p-branes are unstable}.
\newblock {\em Phys. Rev. Lett.}, 70:2837--2840, 1993.

\bibitem{Banks:1998dd}
Tom Banks, Michael~R. Douglas, Gary~T. Horowitz, and Emil~J. Martinec.
\newblock {AdS dynamics from conformal field theory}.
\newblock 1998.

\bibitem{Horowitz:1999uv}
Gary~T. Horowitz.
\newblock {Comments on black holes in string theory}.
\newblock {\em Class. Quant. Grav.}, 17:1107--1116, 2000.

\bibitem{Gubser:2000ec}
Steven~S. Gubser and Indrajit Mitra.
\newblock {Instability of charged black holes in Anti-de Sitter space}.
\newblock 2000.

\bibitem{Gubser:2000mm}
Steven~S. Gubser and Indrajit Mitra.
\newblock {The Evolution of unstable black holes in anti-de Sitter space}.
\newblock {\em JHEP}, 08:018, 2001.

\bibitem{Friess:2005zp}
Joshua~J. Friess, Steven~S. Gubser, and Indrajit Mitra.
\newblock {Counter-examples to the correlated stability conjecture}.
\newblock {\em Phys. Rev.}, D72:104019, 2005.

\bibitem{Buchel:2010wk}
Alex Buchel and Chris Pagnutti.
\newblock {Correlated stability conjecture revisited}.
\newblock {\em Phys. Lett.}, B697:168--172, 2011.

\bibitem{Buchel:2011ra}
Alex Buchel and Alexander Patrushev.
\newblock {Can the correlated stability conjecture be saved?}
\newblock {\em JHEP}, 06:090, 2011.

\bibitem{Brown:1986nw}
J.~David Brown and M.~Henneaux.
\newblock {Central Charges in the Canonical Realization of Asymptotic
  Symmetries: An Example from Three-Dimensional Gravity}.
\newblock {\em Commun. Math. Phys.}, 104:207--226, 1986.

\bibitem{Aharony:1999ti}
Ofer Aharony, Steven~S. Gubser, Juan~Martin Maldacena, Hirosi Ooguri, and Yaron
  Oz.
\newblock {Large N field theories, string theory and gravity}.
\newblock {\em Phys. Rept.}, 323:183--386, 2000.

\bibitem{Green:1987sp}
Michael~B. Green, J.~H. Schwarz, and Edward Witten.
\newblock {\em {SUPERSTRING THEORY. VOL. 1: INTRODUCTION}}.
\newblock Cambridge Monographs on Mathematical Physics. 1988.

\bibitem{Green:1987mn}
Michael~B. Green, J.~H. Schwarz, and Edward Witten.
\newblock {\em {SUPERSTRING THEORY. VOL. 2: LOOP AMPLITUDES, ANOMALIES AND
  PHENOMENOLOGY}}.
\newblock 1988.

\bibitem{Polchinski:1998rq}
J.~Polchinski.
\newblock {\em {String theory. Vol. 1: An introduction to the bosonic string}}.
\newblock Cambridge University Press, 2007.

\bibitem{Polchinski:1998rr}
J.~Polchinski.
\newblock {\em {String theory. Vol. 2: Superstring theory and beyond}}.
\newblock Cambridge University Press, 2007.

\bibitem{Becker:2007zj}
K.~Becker, M.~Becker, and J.~H. Schwarz.
\newblock {\em {String theory and M-theory: A modern introduction}}.
\newblock Cambridge University Press, 2006.

\bibitem{Horowitz:1991cd}
Gary~T. Horowitz and Andrew Strominger.
\newblock {Black strings and P-branes}.
\newblock {\em Nucl. Phys.}, B360:197--209, 1991.

\bibitem{Nastase:2007kj}
Horatiu Nastase.
\newblock {Introduction to AdS-CFT}.
\newblock 2007.

\bibitem{Skenderis:2002wp}
Kostas Skenderis.
\newblock {Lecture notes on holographic renormalization}.
\newblock {\em Class. Quant. Grav.}, 19:5849--5876, 2002.

\bibitem{Witten:1998qj}
Edward Witten.
\newblock {Anti-de Sitter space and holography}.
\newblock {\em Adv. Theor. Math. Phys.}, 2:253--291, 1998.

\bibitem{Gubser:1998bc}
S.~S. Gubser, Igor~R. Klebanov, and Alexander~M. Polyakov.
\newblock {Gauge theory correlators from noncritical string theory}.
\newblock {\em Phys. Lett.}, B428:105--114, 1998.

\bibitem{Skenderis:2008dh}
Kostas Skenderis and Balt~C. van Rees.
\newblock {Real-time gauge/gravity duality}.
\newblock {\em Phys. Rev. Lett.}, 101:081601, 2008.

\bibitem{Skenderis:2008dg}
Kostas Skenderis and Balt~C. van Rees.
\newblock {Real-time gauge/gravity duality: Prescription, Renormalization and
  Examples}.
\newblock {\em JHEP}, 05:085, 2009.

\bibitem{Son:2002sd}
Dam~T. Son and Andrei~O. Starinets.
\newblock {Minkowski space correlators in AdS / CFT correspondence: Recipe and
  applications}.
\newblock {\em JHEP}, 09:042, 2002.

\bibitem{Gibbons:1976ue}
G.~W. Gibbons and S.~W. Hawking.
\newblock {Action Integrals and Partition Functions in Quantum Gravity}.
\newblock {\em Phys. Rev.}, D15:2752--2756, 1977.

\bibitem{Emparan:1999pm}
Roberto Emparan, Clifford~V. Johnson, and Robert~C. Myers.
\newblock {Surface terms as counterterms in the AdS / CFT correspondence}.
\newblock {\em Phys. Rev.}, D60:104001, 1999.

\bibitem{Skenderis:2000in}
Kostas Skenderis.
\newblock {Asymptotically Anti-de Sitter space-times and their stress energy
  tensor}.
\newblock {\em Int. J. Mod. Phys.}, A16:740--749, 2001.
\newblock [,394(2000)].

\bibitem{Maldacena:1998im}
Juan~Martin Maldacena.
\newblock {Wilson loops in large N field theories}.
\newblock {\em Phys. Rev. Lett.}, 80:4859--4862, 1998.

\bibitem{Berenstein:1998ij}
David~Eliecer Berenstein, Richard Corrado, Willy Fischler, and Juan~Martin
  Maldacena.
\newblock {The Operator product expansion for Wilson loops and surfaces in the
  large N limit}.
\newblock {\em Phys. Rev.}, D59:105023, 1999.

\bibitem{Witten:1998zw}
Edward Witten.
\newblock {Anti-de Sitter space, thermal phase transition, and confinement in
  gauge theories}.
\newblock {\em Adv. Theor. Math. Phys.}, 2:505--532, 1998.

\bibitem{Kanitscheider:2008kd}
Ingmar Kanitscheider, Kostas Skenderis, and Marika Taylor.
\newblock {Precision holography for non-conformal branes}.
\newblock {\em JHEP}, 09:094, 2008.

\bibitem{Lehner:2001wq}
Luis Lehner.
\newblock {Numerical relativity: A Review}.
\newblock {\em Class. Quant. Grav.}, 18:R25--R86, 2001.

\bibitem{Chesler:2013lia}
Paul~M. Chesler and Laurence~G. Yaffe.
\newblock {Numerical solution of gravitational dynamics in asymptotically
  anti-de Sitter spacetimes}.
\newblock {\em JHEP}, 07:086, 2014.

\bibitem{Horowitz:1999jd}
Gary~T. Horowitz and Veronika~E. Hubeny.
\newblock {Quasinormal modes of AdS black holes and the approach to thermal
  equilibrium}.
\newblock {\em Phys. Rev.}, D62:024027, 2000.

\bibitem{Arnowitt:1959ah}
Richard~L. Arnowitt, Stanley Deser, and Charles~W. Misner.
\newblock {Dynamical Structure and Definition of Energy in General Relativity}.
\newblock {\em Phys. Rev.}, 116:1322--1330, 1959.

\bibitem{Misner:1974qy}
Charles~W. Misner, K.~S. Thorne, and J.~A. Wheeler.
\newblock {\em {Gravitation}}.
\newblock W. H. Freeman, San Francisco, 1973.

\bibitem{Wu:2012rib}
Bin Wu.
\newblock {On holographic thermalization and gravitational collapse of massless
  scalar fields}.
\newblock {\em JHEP}, 10:133, 2012.

\bibitem{Bhattacharyya:2009uu}
Sayantani Bhattacharyya and Shiraz Minwalla.
\newblock {Weak Field Black Hole Formation in Asymptotically AdS Spacetimes}.
\newblock {\em JHEP}, 09:034, 2009.

\bibitem{Balasubramanian:2013oga}
V.~Balasubramanian, A.~Bernamonti, J.~de~Boer, B.~Craps, L.~Franti, F.~Galli,
  E.~Keski-Vakkuri, B.~Müller, and A.~Schäfer.
\newblock {Inhomogeneous holographic thermalization}.
\newblock {\em JHEP}, 10:082, 2013.

\bibitem{Chen:1995ena}
Lin-Yuan Chen, Nigel Goldenfeld, and Y.~Oono.
\newblock {The Renormalization group and singular perturbations: Multiple
  scales, boundary layers and reductive perturbation theory}.
\newblock {\em Phys. Rev.}, E54:376--394, 1996.

\bibitem{Craps:2014vaa}
Ben Craps, Oleg Evnin, and Joris Vanhoof.
\newblock {Renormalization group, secular term resummation and AdS
  (in)stability}.
\newblock {\em JHEP}, 10:048, 2014.

\bibitem{Craps:2014jwa}
Ben Craps, Oleg Evnin, and Joris Vanhoof.
\newblock {Renormalization, averaging, conservation laws and AdS
  (in)stability}.
\newblock {\em JHEP}, 01:108, 2015.

\bibitem{Craps:2015jma}
Ben Craps and Oleg Evnin.
\newblock {AdS (in)stability: an analytic approach}.
\newblock {\em Fortsch. Phys.}, 64:336--344, 2016.

\bibitem{Bhaseen:2013ypa}
M.~J. Bhaseen, Benjamin Doyon, Andrew Lucas, and Koenraad Schalm.
\newblock {Far from equilibrium energy flow in quantum critical systems}.
\newblock 2013.
\newblock [Nature Phys.11,5(2015)].

\bibitem{Lucas:2015hnv}
Andrew Lucas, Koenraad Schalm, Benjamin Doyon, and M.~J. Bhaseen.
\newblock {Shock waves, rarefaction waves, and nonequilibrium steady states in
  quantum critical systems}.
\newblock {\em Phys. Rev.}, D94(2):025004, 2016.

\bibitem{Balasubramanian:2011ur}
V.~Balasubramanian, A.~Bernamonti, J.~de~Boer, N.~Copland, B.~Craps,
  E.~Keski-Vakkuri, B.~Muller, A.~Schafer, M.~Shigemori, and W.~Staessens.
\newblock {Holographic Thermalization}.
\newblock {\em Phys. Rev.}, D84:026010, 2011.

\bibitem{Ziogas:2015aja}
Vaios Ziogas.
\newblock {Holographic mutual information in global Vaidya-BTZ spacetime}.
\newblock {\em JHEP}, 09:114, 2015.

\bibitem{Bizon:2011gg}
Piotr Bizon and Andrzej Rostworowski.
\newblock {On weakly turbulent instability of anti-de Sitter space}.
\newblock {\em Phys. Rev. Lett.}, 107:031102, 2011.

\bibitem{Craps:2015iia}
Ben Craps, Oleg Evnin, and Joris Vanhoof.
\newblock {Ultraviolet asymptotics and singular dynamics of AdS perturbations}.
\newblock {\em JHEP}, 10:079, 2015.

\bibitem{Craps:2015xya}
Ben Craps, Oleg Evnin, Puttarak Jai-akson, and Joris Vanhoof.
\newblock {Ultraviolet asymptotics for quasiperiodic AdS$_{4}$ perturbations}.
\newblock {\em JHEP}, 10:080, 2015.

\bibitem{Evnin:2015wyi}
Oleg Evnin and Rongvoram Nivesvivat.
\newblock {AdS perturbations, isometries, selection rules and the Higgs
  oscillator}.
\newblock {\em JHEP}, 01:151, 2016.

\bibitem{Matschull:1998rv}
Hans-Jurgen Matschull.
\newblock {Black hole creation in (2+1)-dimensions}.
\newblock {\em Class. Quant. Grav.}, 16:1069--1095, 1999.

\bibitem{Kovtun:2004de}
P.~Kovtun, Dan~T. Son, and Andrei~O. Starinets.
\newblock {Viscosity in strongly interacting quantum field theories from black
  hole physics}.
\newblock {\em Phys. Rev. Lett.}, 94:111601, 2005.

\bibitem{Cremonini:2011iq}
Sera Cremonini.
\newblock {The Shear Viscosity to Entropy Ratio: A Status Report}.
\newblock {\em Mod. Phys. Lett.}, B25:1867--1888, 2011.

\bibitem{PhysRevC.78.034915}
Matthew Luzum and Paul Romatschke.
\newblock Conformal relativistic viscous hydrodynamics: Applications to rhic
  results at $\sqrt{{s}_{\mathit{NN}}}=200$ gev.
\newblock {\em Phys. Rev. C}, 78:034915, Sep 2008.

\bibitem{Bhattacharyya:2008jc}
Sayantani Bhattacharyya, Veronika~E Hubeny, Shiraz Minwalla, and Mukund
  Rangamani.
\newblock {Nonlinear Fluid Dynamics from Gravity}.
\newblock {\em JHEP}, 02:045, 2008.

\bibitem{Chesler:2015wra}
Paul~M. Chesler and Laurence~G. Yaffe.
\newblock {Holography and off-center collisions of localized shock waves}.
\newblock {\em JHEP}, 10:070, 2015.

\bibitem{Chesler:2008hg}
Paul~M. Chesler and Laurence~G. Yaffe.
\newblock {Horizon formation and far-from-equilibrium isotropization in
  supersymmetric Yang-Mills plasma}.
\newblock {\em Phys. Rev. Lett.}, 102:211601, 2009.

\bibitem{vanderSchee:2013pia}
Wilke van~der Schee, Paul Romatschke, and Scott Pratt.
\newblock {Fully Dynamical Simulation of Central Nuclear Collisions}.
\newblock {\em Phys. Rev. Lett.}, 111(22):222302, 2013.

\bibitem{Sakai:2004cn}
Tadakatsu Sakai and Shigeki Sugimoto.
\newblock {Low energy hadron physics in holographic QCD}.
\newblock {\em Prog. Theor. Phys.}, 113:843--882, 2005.

\bibitem{Sakai:2005yt}
Tadakatsu Sakai and Shigeki Sugimoto.
\newblock {More on a holographic dual of QCD}.
\newblock {\em Prog. Theor. Phys.}, 114:1083--1118, 2005.

\bibitem{Craps:2013iaa}
Ben Craps, Elias Kiritsis, Christopher Rosen, Anastasios Taliotis, Joris
  Vanhoof, and Hong-bao Zhang.
\newblock {Gravitational collapse and thermalization in the hard wall model}.
\newblock {\em JHEP}, 02:120, 2014.

\bibitem{Ishii:2015gia}
Takaaki Ishii, Elias Kiritsis, and Christopher Rosen.
\newblock {Thermalization in a Holographic Confining Gauge Theory}.
\newblock {\em JHEP}, 08:008, 2015.

\bibitem{citeulike:4226570}
Abraham Savitzky and M.~J.~E. Golay.
\newblock {Smoothing and Differentiation of Data by Simplified Least Squares
  Procedures.}
\newblock {\em Anal. Chem.}, 36(8):1627--1639, July 1964.

\bibitem{Bizon:2013xha}
Piotr Bizoń and Joanna Jałmużna.
\newblock {Globally regular instability of $AdS_3$}.
\newblock {\em Phys. Rev. Lett.}, 111(4):041102, 2013.

\bibitem{Gursoy:2007cb}
U.~Gursoy and E.~Kiritsis.
\newblock {Exploring improved holographic theories for QCD: Part I}.
\newblock {\em JHEP}, 02:032, 2008.

\bibitem{Gursoy:2007er}
U.~Gursoy, E.~Kiritsis, and F.~Nitti.
\newblock {Exploring improved holographic theories for QCD: Part II}.
\newblock {\em JHEP}, 02:019, 2008.

\bibitem{Hartnoll:2008kx}
Sean~A. Hartnoll, Christopher~P. Herzog, and Gary~T. Horowitz.
\newblock {Holographic Superconductors}.
\newblock {\em JHEP}, 12:015, 2008.

\bibitem{Horowitz:2010gk}
Gary~T. Horowitz.
\newblock {Introduction to Holographic Superconductors}.
\newblock {\em Lect. Notes Phys.}, 828:313--347, 2011.

\bibitem{Cai:2015cya}
Rong-Gen Cai, Li~Li, Li-Fang Li, and Run-Qiu Yang.
\newblock {Introduction to Holographic Superconductor Models}.
\newblock {\em Sci. China Phys. Mech. Astron.}, 58(6):060401, 2015.

\bibitem{Nishioka:2009zj}
Tatsuma Nishioka, Shinsei Ryu, and Tadashi Takayanagi.
\newblock {Holographic Superconductor/Insulator Transition at Zero
  Temperature}.
\newblock {\em JHEP}, 03:131, 2010.

\bibitem{Ryu:2006bv}
Shinsei Ryu and Tadashi Takayanagi.
\newblock {Holographic derivation of entanglement entropy from AdS/CFT}.
\newblock {\em Phys. Rev. Lett.}, 96:181602, 2006.

\bibitem{Hubeny:2007xt}
Veronika~E. Hubeny, Mukund Rangamani, and Tadashi Takayanagi.
\newblock {A Covariant holographic entanglement entropy proposal}.
\newblock {\em JHEP}, 07:062, 2007.

\bibitem{Emparan:2013moa}
Roberto Emparan, Ryotaku Suzuki, and Kentaro Tanabe.
\newblock {The large D limit of General Relativity}.
\newblock {\em JHEP}, 06:009, 2013.

\bibitem{Deser:1983tn}
Stanley Deser, R.~Jackiw, and Gerard 't~Hooft.
\newblock {Three-Dimensional Einstein Gravity: Dynamics of Flat Space}.
\newblock {\em Annals Phys.}, 152:220, 1984.

\bibitem{Banados:1992wn}
Maximo Banados, Claudio Teitelboim, and Jorge Zanelli.
\newblock {The Black hole in three-dimensional space-time}.
\newblock {\em Phys. Rev. Lett.}, 69:1849--1851, 1992.

\bibitem{Banados:1992gq}
Maximo Banados, Marc Henneaux, Claudio Teitelboim, and Jorge Zanelli.
\newblock {Geometry of the (2+1) black hole}.
\newblock {\em Phys. Rev.}, D48:1506--1525, 1993.
\newblock [Erratum: Phys. Rev.D88,069902(2013)].

\bibitem{Israel:1966rt}
W.~Israel.
\newblock {Singular hypersurfaces and thin shells in general relativity}.
\newblock {\em Nuovo Cim.}, B44S10:1, 1966.
\newblock [Nuovo Cim.B44,1(1966)].

\bibitem{Musgrave:1995ka}
Peter Musgrave and Kayll Lake.
\newblock {Junctions and thin shells in general relativity using computer
  algebra. 1: The Darmois-Israel formalism}.
\newblock {\em Class. Quant. Grav.}, 13:1885--1900, 1996.

\bibitem{Barrabes:1991ng}
C.~Barrabes and W.~Israel.
\newblock {Thin shells in general relativity and cosmology: The Lightlike
  limit}.
\newblock {\em Phys. Rev.}, D43:1129--1142, 1991.

\bibitem{Musgrave:1997sfw}
Peter Musgrave and Kayll Lake.
\newblock {Junctions and thin shells in general relativity using computer
  algebra: II. The null formalism}.
\newblock {\em Class. Quant. Grav.}, 14(5):1285--1294, 1997.

\bibitem{Banados:1998gg}
Maximo Banados.
\newblock {Three-dimensional quantum geometry and black holes}.
\newblock pages 147--169, 1998.
\newblock [AIP Conf. Proc.484,147(1999)].

\bibitem{Kraus:2006wn}
Per Kraus.
\newblock {Lectures on black holes and the AdS(3) / CFT(2) correspondence}.
\newblock {\em Lect. Notes Phys.}, 755:193--247, 2008.

\bibitem{Barnich:2016lyg}
Glenn Barnich and Cédric Troessaert.
\newblock {Finite BMS transformations}.
\newblock {\em JHEP}, 03:167, 2016.

\bibitem{Erdmenger:2012xu}
Johanna Erdmenger and Shu Lin.
\newblock {Thermalization from gauge/gravity duality: Evolution of
  singularities in unequal time correlators}.
\newblock {\em JHEP}, 10:028, 2012.

\bibitem{Keranen:2015fqa}
Ville Keranen, Hiromichi Nishimura, Stefan Stricker, Olli Taanila, and Aleksi
  Vuorinen.
\newblock {Gravitational collapse of thin shells: Time evolution of the
  holographic entanglement entropy}.
\newblock {\em JHEP}, 06:126, 2015.

\bibitem{Anous:2016kss}
Tarek Anous, Thomas Hartman, Antonin Rovai, and Julian Sonner.
\newblock {Black Hole Collapse in the 1/c Expansion}.
\newblock {\em JHEP}, 07:123, 2016.

\bibitem{Gross:1988ue}
David~J. Gross.
\newblock {High-Energy Symmetries of String Theory}.
\newblock {\em Phys. Rev. Lett.}, 60:1229, 1988.

\bibitem{Klebanov:2002ja}
I.~R. Klebanov and A.~M. Polyakov.
\newblock {AdS dual of the critical O(N) vector model}.
\newblock {\em Phys. Lett.}, B550:213--219, 2002.

\bibitem{Sezgin:2002rt}
E.~Sezgin and P.~Sundell.
\newblock {Massless higher spins and holography}.
\newblock {\em Nucl. Phys.}, B644:303--370, 2002.
\newblock [Erratum: Nucl. Phys.B660,403(2003)].

\bibitem{Giombi:2009wh}
Simone Giombi and Xi~Yin.
\newblock {Higher Spin Gauge Theory and Holography: The Three-Point Functions}.
\newblock {\em JHEP}, 09:115, 2010.

\bibitem{Giombi:2010vg}
Simone Giombi and Xi~Yin.
\newblock {Higher Spins in AdS and Twistorial Holography}.
\newblock {\em JHEP}, 04:086, 2011.

\bibitem{Bekaert:2012ux}
Xavier Bekaert, Euihun Joung, and Jihad Mourad.
\newblock {Comments on higher-spin holography}.
\newblock {\em Fortsch. Phys.}, 60:882--888, 2012.

\bibitem{Bekaert:2010hw}
Xavier Bekaert, Nicolas Boulanger, and Per Sundell.
\newblock {How higher-spin gravity surpasses the spin two barrier: no-go
  theorems versus yes-go examples}.
\newblock {\em Rev. Mod. Phys.}, 84:987--1009, 2012.

\bibitem{Vasiliev:1990en}
Mikhail~A. Vasiliev.
\newblock {Consistent equation for interacting gauge fields of all spins in
  (3+1)-dimensions}.
\newblock {\em Phys. Lett.}, B243:378--382, 1990.

\bibitem{Vasiliev:2003ev}
M.~A. Vasiliev.
\newblock {Nonlinear equations for symmetric massless higher spin fields in
  (A)dS(d)}.
\newblock {\em Phys. Lett.}, B567:139--151, 2003.

\bibitem{Gutperle:2011kf}
Michael Gutperle and Per Kraus.
\newblock {Higher Spin Black Holes}.
\newblock {\em JHEP}, 05:022, 2011.

\bibitem{Ammon:2012wc}
Martin Ammon, Michael Gutperle, Per Kraus, and Eric Perlmutter.
\newblock {Black holes in three dimensional higher spin gravity: A review}.
\newblock {\em J. Phys.}, A46:214001, 2013.

\bibitem{Perez:2014pya}
Alfredo Perez, David Tempo, and Ricardo Troncoso.
\newblock {Higher Spin Black Holes}.
\newblock {\em Lect. Notes Phys.}, 892:265--288, 2015.

\bibitem{Bunster:2014mua}
Claudio Bunster, Marc Henneaux, Alfredo Perez, David Tempo, and Ricardo
  Troncoso.
\newblock {Generalized Black Holes in Three-dimensional Spacetime}.
\newblock {\em JHEP}, 05:031, 2014.

\bibitem{Achucarro:1987vz}
A.~Achucarro and P.~K. Townsend.
\newblock {A Chern-Simons Action for Three-Dimensional anti-De Sitter
  Supergravity Theories}.
\newblock {\em Phys. Lett.}, B180:89, 1986.

\bibitem{Witten:1988hc}
Edward Witten.
\newblock {(2+1)-Dimensional Gravity as an Exactly Soluble System}.
\newblock {\em Nucl. Phys.}, B311:46, 1988.

\bibitem{van1981supergravity}
P.~Van~Nieuwenhuizen.
\newblock {\em Supergravity}.
\newblock Physics reports. North-Holland, 1981.

\bibitem{Coussaert:1995zp}
Oliver Coussaert, Marc Henneaux, and Peter van Driel.
\newblock {The Asymptotic dynamics of three-dimensional Einstein gravity with a
  negative cosmological constant}.
\newblock {\em Class. Quant. Grav.}, 12:2961--2966, 1995.

\bibitem{Regge:1974zd}
Tullio Regge and Claudio Teitelboim.
\newblock {Role of Surface Integrals in the Hamiltonian Formulation of General
  Relativity}.
\newblock {\em Annals Phys.}, 88:286, 1974.

\bibitem{Banados:1994tn}
Maximo Banados.
\newblock {Global charges in Chern-Simons field theory and the (2+1) black
  hole}.
\newblock {\em Phys. Rev.}, D52:5816--5825, 1996.

\bibitem{Benguria:1976in}
R.~Benguria, P.~Cordero, and C.~Teitelboim.
\newblock {Aspects of the Hamiltonian Dynamics of Interacting Gravitational
  Gauge and Higgs Fields with Applications to Spherical Symmetry}.
\newblock {\em Nucl. Phys.}, B122:61--99, 1977.

\bibitem{Campoleoni:2010zq}
Andrea Campoleoni, Stefan Fredenhagen, Stefan Pfenninger, and Stefan Theisen.
\newblock {Asymptotic symmetries of three-dimensional gravity coupled to
  higher-spin fields}.
\newblock {\em JHEP}, 11:007, 2010.

\bibitem{Henneaux:2010xg}
Marc Henneaux and Soo-Jong Rey.
\newblock {Nonlinear $W_{infinity}$ as Asymptotic Symmetry of Three-Dimensional
  Higher Spin Anti-de Sitter Gravity}.
\newblock {\em JHEP}, 12:007, 2010.

\bibitem{Banados:2015tft}
Máximo Bañados, Alejandra Castro, Alberto Faraggi, and Juan~I. Jottar.
\newblock {Extremal Higher Spin Black Holes}.
\newblock {\em JHEP}, 04:077, 2016.

\bibitem{Henneaux:2013dra}
Marc Henneaux, Alfredo Perez, David Tempo, and Ricardo Troncoso.
\newblock {Chemical potentials in three-dimensional higher spin anti-de Sitter
  gravity}.
\newblock {\em JHEP}, 12:048, 2013.

\bibitem{Castro:2016ehj}
Alejandra Castro, Nabil Iqbal, and Eva Llabrés.
\newblock {Eternal Higher Spin Black Holes: a Thermofield Interpretation}.
\newblock {\em JHEP}, 08:022, 2016.

\end{thebibliography}
\addcontentsline{toc}{chapter}{References}
\bibliographystyle{unsrt}




\end{document}